%% file: xgt2.tex
\makeindex
\documentclass[12 pt,oneside]{report}
\usepackage{epsfig}
\usepackage{amsmath}
\usepackage{rotating}
\usepackage{caption}
\usepackage{subfig}
\usepackage{graphicx}
\usepackage{booktabs}
\usepackage{epstopdf}
\usepackage{listings}
\usepackage{cite}
\usepackage{upquote}
\usepackage{xcolor}
\usepackage{pdfpages}
\usepackage[section]{placeins}

\usepackage{hyperref}
\usepackage{url} 
\usepackage{breakurl} 

\definecolor{dkgreen}{rgb}{0,0.6,0}
\definecolor{gray}{rgb}{0.5,0.5,0.5}
\definecolor{mauve}{rgb}{0.58,0,0.82}
\newcommand*{\TitleFont}{%
      \usefont{\encodingdefault}{\rmdefault}{b}{n}%
      \fontsize{16}{20}%
      \selectfont}

\include{pageshape}

\begin{document}

\input title.tex

\input abstract.tex
\input acknowledgement.tex

\tableofcontents
\listoffigures
\listoftables

\input intro/intro.tex

\input intro/src.tex

\input setup/setup.tex

\input analysis/analysis.tex

\input cross_section/cross_section.tex

\input cross_section/results.tex

\appendix
\input{append/append_trigger.tex}
\input{append/append_xemc.tex}
\input{append/append_deltap.tex}
\input{append/append_target.tex}
\input{append/append_xs.tex}
\renewcommand{\baselinestretch}{1}\normalsize
\bibliographystyle{h-physrev3.bst}
\bibliography{xgt2}

\end{document}

%% file: pageshape.tex
\usepackage[top=1in,right=1in,left=1.5in,bottom=1in]{geometry}
\usepackage{graphicx}
	\graphicspath{
		{./Analysis/dilFac/figures/}
		{./Analysis/pairSym/figures/}
		{./Analysis/polN/figures/}
		{./Experiment/targetApparatus/figures/}
	}
\usepackage{array}
\usepackage{url}
\usepackage{longtable}
\usepackage{setspace}
\usepackage{fancyhdr}

	\fancypagestyle{plain}{
	\fancyhf{} 
	\rhead{\thepage}
	}

\def\be{\begin{equation}}  
\def\ee{\end{equation}}

%% file: title.tex
\thispagestyle{empty}
\begin{center}

{\TitleFont Short Range Correlations in Nuclei at Large $x_{bj}$ through Inclusive Quasi-Elastic Electron Scattering}

\vspace{0.80in}

Zhihong Ye\\
Zhangzhou, Fujian, China

\vspace{0.5 in}
Bachelor of Science, Lanzhou University, 2005
\vspace{0.5 in}

{\small
A Dissertation presented to the Graduate Faculty

of the University of Virginia in Candidacy for the Degree of

Doctor of Philosophy

\vspace{0.35in}

Department of Physics

\vspace{0.35in}

University of Virginia

December, 2013
}

\vspace{0.5in}

\hfill \line(1,0){190}
\vspace{0.20in}

\hfill \line(1,0){190}
\vspace{0.20in}

\hfill \line(1,0){190}
\vspace{0.20in}

\hfill \line(1,0){190}
\vspace{0.20in}

\hfill \line(1,0){190}

\end{center}

%% file: abstract.tex
\begin{abstract}
   The experiment, E08-014, in Hall-A at Jefferson Lab aims to study the short-range correlations (SRC) which are necessary to explain the nuclear strength absent in the mean field theory. The cross sections for $\mathrm{^{2}H}$, $\mathrm{^{3}He}$, $\mathrm{^{4}He}$, $\mathrm{^{12}C}$, $\mathrm{^{40}Ca}$ and $\mathrm{^{48}Ca}$, were measured via inclusive quasielastic electron scattering from these nuclei in a $\mathrm{Q^{2}}$ range between 0.8 and $\mathrm{2.8~(GeV/c)^{2}}$ for $x_{bj}>1$. The cross section ratios of heavy nuclei to $\mathrm{^{2}H}$ were extracted to study two-nucleon SRC for $1<x_{bj}<2$, while the study of three-nucleon SRC was carried out from the cross section ratios of heavy nuclei to $\mathrm{^{3}He}$ for $x_{bj}\ge 2$. Meanwhile, the isospin dependence in SRCs has also been examined through the cross section ratio of $\mathrm{^{48}Ca}$ and $\mathrm{^{40}Ca}$.
      
\end{abstract}

%% file: acknowledgement.tex
\newpage

 \vspace*{1in}

 \emph{Dedicated to my mother, Meishen and my father, Kunlong}

\newpage
\chapter*{Acknowledgements}
  First of all, I would like to thank my advisor Prof. Donal Day for his great support and guidance. He brought me into his group to work on this exciting field. He was always available when I needed his help, standing out for me every time when I was in trouble, encouraging  me to seek more challenges and develop my own academic career. I was also very fortunate to join the E08-014 collaboration and closely work with three other spokespeople: John Arrington, Doug Higinbotham and Patricia Solvignon-Slifer. I have learned a lot from John, not only the knowledge in nuclear physics but also the techniques of analysis and problem-solving skills. Doug is always very supportive, and with his trust and encouragement, I have earned many opportunities to improve my abilities, which greatly boosted my confidence and made me become more versatile in physics. Patricia used her experience to help me not only on the detailed research work, but also on how to become a successful graduate student and how to build rapport with other colleagues. She pointed out my mistakes in the most comfortable way and cared about me and my families. I also received many help from other people in this collaboration, especially Nadia Fomin, who gave me a lot of help on the data analysis. I always feel lucky and grateful that I can own such a supportive team in my career.
  
  Thank to my colleagues and friends in the E08-014 collaboration and other four collaborations in the SRC family during run period of the spring 2011. I'd like to mention these hard working graduate students: David Anez, Navaphon Muangma and Lawrence Selvy; and also our post-docs: Aidan Keller, Charles Hanretty and Vince Sulkosky. Your big contributions were the main reason that our experiments had been so successful. I would like to thank Alexander Camsonne, Jian-Ping Chen, Javier Gomez, Ole Hanson, John LeRose, Bob Michel, Bogdan Wojtsekhowski and all other Hall-A staffs for supporting our experiments. I want to specially thank Vince and Alexandria who patiently taught me the knowledge of the Hall-A detectors and electronics, worked side-by-side with me to prepare the online data checkout and continued to offer guidance on my offline data analysis. Thank Ed. Folts, Jack Segal and all other Hall-A technicians for installing and maintaining the instruments. I would like to thank Dave Meekins and the target group on the support of our target system. I also want to mention the hard working of the staffs in the accelerator division. I am grateful for those colleagues who acted as run coordinators, filled our shifts, monitored the data quality and completed our run plans.
   
   I would like to thank my advisor at Hampton University, Dr. Liguang Tang, who brought me into the field of experimental nuclear physics and guided me to work on the Hypernuclear program in Hall-C. I won't forget our post-doc, Lulin Yuan who showed great patience in teaching me the knowledge of data analysis and earnestly answered every question. He set up a good example of how to work with beginners. I also want to thank Chunhua Chen for working so hard with me to prepare the HES experiment. I am grateful to colleagues in the Hypernuclear collaboration: Prof. Jorg Reinhard, Prof. Nakamura and Prof. Hashimoto, and all other graduate students and post-docs. I especially want to mention Prof. Hashimoto, who played the leading role in the Hypernuclear physics. I still remembered that when he heard I decided to leave the collaboration, he had a long conversation with me and encouraged me to make the best choice for the good of my academic career. He passed away in 2011 before I had a chance to visit his family but I will never forget him, a respectful scientist and a gentleman.

   I am grateful to Prof. Blaine Norum and Prof. Liyanaga who accepted me to continue studying at UVa. I also want to thank these lovely ladies in the office of our department: Dawn, Tammie, Helen, Beth and so on. 
   
   I also want to thank Prof. Bitao Hu, the thesis advisor of my Bachelor degree at Lanzhou University and also a very good friend. He shared his experience and knowledge with me, introduced to me the JLab physics, encouraged and helped me to apply for a Ph.D program in US. My life would have been in a completely different track if I did not meet him in the college.
   
   I would like to thank Chunhua Chen, Xin Qian, Yi Qiang, Huan Yao, Xiaohui Zhan and others, who have been my good friends for many years and take care of each other. Thank Lulin Yuan and Xiaoyan Lin, Xiaofeng Zhu and Change Ma, Bo Zhan and Yangping Yang, for taking care of me when I just came to US and helping me to fit in here. I also want to thank all other friends living in Newport News and Charlottesville. Your friendship is very important to me.
   
   My families are always the most important part of my life, although I haven't spent too much time with them these years. My parents did a great job in raising me, despite the fact that they have sacrificed a lot. For years they had to live a hard life to support my education. Instead of forcing me to choose other careers that I can make more money to support the family, they encourage me to pursue and persist my dream. Although I have been trying to make them proud, I would never be able to repay their selflessness.
   
   Finally, thank you, my dear wife, Zongwen Yang, for all the wonderful moments we have been sharing together. You take care of me and our families without any complains. You tolerate my bad temper and strong character. I feel so lucky to have you in my live and without you, I would have been a completely different person and live a different and difficult life. Thank you for loving me, putting up with me and helping me to write this thesis. 

%% file: intro/intro.tex
\chapter{Introduction}
The basic structure of an atom is understood to be electrons orbiting around a dense central nucleus due to the attractive electromagnetic force. Scattering experiments discovered that the nucleus is further composed of nucleons which include protons with positive charges and electrically neutral neutrons. Due to the Pauli Principle and the long range property of the nucleon-nucleon (NN) interactions, nucleons act like independent particles moving in a mean field inside the nucleus, and their features, e.g. their ground-states, were successfully predicted by the independent particle shell model (IPSM). However, it became clear in the early 1970s via electron scattering experiments that to completely explain high momentum components in the nuclear wave-functions, the short-range NN interactions must be accounted for. Short range correlations (SRC) arise from the tensor component and the repulsive hard-core in the NN interactions, and are essentially important to fully understand the nuclear structure and the properties of nucleons. 

  High energy electron scattering on nuclear targets is used as a probe to unveil the structure of nuclei and nucleons. In this chapter, a brief review of nuclear structure will be given, followed by a discussion of quasielastic (QE) electron scattering.
 
\input intro/nuclear_structure.tex
\input intro/quasielastic.tex

%% file: intro/nuclear_structure.tex
\section{Overview of Nuclear Structure}
  A nucleus is $\mathrm{10^{-6}}$ the size of an atom. However, it is a complicated many-body system where nucleons are bound by the strong interaction. To fully understand the detailed structure of the nucleus, one needs to have the complete knowledge of each nucleon's wave-function. For a light nucleus with only few nucleons, the wave functions can be directly calculated~\cite{PhysRevLett.87.172502}. However, for medium and heavy nuclei ($\mathrm{A\geq 12}$), the explosion of degrees of freedom in the Hamiltonian makes a solution extremely difficult to obtain.
\begin{figure}[!ht]
  \begin{center}
    \includegraphics[type=pdf,ext=.pdf,read=.pdf,width=0.80\linewidth]{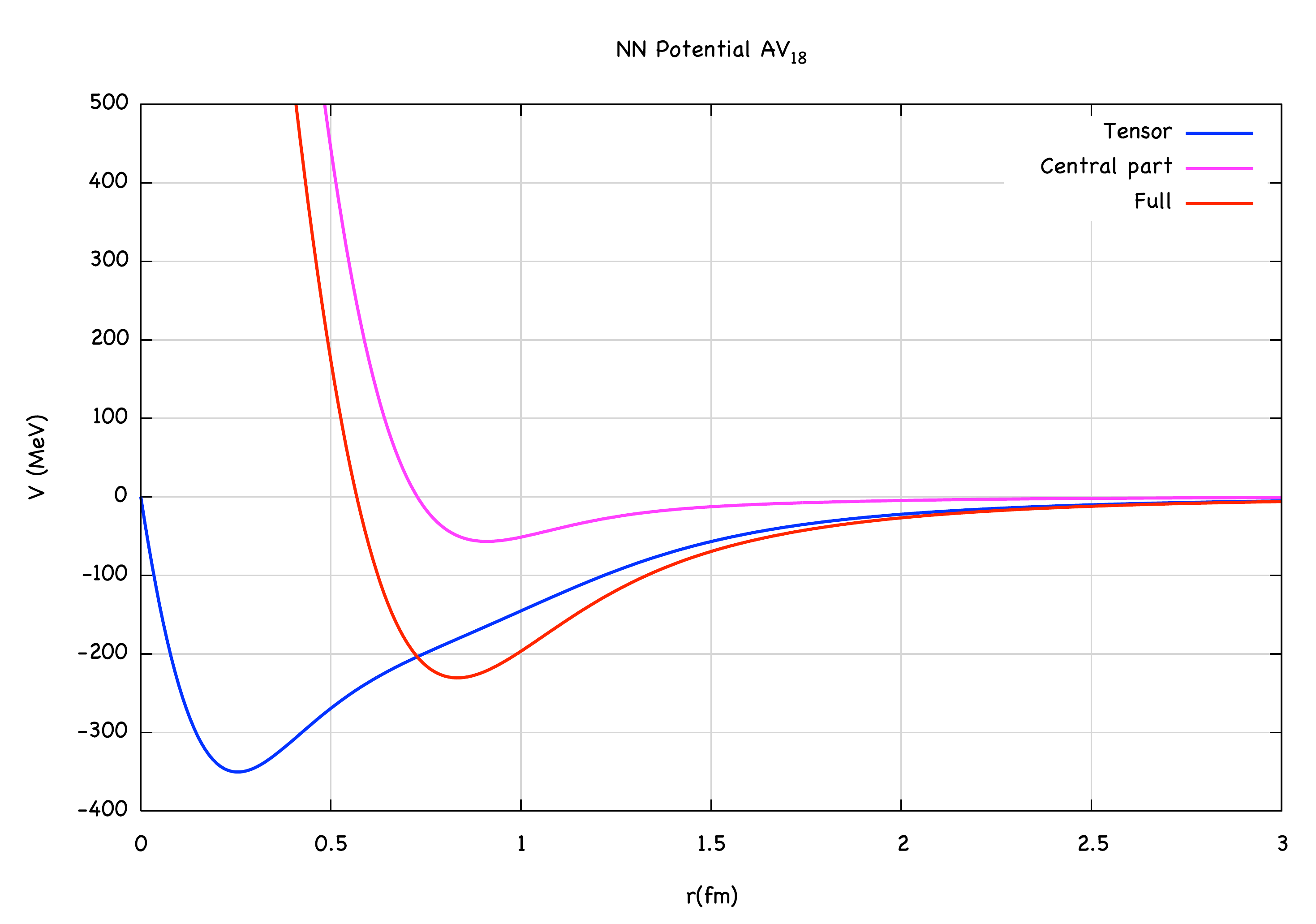}
    \caption[Two-nucleon interactions]{\footnotesize{Two-nucleon interactions calculated from the Argonne V14 potential~\cite{PhysRevC.51.38}, where the blue line represents the tensor force component, the cyan line is the central part and the red line is the total of these two combined. Figure is provided by Ref.~\cite{donal_prvt}.}}
    \label{potential_well}
  \end{center}
\end{figure}

Furthermore, the particular behaviour of the interaction potential between nucleons also increases the complexity of the nuclear system. A specific NN potential is shown in Fig.~\ref{potential_well}~\cite{PhysRevC.51.38}. The weak attractive interaction at moderate distance is generated by the exchange of virtual pions between nucleons. At short distance (e.g. $\mathrm{r\leq 1.5}$ fm), the interaction becomes strongly attractive on account of the tensor components of spin and isospin channels. At much shorter distance, the repulsive hard-core interaction between nucleons prevents the nucleus from collapsing. The Coulomb force between protons and potential three-body forces contribute but play a small role.

Despite the complexity of the nucleus, studies have revealed that nucleons behave like independent particles in the nuclear medium due to the collective effects of the Pauli principle and the average interaction with surrounding nucleons. In this picture, nucleons weakly interact with each other at short distance, and nucleons tended to occupy discrete energy states similar to the arrangement of electrons orbiting around the nucleus. It was also found that some nuclei have much large binding energies when they are composed of certain numbers of nucleons, namely magic numbers. 

These phenomena have been successfully described by the independent particle shell model (IPSM), also called the mean field theory. In this theory, the nucleon is treated as a non-relativistic object and moves in an average field generated by surrounding nucleons, and the NN interactions between nucleons are ignored. Nucleons occupy the lowest energy states first. The momentum and energy of the last occupied state are called the Fermi momentum and Fermi energy, respectively, and the whole set of occupied energy levels is called the Fermi sea. The energy state of each nucleon can be individually obtained by solving the Schrodinger equation with the mean field potential. Combined with the spin-orbit coupling, the IPSM successfully predicts the ground state properties, the excitation of nuclei at low energy, nuclear spins and parities, as well as the magic numbers. 
  
 \begin{figure}[!ht]
  \begin{center}
    \includegraphics[type=pdf,ext=.pdf,read=.pdf,width=0.95\linewidth]{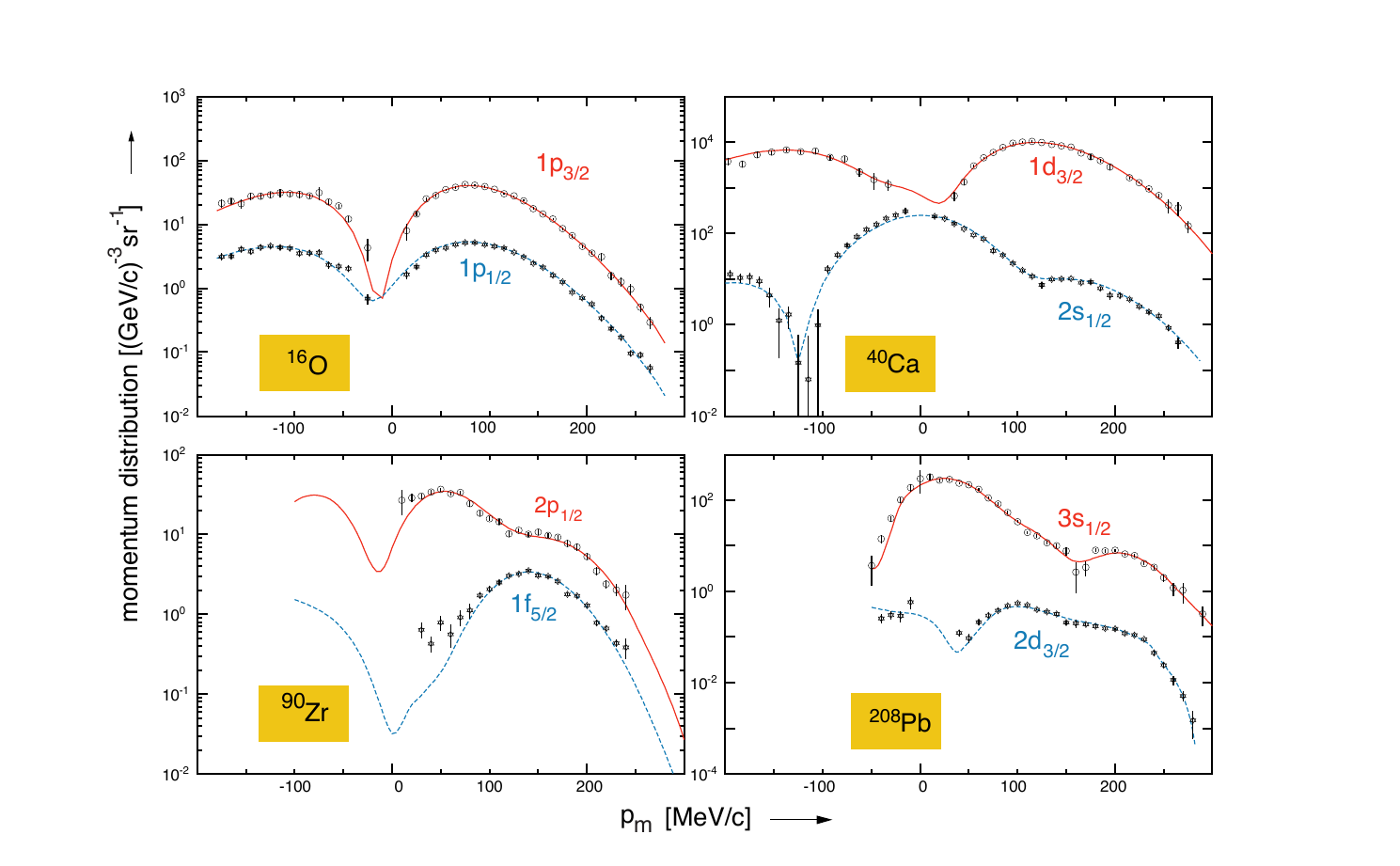}
    \caption[Momentum distribution from NIKNEF data]{\footnotesize{Momentum distribution from NIKNEF data, where each plot denotes the momentum distribution of a nucleon in different shell inside a nucleus. Dots are the results of electron proton-knock-out measurements in NIKNEF and the lines are the theoretical calculation with DWIA. To agree with the data, a factor of 0.65 was applied to the DWIA calculation. Plot is from Ref.~\cite{VanDerSteenhoven1988547}}}
    \label{niknef_mom_dis}
  \end{center}
\end{figure} 
 Nevertheless, the IPSM shows its limitations in predicting the nuclear magnetic moments and highly excited energy states~\cite{review_mft}. Furthermore, discrepancies had been observed in high resolution medium-energy proton-knock-out reactions in the early 1970s. The electron scattering cross section measurements at NIKNEF~\cite{VanDerSteenhoven1988547} studied the momentum distribution for various shell model orbits. In the IPSM, the nucleon wave function was calculated through the impulse plane wave approximation (PWIA)~\cite{DeForest1983} with a simplified optical potential. Taking account of the medium effect, the distorted wave impulse approximation (DWIA)\footnote{The distorted wave impulse approximation (DWIA) is a non-relativistic model used at intermediate energies to compensate for the effects of a mean nuclear potential. Basic scattering reaction calculations often assume that the incident nucleon behaves as a plane wave till it interacts with the a nucleon in the nucleus. In fact, the potential field of the nucleus, which is usually given by an optical potential, will distort the nucleon wave function.} was used with a more complex potential, e.g. the Woods-Saxon potential. However, from the data shown in Fig.~\ref{niknef_mom_dis}, to agree with the experimental observations, the momentum distribution calculated from DWIA had to be scaled by a factor of 0.65, although the shape was well reproduced. Other advanced Hartree-Fock calculations attempted to resolve this discrepancy by involving the long range NN interactions but still overestimated the nuclear strength~\cite{hartree_fock_book}.

\begin{figure}[!ht]
  \begin{center}
    \includegraphics[type=pdf,ext=.pdf,read=.pdf,width=0.45\linewidth]{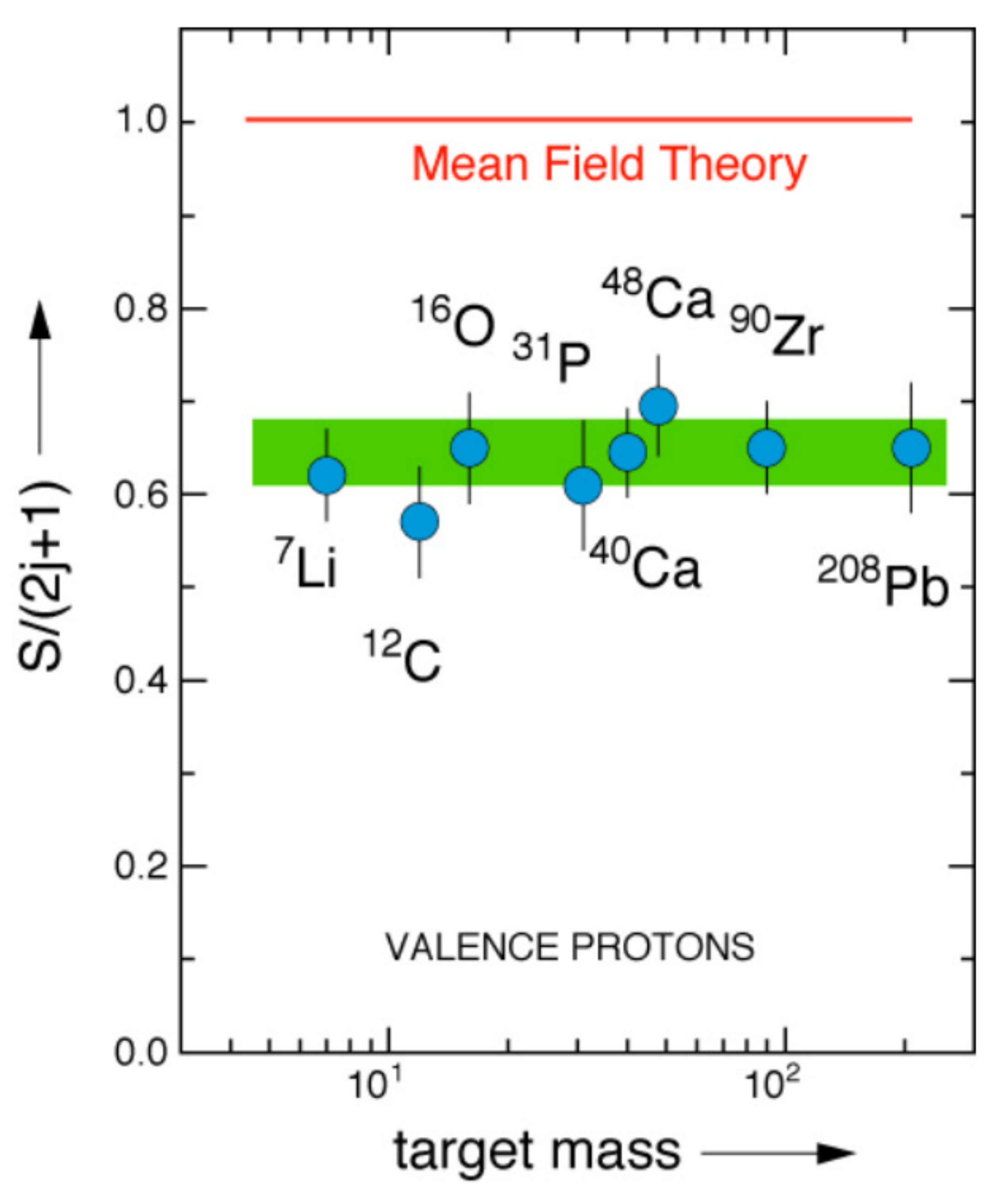}
    \caption[Measurements of spectroscopy factors]{\footnotesize{Measurements of spectroscopy factors for different nuclei deviate from one, where the y-axis denotes the ratio of the measured occupation number of a nucleon in a nucleus compared with the mean field prediction. The plot suggests that the mean field theory overestimates the occupation of a nucleon's states. Figure is from Ref.~\cite{Lapikas1993297}.}}
    \label{spec_fac_exp}
  \end{center}
\end{figure} 
 From these cross section measurements, one obtained the spectroscopic factors or the occupation numbers, and discovered that the occupancies of the same orbits were 30\%-40\% below the expected values~\cite{Lapikas1993297,Kelly:1996hd}, as shown in Fig.~\ref{spec_fac_exp}. 
 
 The missing nuclear strength can be understood by considering the short range NN interactions. The IPSM restricts nucleons in their energy states below the Fermi surface. However, as shown in Fig.~\ref{potential_well}, two nucleons do interact at short distance as a result of the attractive potential, and at much shorter distance, the repulsive force will push them apart. Taken together, it is realized that nucleons must carry high relative momenta which significantly exceed the Fermi momentum. Knock-out measurements revealed that the nuclear strength is beyond the predictions of the mean field. 
 
 Nucleons with high energies and momenta are generally referred to as short range correlations (SRC), which will be discussed in the next chapter.

%% file: intro/quasielastic.tex
\section{Quasi-elastic Scattering}
\begin{figure}[!ht]
  \begin{center}         
    \includegraphics[type=pdf,ext=.pdf,read=.pdf,width=0.60\linewidth]{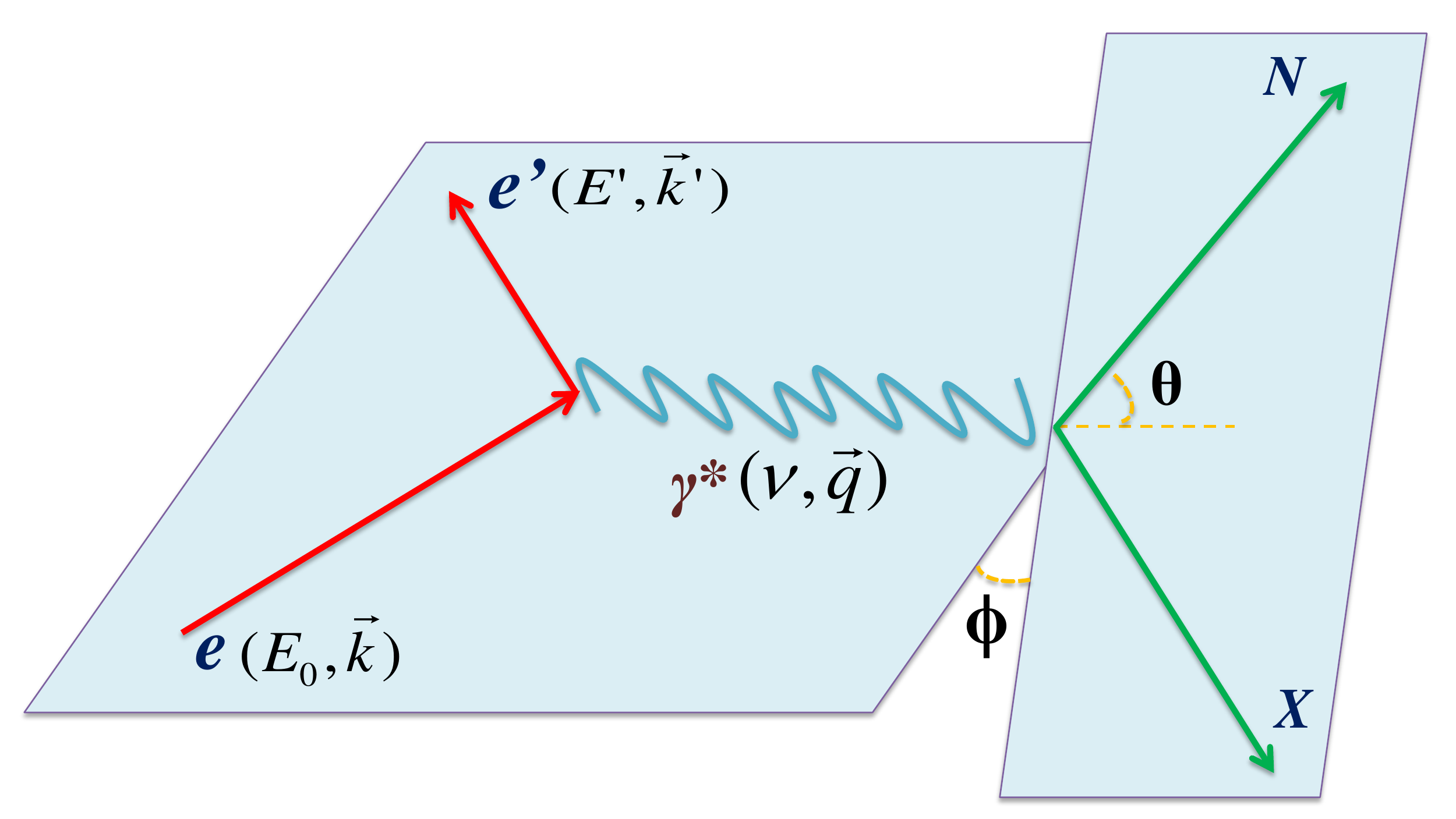}
    \caption[Schematic of electron-scattering]{\footnotesize{Schematic of electron-scattering.}}
    \label{e_scatt}
  \end{center}
\end{figure}

\begin{figure}[!ht]
  \begin{center}
    \includegraphics[type=pdf,ext=.pdf,read=.pdf,width=0.60\linewidth]{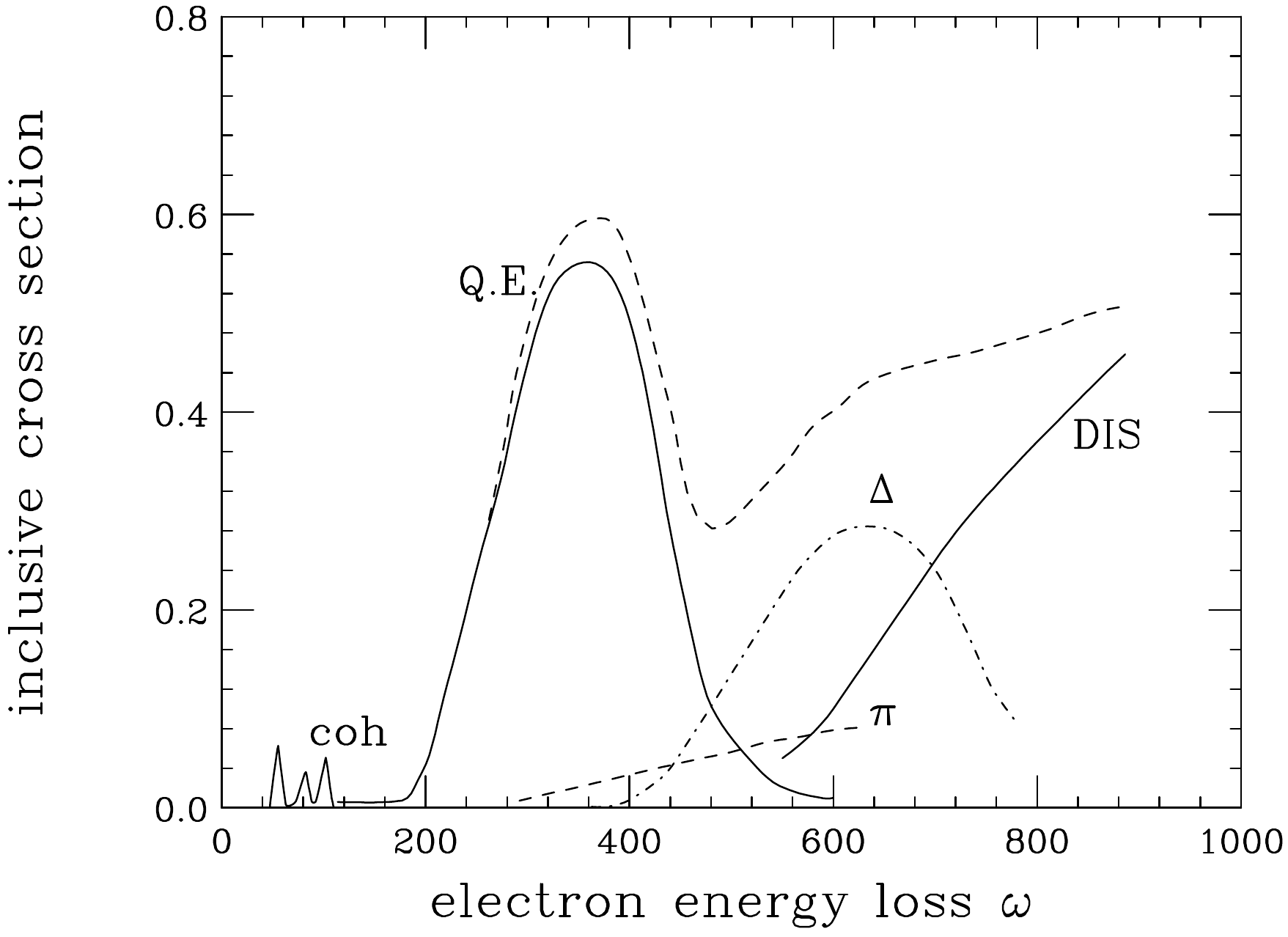}
    \caption[Electron energy transfer]{\footnotesize{Inclusive cross section on the y-axis versus the energy loss $\omega\equiv\nu=E_{0}-E'$ on the x-axis. The Figure is obtained from Ref.~\cite{qe_donal}.}}
    \label{e_trans}
  \end{center}
\end{figure}

 In Fig.~\ref{e_scatt}, an electron with known initial energy, $E_{0}$, and final energy, $E'$, interacts with a charged nucleus by exchanging a virtual photon with the four momentum $q=(\nu,\vec{q})$, where the energy transfer $\nu = E_{0}-E'$ and the momentum transfer $\vec{q}=\vec{k}-\vec{k'}$. One can probe the nucleus at different scales by varying $q$. Elastic scattering denotes the process of electrons interacting with the entire nucleus while the nucleus remains intact. Quasielastic (QE) scattering refers to the case where electrons scatter off individual moving nucleons which are ejected from the nucleus thereafter. Fig.~\ref{e_trans} is a schematic of the inclusive electron scattering cross sections as a function of $\nu$~\cite{qe_donal}, where a broad peak is seen for the QE process due to the interval motion of the nucleons. Nucleons can be excited at even larger $\nu$ and resonances start to contribute to the cross section through inelastic scattering. Electrons directly probe the quark properties in the nucleons at very large $\nu$ through deep inelastic scattering (DIS). 

A convenient kinematic parameter for identifying these different scattering processes is the Bjorken variable, $x_{bj}=Q^{2}/(2m_{N}\nu)$, which was firstly proposed as a scaling variable to describe DIS from nucleons. It is then interpreted as the momentum fraction of the struck quark. For scattering off a nucleon, $x_{bj}$ goes from 0 to 1, and the elastic peak is found at $x_{bj}=1$. In the case of scattering off a nucleus in QE process, $x_{bj}$ extends to the region of $0<x_{bj}<M_{A}/m_{N}$, where $x_{bj}=1$ is now the location of the QE peak and the nuclear elastic peak moves to $x_{bj}=M_{A}/m_{N}\simeq A$. For convenience, $m_{N}$ is usually replaced by the proton mass during the experimental data analysis, and will be used in the rest of this thesis:
\begin{equation}
  x_{bj} = Q^{2}/(2m_{p}\nu).
  \label{xbj_define}
\end{equation}
  
 In QE scattering, an electron knocks a nucleon out from the nucleus. This provides an opportunity to study the nucleon's original ground state. One can detect the scattered electrons in coincidence with the struck nucleon, and extract the exclusive cross section which contains the initial properties of the nucleon inside the nuclear medium. In the PWIA, the exclusive cross section is the sum of the cross sections of the individual nucleons weighted by the spectral function~\cite{DeForest1983,john_thesis}:
\begin{equation}
  \frac{d^{5}\sigma}{dE'd\Omega d^{3}\vec{p'}} = \sum_{nucleons}\sigma_{eN}\cdot S'_{N}(E_{0},\vec{p}_{0}),
  \label{quasi_xs_spectral_function}
\end{equation}
where $\sigma_{eN}$ is the cross section of electron-nucleon scattering, and $S'_{N}(E_{0},\vec{p}_{0})$ is the nuclear spectral function and gives the probability of removing a nucleon with initial energy $E_{0}$, and momentum $\vec{p}_{0}$ from the target nucleus~\cite{qe_donal}. Within the IPSM, a nucleon moves independently in a mean field, and the spectral function can be written as~\cite{qe_donal}:
\begin{equation}
 S'_{N}(E_{0},\vec{p}_{0}) = \sum_{n\in \lbrace F\rbrace}\vert\phi_{n}(\vec{p}_{0})\vert^{2}\delta(E_{0}-E_{n}),
\end{equation}
where $\phi_{n}(\vec{p}_{0})$ is the wave-function of the nucleon when it is in an eigenstate with the eigen-energy $E_{n}$. The sum is extended to all occupied states within the Fermi sea $\lbrace F\rbrace$. 

\subsection{Inclusive Cross Section}
\begin{figure}[!ht]
  \begin{center}
    \includegraphics[type=pdf,ext=.pdf,read=.pdf,width=0.60\linewidth]{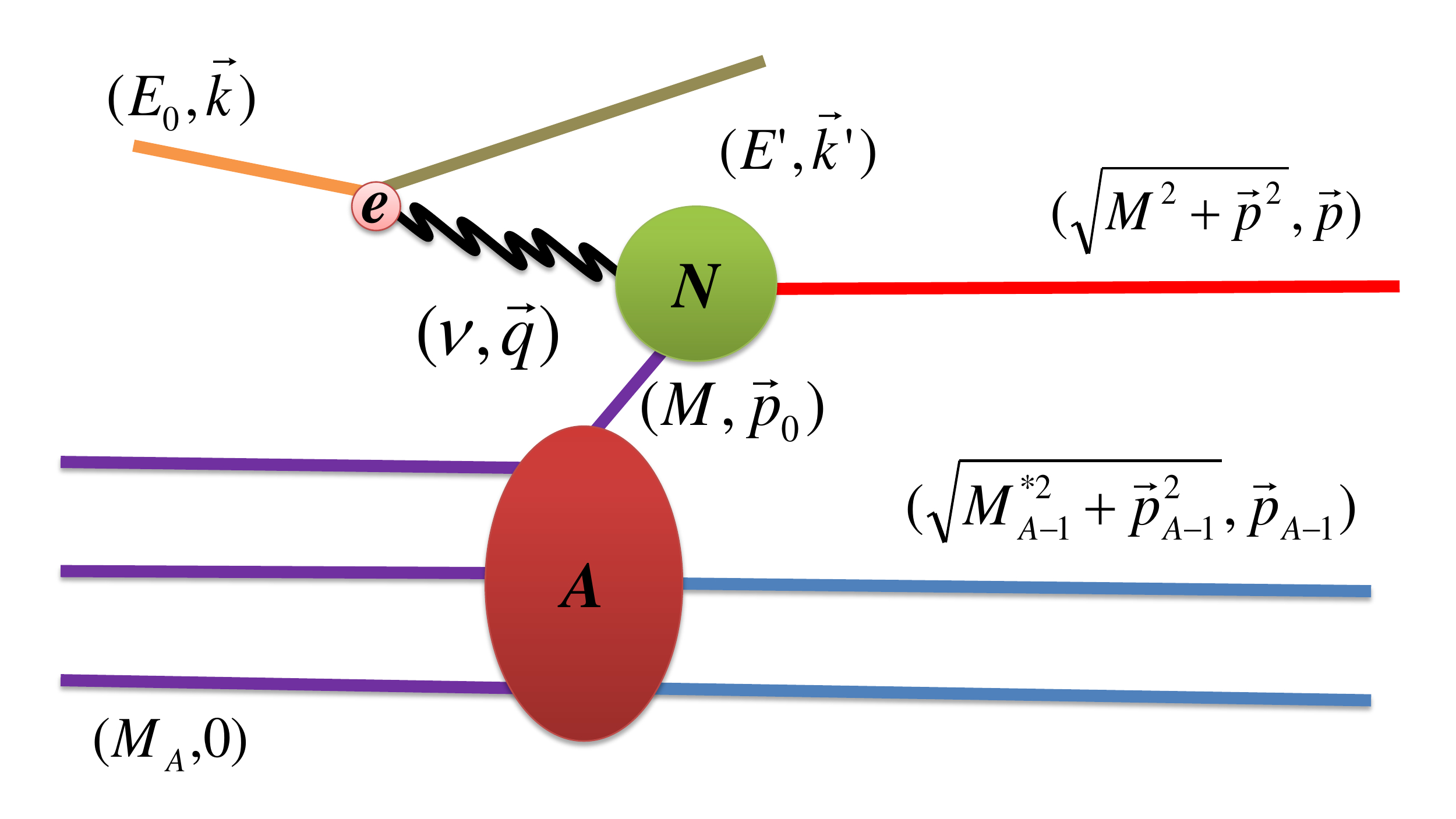}
    \caption[Schematic of QE electron scattering]{\footnotesize{Schematic of QE electron scattering where $\vec{p}_{A-1} = -\vec{p}_{0}$ for fixed targets.}}
    \label{qe_scatt_diag}
  \end{center}
\end{figure}
 In inclusive electron-nucleus QE scattering (Fig.~\ref{qe_scatt_diag}), the cross section is determined by only detecting the scattered electron with the four momentum, ($E'$, $\vec{k'}$). The knocked out nucleons with ($\sqrt{M^{2}+ \vec{p}^{2}}$, $\vec{p}$), and the potentially excited ($A-1$) recoil system in the final state ($\sqrt{M_{A-1}^{*2}+ \vec{p}_{A-1}^{2}}$, $\vec{p}_{A-1}$) are undetected. To obtain the inclusive cross section from Eq.~\eqref{quasi_xs_spectral_function}, one separates the contributions from protons and neutrons, and integrates over the final states of the nucleons:
\begin{equation}
  \frac{d^{3}\sigma}{dE'd\Omega} = \int (Z\sigma_{ep}S'_{p}(E_{0},\vec{p}_{0})+N\sigma_{en}S'_{n}(E_{0},\vec{p}_{0})) d^{3}\vec{p},
\end{equation}
where the subscripts in $dE'_{e}d\Omega_{e}$ have been omitted since only electrons are measured.

Assuming the spectral function is spherically symmetric and the difference in the spectral function of protons and neutrons can be ignored, a more general form $S'(E_{0},p_{0})$ can be factored out from the equation. Since $\vec{p} = \vec{p}_{0}+\vec{q}$, where $\vec{q}$ is fixed when measuring $E_{0}$ and $E'$, one can replace $d^{3}\vec{p}$ by $d^{3}\vec{p}_{0}$. In spherical coordinates, $d^{3}\vec{p}_{0}=p_{0}^{2} dp_{0}d(cos\theta) d\phi$, and the cross section becomes:
\begin{equation}
  \frac{d^{3}\sigma}{dE'd\Omega} = 2\pi \int \tilde{\sigma}_{0}\cdot S'(E_{0},p_{0})\cdot p_{0}^{2}dp_{0}d(cos\theta),
  \label{in_xs_org}
\end{equation}
where
\begin{equation}
  \tilde{\sigma}_{0} = \frac{1}{2\pi}\int_{0}^{2\pi} \left( Z\sigma_{ep}+N\sigma_{en} \right)d\phi.
\end{equation}

Eq.~\eqref{in_xs_org} can be further simplified by considering energy conservation. From Fig.~\ref{qe_scatt_diag}, for a fixed target, $\vec{p}_{A-1} = - \vec{p}_{0}$, which gives: 
\begin{equation}
  M_{A}= E_{0}+ \sqrt{M_{A-1}^{*2}+p_{0}^{2}},
  \label{eqma_e0}
\end{equation}
and,
\begin{equation}
  \quad  M_{A}+\nu = \sqrt{M^{2}+(\vec{p}_{0}+\vec{q})^{2}}+\sqrt{M_{A-1}^{*2}+p_{0}^{2}},
  \label{ene_mom_cons}
\end{equation}
where $M$ and $M_{A}$ are the mass of the ejected nucleon and the target nucleus, respectively. $M_{A-1}^{*}$ is the mass of the recoiling $(A-1)$ system, where the superscript * denotes that it could be in an excited state. Combining Eq.~\eqref{eqma_e0} and Eq.~\eqref{ene_mom_cons}, one has:
\begin{equation}
 E_{0} + \nu = \sqrt{M^{2}+p_{0}^{2}+q^{2}+2p_{0}qcos\theta}.
\end{equation}
Since $\vec{q}$ and $\nu$ are fixed, $E_{0}$ can be determined by $p_{0}$ and $cos\theta$. One defines a $\delta$-function, $\delta(E_{0}+\nu-\sqrt{M^{2}+p_{0}^{2}+q^{2}+2p_{0}qcos\theta})$, and inserts it into the integral:
\begin{equation}
  \frac{d^{3}\sigma}{dE'd\Omega} = 2\pi \int \tilde{\sigma}_{0}\cdot S'(E_{0},p_{0})\cdot \delta \cdot p_{0}^{2}dp_{0}d(cos\theta)dE_{0}.
  \label{in_xs_org2}
\end{equation}
Integrating over $cos\theta$, one rewrites the formula as follows~\cite{john_thesis}:
\begin{equation}
  \frac{d^{3}\sigma}{dE'd\Omega} = 2\pi \int \tilde{\sigma}_{0}\cdot \frac{E_{N}}{p_{0}q} \cdot S'(E_{0},p_{0})\cdot p_{0}^{2}dp_{0}dE_{0},
\end{equation}
where $E_{N}=\sqrt{M^{2}+p^{2}}$ denotes the energy of the struck nucleon.

One can define the separation energy, $E_{s}\equiv M_{A-1}^{*}+M-M_{A}$, and considering Eq.~\eqref{eqma_e0}, the spectral function becomes $S(E_{s},p_{0})dE_{s}\equiv - S'(E_{0},p_{0})dE_{0}$ through the Jacobian transformation~\cite{john_thesis}. By defining $\tilde{\sigma}=\tilde{\sigma_{0}}\cdot E_{N}/q$, the cross section is rewritten as:
\begin{equation}
  \frac{d^{3}\sigma}{dE'd\Omega} = 2\pi \int_{E_{s}^{min}}^{E_{s}^{max}} \int_{p_{0}^{min}}^{p_{0}^{max}}\tilde{\sigma}\cdot S(E_{s},p_{0})\cdot p_{0}dp_{0}dE_{s},
  \label{in_qe_xs1}
\end{equation}
where $p_{0}^{min}$ and $p_{0}^{max}$ can be obtained from Eq.~\eqref{ene_mom_cons} when $\vec{p}_{0}$ and $\vec{q}$ are parallel:
\begin{equation}
  \quad  M_{A}+\nu = \sqrt{M^{2}+y^{2}+2yq+q^{2}}+\sqrt{M_{A-1}^{*2}+y^{2}}.
  \label{ene_mom_cons_y}
\end{equation}	
Two solutions, $y_{1}$ and $y_{2}$ ($y_{1}<y_{2}$), give the values of $p_{0}^{min}$ and $p_{0}^{max}$, respectively. $E_{s}^{min}$ corresponds to the minimum separation energy when the recoil nucleus is in its ground state. $E_{s}^{max}$ is the maximum separation energy when the struck nucleon is at rest in the final state, i.e. $p_{0}^{min}(E_{s}^{max}) = p_{0}^{max}(E_{s}^{max})$, and it can be given as:
\begin{equation}
  E_{s}^{max}=\sqrt{(M_{A}+\nu)^{2}-q^{2}}-M_{A}.
\end{equation}
\subsection{Momentum Distribution and y-Scaling}
 The integral of the spectral function over the separation energy leads to the momentum distribution:
\begin{equation}
  n(p_{0}) = \int_{E_{s}^{min}}^{\infty} S(E_{s},p_{0})dE_{s},
  \label{np_mom_eq}
\end{equation}
which is one of the two main properties of nuclei, along with the nuclear density, and is directly connected to the many-body wave-function. By understanding the momentum distribution, one has an opportunity to examine the effect of the nuclear medium and the NN interactions. For example, one can study the momentum distribution above the Fermi momentum, $k>k_{F}$, where the mean field contribution vanishes, and probe the short-distance properties of the NN interaction, i.e. SRC.

\begin{figure}[!ht]
  \begin{center}
    \includegraphics[type=pdf,ext=.pdf,read=.pdf,width=0.70\linewidth]{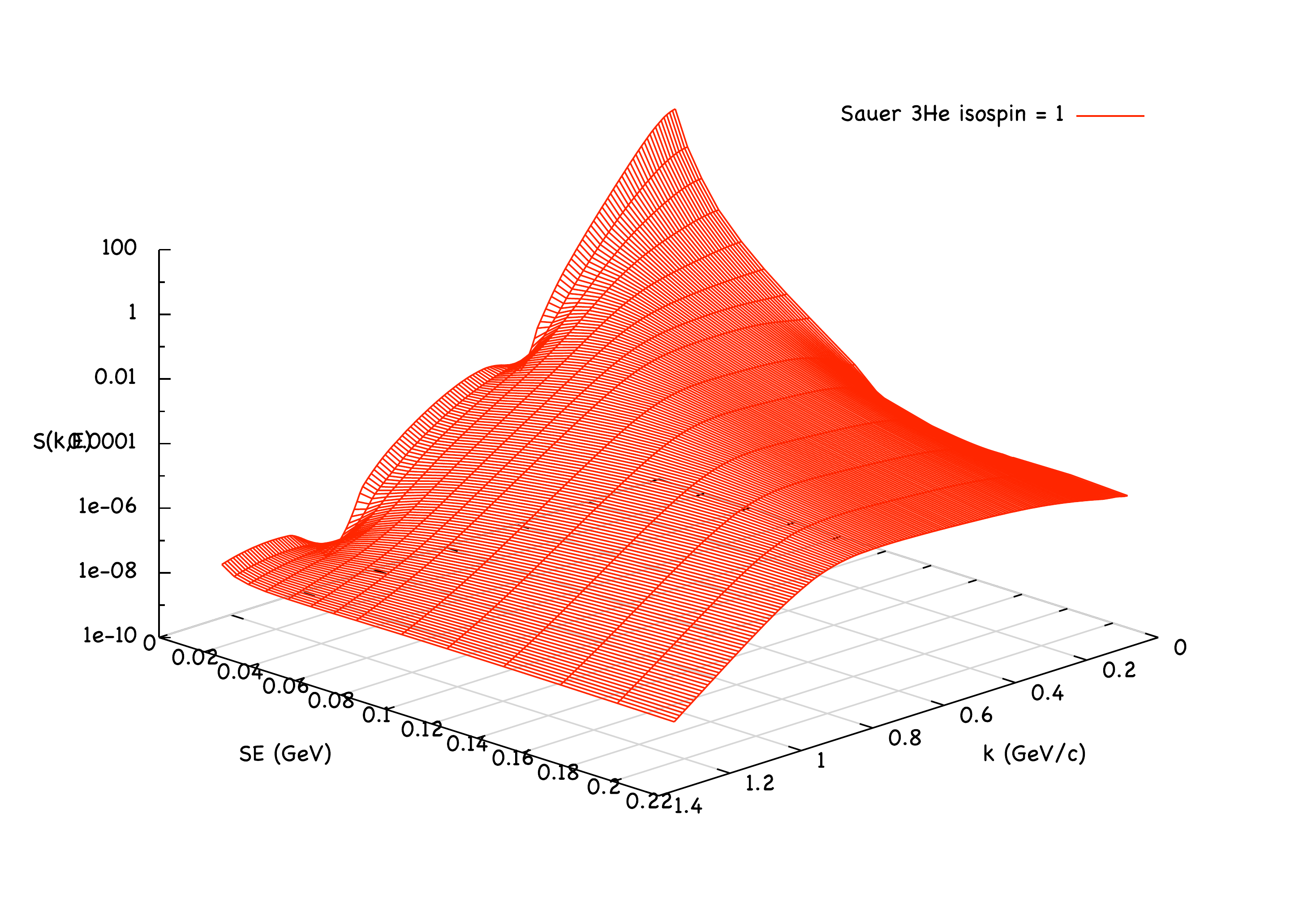}
    \caption[Spectral function for $\mathrm{^{3}He}$]{\footnotesize{Spectral function for $\mathrm{^{3}He}$ as a function of the separation energy $E_{s}$ (given as $SE$ here) and the initial momentum $p_{0}$ (given as $k$ here) for the process of knocking out a nucleon from the nucleus. The magnitude of the spectral function decreases at large values of $E_{s}$ and $p_{0}$, so the upper limits of the double integrals in Eq.~\eqref{in_qe_xs2} can be approximately extended to infinity. Plot was provided by Ref.~\cite{donal_prvt}.}}
    \label{pkem}
  \end{center}
\end{figure}
 However, as the spectral function, the momentum distribution is also not an experimental observable and can not be directly measured from the electron-scattering experiments. Instead, one studies the y-scaling behaviour~\cite{West1975263,day_arns, PhysRevC.41.R2474, Boffi19931} of inclusive QE scattering and extracts the momentum distribution from the scaling function which is directly related to the measured cross sections.

 Because the spectral function decreases rapidly by orders of magnitude toward $E_{s}^{max}$ and $p_{0}^{max}$ (see Fig.~\ref{pkem}), the upper limits of the two integrals in Eq.~\eqref{in_qe_xs1} can be extended to infinity. Meanwhile, $\tilde{\sigma}$ changes very slowly as a function of $E_{s}$ and $p_{0}$, so it can be factored out from the integral and evaluated at the maximum value of the spectral function at $E_{s}=E_{s}^{0}$. Hence Eq.~\eqref{in_qe_xs1} can be rewritten as:
\begin{equation}
  \frac{d^{3}\sigma}{dE'd\Omega} = 2\pi \bar{\sigma}\int_{E_{s}^{min}}^{\infty} \int_{p_{0}^{min}}^{\infty}S(E_{s},p_{0})\cdot p_{0}dp_{0}dE_{s},
  \label{in_qe_xs2}
\end{equation}
where $\bar{\sigma} \propto \tilde{\sigma}(E_{s}^{0},p_{0}^{min})$~\cite{PhysRevB.36.1208}. 

 The scaling function can be defined as:
\begin{equation}
  F(y,q) = 2\pi\int_{E_{s}^{min}}^{\infty} \int_{|y|}^{\infty}S(E_{s},p_{0})\cdot p_{0}dp_{0}dE_{s},
  \label{fy_scaling_eq}
\end{equation}
where the new variable, $y$, is defined as the minimum values of momentum in Eq.~\eqref{ene_mom_cons}, $p_{0}^{min}=|y|$, when the $A-1$ system is in its ground state:
\begin{equation}
  M_{A}+\nu = \sqrt{M^{2}+q^{2}+y^{2}+2yq}+\sqrt{M_{A-1}^{2}+y^{2}}.
  \label{y_enegy_conserv}
\end{equation} 

 The validity of the y-scaling in QE region relies on several assumptions: (i) the final state interaction (FSI, see next section) is small at large momentum transfer; (ii) the DIS contribution can be subtracted or ignored; (iii) since $\mathrm{E_{s}}$ and $\mathrm{p_{0}}$ are limited to the finite ranges during the cross section measurements , the error made by extending the upper limit of Eq.~\eqref{fy_mom_eq} to infinity can be neglected; (v) the (A-1) recoil system has to be minimally excited so that the lower limit ($p_{0}^{min}=|y|$) can be treated as independent of $E_{s}$.

If the assumptions above are valid and only the nucleonic degrees of freedom are considered, the scaling function can be treated as independent of $q$ at large momentum transfer~\cite{Boffi19931}, i.e. $F(y,q)\equiv F(y)$. From Eq.~\eqref{in_qe_xs2}, the scaling function can be extracted from the experimental electron-nucleus scattering cross section, $\sigma_{EX}$:
\begin{equation}
  F(y)=\frac{d^{3}\sigma_{EX}}{dE' d\Omega } \frac{1}{Z\sigma_{p}+N\sigma_{n}} \frac{q}{\sqrt{M^{2}+(y+q)^{2}}}.
  \label{fy_scaling_eq2}
\end{equation}
\begin{figure}[!ht]
  \begin{center}
    \includegraphics[type=pdf,ext=.pdf,read=.pdf,width=0.80\linewidth]{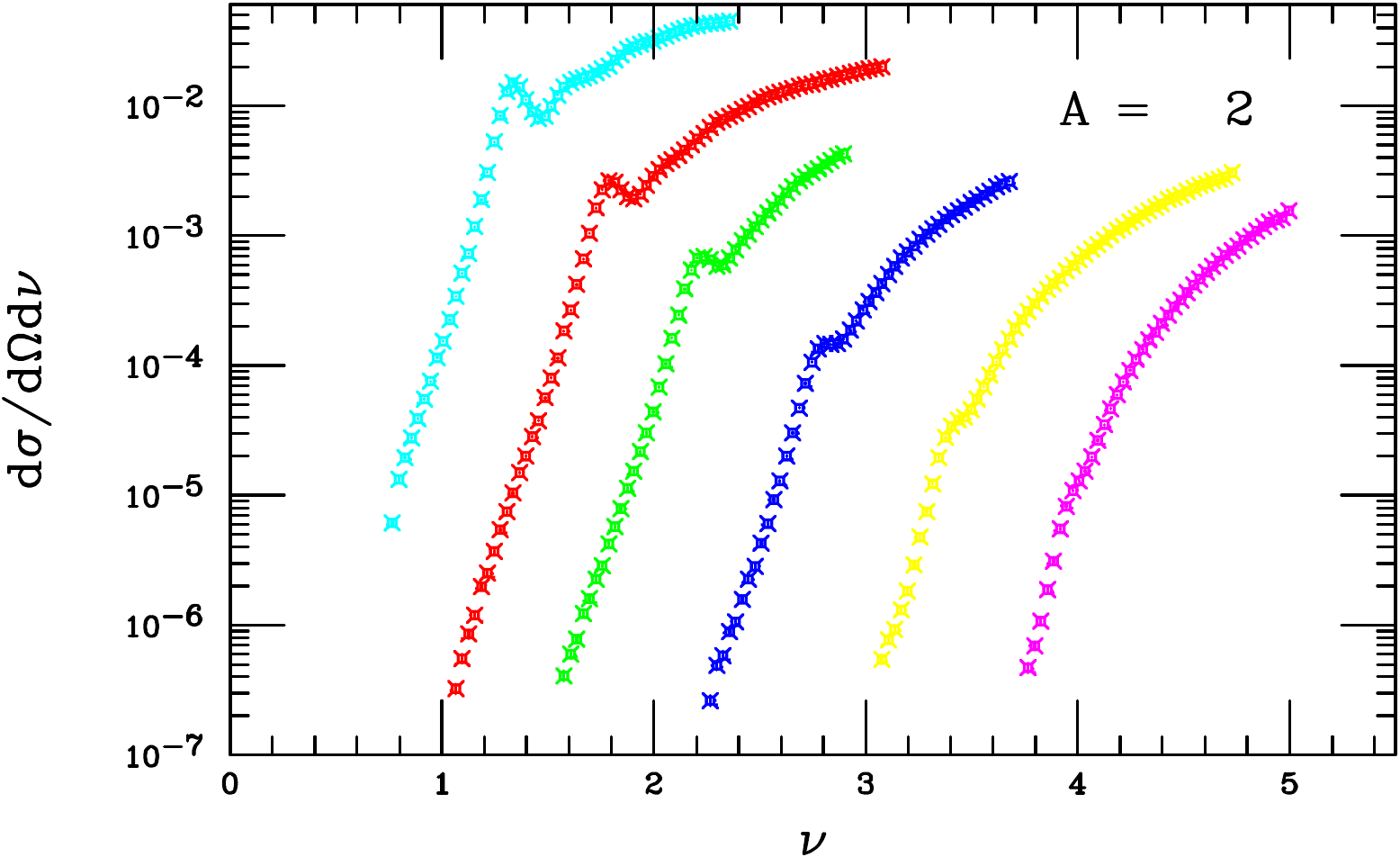}
    \caption[Deuteron inclusive cross section]{\footnotesize{Deuteron inclusive cross sections~\cite{PhysRevLett.108.092502}. Symbols are experiment data from the E02-019~\cite{nadia_thesis}. From left to right, the $\mathrm{Q^{2}}$ values for different distributions are 2.5 (light-blue), 3.3 (red), 4.1 (green), 5.2 (blue), 6.5 (yellow) and 7.4 (purple) $\mathrm{GeV^{2}}$, respectively. The cross section distributions clearly show the strong $\mathrm{Q^{2}}$ dependence.}}
    \label{nadia_cs_deut}
  \end{center}
\end{figure}
 \begin{figure}[!ht]
  \begin{center}
    \includegraphics[type=pdf,ext=.pdf,read=.pdf,width=0.80\linewidth]{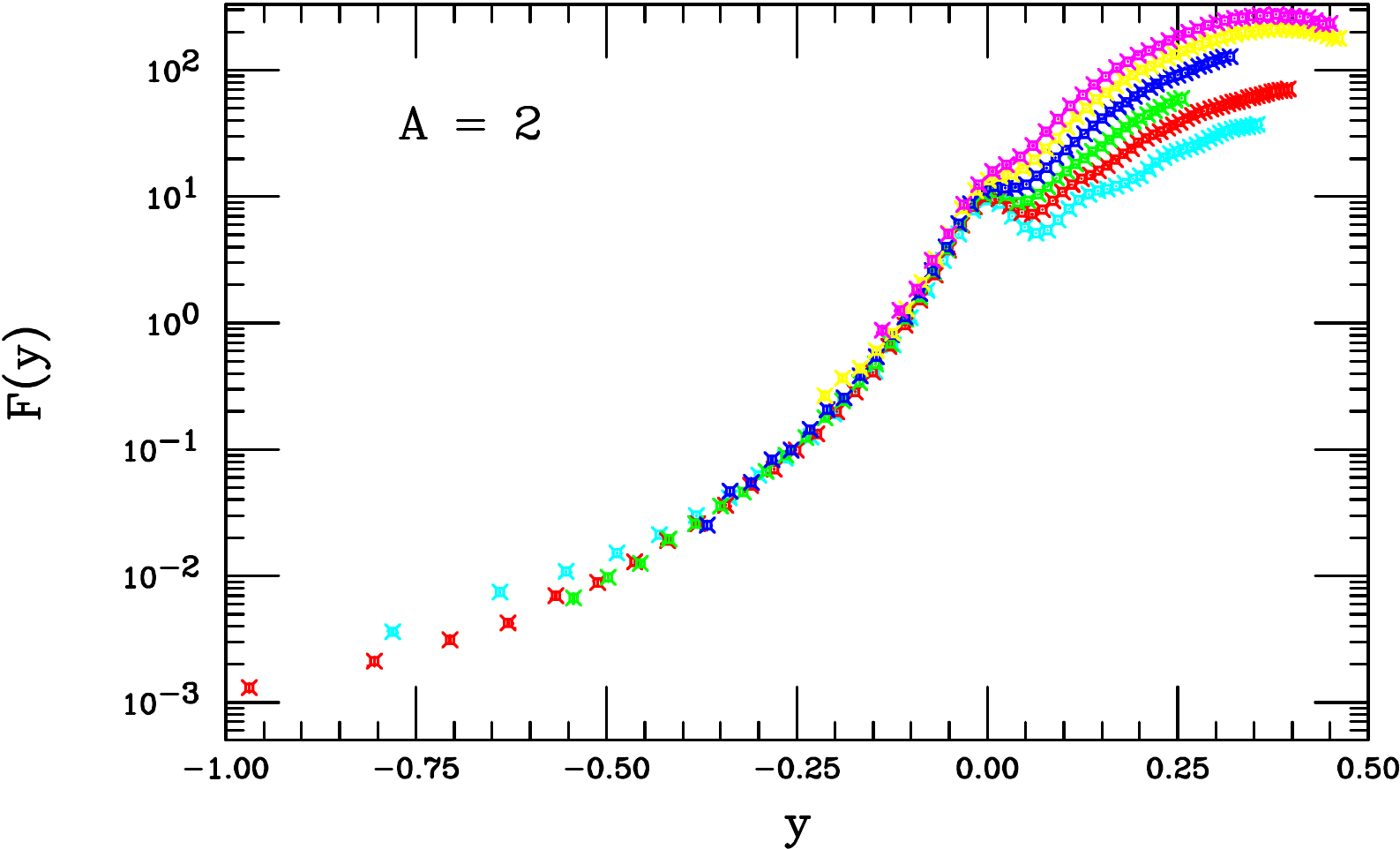}
    \caption[Scaling function $F(y)$ for deuteron]{\footnotesize{Scaling function $F(y)$ for deuteron~\cite{PhysRevLett.108.092502}. Symbols are experiment data from the E02-019~\cite{nadia_thesis}. The $\mathrm{Q^{2}}$ values for different distributions are 2.5 (light-blue), 3.3 (red), 4.1 (green), 5.2 (blue), 6.5 (yellow) and 7.4 (purple) $\mathrm{GeV^{2}}$, respectively. At $y<0$, the $F(y)$ distributions show much less $\mathrm{Q^{2}}$ dependence for data with high $\mathrm{Q^{2}}$, where the FSI is small. At $y\ge 0$, the distributions are dominated by the DIS processes so the y-scaling is largely violated. Figure is from Ref.~\cite{nadia_thesis}.}}
    \label{y_scaling_deut}
  \end{center}
\end{figure}

As shown in Fig.~\ref{nadia_cs_deut}, the deuteron inclusive cross sections, measured in the E02-019~\cite{nadia_thesis}, reveal a strong $\mathrm{Q^{2}}$ dependence. However, in Fig.~\ref{y_scaling_deut}, the $F(y)$ distributions extracted from these cross sections are only modestly dependent on $\mathrm{Q^{2}}$ at $y<0$, especially at large $\mathrm{Q^{2}}$ ($>3~GeV^{2}$). At modest $\mathrm{Q^{2}}$ ($< 2 GeV^{2}$), the y-scaling starts to be violated due to the FSI. The $F(y)$ fails to scale at $y\ge 0$ where the DIS contributions dominate. The result supports the assumption of y-scaling in QE region.

By reversing the order of the integration in Eq.~\ref{fy_scaling_eq} and based on Eq.~\eqref{np_mom_eq}, $F(y)$ can be rewritten as:
\begin{equation}
  F(y) = 2\pi\int_{|y|}^{\infty}n(p_{0})\cdot p_{0}dp_{0}.
  \label{fy_mom_eq}
\end{equation} 
Hence the momentum distribution can be extracted experimentally from the $F(y)$ distribution~\cite{qe_donal}:
\begin{equation}
  n(p_{0}) = \frac{-1}{2\pi p_{0}}\frac{dF(p_{0})}{dp_{0}} \mid_{p_{0}=|y|}.
  \label{mom_dis_fy}
\end{equation}

\begin{figure}[!ht]
  \begin{center}
    \includegraphics[type=pdf,ext=.pdf,read=.pdf,angle=270,width=0.80\linewidth]{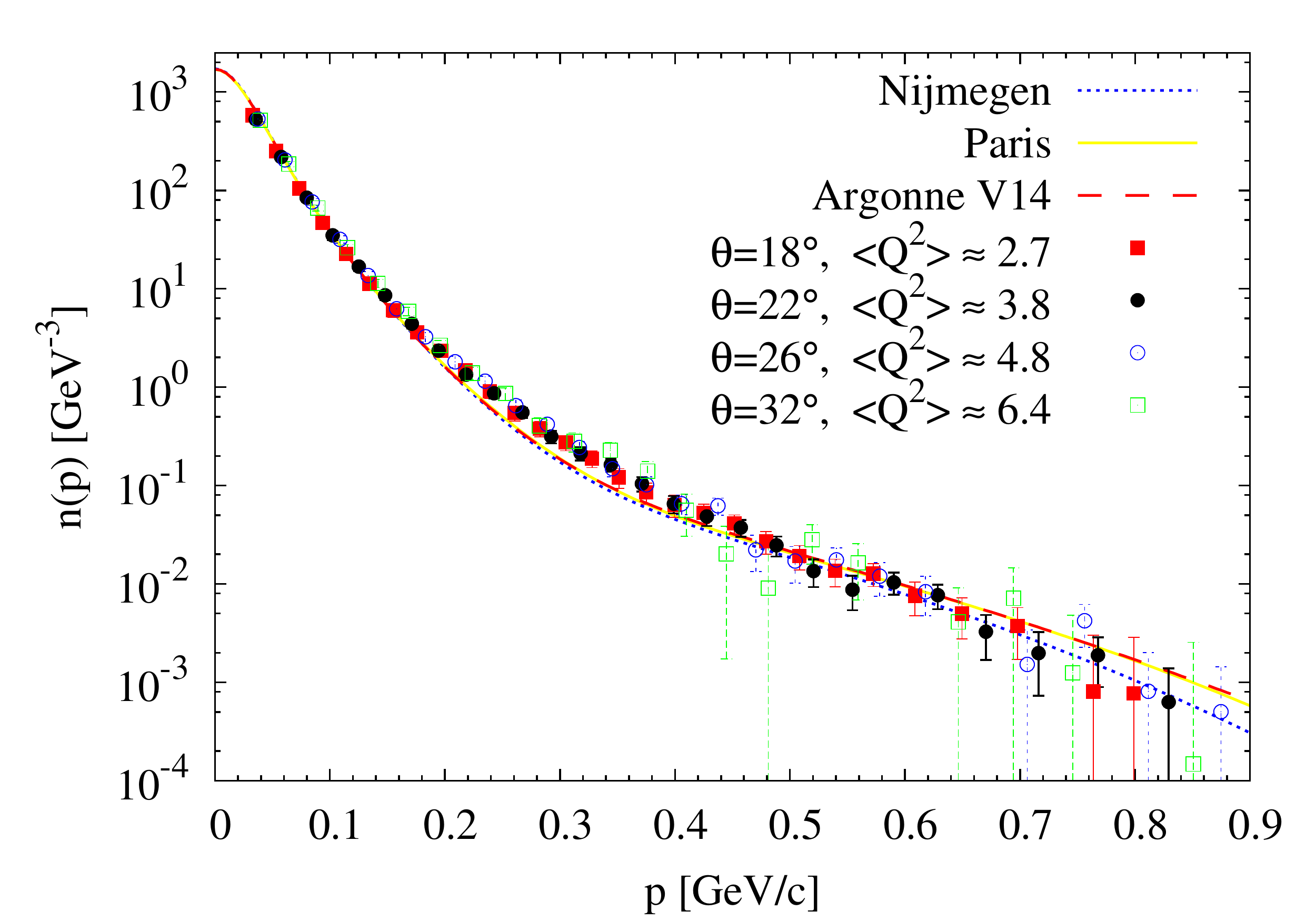}
    \caption[Momentum distribution $n(p_{0})$ for deuteron]{\footnotesize{Momentum distribution $n(p_{0})$ for deuteron extracted from the E02-019 data~\cite{nadia_thesis,PhysRevLett.108.092502}, where in this figure, $p_{0}\rightarrow p$. Symbols show the data at different $\mathrm{Q^{2}}$ where the $\mathrm{Q^{2}}$ values are different with ones in the previous two plots since they were quote at different $x_{bj}$ value for other purposes. Lines give the calculations with three different potential models~\cite{PhysRevC.49.2950,Lacombe1981139, PhysRevC.51.38}. Figure is from Ref.~\cite{PhysRevLett.108.092502}.}}
    \label{mom_dis_deut}
  \end{center}
\end{figure}
Eq.~\eqref{mom_dis_fy} provides a way to obtain the nucleon's momentum distribution in the nucleus since $F(y)$ can be directly extracted from the inclusive QE scattering cross section. Fig.~\ref{mom_dis_deut} shows the momentum distribution for the deuteron extracted from the experiment data taken in the $x_{bj}>1$ region along with theoretical calculations derived from various NN potentials~\cite{nadia_thesis,PhysRevLett.108.092502}. From this plot, one can draw the conclusion that studying y-scaling at high $\mathrm{Q^{2}}$ allows the experimental data to be compared with theory in a productive way.

\subsection{Final State Interaction}
\begin{figure}[!ht]
  \begin{center}
    \includegraphics[type=pdf,ext=.pdf,read=.pdf,width=0.80\linewidth]{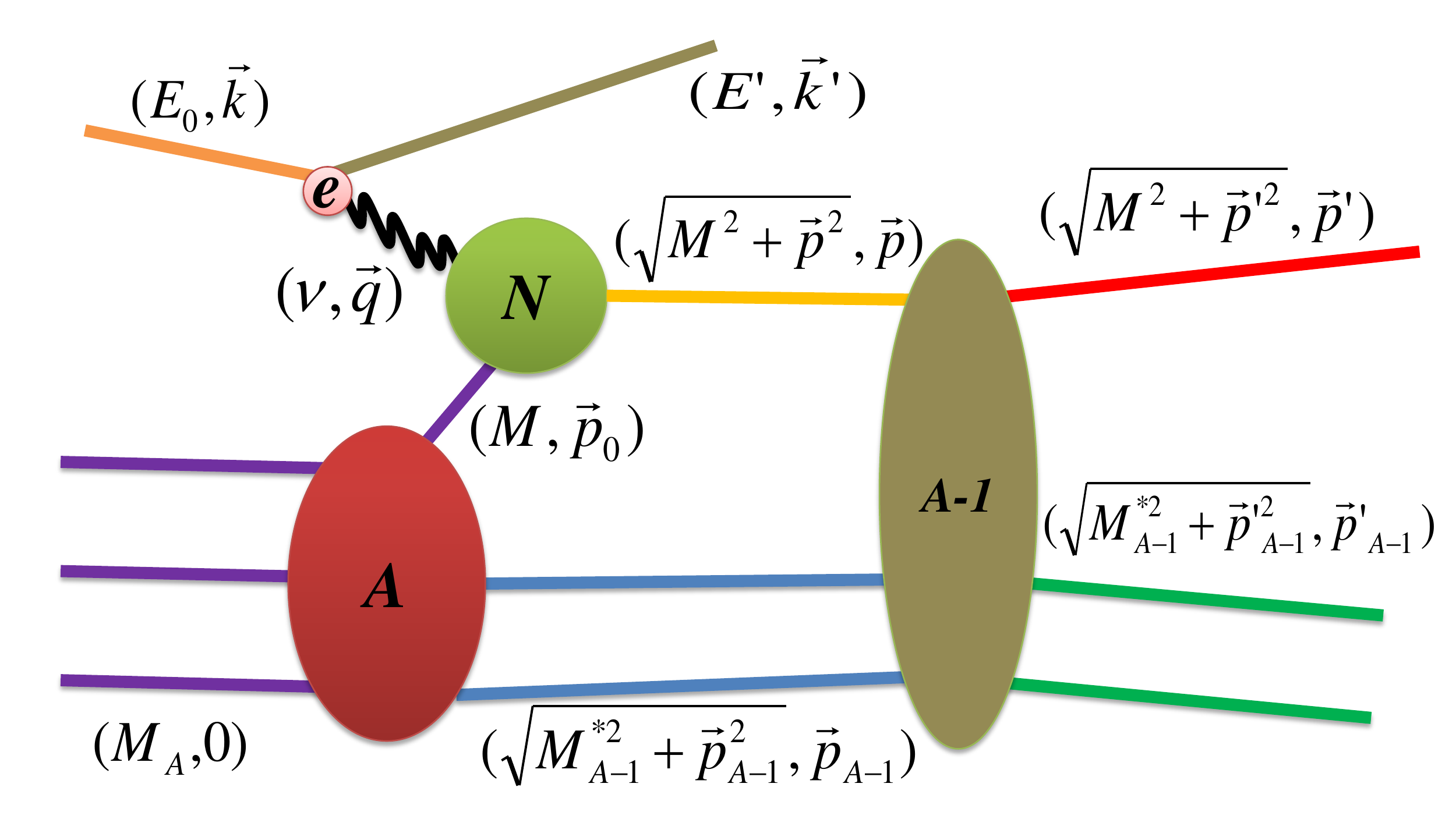}
    \caption[Final state interaction]{\footnotesize{General diagram of final state interaction. The struck nucleon is re-scattered by the $A-1$ system and its final momentum is modified.}}
    \label{fsi}
  \end{center}
\end{figure}
 The FSI effect in the y-scaling has been mentioned above. The FSI refers to the effect of the struck nucleon being re-scattered by the $A-1$ recoil system as it escapes from the nucleus. In the PWIA, nucleons in the nucleus are treated as individual constituents and the spatial resolution of the electron probe is approximately $1/q$. Hence, in the inclusive cross section measurement at large $\mathrm{Q^{2}}$, the FSI is expected to be small (Fig.~\ref{y_scaling_deut}), based on the fact that the interaction time between the virtual photon and the struck nucleon is significantly smaller than the one between the struck nucleon and the recoil system.

However, comparisons between the theoretical calculations and experimental results~\cite{PhysRevC.46.1045,PhysRevC.87.024606} suggest that violations of y-scaling for heavy targets exist. It indicates that the FSI still plays a significant role even at large $\mathrm{Q^{2}}$. Thus, further studies of the FSI contributions for QE scattering at large $\mathrm{Q^{2}}$ are still important.

%% file: intro/src.tex
\chapter{Short Range Correlations in Nuclei}
 As discussed in the previous chapter, the independent particle shell model (IPSM) has its great success in the description of the nuclear structure as nucleons occupying discrete energy states with their energies and momenta limited to the Fermi energy and the Fermi momentum, i.e. $\epsilon < \epsilon_{F}$ and $k<k_{F}$. However, this theory arises from a picture of a nucleons interacting only with the mean field potential generated by surrounding nucleons and it does not take into account the two-body (NN) interactions. Thus, IPSM is incapable of describing the short range properties of the NN interactions and fails to describe the structure of nuclear matter beyond the saturation density. Furthermore, measurements of the spectroscopic factors for the nuclear valence orbits via the proton knock-out experiments revealed that 30-40\% of the nuclear strength was missing compared to the predictions made by the mean field theory.
 
 Short range correlations (SRC) provide a successful explanation for the missing strength in the IPSM by considering the short distance behaviour of the NN interactions beyond the mean field, and reveal the importance of the high momentum components in the nucleon momentum distribution at $k>k_{F}$. At short distance ($\mathrm{\leq 1.0}$ fm), the attractive potential and repulsive force between nucleons excite the nucleons from their single shells and generate significant strength in the nuclear spectral functions at high momenta and energies. 
 
 To experimentally study the features of the SRC, one can use high-energy probes to directly measure the high-momentum nucleons and examine their correlations inside the nuclei. Early experiments at SLAC produced the first evidence of the SRC~\cite{SLAC_Measurement_PRC.48.2451} in inclusive electron-nucleus scattering. Recent experiments at JLab extended the study to map out the strength of the SRC in a wider range of nuclei and further examined the isospin dependence of the SRC~\cite{PhysRevLett.96.082501,PhysRevLett.99.072501,Subedi:2008zz,PhysRevLett.108.092502}. The new experiment in Hall-A at JLab, E08-014, was designed to study the structure of the SRC via inclusive electron-nucleus scattering and also to examine the isospin dependence of the SRC. 
 
 In this chapter, the features of the SRC will be discussed and the experimental techniques to explore the SRC will be briefly reviewed. 

\section{The Features of SRC}
\begin{figure}[!ht]
  \begin{center}
    \includegraphics[type=pdf,ext=.pdf,read=.pdf,width=0.80\linewidth]{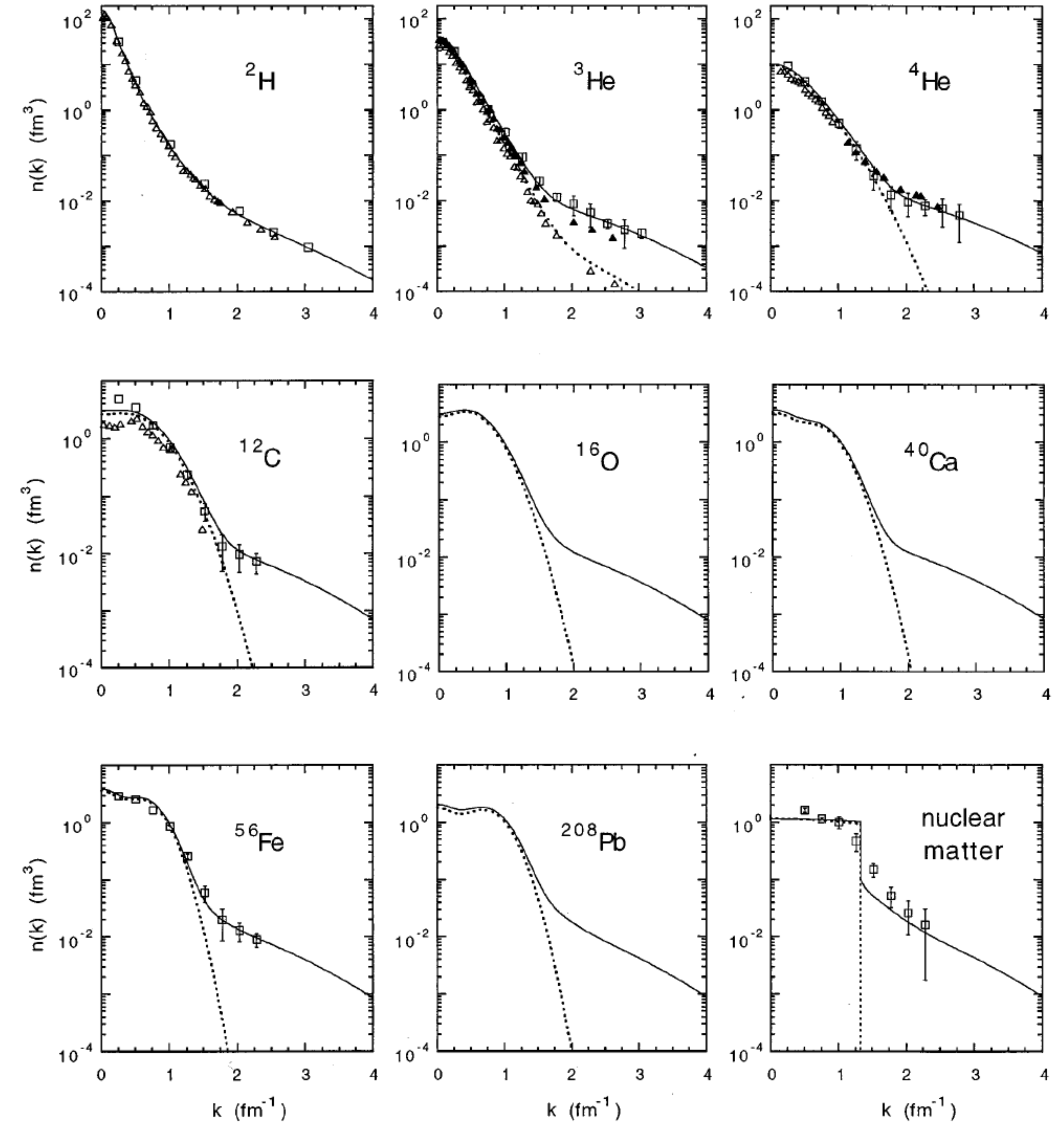}
    \caption[Nucleon momentum distribution]{\footnotesize{Nucleon momentum distribution for various nuclei~\cite{PhysRevC.53.1689}, where dotted lines are from a mean field calculation, solid lines are the calculations including the SRC. Symbols are from experimental data. The unit of the momentum is $\mathrm{fm^{-1}}$ ($\mathrm{1~fm^{-1}\simeq 197.3~MeV/c}$). Figures were taken from Ref.~\cite{PhysRevC.53.1689}.}}
    \label{mom_dis_ox}
  \end{center}
\end{figure}

To understand how the IPSM fails to predict the nuclear strength and how the SRC contributes to an understanding of the missing nuclear strength, one needs to examine the modern theoretical nucleon momentum distributions. While the prediction made by the mean field theory gives a rapid fall-off at momenta approaching $k_{F}$, experimental results from SLAC~\cite{PhysRevC.53.1689} and elsewhere show that each nucleus has a momentum tail falling off much slower at $k>k_{F}$. The tails for all nuclei are similar from deuteron to nuclear matter, as shown in Fig.~\ref{mom_dis_ox}. 

 These results strongly contrast with the mean field prediction, but can be easily understood if these high momentum tails are generated by the short-range part of the NN interactions. In Fig.~\ref{potential_well}, nucleons interact weakly at long distance where the mean field effect dominates. The attractive force at short distance is much stronger so that the nucleons can be bound together and their wave-functions largely overlap. When the nucleons become much closer, the strong repulsive hard-core dominates the NN interactions. These short distance components of the interactions generate highly correlated nucleons in the ground states with momenta significantly larger than the Fermi momentum ($k_{F}$), which is prohibited in the IPSM. However, the total momentum of these correlated nucleons is still very small and the nucleus remains in its ground state~\cite{src_john}.

\begin{figure}[!ht]
  \begin{center}
    \includegraphics[type=pdf,ext=.pdf,read=.pdf,width=0.80\linewidth]{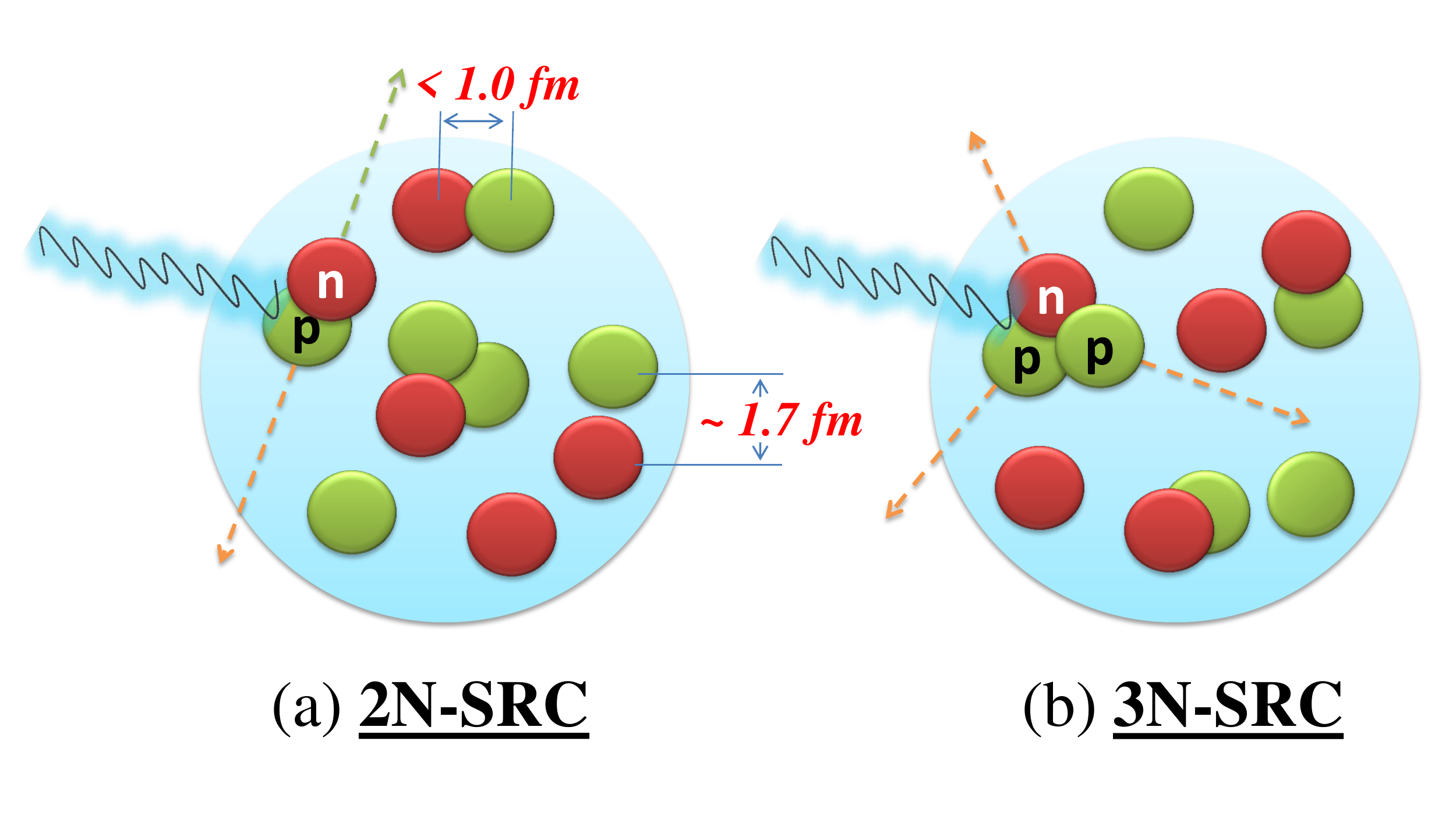}
    \caption[Diagrams of the 2N- and 3N-SRC]{\footnotesize{Diagram of the 2N- and 3N-SRC. On the left diagram the virtual phone breaks up the 2N-SRC pair in back-to-back ejection, and on the right diagram the break-up of  the 3N-SRC configuration results in correlated nucleon ejecting in different direction so their total momentum remains at zero.}}
    \label{2nsrc3nsrc}
  \end{center}
\end{figure}
 The asymptotic form of the momentum distribution can be broken down into several regions. At $k\leq k_{F}$, the strength is mainly contributed by the mean field potential. At large momentum, e.g. $k$ $>$ 300 MeV/c, the contribution of the mean field effect vanishes and the effect of two-nucleon short range correlation (2N-SRC) becomes dominant. These two nucleons in the configuration carry large and back-to-back momenta (Fig.~\ref{2nsrc3nsrc}.(a)), while their total center of mass momentum is modest. 

\begin{figure}[!ht]
  \begin{center}
    \includegraphics[type=pdf,ext=.pdf,read=.pdf,width=0.65\linewidth]{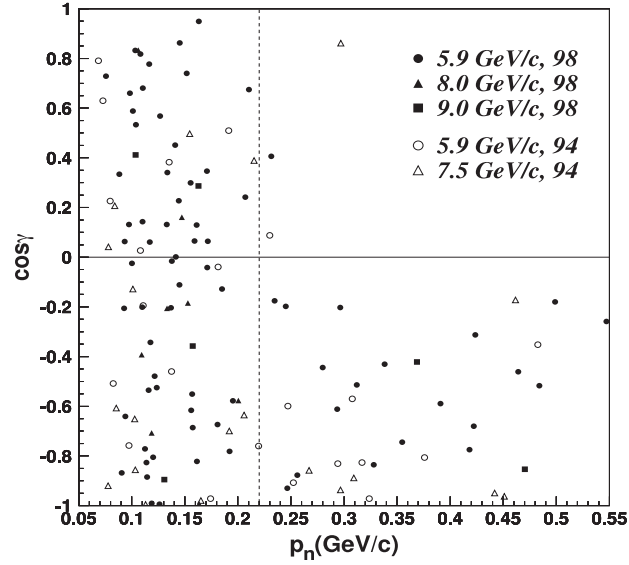}
    \caption[Correlation between the recoil momentum and direction of neutrons in the 2N-SRC]{\footnotesize{Correlation between the recoil momentum and direction of neutrons in the 2N-SRC where $cos\gamma$ is the cosine of the opening angle between the struck nucleon and the spectator proton. Data was from the E850 at BNL with the $\mathrm{^{12}C(p,p'pn)}$ reaction. The dash line indicates the location of the Fermi momentum($k{F}$). The recoil neutrons with $p_{n}>k_{F}$ show up in the opposite direction while recoil neutrons with momenta below the Fermi momentum have no angular correlation. Plot is adopted from Ref.~\cite{PhysRevLett.90.042301}.}}
    \label{bnl_cos}
  \end{center}
\end{figure}
\begin{figure}[!ht]
  \begin{center}
    \includegraphics[type=pdf,ext=.pdf,read=.pdf,width=0.65\linewidth]{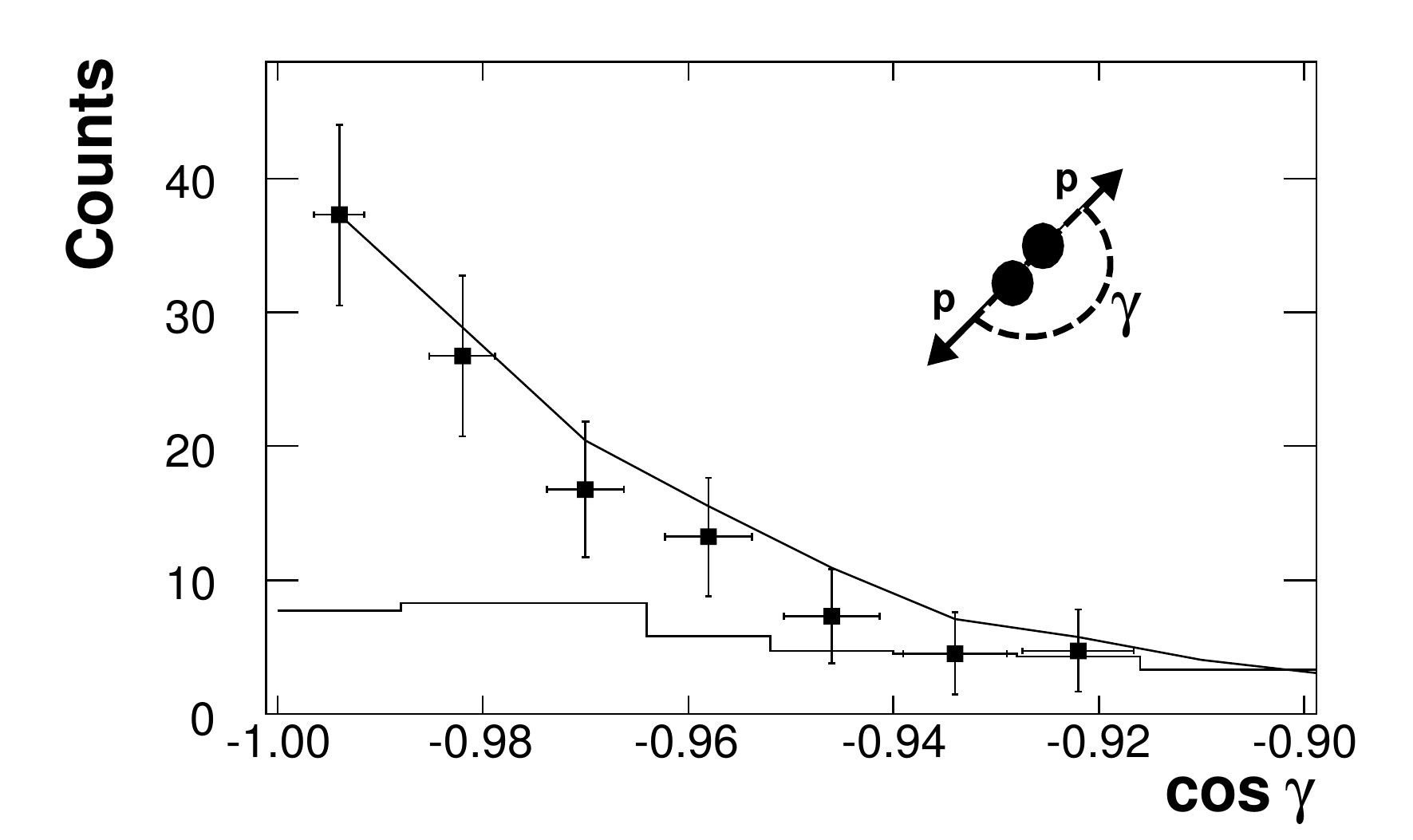}
    \caption[Angular correlation between nucleons in the 2N-SRC]{\footnotesize{Angular correlation between nucleons in the 2N-SRC, where the x-axis is the cosine of the opening angle between the struck nucleon with $k>k_{F}$ and the spectator nucleon in the $\mathrm{^{12}C(e,e'pp)}$ reaction. Figure is adopted from Ref.~\cite{PhysRevLett.99.072501}.}}
    \label{triple_src_cos}
  \end{center}
\end{figure}

  
 The break-up of a 2N-SRC pair must result in the strong angular correlation between these two nucleons. Such a correlation has been first observed in the E850 at Brookhaven National Lab (BNL)~\cite{PhysRevLett.90.042301} with the $\mathrm{^{12}C(p,p'pn)}$ reaction. The recoil neutron was measured in coincidence with the knocked-out proton. The opening angle (in $cos\gamma$) between the recoil neutron's momentum ($\vec{p_{n}}$) and the knock-out proton's initial momentum ($\vec{p_{i}}$) was correlated with $p_{n}$, as shown in Fig.~\ref{bnl_cos}. The result gives a uniform distribution of $cos\gamma$ for $p_{n}$ below the Fermi momentum. For $p_{n}>k_{F}$, only neutrons with $cos\gamma<0$ were observed, indicating that $\vec{p_{n}}$ is opposite to $\vec{p_{i}}$. The E01-015 in Hall-A at JLab~\cite{PhysRevLett.99.072501} used the $\mathrm{^{12}C(e,e'pn)}$ reaction and gave similar results. As shown in Fig.~\ref{triple_src_cos}, the opening angle between the knock-out proton and the recoil nucleon clearly peaks at $\mathrm{180^{o}}$ when the small center of mass motion is ignored. In addition, this experiment also discovered that these 2N-SRC pairs are dominated by $np$ configurations~\cite{Subedi:2008zz}.
 
 At much higher momentum (k $\geq$ 600 MeV/c), the dominance of $np$ pairs in the SRC should be broken down since the isospin-independent repulsive core begins to prevail, and the inclusion of three-nucleon short range correlations (3N-SRC) (Fig.~\ref{2nsrc3nsrc}.(b)) may become important~\cite{src_john}. Compared with the 2N-SRC, the 3N-SRC is more difficult to observe and its configuration is much more complicated. The observation of possible 3N-SRC configurations was one of the major goals for the E08-014 and will be discussed in more details in the next section.
 
  In the limit of extremely high $k$ where the nucleon kinetic energy is comparable with the excitation energy of nucleons, non-nucleonic degree-of-freedom may need to be considered.

\section{Probing SRC with Electron Scattering}
  The existence of high energy electron accelerators, e.g. NIKHEF, SLAC and JLab, provide a good opportunity to directly probe the NN short distance interactions. 
  
  The most complete experimental technique to study the 2N-SRC is the triple-coincidence measurement which not only detects the scattered electron but also maps out the momentum and angular correlations of the struck nucleon and the spectator nucleon. Due to the low counting rate, such a measurement requires a high luminosity electron beam and large acceptance spectrometers. Two experiments at JLab ~\cite{PhysRevLett.99.072501,Subedi:2008zz,E07006_pr} have successfully performed these types of measurements (Fig.~\ref{triple_src_cos}).

\begin{figure}[!ht]
  \begin{center}
    \includegraphics[type=pdf,ext=.pdf,read=.pdf,width=0.60\linewidth]{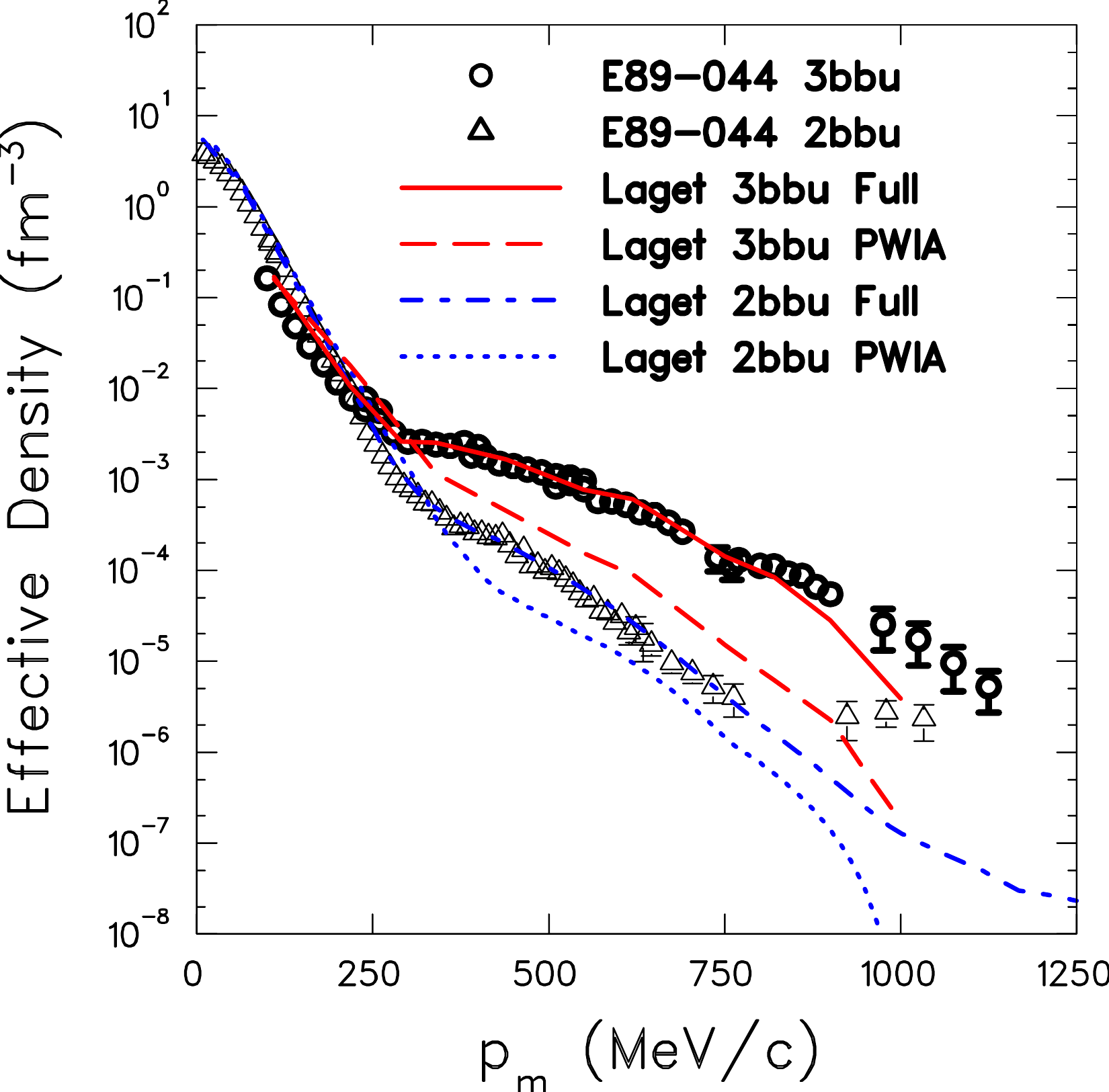}
    \caption[Proton effective momentum distribution in $\mathrm{^{3}He}$.]{\footnotesize{Proton effective momentum distribution in $\mathrm{^{3}He}$~\cite{PhysRevLett.94.082305}, where circles and triangles are experimental data in $\mathrm{^{3}He(e,e'p)pn}$ three-body break-up and $\mathrm{^{3}He(e,e'p)d}$ two-body break-up, respectively. Lines are theoretical calculations from~\cite{Laget200549}. Above the Fermi momentum (250 MeV), the momentum distribution of three-body break-up is much larger than the one of two-body break-up, and was explained as the combined contribution of FSI, MEC and SRC. Figure is adopted from Ref.~\cite{PhysRevLett.94.082305}}}
    \label{10yrSRC_fig3}
  \end{center}
\end{figure} 
\begin{figure}[!ht]
  \begin{center}
    \includegraphics[type=pdf,ext=.pdf,read=.pdf,width=0.80\linewidth]{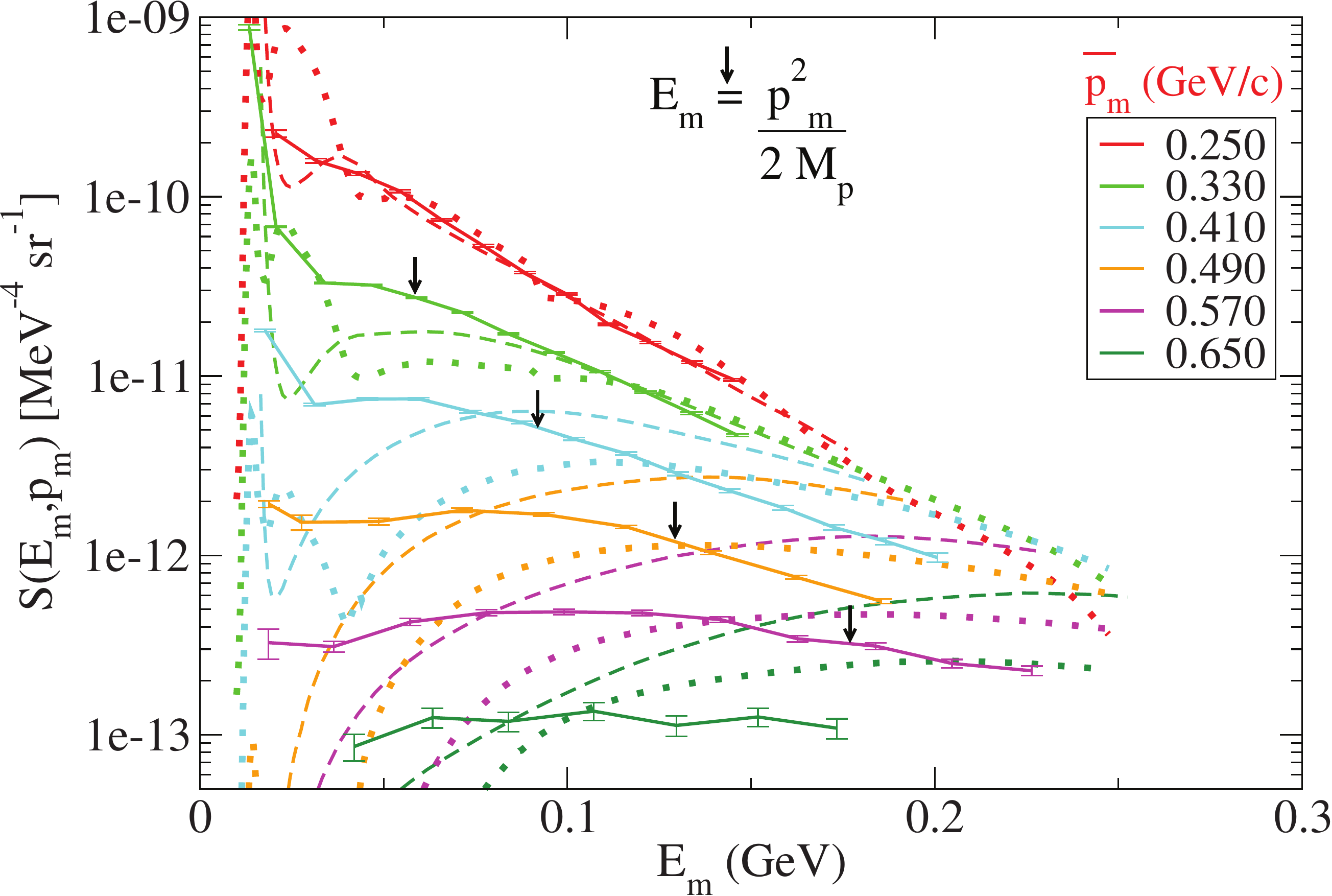}
    \caption[Measured distorted spectral functions for $\mathrm{^{12}C}$ from E97-006]{\footnotesize{Measured distorted spectral functions for $\mathrm{^{12}C}$ from E97-006. The data was taken in parallel kinematics to minimize the FSI. The results were compared with the theoretical predictions of the CBF~\cite{Benhar:1994hw} (dashed) and the Green's function approach~\cite{Muther:1995bk} (dotted). Figure is adopted from Ref.~\cite{PhysRevLett.93.182501}.}}
    \label{hallc_e97006}
  \end{center}
\end{figure} 
   The semi-exclusive measurement, which only detects the scattered electrons and the knocked-out protons~\cite{PhysRevLett.94.082305}, can fully probe the nuclear spectral functions (Eq.~\eqref{quasi_xs_spectral_function}) and study the dominance of the SRC and other competing processes in different kinematic regions, e.g. final-state interactions (FSI) and meson-exchange currents (MEC). NIKHEF studied the spectral function in $\mathrm{^{4}He(e,e'p)}$ data at $\mathrm{Q^{2}=0.34~GeV^{2}}$ but observed that the cross section was dominated by the long range NN interactions~\cite{vanLeeuwe1998593,Leeuwe20016}. The E89-044 in Hall-A at JLab~\cite{PhysRevLett.94.082305} measured the semi-exclusive cross sections of $\mathrm{{^{3}He}}$ at $\mathrm{Q^{2}=1.5~GeV^{2}}$. The experiment result in Fig.~\ref{10yrSRC_fig3} displays a great increase of strength in the high momentum tail which was explained as an interference between the SRC and the FSI~\cite{src_john}. The Hall-C experiment, E97-006~\cite{PhysRevLett.93.182501}, was designed to measure the spectral function at high initial energy and momentum through A(e,e'p). Special efforts were made to minimize the role of FSI by working in parallel kinematics. A sample of the results, as shown in Fig.~\ref{hallc_e97006}, agrees well with theoretical predictions.

   An inclusive measurement, where only the scattered electrons are measured, provides a powerful tool to study the feature of the SRC and probe the momentum distribution of the struck nucleon. Besides, the 3N-SRC can only be measured by the inclusive method in the current stage because of their much lower production rate and the complexity of the configurations.

\subsection{Kinematics}
  One of the key requirements to perform these experimental studies is to carefully determine the kinematic conditions and reactions, which can provide a clean measurement of high momentum nucleons and meanwhile suppress other processes such as FSI and MEC. It is also crucial to distinguish the processes of scattering off the high-momentum nucleons originally in the 2N- or 3N-SRC by varying the kinematic conditions.
    
 Although there are different kinds of reactions for probing the SRC as discussed above, these reactions share the common kinematic conditions to provide a clean study. Overall, the desire to instantly remove the nucleon from the SRC can be achieved by requiring sufficiently large energy and momentum transfer scales which significantly exceed the excitation scale of the nucleus~\cite{Frankfurt1981215,Frankfurt_misak}
\begin{equation}
  \nu >> V_{NN}, \qquad |\mathbf{q}| >> m_{N}/c,
  \label{src_condition1}
\end{equation}
where $V_{NN}$ is the characteristic potential of the NN interactions and $m_{N}$ is the nucleon mass. A reaction removing a nucleon from the nucleus under this condition allows the residual system to remain intact at the time when the nucleon in the SRC is removed, so that the properties of the SRC can be directly studied in this process.

The contribution of long range interactions, such as MEC, is suppressed by a factor of $Q^{-4}$ with respect to the production of the SRC, so they can be generally removed by requiring~\cite{M_Sargsian_JPG_29_2003}:
\begin{equation}
  Q^{2} > 1.0~GeV^{2} >> m_{meson}^{2}.
  \label{src_condition2}
\end{equation}
In this condition, intermediate state resonances still have sizeable contributions. For example, for $\mathrm{1~GeV^{2}<Q^{2}<4~GeV^{2}}$, $\gamma N\rightarrow \Delta$ transition is comparable with $\gamma N\rightarrow N$. Those resonance states are generally restricted to the region of $0<x_{bj}<1$, and their contributions can be suppressed by working at the region above the quasi-elastic (QE) peak:
\begin{equation}
  x_{bj} > 1.
  \label{src_condition3}
\end{equation}

\begin{figure}[!ht]
  \begin{center}
    \includegraphics[type=pdf,ext=.pdf,read=.pdf,width=0.60\linewidth]{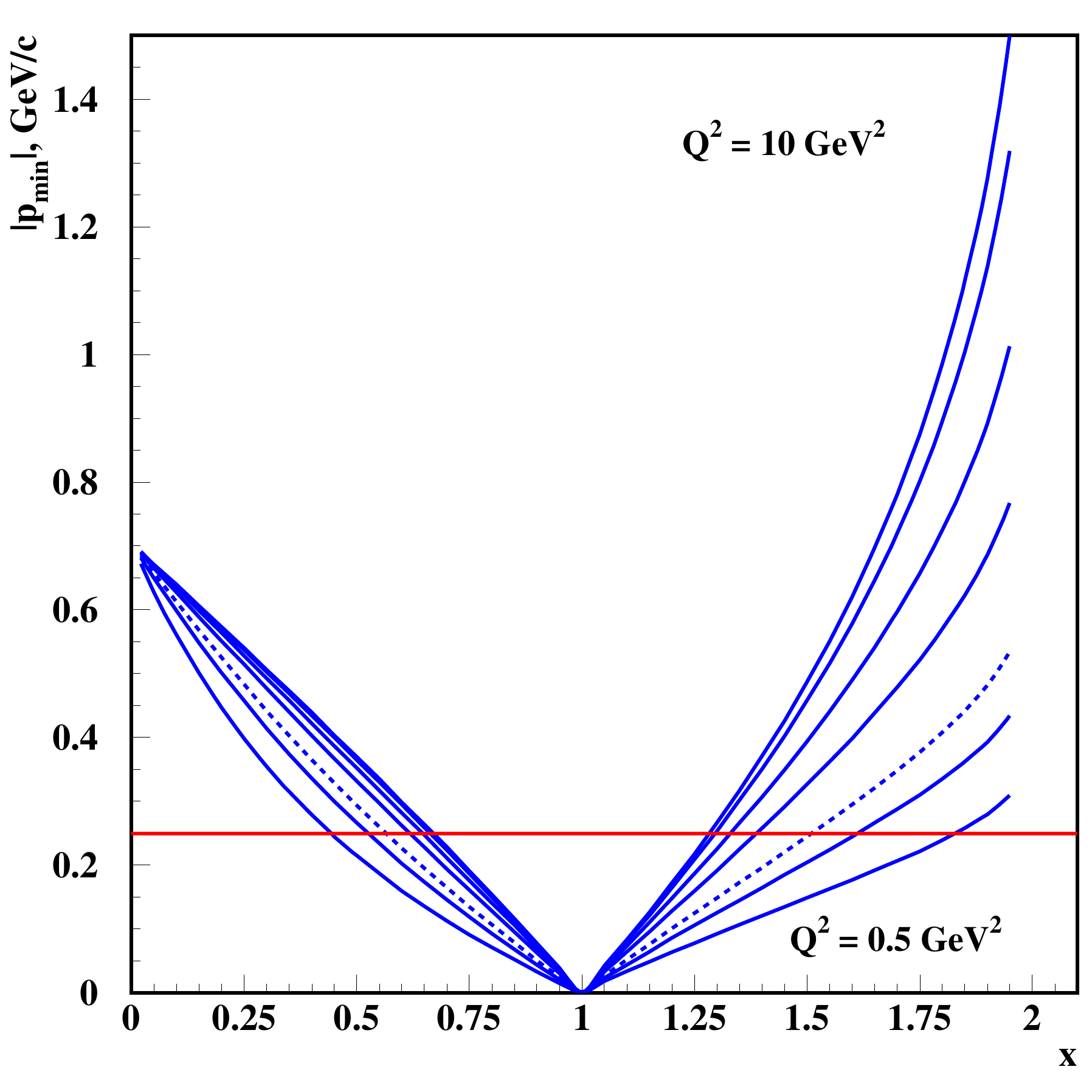}
    \caption[Minimum momentum of the struck nucleon as function of $x_{bj}$ and $\mathrm{Q^{2}}$]{\footnotesize{Minimum momentum of the struck nucleon as function of $x_{bj}$ and $\mathrm{Q^{2}}$ for scattering off a nucleon from the nucleus, where the values of $\mathrm{Q^{2}}$ from bottom to top are 0.5, 1.5, 3.0 and 10 $GeV^{2}$, respectively~\cite{Frankfurt_misak, src_john}. The red line sets the value of the Fermi momentum ($k_{F}$). Figure is adopted from Ref.~\cite{Frankfurt_misak}.}}
    \label{kin_cond_q2_xbj}
  \end{center}
\end{figure}

 The combination of kinematic conditions (Eq.~\eqref{src_condition2} and Eq.~\eqref{src_condition3}) is demonstrated in Fig.~\ref{kin_cond_q2_xbj}. The plot is based on the energy and momentum conservation of the struct nucleon with different initial momenta, and the different lines represent different $\mathrm{Q^{2}}$. For scattering off nucleons in the nucleus at low $\mathrm{Q^{2}}$ (e.g. $\mathrm{0.5~GeV^{2}}$), it requires one to measure the struck nucleons at very high $x_{bj}$ (e.g. 1.8) to reach the minimum momentum requirement ($k>k_{F}$) and suppress the mean field contribution. When the $\mathrm{Q^{2}}$ is sufficiently high (e.g. $\mathrm{10~GeV^{2}}$), one can detect the struck nucleons with $k>k_{F}$ at relatively low $x_{bj}$ (e.g. 1.3). Those kinematic conditions enable a clean measurement of high momentum nucleons from the SRC.

 To separate the 2N- and 3N-SRC during the scattering process, one can study the light-cone (LC) variable in the relativistic regime. A relativistic projectile moving along the z-direction probes the LC wave-function of the nucleus, $\psi_{A}(\alpha_{1},k_{1,t},...,\alpha_{i},k_{i,t},...,\alpha_{A},k_{A,t})$, where the LC variable is defined as~\cite{Frankfurt_misak}:
\begin{equation}
  \alpha_{i} = A\left(\frac{E_{i}-p_{i,z}}{E_{A}-p_{A,z}}\right)=A\left(\frac{E_{i}^{lab}-p_{i,z}^{lab}}{M_{A}}\right),
\end{equation}  
where $E_{i}$ and $p_{i,z}$ ($E_{A}$ and $p_{A,z}$) are the initial energy and longitudinal momentum of the constituent nucleon (the target nucleus $A$), respectively. $\alpha_{i}$ is invariant under Lorentz boosts in the z-direction. In the rest frame of the nucleus, $E_{A}-p_{A,z}=M_{A}$, where $M_{A}$ is the nuclear mass. 

 Similar to the definition of $x_{bj}$ in inelastic scattering (Eq.~\ref{xbj_define}), $\alpha_{i}$ denotes the LC fraction of the nucleus momentum carried by the nucleon, hence $\sum_{i}^{A}\alpha_{i}=A$. While $\alpha_{i}\leq 1$ limits the momentum fraction of the nucleon carried by the quark, to have $\alpha_{i}>1$ requires at least two nucleons involved in the scattering. Furthermore, three nucleons are required to share their momentum to have $\alpha_{i}>2$. Consequently, $\alpha_{i}$ becomes an ideal variable to distinguish the 2N-SRC from the 3N-SRC. Considering the energy and momentum conservation law for the nucleon knock-out with a virtual photon from the nucleus, one can rewrite the LC variable as~\cite{Frankfurt_misak}:
\begin{equation}
  \alpha_{i}=x_{bj}\left(1+\frac{2p_{i,z}}{\nu+|\mathbf{q}|}\right)+\frac{W_{N}^{2}-m_{i}^{2}}{2m_{i}\nu},
  \label{alpha_xbj}
\end{equation}
where $\nu$ and $\mathbf{q}$ is the energy and momentum transfer of the virtual photon, respectively, and $W_{N}^{2}=(\mathbf{p_{i}}+\mathbf{q})^{2}$. For the QE process, $W_{N}\simeq m_{i}$ yields a simple connection between $\alpha_{i}$ and $x_{bj}$. At sufficiently large $\mathrm{Q^{2}}$, $\alpha_{i}$ is usually replaced by $x_{bj}$: 
\begin{equation}
  \alpha_{i} (Q_{2}\rightarrow \infty)\rightarrow x_{bj}. 
\end{equation}
However, these two variables are different for $\mathrm{Q^{2}}$ values in few $\mathrm{GeV^{2}}$ range. One needs to examine the different scaling behaviours of the SRC as a function of $x_{bj}$ and $\alpha_{i}$ at the low $\mathrm{Q^{2}}$ region. 	

\subsection{Inclusive Measurements}
 The inclusive cross section measurement of $A(e,e')$ reaction in QE region was the first method used to isolate the SRC and currently is the only reaction to study the 3N-SRC. The cross section for $x>1.3$ and $\mathrm{Q^{2}>1~GeV^{2}}$ can be written as~\cite{Frankfurt1988235}:
\begin{eqnarray}
  \sigma_{A}(x_{bj},Q^{2}) &=& \sum_{j=2}^{A}\frac{A}{j} a_{j}(A) \sigma_{j}(x_{bj},Q^{2}) \nonumber \\
  &=& \frac{A}{2}a_{2}(A)\sigma_{2}(x_{bj},Q^{2})+\frac{A}{3}a_{3}(A)\sigma_{3}(x_{bj},Q^{2})+...,
  \label{xs_src_inclusive}
\end{eqnarray}
where $\sigma_{j}$ is the cross section for scattering off a $j$-nucleon correlation and $a_{j}(A)$ denotes the probability of finding the number of j-nucleon correlations in the nucleus. The first two terms represent the contributions from the 2N- and 3N-SRC. The 2N-SRC is expected to dominate at $1.3<x_{bj}<2$, and it generally vanishes at $x_{bj}>2$ where the 3N-SRC becomes more important.
 \begin{figure}[!ht]
  \begin{center}
    \includegraphics[type=pdf,ext=.pdf,read=.pdf,width=0.60\linewidth]{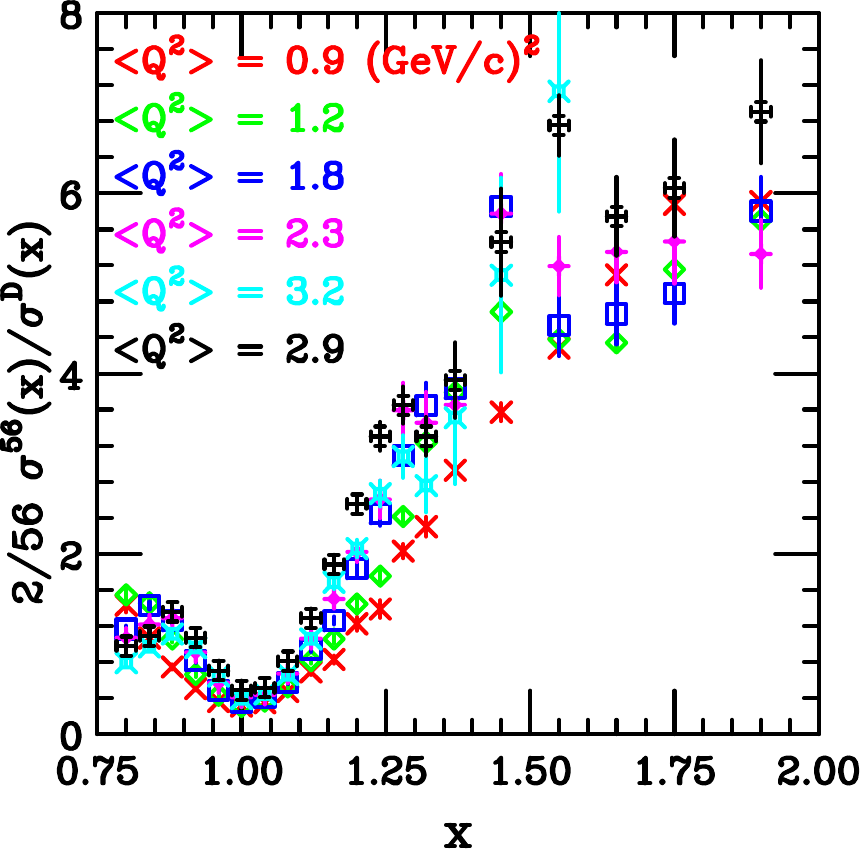}
    \caption[Evidence of the 2N-SRC from SLAC]{\footnotesize{Evidence of the 2N-SRC from SLAC~\cite{SLAC_Measurement_PRC.48.2451}. Each plot corresponds to the different $\mathrm{Q^{2}}$. Y-axis is the cross section ratio of $\mathrm{^{56}Fe}$ to $^{2}H$ for $\mathrm{Q^{2}=0.9-3.2~GeV^{2}}$ and x-axis is $x_{bj}$. The 2N-SRC plateau shows up at $x_{bj}>1.3$ and becomes more clear at larger $\mathrm{Q^{2}}$. Figure from Ref.~\cite{SLAC_Measurement_PRC.48.2451}.}}
    \label{SLAC_2NSRC_xbj}
  \end{center}
\end{figure}

 In the region of the 2N-SRC (3N-SRC), one expects that electron scattering off a heavy nucleus $A$ is identical to electron scattering off the deuteron-like ($\mathrm{^{3}He}$-like) configuration inside the nucleus. Hence, the inclusive cross section of the heavy nucleus $A$ is expected to scale to the one of the deuteron ($\mathrm{^{3}He}$). In the region of $1.3<x_{bj}<2.0$ where the 2N-SRC dominates, the scaling factor, $a_{2}(A)$, is given by the cross section ratio:
\begin{equation}
  a_{2}(A) = \frac{2}{A}\frac{\sigma_{A}(x_{bj},Q^{2})}{\sigma_{D}(x_{bj},Q^{2})},
  \label{src_a2}
\end{equation}
where $\sigma_{D}(x_{bj},Q^{2})$ is the inclusive cross section of electron scattering of the deuteron. Implicated in Eq.~\ref{src_a2}, $a_{2}$ in the 2N-SRC region is independent of $x_{bj}$ and $\mathrm{Q^{2}}$, and only depends on A. The value of the ratio in the scaling plateau directly gives the relative number of the 2N-SRC pairs in the nucleus compared to the deuteron.

\begin{figure}[!ht]
  \begin{center}
    \includegraphics[type=pdf,ext=.pdf,read=.pdf,width=0.70\linewidth]{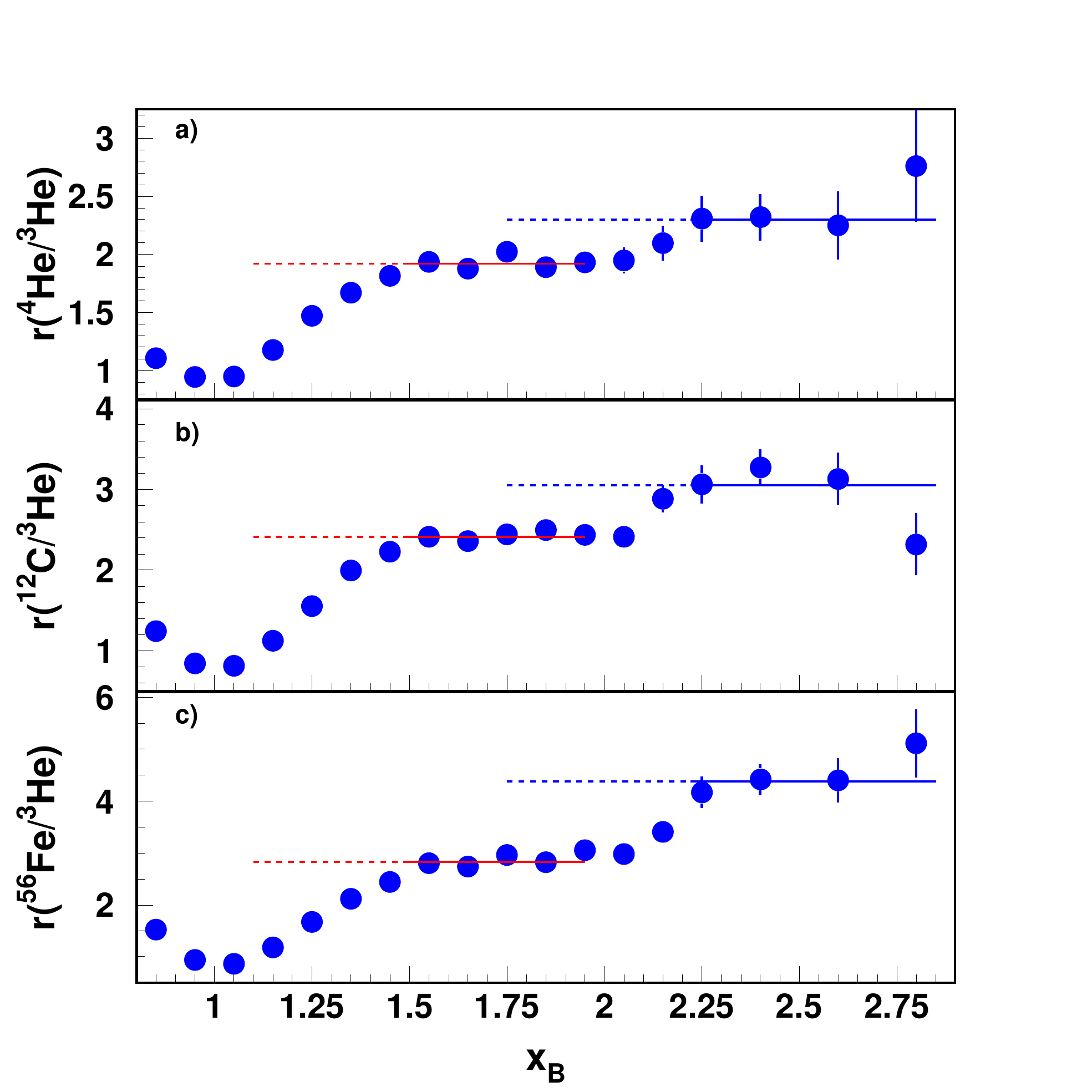}
    \caption[2N- and 3N-SRC results from CLAS in Hall-B]{\footnotesize{2N- and 3N-SRC results from CLAS Hall-B~\cite{PhysRevLett.96.082501}. The top, middle and bottom plots give the cross section ratios of $\mathrm{^{4}He}$, $\mathrm{^{12}C}$ and $\mathrm{^{56}Fe}$ to $\mathrm{^{3}He}$, respectively. In each plot, both the 2N-SRC plateau (in $1.5<x_{bj}<2$) and the 3N-SRC plateau (in $x_{bj}>2$) can be observed. Figure are reproduced from the original plots in Ref.~\cite{PhysRevLett.96.082501}.}}
    \label{CLAS_2NSRC_3NSRC}
  \end{center}
\end{figure}

\begin{figure}[!ht]
  \begin{center}
    \includegraphics[type=pdf,ext=.pdf,read=.pdf,angle=270.,width=0.80\linewidth]{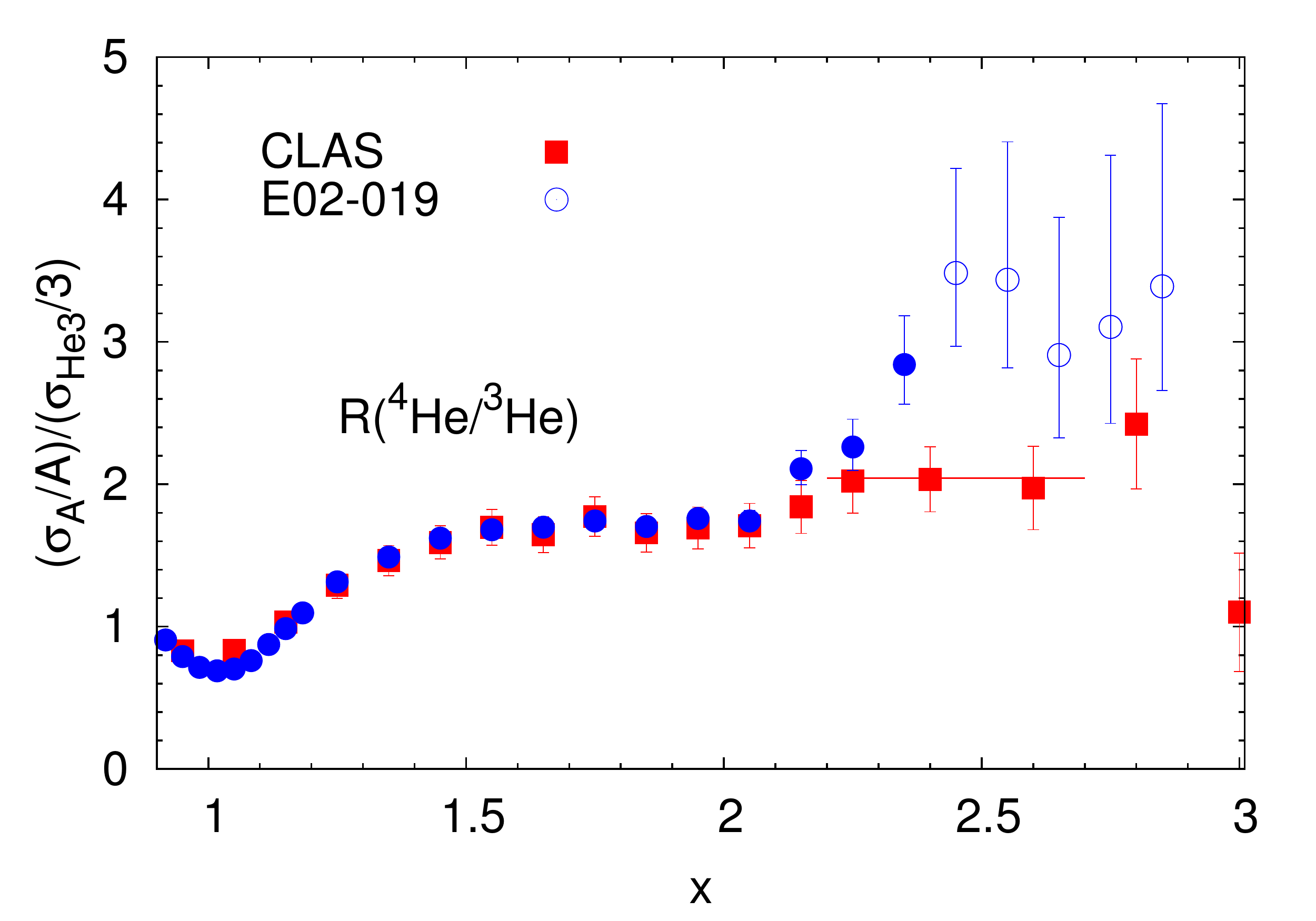}
    \caption[2N- and 3N-SRC in $\mathrm{^{4}He/^{3}He}$ from the E02-019]{\footnotesize{2N- and 3N-SRC in $\mathrm{^{4}He/^{3}He}$ from the E02-019 in Hall-C compared with the result from Hall-B, where the blue dots are the E02-019 data and red dots are the CLAS data. Figure is adopted from Ref.~\cite{PhysRevLett.108.092502}.}}
    \label{E02019_2NSRC_3NSRC}
  \end{center}
\end{figure}

Fig.~\ref{SLAC_2NSRC_xbj} shows the results from SLAC~\cite{SLAC_Measurement_PRC.48.2451}, which for the first time observed such a plateau with the cross section ratio of $\mathrm{^{56}Fe}$ to $\mathrm{^{2}H}$ at $x_{bj}>1.5$ and $\mathrm{Q^{2}=0.9-3.2~GeV^{2}}$. Other targets, $\mathrm{^{4}He}$, $\mathrm{^{27}Al}$ and $\mathrm{^{64}Cu}$ were also studied and all showed clear evidences of the 2N-SRC plateaus. However, Fig.~\ref{SLAC_2NSRC_xbj} also suggests a $\mathrm{Q^{2}}$ dependence of $a_{2}$. The statistics were limited and the deuteron data was taken at different kinematics, so the result was extracted with nontrivial extrapolations. Recent Jefferson Lab results from the E89-008~\cite{Arrington:2003tw, Arrington:2006pn} and the E02-019~\cite{PhysRevLett.108.092502} in Hall-C, and Large Acceptance Spectrometer (CLAS)~\cite{PhysRevLett.96.082501} in Hall-B extracted the values of $a_{2}$ from various nuclei with higher statistics and better resolution, and their results indicate a sound agreement in the 2N-SRC region (Fig.~\ref{CLAS_2NSRC_3NSRC} and Fig.~\ref{E02019_2NSRC_3NSRC}). 

Similarly, one can study the scaling factor, $a_{3}(A)$, in the 3N-SRC at $2<x_{bj}<3$ with the cross section ratio of the heavy nucleus to $\mathrm{^{3}He}$:
\begin{equation}
  a_{3}(A) = K\cdot\frac{3}{A}\frac{\sigma_{A}(x_{bj},Q^{2})}{\sigma_{^{3}He}(x_{bj},Q^{2})},
  \label{src_a3}
\end{equation}
which denotes the number of $\mathrm{^{3}He}$-like 3N-SRC configuration in the nucleus. $K$ corrects for the difference of the electron-proton and electron-neutron cross sections:
\begin{equation}
  K = \frac{\sigma_{ep}+\sigma_{en}}{Z\sigma_{ep}+(A-Z)\sigma_{en}}.
\end{equation}

In the CLAS data, the kinematics was for the first time extended into the region of 3N-SRC and a plateau at $x_{bj}>2.3$ was observed. The E02-019 data, however, yields a different result in this region. From Fig.~\ref{E02019_2NSRC_3NSRC}, the cross section ratio of $\mathrm{^{4}He/^{3}He}$ reaches the scaling region at $x_{bj}>2.5$, slightly later than the CLAS result, and the scaling plateau can not be clearly identified because of the large error bars. An explanation of the discrepancy is not straightforward since these two experiments ran at very different $\mathrm{Q^{2}}$ ranges ($\mathrm{Q^{2}\sim 1.6~GeV^{2}}$ for CLAS and $\mathrm{Q^{2}\sim 2.7~GeV^{2}}$ for the E02-019). It is unclear whether these measurements isolated the 3N-SRC contributions or not. The E08-014 in Hall-A focuses on studying the scaling of 3N-SRC at $x_{bj}>2$ with much better accuracy, and the new preliminary results will be presented later. 

\begin{figure}[!h]
  \begin{center}
        \includegraphics[type=pdf,ext=.pdf,read=.pdf,width=0.6\linewidth]{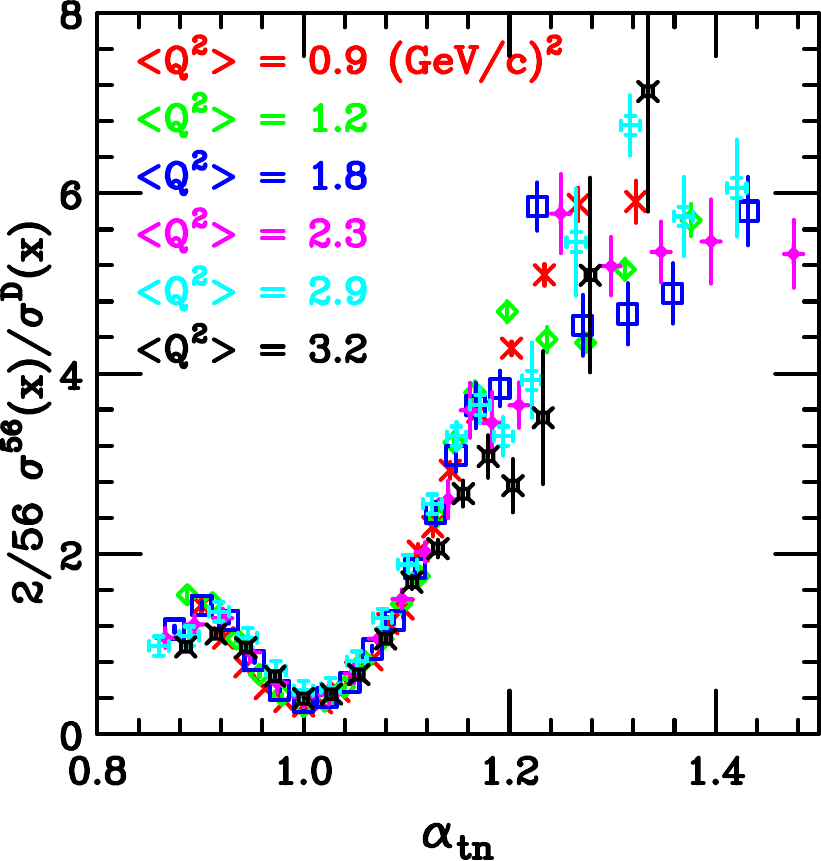}
        \caption[Ratio of $\mathrm{^{56}Fe/^{2}H}$ as a function of $\alpha_{2N}$ from SLAC]{\footnotesize{Ratio of $\mathrm{^{56}Fe/^{2}H}$ as a function of $\alpha_{2N}$ for different $\mathrm{Q^{2}}$ values~\cite{SLAC_Measurement_PRC.48.2451}. $\alpha_{2N}$ (labelled as $\alpha_{tn}$ in this plot) is an approximation of the LC variable by assuming the total momentum of the nucleons in the 2N-SRC is zero. Compared with Fig.~\ref{SLAC_2NSRC_xbj}, $\alpha_{2N}$ provides better scaling behaviour and indicates less $\mathrm{Q^{2}}$ dependence.}}
    \label{SLAC_2NSRC_alpha}
  \end{center}
\end{figure}
The other important element in the study of the SRC in inclusive measurements is the difference between two scaling variables, $x_{bj}$ and $\alpha_{i}$. From Eq.~\eqref{alpha_xbj}, the LC variable $\alpha_{i}$ is approximately equal to $x_{bj}$ at large $\mathrm{Q^{2}}$. At the few $\mathrm{GeV^{2}}$ level, the approximation is invalid and the difference in the scaling behaviour of the SRC ratios as a function of $x_{bj}$ and $\alpha_{i}$ must be carefully examined. 

Although $\alpha_{i}$ can not be reconstructed in inclusive scattering, one can assume that in the PWIA the virtual photon interacts with the nucleon in a 2N-SRC pair at rest. This assumption leads to a new expression of the LC variable specifically for the 2N-SRC:
\begin{equation}
  \alpha_{2N} = 2-\frac{q_{-}+2m}{2m}\frac{\sqrt{W^{2}-4m^{2}}+W}{W},  
\end{equation}
where $q_{-}$ is the initial longitudinal momentum of the struck nucleon and $W^{2}=4m_{N}^{2}+4m_{N}\nu-Q^{2}$. 

The analysis of SLAC data~\cite{SLAC_Measurement_PRC.48.2451} reveals that compared with $x_{bj}$ in Fig.~\ref{SLAC_2NSRC_xbj}, $\alpha_{2N}$ can better isolate the 2N-SRC (Fig.~\ref{SLAC_2NSRC_alpha}) and allow one to examine the transition region from the 2N-SRC to the 3N-SRC. A more general expression for all $\alpha_{i}$~\cite{e08014_pr} in the inclusive measurement can be obtained from:
\begin{equation}
  q_{-}\cdot\alpha_{jN}m_{N}+q_{+}\cdot\left(M_{A}-\frac{M_{r}^{2}}{m_{N}(j-\alpha_{jN})}\right)=m_{N}^{2},
\end{equation}
where $j=2,3,...$. $q_{+}$ is the initial transverse momentum of the struck nucleon and $M_{r}$ is the mass of the residual system. Taking $j=3$, one can solve for $\alpha_{3N}$, but the exact expression depends on the value of $M_{r}$ which is difficult to specify since the 3N-SRC is a more complicated configuration.

\begin{figure}[h]
	  \begin{center}
	    \includegraphics[type=pdf,ext=.pdf,read=.pdf,width=0.80\linewidth]{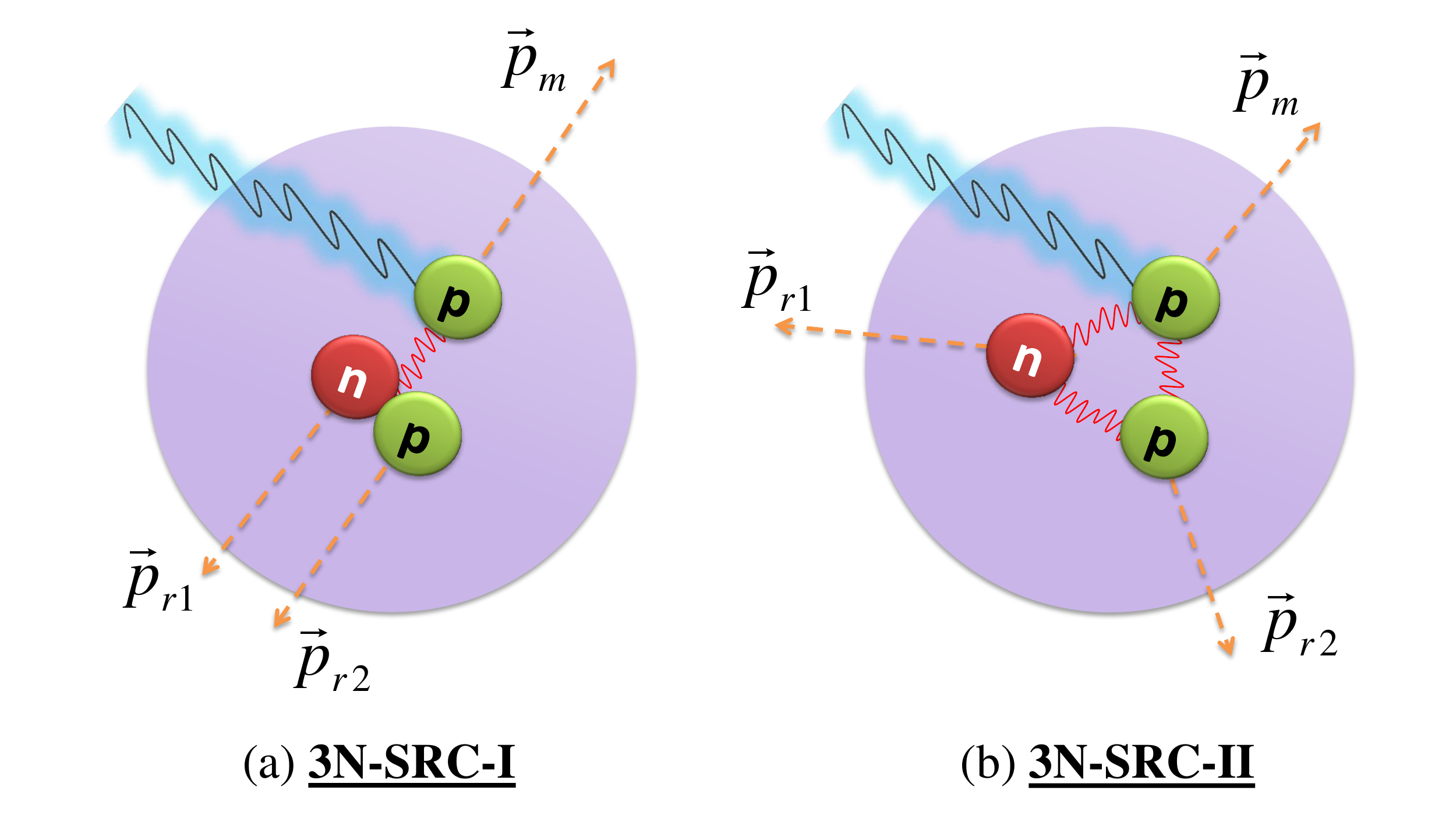}
	    \caption[Two types of 3N-SRC configuration]{\footnotesize{Two types of 3N-SRC configuration~\cite{Frankfurt_misak}.}}
	  \label{3nsrc_two_types}
	  \end{center}
\end{figure}

 In general, there are two types of 3N-SRC, as shown in Fig.~\ref{3nsrc_two_types}. In the first type, namely 3N-SRC-I, the total initial momentum of two nucleons is equal to that of the struck nucleon but in the opposite direction. This configuration is similar to the 2N-SRC but involving in three nucleons. The second type, 3N-SRC-II, refers to the configuration of three nucleons carrying momenta all exceeding the Fermi momentum in three different directions. Although the 3N-SRC-II configuration is easier to be observed experimentally via semi-exclusive reactions, it is less likely to occur than the 3N-SRC-I one since it requires a larger separation energy~\cite{Frankfurt_misak}.
 
 The situation of the 3N-SRC-I, where $M_{r}=2m_{N}$, gives:
\begin{equation}
  \alpha_{3N-I} = \frac{3}{2}+\frac{1}{2}[\sqrt{ (3+b_{1})^{2}-b_{2}}-b_{1}],
\end{equation} 
where,
\begin{equation}
  b_{1} = \frac{q_{+}}{q_{-}} \frac{M_{A}}{m_{N}}-\frac{m_{N}}{q_{-}},\qquad b_{2} = 16 \frac{q_{+}}{q_{-}}.
\end{equation} 
The value of $M_{r}$ becomes larger for the 3N-SRC-II, and its LC variable, $\alpha_{3N-II}$, has a more complicated form. Examining the scaling as a function of $\alpha_{3N-I}$ and $\alpha_{3N-II}$ by varying the value of $M_{r}$ provides a sensitive probe to the detailed structure of 3N-SRC~\cite{e08014_pr}.
\subsection{Isospin Dependence}
\begin{figure}[!ht]
  \begin{center}
    \includegraphics[type=pdf,angle=270,ext=.pdf,read=.pdf,width=0.80\linewidth]{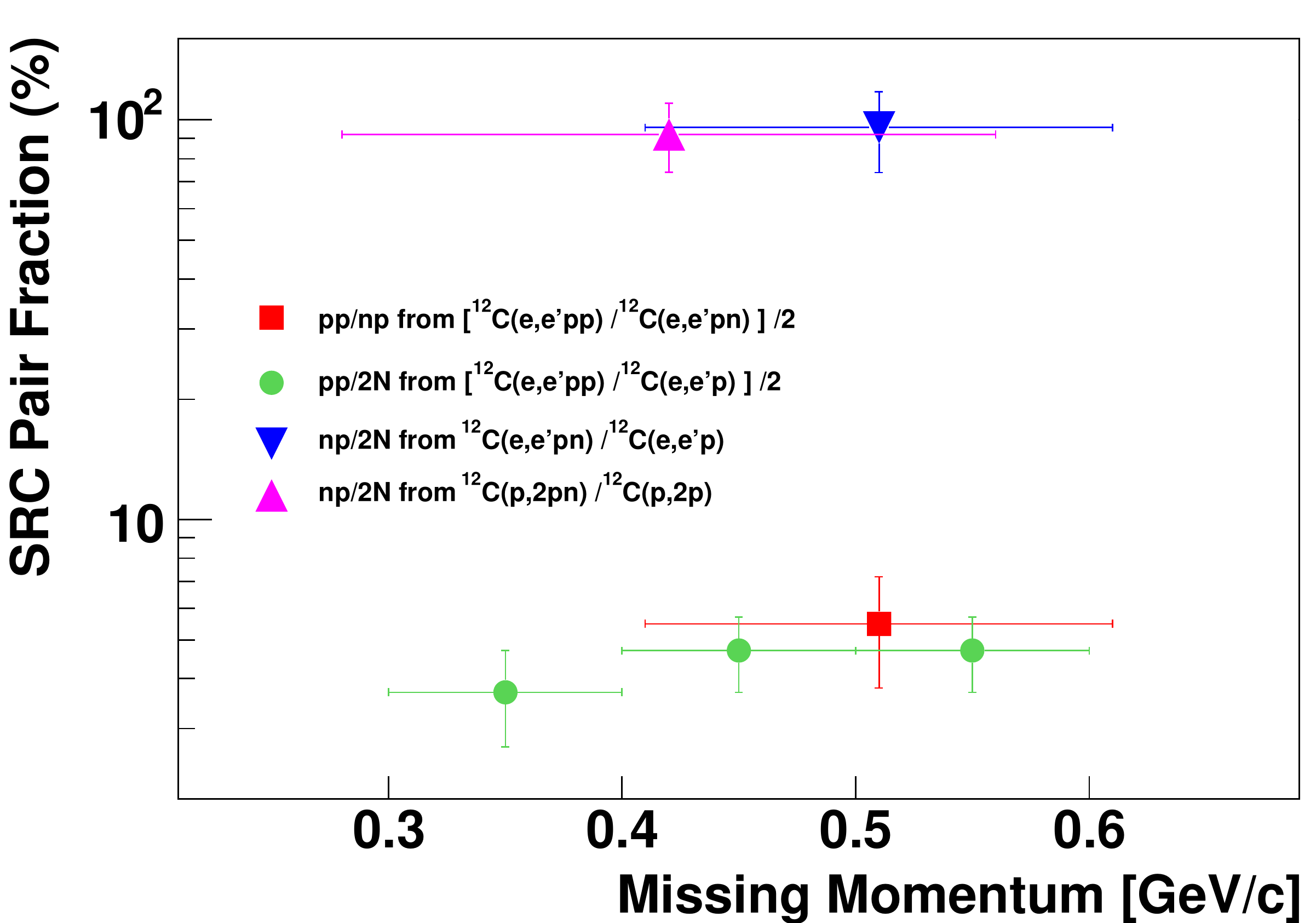}
    \caption[The fraction of $np$ pairs to $pp$ pairs in the 2N-SRC]{\footnotesize{The fraction of $np$ pairs to $pp$ pairs in the 2N-SRC in carbon from the triple-coincidence experiment in Hall-A. Figure is adopted from Ref.~\cite{Subedi:2008zz}.}}
    \label{triple_src_np}
  \end{center}
\end{figure} 
  In the early analysis of inclusive measurements where the struck nucleon was not identified,  isospin-independence was assumed and the ratio of neutrons to protons in the SRC configurations was treated to be equal to the $N/Z$ ratio. 

  Triple-coincidence experiments at JLab~\cite{PhysRevLett.90.042301,PhysRevLett.99.072501,Subedi:2008zz} studied the isospin effect by measuring the ratio of $pn$ and $pp$ in the 2N-SRC with the $\mathrm{^{12}C(e,e'pp)}$ and $\mathrm{^{12}C(e,e'pn)}$ reactions. As shown in Fig.~\ref{triple_src_np}, the result~\cite{Subedi:2008zz} revealed that the ratio of $np/pp$ pairs is around $\mathrm{18\pm 5}$. The dominance of $np$ pairs indicates that the assumption of the isospin independence is invalid.  
  
  Numerical studies~\cite{PhysRevC.72.054310} suggest that because of the tensor interaction, the 2N-SRC pairs should be mainly in iso-singlet ($np$ with T=0) states. The iso-triplet ($pp$,$np$ and $nn$ with $T=1$) pairs experience a much smaller attractive component in the NN potential till they are close enough to interact via the repulsive core.
 
\begin{figure}[!ht]
  \begin{center}
    \includegraphics[type=pdf,ext=.pdf,read=.pdf,width=0.90\linewidth]{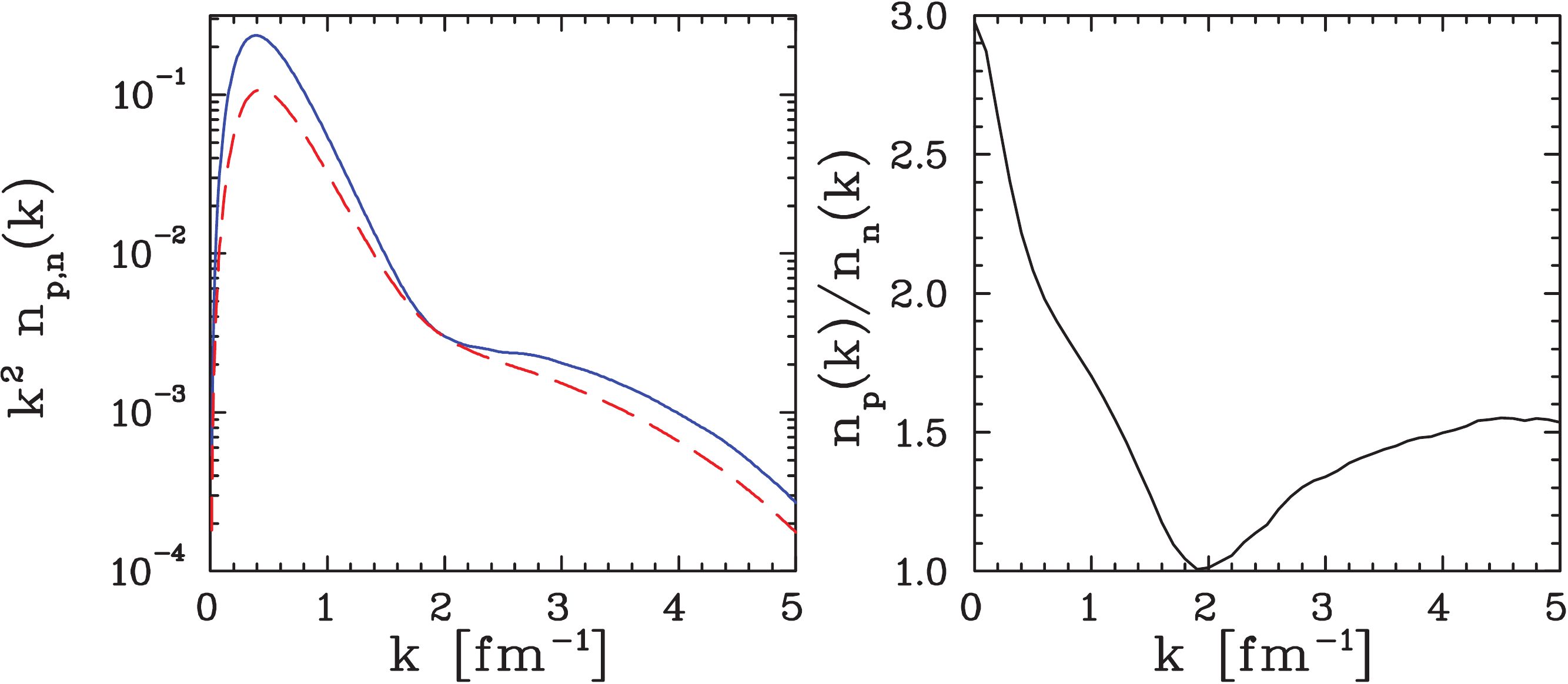}
    \caption[Momentum distribution for proton and neutron and their ratio]{\footnotesize{Left: Momentum distribution for proton (solid) and neutron (dashed) in $\mathrm{^{3}He}$; Right: Ratio of proton to neutron momentum distribution. Plots were originally from Ref.~\cite{Pieper_Wiringa}.}}
    \label{mom_dis_np}
  \end{center}
\end{figure}  
 
\begin{figure}[!ht]
  \begin{center}
    \includegraphics[type=pdf,ext=.pdf,read=.pdf,width=0.60\linewidth]{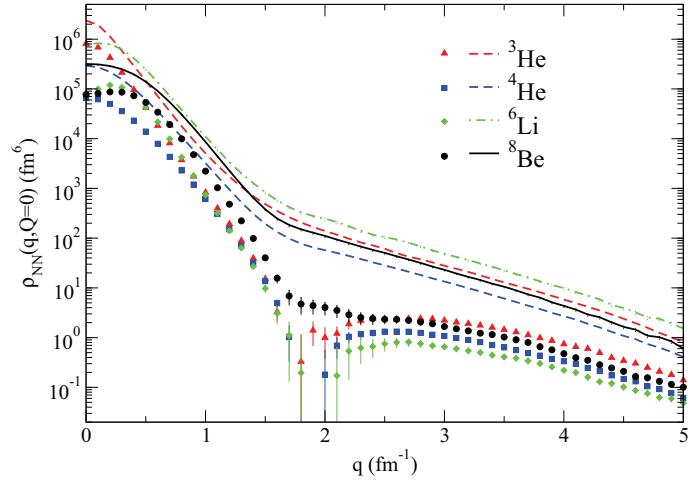}
    \caption[Isospin effect in momentum distribution]{\footnotesize{Isospin effect in momentum distribution, where lines represent the momentum distribution in $np$ and dots represent the momentum distribution in $pp$. The unit of the momentum is $\mathrm{fm^{-1}}$ ($\mathrm{1~fm^{-1}\simeq 0.1973~GeV/c}$). Figure is adopted from Ref.~\cite{PhysRevLett.98.132501}.}}
    \label{isospin_src}
  \end{center}
\end{figure} 

Fig.~\ref{mom_dis_np} presents a calculation of the momentum distribution for protons and neutrons and their ratio in $\mathrm{^{3}He}$~\cite{Pieper_Wiringa}. In the assumption of isospin-independence, the momentum ratio of protons to neutrons should be equal to two, but if the SRC is isospin-dependent, the ratio becomes one when the SRC dominates at $k>k_{F}$. The right plot in Fig.~\ref{mom_dis_np} gives a ratio at $k>k_{F}$ roughly equal to 1.5, which suggests that the isospin effect plays a large role in the SRC.

 The reason why $np_{T=0}$ configuration does not totally dominate is that the $T=1$ channels are not completely suppressed, especially at very large momentum, where the 3N-SRC configuration is more complicated. Another calculation extended the study to other nuclei provides similar results~\cite{PhysRevLett.98.132501}. As shown in Fig.~\ref{isospin_src}, the momentum distribution of $np$ pairs is much larger than the one of $pp$ pairs at $300<k<600$ $MeV/c$.

 Inclusive cross sections are also sensitive to the role of isospin. One can examine the isospin-dependence by measuring the cross section ratio of two isotopes at different $\mathrm{Q^{2}}$ in the SRC region. For example, assuming the SRC is independent of the isospin, the cross section of protons to neutrons at large momenta is proportional to N/Z. Considering that the cross section of scattering off a proton is approximately three times larger than the cross section of scattering off a neutron (i.e. $\sigma_{p}\simeq 3\sigma_{n}$), the per nucleon cross section ratio of $\mathrm{^{48}Ca}$ to $\mathrm{^{40}Ca}$ can be written as~\cite{e08014_pr}:
 \begin{equation}
  \frac{\sigma_{^{48}Ca}/48}{\sigma_{^{40}Ca}/40}=\frac{\left(20\sigma_{p}+28\sigma_{n}\right)/48}{\left(20\sigma_{p}+20\sigma_{n}\right)/40}\simeq\frac{\left(20\sigma_{p}+28\sigma_{p}/3\right)/48}{\left(20\sigma_{p}+20\sigma_{p}/3\right)/40}=0.916.
  \label{eq_ca_ratio1}
 \end{equation}

 On the contrary, if the $np$ pairs dominate in the SRC region, one only can compare the cross sections for scattering off nucleons in the $np$ correlations. For $\mathrm{^{40}Ca}$, the maximum number of $np$ pairs is $\mathrm{20\times 20}$, while the number becomes $\mathrm{20\times 28}$ for $\mathrm{^{48}Ca}$. The ratio in Eq.~\eqref{eq_ca_ratio1} becomes:
 \begin{equation}
  \frac{\sigma_{^{48}Ca}/48}{\sigma_{^{40}Ca}/40}=\frac{\left(20\times 28\right)/48}{\left(20\times 20\right)/40}=1.17,
  \label{eq_ca_ratio2}
 \end{equation} 
which is 28\% larger than the ratio with the assumption of isospin independence. 

 However, Eq.~\eqref{eq_ca_ratio2} is a naive calculation since it does not consider the fact that nucleons can only form SRC pairs with their neighbors on account of their short distance properties of NN interactions. The calculations in Ref.~\cite{PhysRevC.84.031302} and \cite{PhysRevC.86.044619} take into account the size effect of $\mathrm{^{48}Ca}$ and $\mathrm{^{40}Ca}$, and predict the ratio in Eq.~\eqref{eq_ca_ratio2} to be near 1.0.
  
  These two Calcium isotopes were used in the E08-014 and the preliminary result will be presented in this thesis. A new proposal~\cite{E12_11_112_pr} in Hall-A at JLab will continue to study the isospin dependence of the SRC with $\mathrm{^{3}He}$ and $\mathrm{^{3}H}$ which have much smaller mass and size differences. The new measurement will provide 40\% deviation of the ratios between the two assumptions about isospin dominance in the SRC. Besides, the ground state wave-functions of $\mathrm{^{3}He}$ and $\mathrm{^{3}H}$ can be calculated exactly, therefore the experimental results will be directly compared with the theoretical models. 

\input ./intro/emc.tex

\section{Final State Interaction in SRC}
A nucleus is a complicated system and the struck nucleon experiences multiple interactions both in its initial and final states. The major problems in the experimental study of the SRC are the FSI, where the momentum and energy of the struck nucleon can be modified during the re-scattering processes with other spectators in the residual system. It is crucial to disentangle the role of the FSI from the SRC in the measurements of electron-nucleon scattering in the nuclei.

 In inclusive electron scattering, the effect from the FSI falls off rapidly at high $\mathrm{Q^{2}}$ as $\mathrm{1/Q^{2}}$~\cite{day_arns,SLAC_Measurement_PRC.48.2451}. At low $\mathrm{Q^{2}}$, the contribution of the FSI is large enough to break down the y-scaling feature of QE scattering in the PWIA~\cite{Day:1987az}. The study of the SRC with inclusive cross section measurements requires sufficiently large $\mathrm{Q^{2}}$ to diminish the FSI contribution. The current results from inclusive data (i.e. in Fig.~\ref{SLAC_2NSRC_xbj}) indicate little dependence of $\mathrm{Q^{2}}$ for the scaling region of the 2N-SRC, which implies that the FSI becomes less important in the kinematic settings of the SRC study ($\mathrm{Q^{2}>1~GeV^{2}}$).

  However, the contribution of the FSI may not completely vanish even at very large $\mathrm{Q^{2}}$. When the electron scatters off a nucleon in a SRC configuration, the struck nucleon may be very close to other correlated nucleons and the probability of re-scattering from the residual system is non-zero. Despite the possible large contributions, they can be removed by taking the cross section ratio if the FSI is localized in the SRC. For example, the FSI contribution in the 2N-SRC pairs in heavy nuclei should be similar to one in $\mathrm{^{2}H}$, and the ratio, $\sigma_{A}/\sigma_{^{2}H}$, should be able to cancel the FSI effect and only yield the clean contribution from the 2N-SRC. 

 Overall, despite that the FSI always exists in the SRC configurations, the effects in a heavy nucleus should be identical to the ones in a light nucleus due to the short distance feature of the SRC configurations. These effects are minimized in the study of the 2N- and 3N-SRC when taking the cross section ratio of heavy nuclei to light nuclei.

\section{E08-014 Experiment}
\begin{figure}[!ht]
  \begin{center}
    \includegraphics[type=pdf,ext=.pdf,read=.pdf,width=0.70\linewidth]{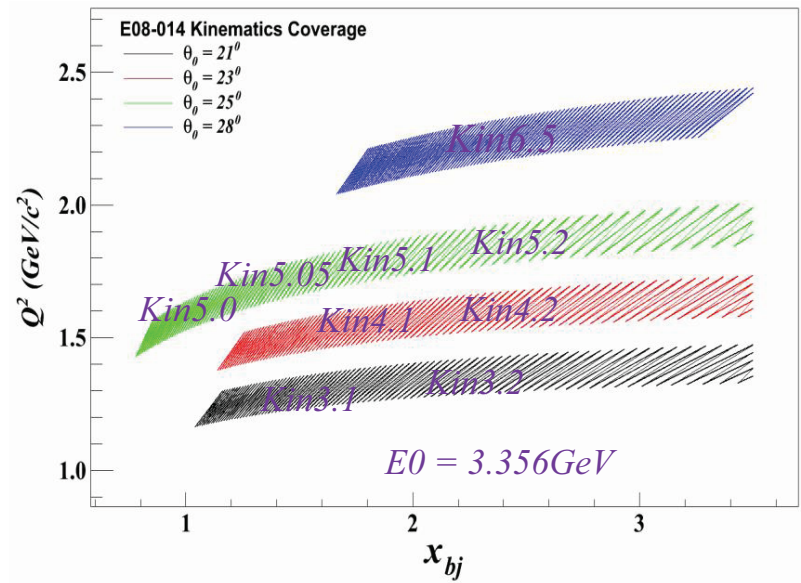}
    \caption[Kinematic coverage of the E08-014]{\footnotesize{Kinematic coverage of the E08-014.}}
    \label{kin_cor}
  \end{center}
\end{figure}

 A new experiment, E08-014~\cite{e08014_pr}, was carried out in 2011 in Hall-A at Jefferson Lab, with an electron beam energy of 3.356 GeV from the continuous electron beam accelerator facility (CEBAF). Utilizing the high resolution spectrometers in their standard configurations, this experiment measured the inclusive cross section of $\mathrm{^{2}H}$, $\mathrm{^{3}He}$, $\mathrm{^{4}He}$, $\mathrm{^{12}C}$, $\mathrm{^{40}Ca}$ and $\mathrm{^{48}Ca}$ at $\mathrm{1.1<Q^{2}<2.5 (GeV/c)^{2}}$, which covered the range of $x_{bj}$ from the QE peak region to above 3.0, as shown in Fig.~\ref{kin_cor}. The absolute cross section results will be used to study the scaling functions and momentum distributions at larger missing momentum, as well as the effect from the FSI. By taking the cross section ratio of heavy targets to $\mathrm{^{2}H}$ or $\mathrm{^{3}He}$, one can examine the $x_{bj}$ and $\mathrm{Q^{2}}$ dependence of the SRC, and measure the values of $a_{2}$ and $a_{3}$. The relatively low $\mathrm{Q^{2}}$ setting allows the study of $\alpha_{2N}$ and $\alpha_{3N}$ in the scaling of SRC. The Calcium isotopes, $\mathrm{^{40}Ca}$ and $\mathrm{^{48}Ca}$, were also used to study the isospin dependence of the 2N- and 3N-SRC. The experimental setup and the data analysis will be described in great details in this thesis and preliminary results will be presented. 

%% file: intro/emc.tex
\section{Medium Modification and EMC Effect}
  The nucleons in the SRC acquire high momenta through the short distance parts of the NN interaction. While the typical radius of a nucleon is roughly 0.85 fm~\cite{xiaohui}, the strong repulsive core in the NN interaction appears below 1 fm, which means that the wave-functions of these nucleons have significant overlap. One may argue that the structures and properties of these nucleons at close distance could have been modified in such localized high density configurations. This is generally referred to as medium modification~\cite{john_src_emc}. The inter-nucleon separation in heavy nuclei is typically smaller than the one in light nuclei, so the medium modification is expected to show a dependence on the average nuclear density.
  
 In Eq.~\eqref{src_a2}, the magnitude of the SRC in $1.3\leq x_{bj}<2.0$ is given by the scale factor, $\mathrm{a_{2}}$, which only depends on the nuclear number A in this region. Studies of the A-dependence of the SRC have been recently performed in Ref.~\cite{PhysRevLett.106.052301, john_src_emc} by combining the results of measurements in SLAC~\cite{SLAC_Measurement_PRC.48.2451}, Hall-B~\cite{PhysRevLett.96.082501} and JLab Hall-C~\cite{PhysRevLett.108.092502}. A different quantity, $\mathrm{R_{2N}}$, which is derived from $\mathrm{a_{2}}$ and includes a center of mass (c.m.) motion correction, is more applicable for this kind of study. While $\mathrm{a_{2}}$ represents the relative strength of the high-momentum tail in the nucleus, $\mathrm{R_{2N}}$ refers to the probability of a nucleon being part of a SRC configuration in a nucleus A compared with a nucleon in a deuteron.
 
  \begin{figure}[!ht]
  \begin{center}
    \includegraphics[type=pdf,ext=.pdf,read=.pdf,angle=270,width=0.54\linewidth]{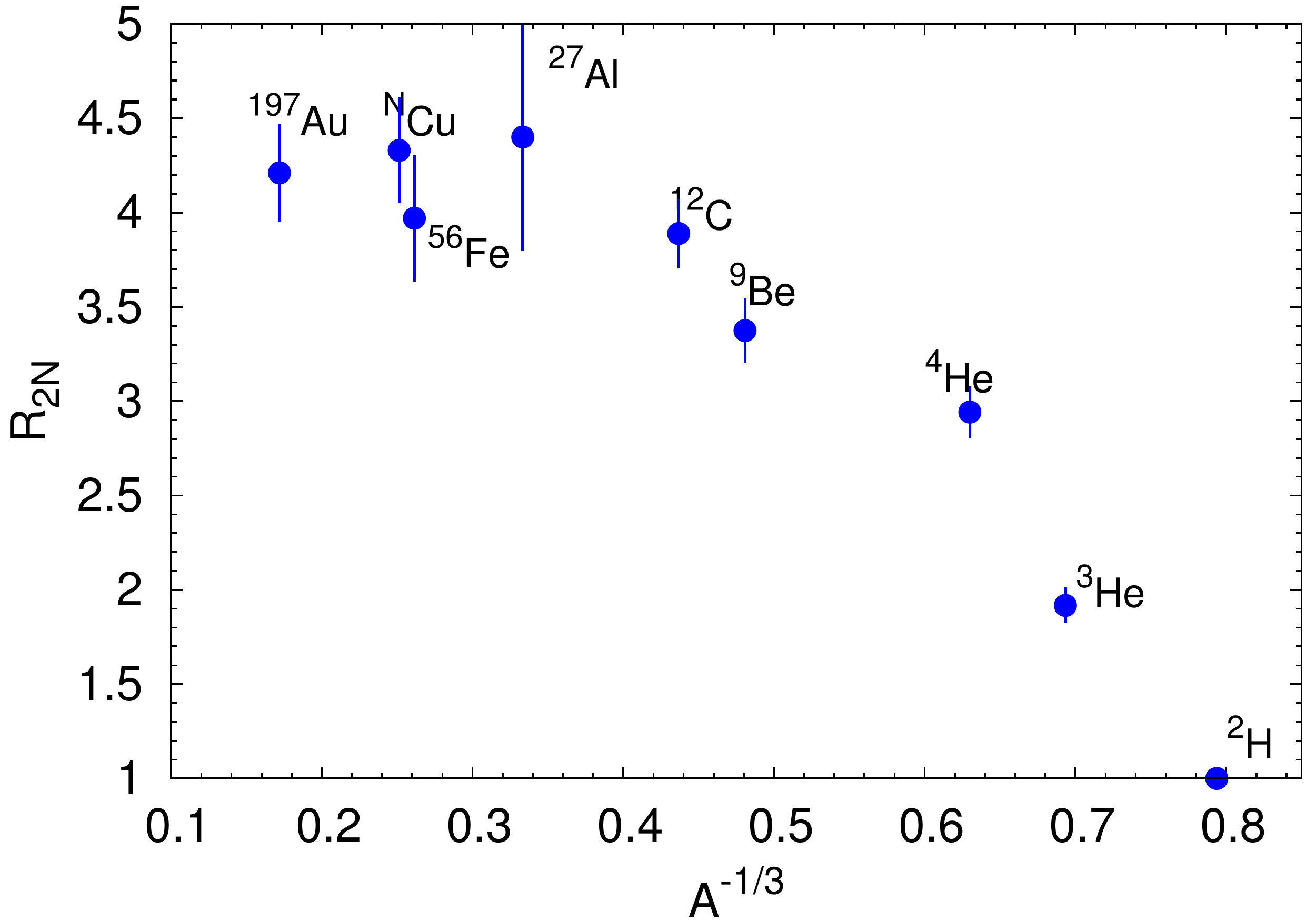}
    \caption[$\mathrm{R_{2N}}$ vs $\mathrm{A^{-1/3}}$]{\footnotesize{$\mathrm{R_{2N}}$ vs $\mathrm{A^{-1/3}}$ where $\mathrm{R_{2N}}$ in the y-axis is the scaling factor of the 2N-SRC defined in Eq.~\eqref{src_a2} with the center of mass correction. Figure is adopted from Ref.~\cite{john_src_emc}.}}
    \label{src_vs_a}
  \end{center}
\end{figure}  
  \begin{figure}[!ht]
  \begin{center}
    \includegraphics[type=pdf,ext=.pdf,read=.pdf,angle=270,width=0.6\linewidth]{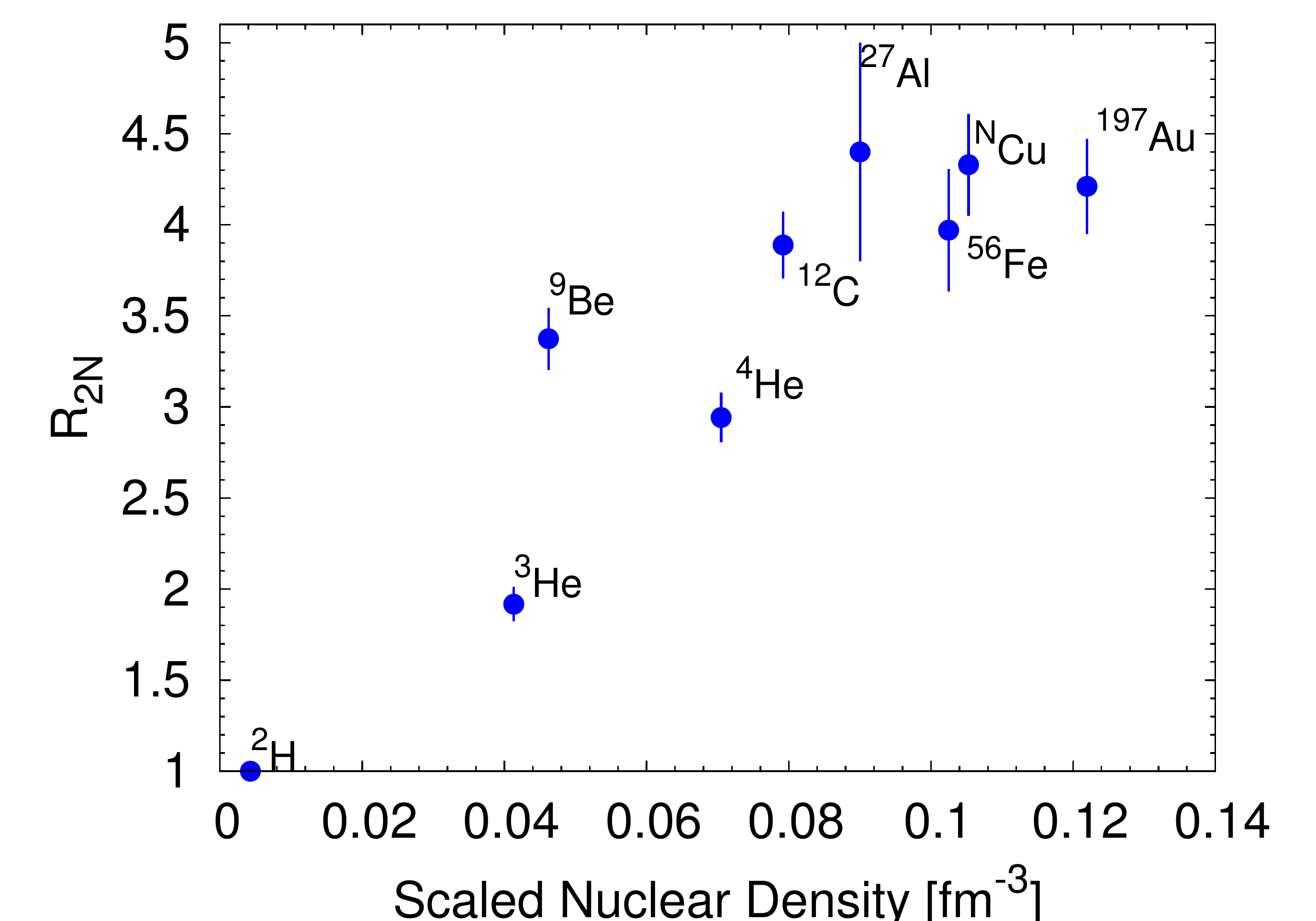}
    \caption[$\mathrm{R_{2N}}$ vs nuclear density]{\footnotesize{$\mathrm{R_{2N}}$ vs the nuclear density where the scaled nuclear density in x-axis is defined in Eq.~\eqref{density_scaled}. Note that the y-axis is $\mathrm{R_{2N}-1}$. Figure is adopted from Ref.~\cite{john_src_emc}.}}
    \label{src_vs_dens}
  \end{center}
\end{figure}    
  Fig.~\ref{src_vs_a} shows $\mathrm{R_{2N}}$ as a function of $\mathrm{A^{-1/3}}$, which appears to saturate at large A. A linear dependence would be expected if the nuclear response was the incoherent sum of scattering from individual nucleons. In fact, one is more interested in the variation of $\mathrm{R_{2N}}$ in the average nuclear density. A scaled nuclear density is defined as~\cite{john_src_emc}:
  \begin{equation}
    \rho_{scaled}(A) = \frac{A-1}{A}\rho(A),
    \label{density_scaled}
  \end{equation}
where $\rho(A)$ is the actual average nuclear density of the nucleus A, and the correction factor of (A-1)/A accounts for the excess nuclear density seen by the struck nucleon. Fig.~\ref{src_vs_dens} shows $\mathrm{R_{2N}}$ as a function of the scaled nuclear density. The figure indicates a nearly linear correlation for most nuclei except for $\mathrm{^{9}Be}$. 

\begin{figure}[!ht]
  \begin{center}
    \includegraphics[type=pdf,ext=.pdf,read=.pdf,angle=270,width=0.60\linewidth]{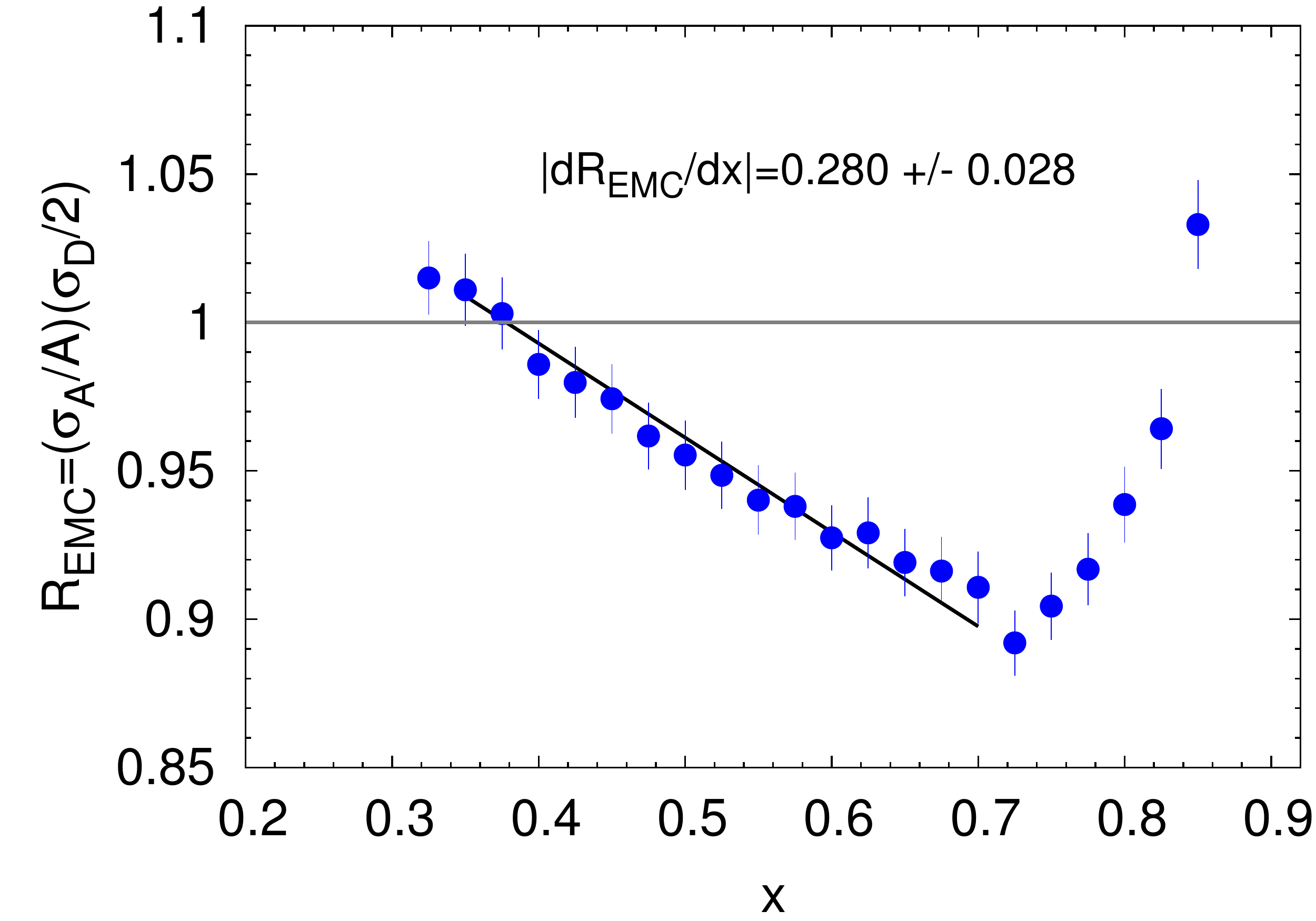}
    \caption[The EMC effect]{\footnotesize{The EMC effect illustrated by the slope of the ratio of the per-nucleon inclusive DIS cross sections of $\mathrm{^{12}C}$ to those of deuteron. Figure is adopted from Ref.~\cite{PhysRevLett.103.202301}.}}
    \label{emc_slop_03013}
  \end{center}
\end{figure}  
 The average density dependence in the SRC shown in Fig.~\ref{src_vs_dens} has also been seen in the analysis of the EMC effect, which refers to the in-medium modification of the nucleon structure, $\mathrm{F_{2}}$~\cite{EMC_Review_1995, EMC_Review_2003}. It was first discovered in the mid 1980s~\cite{EMC_first} and confirmed by many other measurements. In these inclusive DIS measurements, the per-nucleon cross sections (proportional to $\mathrm{F_{2}}$) in nuclei were compared to the deuteron at $\mathrm{Q^{2}\geq 2~GeV^{2}}$ and $0.35\leq x_{bj} \leq 0.7$ and turned out to be smaller. As an example, Fig.~\ref{emc_slop_03013} from Ref.~\cite{PhysRevLett.103.202301} shows an $x_{bj}$-dependence of the ratio of the $\mathrm{^{12}C}$ to deuteron deviates from one and decreases for an increasing $x_{bj}$. Many different models have been proposed to explain the EMC effect~\cite{EMC_Review_1995, EMC_Review_2003}. One common assumption is that the modification of the nucleon structure is driven by the average nuclear density. 
 
 \begin{figure}[!ht]
  \begin{center}
    \includegraphics[type=pdf,ext=.pdf,read=.pdf,angle=270,width=0.54\linewidth]{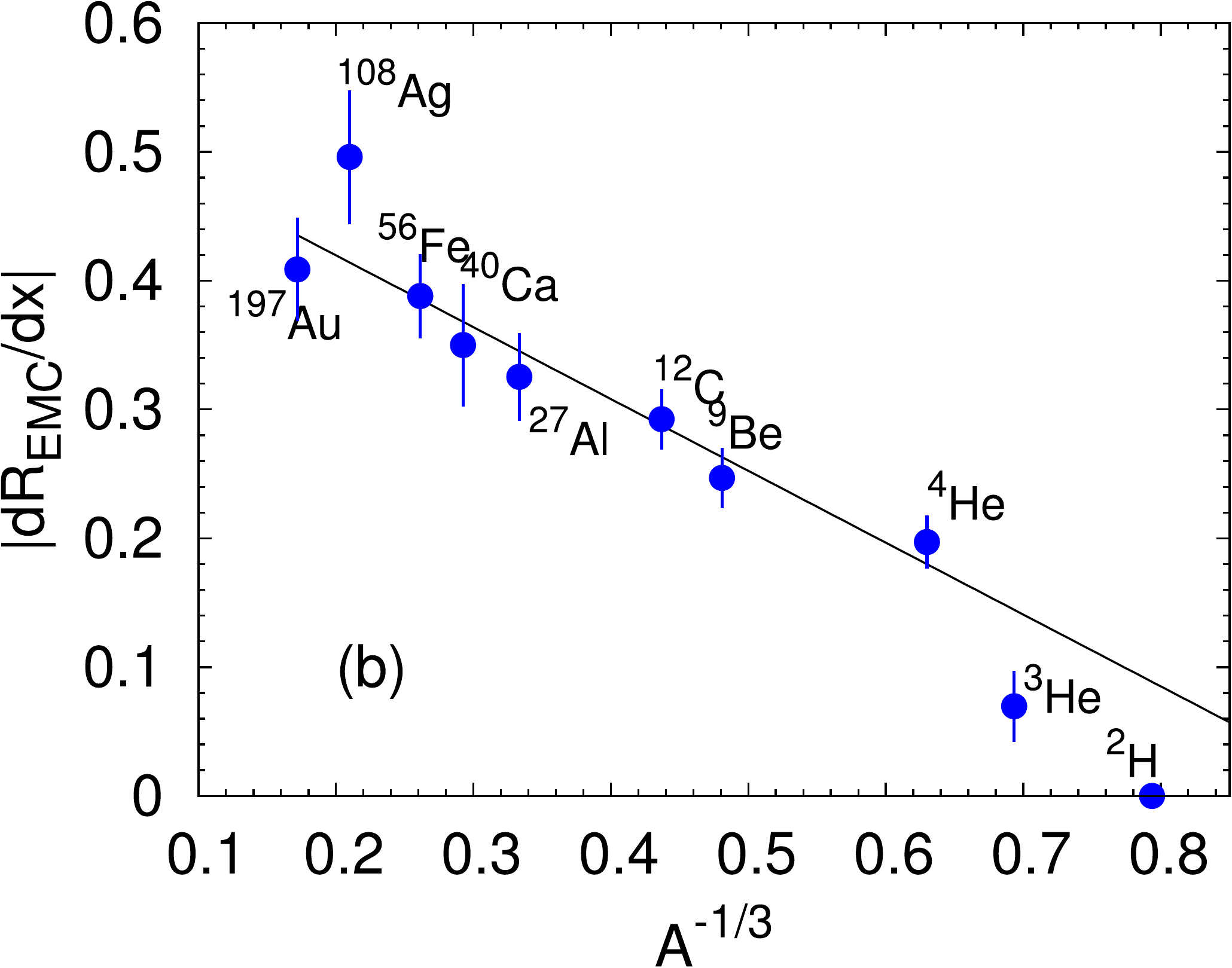}
   \caption[$\mathrm{dR_{EMC}/dx}$ vs $\mathrm{A^{-1/3}}$]{\footnotesize{$\mathrm{dR_{EMC}/dx}$ vs $\mathrm{A^{-1/3}}$. Figure is adopted from Ref.~\cite{john_src_emc}.}}
    \label{emc_vs_a}
  \end{center}
\end{figure}   
  The magnitude of the EMC effect can be characterized by the slope of a linear fit to this region, i.e., $\mathrm{dR_{EMC}/dx}$, shown in Fig.~\ref{emc_slop_03013}. The values of $\mathrm{dR_{EMC}/dx}$ for a wide range of nuclei are correlated with $\mathrm{A^{-1/3}}$, and the result given in Fig.~\ref{emc_vs_a} does indicate a linear connection. This result would be expected if the "surface" density distribution is independent of A. However, for light nuclei ($A\leq 12$), this assumption is invalid~\cite{john_src_emc}, so this simple A-dependence study of the EMC effect may be inappropriate. 

   \begin{figure}[!ht]
  \begin{center}
    \includegraphics[type=pdf,ext=.pdf,read=.pdf,angle=270,width=0.6\linewidth]{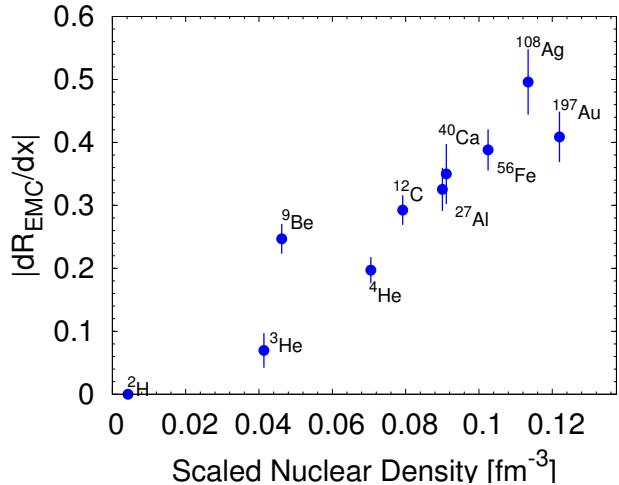}
   \caption[$\mathrm{dR_{EMC}/dx}$ vs nuclear density]{\footnotesize{$\mathrm{dR_{EMC}/dx}$ vs the nuclear density where in x-axis the scaled nuclear density defined in Eq.~\eqref{density_scaled}. Figure is adopted from Ref.~\cite{john_src_emc}.}}
    \label{emc_vs_dens}
  \end{center}
\end{figure} 
   The plot of $\mathrm{dR_{EMC}/dx}$ as a function of the scaled average nuclear density in Fig.~\ref{emc_vs_dens} also presents a linear correlation for most nuclei, but $\mathrm{^{9}Be}$, similar to the $\mathrm{R_{2N}}$ distribution in Fig.~\ref{src_vs_dens}, significantly deviates from the linear pattern. It was suggested that, since $\mathrm{^{9}Be}$ is composed of two $\alpha$-like clusters surrounded by a neutron, the local density is more appropriate for studying the EMC effect~\cite{PhysRevLett.103.202301}. 

\begin{figure}[!ht]
  \begin{center}
    \subfloat[EMC and SRC vs A]{
      \includegraphics[type=pdf,ext=.pdf,read=.pdf,angle=270,width=0.6\textwidth]{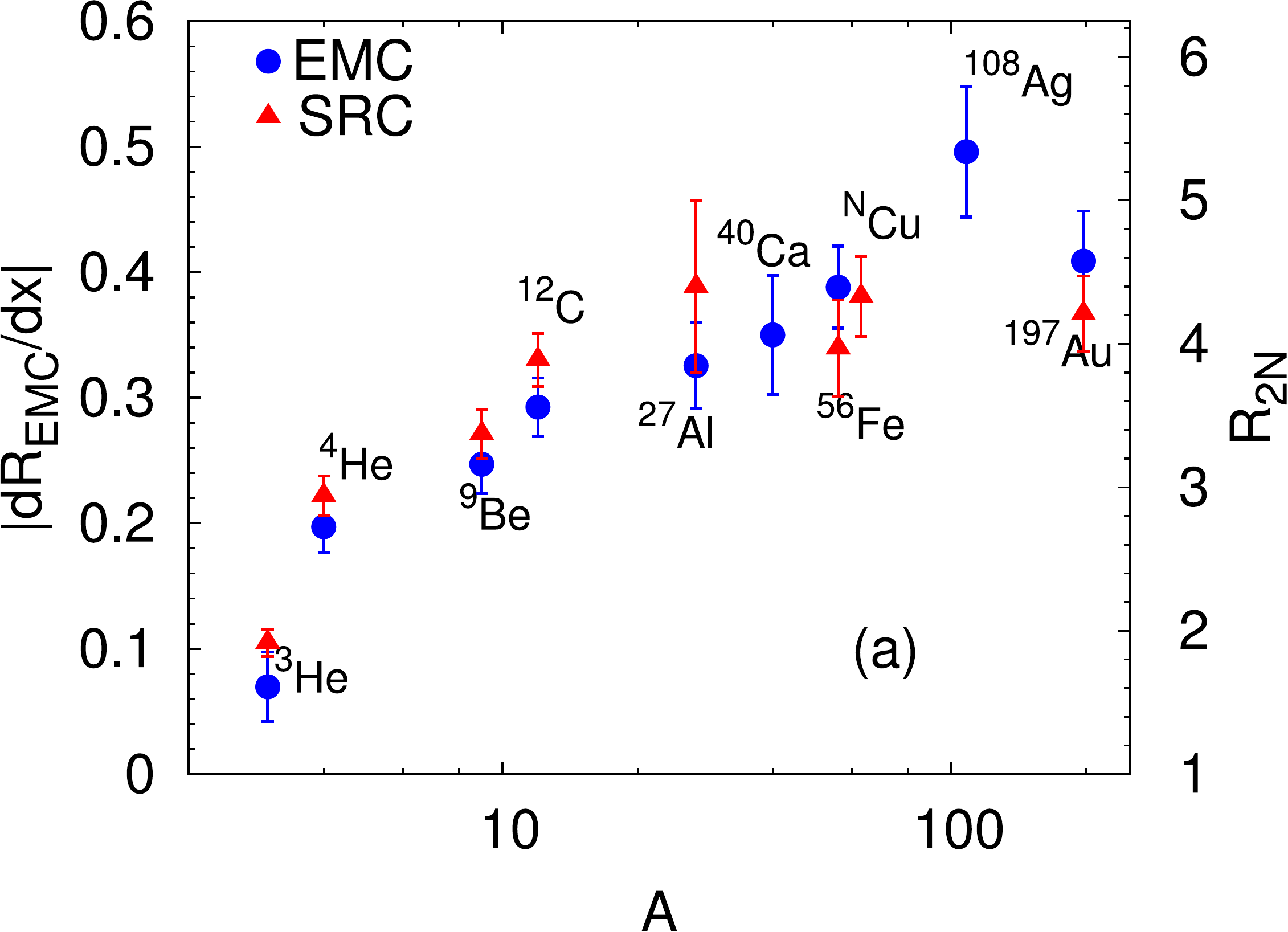}
      \label{emc_src_a_all}
    }
    \\ 
    \subfloat[EMC and SRC vs nuclear density]{
      \includegraphics[type=pdf,ext=.pdf,read=.pdf,angle=270,width=0.6\textwidth]{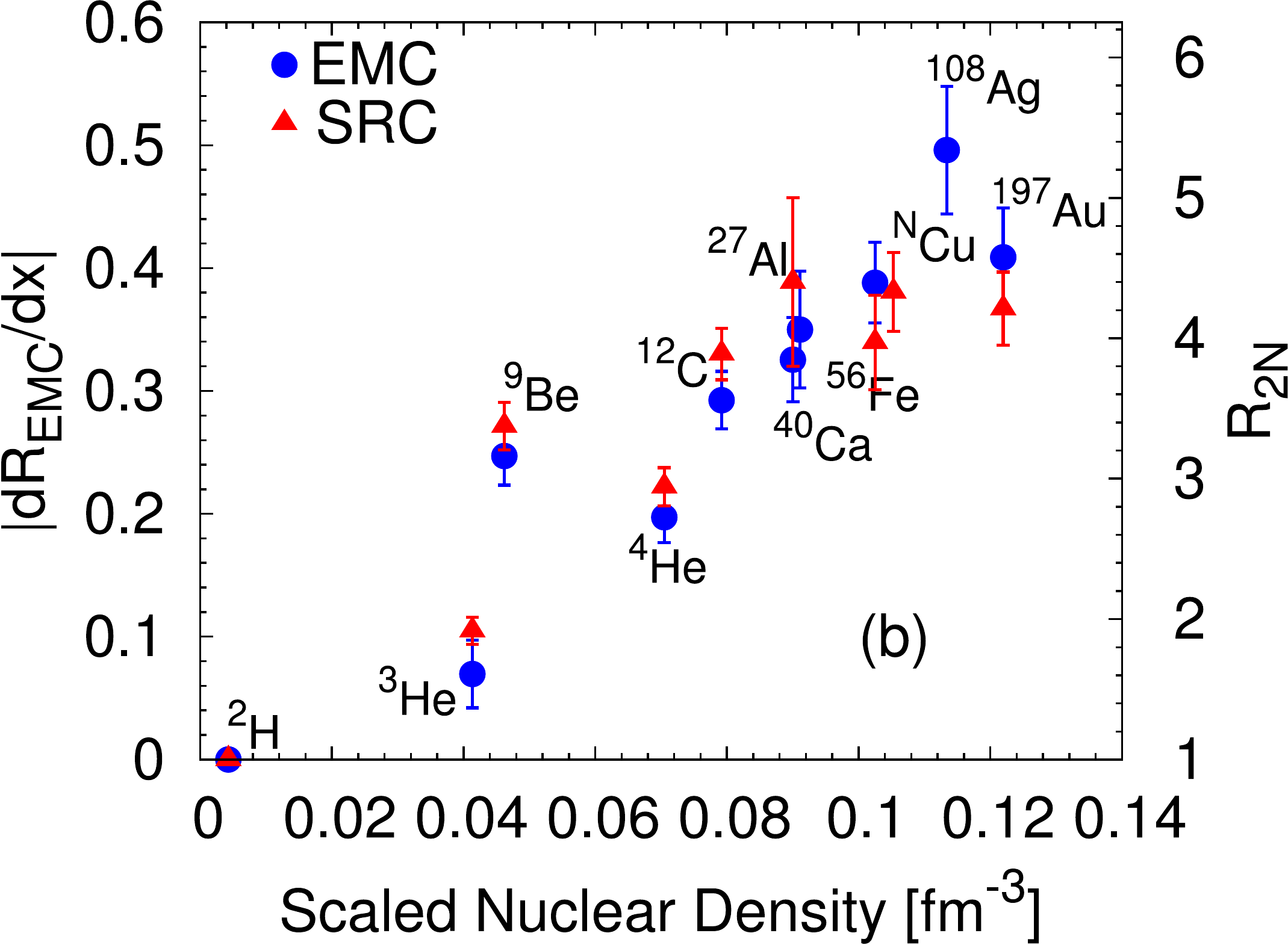}
    \label{emc_src_dens_all}    
    }
    \caption[Connecting EMC and SRC effects]{\footnotesize{Connecting EMC and SRC effects, where $\mathrm{dR_{EMC}/dx}$ and $\mathrm{R_{2N}}$ are shown as a funciton of A (top) and the scaled nuclear density (bottom). Figure is adopted from Ref.~\cite{john_src_emc}.}}
    \label{emc_src_all}
  \end{center}
\end{figure}
  In Fig.~\ref{emc_src_a_all} and Fig.~\ref{emc_src_dens_all}, the A- and density-dependence of $\mathrm{dR_{EMC}/dx}$ and $\mathrm{R_{2N}}$ are directly compared. Remarkably, the same pattern for all nuclei, including $\mathrm{^{9}Be}$, are seen in both plots. Since the measurements of the SRC directly probe the high density configurations inside the nucleus, the results strongly suggest that the local density is driving both of these disparate effects. 
  
    \begin{figure}[!ht]
  \begin{center}
    \includegraphics[type=pdf,ext=.pdf,read=.pdf,angle=270,width=0.60\linewidth]{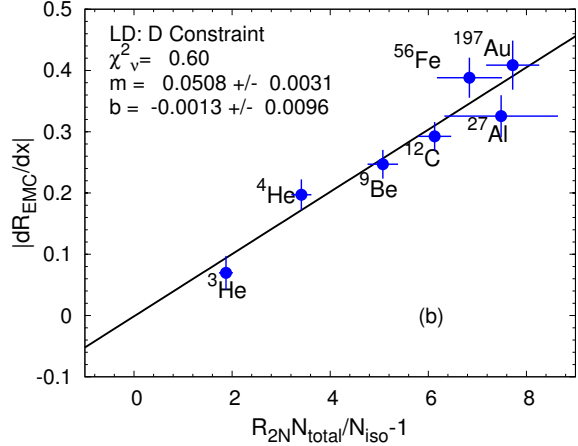}
   \caption[$\mathrm{dR_{EMC}/dx}$ vs $\mathrm{R_{2N}}$]{\footnotesize{$\mathrm{dR_{EMC}/dx}$ vs $\mathrm{R_{2N} N_{total}/N_{iso}-1}$, where $N_{total}=A(A-1)/2$ and $N_{iso}=Z(A-Z)$~\cite{john_src_emc}. The plot clearly shows the strong linear correlation between the EMC and the SRC effects. Figure is provided by Ref.~\cite{john_src_emc}.}}
    \label{emc_vs_src}
  \end{center}
\end{figure} 
 The linear connection between the EMC effect and the SRC can be seen in Fig.~\ref{emc_vs_src} where $\mathrm{dR_{EMC}/dx}$ is plotted against $\mathrm{R_{2N}N_{total}/N_{iso}-1}$ with $N_{total}=A(A-1)/2$ and $N_{iso}=Z(A-Z)$. This strong correlation provides a new constraint when modeling these two phenomena. The discussion above clearly demonstrates that the local density configuration plays an essential role in driving both the EMC effect and the SRC. Another hypothesis~\cite{PhysRevLett.106.052301} explains the linear connection between these two effects is due to the high virtuality~\cite{M_Sargsian_JPG_29_2003} of the high momentum nucleon. 
  
 Systematically understanding the connection between the SRC and EMC effects is still desirable, and will be of greatest interest in the 12-GeV era at JLab. Several new experiments have been approved to map out the nuclear dependence of these two effects~\cite{E12_10_103_pr,E12_06_105_pr,E12_10_008_pr,E12_11_107_pr,E12_10_003_pr}.

%% file: setup/setup.tex
\chapter{Experiment Setup}
\section{Overview}
 
\begin{figure}[!ht]
 \begin{center}
  \includegraphics[width=0.55\textwidth]{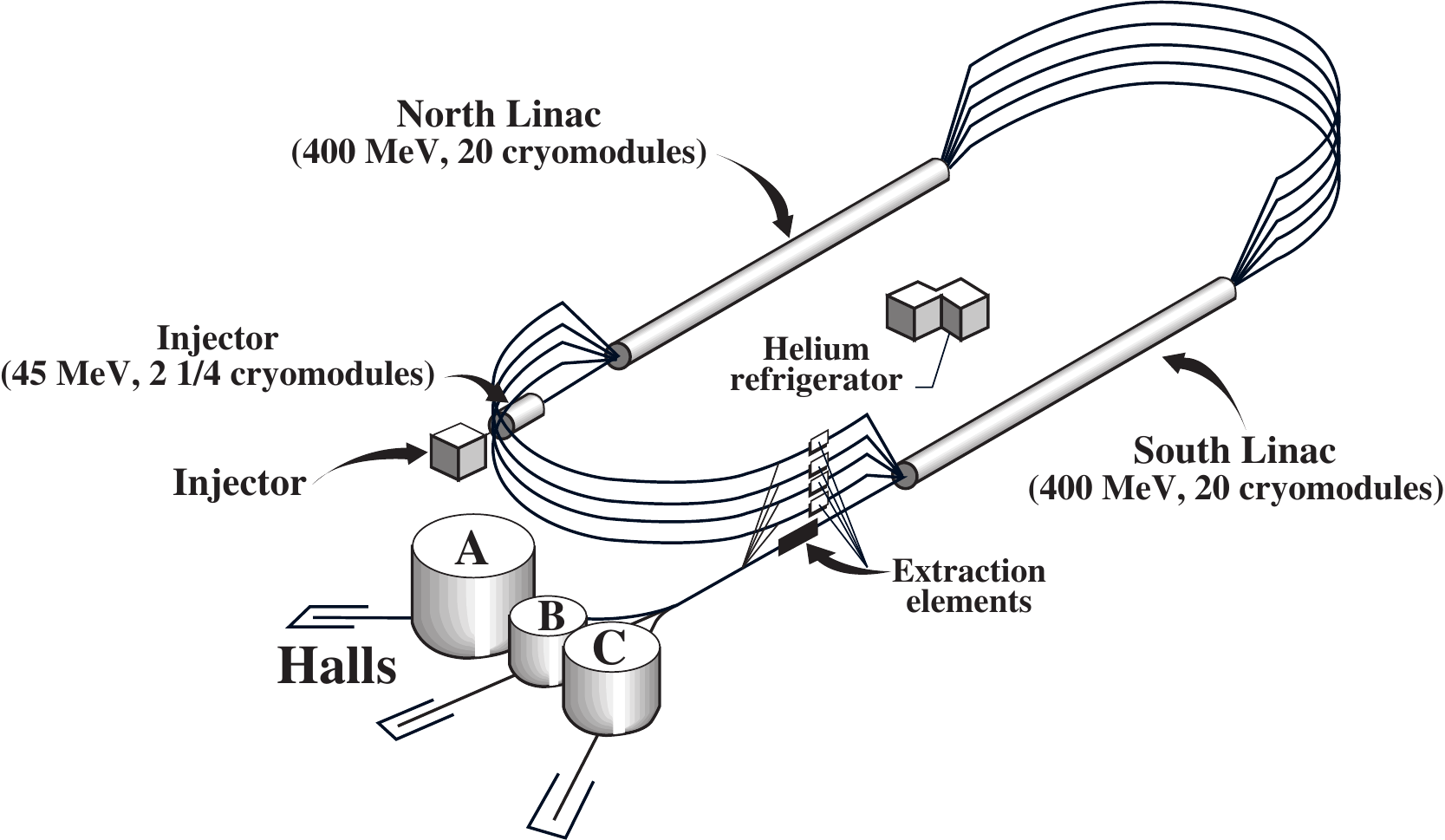}
  \caption[The Accelerator (CEBAF) at JLab]{The Accelerator (CEBAF) at JLab. Figure is from Ref.~\cite{halla_nim}.}
  \label{cebaf}
 \end{center}
\end{figure}
Thomas Jefferson Lab (JLab) is the world's leading medium energy electron scattering laboratory, consisting of a continuous electron beam accelerator facility (CEBAF), three experimental halls (A, B and C), a free electron laser facility and several applied research centers (Fig.~\ref{cebaf}). An upgrade project has been proceeding to increase the beam energy from 6 GeV to 12 GeV, and a complete new experimental hall, Hall D, is currently under construction and data taking is expected to begin by late 2014.

 CEBAF uses the radio frequency (RF) technique to deliver the polarized continuous-wave (CW) electron beam simultaneously to all three experimental halls. An injector provides electrons with polarization up to 85\% and a maximum current of $\mathrm{200~\mu A}$. The electron beam gains 400$\sim$ 600~MeV when passing through each of two super-conducting linear accelerators (linac), so the energy of the electron can be in the range of 0.8~GeV and 6.0~GeV within a maximum of 5 passes. Two arcs connect the linacs and provide $\mathrm{180^{\circ}}$ bending. The electron beam can be delivered to three halls at the same time with different energy and current. During the E08-014, the 3.356 GeV electron beam was delivered into Hall A with the current up to 150 $\mathrm{\mu A}$. Polarization was not required.

\begin{figure}[!ht]
 \begin{center}
  \includegraphics[width=0.75\textwidth]{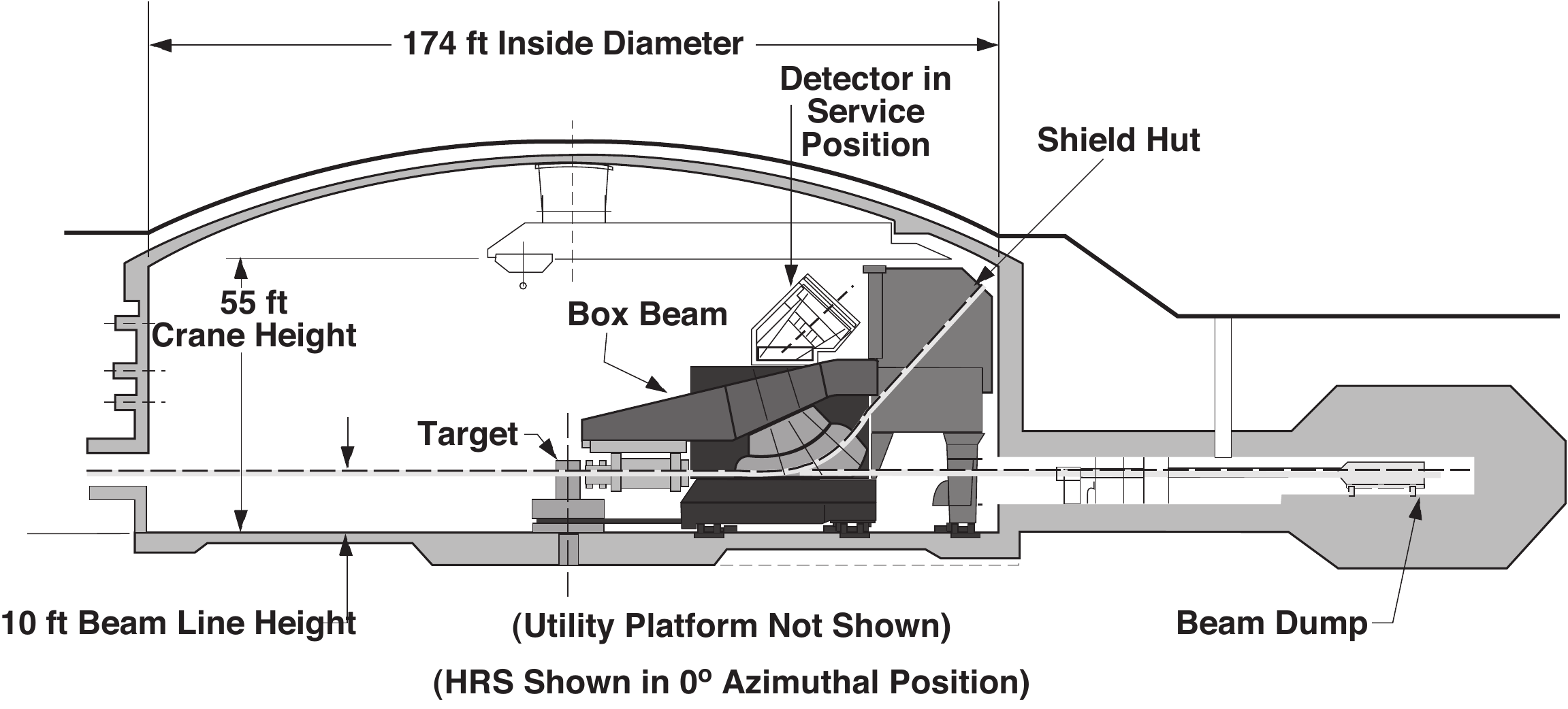}
  \caption[Side view of Hall-A]{Side view of Hall-A, which is a cylinder with 53 m in diameter and 17 m deep in the ground. The main instruments are the beamline components, the target system, two spectrometers and the beam dump. Figure is from Ref.~\cite{halla_nim}.}
  \label{sideview}
 \end{center}
\end{figure}
\begin{figure}[!ht]
 \begin{center}
  \includegraphics[width=0.75\textwidth]{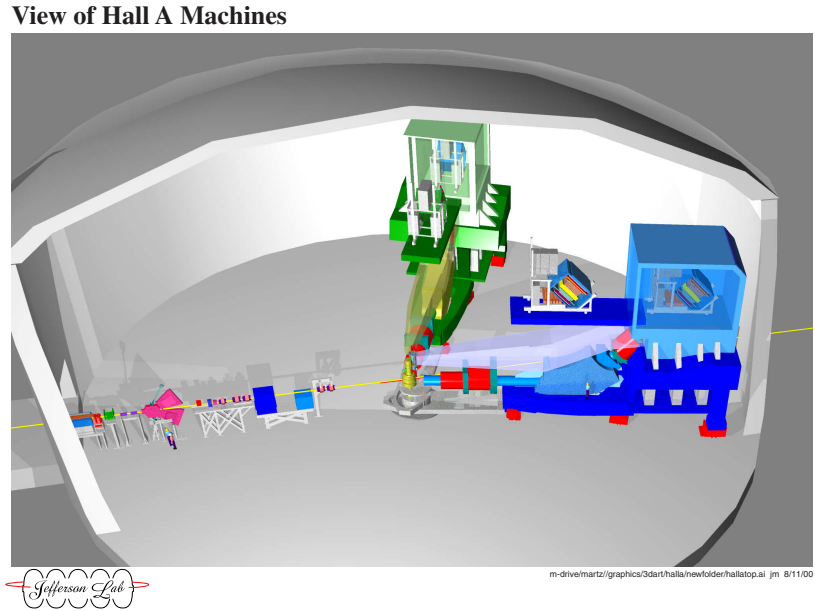}
  \caption[Top view of Hall-A]{Top view of Hall-A. Two high resolution spectrometers are on each side of the beam line and can be rotated around the central pivot where the target chamber is installed. The detectors are located on top of each spectrometer and shielded by the detector hut.  Figure is from Ref.~\cite{halla_main}.}
  \label{topview}
 \end{center}
\end{figure}

 Hall-A is a circular bulk (Fig.~\ref{sideview}) with a diameter of 53 m and a height of 17 m. The entire hall is buried underground and covered with concrete and earth. As shown in Fig.~\ref{topview}, the central elements in the hall include beamline components, a target system, and two identical high resolution spectrometers (HRSs). A detector package is stationed within a concrete shielding, called the detector hut, at the top of each HRS. The detector hut is designed to reduce the background in the detectors and protect them from radiation damage. Besides, it also stores electronic modules which collect signal outputs from detectors and the beamline, generate triggers, and provide the front end of the CEBAF Online Data Acquisition system (CODA). A detailed discussion of the Hall-A instrumentation is presented in the reference~\cite{halla_nim}.

\section{Beam}
\begin{figure}[!ht]
 \begin{center}
  \includegraphics[width=0.6\textwidth]{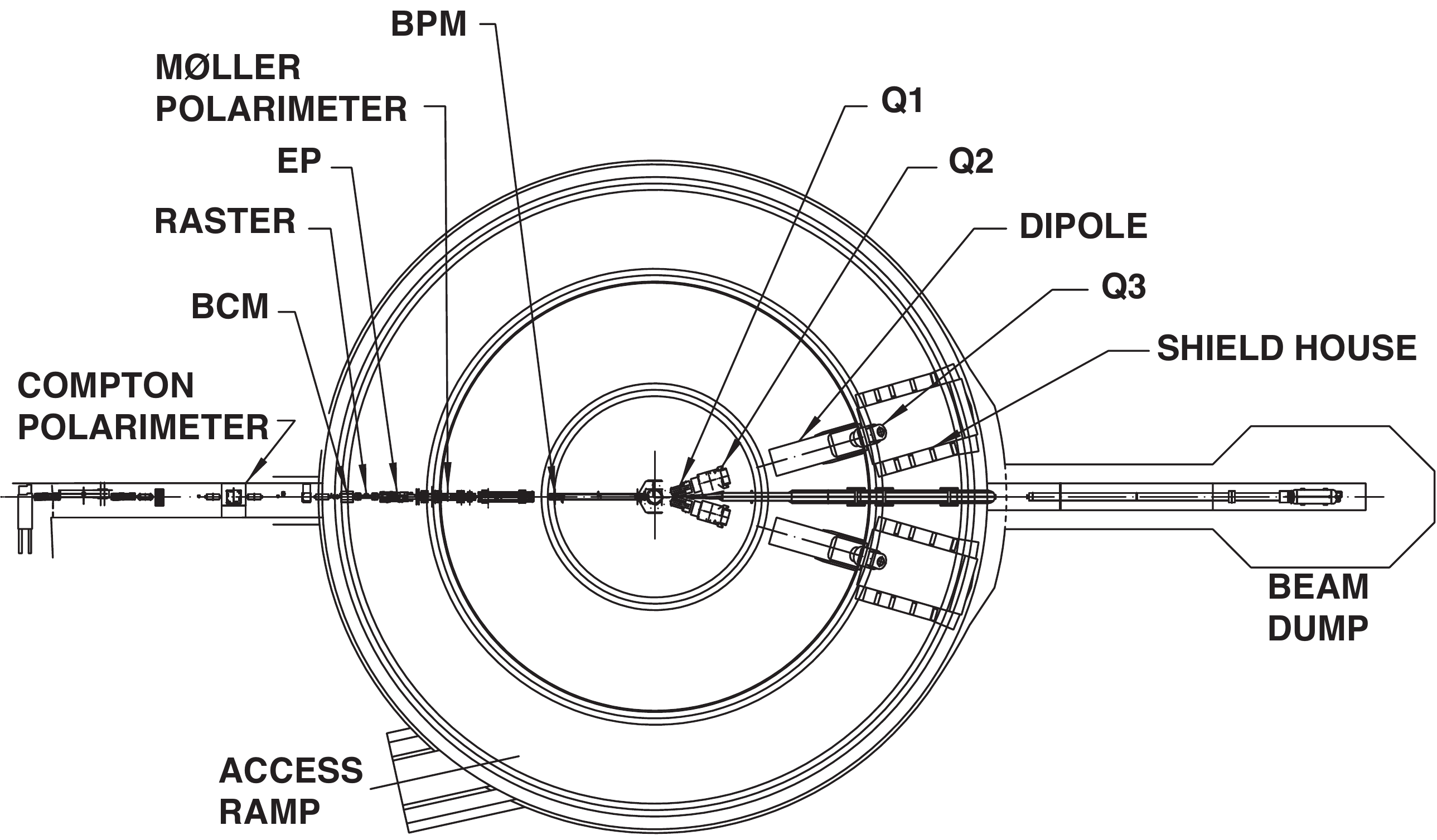}
  \caption[Schematic layout of beam instruments and spectrometers in Hall-A]{\footnotesize{Schematic layout of beam instruments and spectrometers in Hall-A, including beamline components, beam diagnotistic elements,and beam dump. Figure is from Ref.~\cite{halla_nim}.}}
  \label{whole_beam}
 \end{center}
\end{figure}

 The electron beam is delivered into Hall-A through a stainless steel tube which is 10 ft above the hall floor and holds a pressure $\mathrm{\leq 10^{-6}}$ Torr. The beam optics elements, including quadrupoles, sextupoles and corrector magnets, focus the beam on the target with spot sizes varying from 100 to 200 $\mathrm{\mu m}$ . A fast-raster system at 23 m upstream of the target position provides a beam spot of several millimeters at the target. As shown in Fig.~\ref{whole_beam}\cite{halla_nim}, the beam passing through the target is sent into the beam dump and spread out by a diffuser consisting of two 6.4~mm thick beryllium foils with water flowing between them. In addition, there are multiple beam diagnostics elements along the beamline to monitor, determine and control the relevant properties of the beam including the beam current, the beam position and direction, and the beam spot size at the target location. The energy and polarization of the electron beam are measured by individual methods and instruments~\cite{halla_nim}. 
\subsection{Beam Position Monitors}
 The position and direction of the beam at the target are determined by two beam position monitors (BPMs) located at 7.524~m and 1.286~m upstream of the target. Each BPM contains 4 antennas orientated orthogonally inside the beam pipe. Each antenna picks up a voltage reading from the beam when the beam current is above 1 $\mathrm{\mu A}$, and the signals from these antennas are used to calculate the beam position with the resolution of 100 $\mathrm{\mu m}$. The BPMs have to be calibrated independently to obtain the absolute position of the beam. When taking the calibration data, two pre-surveyed super-harps adjacent to BPMs are used to determine the absolute beam position~\cite{bpm_cali}. Event-by-event information from the BPMs is injected into the data stream, while the position average over every 0.3~s is also injected into the data stream every 3-4~s.

\subsection{Beam Charge Monitors}
The beam current monitor (BCM) is installed 25 m upstream of the target location and provides a non-interfering measurement of beam current. It consists of an Unser monitor, two RF cavities, several electronic modules and an associated DAQ system~\cite{halla_nim}. Two cavities on each side of the Unser monitor are high frequency wave-guides. The signal strength is proportional to the beam current when the cavities are tuned to the frequency of the beam. Before being sent into the RMS-to-DC converter~\cite{halla_nim}, each BCM output signal is split into three copies, two of which are amplified by 3 times and 10 times, respectively. Hence there are a total of six digital signals, $\mathrm{U_{1},~U_{3},~U_{10},~D_{1},~D_{3}}$ and $\mathrm{D_{10}}$, each of which is further divided into two copies and fed separately into scalers in HRSs. These digital signals are recorded by scalers in counts.

 During the data analysis, a BCM calibration is required to obtain the parameters to convert the scaler counts into electron charge. The procedure and result of the BCM calibration for this experiment can be found in Ref.~\cite{bcm_patricia}.
 
\subsection{Beam Energy}
The absolute energy of the beam can be determined by measuring the bend angle of the beam in the arc section of the beamline~\cite{beam_energy1,beam_energy2}. The momentum of the beam is related to the field integral of the eight dipoles and the bend angle:
\begin{equation}
  p = k \frac{\vec{B}\cdot \vec{dl}}{\theta},
\end{equation}
where $k\mathrm{=0.299792~GeV\cdot rad\cdot T^{-1}m^{-1}/c}$ and $\theta$ is the bend angle. The magnetic field integral of the eight dipoles are measured with respect to a reference dipole, the 9th dipole. The value of the bend angle is measured with a set of wire scanners.
 
\section{Target System}
 \begin{figure}[!ht]
 \begin{center}
  \includegraphics[angle=270,width=0.3\textwidth]{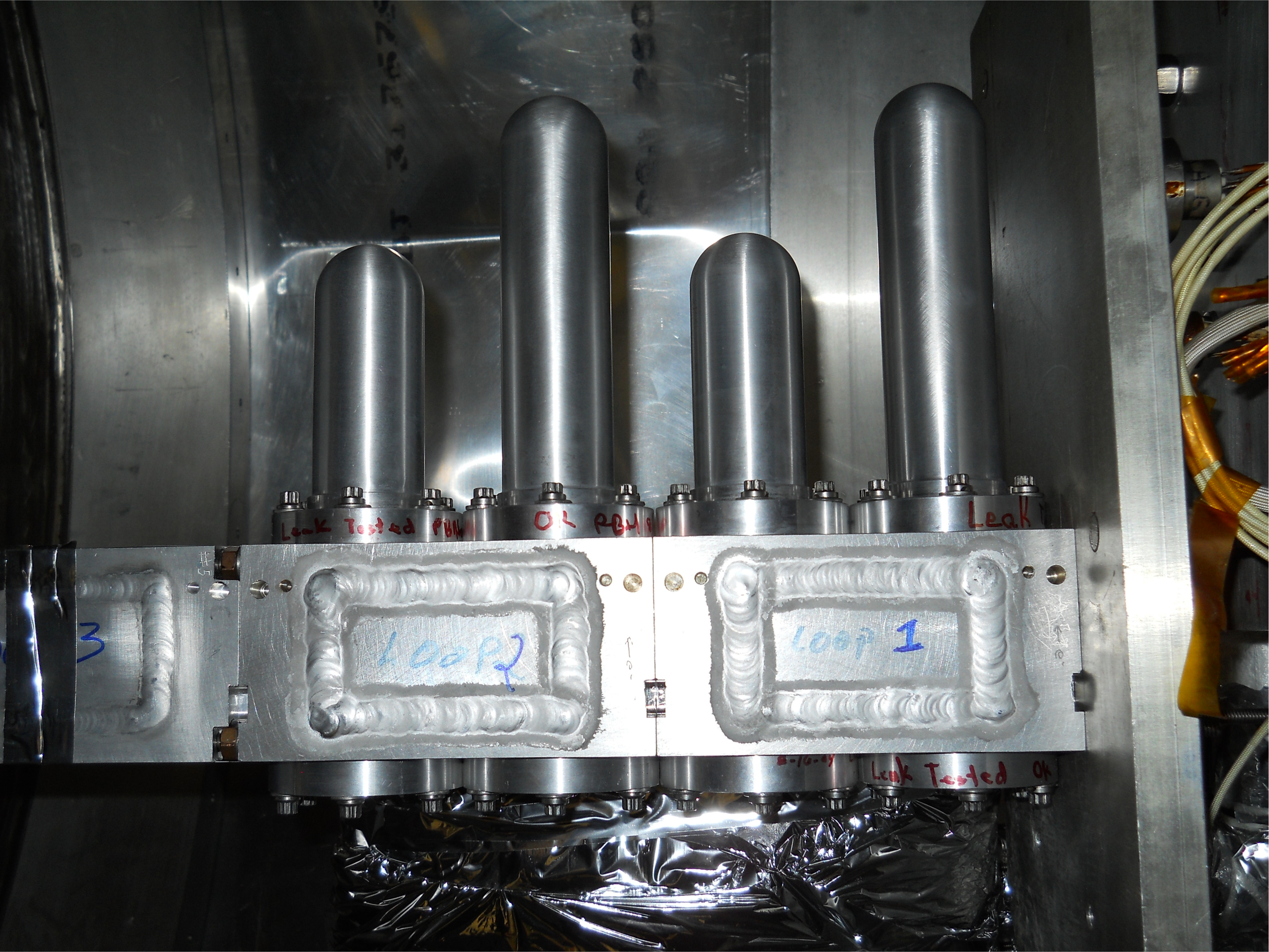}
  \caption[Picture of cryogenic target loops]{\footnotesize{Picture of cryogenic target loops, where Loop-1 and Loop-2 include 10 cm and 20 cm aluminium cells, respectively. Loop-3, which has two 20 cm cells, is not shown in this picture.}}
  \label{cryotarget_loop}
 \end{center}
\end{figure}
  The targets are located in a scattering vacuum chamber, which is supported by a 607~mm diameter central pivot connected to two HRSs. The main element within the scattering chamber is a cryogenic target system, which includes three loops of cryogenic targets, a target ladder to support solid targets, sub-systems for cooling and gas handling, temperature and pressure monitors, and target control and motion systems~\cite{halla_nim}. Three target loops are called Loop-1, Loop-2 and Loop-3. Both Loop-1 and Loop-2 contain two aluminium target cells with lengths of 10~cm and 20~cm, and Loop-3 has two 20~cm cells (Fig.~\ref{cryotarget_loop}).
\begin{figure}[!ht]
  \begin{center}
    \subfloat[First run period]{
      \includegraphics[type=pdf,ext=.pdf,read=.pdf,width=0.7\textwidth]{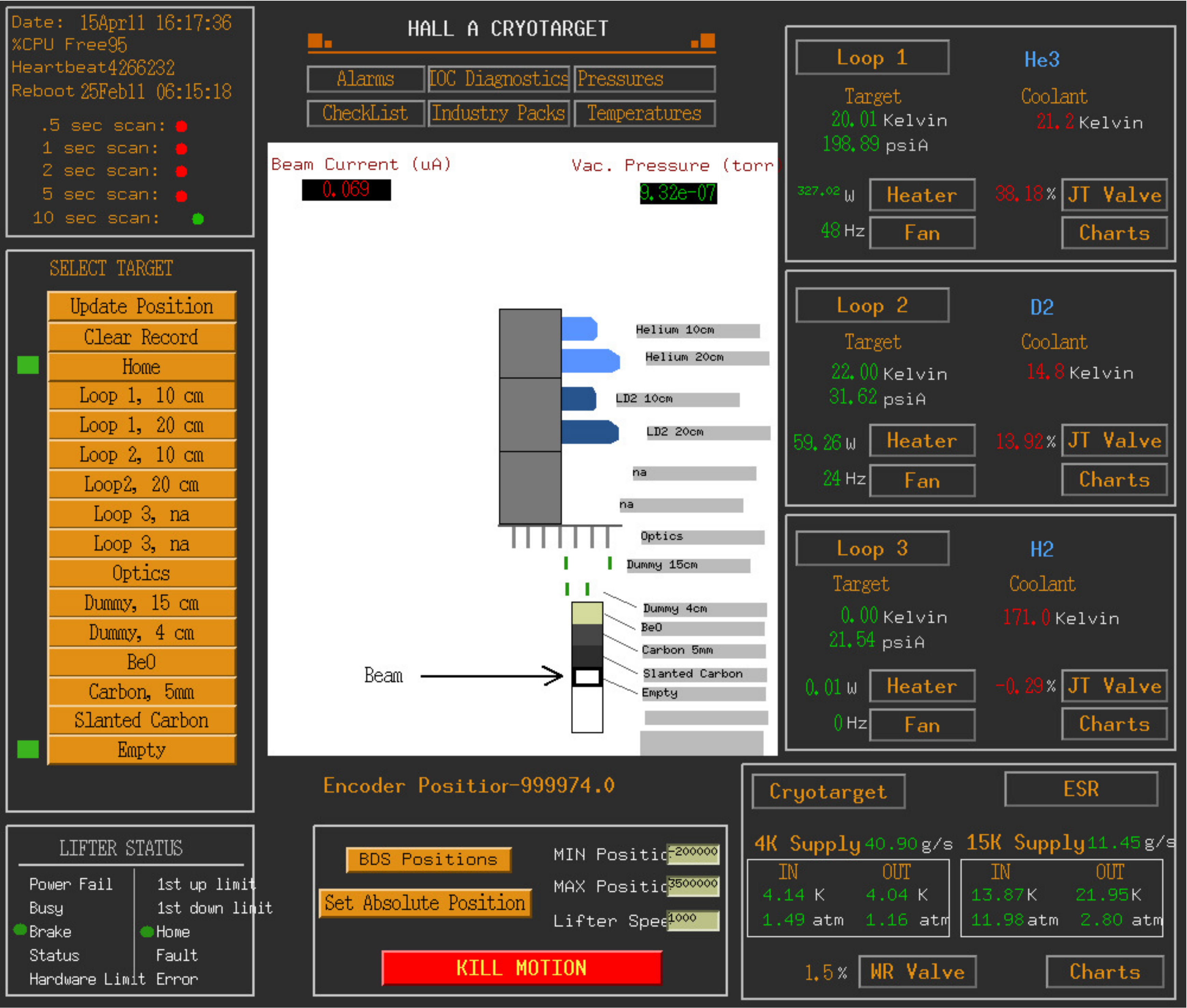}
      \label{target_run1}
    }
    \\
    \subfloat[Second run period]{
      \includegraphics[type=pdf,ext=.pdf,read=.pdf,width=0.7\textwidth]{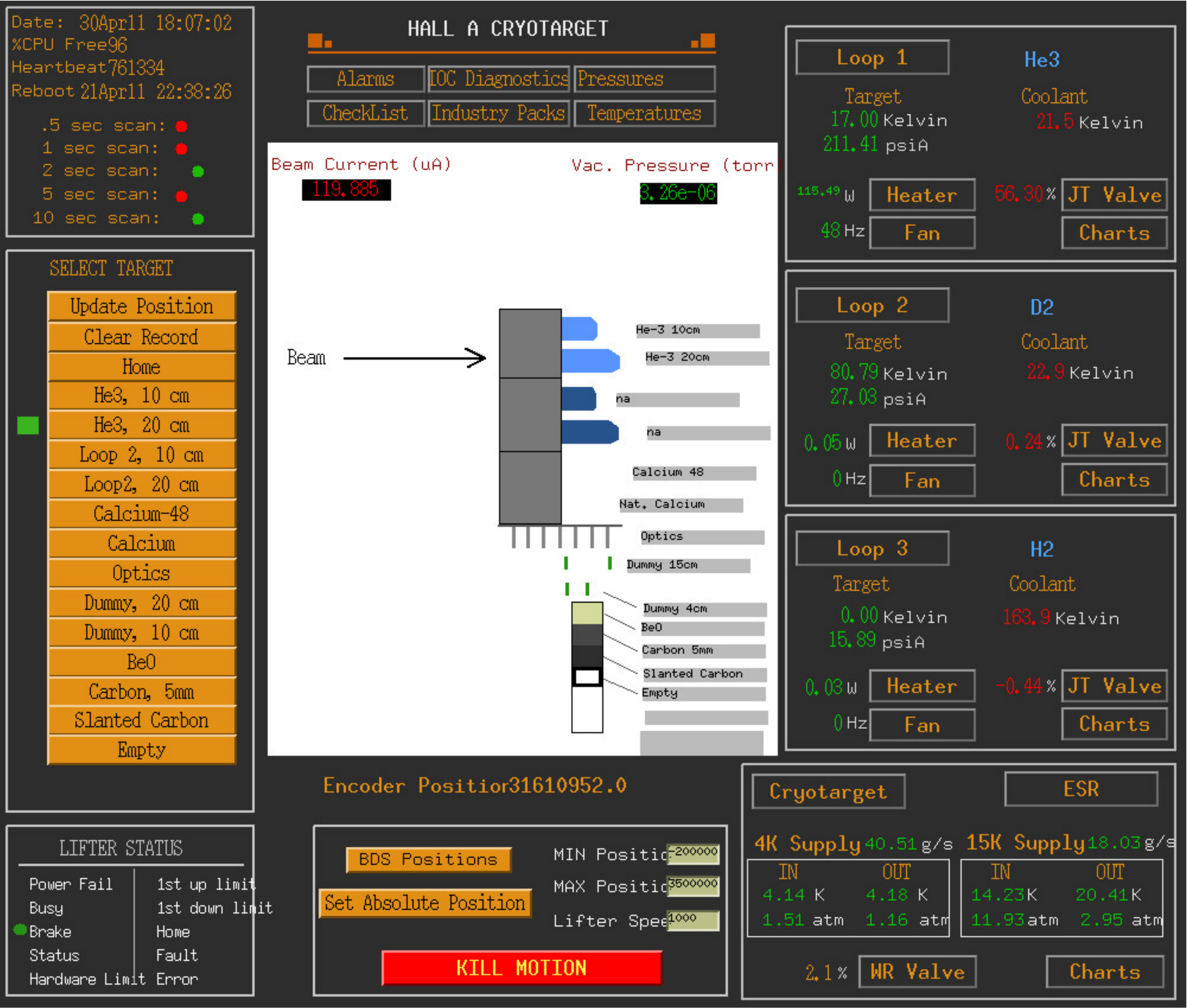}
      \label{target_run2}
    }
    \caption[Target control screens]{\footnotesize{Target control screen. (a) shows the targets installed in the first run period and (b) shows the targets installed in the second run period. Figures were obtained during this experiment.}}
  \end{center}
\end{figure}

  The cryogenic targets used in this experiment were liquid deuterium ($\mathrm{LD_{2}}$), gaseous $\mathrm{^{3}He}$ and $\mathrm{^{4}He}$. In the first run period of this experiment (from April 15th 2011 to April 19th 2011), the 20~cm cells of Loop-1 and Loop-2 were filled with $\mathrm{^{4}He}$ and $\mathrm{LD_{2}}$, respectively. $\mathrm{^{4}He}$ was then replaced by $\mathrm{^{3}He}$ in the second run period (from April 21st 2011 to May 15th 2011), and $\mathrm{LD_{2}}$ was evacuated from Loop-2 (Fig.~\ref{target_run2}). Two 20 cm cells in Loop3 were used to store a $\mathrm{^{40}Ca}$ foil and a $\mathrm{^{48}Ca}$ foil which could not be directly exposed to the air. The temperature (pressure) of $\mathrm{LD{2}}$, $\mathrm{^{3}He}$ and $\mathrm{^{4}He}$ was maintained at 22~K (30.5 psia), 17~K (211 psia) and 20~K (202 psia), respectively. The cooling power was provided by the end station refrigerator (ESR)~\cite{cryo_grp} operated by the JLab cryogenic group. 
  \begin{table}[htbp]
   \begin{tabular}{lccccc}
   \toprule
   Target       &$\rho$ ($\mathrm{g/cm^{3}}$)& Length (cm)   & $\mathrm{\delta\rho}$ ($\mathrm{g/cm^{2}}$)& I ($\mathrm{\mu A}$)& Comment   \\
   \midrule
   $\mathrm{LD_{2}}$& 0.1676                 & 20.0          &     N/A                & 40          &Loop2      \\
   Al can (Loop-2)  & 2.7                    & 0.0272        &     0.0001             &             &  Entrance\\
                    & 2.7                    & 0.0361        &     0.0011             &             &  Exit     \\
                    & 2.7                    & 0.0328        &     0.0002             &             &  Wall     \\
   $\mathrm{^{3}He}$& 0.0296                 & 20.0          &     N/A                &120         &  Loop1    \\
   $\mathrm{^{4}He}$& 0.0324                 & 20.0          &     N/A                &90           &  Loop1    \\
   Al-can (Loop-1)  & 2.7                    & 0.0272        &     0.0002             &             &  Entrance \\
                    & 2.7                    & 0.0361        &     0.0006             &             &  Exit     \\
                    & 2.7                    & 0.0328        &     0.0005             &             &  Wall     \\
   $\mathrm{^{12}C}$&      2.265             & 0.3937        &     0.0008             &120          &           \\
   $\mathrm{^{40}C}$&      1.55              & 0.5735        &     0.01               &40           &  Loop3    \\
   $\mathrm{^{48}C}$&      1.55              & 0.5284        &     0.01               &40           &  Loop3    \\
   Al-can (Loop-3)  & 2.7                    & 0.0272        &     0.0001             &             &  Entrance \\
                    & 2.7                    & 0.0361        &     0.001             &              &  Exit     \\
                    & 2.7                    & 0.0328        &     0.0002             &             &  Wall     \\
   Dummy-20cm       &      2.7               & 0.1581        &     0.0005             &40           & Upstream  \\
                    &      2.7               & 0.1589        &     0.0005             &             & Downstream\\
   Dummy-10cm       &      2.7               & 0.1019        &     0.0003             &40           & Upstream  \\
                    &      2.7               & 0.1000        &     0.0003             &             & Downstream\\        
   \bottomrule
   \end{tabular}
  \caption[Targets in the E08-014]{\footnotesize{Targets in the E08-014, where BeO target and optics target are not listed. The detailed report is in Ref.~\cite{target_report}. The uncertainties of three cryo-targets are needed to be conformed so they are listed temporarily as "N/A". }}
  \label{target_table}
  \end{table}
  
   A 30~cm long optics target was installed right below Loop-3 for taking optics calibration data. The optics target contains 7 carbon foils located at -15 cm, -10 cm, -5 cm, 0 cm, 5 cm, 10 cm, and 15 cm, respectively. Two dummy targets, Dummy-20cm and Dummy-10cm, were installed below the optics target to measure contributions from the endcaps of the cryogenic target cells. Each of them contains two thick aluminium foils separated by 10~cm for Dummy-10cm and 20~cm for Dummy-20cm. There were three other targets, BeO, $\mathrm{^{12}C}$ and an empty target, installed on the target ladder below Dummy-10cm. 
 
  The list of targets used in this experiment is given in Table~\ref{target_table}. Detailed information of targets and related systems can be found in Ref.~\cite{target_report}. The target positions are typically surveyed during the preparation of the experiment. However, survey reports were only available for experiments that ran before this experiment, and the targets installed in the second run period were not surveyed. Their positions were extracted by comparing their positions, e.g. the positions of endcaps for cryo-targets, and the central foil of the optics target.
  
\section{High Resolution Spectrometers}
\begin{figure}[!ht]
 \begin{center}
  \includegraphics[width=0.8\textwidth]{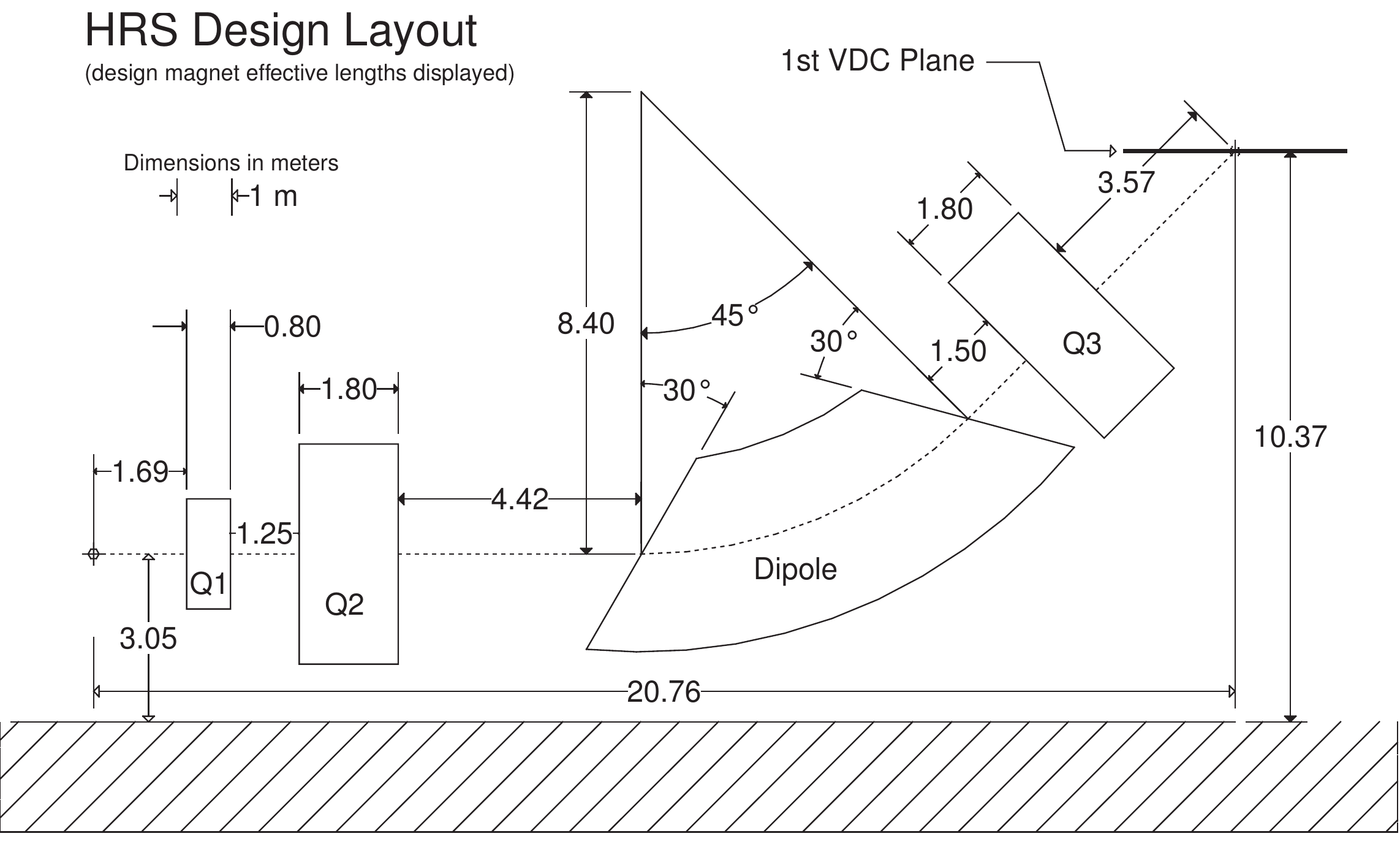}
  \caption[Schematic layout of HRS]{\footnotesize{Schematic layout of HRS, which shows the sizes and locations of the dipole and the three quadrupoles. Figure is from Ref.~\cite{halla_nim}.}}
  \label{halla_spectro}
 \end{center}
 \end{figure}
The essential equipments in Hall-A are two identical HRSs which provide high momentum resolution at the $\mathrm{10^{-4}}$ level over the range from 0.8 to 4.0~GeV/c, high position and angular resolution in the scattering plane, and large angular acceptance. As shown in Fig.~\ref{whole_beam}, the spectrometer on the left of the beam direction (to the beam dump) is called HRS-L, and the one on the right is called HRS-R. The basic layout of a HRS is given in Fig.~\ref{halla_spectro}. The magnet configuration of each HRS is QQDQ, including a dipole and three superconducting quadrupoles~\cite{halla_nim}. Two quadrupoles, Q1 and Q2, are installed in front of the dipole to achieve the desired angular acceptance and maximize the resolving power for the bend angle. The dipole performs a $\mathrm{45^{\circ}}$ vertical bending of the charged particles, and additionally, accommodates the extended targets and focuses the parallel beam. The third quadrupole, Q3, is behind the dipole to enhance the position and angular resolutions. Some important characteristics of the HRSs are listed in Table~\ref{hrs_table}.
 \begin{table}[htbp]
   \begin{tabular}{lcc}
   \toprule
   Bend Angle:           &$45^{\circ}$    \\   
   Optical Length:       &23.4 m      \\
   Momentum Range:       &0.3-4.0 GeV/c \\
   Momentum Acceptance:  &$-4.5\%<\delta p/p<+4.5\%$ \\
   Momentum Resolution:  &$1\times 10^{-4}$ \\
   Angular Range         &$12.5-150^{\circ}$ (HRS-L),12.5-130$^{\circ}$ (HRS-R) \\
   Angular Acceptance:   &$\pm$ 30~mrad (Horizontal), $\pm$ 60~mrad (Vertical)\\
   Angular Resolution:   &0.5 mrad (Horizontal), 1.0 mrad (Vertical)\\
   Solid Angle:          & 6 msr at $\delta p/p=0,y_{0}=0$\\
   Transverse Length Acceptance: &$\pm$ 5~cm \\
   Transverse Position Resolution: & 1 mm\\
   \bottomrule
   \end{tabular}
  \caption[Design characteristics of HRSs]{\footnotesize{Design characteristics of HRSs, where the resolution values are for the FWHM~\cite{halla_nim}.}}
  \label{hrs_table}
  \end{table}

 The power supply for the Q3 on HRS-R (RQ3) was not working properly during the experiment and limited the maximum central momentum setting to 2.876 GeV/c, but the experiment was planned to reach the maximum central momentum to 3.055~GeV/c. The RQ3 magnetic field was scaled down to 87.72\% at each kinematic setting. Accordingly, a new optics matrix was needed to match this new magnetic setting. An optics calibration to obtain the new matrix will be discussed in the next chapter.

\input ./setup/setup_detector.tex

\input ./setup/setup_daq.tex

%% file: setup/setup_detector.tex
\section{Detector Packages}
\begin{figure}[!ht]
 \begin{center}
  \includegraphics[width=0.7\textwidth]{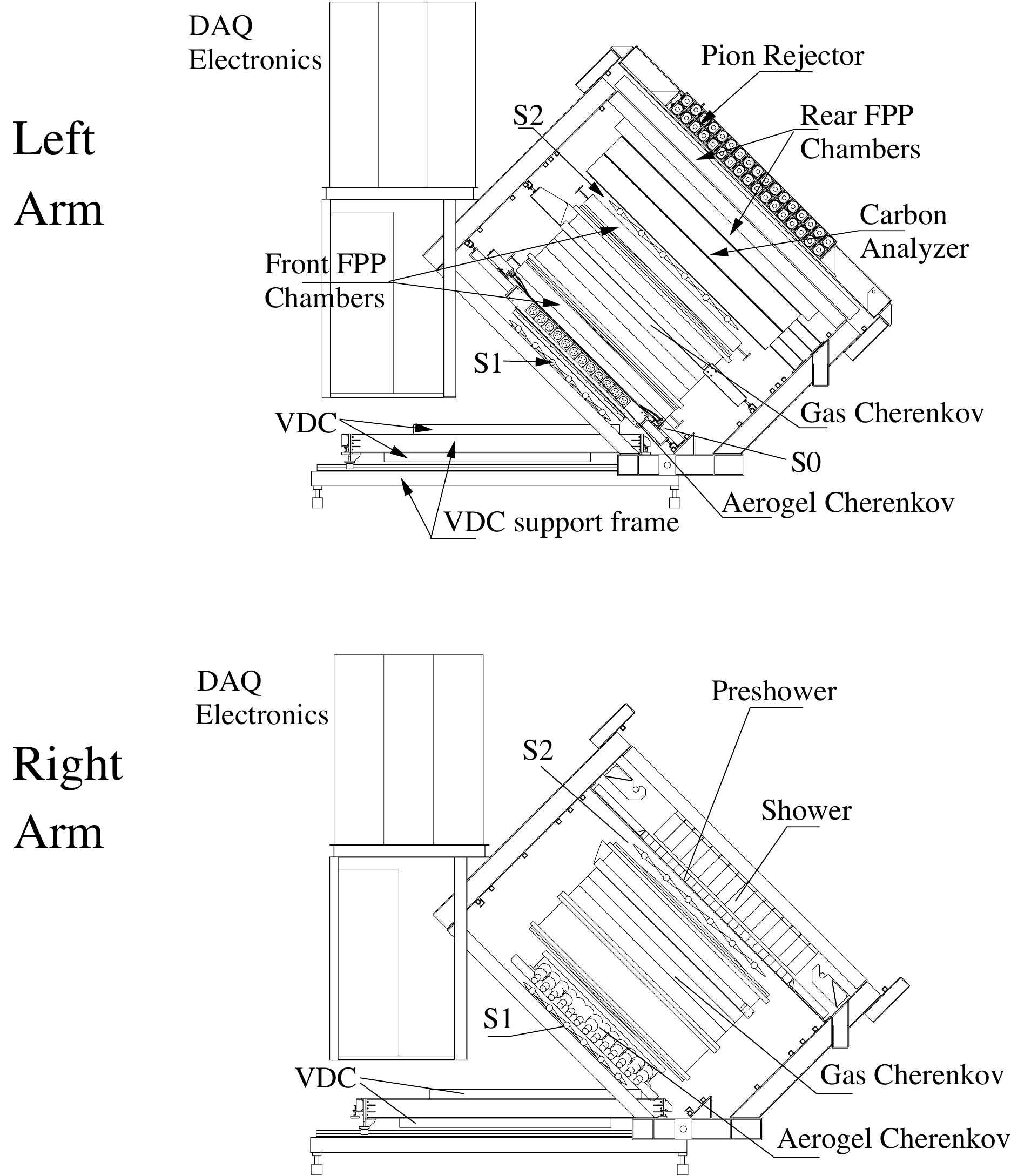}
  \caption[HRS detector stack]{\footnotesize{HRS detector stack, where all detectors available for different experiments are shown. VDCs, S1, S2m, Gas Cherekov detectors and Calorimeters were installed in the E08-014. In addition to these standard detectors, a long single-bar scintillator (S0) and an Aerogel \v{C}erenkov detector (AC) are included in each HRS, and a focal plane polarimeter (FPP) is available in HRS-L. S0, AC and FPP were not used during this experiment. Figure is from Ref.~\cite{halla_nim}.}}
  \label{detecor_hut}
 \end{center}
\end{figure}
 
 Particles coming through the HRS are fully characterized by the detector package and their signal outputs are delivered to the front-end electronics to form trigger signals and to be recorded by the data acquisition (DAQ) system. As shown in Fig.~\ref{detecor_hut}, the detector package in each arm includes two vertical drift chambers (VDCs), two scintillator planes (S1 and S2m), a gas \v{C}erenkov detector (GC), and a calorimeter. 
  
 Signals from VDCs are converted into digital types by the discriminator cards attached on the VDCs and then sent directly into the front-end of the time-to-digital converters (TDC) on the FastBus crate. For all other detectors, each analog signal from the corresponding photomultiplier tube (PMT) is split into two copies. One is properly delayed through a long cable before it is fed into the front-end of an analog-to-digital converter (ADC), and the other one, at the same time, goes through the discriminator module (DIS). If the amplitude of the analog signal is over the threshold value, a digital signal will be created and further used to form trigger signals or be recorded by the TDC front-end. 
 
 In the following sections, the detectors used in this experiment will be individually introduced.

\subsection{Vertical Drift Chambers}
\begin{figure}[!ht]
 \begin{center}
  \includegraphics[width=0.6\textwidth]{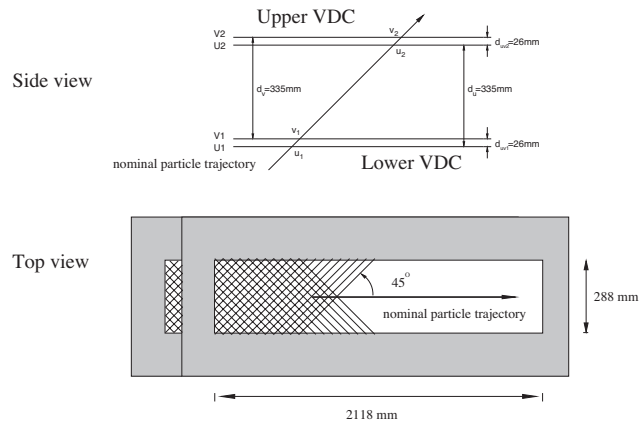}
  \caption[Layout of Vertical Drift Chambers]{\footnotesize{Layout of Vertical Drift Chambers. Figure is from Ref.~\cite{halla_nim}.}}
  \label{vdc}
 \end{center}
\end{figure}
 The trajectory of a particle after the Q3 exit is tracked by two identical VDCs, which are placed vertically 335~mm apart and lay horizontally $\mathrm{45^{\circ}}$ from the normal particle trajectory~\cite{halla_nim}, as shown in Fig.~\ref{vdc}. There are two wire planes (U and V) in each VDC oriented at $\mathrm{90^{\circ}}$ for one another, and each plane contains 368 wires. Two gold-plated Mylar planes are placed below and above each wire plane, and a high electric field is generated by applying the high voltage (-4~kV) between the wire plane and each Mylar plane. Both VDCs are filled with argon (62\%) and ethane (38\%) with a flow rate of 10~liter/hr. 
 
 When a particle goes through the VDC, the gas molecules are ionized and create a bunch of electrons and ions on the trajectory of the particle. The electrons are accelerated by the high field toward the closest wires, and the signal collected by each wire is amplified and read out by a pre-amplifier TDC card. On average, five sense wires have read-out signals when a particle passes through each wire plane. The exact location where the particle hits on the plane can be reconstructed by those TDC signals. Four locations provided by the four wire planes are used to fit the trajectory of the particle. The position resolution in the focal plane is about 100 $\mathrm{\mu m}$ and the angle resolution is near 100~mrad.

\subsection{Scintillator Counters}
 Two scintillator planes, S1 and S2m, are placed after the VDCs and separated by 2 m. S1 is composed of 6 overlapping thin plastic paddles, and S2m has 16 smaller paddles. When a charged particle passes through a paddle, it creates light which travels toward both ends of the paddle. A PMT attached to each end of the paddle collects the light and converts it into an analog signal. Scintillators have very fast time-response with very good resolution ($\mathrm{\sim}$30 ns), so their signals are the major source of generating triggers for the DAQ system. The traditional production trigger in Hall-A is generated by requiring both S1 and S2m to be fired within a narrow time window. A detailed discussion of trigger system is given in Section 3.7 and Appendix A.

\subsection{Gas \v{C}erenkov Detectors}
\begin{figure}[!ht]
 \begin{center}
  \includegraphics[width=0.6\textwidth]{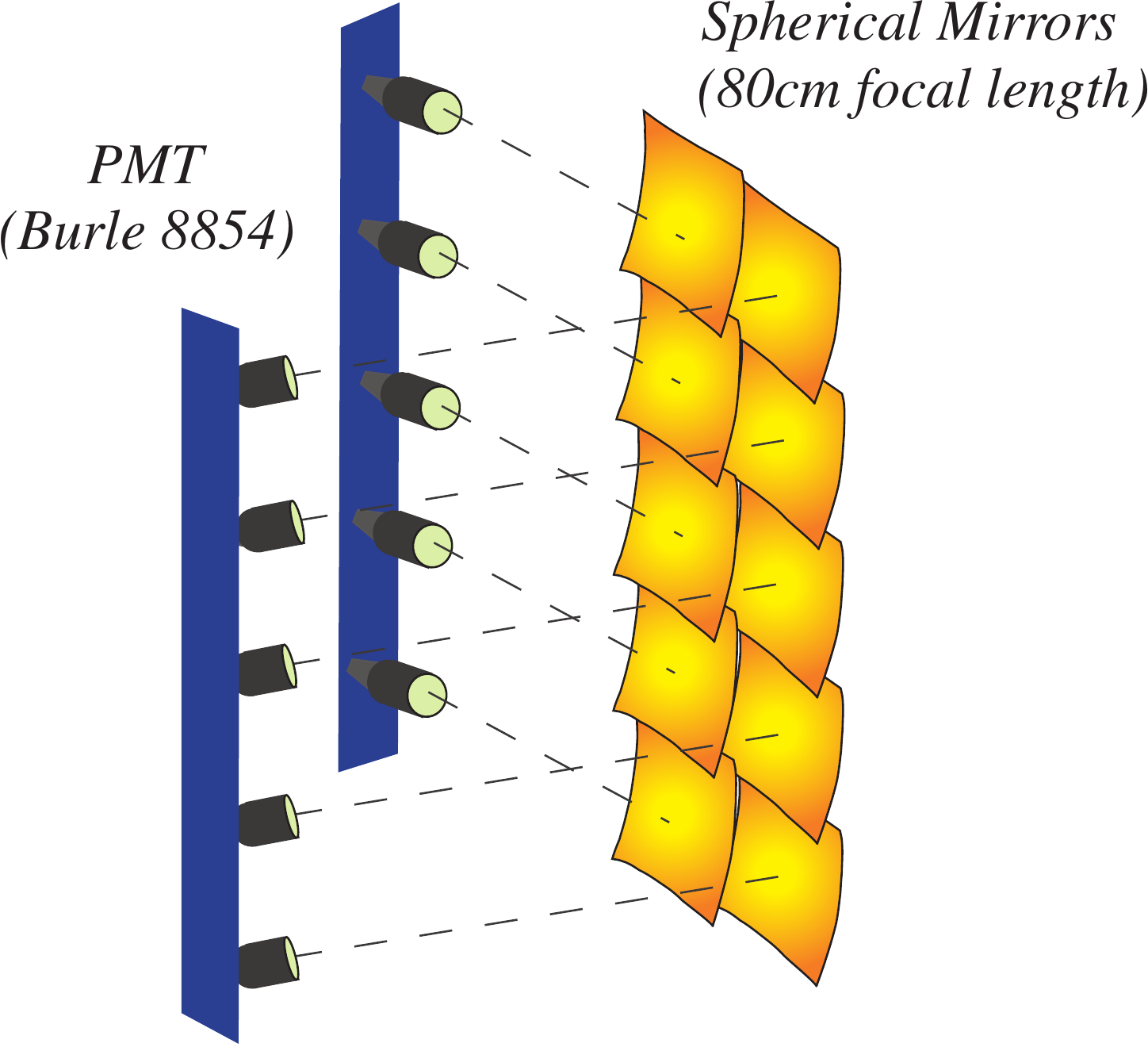}
  \caption[Design of the gas \v{C}erenkov detector]{\footnotesize{Design of the gas \v{C}erenkov detector. Ten spherical mirrors were carefully arranged to collect the \v{C}erenkov light and focus it into their corresponding PMTs.}}
  \label{gc_pmt}
 \end{center}
\end{figure}
 A high energy charged particle radiates \v{C}erenkov light when it travels in a medium with its speed faster than that of light. The basic mechanism of \v{C}erenkov radiation is that atoms along the track of the particle are polarized and become dipoles, and the variation of these dipole moments emits electromagnetic light\cite{R_Bock}. 
 
 The angle between the direction of \v{C}erenkov light and the track of the charged particle is given by:
\begin{equation}
 cos \theta = \frac{1}{\beta n},
\end{equation}
where $n$ is the index of reflection of the medium. $\beta=v/c$ where $v$ is the charged particle's velocity in the medium and $c$ is the speed of light. The velocity-dependence property of \v{C}erenkov radiation provides an effective tool to discriminate particles with different masses, since the momentum threshold to emit \v{C}erenkov light depends on the mass of the particle:
\begin{equation}
 P_{threshold} = \frac{mc}{\sqrt{n^{2}-1}}.
\end{equation}

 A gas \v{C}erenkov detector (GC), made up of a steel box with thin entry and exit window, is mounted between S1 and S2m on each HRS. Within the box ten light-weight spherical mirrors with very small thickness (0.23~$\mathrm{g\cdot cm^{-2}}$) are positioned in a 2 (horizontal) $\times$ 5 (vertical) array, as shown in Fig.~\ref{gc_pmt}. These mirrors are carefully arranged to efficiently reflect and focus the \v{C}erenkov light on the associated ten PMTs.
 
 The GC box is filled with atmospheric pressure $\mathrm{CO_{2}}$, which gives the index of refraction to be 1.00041. The momentum threshold for electrons to radiate \v{C}erenkov light in this detector is about 18 MeV/c, while the threshold for pions is as high as 4.9 GeV/c. Since the momentum coverage of HRS is from 0.5 GeV/c to 4.3 GeV/c, only electrons can emit \v{C}erenkov light in the detectors. Pions may still be able to produce signals in the GC when they interact with the gas and create low-energy electrons, i.e. $\mathrm{\delta}$-electrons~\cite{pdg}. However, the probability of such process is relatively low and the amplitude of the signal is comparable to the background signal. The path length of the GC on HRS-L is 80~cm which yields an average of 7 photon-electrons, while on HRS-R, the path length for the GC is 130~cm, leading to 12 photon-electron on average~\cite{halla_nim}. The design of GCs provides an excellent electron detection with efficiencies normally above 99\%.

The signal from each PMT is amplified 10 times by an amplifier and divided into two copies. One copy is directly sent to the front-end ADC for offline analysis. The other copy is further split into two pieces, where one is converted into a digital signal and sent to the TDC, while the other one is added together with the similar signals from the other 9 PMTs. The sum of the ten signals is then converted into a digital signal which is used for the design of online triggers, such as the efficiency triggers. During the E08-014, GCs were also included in the production triggers to suppress pion events during the data recording.

\subsection{Lead Glass Calorimeters}
\begin{figure}[!ht]
 \begin{center}
  \includegraphics[width=0.7\textwidth]{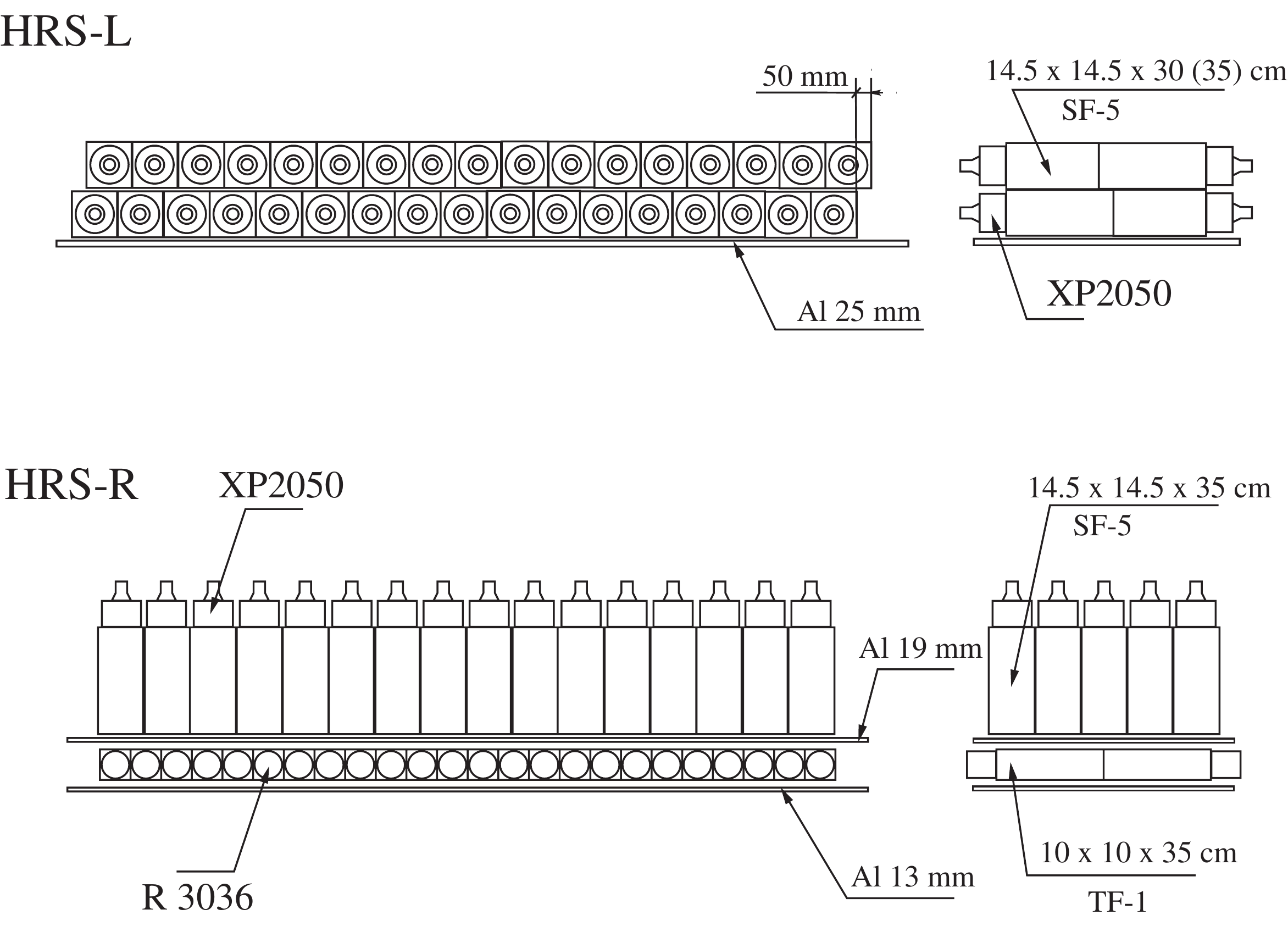}
  \caption[Schematic layout of calorimeters in HRS-L and HRS-R]{\footnotesize{Schematic layout of calorimeters in HRS-L and HRS-R. Figure is from Ref.~\cite{halla_nim}.}}
  \label{shower}
 \end{center}
\end{figure}

 In each HRS, a calorimeter is placed behind S2m for the energy measurement of charged particles. Each calorimeter is composed of two layers of lead glass blocks and associated PMTs (Fig.~\ref{shower}). The gaps between blocks in the first layer are covered by the blocks in the second layer. The two layers of the calorimeter in HRS-L are called Pion-Rejector-1 (PRL1) and Pion-Rejector-2 (PRL2), respectively, and each layer consists of two columns of 17 lead glass blocks. In HRS-R, the first layer of the calorimeter, also named as PreShower (PS), is formed by two columns of 24 lead glass blocks, while the second layer, called Shower (SH), has five columns of 16 lead glass blocks. 

 When propagating through the dense material, a high energy charged particle loses its energy exclusively through Bremsstrahlung radiation. The emitted photons sequentially create electron-positron pairs which generate secondary Bremsstrahlung radiation. Along the path in the material, an electromagnetic cascade is developed in the direction of incident particle. At the GeV energy scale, only electrons are able to develop such a cascade in the HRS calorimeter. Since heavier particles require a much longer path length, the calorimeter provides a useful substantial particle identification in addition to the GC. PS and SH are arranged to be a total absorber, while PRL1 and PRL2 still provide powerful capability of electron identification even though they don't form a total absorber.

%% file: setup/setup_daq.tex
\section{Data Acquisition System}
 The data-acquisition (DAQ) system in Hall-A is composed of the CEBAF online data acquisition system (CODA) developed by the JLab CODA group, and the associated hardware components. CODA is a tool-kit of software components, including read-out controllers (ROCs), the event builder (EB), the event recorder (ER) and the event-transfer (ET). The other major component is the Run-Control (RC) which is a graphical user interface to choose experimental configurations, start and stop runs, and monitor and reset CODA components~\cite{halla_nim}. The hardware elements are basically composed of front-end Fastbus crates, VME devices (ADCs, TDCs and scalers), VME-Fastbus interfaces, single-board VME computers, trigger supervisors (TS) and network components. CODA is operated on a Linux based workstation which stores the recorded data (called raw data) in the local hard-drive. The data is subsequently transferred to a mass storage tape silo (MSS) for long term storage. Data in the local hard-drive will be deleted when the hard-drive runs out of space. 
 
  The E08-014 ran consecutively with four other experiments during the spring of 2011. Besides the HRSs, the BigBite spectrometer and a neutron detector were installed in the hall for double-coincidence and triple-coincidence experiments. Triggers from four devices were sent to the same TS located in the electronic hut on the floor. When a trigger was accepted by the TS, a Level-One-Accept (L1A) signal was generated and sent back to each spectrometer. The leading-edge of the L1A signal was then adjusted by the strobe signal in a retiming-module (RT) installed in the local front-end crate. The signal from RT was fed to the Transition Module (TM)~\cite{ts_tm}, where an ADC gate, a TDC Start/Stop signal and control signals were generated and distributed to the front-end electronics on Fastbus crates and VME crates where ADCs, TDCs and scalers start to record data when these signals arrive. An event number associated with this trigger was registered in the DAQ system and all signals associated with this events were recorded.
  
   Limited by the dead-time and data size, not all triggers were accepted by the TS. A pre-scale factor was assigned to each trigger type to control the total event rate before CODA starts taking data. For example, a pre-scale factor "3" represents the case that only the first one can be accepted for every three consecutive events from the trigger, and a pre-scale factor "0" means that no event from the trigger will be recorded. Each time when CODA starts to take data, a unique run number is given to the raw data file which stores all events coming after the start of the run. To control the total size of the data file and to prevent the data file from being damaged by any errors during the data taking, CODA will be stopped when each run reaches a pre-defined length of time or a certain number of events, and then a new run will be started with a new run number. 

   Scaler events are read every 1-4 seconds and stored in the data stream. Meanwhile, data from the Experimental Physics and Industrial Control System (EPICS), such as the beam energy, the BPM and BCM readings, the information of the target system, the angles and magnet fields spectrometers, etc., are also inserted into the data stream for every few seconds.

\section{Trigger Design}
\begin{figure}[!ht]
 \begin{center}
  \includegraphics[angle=90,width=0.95\textwidth]{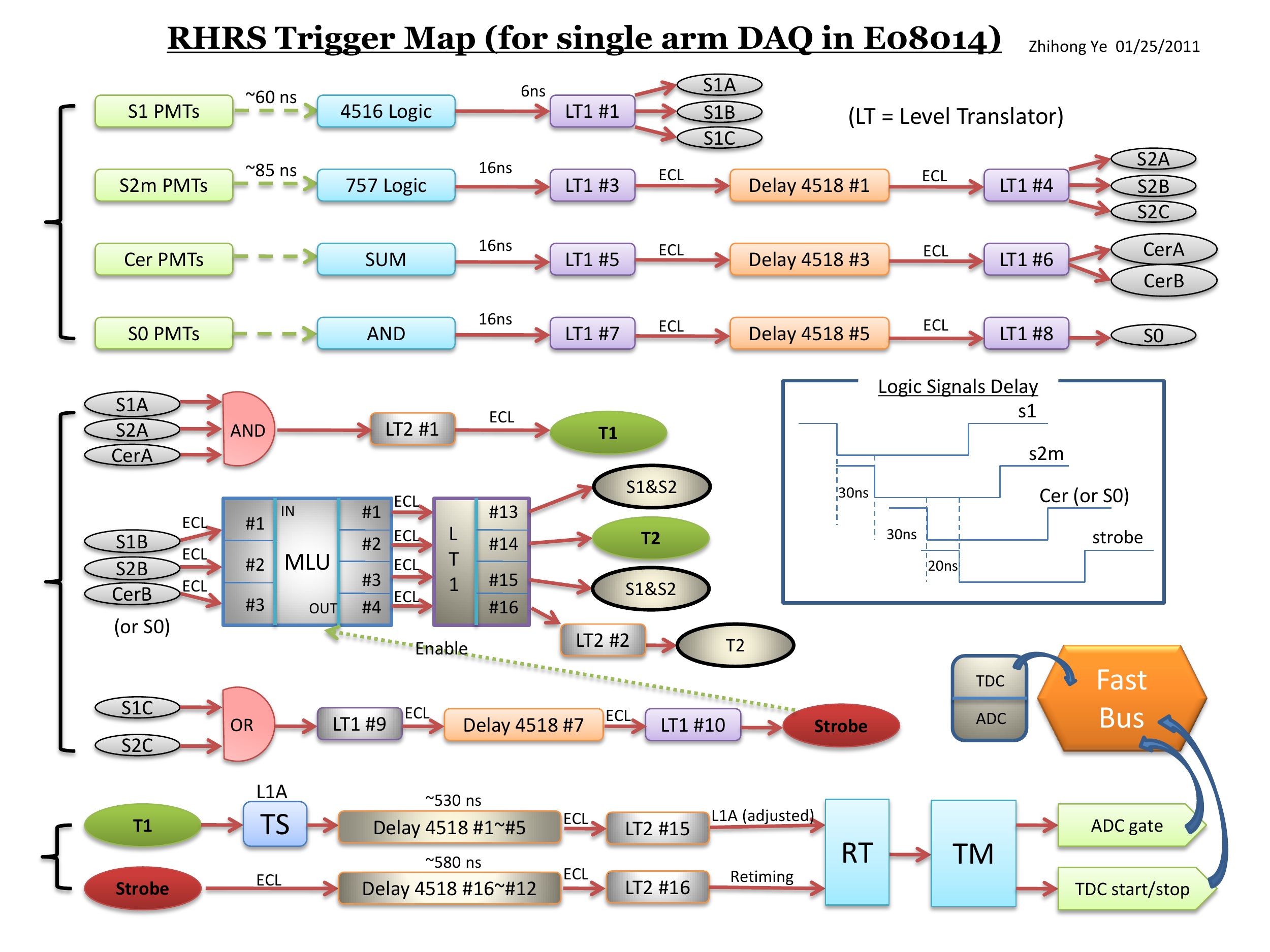}
  \caption[Single arm trigger design]{\footnotesize{Single arm trigger design on HRS-R, where the logic signal from the GC was also included to form the TS1 signal which produces the T1 trigger. The HRS-L trigger has the similar layout except some electronic modules were different.}}
  \label{trigger_design}
 \end{center}
\end{figure}

  During the E08-014 the scattered electrons were measured by both HRSs simultaneously. The BigBite and the neutron detector were turned off and their triggers were ignored. Both HRSs shared the similar design of the trigger system which is illustrated in Fig.~\ref{trigger_design}. Three detector planes, S1, S2m and GC were included in the trigger design. A logic signal was created when one or more scintillator bar in S1 or S2m was fired. The logic signal of the GC was the digital signal converted from the sum of ten PMT signals. The coincidence of logic signals from S1, S2m and GC created T1 (T3) trigger in HRS-R (HRS-L), which was the production trigger in this experiment. T2 (T4) was formed in HRS-R (HRS-L) by the coincident signal of the GC logic signal and only one of the S1 or S2m logic signal. T2 (T4) was designed to evaluate the trigger efficiency of T1 (T3). T6 (T7) was generated from the overlapped signal of S1 and S2m, and is the traditional HRS main trigger. Events from T6 and T7 were used for particle identification study since pions were also recorded. T5 is the coincident signal of T1 and T3, and was disabled in this experiment. A discussion of triggers during the data analysis is given in Appendix A.
  
%% file: analysis/analysis.tex
\chapter{Calibration}

\section{Overview}
 Experimental raw data collected by the DAQ system is stored in individual data files. Each file is associated with a unique run number and hence is also called "a run". The raw data contains plentiful information on each event, including the experimental settings during the run and all the signal readouts from experimental instruments. However, those information can not be directly read out for offline analysis. The Hall A C++ Analyzer~\cite{analyzer}, an object-oriented framework on top of ROOT~\cite{cern_root} and developed by the Hall A software group, is used to replay the raw data, extract and calculate important quantities, and store these quantities in ROOT files which can then be directly accessed through the ROOT interface or C/C++ subroutines. Each ROOT file contains several subdirectories which are called  "trees". The event-by-event detector readouts, including both the uncalibrated and calibrated signals, are stored in the \emph{\bf{T}} tree. The EPICS readings are put in the \emph{\bf{E}} tree, and the \emph{\bf{RIGHT}} tree and \emph{\bf{LEFT}} tree store signal readouts from scalers in HRS-R and HRS-L, respectively.
 
   During the data replay, each quantity must be correctly linked to the corresponding readout-signal with an up-to-date map which contains the front-end crate number, the model of the electronic module and the slot ID in the FastBus crate, and the channel number which the signal cable connects to. Such a map is given in an individual file associated with the instrument. The Analyzer's data base (DB) stores these files for all Hall-A instruments. The parameters to convert the raw signals into calibrated quantities are also stored in the DB.  
   
  The first step of the data analysis is to calibrate the parameters for each instrument with the calibration data, which will be discussed in this chapter. After these parameters are updated, the raw data will be replayed again and the new ROOT files can be used to extract useful physics quantities, for example, inclusive cross sections, as given in the next chapter. 
  
  In this experiment, the calibration is composed of three major parts: beam instruments, detector packages and optics matrices of the HRSs. 
 
 The calibration of beam instruments aims to obtain the parameters of the beam position monitors (BPMs) and the raster system which determines the event-by event beam position, and to calculate the accumulated beam charge from the scaler readings of the beam charge monitors (BCMs). The beam position calibration has been processed during the experiment by using the Harp scan data~\cite{bpm_runs}. The detailed calibration procedure of the BPMs and the raster system can be found in Ref.~\cite{bpm_cali}. The result of BCM calibration is given in Ref.~\cite{bcm_patricia}, and the calculation of beam charge will be presented in Section 5.2.
 
 Each detector in the HRS can be individually calibrated, while the calibration of HRS optics requires a good determination of the beam position and an updated reference time ($\mathrm{T_{0}}$) for each VDC wire. $\mathrm{T_{0}}$ can be changed when the parameters of the TDC signals from S1 and S2m are updated. In this experiment, S1 and S2m were unable to be calibrated because several TDC channels showed multiple peaks in each TDC spectrum and the real signals could not be identified. The values of $\mathrm{T_{0}}$ were calculated with old S1 and S2m parameters. The detailed calibration of the gas \v{C}erenkov detectors, calorimeters and the HRS optics will be given in next two sections.

\input analysis/analysis_detector.tex

\input analysis/analysis_optics.tex

%% file: analysis/analysis_detector.tex
\section{Detector Calibration}
   The gas \v{C}erenkov detector (GC) in each HRS contains 10 PMTs. The calorimeter in HRS-L contains two layers, Pion-Rejector-1 (PRL1) and Pion-Rejector-2 (PRL2), each of which has 34 PMTs. The calorimeter in HRS-R has 48 and 80 PMTs in Pre-Shower (PS) and Shower (SH), respectively. These PMTs collect the signals created by a particle passing through the detectors. The read-out signal from each PMT is split into two copies which are then recorded in the TDC and the ADC front-ends, respectively.
   
   For a common-stop TDC module, the channel numbers in the TDC spectrum represent the time difference between when the event triggers and when the STOP signal arrives. The channel numbers in the ADC spectrum, on the other hand, are directly related to the strength of the PMT signals. However, the PMT signal not only is proportional to the photon energy, but also depends on the high voltage on the PMT as well as the amplitude of the background signal. Hence, when collecting the signals with the same strength, different PMTs in the same detector may give different channel numbers in their ADC spectra. In the DB, each detector is associated with a group of parameters, or called gain factors, which can convert the channel number of each ADC spectrum into a common energy unit. These gain factors have to be calibrated each time the detector configuration is modified. After updating the gain factors in the DB and performing the new data replay, the calibrated ADC spectra can be added up together to obtain the total energy deposited by the particles in the detector.
 
\subsection{Gas \v{C}erenkov Detectors}
\begin{figure}[!ht]
  \begin{center}
    \subfloat[Before alignment]{
      \includegraphics[type=pdf,ext=.pdf,read=.pdf,width=1.0\textwidth]{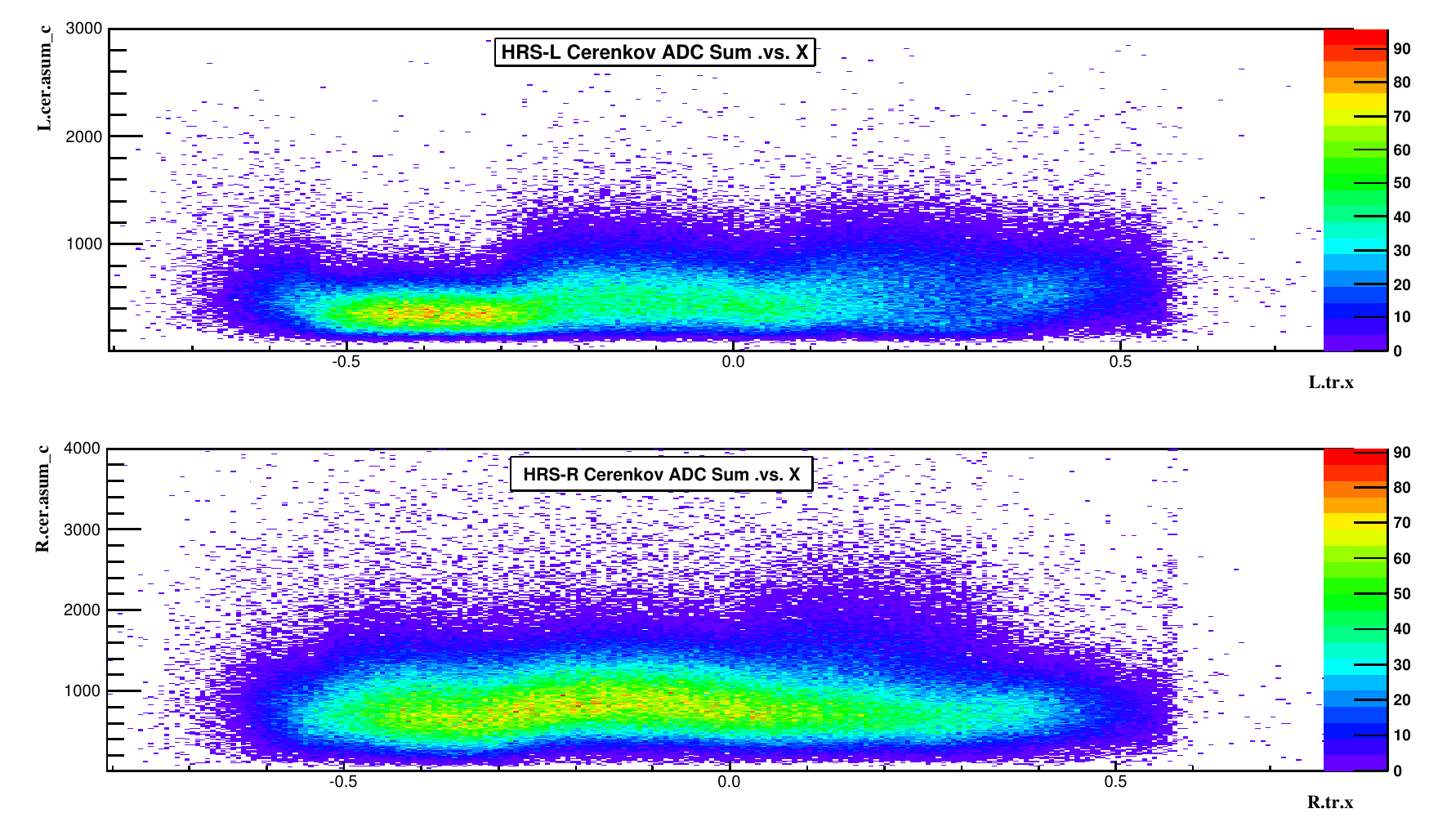}
    }
    \\ 
    \subfloat[After alignment]{
      \includegraphics[type=pdf,ext=.pdf,read=.pdf,width=1.0\textwidth]{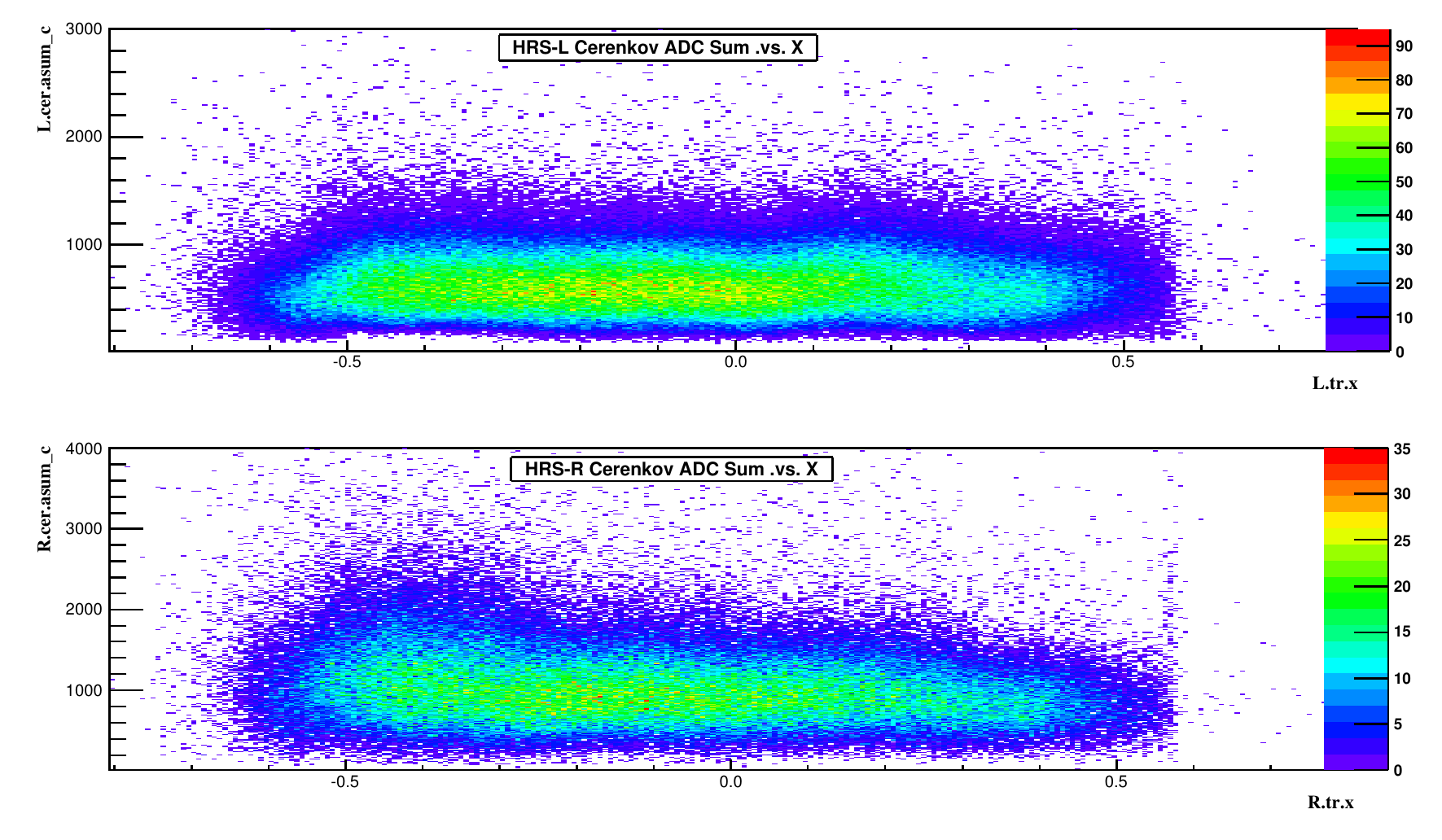}
    }
    \caption[Alignment of gas \v{C}erenkov detectors]{\footnotesize{Alignment of gas \v{C}erenkov detectors (GC). Each 2-D histogram gives the distribution of the sum of the GC ADC spectra along the detector plane. Plots in (a), top for HRS-L and bottom for HRS-R, show that the ADC peaks are off by certain channels before the calibration. Plots in (b) demonstrate that those peaks are nearly at the same channel number after the alignment.}}
    \label{cer_align}
  \end{center}
\end{figure}

 The energy of a single photon which causes the emission of the single photo-electron (SPE) is only determined by the material of the photocathode in the PMT. If all PMTs for the detector are from the same model, the SPE peaks should represent the same photon energy in their ADC spectra. A calibration procedure of the GC aims to adjust the single photon electron (SPE) peak in each ADC spectrum to appear at channel 100. The gain factor for the $ith$ PMT is defined as:
\begin{equation}
  C_{i} = \frac{100}{M_{i}^{SPE}-M_{i}^{ped}},
  \label{eq_cer_gain}
\end{equation}
where $M_{i}^{SPE}$ and $M_{i}^{ped}$ are the mean values of the SPE peak and the pedestal peak in the $ith$ ADC spectrum. 
\begin{figure}[!ht]
  \begin{center}
    \includegraphics[type=pdf,ext=.pdf,read=.pdf,width=0.60\linewidth]{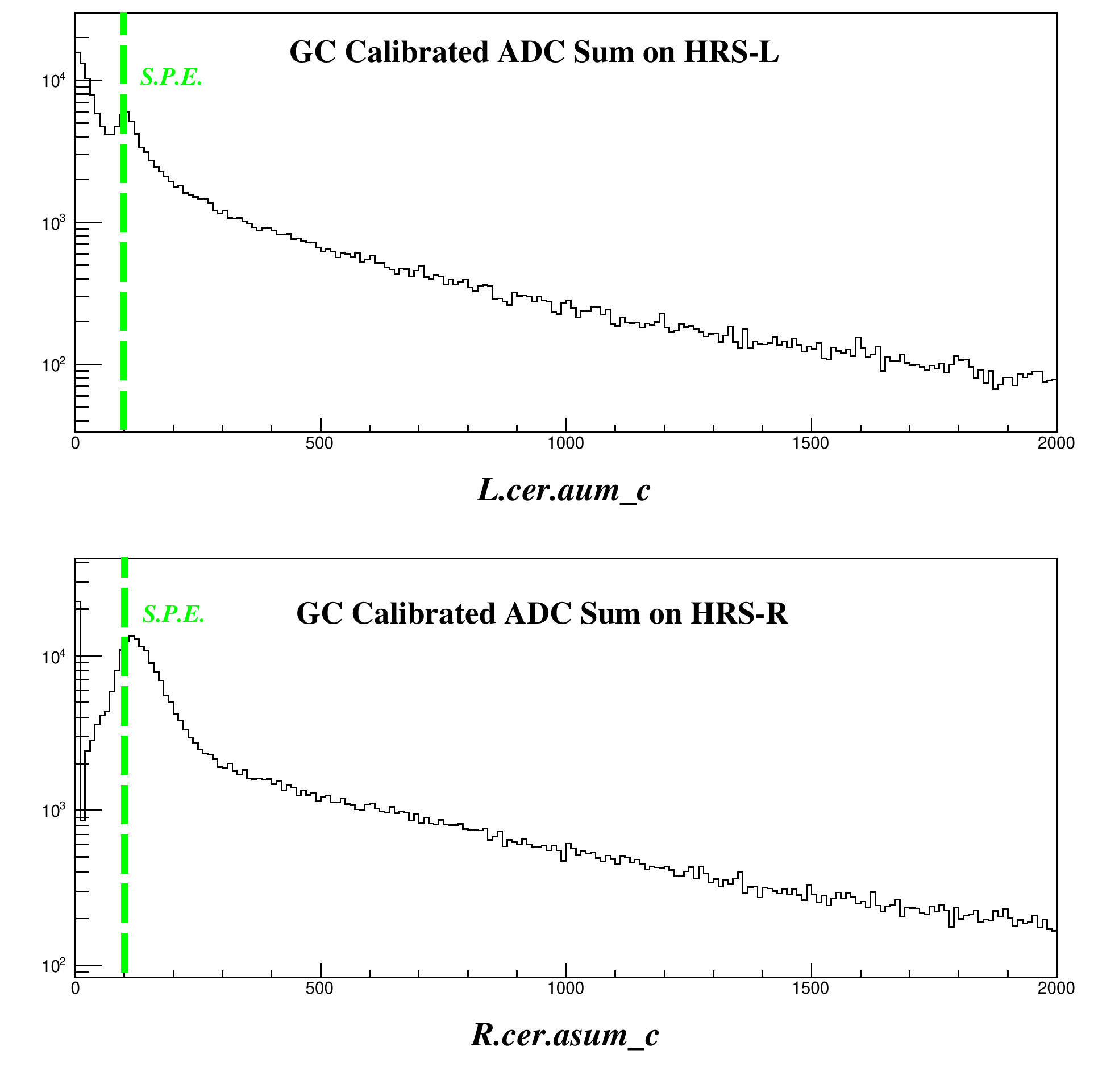}
    \caption[Single electron peaks on both GCs]{\footnotesize{Single photon electron peaks (SPE) in the sum of the GC ADC spectra. Top for HRS-L and Bottom for HRS-R. The green lines indicate that the SPE peaks have been properly aligned at 100 ADC channels on each GC. (Narrow the range to 500!)}}
  \label{cer_spe}
  \end{center}
\end{figure}

 The SPE peaks didn't show on the ADC spectra when plotting events from the main production triggers (T1 for HRS-R and T3 for HRS-L). A threshold was set on the GC to form these triggers, and it excluded most of weak signals, including SPEs. The calibration was performed with events from the T6 trigger for HRS-R and the T7 trigger for HRS-L which didn't include the GCs. The raw ADC spectrum of each PMT was plotted and the channel numbers of the pedestal peak and the SPE peak were identified and recorded. The gain factors for all ten PMTs in the GC were calculated with Eq.~\eqref{eq_cer_gain} and their values were updated in the DB. The data was replayed again with these new parameters and then the calibrated ADC spectra for all PMTs had the same energy scale.
 
 Fig.~\ref{cer_align} shows that the calibrated ADC spectra were well aligned. The sum of ten calibrated ADC spectra clearly shows the SPE peak located at channel 100, as shown in Fig.~\ref{cer_spe}, and can now be directly used for the particle identification.

\subsection{Electromagnetic Calorimeters}
  The Hall-A calorimeters are able to measure the energy of few GeV electrons deposited exclusively in the detectors. The resolution of the energy measurement is determined by the design of calorimeters, as well as the energy range and the type of charge particles. In general, the energy resolution of a calorimeter can be parametrized by\cite{R_Bock}:
\begin{equation}
  \frac{\sigma(E)}{E} = a \oplus \frac{b}{\sqrt{E}},
\end{equation}
where $\oplus$ represents two terms added in quadrature. The first term is mainly contributed by systematic errors, such as intrinsic shower fluctuations, which should be small for homogeneous calorimeters, such as total absorbers. The value of second term is determined by the uniformity of calorimeters as well as uncertainty of detector calibration. It is typically $\mathrm{5\%/\sqrt{E}}$ for lead glass calorimeters. 

  The electron creates a track during the cascade and the lead glass blocks along the track collect these photon signals. During the data replay, these blocks can be identified by using the VDC tracking information, and within the same layer the group of these blocks is called a cluster. The sum of their ADC spectra after the calibration denotes the energy deposited in this cluster, and should be distinguished from the sum of all calibrated ADC spectra since blocks outside the cluster only pick up the background signals. Once the individual ADC spectra are calibrated, the sum of the energy deposited in both layers should be equal to the energy (or equivalently the momentum) of scattered electrons. A new variable, E/P, is defined as the ratio of the energy sum of two clusters to the electron's momentum, and should be centered at one if the gain factors are properly calibrated. Different procedures were applied on each calorimeter. 
 
 \begin{figure}[!ht]
  \begin{center}
     \subfloat[Pion Rejectors on HRS-L]{
      \includegraphics[type=pdf,ext=.pdf,read=.pdf,width=0.75\linewidth]{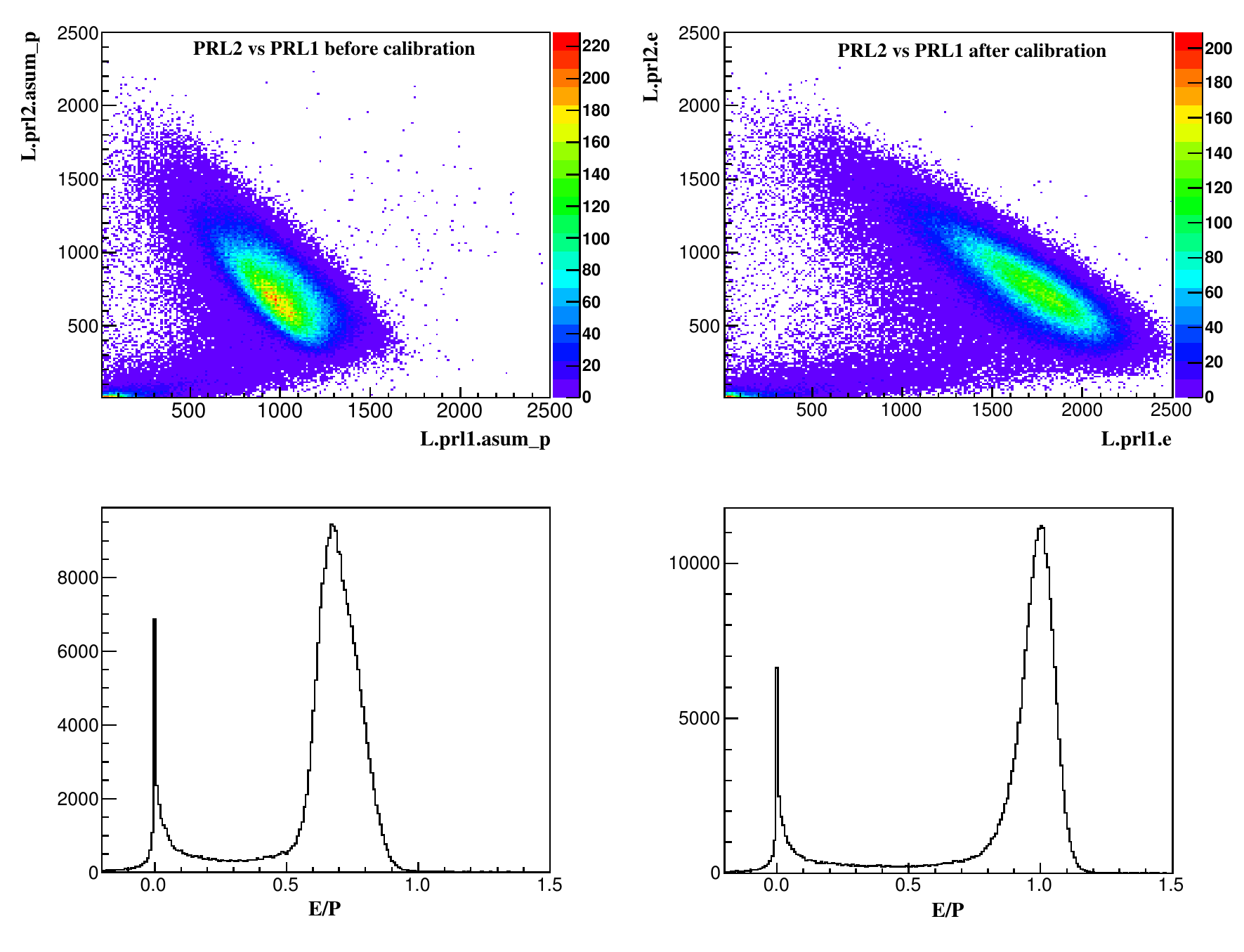}
      \label{prl_cali}
   }
   \\
    \subfloat[Pre-Shower and Shower on HRS-R]{
      \includegraphics[type=pdf,ext=.pdf,read=.pdf,width=0.75\linewidth]{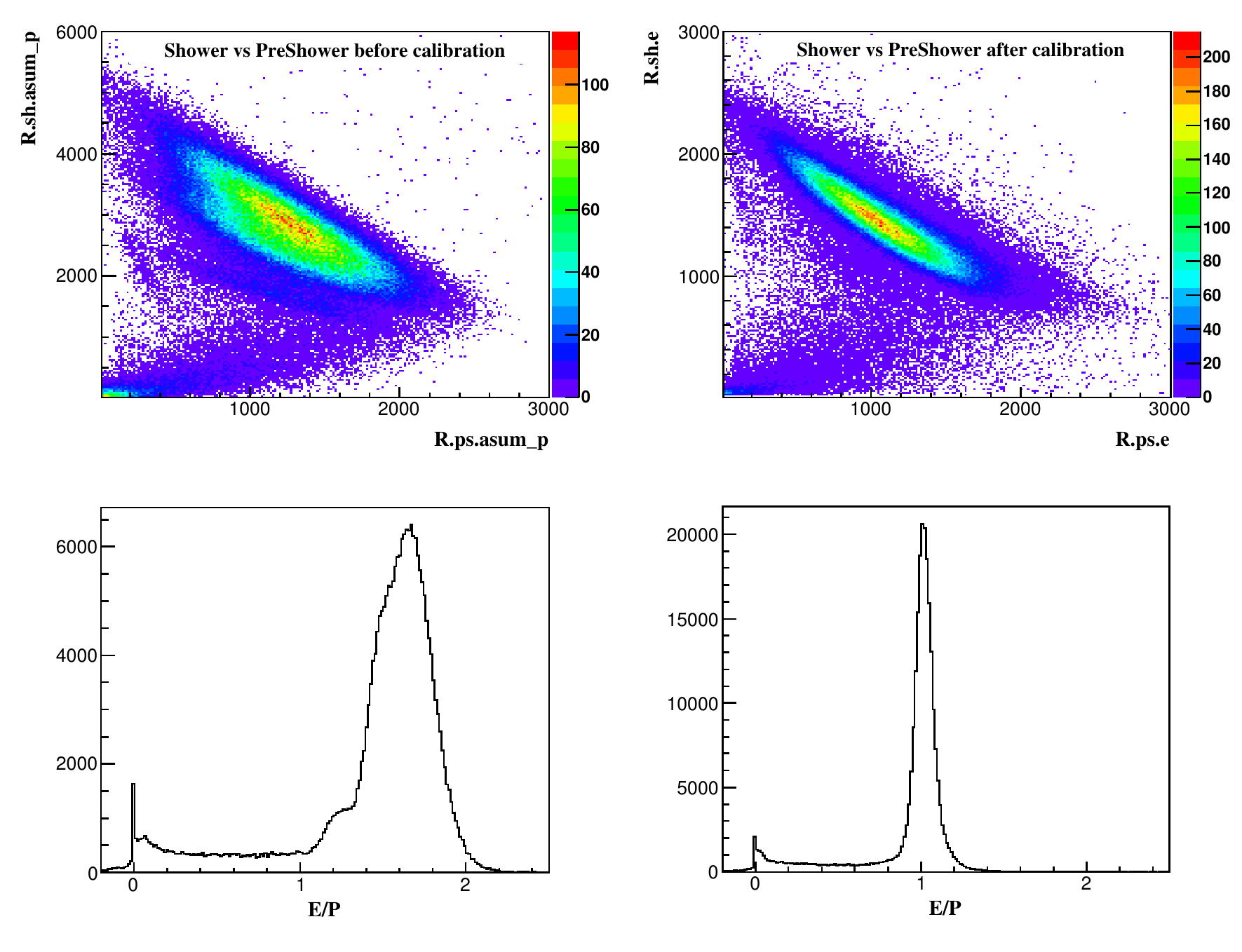}
      \label{sh_cali}
      }
   \caption[Calibration of calorimeters]{\footnotesize{Calibration of calorimeters. In each figure, the top two plots are the 2-D histograms of the PRL1 (PS) ADC sum versus the PRL2 (SH) ADC sum before and after the calibration. The electron band is clearly isolated after the calibration. The bottom two 1-D histograms are the distributions of E/P before and after the calibration. The peak becomes sharp and locates at one with new gain factors.}}
   \label{calo_cali}
  \end{center}
\end{figure}
 A minimization method was used to calibrate the calorimeter on HRS-R, composed of two layers, Pre-Shower (PS) and Shower (SH)\cite{shower_ak}. The Chi-Square was defined as: 
\begin{equation}
  \chi^{2} = \sum_{i=1}^{N}\left[\sum_{j\in M_{ps}^{i}}C_{j}\cdot (ADC_{j}^{i}-Ped_{j})+\sum_{k\in M_{sh}^{i}}C_{k}\cdot (ADC_{k}^{i}-Ped_{k})-P_{kin}^{i}\right]^{2},
\end{equation}
where \emph{i} is the \emph{ith} event; \emph{j} is the \emph{jth} PS block; \emph{k} is the \emph{kth} SH block; $M_{ps}^{i}$ and $M_{sh}^{i}$ are sets of PS and SH blocks included in the reconstructed cluster for the \emph{ith} event; $ADC_{j/k}^{i}$ and $Ped_{j/k}$ represent the ADC channel number of the event and mean pedestal value in the ADC spectrum, respectively; $P_{kin}^{i}$ is the particle momentum of the \emph{ith} event; and $C_{j/k}$ is the gain factor of the ADC spectrum used as a fitting parameter during the minimization.

 To obtain the best fitting result, electron samples were selected from data taken in the QE tail ($x_{bj}>1$) where scattered electrons were uniformly distributed among all lead glass blocks. A minimization package \cite{shower_luhj} was called to minimize $\chi^{2}$, and the gain factors obtained from the fitting parameters were stored in the database.

  The calorimeter on HRS-L, composed of the layers of Pion-Rejector-1 (PRL1) and Pion-Rejector-2 (PRL2), is not a total absorber and the minimization method used on the PS and SH is not applicable. Instead, it was calibrated by aligning the minimum ionization peak of each ADC spectrum to a common channel number, similar to the GC calibration. The cosmic ray events were used during the calibration since they were uniformly distributed along the entire blocks. Furthermore, the particles in cosmic ray are mostly muons which have small energy spread. The pedestal peak ($ADC_{i}^{ped}$) and muon peak ($ADC_{i}^{muon}$) in the ADC spectrum of the $ith$ PMT were located and their distance were aligned to 100, by applying a gain factor defined as:
\begin{equation}
  C_{i} = \frac{100}{ADC_{i}^{muon}-ADC_{i}^{ped}}.
\end{equation}
 The gain factors for all PMTs in PRL1 and PRL2 were calculated similarly. With these updated gain factors in the data base, the E/P was calculated at the new data replay. To shift the peak of the E/P distribution to one, the gain factors were further adjusted:
\begin{equation}
  C_{i}^{real} = C_{i} \times \frac{1}{M_{E/P}},
\end{equation}
where $M_{E/P}$ represents the mean value of the E/P peak before the adjustment. The adjusted gain factors were then updated in the data base and the data was replayed again.
\begin{figure}[!ht]
  \begin{center}
    \subfloat[on HRS-L]{
      \includegraphics[type=pdf,ext=.pdf,read=.pdf,width=0.9\textwidth]{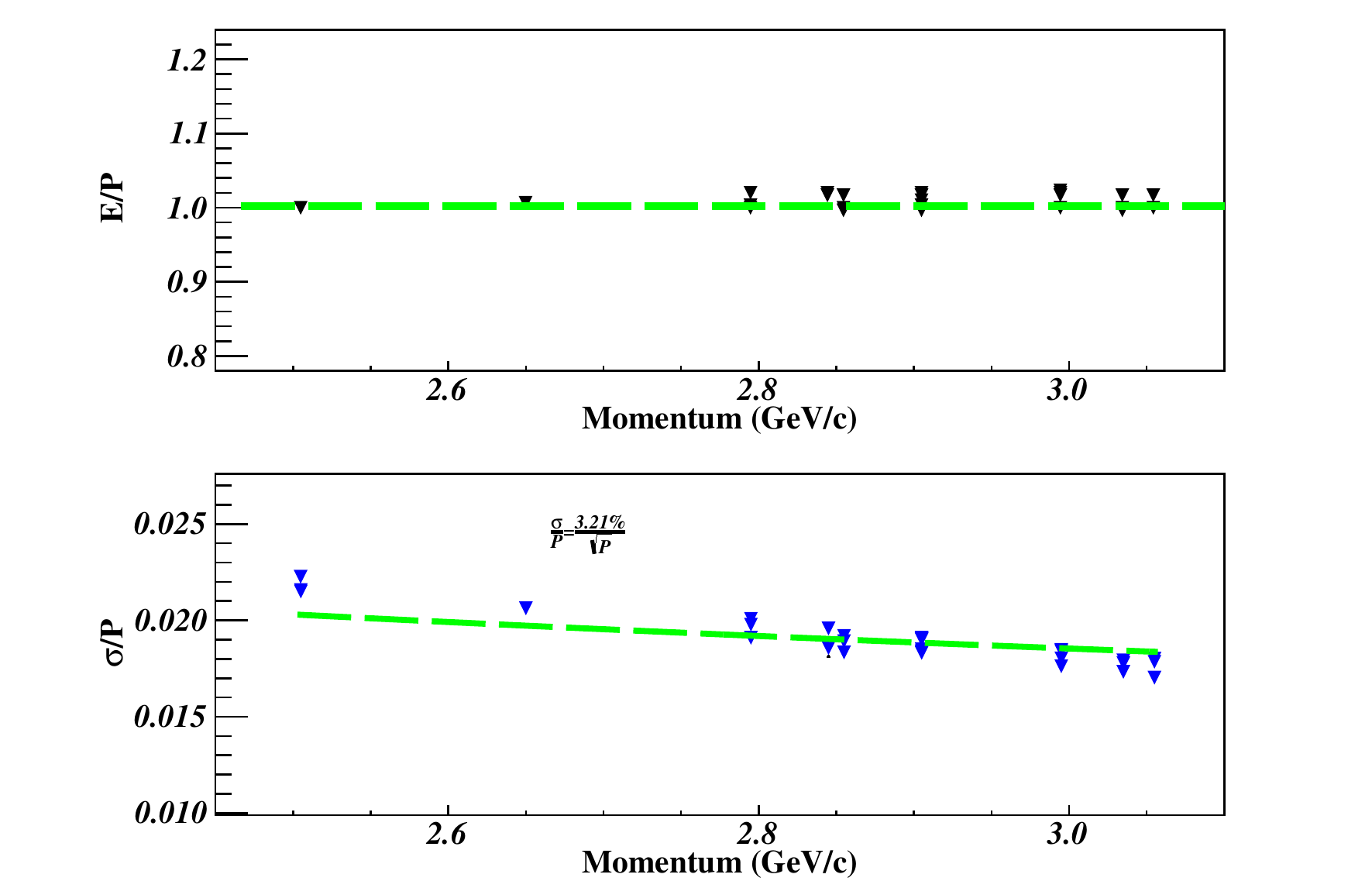}
    }
    \\ 
    \subfloat[on HRS-R]{
      \includegraphics[type=pdf,ext=.pdf,read=.pdf,width=0.9\textwidth]{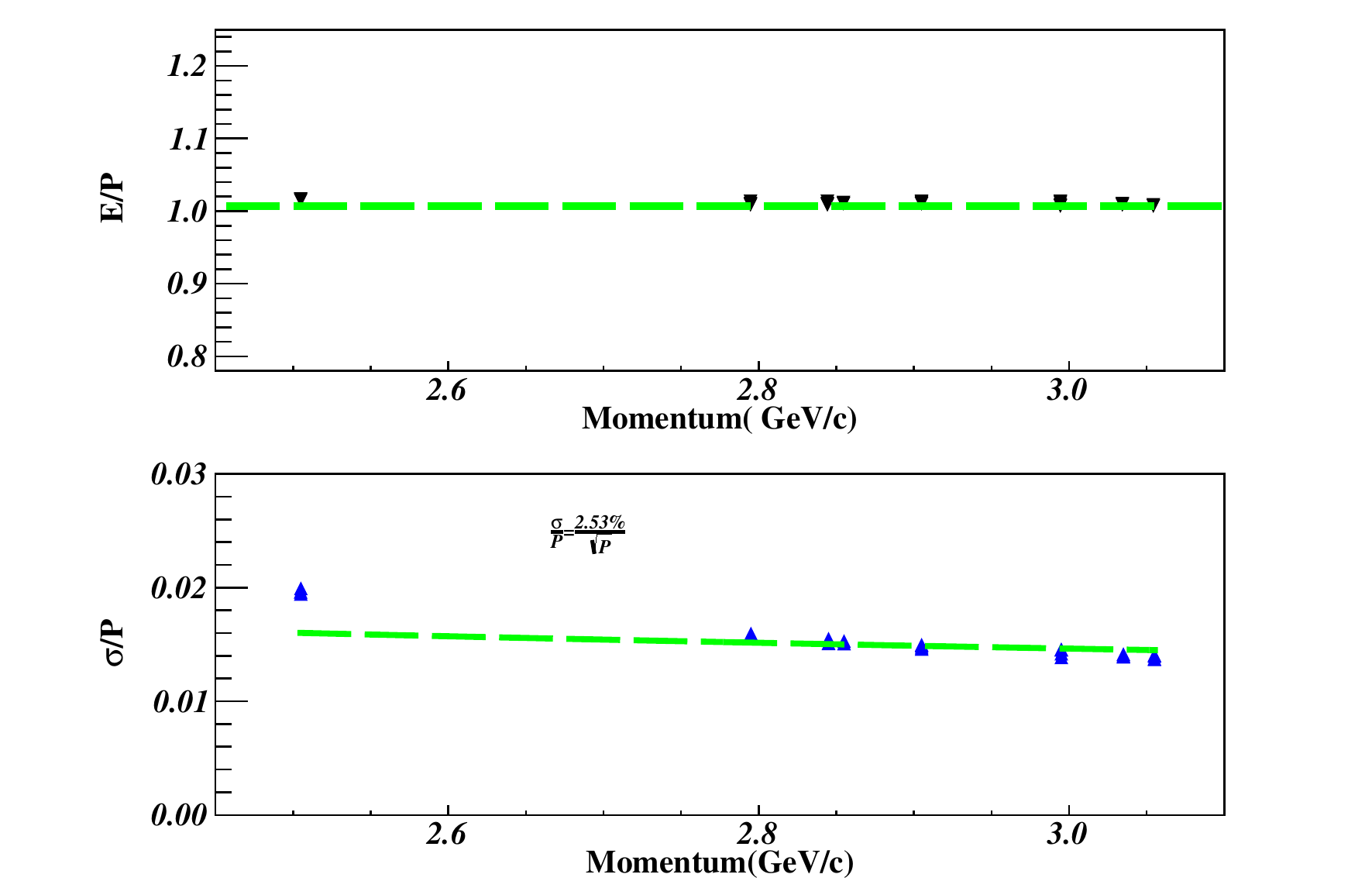}
    }
    \caption[Calibration performance and resolution of calorimeters]{\footnotesize{Calibration performance and resolution of calorimeters. The top plot in each figure reveals the performance of calibration at different momentum setting. The two bottom plots give the resolution of calorimeters, which are 3.21\%$\mathrm{/\sqrt{GeV}}$ on HRS-L and 2.53\%$\mathrm{/\sqrt{GeV}}$ on HRS-R.} }
    \label{calo_resol}
  \end{center}
\end{figure}

 The calibration results of both calorimeters are shown in Fig.~\ref{calo_cali}, where electrons are better separated from backgrounds and E/P is well centered at one. The locations of E/P peaks at different momentum settings are shown in Fig.~\ref{calo_resol}. The energy resolution was also given by fitting the spread of \emph{E/P} peaks as the function of the momentum. The overall resolution of PS and SH is 2.53\%$\mathrm{/\sqrt{GeV}}$. The resolution of Pion Rejectors is 3.21\%$/\mathrm{\sqrt{GeV}}$, slightly worse as they are not total absorbers.

%% file: analysis/analysis_optics.tex
\section{HRS Calibration}
 After exiting the target chamber, a charged particle travels a long distance within the magnets of the HRS, and its trajectory after the Q3 exit is determined by two VDCs placed at the focal plane of the HRS. By using the focal plane quantities, an optics matrix reconstructs the particle's position and direction in the target plane where the electron interacts with the target.

 The standard HRS optics matrices have already been optimized in previous Hall-A experiments. However, the absolute positions of the target, the HRS and detectors change from time to time, so these offsets should be taken into account in the optics matrices. Furthermore, during the E08-014, the momentum of the third quadrupole in HRS-R (RQ3) was limited to 2.8273 GeV/c due to a power supply issue, while our maximum momentum setting was 3.055 GeV/c. The RQ3 field had to be scaled down to 87.72\% of the dipole field for each setting. Therefore, the previous HRS-R optics matrix was not applicable. 
 
 In this section, a calibration procedure to obtain new optics matrices for this experiment will be introduced.

\subsection{Coordinate Systems}
  \begin{figure}[!ht]
    \begin{center}
      \includegraphics[type=pdf,ext=.pdf,read=.pdf,angle=270,width=0.60\textwidth]{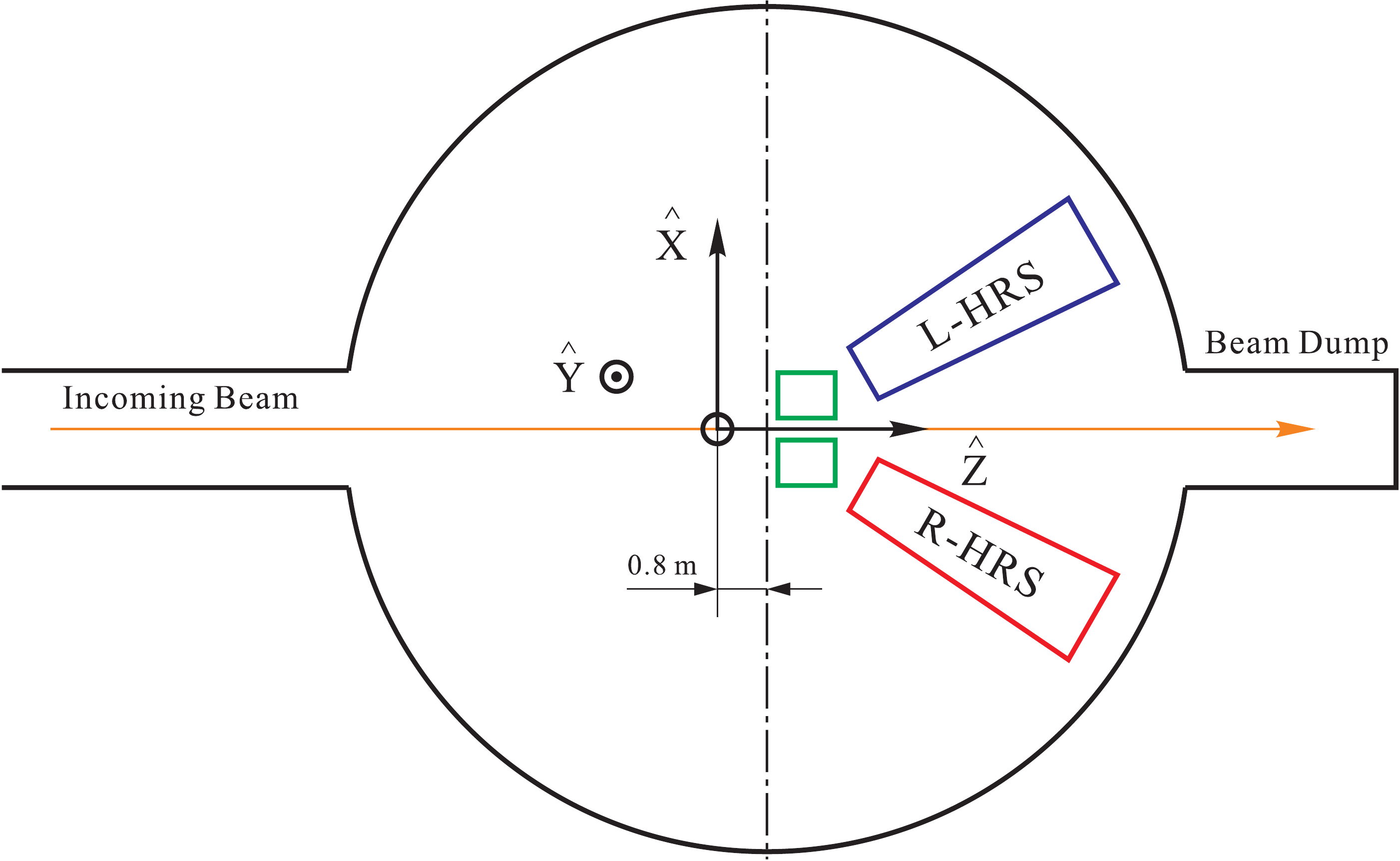}
      \caption[Hall coordinate system (HCS)]{\footnotesize{Hall coordinate system (HCS), which defines the beam position and the target location with respect to the hall center which is given as the intersection of the beam and the vertical axis of the target system. $\hat{x}$ is to the left of the beam direction, $\hat{z}$, and $\hat{y}$ is vertically up.}}
      \label{HCS}
    \end{center}
  \end{figure}
 The coordinates used during data analysis are briefly presented here. A more detailed description of Hall A coordinate systems and the translation between coordinates is given in Ref.~\cite{nilanga_optics}. Notes that angles defined in all coordinates are the tangent of their values.
\begin{itemize}
\item \textbf{Hall Coordinate System (HCS)} \\
  The center of the HCS is defined as the intersection of the beam and the vertical axis of the target system. $\hat{z}$ is along the direction of the beam, $\hat{x}$ is to the left of $\hat{z}$ and $\hat{y}$ is vertically up (Fig.\ref{HCS}).
  \begin{figure}[!ht]
    \begin{center}
      \includegraphics[type=pdf,ext=.pdf,read=.pdf,angle=270,width=0.85\textwidth]{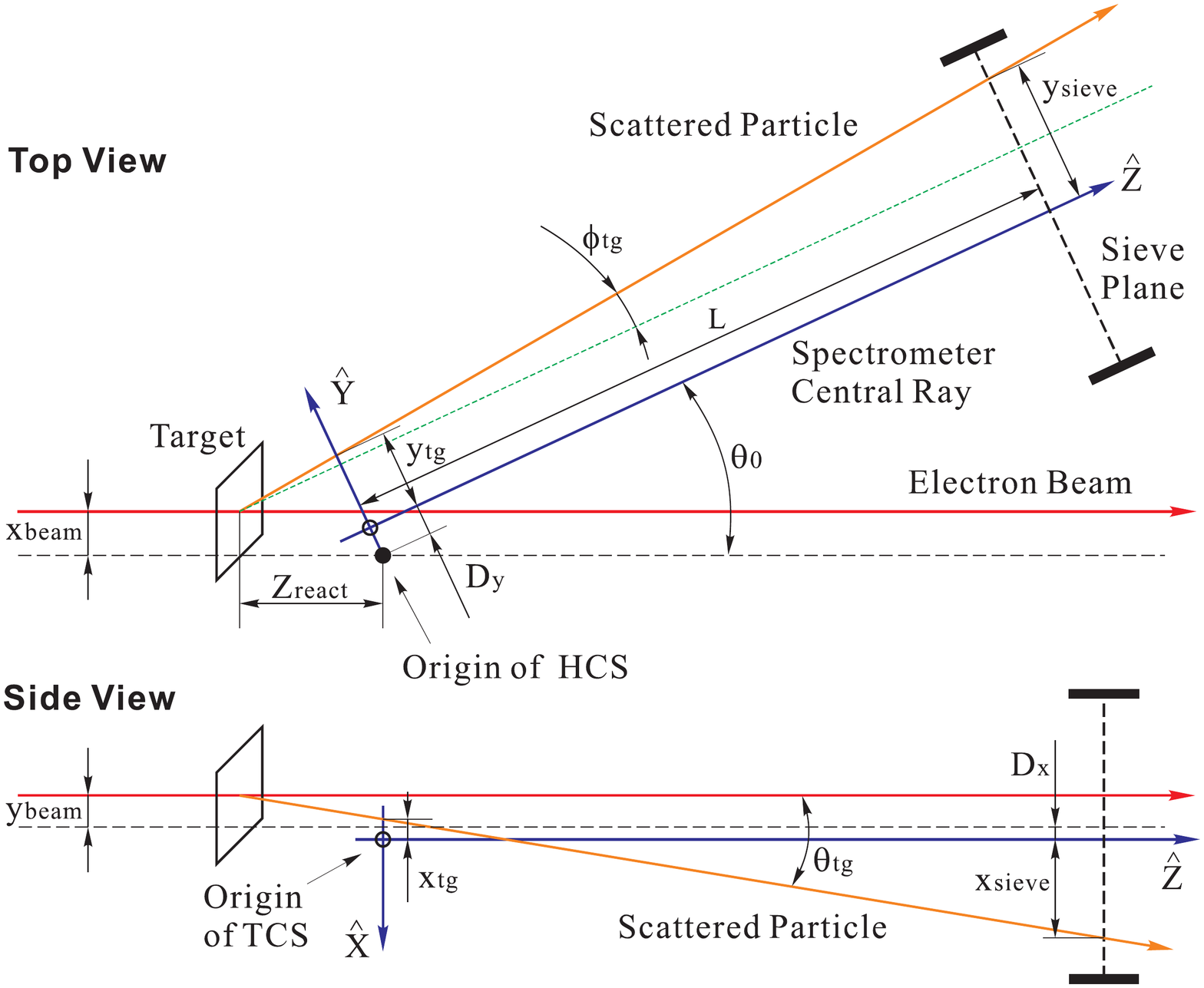}
      \caption[Target coordinate system (TCS)]{\footnotesize{Target coordinate system (TCS). $\hat{z}$ goes from the target system perpendicularly to the center hole of the sieve slit plane attached to the Q1 entrance of each HRS. The intersection of $\hat{z}$ and the vertical axis of the target system defines the origin, hence there is a potential offset between the hall center and the origin of TCS. $\hat{x}$ is normal to $\hat{z}$ and points down, and $\hat{y}$ is to the left of $\hat{z}$.}}
      \label{TCS}
    \end{center}
  \end{figure}
    \begin{figure}[!ht]
    \begin{center}
      \includegraphics[type=pdf,ext=.pdf,read=.pdf,angle=270,width=0.7\textwidth]{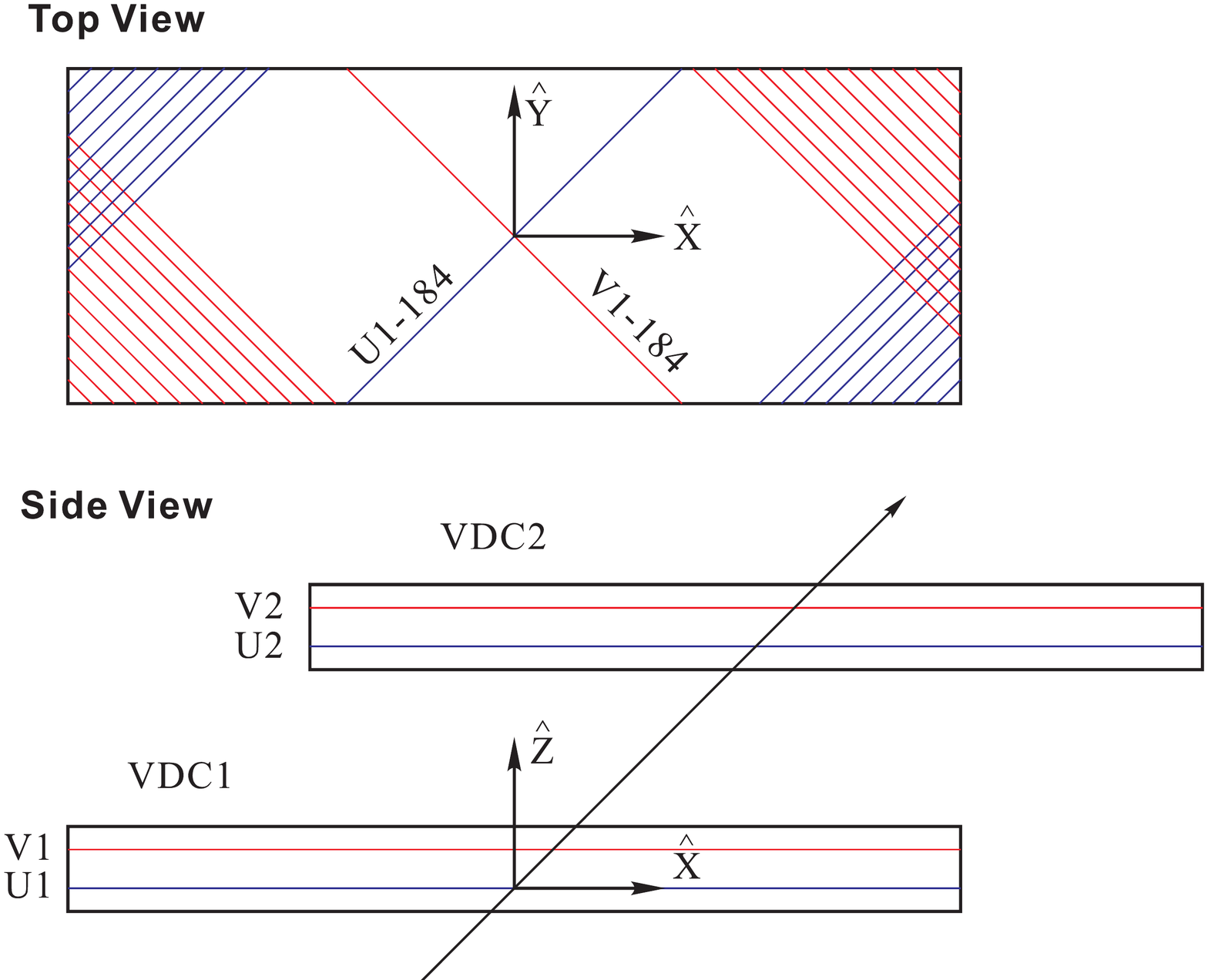}
      \caption[Detector coordinate system (DCS)]{\footnotesize{Detector coordinate system (DCS). Its origin is given by the intersection of wire-184 of U1 and wire 184 of V1. $\hat{z}$ is perpendicular to the VDCs, $\hat{x}$ is horizontally along the long edge of the VDC and points away from the hall center, and $\hat{y}$} is along the short edge of the VDC and vertically up.}
      \label{DCS}
    \end{center}
  \end{figure}
\item \textbf{Target Coordinate System (TCS)}   \\
  As shown in Fig.\ref{TCS}, $\hat{z_{tg}}$ is the direction from the target system perpendicular through the center hole in the sieve slit plane on each spectrometer. The origin of TCS is given by the intersection of $\hat{z}_{tg}$ and the vertical axis of the target system, and $L$ is a constant length from the origin of TCS to the sieve slit plane. $\hat{x}_{tg}$ is parallel to the sieve slit plane and vertically down, and $\hat{y}_{tg}$ is to the left of $\hat{z}_{tg}$. $\hat{\theta}_{tg}$ (the out-of-plane angle) and $\hat{\phi}_{tg}$ (the in-plane angle) are taken to be $dx_{sieve}/L$ and $dy_{sieve}/L$. The origins of HCS and TCS are not necessarily in the same location and the value of \emph{D},  the offset between two positions, changes when moving HRSs to different angles. Surveys are required during the experiment running to obtain the offset value.
    \begin{figure}[!ht]
    \begin{center}
      \includegraphics[type=pdf,ext=.pdf,read=.pdf,angle=270,width=0.60\textwidth]{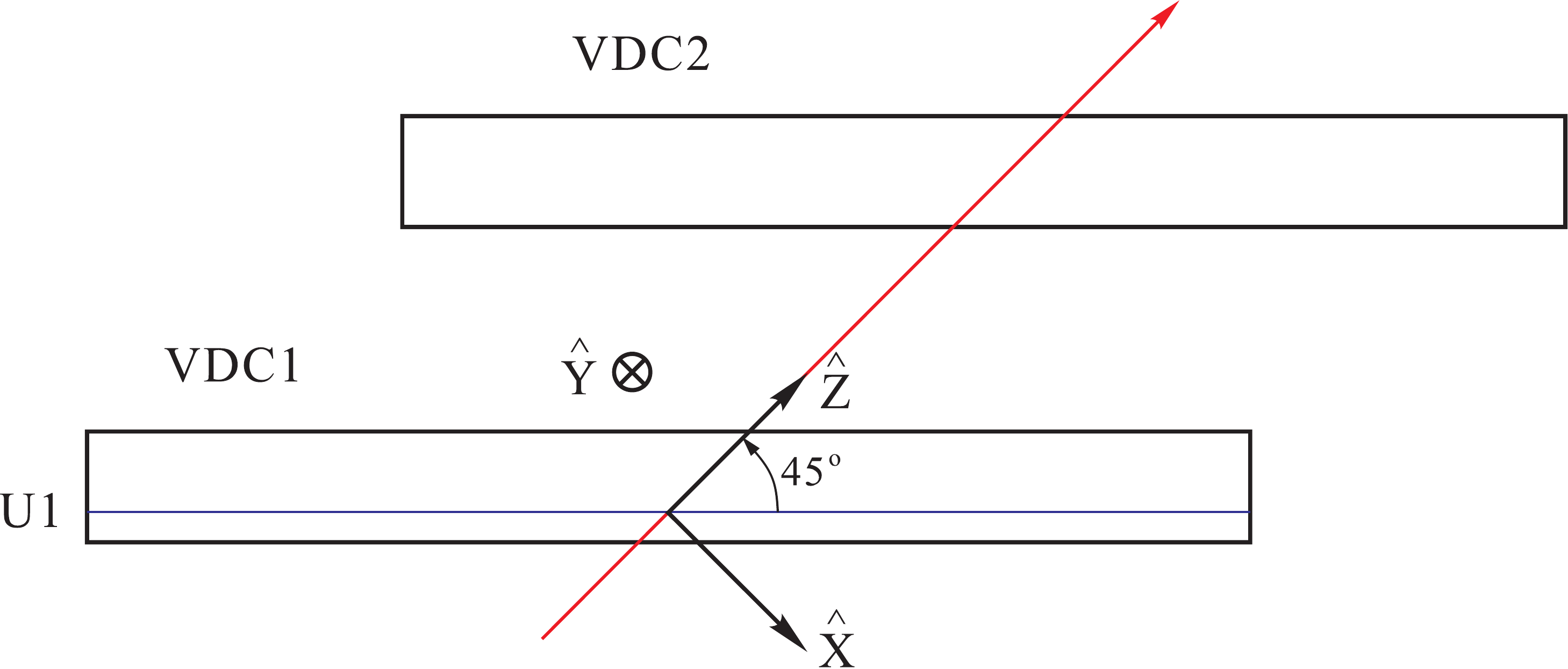}
      \caption[Transport coordinate system (TRCS)]{\footnotesize{Transport coordinate system (TRCS), which is generated by rotating the DCS clockwise around  $\hat{y_{det}}$ by $45^{\circ}$.}}
      \label{TRCS}
    \end{center}
  \end{figure}
    \begin{figure}[!ht]
    \begin{center}
      \includegraphics[type=pdf,ext=.pdf,read=.pdf,angle=270,width=0.65\textwidth]{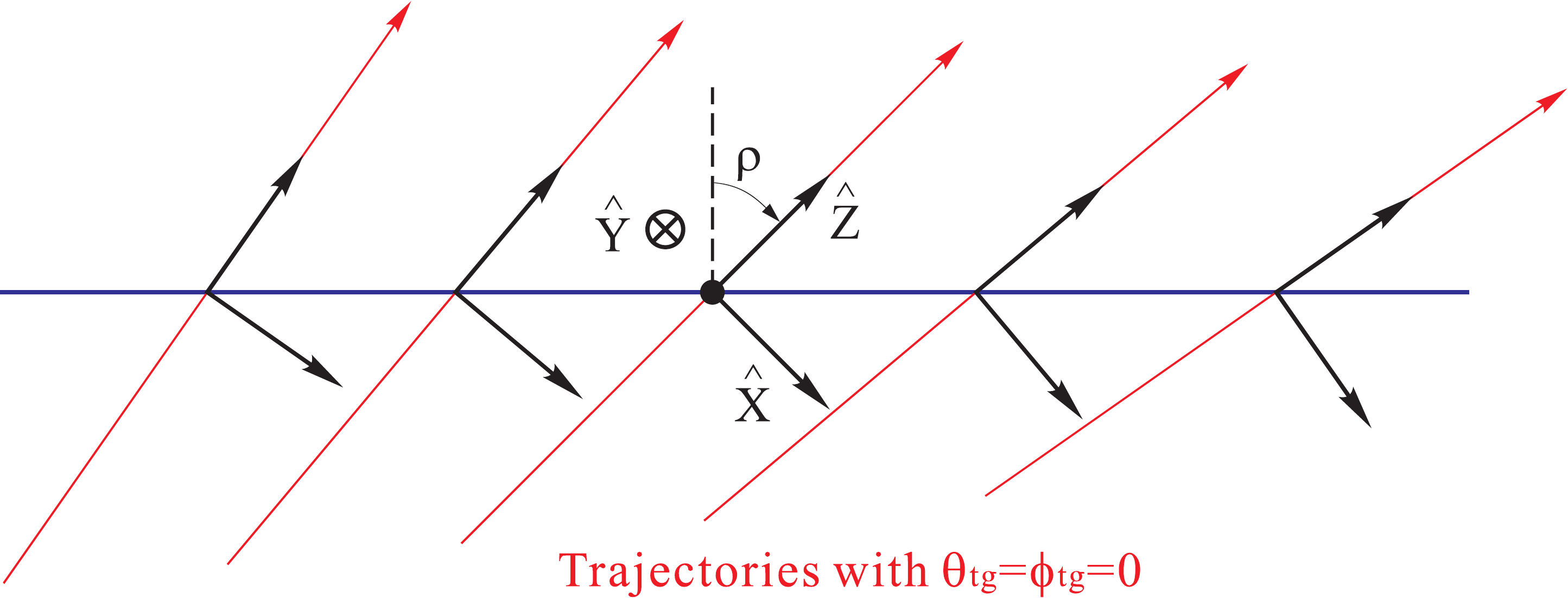}
      \caption[Focal plane coordinate system (FCS)]{\footnotesize{Focal plane coordinate system (FCS), obtained from rotating DCS around its $\hat{y}$ by an angle $\rho$ so $\hat{z}$ is parallel to the central ray with $\hat{\theta}_{tg}$=$\hat{\phi}_{tg}$=0 and $\delta p=(p-p_{0})/p_{0}$.}}
      \label{FCS}
    \end{center}
  \end{figure}
\item \textbf{ Detector Coordinate System (DCS)} \\
  The origin of DCS can be defined as the intersection point of the wire 184 of U1 plane and the wire 184 of V1 plane on the first VDC (VDC1). $\hat{z}_{det}$ is perpendicular to the VDC planes away from HRS, $\hat{x}_{det}$ is horizontally along the long symmetry axis of VDC1 pointing away from the hall center, and $\hat{y}_{det}$ is vertically up toward $\hat{z}_{det}$ (Fig.~\ref{DCS}).

\item \textbf{Transport Coordinate System (TRCS)} \\ 
  The TRCS is generated by rotating the DCS clockwise around  $\hat{y_{det}}$ by $45^{\circ}$ (Fig.~\ref{TRCS}).

\item \textbf{Focal Plane Coordinate System (FCS)} \\
  The FCS is obtained by rotating DCS around its $\hat{y}_{det}$ axis by an angle $\rho$, which is the angle between $\hat{z}_{det}$ axis and the local central ray with $\hat{\theta}_{tg}$=$\hat{\phi}_{tg}$=0 for the corresponding relative  momentum $\delta p=(p-p_{0})/p_{0}$ (Fig.\ref{FCS}).

\end{itemize}

\subsection{Optics Optimization}
The optics calibration follows the procedure described in Ref.~\cite{nilanga_optics}. An optics matrix for HRS is a set of polynomial functions to calculate the target plane quantities, $\delta p$, $y_{tg}$, $\theta_{tg}$ and $\phi_{tg}$, by using the focal plane quantities, $x_{fp}$, $y_{fp}$, $\theta_{fp}$ and $\phi_{fp}$. The functions are given as: 
\begin{eqnarray}
  & &\delta p    = \sum_{i,j,k,l} C^{D}_{ijkl}x^{i}_{fp} \theta^{j}_{fp}y^{k}_{fp}\phi^{l}_{fp}, \\
  & &y_{tg}      = \sum_{i,j,k,l} C^{Y}_{ijkl}x^{i}_{fp} \theta^{j}_{fp}y^{k}_{fp}\phi^{l}_{fp}, \\
  & &\theta_{tg} = \sum_{i,j,k,l} C^{T}_{ijkl}x^{i}_{fp} \theta^{j}_{fp}y^{k}_{fp}\phi^{l}_{fp}, \\
  & &\phi_{tg}   = \sum_{i,j,k,l} C^{P}_{ijkl}x^{i}_{fp} \theta^{j}_{fp}y^{k}_{fp}\phi^{l}_{fp},
  \label{optics_eq}
\end{eqnarray}
where D-terms ($C^{D}_{jkl}$), Y-terms ($C^{Y}_{jkl}$), T-terms ($C^{T}_{jkl}$), and P-terms ($C^{P}_{jkl}$) represent the matrix elements of $\delta p$, $y_{tg}$, $\theta_{tg}$ and $\phi_{tg}$, respectively. An optics calibration procedure is set to determine the matrix elements by using the optics data taken during the experiment.

There are three new variables in HCS which are more practical for long targets and foil targets with known offsets from the hall center:
\begin{eqnarray}
  & & z_{react} = \frac{-\left(y_{tg} + D_{y}\right)+x_{beam}\left(cos\left(\Theta_{0}\right)-\phi_{tg} sin\left(\Theta_{0}\right)\right)}{cos(\Theta_{0})\phi_{tg}+sin(\Theta_{0})}, \\
  & & x_{sieve} = x_{tg} + L\cdot \theta_{tg},\\
  & & y_{sieve} = y_{tg} + L\cdot \phi_{tg},
\end{eqnarray}
where $x_{beam}$ is the horizontal position of the beam, $\theta_{0}$ is the central angle of the spectrometer, and \emph{L} and \emph{D} are defined in TCS. $z_{react}$ is the reaction location along the beam direction and also provides the target position in HCS. $x_{sieve}$ and $y_{sieve}$ represent the vertical and horizontal positions at the sieve slit plane. Table~\ref{optics_offset_table} and Table~\ref{sieve_offset_table} give the values of \emph{D}, $x_{sieve}$ and $y_{sieve}$ from survey reports. During the experiment, the beam position was locked at (-2.668 mm, 3.022 mm).

\begin{table}
  \centering 
  \begin{tabular}{|c||c|c|c|c|c|c|}
    \hline
    &   Angle   &    $D_{x} (mm)$ & $D_{y}$ (mm) & $D_{z} (mm)$ & Survey Report \\
    \hline \hline
    HRS-L & 21.480   & 1.78 & 1.25 & -0.70  & DVCS~\cite{survey_dvcs_2010}\\
    \hline 
    HRS-R & -20.022 & -2.91 & 0.73 & -1.06 & PVDIS~\cite{survey_pvdis_2009}\\
    \hline 
  \end{tabular}
  \caption[Spectrometer offsets from survey reports]{\footnotesize{ Spectrometer offsets for survey reports, where \emph{D}, the offset between the origins of HCS and TCS, is given in term of three components in HCS.}}
  \label{optics_offset_table}
\end{table}

\begin{table}
  \centering
  \begin{tabular}{|c||c|c|c|c|c|c|}
    \hline
    &   L  (mm) &    $x_{sieve} (mm)$ &  $y_{sieve} (mm)$  &  Survey Report\\
    \hline \hline
    HRS-L & 1182.3  & -1.05 & 0.20  & DVCS~\cite{survey_dvcs_2010}\\
    \hline 
    HRS-R & 1175.9  & 1.04  & 0.05 & A1n~\cite{survey_a1n_2009} \\
    \hline 
  \end{tabular}
  \caption[Sieve slit plates offsets from survey reports.]{\footnotesize{ Sieve slit plates offsets from survey reports. The values were measured in HCS.}}
  \label{sieve_offset_table}
\end{table}

\begin{figure}[!ht]
  \begin{center}
    \includegraphics[type=pdf,ext=.pdf,read=.pdf,width=0.85\textwidth]{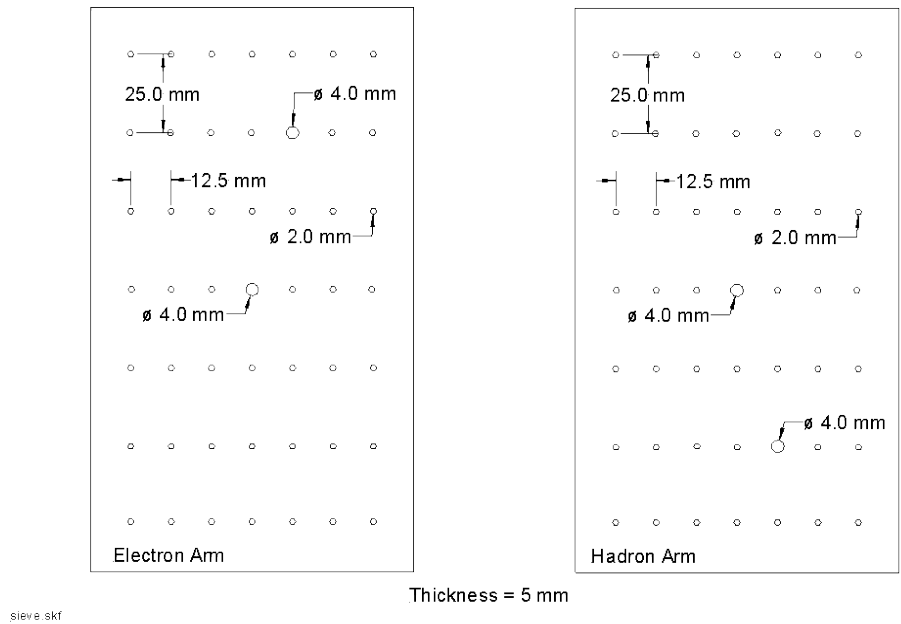}
    \caption[Design of sieve slit plates]{\footnotesize{The design of sieve slit plates. Both arm have the identical plates but different mounting system. The graphic is taken from Hall A web-page.}}
    \label{sieve_slit}
  \end{center}
\end{figure}

As given in table~\ref{optics_data_table}, a set of optics data has been taken during the experiment with the optics target for $y_{tg}$ calibration. When taking angular calibration data, a sieve slit plate (Fig.~\ref{sieve_slit}) was installed at the entrance at Q1 for each HRS. The data was taken in the QE region to ensure each hole of the sieve slit plate had enough events. 
\begin{table}[!ht]
  \centering
  \begin{tabular}{|c||c|c|c|c|c|c|}
    \hline
    Run Number & Target   & Angle    & $P_{0} / P_{0}^{RQ3}$ (GeV/c)  & Raster & Sieve & Comment\\
    \hline \hline
    3695    & Dummy4cm & $23^{\circ}$ & 2.678/2.3492  &  Off   &  Out  & $\delta p$ +3\% \\
    \hline
    3698    & Dummy4cm & $23^{\circ}$ & 2.600/2.2808  &  Off   &  Out  & $\delta p$ 0\% \\
    \hline
    3704    & Dummy4cm & $23^{\circ}$ & 2.522/2.2124  &  Off   &  Out  & $\delta p$ -3\% \\
    \hline
    3700    & Multi-C  & $23^{\circ}$ & 2.600/2.2808  &  Off   &  Out  & \\
    \hline
    3701    & Multi-C  & $23^{\circ}$ & 2.600/2.2808  &  On    &  Out  & \\
    \hline
    4201-4205& Multi-C    & $25^{\circ}$ & 2.505/2.1975 & Off    &  In & \\
    \hline 
  \end{tabular}
  \caption[Run list of optics data.]{Run list of optics data, where Dummy4cm means two dummy foils separated by 4 cm, and Multi-C means the optics target with seven carbon foils. Two HRSs took data simultaneously with the same settings.}
  \label{optics_data_table}
\end{table}
\begin{figure}[!ht]
  \begin{center}
    \subfloat[Before optics calibration]{
      \includegraphics[type=pdf,ext=.pdf,read=.pdf,width=1.0\textwidth]{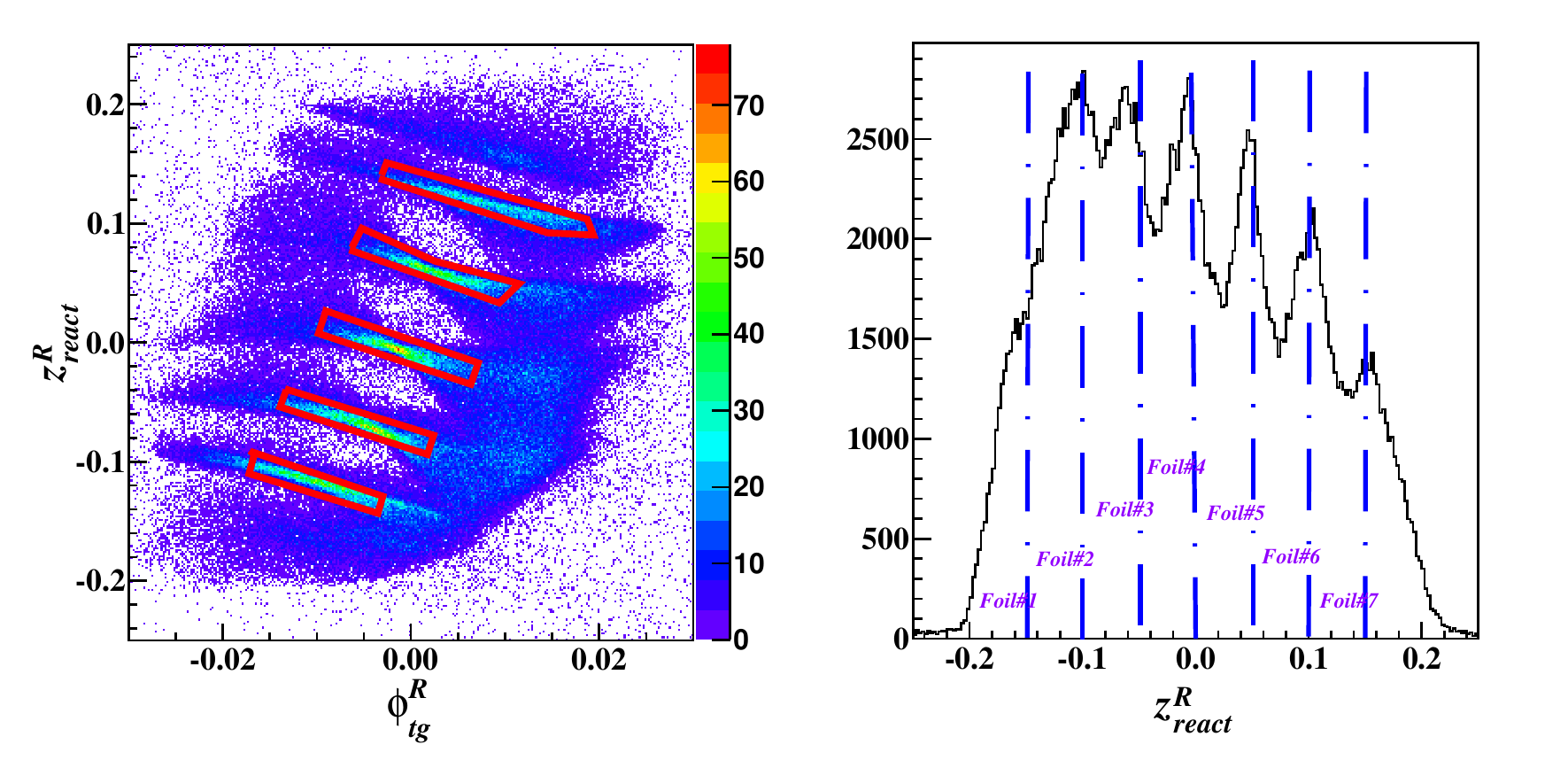}
      \label{vz_before}
    }
    \\ 
    \subfloat[After optics calibration]{
      \includegraphics[type=pdf,ext=.pdf,read=.pdf,width=1.0\textwidth]{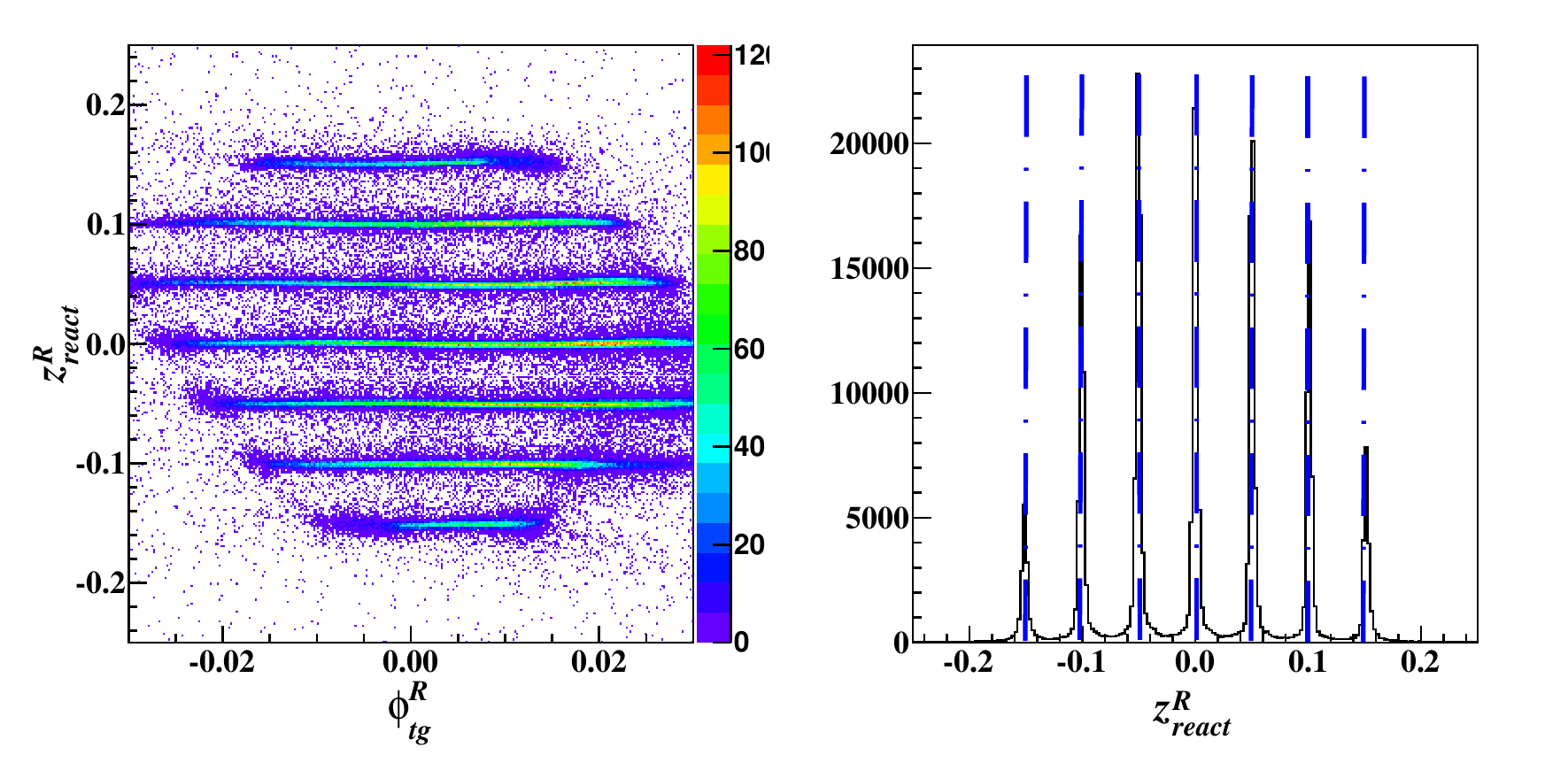}
      \label{vz_after}
    }
    \caption[$Z_{react}$ distribution before and after the optics calibration]{\footnotesize{$Z_{react}$ distribution before and after the optics calibration. The 2-D plots reveal that each strip represents electrons scattered from the corresponding foil indicated in the 1-D plots. The red boxes in the first 2-D plot represent graphic cuts applied to selected good electron samples during the first iteration of the Y-terms calibration.}}
    \label{optics_vertex}
  \end{center}
\end{figure}

\begin{figure}[!ht]
  \begin{center}
    \subfloat[Before optics calibration]{
      \includegraphics[type=pdf,ext=.pdf,read=.pdf,width=1.0\textwidth]{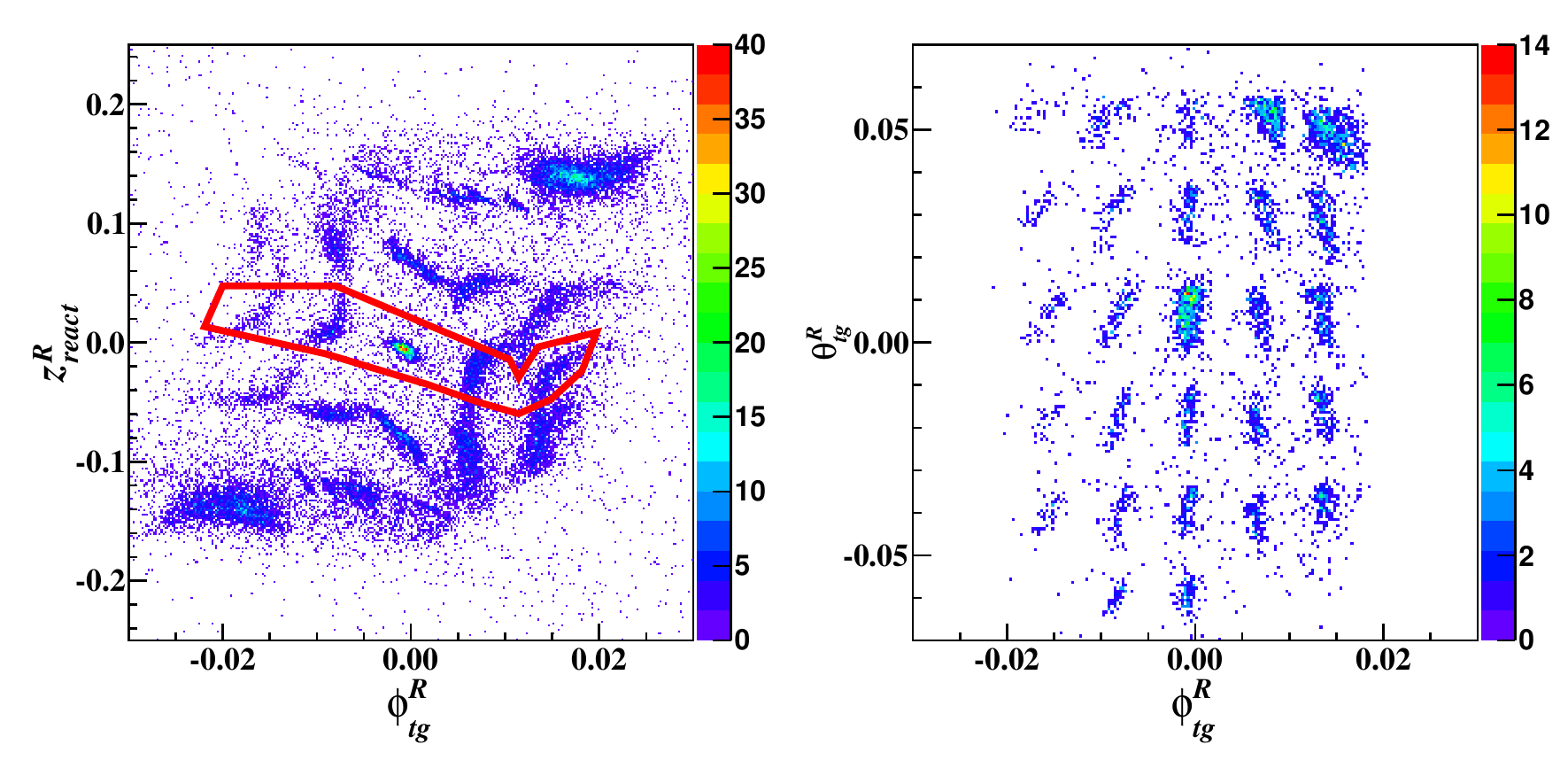}
      \label{sieve_before}
    } 
    \\
    \subfloat[After optics calibration]{
      \includegraphics[type=pdf,ext=.pdf,read=.pdf,width=1.0\textwidth]{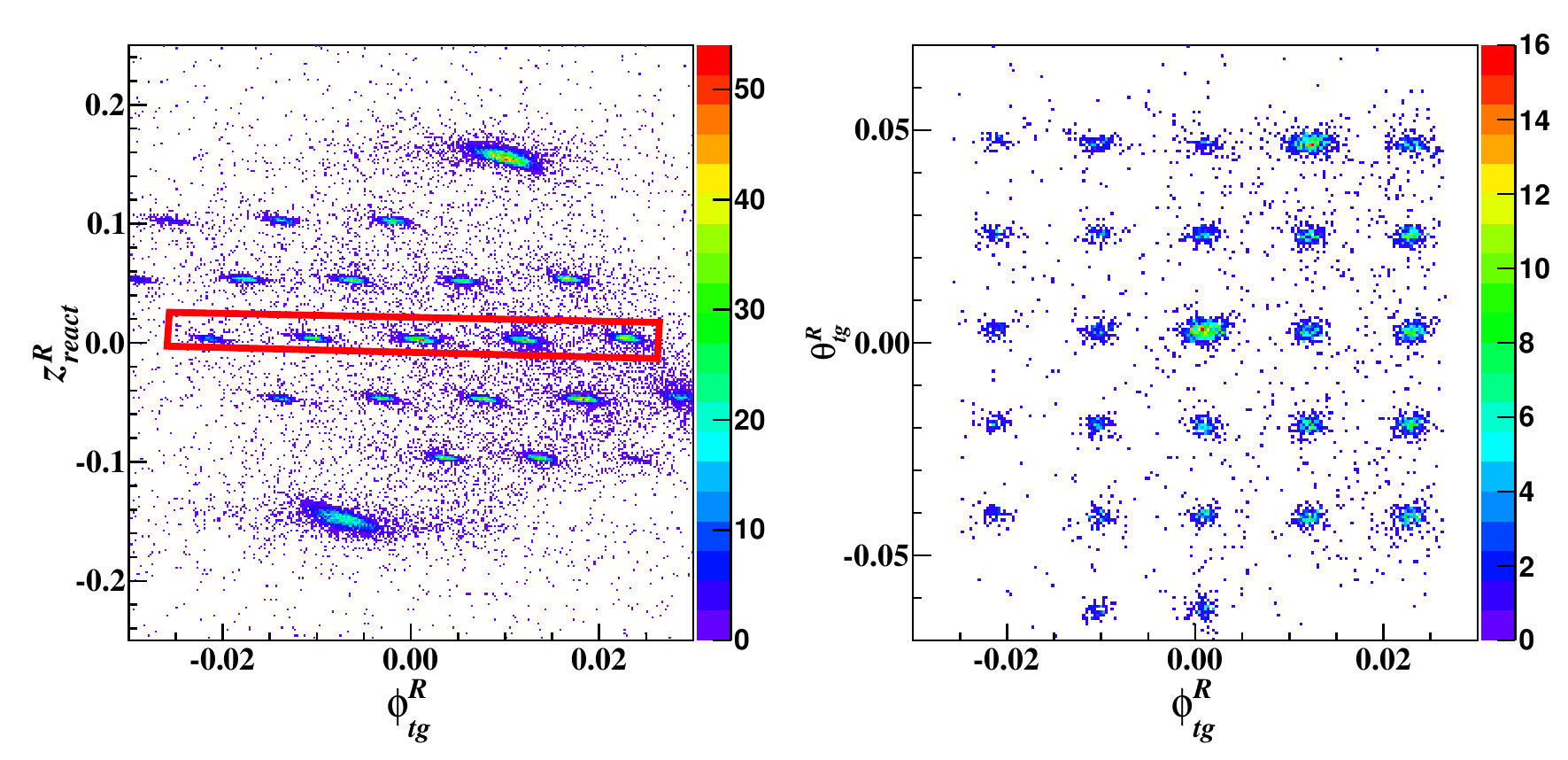}
      \label{sieve_after}
    }
    \caption[Sieve slit pattern before and after the optics calibration]{\footnotesize{$Z_{react}$ Sieve slit pattern before and after the optics calibration. Cutting on a single foil (the red box) is required to see the clear sieve slit pattern. Events from each hole are individually extracted and assigned with the values of $\theta_{tg}$ and $\phi_{tg}$ at the center of the hole.} }
    \label{optics_sieve}
  \end{center}
\end{figure}

 The optics matrices used to replay optics calibration data were from previous experiments which shared similar spectrometer settings as this experiment. The initial HRS-L matrix taken from E05-102~\cite{jinge_thesis} was able to reconstruct the target plane quantities with good accuracy. The matrix was refitted by using the calibration data and the updated offsets from survey reports. The HRS-R optics matrix used by previous experiments, however, performed poor reconstruction of the target plane quantities because of the mis-tuned RQ3 field, as shown in Fig.~\ref{vz_before} and Fig.~\ref{sieve_before}. The detailed procedure to calibrate the new HRS-R optics matrix will be discussed. 

 Due to the change of the RQ3 field, the higher order effect of the HRS-R optics may be different from the one with normal RQ3 field and could change the number of elements and their values in the matrix. In addition to the old matrix elements, new matrix elements were added in the optics terms to form a complete set of polynomials up to the 5th-order, but their values were initially set to zero. After the calibration data was replayed with this optics matrix, events selection at the focal plane was performed to select event from main trigger (T1). The one-track cut on VDCs and PID cuts on the GC and calorimeters were applied to select electrons only. Events at the edge of HRS acceptance were eliminated by choosing only the flat regions of the focal plane quantities. 

Events from Run-3700 were used to calibrate the matrix elements in Y-terms (Eq.~\eqref{optics_eq}). When one plots the 2-D histogram of $z_{react}$ versus $\phi_{tg}$ (Fig.~\ref{vz_before}), events scattered from a specific foil form a strip where both ends  were smeared due to the defocusing effect from RQ3. The first iteration was to select electron samples which clearly belong to a certain strip (e.g. near the center of each strip as included in the red boxes in Fig.~\ref{vz_before}), while events in the overlap regions were discarded. For each sample, its $z_{react}$ value was assigned with the position of the foil it belonged to. After the offsets in Table~\ref{optics_offset_table} and Table~\ref{sieve_offset_table} were applied in the calibration, the matrix elements in Y-terms were fitted by an optimizer based on the Minuit minimization method~\cite{jin_huang_optics}. The high order elements in Y-terms were removed one by one until the minimization started to fluctuate. The data was replayed with these new matrix elements and the big improvement of $z_{react}$ distribution is demonstrated in Fig.~\ref{vz_after}.

The procedure of calibrating T-terms and P-terms was similar but used calibration data taken with a sieve slit plate (Runs 4201-4205). Since Y-terms have been optimized, these sieve slit runs were first replayed with new elements of Y-terms updated in the DB. The sieve slit patterns shown in a 2-D plot of target plane quantities $\theta_{tg}$ and $\phi_{tg}$ were compared with the design of the sieve slit plate in Fig.~\ref{sieve_slit}. On the right plots of Fig.~\ref{optics_sieve}, each spot corresponds to the one sieve slit hole. For an electron sample that could be clearly identified from a spot, the coordinate of this event on the sieve slit plane, ($x_{ss}$, $y_{ss}$), was set at the center of the hole since the diameter of the hole is very small (2 mm). The values of $\theta_{tg}$ and $\phi_{tg}$ can be directly calculated from the values $x_{ss}$ and $y_{ss}$.

 The matrix elements of T-terms and P-terms were fitted separately with the same optimizer and unnecessary matrix elements were removed by checking the variation of the minimization Chi-Square. Fig.~\ref{sieve_after} shows that the sieve slit holes were well aligned after the calibration of angular terms. 

 With updated Y-terms, T-terms and P-terms in the data base, the calibration runs were replayed again. As the target plane quantities were reconstructed with higher resolutions and the events were better separated, the second iteration was processed with more good samples. The calibration was completed when the minimization Chi-Square started to fluctuate after several iterations. 

 The calibration of D-terms requires data to be taken with the central momentum intentionally shifted by small values, for example, by $\mathrm{\pm 3\%}$. However, the experiment was running in the QE region and the peak of the momentum distribution was too broad and insensitive to these small offsets. Without the elastic data, the D-terms could not be calibrated. In Appendix C, a different method was discussed to obtain the correct $\delta p$ reconstruction.

%% file: cross_section/cross_section.tex
\chapter{Cross Section Extraction}

\section{Overview}
 Assuming the data is binned in the energy of scattered electrons, $E'$, the experimental raw cross section can be written as:
\begin{equation}
  \frac{d\sigma^{raw}_{EX}}{dE'd\Omega} (E_{0},E'_{i}, \theta_{0}) = \frac{N^{i}_{EX}\cdot \epsilon_{e-\pi}}{N_{e} \cdot \eta_{tg} \cdot \epsilon_{eff}\cdot (\Delta E'_{EX}\Delta\Omega_{EX})},
  \label{eqxs_org}
\end{equation}
where the superscript $i$ denotes the $ith$ bin. $E_{0}$ is the incident energy set at 3.356 GeV in the E08-014, $E'_{i}$ is the scattered energy at the center of the bin, and $\theta_{0}$ is the central scattering angle. $\Delta E'$ and $\Delta\Omega=\Delta\theta_{tg}\cdot \Delta\phi_{tg}$ are the momentum acceptance and the solid angle acceptance of the spectrometer; $N^{i}_{EX}$ is the number of scattered electron events in this bin; $\eta_{tg}$ is the areal density of scattering centers; $N_{e}$ is the total number of electrons in the beam; and $\epsilon_{eff}$ is the total efficiency of all detectors combined, including the detection efficiency and the cut efficiency. $\epsilon_{e-\pi}$ corrects for the pion-contamination in the electrons after the PID cuts. In the rest of this chapter, the differential form of the cross section, $\frac{d\sigma}{dE'd\Omega}(E_{0},E'_{i}, \theta_{0})$, is abbreviated to $\sigma(E'_{i}, \theta_{0})$.

\begin{figure}[!ht]
 \begin{center}
  \includegraphics[angle=90, type=pdf, ext=.pdf,read=.pdf,width=1.\textwidth]{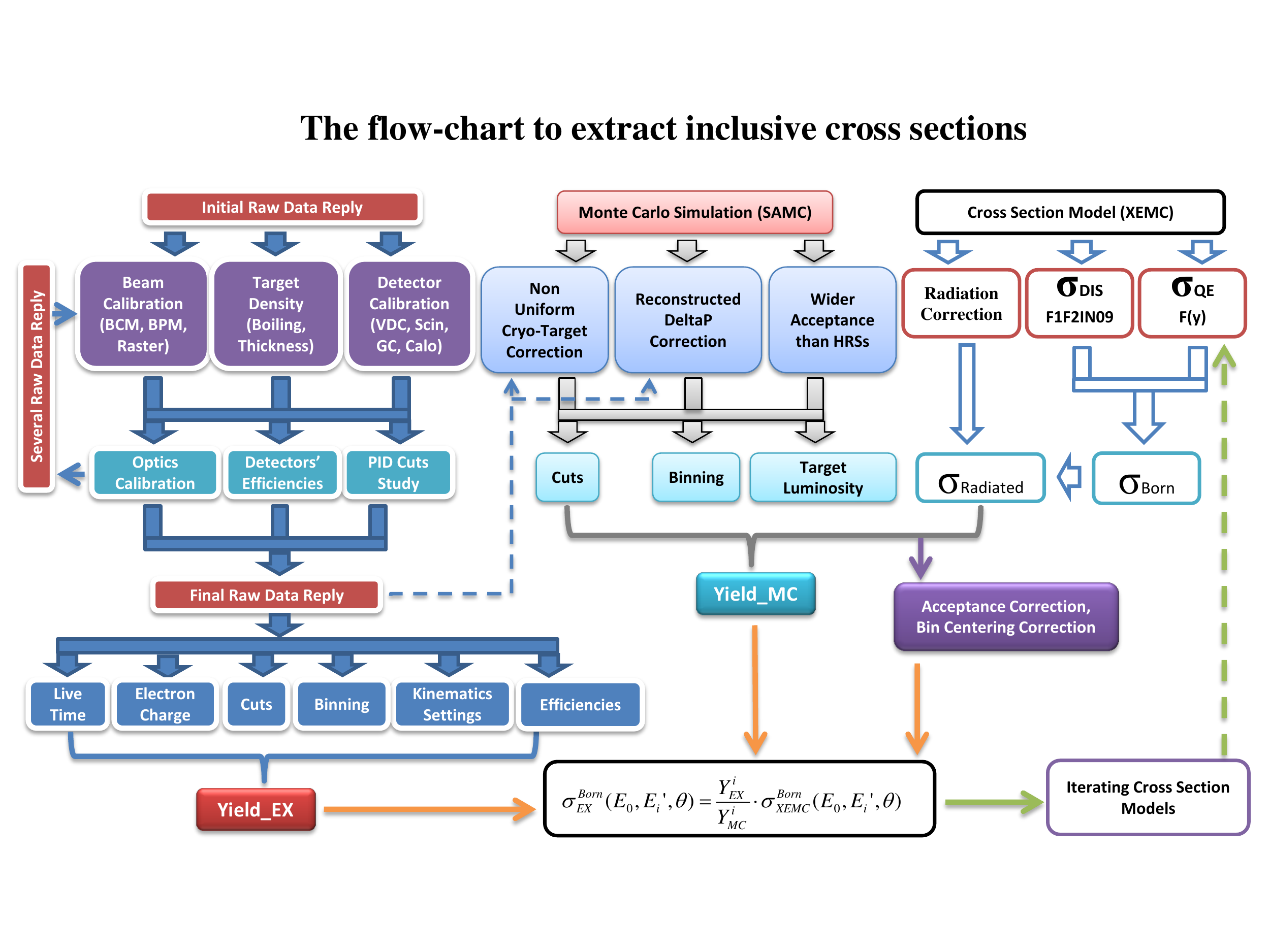}
  \caption[Flow-chart illustrating cross section extraction]{Flow-chart illustrating cross section extraction}
  \label{xs_flow_chart}
 \end{center}
\end{figure}
The raw cross section in Eq.~\eqref{eqxs_org} requires additional corrections to remove the effects from the spectrometer acceptance. Also, $E_{0}$ and $E'_{i}$ are altered when the electron loses its energy as it passes through the target due to the radiative effects before and after the scattering (see Appendix B.5). The experimental cross section, usually called the radiated cross section, has to be further corrected for radiative effects. The final cross section is the Born cross section, which can be directly compared with theoretical calculations. 

 The basic procedure of extracting cross sections from experimental data is demonstrated in Fig.~\ref{xs_flow_chart}. First of all, the signals from detectors and electronics were stored in the raw data in the form of TDC channels, ADC channels and scaler counts. These signals have to be properly calibrated and converted into applicable quantities. The calibrated HRS optics matrix reconstructs the scattered electron's momentum, scattering angle and reaction point at the target plane. The full set of raw data was replayed with updated parameters in the data base. The calibration of detectors and the HRS optics matrices have been introduced in the previous chapter. 
 
 Secondly, the results of the beam charge monitor (BCM) calibration convert the BCM scaler counts into electron beam charge. The dead-time associated with the DAQ system needs to be evaluated to recover the events lost during the data acquisition. $\eta_{tg}$ is determined by the target thickness after the boiling study. Good electrons are identified by applying cuts on calibrated detector signals, and the efficiencies of the event selection can be individually determined. By binning the data with the kinematic variable, e.g. $E'$, in its proper acceptance range, one can extract the experiment yield in each bin. A description of all the procedures will be given in this chapter.
 
 In addition, the single arm Monte Carlo simulation (SAMC) generates simulation events with the same kinematic settings but with a wider acceptance range to correct the acceptance effect of the HRSs. After weighting the simulation events with the cross sections calculated from model (e.g. XEMC in this experiment), the simulation yields were extracted with the same acceptance cuts and binning method. The Monte Carlo simulation and cross section models will also be discussed in this chapter.

 Finally, the yield ratio method used to extract the cross sections will be introduced, followed by a discussion of errors.

\input{ cross_section/analysis_charge.tex}

\input{ cross_section/analysis_dt.tex}

\input{ cross_section/analysis_target.tex}

\input{ cross_section/analysis_effi.tex}

\input{ cross_section/analysis_mc.tex}

\input{ cross_section/analysis_xs.tex}

%% file: cross_section/analysis_charge.tex
\section{Beam Charge}
The accumulated electron charge from the beam was monitored by BCMs, where signals were recorded in scalers. The scalers signals, in term of number of counts, have been calibrated~\cite{bcm_patricia} to correctly reflect the accumulated electron charge. When the beam is stable during one run, the total electron charge is simply the product of the beam current and the total run time, and should be directly proportional to the total number of scaler counts. However, events taken during the beam trips must be removed by applying a cut on the electron beam current.

 The average electron beam current in between two consecutive scaler events, called the real-time current, is calculated from the total electron charge collected between these events divided by the time gap. For example, between the $ith$ and the $ith+1$ event, the real-time current measured by the upstream BCM scaler, $\mathrm{U_{1}}$, is given by:
\begin{equation}
  I_{i}^{U_{1}} = \Delta C_{i}^{U_{1}}/\Delta T_{i}, 
\end{equation}
where $\mathrm{\Delta C_{i}^{U_{1}} = C_{i+1}^{U_{1}} - C_{i}^{U_{1}}}$ gives the charge accumulated between two scaler events with the time gap, $\mathrm{\Delta T_{i}=T_{i+1}-T_{i}}$. Similarly, the real-time current measured by the downstream BCM scaler, $\mathrm{D_{1}}$, is also calculated. There are other BCM scaler signals, $\mathrm{U_{3}}$ and $\mathrm{U_{10}}$ ($\mathrm{D_{3}}$ and $\mathrm{D_{10}}$), which basically measure the same charge signal as $\mathrm{U_{1}}$ ($\mathrm{D_{1}}$) but with 3 times and 10 times amplification, respectively. Only $\mathrm{U_{1}}$ and $\mathrm{D_{1}}$ were used since this experiment required very high currents.

 The beam trip cut is applied on the average of these two real-time current values:
\begin{equation}
\frac{1}{2}(I_{i^{*}}^{U_{1}}+I_{i^{*}}^{D_{1}})>I_{beam\_trip\_cut},
\end{equation}
where the cut value can be any value between zero (when beam is tripped) and the value slightly below the maximum current. In this analysis, the beam trip cut was chosen to be 50\% of the normal beam current. The total charge after the beam trip cut is given as:
\begin{equation}
   Q_{e} = \frac{1}{2}\sum_{i^{*}}(\Delta C_{i^{*}}^{U_{1}}+\Delta C_{i^{*}}^{D_{1}}), 
  \label{eq_qe}
\end{equation}
where $i^{*}$ means summing over scaler events with beam current $I_{i^{*}}$ higher than the cut. And the number of electrons in the beam can be calculated as follows:
\begin{equation}
   N_{e} = Q_{e}/e, 
  \label{eq_ne}
\end{equation}
with the electron charge, $\mathrm{e=1.602\times 10^{-19}~C}$.

 After the data replay, scaler events are stored in the scaler trees, \emph{\bf{RIGHT}} for HRS-R and \emph{\bf{LEFT}} for HRS-L, respectively, and they are synchronized with trigger events in the \emph{\bf{T}} tree. There are certain number of trigger events recorded between two consecutive scaler events, and these events are assigned the same value of the real-time beam current evaluated between these two scaler events. Consequently, a beam trip cut removes all trigger events in between two scaler events if the real-time current is lower than the cut.

 During this experiment, BCM scalers on HRS-L did not work properly. Due to the fact that the scalers on both HRSs recorded the same BCM signals, the real-time current for data taken in HRS-L was calculated with scaler events in HRS-R.

%% file: cross_section/analysis_dt.tex
\section{Dead-Time}
  There are two types of dead-time that can cause the loss of events, the electronic dead-time and the computer dead-time. The electronic dead-time comes from the front-end electronics of the DAQ system, which can discard the incoming trigger events while they are busy with processing the current trigger. The computer dead-time is caused by the limitation of computer speed which can lead to the loss of new events when the computer is still writing the current event into the hard disk. Unless the computer is overloaded by processes other than the DAQ system, the computer dead-time is negligible due to the application of high performance computer hardware. 
 
 One evaluates the dead-time as the percentage of the trigger events being discarded to the total trigger events in a certain period of time. The value of the dead-time is directly related to the performance of electronics and computers, but also strongly depends on the total trigger rate. Rather than increasing the hardware performance, a typical method to reduce the dead-time is to limit the total trigger rate below a reasonable value by assigning a pre-scale factor to each trigger. 
 
 The online dead-time during data taking is monitored by using the electron dead-time monitor module (EDTM) which mixes pulse signals with fixed frequency into TDC signals. Within a certain amount of time, the total number of the pulse signals is known and the dead-time value can be given by calculating the percentage of the pulse signals which are not recorded by the DAQ system. By changing the pre-scale factors before the start of the each run, this value was kept under 30\% in this experiment.
 
 The average value of dead-time in each run for the main production triggers was calculated individually during the offline analysis. Although the total number of events recorded by the DAQ system was scaled by the pre-scale factor, their total triggers were counted by scalers, hence the average dead-time for the $ith$ trigger can be given by:
\begin{equation}
  DT_{T_{i}} = 1 - \frac{PS_{T_{i}}\cdot N_{T_{i}}^{DAQ} }{N_{T_{i}}^{Scaler}},
  \label{eq_dt}
\end{equation}
where $PS_{T_{i}}$ is the pre-scale factor of the trigger. $N_{T_{i}}^{Scaler}$ and $N_{T_{i}}^{DAQ}$ are the total number of scaler counts (in $\mathbf{RIGHT}$ tree for $i=1$ or $\mathbf{LEFT}$ tree for $i=3$) and trigger events (in $\mathbf{T}$ tree) for each run, respectively. The beam trip cut was applied when calculating $N_{T_{i}}^{Scaler}$ and $N_{T_{i}}^{DAQ}$.

  A different quantity, live-time ($LT_{T_{i}} = 1 -DT_{T_{i}}$), is more commonly used to correct the total number of good events in each run:
 \begin{equation}
  N^{r}_{T_{i},EX} = PS^{r}_{T_{i}}\cdot \frac{N^{r,recorded}_{T_{i}}}{LT^{r}_{T_{i}}},
  \label{eq_lt}
 \end{equation}
where $r$ denotes the run number; $PS^{r}_{T_{i}}=PS1^{r}$ for the $T_{1}$ trigger on HRS-R and $PS^{r}_{T_{i}}=PS3^{r}$ for the $T_{3}$ trigger on HRS-L; $N^{r}_{T_{i},EX}$ and $N^{r,recorded}_{T_{i}}$ are the number of selected events which create triggers and the number of those events which are recorded by the DAQ system after pre-scaling, respectively. Note that without event selection, e.g. PID cuts, $N^{r,recorded}_{T_{i}}=N^{r,DAQ}_{T_{i}}$.

 In this experiment, since only events from $T_{1}$ ($T_{3}$) were used for data analysis on HRS-R (HRS-L), the subscript, $T_{i}$, is omitted in any future discussion.

%% file: cross_section/analysis_target.tex
\section{Targets}
 The areal density of scattering centers ( in $\mathrm{cm^{-2}}$) in Eq.~\eqref{eqxs_org} is calculated from the known target thickness:
\begin{equation}
  \eta_{tg} = \frac{\rho\cdot l \cdot N_{a}}{A},
  \label{eq_ntg}
\end{equation}
where $\rho$ is the density of the target material in $\mathrm{g/cm^{3}}$, $l$ is the effective target length in \emph{cm}, $N_{a}$ is the Avogadro's number and A is the nuclear number of the target.
behaviour
\subsection{Cryo-Target Boiling Effect}
\begin{figure}[!ht]
  \begin{center}
    \subfloat[$^{2}H$]{
      \includegraphics[type=pdf, ext=.pdf,read=.pdf,width=0.62\textwidth]{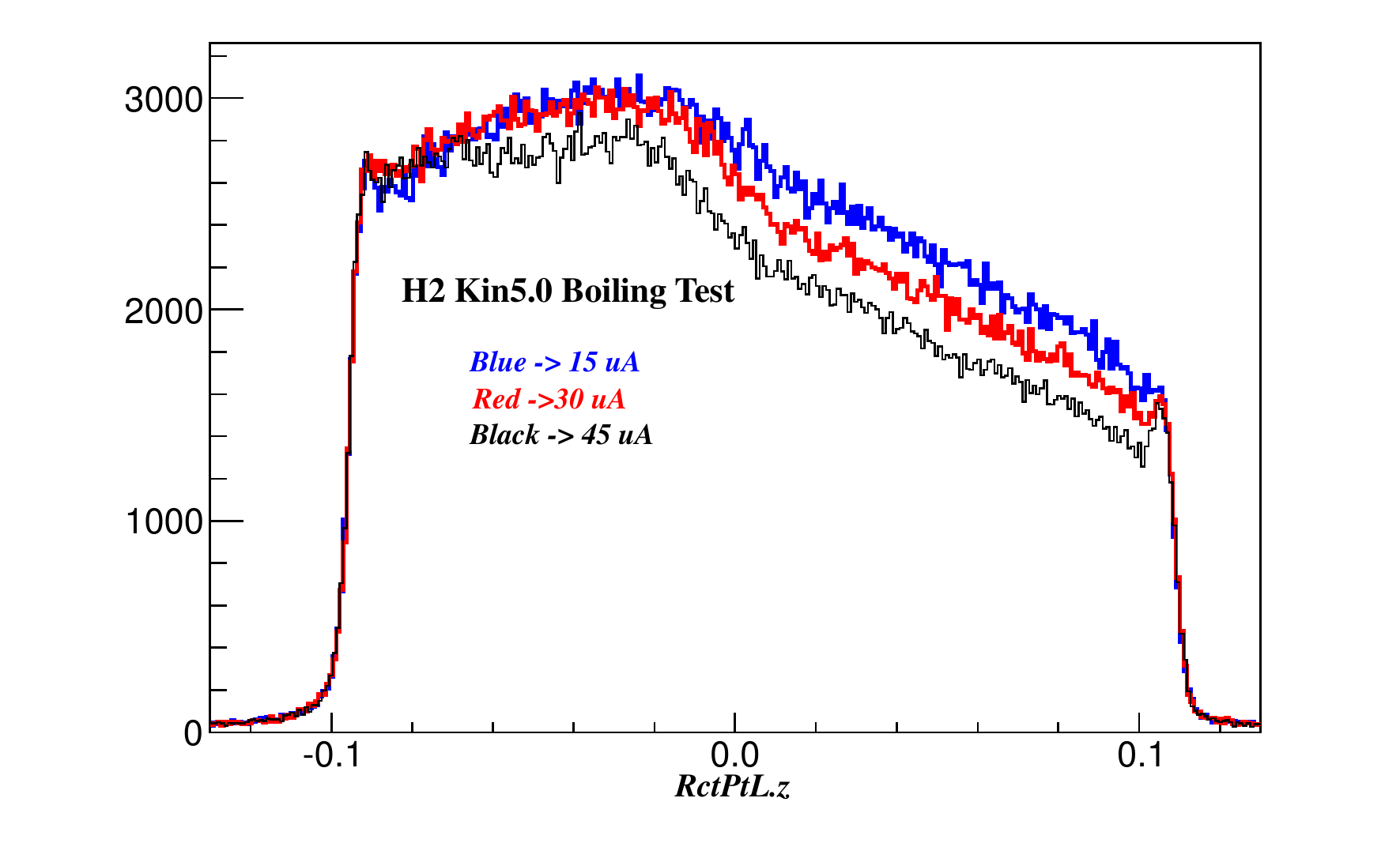}
    }
    \\
    \subfloat[$^{3}He$]{
      \includegraphics[type=pdf, ext=.pdf,read=.pdf,width=0.62\textwidth]{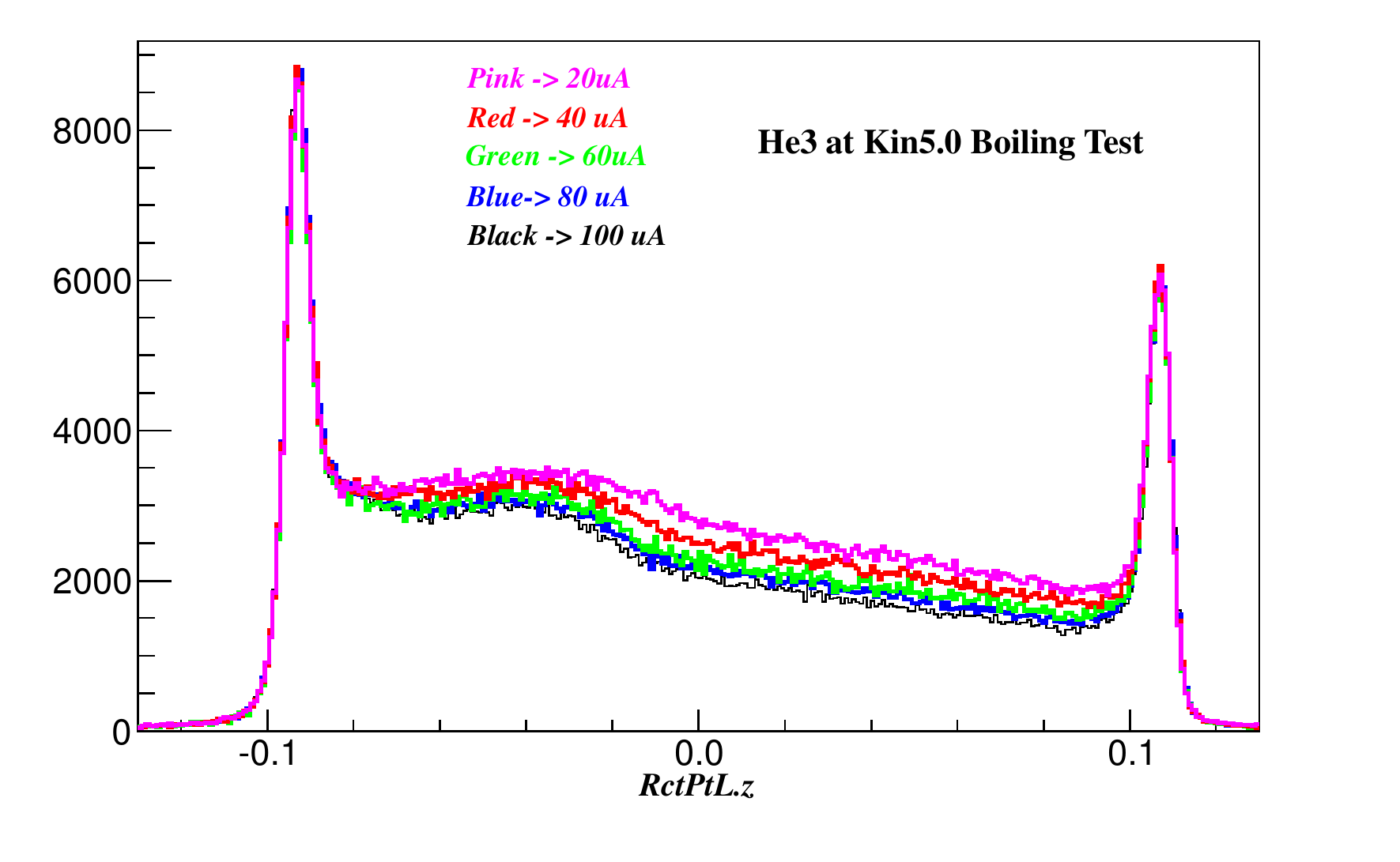}
    }
    \\
    \subfloat[$^{4}He$]{
      \includegraphics[type=pdf, ext=.pdf,read=.pdf,width=0.62\textwidth]{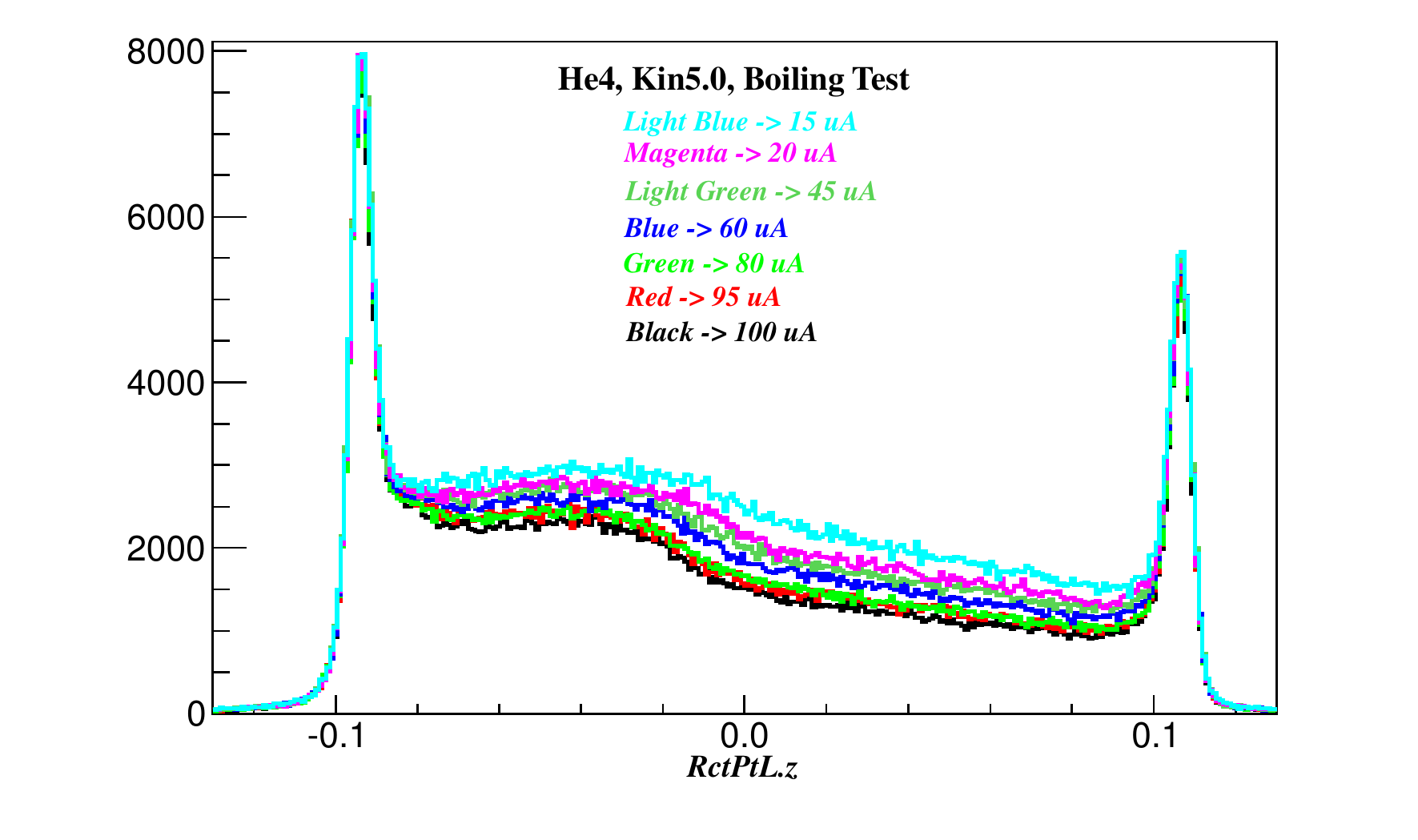}
    }
    \caption[Cryo-target bumps]{\footnotesize{Cryo-target bumps which appear on the $z_{react}$ distributions because of the non-uniform density of cryo-targets. Due to the boiling effect, the bumps become more significant when the beam current is larger.}}
    \label{bump_current}
  \end{center}
\end{figure}
 When the electron beam passes through the target, the local temperature fluctuates and causes the target density to vary with the beam current. This phenomenon is called the boiling effect. While the density variation of solid targets is usually negligible, liquid and gas targets have significant boiling effects and their densities correlate to the beam current as follow:
\begin{equation}
  \rho = \rho_{0} \cdot (1.0 - B \cdot I /100),
  \label{eq_tgrho}
\end{equation}
where $I$ and $B$ are the values of the beam current and the boiling factor for the target, respectively. $\rho_{0}$ is the nominal target density at $I=0$ and $\rho$ is the actual density with the boiling effect.

 In the E08-014, three cryogenic targets (cryo-targets), $\mathrm{^{2}H}$, $\mathrm{^{3}He}$ and $\mathrm{^{4}He}$, were held in 20 cm long aluminium cells. The cryogenic coolant flowed from the upstream to the downstream of a target cell, and the variation of temperature among different parts of the target leaded to a non-uniform density distribution. When the beam was on, the temperature fluctuation became more significant with higher current. The boiling effect was different along the cryo-target and further increased the non-uniformity of the target densities. Fig.~\ref{bump_current} shows the irregular density distribution and the strong correlation between the density and the beam current. The areal density, $\eta_{tg}$, for these cryo-targets could not be simply calculated from Eq.~\eqref{eq_ntg}.

  A boiling study was performed by dividing each target into several sections along the cell, where the boiling effect was individually evaluated. The relative density distributions were extrapolated from the boiling study results and the absolute target densities were calculated with the survey report of the target system~\cite{target_report}. A detailed discussion is given in Appendix D. 
 

%% file: cross_section/analysis_effi.tex
\section{Detector Efficiencies}
  To extract the electron-scattering cross section, one needs to know the number of scattered electrons coming out from the reaction plane (i.e. the target plane). Every detector is designed to be sensitive to certain types of particles within the known energy ranges. In practice, the detector may not be able to detect everyone of these particles passing through. Each detector has a detection efficiency ($\epsilon_{det}$) which is given as the portion of particles detected to the total. In addition, during the offline analysis, one applies cuts on the reconstructed quantities of the detectors to remove background and select good events, e.g. to identify pure electron events from the target. However, depending on the range of the cut, each cut may also unintentionally discard some good events. The cut efficiency ($\epsilon_{cut}$) denotes the percentage of good events remaining after applying a cut and has to be evaluated when one chooses the value of the cut. In other words, the detection efficiency denotes the survival rate of particles at the hardware level and the cut efficiency represents the level of confidence when selecting good particles at the software level, respectively. In this section, the efficiencies of the HRS detectors will be individually evaluated.

\subsection{Trigger Efficiency}
 The traditional HRS production trigger is generated by the coincidence of logic signals from two scintillator planes (S1 and S2m), so the trigger efficiency is equal to the product of the detection efficiency of these two scintillators. An inefficiency arises when either S1 or S2m does not fire when a particle passes through. As discussed in Section 3.7, T2 (T4) is the trigger generated when only one of S1 and S2m signals coincides with the gas \v{C}erenkov (GC) signal on HRS-R(-L). Using the events from T2 (T4), one can calculate the trigger efficiency of T1 (T3), or equivalently T6 (T7) in the E08-014, as follow: 
\begin{equation}
  \epsilon_{trig} = \frac{PS1(3)\cdot N_{T1(3)}}{PS1(3) \cdot N_{T1(3)}+PS2(4)\cdot N_{T2(4)}},
  \label{trigger_eff}
\end{equation}
where $N_{T1(2,3,4)}$ is number of events triggered by T1(2,3,4) and PS1(2,3,4) is the prescale factor of the trigger.

 Note that Eq.~\eqref{trigger_eff} is only valid when the GC has 100\% detection efficiency. Particles creating T1 (T3) Events ($N_{T1(3)}$) may not necessarily fire the GC, but events from T2 (T4) are recorded when the GC is fired, so $N_{T2(4)}$ has to be corrected by the detection efficiency of the GC. The trigger efficiency should be given by:
 \begin{equation}
  \epsilon_{trig} = \frac{PS1(3)\cdot N_{T1(3)}}{PS1(3) \cdot N_{T1(3)}+PS2(4)\cdot N_{T2(4)}/\epsilon^{GC}_{det}} 
  \label{trigger_eff2},
\end{equation}
where $\epsilon^{GC}_{det}$ is the detection efficiency of the GC. The HRS GCs usually have very high efficiency for detecting electrons, so Eq.~\eqref{trigger_eff} is still valid. However, when the efficiency of the GC falls, the trigger efficiency has to be corrected by the detection efficiency of the GC which is evaluated independently.

 In the E08-014, as the design of T1 and T3 involved S1, S2m and the GC, hence the trigger efficiency does not depend on the detection efficiency of the GC, which cancels in Eq.~\eqref{trigger_eff}:
 \begin{eqnarray}
 \epsilon_{trig} &=& \frac{PS1(3)\cdot N_{T1(3)}/\epsilon^{GC}_{det}}{PS1(3) \cdot N_{T1(3)}/\epsilon^{GC}_{det}+PS2(4)\cdot N_{T2(4)}/\epsilon^{GC}_{det}} \nonumber \\
                 &=& \frac{PS1(3)\cdot N_{T1(3)}}{PS1(3) \cdot N_{T1(3)}+PS2(4)\cdot N_{T2(4)}}.
  \label{trigger_eff3}
 \end{eqnarray}

 In summary, the assumption that the trigger efficiency is equivalent to the detection efficiency of S1 and S2m is valid only when both T1 (T3) and T2 (T4) involve the logic signal from the GC. 
\begin{figure}[!ht]
  \begin{center}
    \includegraphics[type=pdf,ext=.pdf,read=.pdf,width=0.9\textwidth]{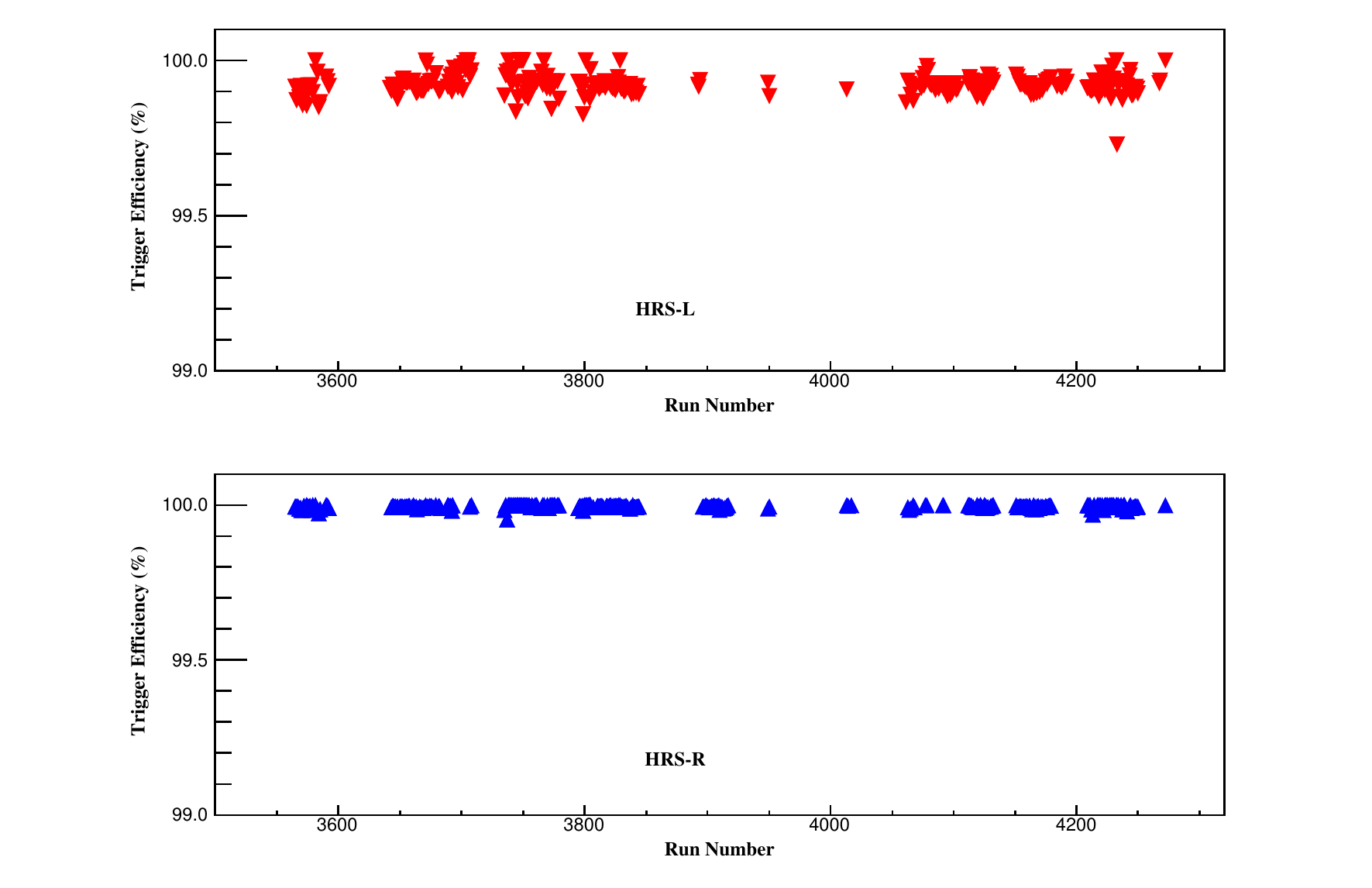}
    \caption[Trigger efficiency vs run number]{\footnotesize{Trigger efficiency vs run number, where the top plot is for T3 trigger on HRS-L and the bottom plot is for T1 trigger on HRS-R.}}
    \label{trig_effi}
  \end{center}
\end{figure}
The trigger efficiencies of T1 and T3 were calculated individually for each run, shown in Fig.~\ref{trig_effi}. The results show that the triggers have very high efficiencies.

\subsection{Vertical Drift Chamber Efficiency}
 The detection efficiency of vertical drift chambers (VDCs) is usually very high and the inefficiency is mainly caused by the mis-reconstruction of particle tracks given by the tracking algorithm. Only events with one track were kept for the data analysis, and other events with zero-track and multi-tracks were discarded by applying a one-track-cut. The cut efficiency is generally called the one-track-cut efficiency, which is defined as:
\begin{equation}
  \epsilon_{vdc} = \frac{N_{Track=1}}{N_{0\leq Tracks\leq 4}},
  \label{eq_vdc_eff}
\end{equation}
where $N_{Track=1}$ is the number of events with only one track and $N_{0\leq Tracks\leq 4}$ is the number of events with tracks less than 4. Events with tracks more than 4 are extremely rare for HRS VDCs.

  To correctly evaluate $\epsilon_{vdc}$, good electrons were sampled by applying cuts on detector quantities. Those quantities that require tracking information were avoided when selecting electrons --- quantities derived from VDCs, the acceptance cuts on the focal plane and the target plane quantities, and the calorimeter's energy sum from the cluster reconstruction. Electrons can be alternately identified by cutting the calibrated ADC sums of the calorimeter and the GC. Events with multi-tracks can also be caused by multiple particles coming in one trigger window, and such events can be eliminated by requiring only one hit in each scintillator plane. Note that because paddles in S1 partially overlap, good events coming through the overlapped region are discarded when applying such a cut. This cut should be avoided for any other parts of data analysis.
  
 Cosmic ray events usually come into the VDC at large angles and give bad tracking reconstruction, and they can be eliminated by cutting on the time-of-flight velocity ($\beta_{TOF}$) calculated from the timing information from S1 and S2m. However, this information was not available in this experiment due to several nonfunctional TDC signals in S1 and S2m, so cosmic ray events were not removed. To suppress the cosmic ray background, data with high trigger rates, such as the carbon target data taken at the kinematic setting at the QE peak, were used to calculate the one-track-cut efficiency. From Table~\ref{vdc_table}, the fraction of one-track and multi-track events are listed, where the one-track efficiency is mostly above 99\%. The detection efficiency is the essential property of the detector and should not depend on the kinematic settings, hence one can conclude that the real value of the one-track-cut efficiency is equal to the values calculated with data taken at high rates.
\begin{table}[!ht]
  \centering
  \begin{tabular}{|c||ccccc|}
    \hline
    \textbf{Number of tracks}  & 0 & 1 & 2 & 3 & 4     \\
    \hline \hline
    HRS-L   & 0.030\% & 99.175\% & 0.743\% & 0.045\% & 0.005\%  \\
    \hline
    HRS-R   & 0.048\% & 99.360\% & 0.545\% & 0.039\% & 0.007\%  \\
    \hline 
  \end{tabular}
  \caption{Fraction of different tracks events from QE data,w/o $\beta$ cut}
  \label{vdc_table}	
\end{table} 
\subsection{Particle Identification Efficiencies}
 \begin{figure}[!ht]
  \begin{center}
    \subfloat[\footnotesize{Pion Rejectors}]{
      \includegraphics[type=pdf,ext=.pdf,read=.pdf,width=0.7\textwidth]{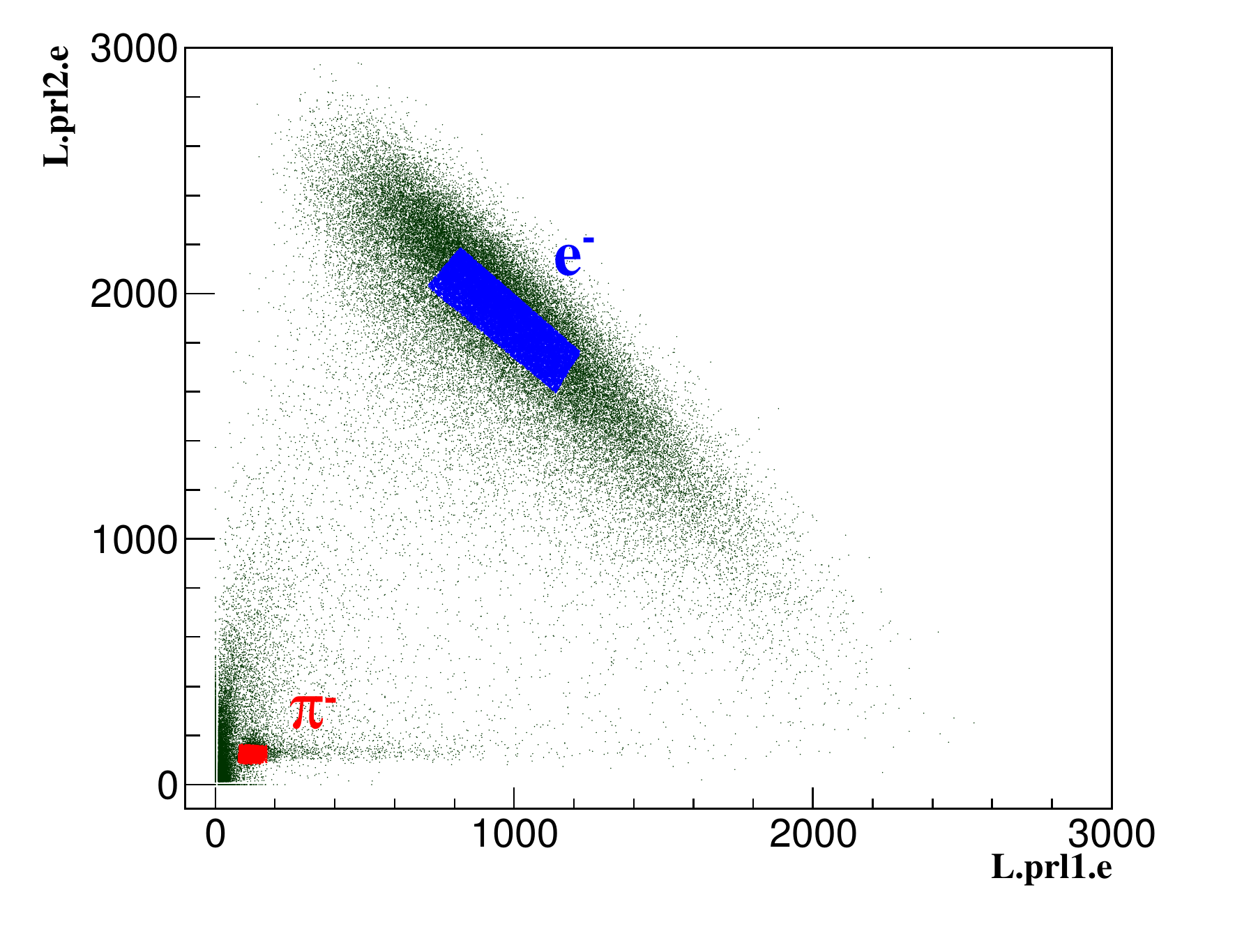}
    }\\
    \subfloat[\footnotesize{Pre-Shower and Shower}]{
      \includegraphics[type=pdf,ext=.pdf,read=.pdf,width=0.7\textwidth]{./figures/pid/L_PID_Calo.Cut_T7}
    }    
    \caption[Electron and pion samples from the calorimeters]{\footnotesize{Electron (blue) and pion (red) samples from the calorimeters. In each plot, the x-axis and the y-axis are the total energies collected by the first layer and the second layer of the calorimeter, respectively. Electrons create large signals either in the first or the second layer during the cascade while the signals created by pions are relatively small in each layer. Graphic cuts were applied on these regions (in color) to select the electrons and pions. } }
    \label{calo_sample}
  \end{center}
\end{figure}

 Electrons are identified by the GC and the calorimeter on each HRS. The GC gives high detection efficiency, since the momentum threshold for electrons to create \v{C}erenkov radiation is only 18 MeV/c, while pions and other heavy particles must have their momenta above 4 GeV/c to fire the detectors. The efficiency is mainly related to the performance of the mirrors in the GC to collect and focus the \v{C}erenkov light. 
\begin{figure}[!ht]
  \begin{center}
    \includegraphics[type=pdf,ext=.pdf,read=.pdf,width=1.00\textwidth]{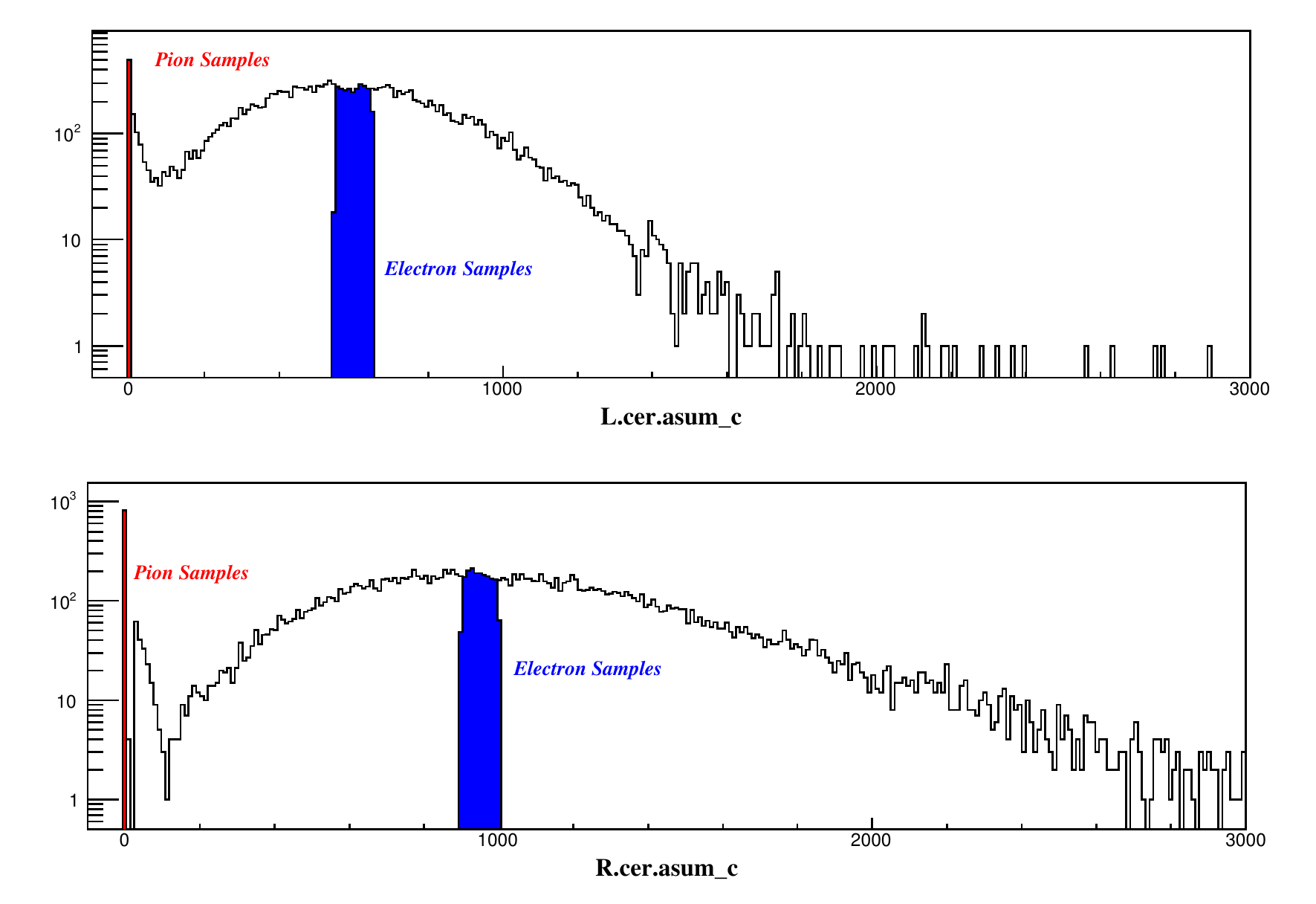}
    \caption[Electron and pion samples from the GC]{\footnotesize{Electron and pion samples from the GC. The x-axis is the sum of calibrated ADC spectra of ten PMTs in the GC in HRS-L (top) or HRS-R (bottom). Electrons were selected by applying cut on the main peak of the spectrum. Pions can not directly create \v{C}erenkov light and they were selected by cutting on low ADC values.}} 
    \label{gc_sample}
  \end{center}
\end{figure}

 The detection efficiencies of the calorimeters are expected to be lower than the GCs. Each calorimeter is composed of many lead glass blocks, so the inefficiency arises when particles go through gaps between blocks or hit the edges of the calorimeter before it creates a shower. 
\begin{figure}[!ht]
  \begin{center}
     \subfloat[GC cut scan on HRS-L]{
      \includegraphics[type=pdf,ext=.pdf,read=.pdf,width=0.7\textwidth]{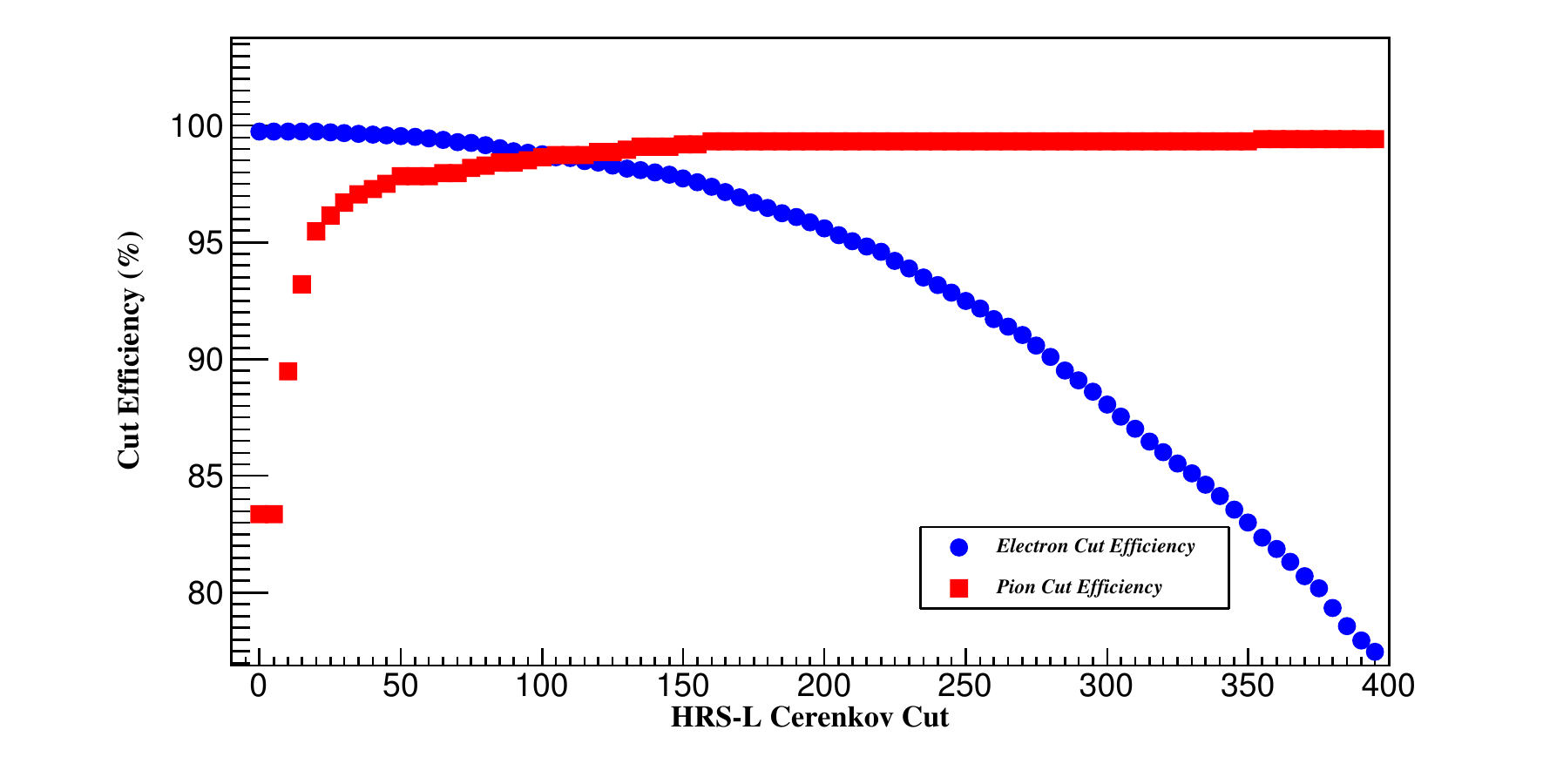}
      } 
      \\
    \subfloat[GC cut Scan on HRS-R]{
      \includegraphics[type=pdf,ext=.pdf,read=.pdf,width=0.75\textwidth]{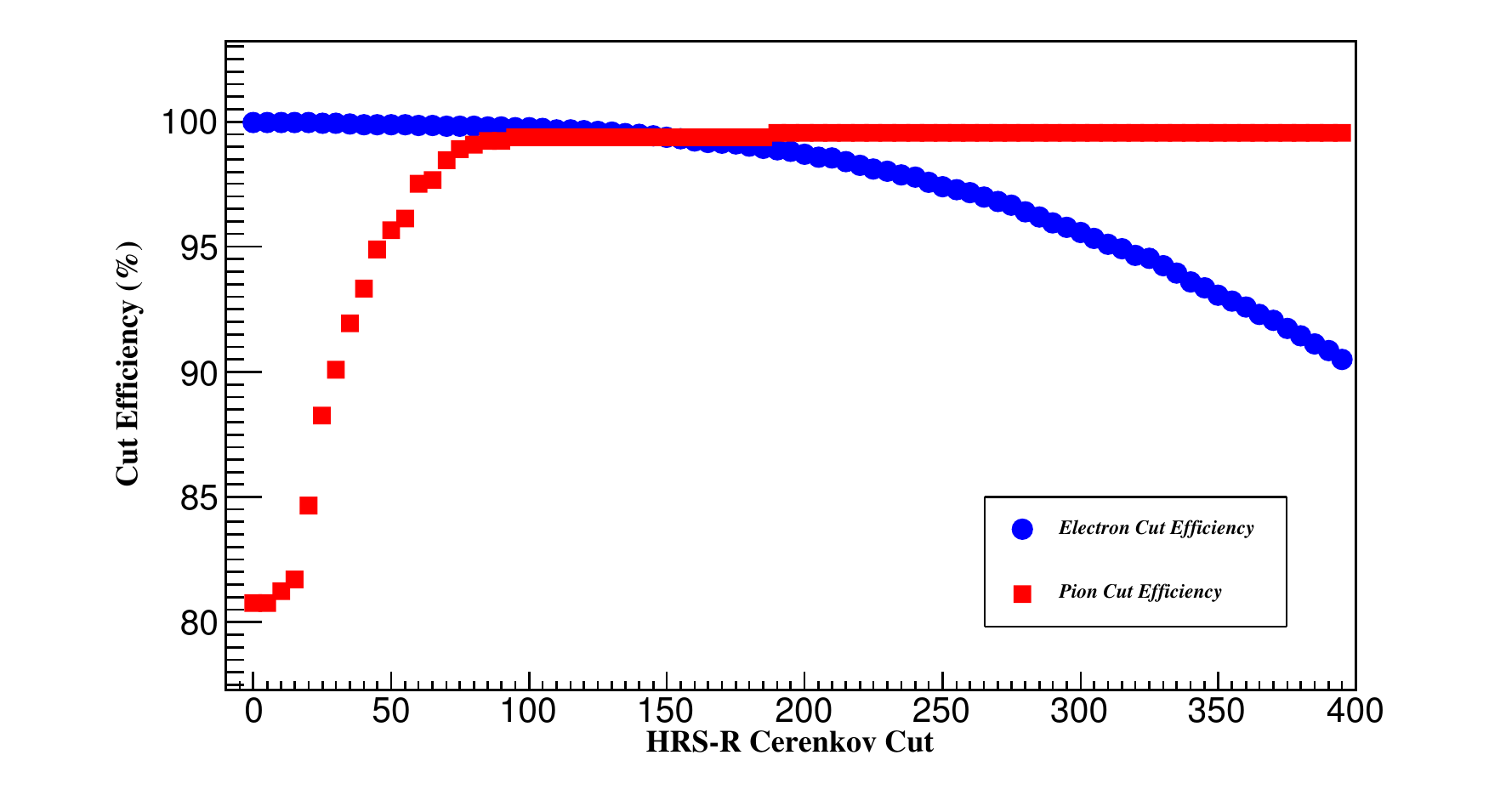}
      }
      \caption[Cut scan of the GCs]{\footnotesize{Cut scan of the GCs on HRS-L (top) and HRS-R (bottom). The x-axis is the channel number of the GC's ADC sum where the cut applies on. The cut efficiencies of pion (red boxes) and electrons (blue dots) were calculated with Eq.~\eqref{cut_eff_pi} and Eq.~\eqref{cut_eff_e} by varying the cut on the GC.} }
     \label{gc_cut_scan}
  \end{center}
\end{figure}

\begin{figure}[!ht]
  \begin{center}
     \subfloat[Calorimeter cut scan on HRS-L]{
      \includegraphics[type=pdf,ext=.pdf,read=.pdf,width=0.75\textwidth]{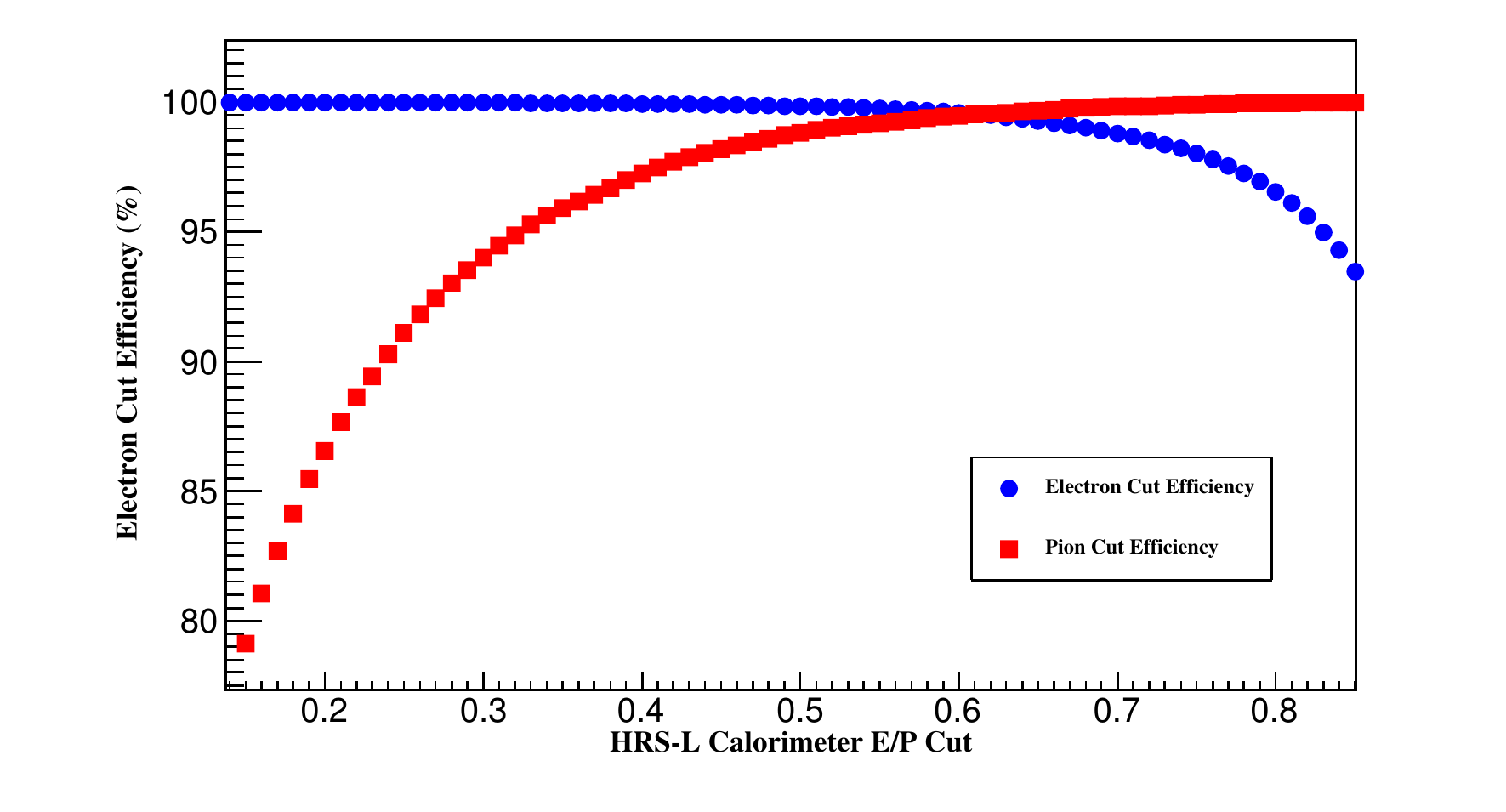}
     } 
     \\
    \subfloat[Calorimeter cut scan on HRS-R]{
      \includegraphics[type=pdf,ext=.pdf,read=.pdf,width=0.75\textwidth]{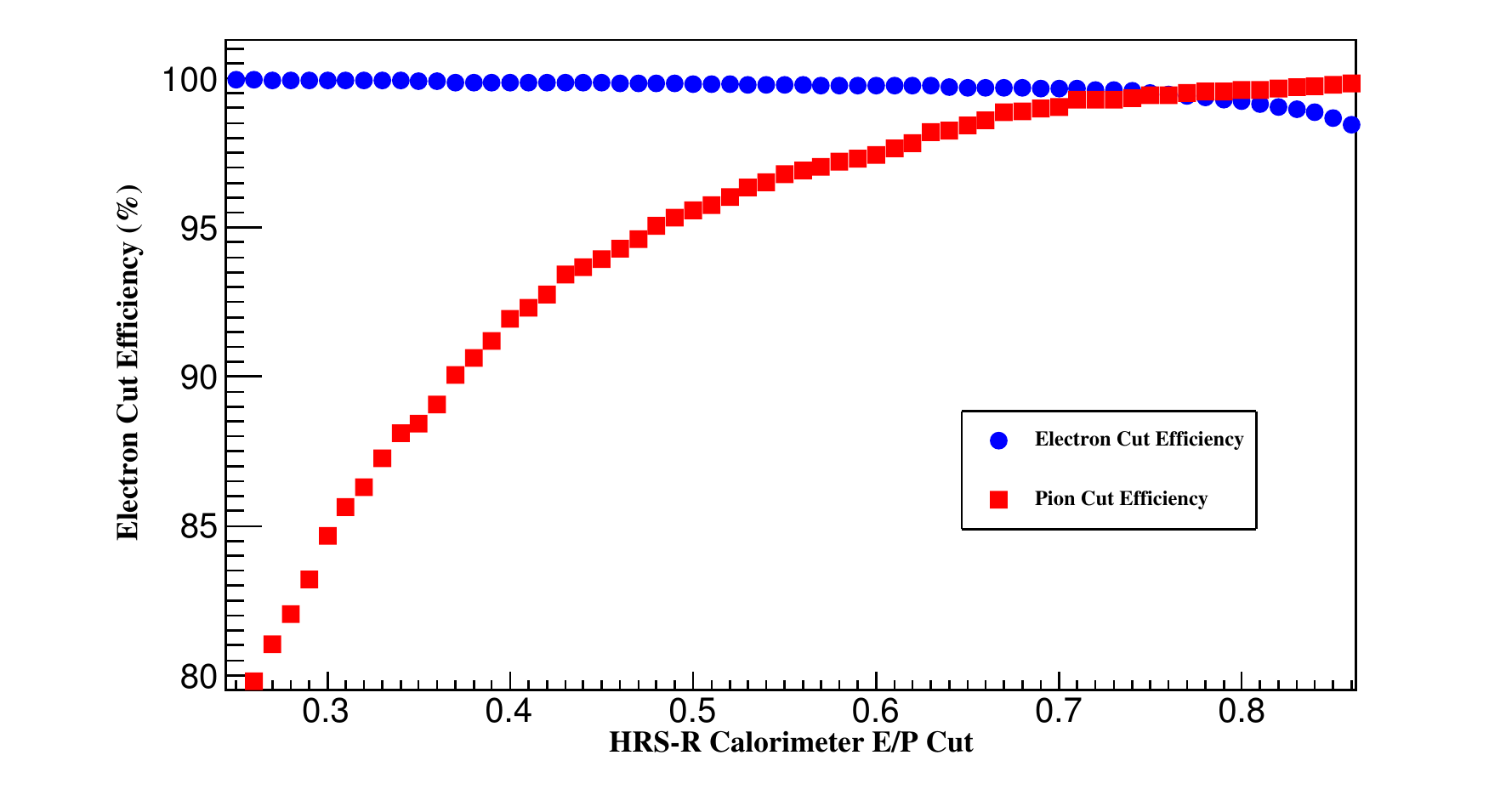}
    }
     \caption[Cut scan of the calorimeters]{\footnotesize{Cut scan of the calorimeters on HRS-L (top) and HRS-R (bottom). The x-axis is the channel number of the calorimeter's ADC sum where the cut applies on. The cut efficiencies of pion (red boxes) and electrons (blue dots) were calculated with Eq.~\eqref{cut_eff_pi} and Eq.~\eqref{cut_eff_e} by varying the cut on the calorimeter.} }
     \label{calo_cut_scan}
  \end{center}
\end{figure}

 The particle identification (PID) for electrons was performed by applying cuts on the calibrated quantities of the GC and the calorimeter. The cuts can reject most unwanted particles, e.g. pions, but on the other hand, they may also accidentally discard good electrons. The PID study aims to obtain the optimized PID cuts on the GC and the calorimeter which can nearly eliminate pions while keeping as many electrons as possible. The cut efficiencies of the GC and the calorimeter have to be individually evaluated to correct the portion of electrons lost during the cuts. 
 
 To evaluate the detection efficiency of the GC (the calorimeter), one first selects electron samples from the calorimeter (the GC) and calculates the percentage of these samples being detected by the GC (the calorimeter), e.g. their signals are slightly larger than the pedestals in the ADC spectrum. Similarly, the evaluation of the cut efficiency for one detector also requires electron samples from the other detectors, but the cut applied on the signals of these samples should be significantly above the pedestals. Hence the calculation of the cut efficiencies for the GCs and the calorimeters should automatically include the detection efficiencies of these detectors.

 In general, for experiments with a large pion background, evaluating the percentage of residual pions mixed into the electron events ($\epsilon_{\pi}=1-\epsilon_{e-\pi}$) is also very crucial. However, compared with the electron rate in the QE region, the pion production rate during the E08-014 was very low. Additionally, the new trigger design had already removed most of pions during online data taking by introducing the GC in the trigger system. Hence the value of $\epsilon_{\pi}$ was expected to be very small. 
\begin{figure}[!ht]
  \begin{center}
    \subfloat[on HRS-L]{
      \includegraphics[type=pdf,ext=.pdf,read=.pdf,width=0.70\textwidth]{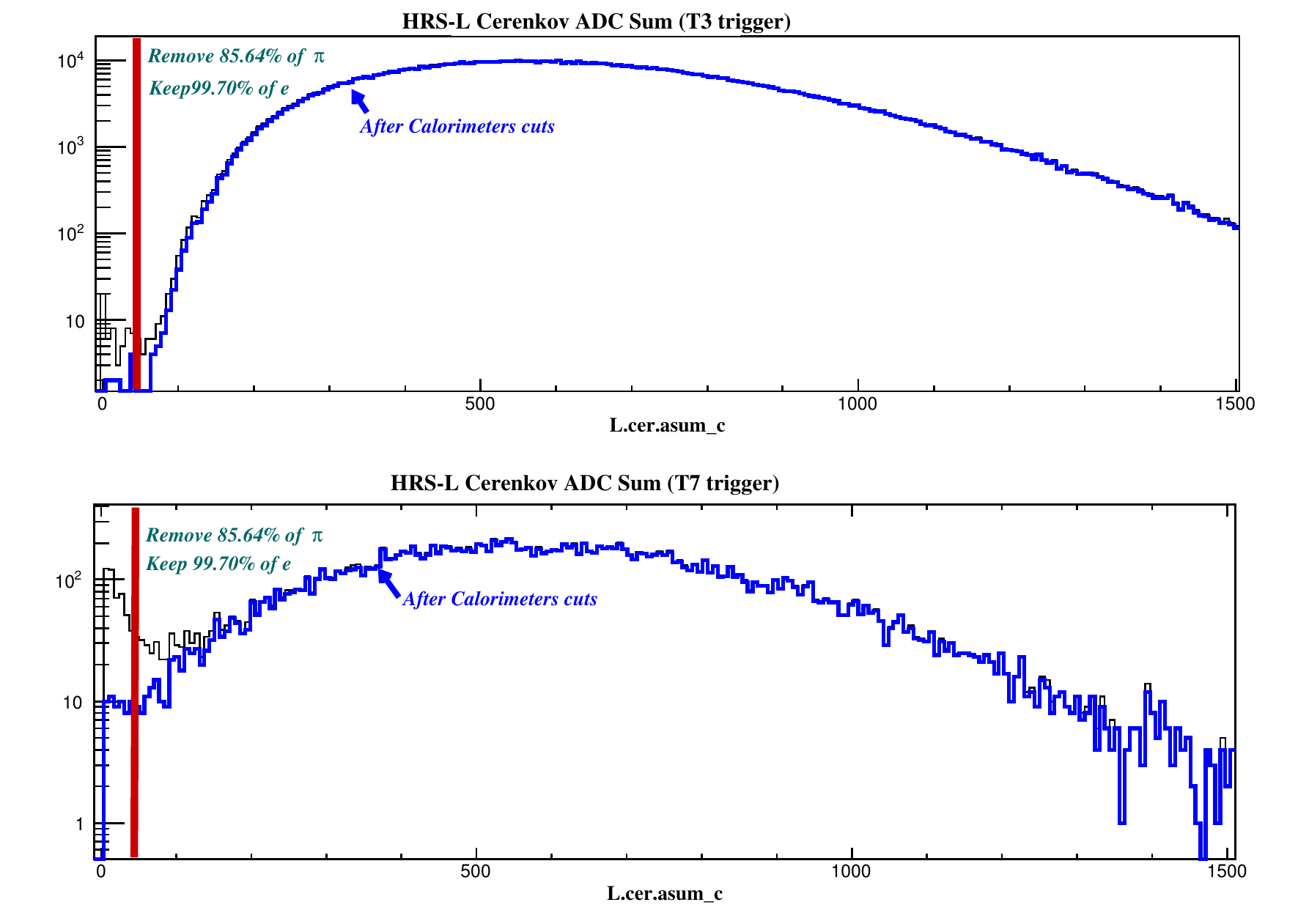}
      \label{Lgc_eff}
    } 
    \\
    \subfloat[on HRS-R]{
      \includegraphics[type=pdf,ext=.pdf,read=.pdf,width=0.70\textwidth]{./figures/pid/L_Cer_PID_Cut}
      \label{Rgc_eff}
    }
    \caption[PID cut on the GCs]{\footnotesize{PID cut on the GCs. In each panel, the top and bottom histograms plot the calibrated ADC sum of events triggered by T1 (T3) and T6 (T7) from HRS-R (HRS-L), respectively. Most of pions have already been rejected in events from T1 and T3 during data taking, so a minimum cut on the GC's ADC spectrum ($\geq 50$) can further remove the rest of pions.}}
    \label{gc_eff}
  \end{center}
\end{figure}

 Events from the T6 and T7 triggers were used to study the PID cut efficiencies since they contained the most of pions. The VDC one-track-cut and the acceptance cuts were applied to select good events. Then pure pion samples and pure electron samples were chosen from the calorimeter (GC) when studying the cut efficiency of the GC (calorimeter). The pion~rejection~efficiency is defined as the percentages of pions removed by applying the PID cuts:
\begin{equation}
\epsilon_{\pi\_rej}^{GC(calo)} = \frac{N_{\pi}^{GC(calo)}}{N_{\pi\_samples}^{calo(GC)}}, 
\label{cut_eff_pi}
\end{equation}
  and the electron cut efficiency can be calculated from:
\begin{equation}
   \epsilon_{e\_cut}^{GC(calo)} = \frac{N_{e}^{GC(calo)}}{N_{e\_samples}^{calo(GC)}},
   \label{cut_eff_e}
\end{equation}
where $N_{\pi\_samples}^{calo(GC)}$ ($N_{e\_samples}^{calo(GC)}$) is the pion (electron) samples from the calorimeter (GC) (Fig.~\ref{calo_sample} and Fig.~\ref{gc_sample}). $N_{\pi}^{GC(calo)}$ is the number of pions rejected and $N_{e}^{GC(calo)}$ is the number of electrons left over after cutting on the GC (calorimeter), respectively. 
\begin{figure}[!ht]
  \begin{center}
    \subfloat[on HRS-L]{
      \includegraphics[type=pdf,ext=.pdf,read=.pdf,width=1.0\textwidth]{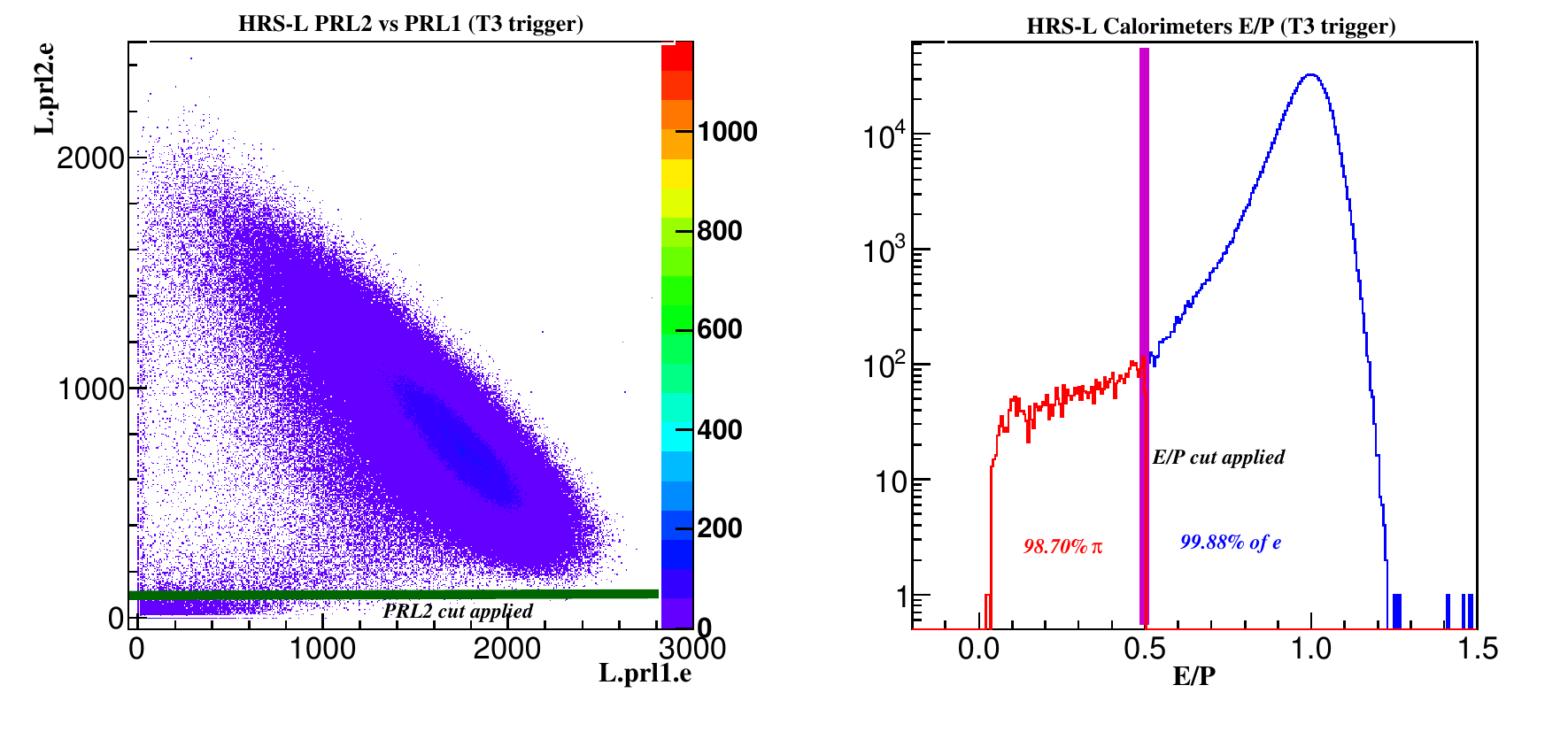}
      \label{Lcalo_eff}
    } 
    \\
    \subfloat[on HRS-R]{
      \includegraphics[type=pdf,ext=.pdf,read=.pdf,width=1.0\textwidth]{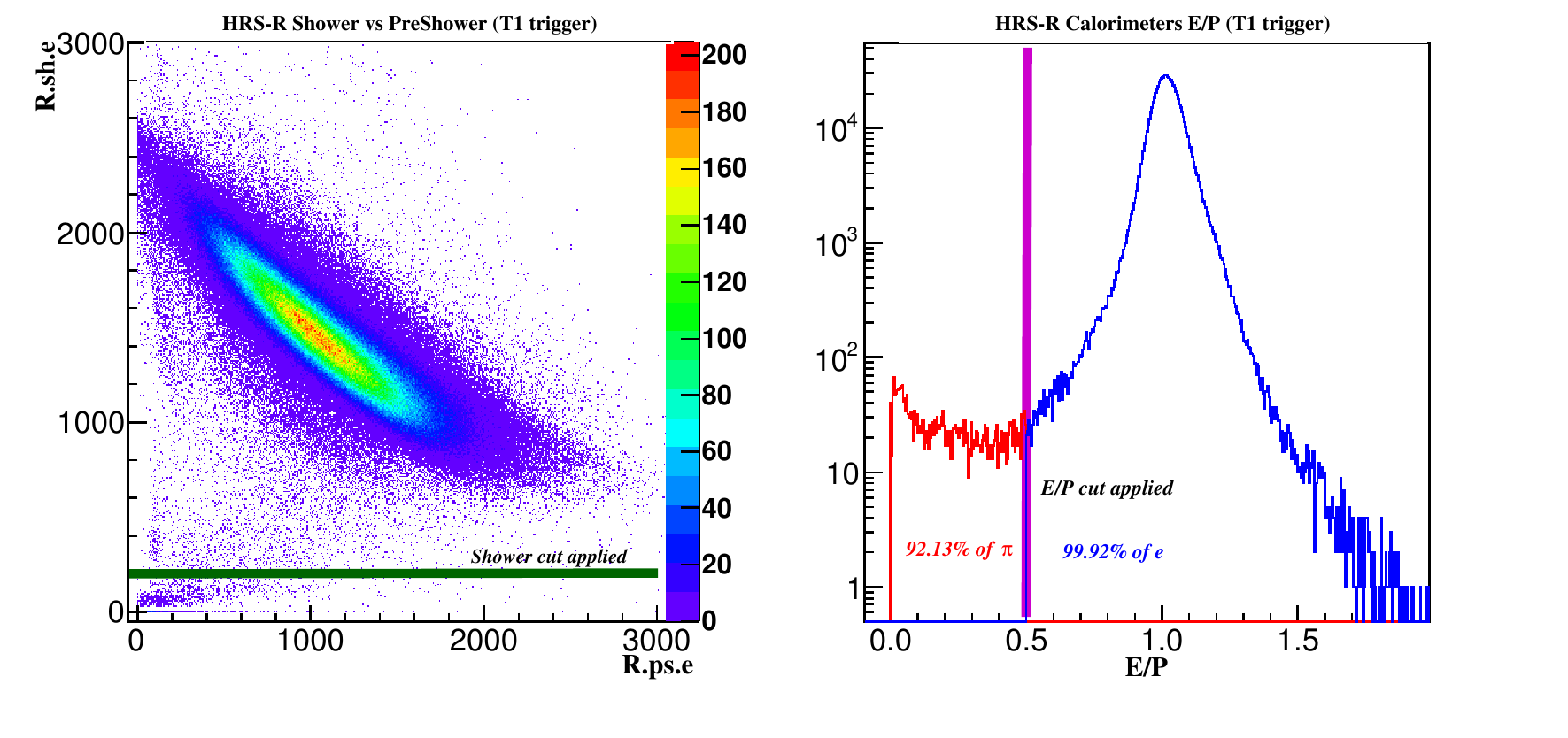}
      \label{Rcalo_eff}
    }
    \caption[PID cut on the calorimeters]{\footnotesize{PID cut on the calorimeters. Most of pions can be removed by the E/P cut ($E/P\geq 0.5$) and the cut on the second layer's ADC spectrum ($PRL2\geq 100$ or $SH\geq 200$).}}
    \label{calo_eff}
  \end{center}
\end{figure}

  A cut scan was performed to study the distributions of the pion rejection efficiencies and the electron cut efficiencies by varying the cuts on the GCs and the calorimeters, shown in Fig.~\ref{gc_cut_scan} and Fig.~\ref{calo_cut_scan}. Fig.~\ref{Lgc_eff} and Fig.~\ref{Rgc_eff} show that for the GC, a cut at the low channel value of the calibrated ADC sum, e.g. $\mathrm{L.cer.asum\_c\geq 50}$ for HRS-L or $\mathrm{R.cer.asum\_c\geq 50}$ for HRS-R, has already remove most of pions and preserve more than 99\% of electrons. The combined cuts on the calorimeter, $\mathrm{E/P\geq 0.5}$ and $\mathrm{L.prl2.e\geq 100}$ ($\mathrm{R.sh.e\geq 200}$), can further remove more than 90\% of pions while remaining more than 99\% of electrons, shown Fig.~\ref{Lcalo_eff} and Fig.~\ref{Rcalo_eff}. In total, on HRS-L (HRS-R), 99.85\% (99.62\%) of pions are eliminated with these combined PID cuts, while 99.58\% (99.86\%) of electrons survive after the cuts. Considering the high electrons rates and low pion production for this experiment, one is not required to specifically correct the pion contamination, and the value of $\epsilon_{e-\pi}$ in Eq.~\eqref{eqxs_org} was set to one.

%% file: cross_section/analysis_mc.tex
\section{Monte Carlo Simulation}
 The Hall-A Single Arm Monte Carlo simulation tool (SAMC) was designed to simulate the transportation of particles from the target plane to the focal plane. SAMC was originally developed in FORTRAN~\cite{A_Duer} and then converted into C++~\cite{hyao_thesis}. The beam position, the spectrometer settings, and the information of the target system can be specified in the code to match the experimental settings. A simulated event has its specified values of the incoming energy, the scattered momentum and the scattering angle, which are defined in the target coordinate system and called the target plane quantities. These quantities are randomly generated with uniform distributions, and with these quantities as inputs, each focal plane quantity is calculated by a set of forward transportation functions which are generated by the SNAKE model~\cite{snack_lerose}. After the focal plane quantities are smeared with the resolution of VDCs, another set of backward transportation functions are used to reconstruct the target plane quantities. During these two processes, events inside and outside the HRS acceptance can be individually identified. Before comparing with the experimental data, the distributions of the target plane quantities are weighted by the radiated cross section values of these simulated events which can be calculated with cross section models embedded in the code. In this analysis, a new cross section model and a special treatment of the no-uniform cryogenic targets have been added in SAMC.

\begin{figure}[!ht]
  \begin{center}
    \subfloat[Target plane quantities on HRS-L]{
      \includegraphics[type=pdf, ext=.pdf,read=.pdf,width=0.92\textwidth]{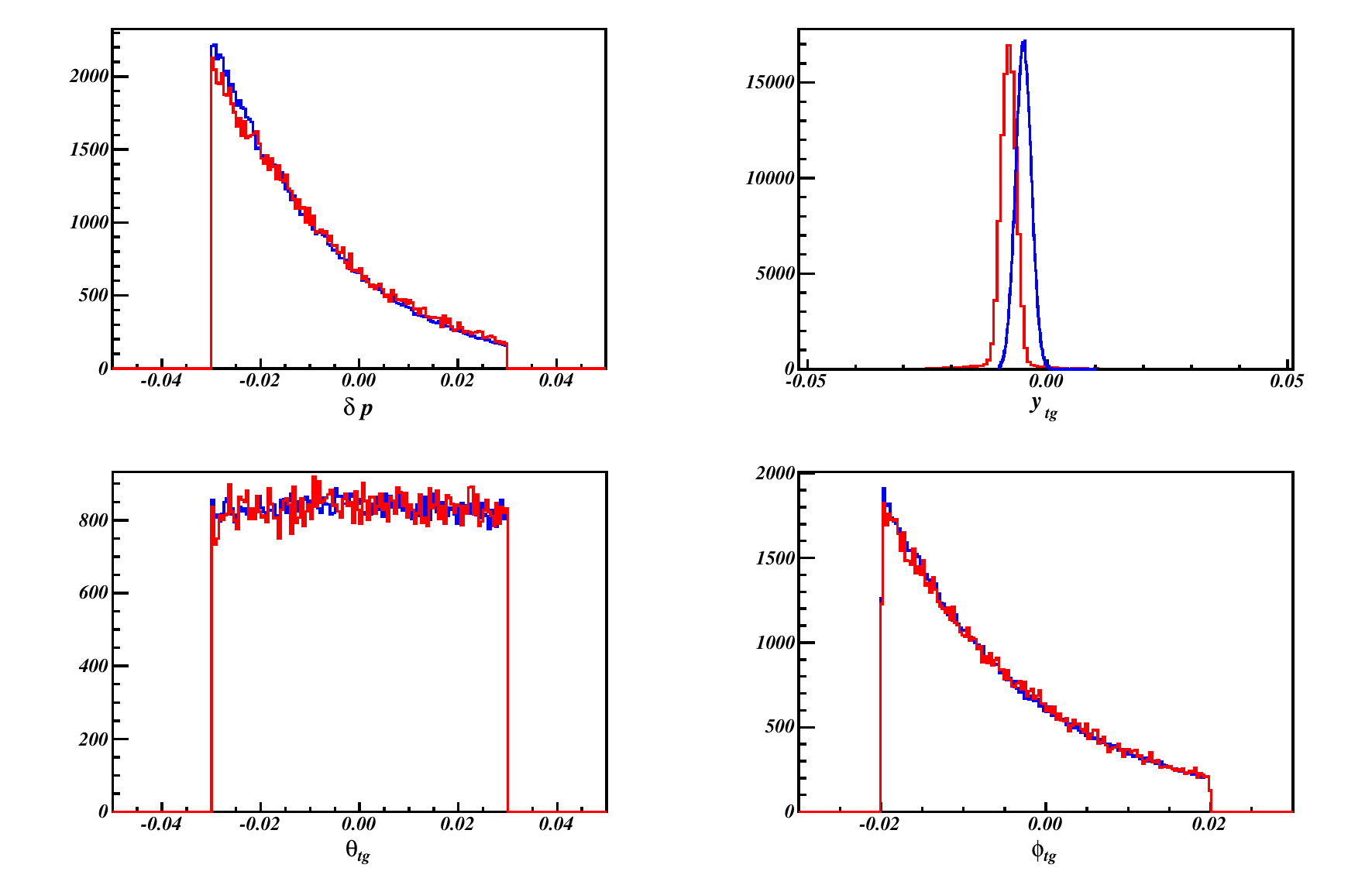}
    }
    \\
    \subfloat[Target plane quantities on HRS-R]{
      \includegraphics[type=pdf, ext=.pdf,read=.pdf,width=0.92\textwidth]{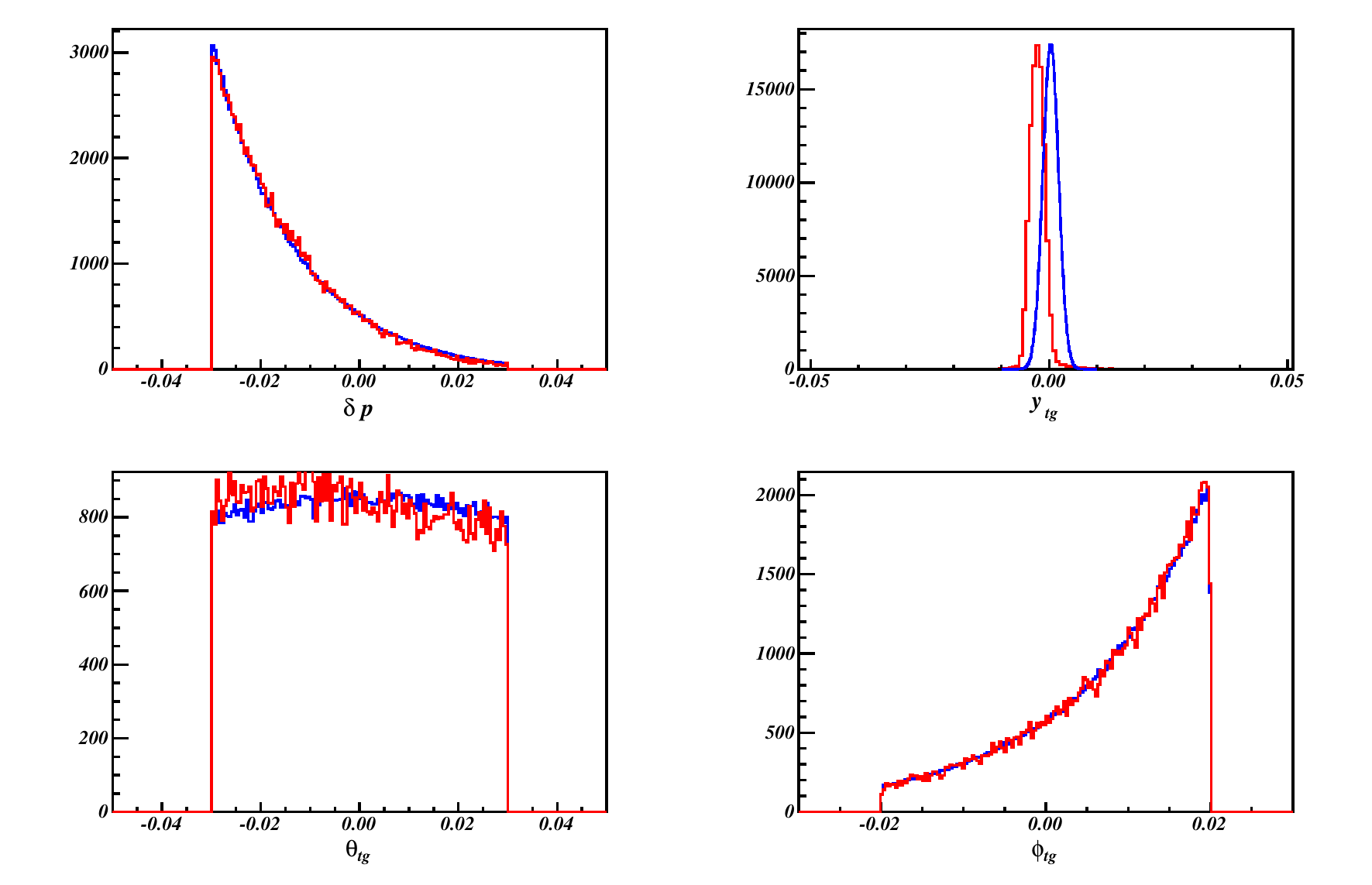}
    }
    \caption[Simulation of $\mathrm{^{12}C}$ target plane quantities]{\footnotesize{Simulation of $\mathrm{^{12}C}$ target plane quantities, where red lines are simulation data from SAMC and blue lines are from the E08-014 data. The offset of $y_{tg}$ between two data is a known issue of SAMC but the offset was too small to affect the acceptance.}}
    \label{samc_tg_c12}
  \end{center}
\end{figure}

\begin{figure}[!ht]
  \begin{center}
    \subfloat[Target plane quantities on HRS-L]{
      \includegraphics[type=pdf, ext=.pdf,read=.pdf,width=0.92\textwidth]{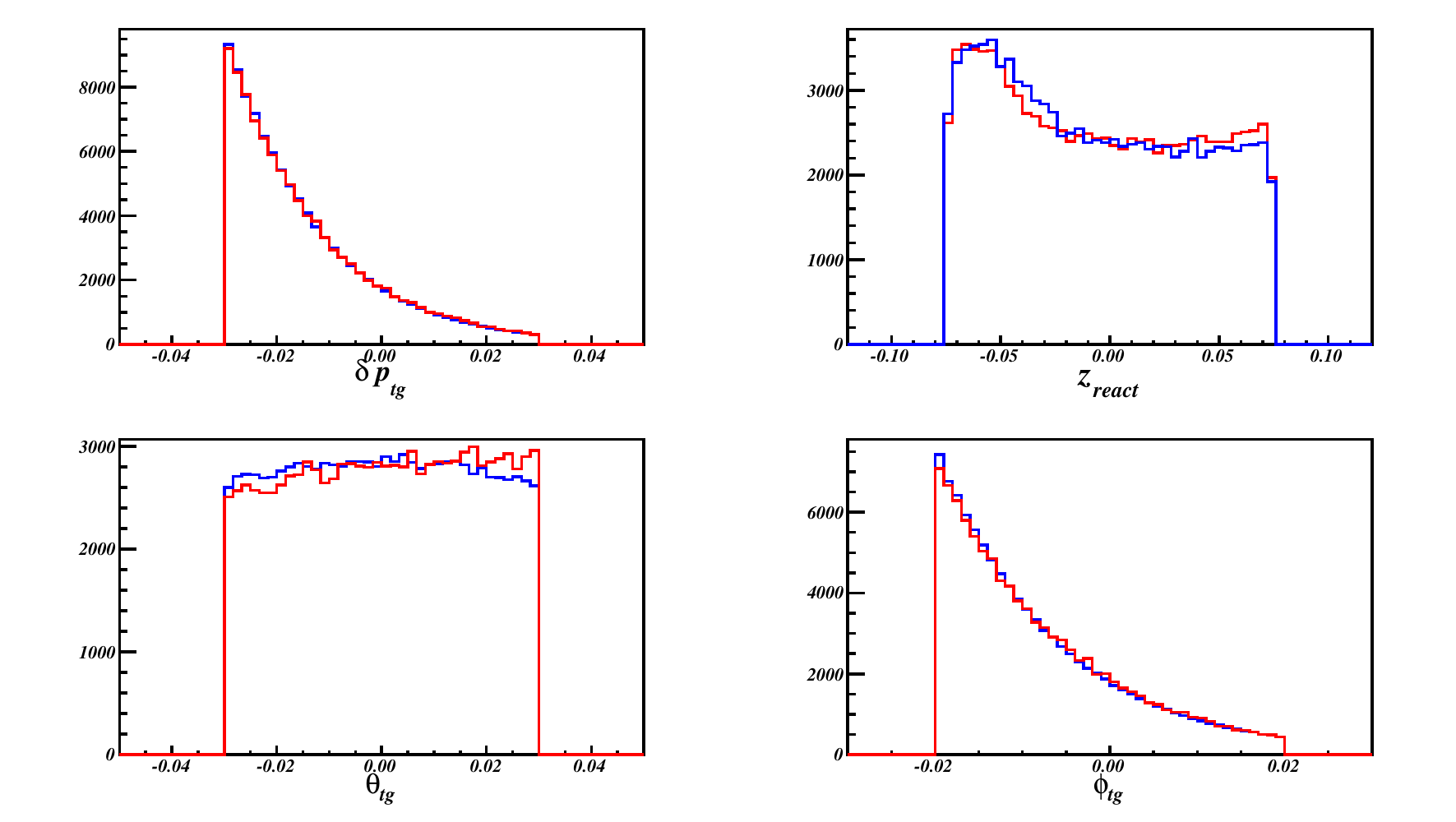}
    }
    \\
    \subfloat[Target plane quantities on HRS-R]{
      \includegraphics[type=pdf, ext=.pdf,read=.pdf,width=0.92\textwidth]{./figures/samc/He3_L_Kin31_TG}
    }
    \caption[Simulation of $\mathrm{^{3}He}$ target plane quantities]{\footnotesize{Simulation of $^{3}$He target plane quantities, where red lines are simulated data from SAMC and blue lines are the experimental data. Instead of $y_{tg}$, the $z_{react}$ distribution is given to compare the real density distribution which was simulated with the function fitted from data (Appendix D).}}
    \label{samc_tg_he3}
  \end{center}
\end{figure}

 There were 20 million events generated for each target in each kinematic setting. Fig.~\ref{samc_tg_c12} and Fig.~\ref{samc_tg_he3} compare the distributions of reconstructed target plane quantities between simulated data and experimental data for $\mathrm{^{12}C}$ and $\mathrm{^{3}He}$. The histograms for simulation data were weighted by the cross sections calculated by XEMC (see next section and Appendix B). The distribution of the same quantity from these two data sets agree nicely with each other. The distribution of $z_{react}$ for the cryogenic target was simulated with the relative density distribution function extracted with the method discussed in Appendix D.

\section{Cross Section Model}
 The inclusive electron scattering cross sections model used in this data analysis is XEMC, a C++ package to compute Born cross sections and radiated cross sections. A brief discussion of the cross section models and radiative correction is given in Appendix B. 
\begin{figure}[!ht]
  \begin{center}
    \includegraphics[type=pdf, ext=.pdf,read=.pdf,width=0.60\textwidth]{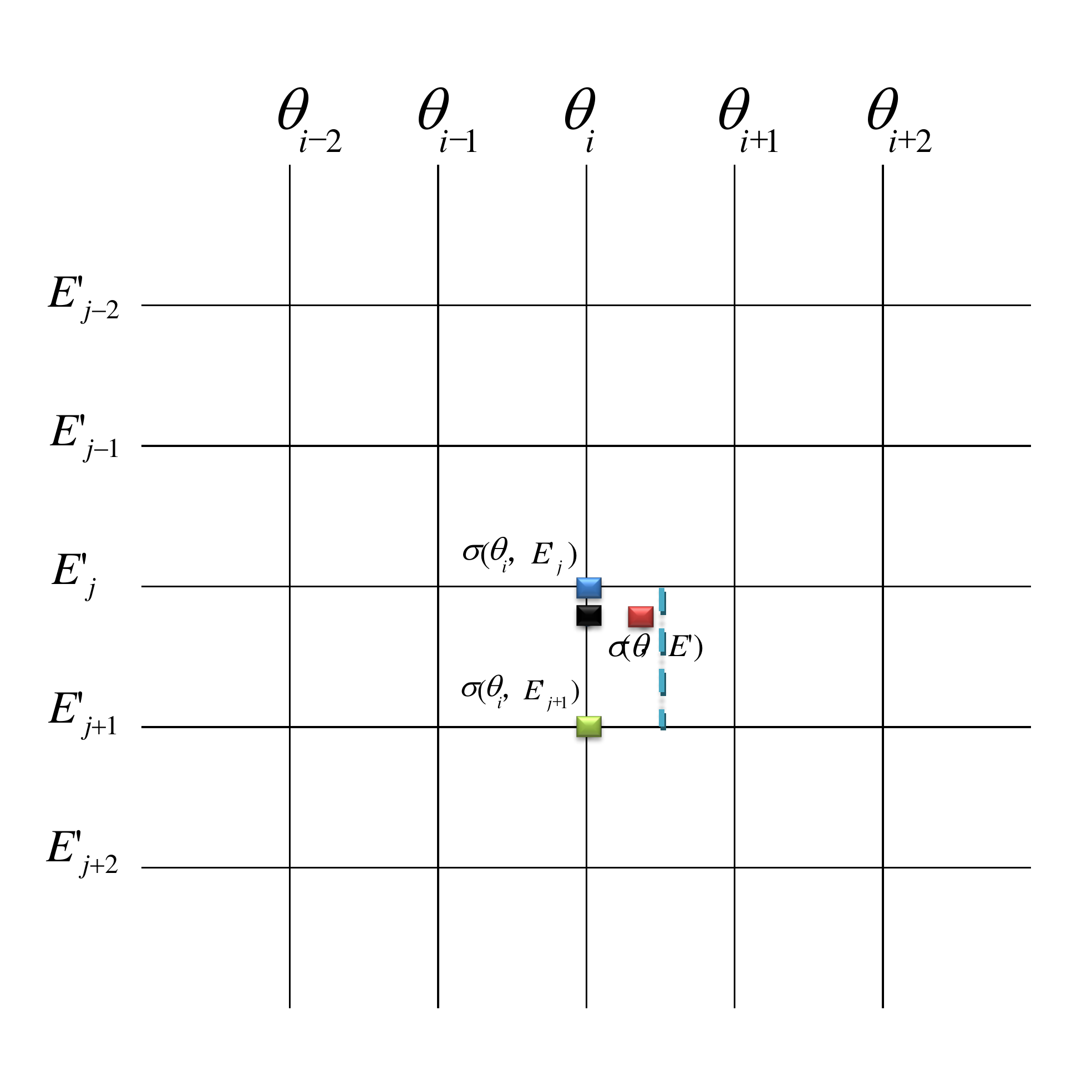}
    \caption[A sketch of cross section lookup tables]{\footnotesize{A sketch of cross section lookup tables. $\sigma(\theta,E') \equiv \sigma(\theta_{i},E')$ when $\theta_{i}\leq \theta < (\theta_{i}+\theta_{i+1})/2$, and $\sigma(\theta,E') \equiv \sigma(\theta_{i+1},E')$ when $(\theta_{i}+\theta_{i+1})/2\leq \theta <\theta_{i+1}$, e.g. from the red point to the black point in this plot. For $E'_{j}<E'<E'_{j+1}$, the cross section is calculated with the linear relationship given in Eq.~\eqref{xs_lookup_ep}.}}
    \label{xs_table}
  \end{center}
\end{figure}

 Calculating radiated cross sections with XEMC usually takes very long time. To generate millions of simulated events, cross section look-up tables were generated for each target in each kinematic setting. When generating each table, the range of the scattering angle, $\Delta\theta$, and the scattered energy, $\Delta E'$, were slightly wider than the actual HRS acceptance. $\Delta\theta$ was divided into 200 bins and $\Delta E'$ was also split into bins of 5 MeV. As shown in Fig.~\ref{xs_table}, the kinematic space for each setting was given as a 2-dimensional lattice where the born cross section and the radiated cross section for each grid, ($\theta_{i}$, $E'_{j}$), were simultaneously calculated. Since the bin sizes are very fine, for fixed momentum, the cross sections at different angles are considered to be equal within one $\theta$ bin, while for a fixed angle, the cross sections are assumed to be proportional to the momentum values inside one $E'$ bin. As illustrated in Fig.~\ref{xs_table}, for a given event, ($\theta,E'$), the value of $\theta$ is replaced by the closest angle bin, e.g. $\theta_{i}$, and when two momentum bins are specified, e.g. $E'_{j}<E'<E'_{j+1}$, the cross section value for this event can be calculated with the linear relationship:
\begin{equation}
  \sigma(E',\theta) = \sigma(E'_{j},\theta^{i}) - \frac{E'-E'_{j}}{E'_{j+1}-E'_{j}}\left(\sigma(E'_{j},\theta^{i})-\sigma(E'_{j+1},\theta^{i})\right).
  \label{xs_lookup_ep}
\end{equation}
For the same event, the difference between the cross section obtained from the look-up table and the cross section directly calculated from XEMC is less than 0.1\%. This method can dramatically reduce the computation time when generating simulation events. Tables were re-generated each time when the model was changed or the experimental details were updated, e.g. the target thickness.

%% file: cross_section/analysis_xs.tex
\section{Event Selection and Corrections}
  The ideal way to extract an experimental cross section is to use the scattered electrons with the same values of $E_{0}$, $E'$ and $\theta$. However, although the beam energy can be easily locked at one value, $E'$ and $\theta$ can vary within the acceptance of the spectrometer, and due to the statistical limitation, the cross sections can only be calculated by allowing the values of $E'$ and $\theta$ to change within finite ranges, e.g. $\Delta E'$ and $\Delta\Omega$ in Eq.~\eqref{eqxs_org}. In practice, the experimental data is divided by binning one or more kinematic variables with known bin sizes, and the cross section is evaluated at the center of each bin. The way to choose the binning method, including the acceptance ranges and the bin sizes, requires additional corrections during the cross section extraction.
   
   For the E08-014, the data was binned in $E'$ only, and the cross sections were calculated in each $E'$ bin with the same scattering angle, $\theta=\theta_{0}$. The determination of the kinematic space, the acceptance correction and the binning correction will be discussed in this section. A list of cuts to select the good electron events is also given.  
  
\subsection{Central Momentum and Angle}
\begin{figure}[!ht]
  \begin{center}
    \includegraphics[type=pdf, ext=.pdf,read=.pdf,width=0.60\textwidth]{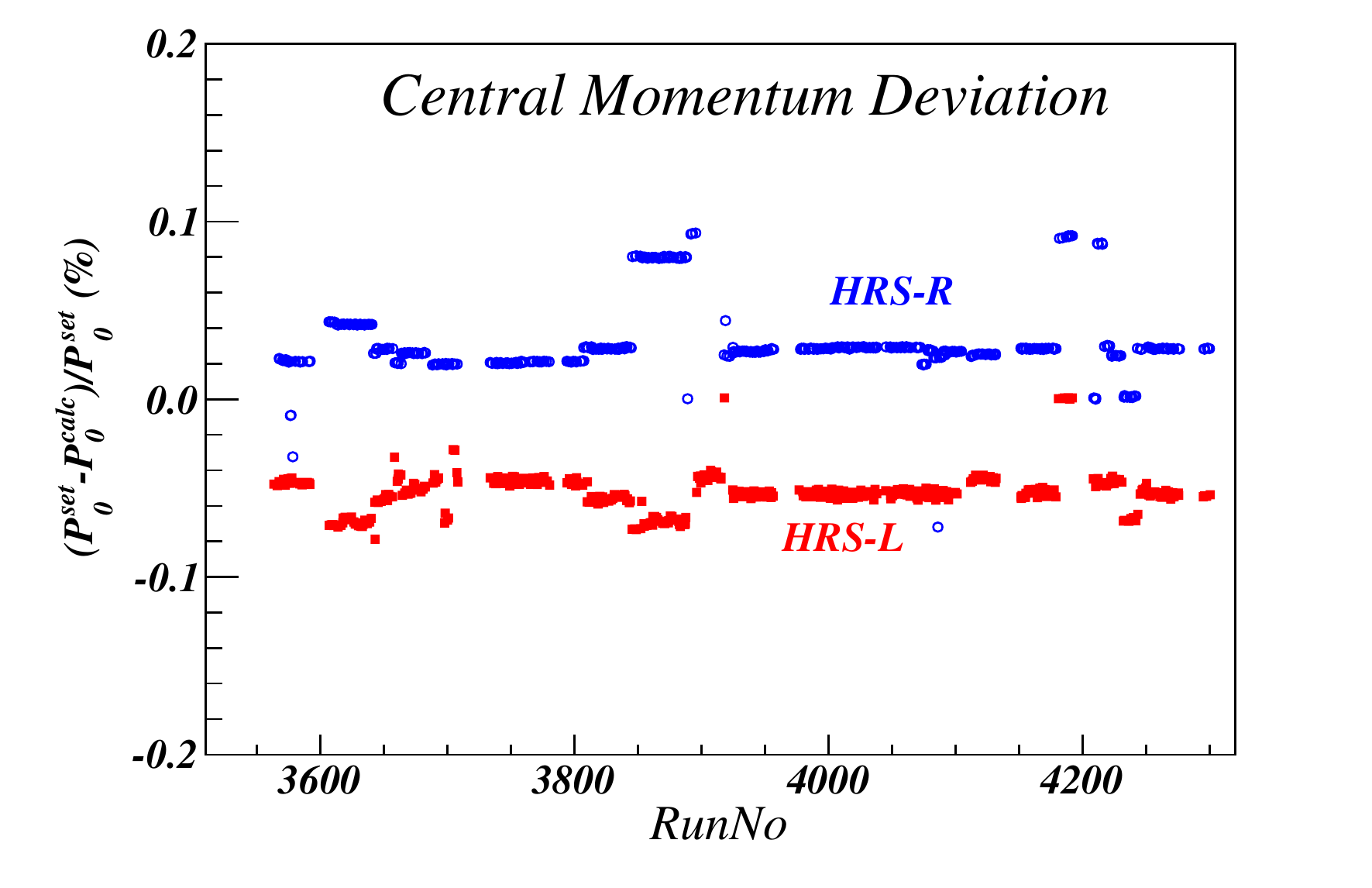}
    \caption[Central momentum deviation]{\footnotesize{Central momentum deviation, where the blue circles and the red boxes are the deviations of the central momentum on HRS-R and HRS-L, respectively. The x-axis is the run number and the y-axis is the deviation in percentage.}}
    \label{point_mom}
  \end{center}
\end{figure}
The kinematic space is determined by the central scattered momentum, the central scattering angle, and the acceptance of the HRS. The central momentum was given by the field values of the HRS magnets which were locked at the setting values by the HRS NMR system during the experiment. The off-line calculation gives the absolute value of the central momentum with the magnetic field of the dipole~\cite{halla_nim}:
\begin{equation}
 P_{0} = \sum_{i=0}^{4} \gamma_{i} \cdot \left( 10\cdot B^{NMR}_{dipole} \right)^{i},
\end{equation}
where $\mathrm{\gamma_{1,2,3,4}=(0, 270.2, 0, -0.0016)}$ for HRS-L and  $\mathrm{\gamma_{1,2,3,4}=(0, 269.8, 0, -0.0016)}$ for HRS-R. $B^{NMR}_{dipole}$ is the field reading from the NMR monitor. Fig.~\ref{point_mom} shows that the actual central momenta were mostly off by $\pm$3\% while few of them were off by $\pm$10\%. During the cross section extraction, the central momenta were assigned to the calculated values instead of the set values.

The central scattering angle was specified during the experiment by moving the HRS to point at the angle marked on the floor. These floor marks were drawn with respect to the hall center and may not accurately reflect the true values. Moreover, the actual central scattering angle also depends on the offsets between the spectrometer center and the hall center which are different when the spectrometer points at different angles. For some extreme cases, when the spectrometer is moved away from one angle and later moved back to the same value, the actual angles may be different between these two periods.
\begin{table}[!ht]
  \centering
  \begin{tabular}{|c||cccc|}
    \hline
    \textbf{RunNo} &$\theta^{set}_{0}(L)$&$\theta^{true}_{0}(L)$&$\theta^{set}_{0}(R)$&$\theta^{true}_{0}(R)$\\
    \hline \hline
    3565$\sim$3656          & 25.00 & 25.00  & 25.00 & 25.00 \\
    \hline
    3657$\sim$3683          & 21.00 & 21.03  & 21.00 & 21.04 \\
    \hline
     3684$\sim$3708         & 23.00 & 23.00  & 23.00 & 23.01 \\
    \hline
     3735$\sim$3891         & 25.00 & 24.99  & 25.00 & 25.00 \\
    \hline
     3892$\sim$3916         & --    &  --    & 21.00 & 21.03 \\
    \hline
     3917$\sim$4071         & 28.00 & 27.98  & 28.00 & 27.99 \\
    \hline
    4073$\sim$4103          & 21.00 & 21.04  & 28.00 & 27.99 \\
    \hline
    4112$\sim$4179          & 23.00 & 23.00  & 23.00 & 23.04 \\
    \hline
    4181$\sim$4241          & 25.00 & 24.98  & 25.00 & 25.00 \\
    \hline
    4242$\sim$4250          & 21.00 & 21.02  & 21.00 & 21.03 \\
    \hline    
    4251$\sim$4299          & 28.00 & 27.98  & 28.00 & 27.99 \\
    \hline
    \end{tabular}
  \caption{Scattering angle correction}
  \label{scat_angle_table}	
\end{table}

 To obtain the actual central scattering angle each time after the spectrometer was moved, a survey would be performed to correct the errors of the floor marks and to measure the offset between the two centers. Unfortunately, the survey could not been done each time the spectrometers were moved. However, the optics target was surveyed at the beginning of this experiment when both HRSs were set at $\mathrm{25^{\circ}}$, and the positions of $z_{react}$ at different angles can be extracted from the data. Combined with the survey reports from earlier experiments which had similar settings, the actual central scattering angles can be calculated with the difference of $z_{react}$ at $\mathrm{25^{\circ}}$ and at the setting angle ($\Delta z_{react}=z_{react}(\theta_{0})-z_{react}(25^{\circ})$), as follow: 
\begin{eqnarray}
 & &\theta_{tg} = \frac{D_{x}+x_{sieve}-y_{beam}}{L-x_{beam} \cdot sin\theta^{set}_{0}-\Delta z_{react} \cdot cos\theta^{set}_{0}},\\
 & &\phi_{tg}   = \frac{D_{y}+y_{sieve}-x_{beam} \cdot cos\theta^{set}_{0}+\Delta z_{react} \cdot sin\theta^{set}_{0}}{L-x_{beam} \cdot sin\theta^{set}_{0}-\Delta z_{react} \cdot cos\theta^{set}_{0}},\\
 & &\theta^{true}_{0} = acos\left( \frac{cos\theta^{set}_{0}-\phi_{tg}sin\theta^{set}_{0}}{\sqrt{1+\theta_{tg}^{2}+\phi_{tg}^{2}}} \right),
\end{eqnarray}
where $D_{x}$, $D_{y}$, $x_{sieve}$, $y_{sieve}$ and L are given in Table~\ref{optics_offset_table} and Table~\ref{sieve_offset_table}. The beam position ($x_{beam}$, $y_{beam}$) was locked at (-2.668 mm, 3.022 mm) during the experiment. $\theta^{set}_{0}$ is the central scattering reading from the floor marks and $\theta^{true}_{0}$ is the actual central scattering angle after the correction. As shown in Table~\ref{scat_angle_table}, the calculation showed that the maximum offset between $\theta^{true}_{0}$ and $\theta^{set}_{0}$ was not larger than $\mathrm{0.04^{o}}$. The value of $\theta^{true}_{0}$ was calculated for runs taken at each run period when the spectrometer was moved to different positions. The cross sections were calculated with these updated values.

\subsection{Acceptance Correction}
\begin{figure}[!ht]
  \begin{center}
    \includegraphics[type=pdf, ext=.pdf,read=.pdf,width=0.60\textwidth]{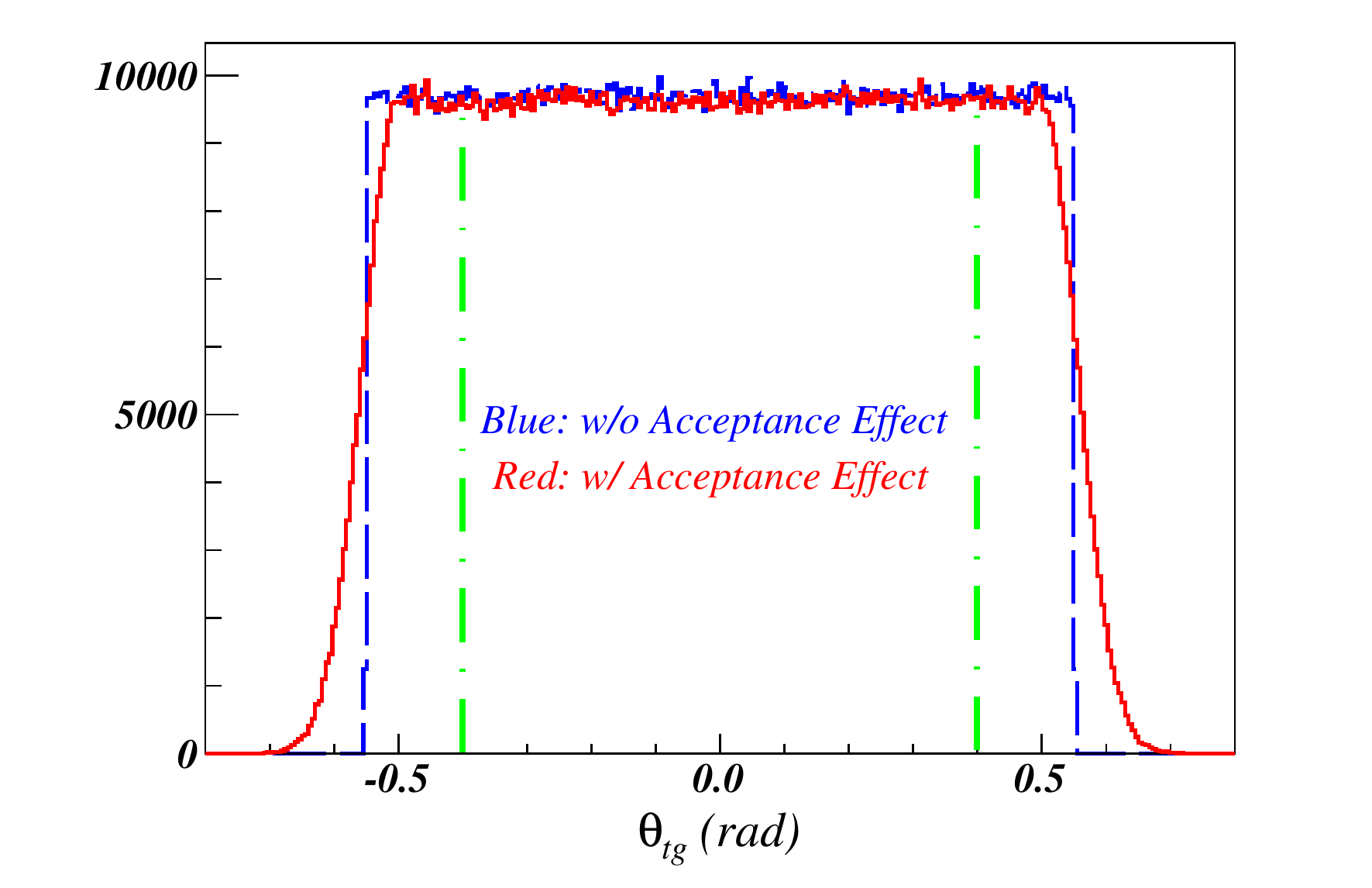}
    \caption[A demonstration of the acceptance effect]{\footnotesize{A demonstration of the acceptance effect, where the distribution of $\theta_{tg}$ is generated by assuming no cross section weighting effect. The blue line shows that the acceptance is flat when the HRS acceptance is perfect, while the red line demonstrates the slow fall-off of the acceptance edges. Such an effect is mainly due to the geometry of the HRS magnets and also contributed by the resolutions of the VDC tracking and the optics reconstruction. Green lines show the cuts to select the flat acceptance region.}}
    \label{accp_demo}
  \end{center}
\end{figure}
 The HRS acceptance includes both the range of momentum dispersion ($\Delta\delta p$) and the total solid angle which is the product of the out-of-plane angle ($\theta_{tg}$) and the in-plane-angle ($\phi_{tg}$). For an extended target, the optics reconstructed reaction point along the beam direction ($z_{react}$) is also affected by the HRS acceptance. These four quantities, called the target plane quantities, are essential to reconstruct the reaction at the target. Due to the geometry of the HRS magnets, the event distributions of these quantities are not cut off immediately at the edge of the acceptance and instead, they fall off relatively slowly with a gaussian tail, as can be seen in Fig.~\ref{accp_demo}. In addition, the resolution of VDC tracking and the accuracy of the optics reconstruction can also smear the distributions of these quantities. 
 
 Choosing the right acceptance ranges of the target plane quantities is crucial in order to obtain the correct cross section results. Tight cuts on the target plane quantities were used to select events at the central region of the HRS acceptance. Cutting out the tails on the edges of the focal plane variables also removes multi-scattering events produced inside the spectrometer. The acceptance cuts will be enlarged to increase the statistics of events in one bin, until the cross section results start to deviate from the results calculated with tighter cuts.
 
  However, good events can be incorrectly discarded when one applies the combination cuts of the four target plane quantities to define a valid acceptance region. Such an effect can be corrected by the HRS simulation for each bin:
 \begin{equation}
  A(E_{0},E_{i}, \theta_{0}) = \frac{N^{gen}_{MC}}{\Delta E'_{MC} \Delta\Omega_{MC}}/\frac{N^{i}_{MC}}{\Delta E'_{bin} \Delta\Omega_{EX}},
 \label{acc_corr}  
  \end{equation}
where $\Delta E'_{bin}$ is the bin size of $E'$ and is fixed in both the simulated data and experimental data, and $\Delta\Omega_{EX}$ is the selected angular acceptance range for the experimental data. $N^{i}_{MC}$ is the number of simulated events in the $ith$ bin, with the same acceptance cuts used for the experimental events ($N^{i}_{EX}$) in this bin.  $N^{gen}_{MC}$ is the total number of simulated events without any cuts. $\Delta E'_{MC}$ and $\Delta \Omega_{MC}$ define the full momentum and angular acceptance in the simulation, respectively, and they are slightly larger than the HRS acceptance. Overall,  $\frac{N^{i}_{MC}}{\Delta E'_{bin} \Delta\Omega_{EX}}$ denotes the average number of events in the unit kinematic space which is limited by the HRS geometry, while the other term, $\frac{N^{gen}_{MC}}{\Delta E'_{MC} \Delta\Omega_{MC}}$, gives the average number of events in the unit kinematic space without any spectrometer limitations. Eq.~\eqref{acc_corr} is usually referred to as the acceptance correction. 

\subsection{Binning Correction}
 The cross section results were calculated by binning the data on $E'$. The binning ranges and step sizes are given in the following table:
\begin{table}[!ht]
  \centering
  \begin{tabular}{|c||ccccccccc|}
    \hline
    \textbf{Kin}        & 3.1 & 3.2 & 4.1 & 4.2 &5.0 &5.05 & 5.1 & 5.2 & 6.5 \\
    \hline \hline
    $E'^{Min}$    & 2.76   & 2.90   & 2.71   & 2.88   &2.38   & 2.52   & 2.66   & 2.85   & 2.70   \\
    \hline
    $E'^{Max}$    & 3.05   & 3.21   & 3.00   & 3.19   &2.63   & 2.78   & 2.94   & 3.14   & 2.99   \\
    \hline
    $\Delta E'$      & 0.01   & 0.01   & 0.01   & 0.01   & 0.01  &0.01    & 0.01   & 0.01   & 0.01  \\
    \hline
  \end{tabular}
  \caption{E' binning size and range}
  \label{bin_table}	
\end{table}

  From Eq.~\eqref{eqxs_org}, when binning on $E'$, the cross section in each bin is given as a function of the central scattering angle ($\theta_{0}$) and the momentum value at the center of the bin ($E'_{i}$). However, events in each bin carry different momenta varying from $E'_{i}-\frac{1}{2}\Delta E'$ to $E'_{i}+\frac{1}{2}\Delta E'$ , while their central scattering angles can deviate from $\theta_{0}$ within the solid angle, $\Delta \Omega_{EX}$. A bin-centering correction is applied to remove the effect with the simulation data and the cross section model:
\begin{equation}
 B(E_{0},E_{i}, \theta_{0}) = \frac{\sigma^{rad}_{XEMC}(E_{0},E'_{i},\theta_{0})}{\sum_{j\in i}\sigma^{rad}_{XEMC}(E_{0},E'_{j},\theta_{j})},
  \label{bin_corr}  
\end{equation}
where $\sum_{j\in i}$ means summation over the radiated cross section values, $\sigma(E'_{j},\theta_{j})$), of all Monte Carlo events in the $ith$ bin. $\sigma^{rad}_{XEMC}(E'_{i},\theta_{0})$ and $\sigma^{rad}_{XEMC}(E'_{j},\theta_{j})$ are calculated from the XEMC model.

\subsection{Cuts}
In addition to cutting on the binning variable, there are several other cuts which were applied to select good scattered electron events:
\begin{enumerate}
\item Cutting on production trigger events (see Appendix A);
\item Removing pulser events generated by EDTM modules;
\item Beam trip cut;
\item Selecting events with only one track in VDCs; 
\item Cuts on the focal plane acceptance;
\item Cuts on the target plane acceptance;
\item PID cuts on the GC and the calorimeter.
\end{enumerate}

 When the extraction of cross sections involves data from more than one run, the total number of events after the cuts defined above is given by:
\begin{equation}
  N_{EX}^{i} = \sum_{r} \frac{PS1(3)^{r}\cdot N_{T_{1(3)}}^{r}}{LT_{T_{1(3)}}^{r}},
  \label{eq_nex}
\end{equation}
where $r$ represents the run number and $N_{T_{1(3)}}^{r}$ is the total number of events from $T_{1}$ on HRS-R ($T_{3}$ on HRS-L) and recorded by DAQ after cutting out the beam trip. Note that events from each run are individually corrected by the Live-Time ($LT_{T_{1(3)}}^{r}$) before they are added together.

\section{From Yields to Cross Sections}
The experimental Born cross section can be calculated from Eq.~\eqref{eqxs_org} after applying the acceptance correction (Eq.~\eqref{acc_corr}) and the bin-centering correction (Eq.~\eqref{bin_corr}): 
\begin{equation}
  \sigma^{Born}_{EX} (E'_{i}, \theta_{0}) =  A(E'_{i}, \theta_{0}) \cdot B(E'_{i}, \theta_{0})  \cdot \sigma^{rad}_{EX} (E'_{i}, \theta_{0}) \cdot RC(E'_{i}, \theta_{0}).
  \label{eqxs_org_corr}
\end{equation}
Note that the initial electron energy, $E_{0}$, is fixed at 3.356 GeV during this experiment so it is omitted from the equation. The last term is the radiative correction factor:
\begin{equation}
 RC(E'_{i}, \theta_{0}) = \frac{\sigma^{Born}_{XEMC}(E'_{i},\theta_{0})}{\sigma^{rad}_{XEMC}(E'_{i}, \theta_{0})}.
 \label{eq_radc_fact}
\end{equation}

 Extraction of cross sections from Eq.~\ref{eqxs_org_corr} largely relies on the performance of the simulation and the cross section model, which, however, can not be directly examined from the cross section results. Two useful quantities, the experimental yield and the Monte Carlo (MC) yield, can be extracted to directly compare their differences. The experimental yield is written as:
\begin{equation}
  Y^{i}_{EX} = \frac{N^{i}_{EX}}{N_{e} \cdot \epsilon_{eff}},
  \label{eqyex}
\end{equation}
where $\epsilon_{eff}=\epsilon_{trig}\cdot\epsilon_{vdc}\cdot\epsilon_{e\_cut}^{GC}\cdot\epsilon_{e\_cut}^{calo}$ which are given in Eq.~\eqref{trigger_eff3}, Eq.~\eqref{eq_vdc_eff} and Eq.~\eqref{cut_eff_e}, respectively. The MC yield is given by:
\begin{equation}
  Y^{i}_{MC} = \eta_{tg}\cdot \sum_{j\in i}\sigma^{rad}_{model}(E'_{j},\theta_{j}) \cdot \frac{\Delta\Omega_{MC} \Delta E'_{MC}}{N_{MC}^{gen}}.
  \label{eqymc}
\end{equation}

 The ratio of the experimental yield to the MC yield should be close to one if the performance of the HRS can be well simulated by the MC data and the XEMC model produces cross sections close to the actual values. The experimental Born cross section from Eq.~\ref{eqxs_org_corr} can be rewritten as:
\begin{equation}
  \sigma^{Born}_{EX}(E'_{i}, \theta_{0}) = \frac{ Y^{i}_{EX}}{Y^{i}_{MC}} \cdot \sigma^{Born}_{XEMC}(E'_{i}, \theta_{0}),
  \label{eqxs_ratio}
\end{equation}

The yield ratio method can largely reduce the bias caused by the choice of different cross section models and Monte Carlo simulation tools. While the experimental yield is completely extracted from the data and remains unchanged, one can iterate the cross section model and apply necessary corrections only on the MC yield until the the yield ratio becomes close to one for all $E'$ bins. Furthermore, the acceptance cuts on the HRS can also be studied by varying the cuts and checking the distribution of the yield ratio as a function of the binning variable. Most of other potential issues, such as junk runs, incorrect input parameters and so on, can also be examined in the yield ratio method.


\section{Calculation of Errors}
 One of the most important tasks in the extraction of experimental cross sections is to calculate the systematic errors and statistical errors. Systematic errors are introduced by the experimental instrumentation, the simulation tools and the cross section model, etc. Statistical errors are related to the number of measurements of one quantity during the experiment. It is very important to properly propagate the errors when extracting new quantities from the existing quantities, and any mistakes such as mis-counting or double-counting should be avoided during the cross section extraction. The detailed explanation of the error calculation and propagation is given in the following subsections.

\subsection{Statistical Errors}
 A detailed propagation of statistical errors is discussed here:
\begin{enumerate}

\item \textbf{$N_{e}$:} From Eq.\ref{eq_ne}, since the charge is obtained from the average of two BCM monitor outputs ($U_{1}$ and $D_{1}$),the error is also averaged:
  \begin{equation}
   \delta N_{e}^{r} = \sqrt{\frac{\left(\delta N_{e}^{r,D_{1}}\right)^{2}+\left(\delta N_{e}^{r,U_{1}}\right)^{2}}{2}}
                    = \sqrt{\frac{N_{e}^{r,D_{1}}+N_{e}^{r,U_{1}}}{2}}
                    = \sqrt{\frac{N_{e}^{r}}{2}},
  \end{equation}
where, $r$ means the run number. Hence,
  \begin{equation}
    \delta N_{e} = \sqrt{\sum_{r}\left(\delta N_{e}^{r}\right)^{2}}=\sqrt{\frac{\sum_{r}N_{e}^{r}}{2}}=\sqrt{\frac{N_{e}}{2}}.
  \end{equation}
  
\item \textbf{Live-Time:} Form Eq.\ref{eq_lt}, when  $PS^{r} = 1$:
  \begin{equation}
    \delta LT^{r} = LT^{r} \cdot \sqrt{\frac{1}{N^{r,Scaler}}+\frac{1}{N^{r,DAQ}}},
  \end{equation}
where $PS=PS1$ for HRS-R and $PS=PS3$ for HRS-L. When  $PS^{r} > 1$, the calculation of $\delta LT^{r}$ is given differently~\cite{vince_thesis}:
 \begin{equation}
   \delta LT^{r} = LT^{r} \cdot \sqrt{\frac{1}{N^{r,Scaler}}-\frac{1}{N^{r,DAQ}}}.
 \end{equation}

\item \textbf{$N_{EX}:$} From  Eq.\ref{eq_nex} and $N_{EX}=\sum_{r}N_{EX}^{r}$ for all runs, one gets:
  \begin{equation}
    \delta N_{EX}^{r} = N_{EX}^{r} \cdot \sqrt{\frac{1}{N_{recorded}^{r}} + \left(\frac{\delta LT^{r}}{LT^{r}}\right)^{2} }, \delta N_{EX}=\sqrt{\sum_{r}\left(\delta N_{EX}^{r}\right)^{2}},
  \end{equation}
where $N_{recorded}^{r}$ is defined in Eq.~\eqref{eq_lt}. 

\item \textbf{$Y_{EX}:$} From Eq.\ref{eqyex},
  \begin{equation}
    \delta Y_{EX} =  Y_{EX} \cdot \sqrt{\left(\frac{\delta N_{EX}}{N_{EX}}\right)^{2}+\left(\frac{\delta N_{e}}{N_{e}}\right)^{2}+\left(\frac{\delta\epsilon_{eff}}{\epsilon_{eff}}\right)^{2}},
  \end{equation}
where $\epsilon_{eff}$ is set to one and its statistic error and systematic error are set to zero and 1\%, respectively.

\item \textbf{$Y_{MC}:$} From Eq.\ref{eqymc},
  \begin{equation}
    \delta Y_{MC} =  Y_{MC} \cdot \sqrt{\left(\frac{\delta\sum_{j\in i}}{\sum_{j\in i}}\right)^{2}+\left(\frac{\delta N_{MC}^{gen}}{N_{MC}^{gen}}\right)^{2}},
  \end{equation}
  where $\delta\sum_{j\in i} = \sum_{j\in i}\cdot\frac{1}{\sqrt{N_{MC}^{i}}}$, since it summarizes the cross section of MC events ($N_{MC}^{i}$) in one bin.

\item \textbf{$\sigma_{EX}^{Born}:$} From Eq.\ref{eqxs_ratio},
  \begin{equation}
    \delta \sigma_{EX}^{Born} = \sigma_{EX}^{Born} \cdot \sqrt{\left(\frac{\delta Y_{EX}}{Y_{EX}}\right)^{2}+\left(\frac{\delta Y_{MC}}{Y_{MC}}\right)^{2}}.
  \end{equation}

\end{enumerate}

\subsection{Systematic Errors}
 The entire list of systematic errors has not been determined in this thesis. Few items are given as follows:
\begin{enumerate}
\item \textbf{$\eta_{tg}$:}  Form Eq.~\eqref{eq_ntg} and Eq.~\eqref{eq_tgrho}, there are three quantities that can introduce errors: beam current measurement and calculation ($\delta I$), accuracy of Boiling Factors ($\delta B$), and the accuracy of target thickness measurement ($\delta \rho$). First two terms were temporarily set to zero, hence:
  \begin{equation}
    \delta \eta_{tg} = \frac{\delta\rho}{\rho} \cdot \eta_{tg}.
  \end{equation} 
The value of $\delta \rho$ for each target can be found in Table~\ref{target_table} and in Ref.~\cite{target_report}.
\item \textbf{$\epsilon_{eff}$:}  1\% systematic errors is assigned to each of VDC One-Track efficiency, trigger efficiency, detection and cut efficiencies of Gas \v{C}erenkov and Calorimeters.
\item \textbf{$\delta p$ correction (HRS-R only):} The error caused by correcting the un-calibrated $\delta p$ on HRS-R as given in Appendix D has to be evaluated. 0.3\% is assigned in this thesis. The value will be updated in the near future.
\item \textbf{Cross section model and radiative correction:} The error from the cross section models and the radiative correction. An estimation of 3\% is given in this thesis. The value will be updated in the near future.
\end{enumerate}

%% file: cross_section/results.tex
\chapter{Results and Discussion}
  The preliminary cross sections for all targets are given in Appendix E. In this chapter, the y-scaling functions, the momentum distributions and cross section ratios will be presented and compared with the existing data from CLAS and the E02-019. The new measurement of the 2N-SRC plateau ($a_{2}$) for $\mathrm{^{40}Ca}$ is included to study the linear correlation between the EMC and the SRC effects. The new results shown in this chapter are preliminary, and a discussion of the remaining analysis work will be given. Note that some systematic errors are not yet included in these results, such as the errors of acceptance correction, bin-centering correction, target densities of cryo-targets and the radiative corrections.

\section{y-Scaling and Momentum Distribution}
 \begin{figure}[!ht]
  \begin{center}
    \includegraphics[type=pdf,ext=.pdf,read=.pdf,width=.90\textwidth]{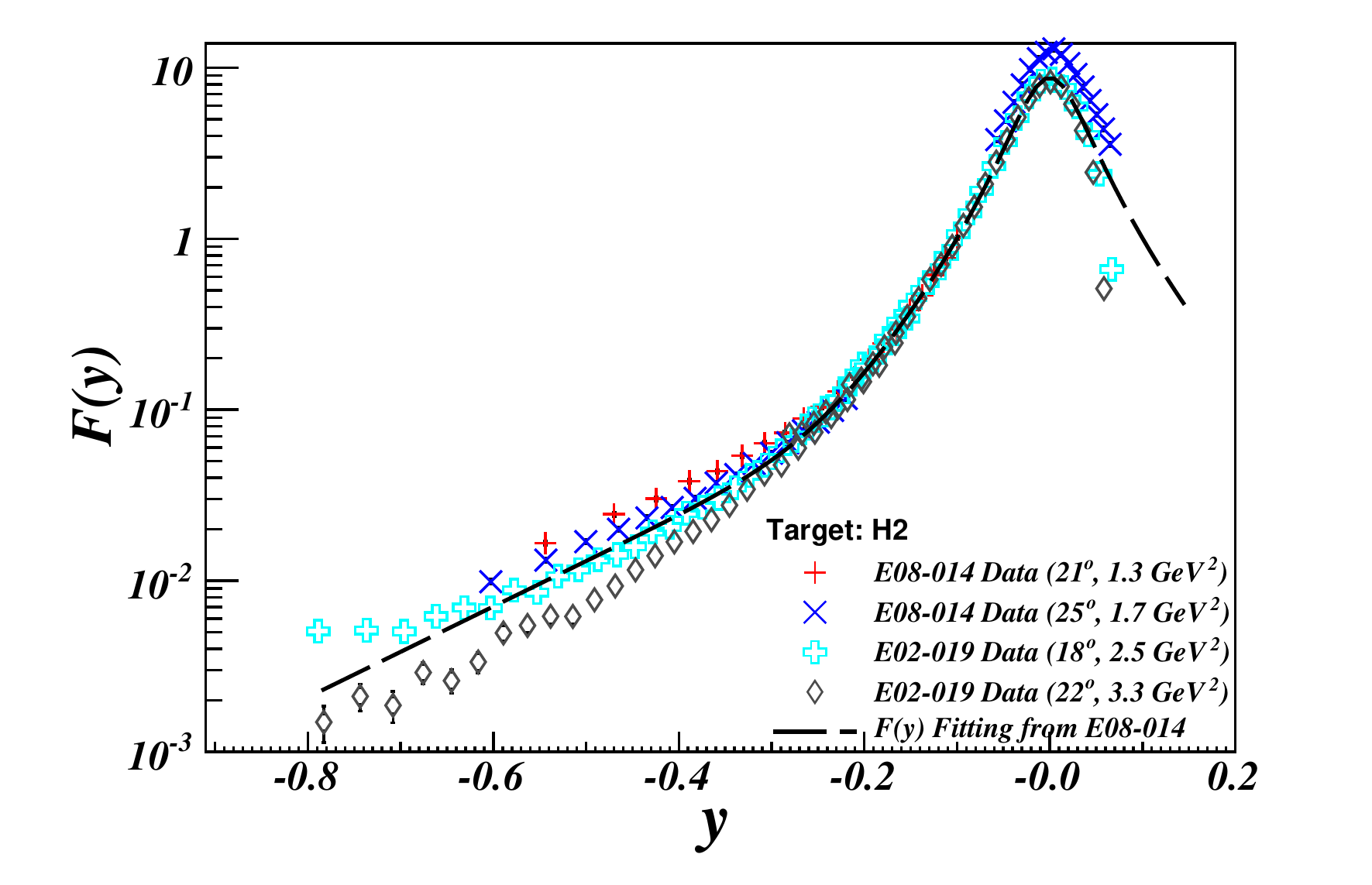}
    \caption[$F(y)$ distribution for $\mathrm{^{2}H}$]{\footnotesize{$F(y)$ distribution for $\mathrm{^{2}H}$. Symbols are experimental results from the E08-014 and the E02-019, where the kinematic settings are as indicated. The dash line is the fit of the new data with Eq.~\eqref{fy_fit_func1}. The new results have better agreement with the lowest $\mathrm{Q^{2}}$ data from the E02-019 at $18^{\circ}$ and deviate from the $22^{\circ}$ data with a higher $\mathrm{Q^{2}}$ value. It could be due to the FSI contribution. At $y\simeq 0$, the $25^{\circ}$ data significantly deviates from the E02-019 results, which might be due to the difference DIS subtraction procedure.}}
    \label{fy_h2_xgt2}
  \end{center}
\end{figure}
 \begin{figure}[!ht]
  \begin{center}
    \includegraphics[type=pdf,ext=.pdf,read=.pdf,width=.90\textwidth]{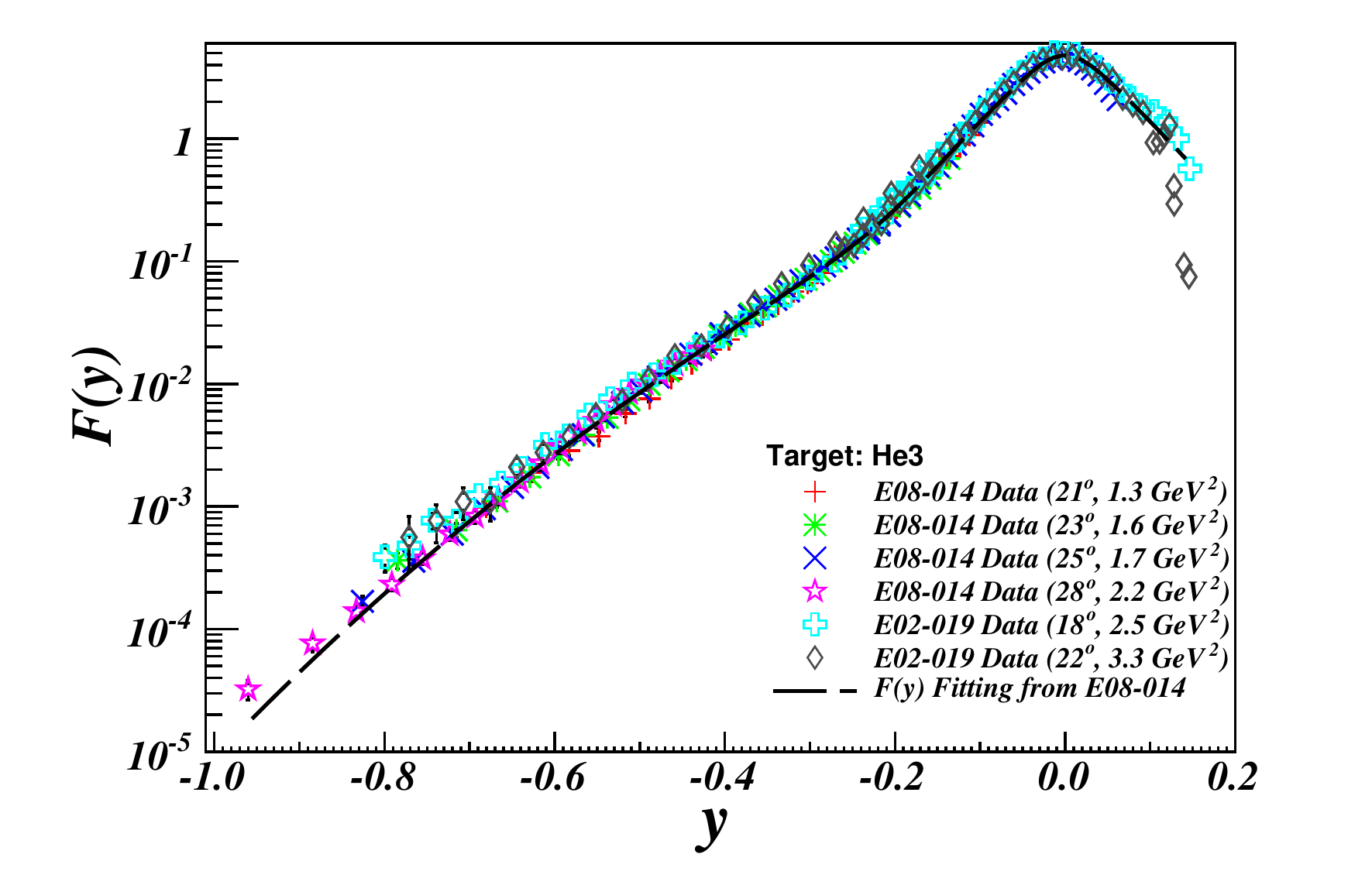}
    \caption[$F(y)$ distribution for $\mathrm{^{3}He}$]{\footnotesize{$F(y)$ distribution for $\mathrm{^{3}He}$. Symbols are experimental results from the E08-014 and the E02-019, where the kinematic settings are as indicated. The dash line is the fit of the new data with Eq.~\eqref{fy_fit_func2}.}}
    \label{fy_he3_xgt2}
  \end{center}
\end{figure}
 \begin{figure}[!ht]
  \begin{center}
    \includegraphics[type=pdf,ext=.pdf,read=.pdf,width=.90\textwidth]{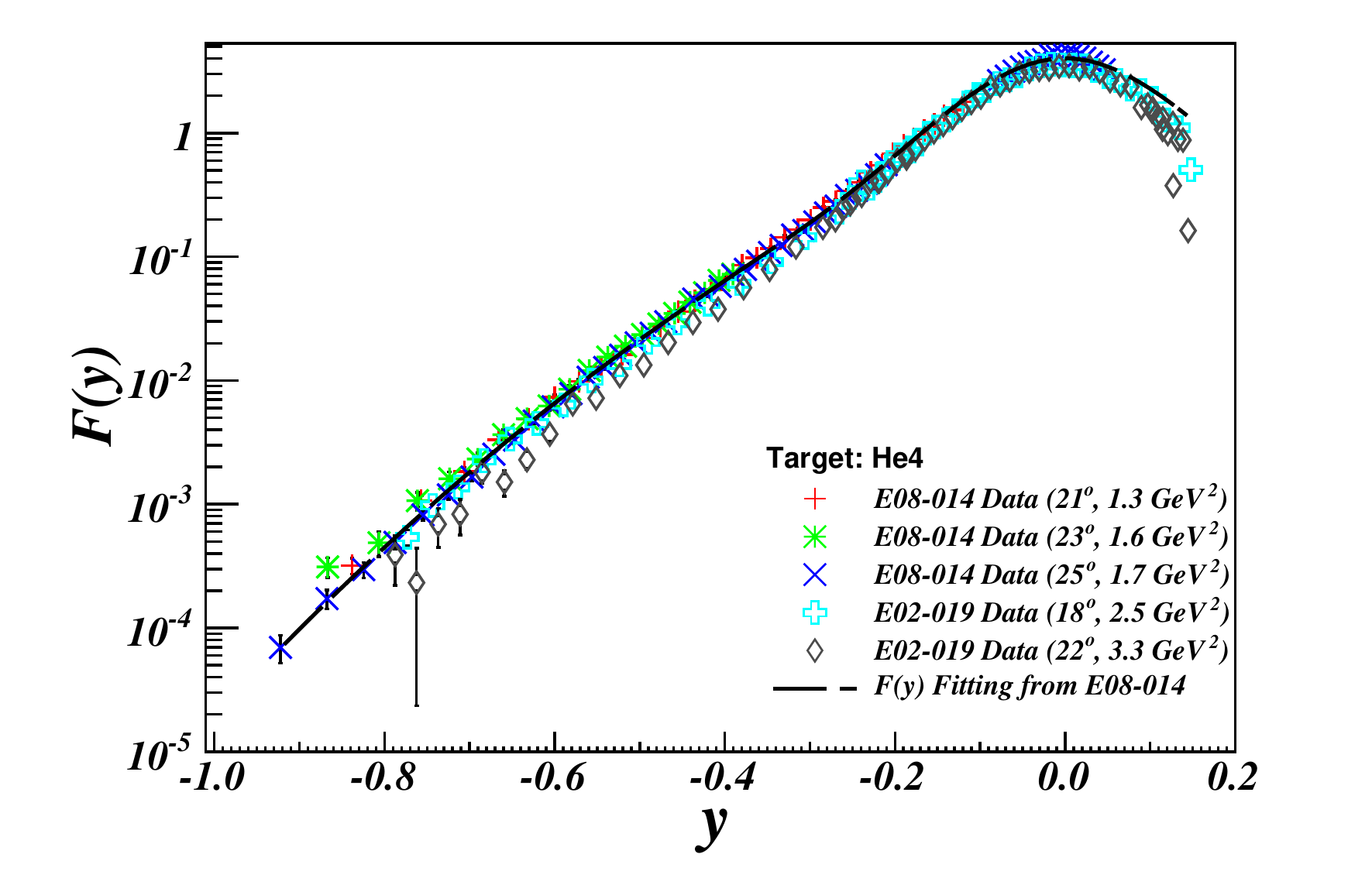}
    \caption[$F(y)$ distribution for $\mathrm{^{4}He}$]{\footnotesize{$F(y)$ distribution for $\mathrm{^{4}He}$. Symbols are experimental results from the E08-014 and the E02-019, where the kinematic settings are as indicated. The dash line is the fit of the new data with Eq.~\eqref{fy_fit_func2}.}}
    \label{fy_he4_xgt2}
  \end{center}
\end{figure}
\begin{figure}[!ht]
  \begin{center}
    \includegraphics[type=pdf,ext=.pdf,read=.pdf,width=.90\textwidth]{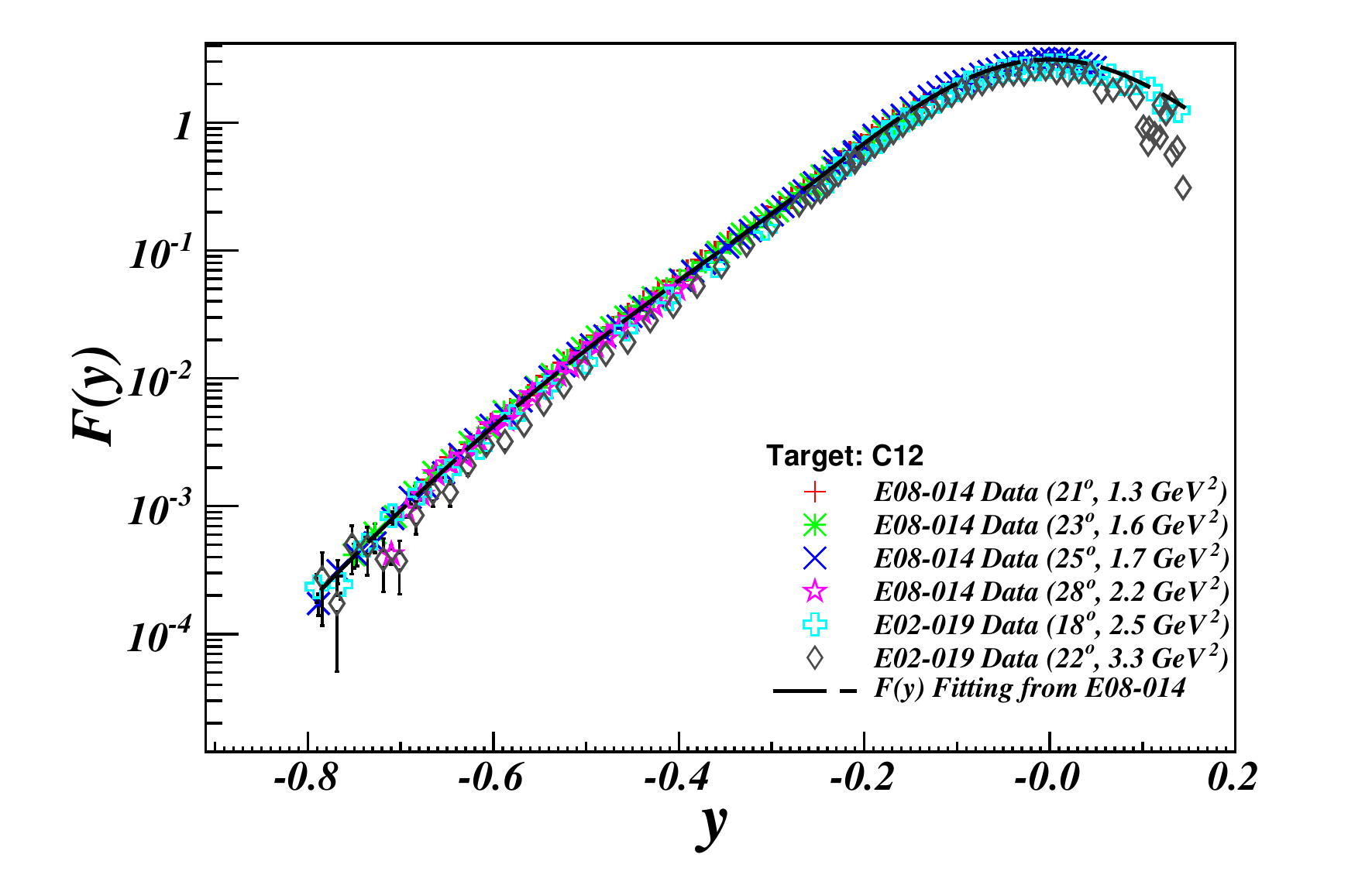}
    \caption[$F(y)$ distribution for $\mathrm{^{12}C}$]{\footnotesize{$F(y)$ distribution for $\mathrm{^{12}C}$. Symbols are experimental results from the E08-014 and the E02-019, where the kinematic settings are as indicated. The dash line is the fit of the new data with Eq.~\eqref{fy_fit_func2}.}}
    \label{fy_c12_xgt2}
  \end{center}
\end{figure}
 \begin{figure}[!ht]
  \begin{center}
    \includegraphics[type=pdf,ext=.pdf,read=.pdf,width=.90\textwidth]{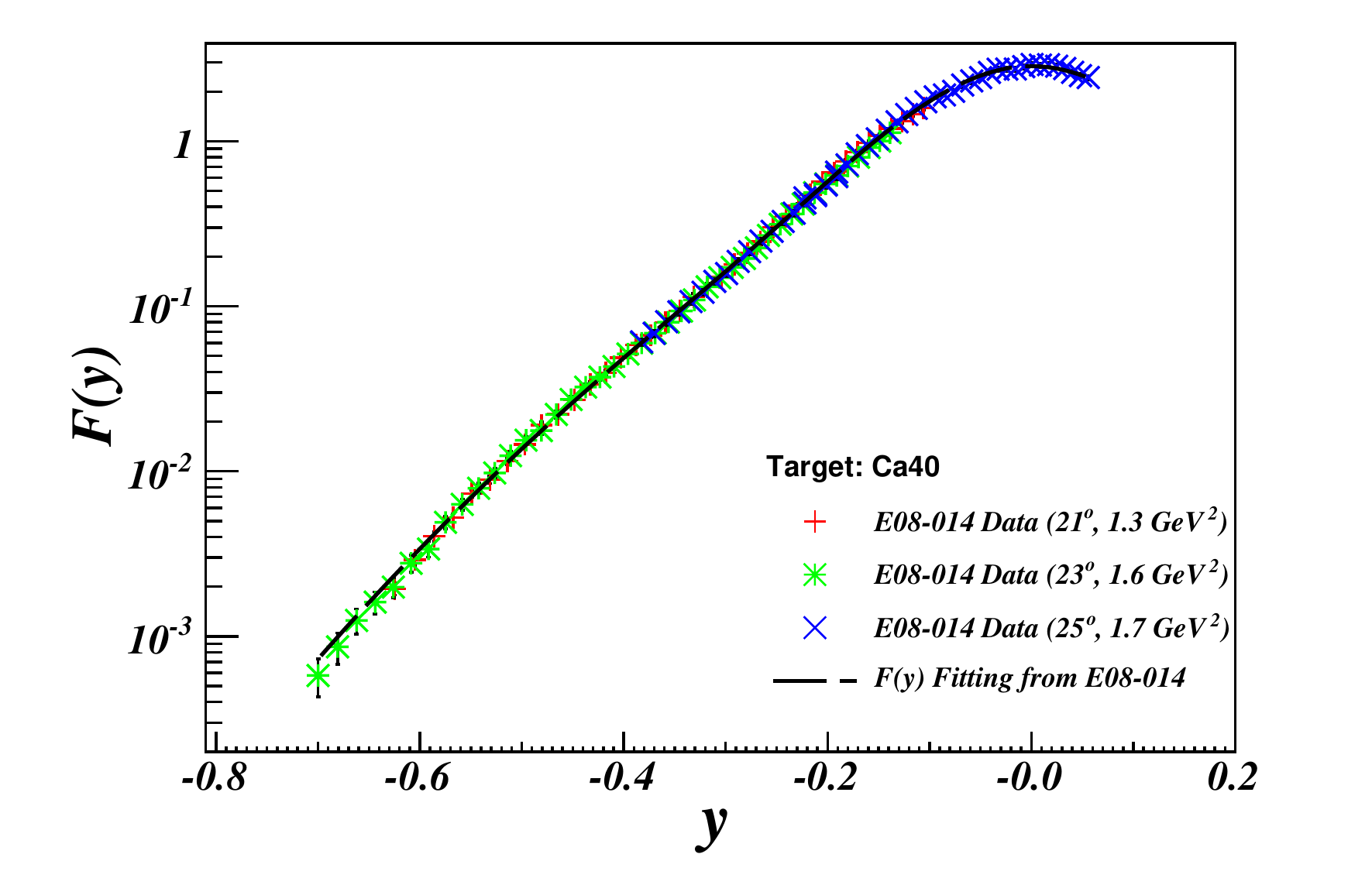}
    \caption[$F(y)$ distribution for $\mathrm{^{40}Ca}$]{\footnotesize{$F(y)$ distribution for $\mathrm{^{40}C}$. Symbols are experimental results from the E08-014, where the kinematic settings are as indicated. The dash line is the fit of the new data with Eq.~\eqref{fy_fit_func2}.}}
    \label{fy_ca40_xgt2}
  \end{center}
\end{figure}
 \begin{figure}[!ht]
  \begin{center}
    \includegraphics[type=pdf,ext=.pdf,read=.pdf,width=.90\textwidth]{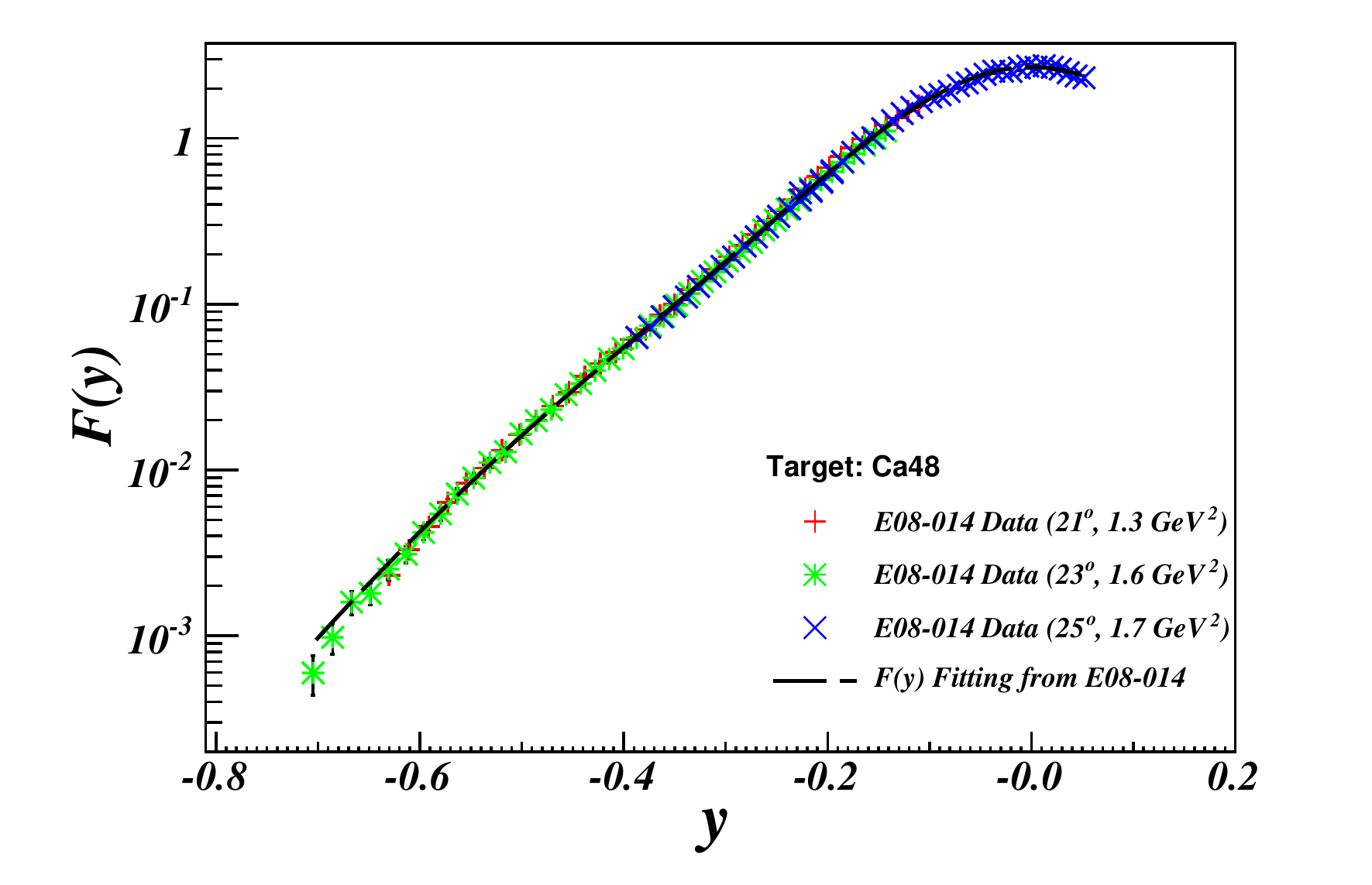}
    \caption[$F(y)$ distribution for $\mathrm{^{48}Ca}$]{\footnotesize{$F(y)$ distribution for $\mathrm{^{48}C}$ extracted from the experimental cross sections. Symbols are experimental results from the E08-014, where the kinematic settings are indicated. The dash line is the fit of the new data with Eq.~\eqref{fy_fit_func2}.}}
    \label{fy_ca48_xgt2}
  \end{center}
\end{figure}
  As discussed in Section 1.2.2, the y-scaling function, $F(y)$, is one of the most important features of Quasielastic scattering (QE) and it can be applied to extract the momentum distribution of the nucleus. $F(y)$ can be obtained from the experimental cross sections with Eq.~\eqref{fy_scaling_eq2}, after removing the deep inelastic scattering (DIS) contributions. 
   
  The $F(y)$ distributions for $\mathrm{^{2}H}$, $\mathrm{^{3}He}$, $\mathrm{^{4}He}$, $\mathrm{^{12}C}$, $\mathrm{^{40}Ca}$ and $\mathrm{^{48}Ca}$ were extracted from the preliminary cross section results given in Appendix E, and they are presented in Fig.~\ref{fy_h2_xgt2}, Fig.~\ref{fy_he3_xgt2}, Fig.~\ref{fy_he4_xgt2}, Fig.~\ref{fy_c12_xgt2}, Fig.~\ref{fy_ca40_xgt2}, and Fig.~\ref{fy_ca48_xgt2}, respectively. The E02-019 data from Hall-C with larger $\mathrm{Q^{2}}$ was also compared with each target, except $\mathrm{^{40}Ca}$ and $\mathrm{^{48}Ca}$ which were firstly measured in this experiment with high $\mathrm{Q^{2}}$.
  
   From these plots, the distributions at different $\mathrm{Q^{2}}$ tend to be nearly identical and generally show very small $Q^{2}$ dependence. The only significant $\mathrm{Q^{2}}$ dependence occurs for deuterium at $y<-0.4~GeV/c$ where there is a clear decrease in $F(y)$ with $\mathrm{Q^{2}}$, as shown in Fig.~\ref{fy_h2_xgt2}. Also in this plot, one notices that at $y\simeq 0$ where the QE peak locates, there is a non-trivial disagreement between the new data at $25^{\circ}$ and the E02-19 data. This might be due to the different DIS subtraction procedures. As mentioned in Section 1.2.2, a cross section model is required to remove the DIS contributions from the experimental cross sections, and two different models were used for these two experiments.
 
 As also shown in these plots, the extracted $F(y)$ distributions for all $\mathrm{Q^{2}}$ settings were fitted with the fitting function, Eq.~\eqref{fy_fit_func1} for deuteron and Eq.~\eqref{fy_fit_func2} for other heavier nuclei, as given in Appendix II.

\begin{figure}[!ht]
  \begin{center}
    \includegraphics[type=pdf,ext=.pdf,read=.pdf,width=1.\textwidth]{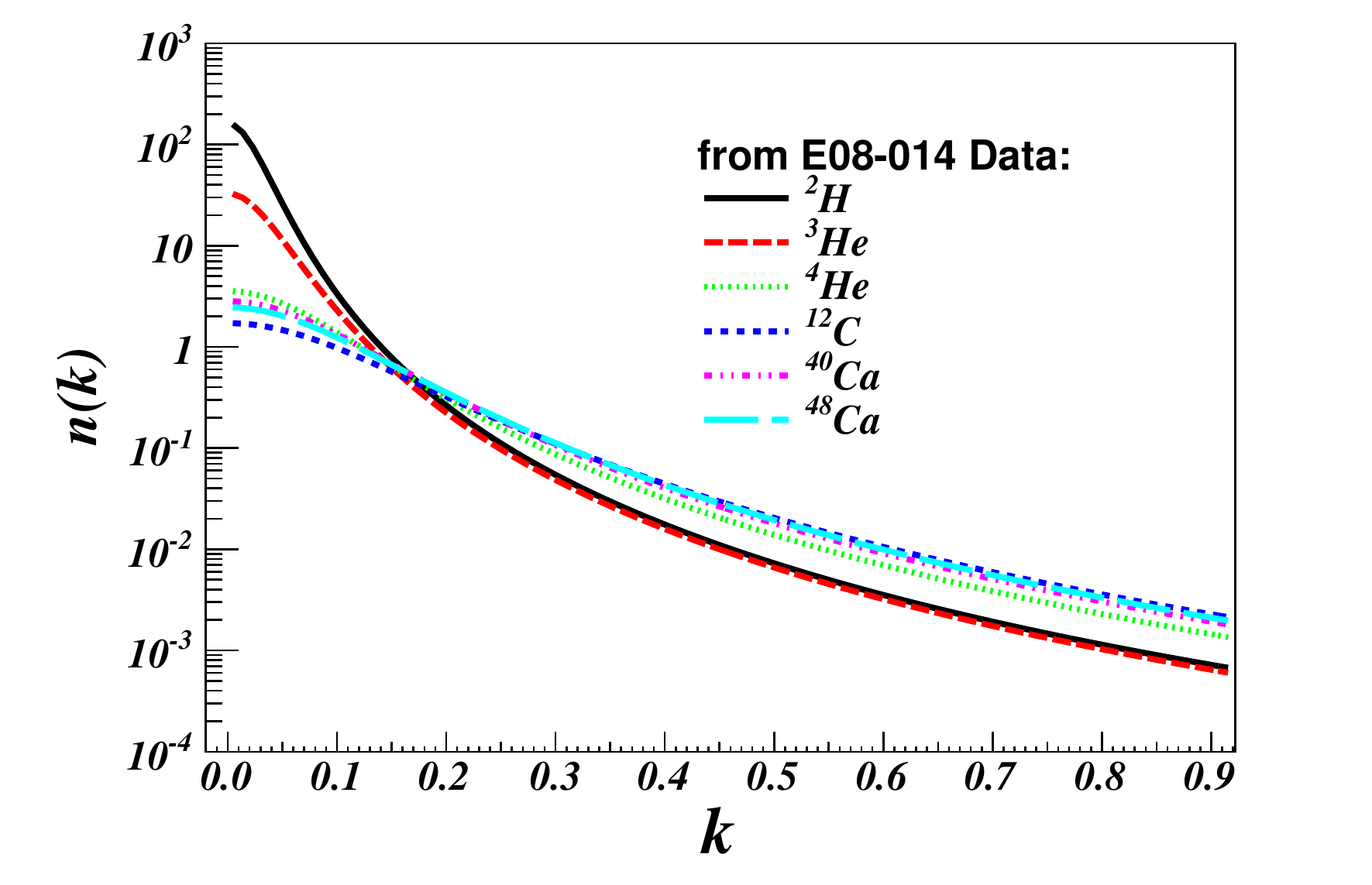}
    \caption[Momentum distribution for all nuclei]{\footnotesize{Momentum distribution for all nuclei, where the distributions were extracted from the $F(y)$ function fitted from the experimental cross sections.}}
    \label{mom_dis_xgt2}
  \end{center}
\end{figure}

 The momentum distribution for each nucleus was also extracted from the fit of the new results with Eq.~\eqref{mom_dis_fy}. The distributions for different nuclei are compared in Fig.~\ref{mom_dis_xgt2}. The absolute strength of each nucleus is comparable with the one given in Fig.~\ref{mom_dis_ox}. As discussed in Chapter 2, the momentum distributions for different nuclei should have similar shapes for $k>k_{F}$ due to the short-distance property of the 2N-SRC pairs in nuclei. The results given in Fig.~\ref{mom_dis_xgt2} show identical curves at high momentum ($k>300~MeV/c$). 

\section{2N-SRC and 3N-SRC Ratios}
\subsection{$\mathrm{^{4}He/^{3}He}$}
 \begin{figure}[!ht]
  \begin{center}
    \includegraphics[type=pdf,ext=.pdf,read=.pdf,width=0.95\textwidth]{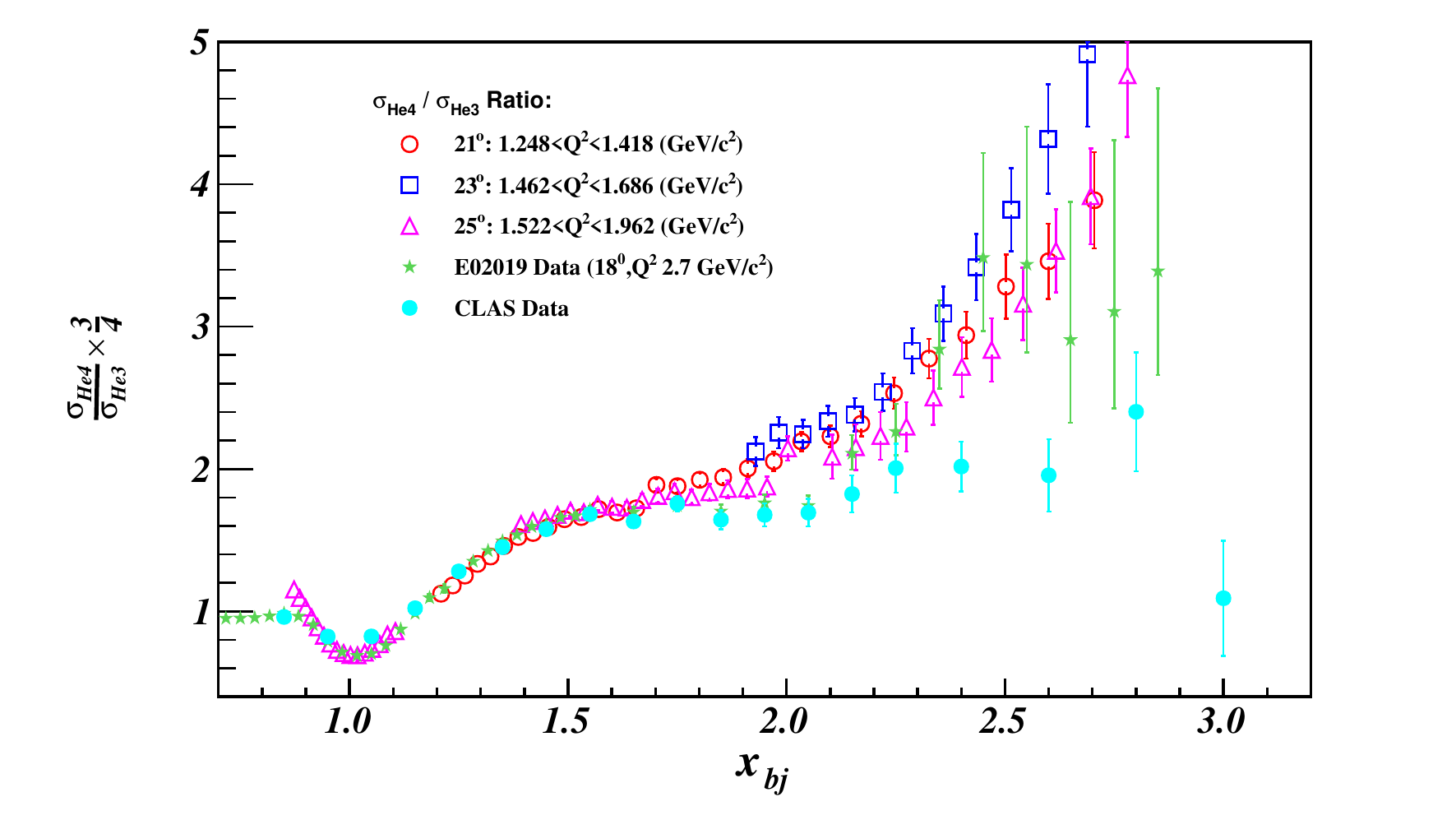}
    \caption[Cross section ratio of $\mathrm{^{4}He}$ to $\mathrm{^{3}He}$]{\footnotesize{Cross section ratio of $\mathrm{^{4}He}$ to $\mathrm{^{3}He}$. Open symbols are the new E08-14 data at $21^{\circ}$ (circle), $23^{\circ}$ (box), and $25^{\circ}$ (triangle). The values of the 2N-SRC ratio from the E02-019 data~\cite{PhysRevLett.108.092502} (solid boxes) at $18^{\circ}$ and CLAS data~\cite{PhysRevLett.96.082501} (solid dishes) are also included from comparison.}}
    \label{ratio_he4_he3}
  \end{center}
\end{figure}
 The preliminary cross section ratio of $\mathrm{^{4}He}$ to $\mathrm{^{3}He}$ is given in Fig.~\ref{ratio_he4_he3}. In the 2N-SRC region ($1.3<x_{bj}<2$), the $21^{\circ}$ data does not reveal a plateau. It is consistent with the CLAS data~\cite{PhysRevLett.96.082501} which found a breakdown of the plateau below $\mathrm{Q^{2}=1.4~GeV^{2}}$. The $\mathrm{^{4}He}$ data at $23^{\circ}$ does not cover the 2N-SRC region. However, the $25^{\circ}$ data does show an obvious plateau in the 2N-SRC region and a fit in $1.55<x_{bj}<1.85$ gives the value of the plateau equal to 1.84$\pm$0.01. Within the same $x_{bj}$ range, the CLAS data gives the plateau at 1.68$\pm$0.02, and the value in the E02-019 data~\cite{PhysRevLett.108.092502} is 1.71$\pm$ 0.02 at $18^{\circ}$. The result from CLAS agrees more with the one from the E02-019, while the preliminary result of this experiment is about 8\% larger. The new data was taken at lower $\mathrm{Q^{2}}$ compared with the E02-019 data, but the CLAS data was taken at the similar $\mathrm{Q^{2}}$ range, so the reason for the disagreement is yet unknown. The E02-019 data has been corrected for the Coulomb effect, and this correction will also be applied to this data later.
 
  In the 3N-SRC region for $x_{bj}>2$, this new data tends to agree with the E02-019 data. Neither data sets reveal a flat region starting from $x_{bj}\simeq 2.3$ as seen in the CLAS data. While the E02-019 result has large error bars in this region, the new data has much smaller error bars. However, one can not draw a firm conclusion from the current results, and more detailed analysis work for the new data must be completed before the results are final.
 
 \subsection{$\mathrm{^{12}C/^{3}He}$}
 \begin{figure}[!ht]
  \begin{center}
    \includegraphics[type=pdf,ext=.pdf,read=.pdf,width=1.\textwidth]{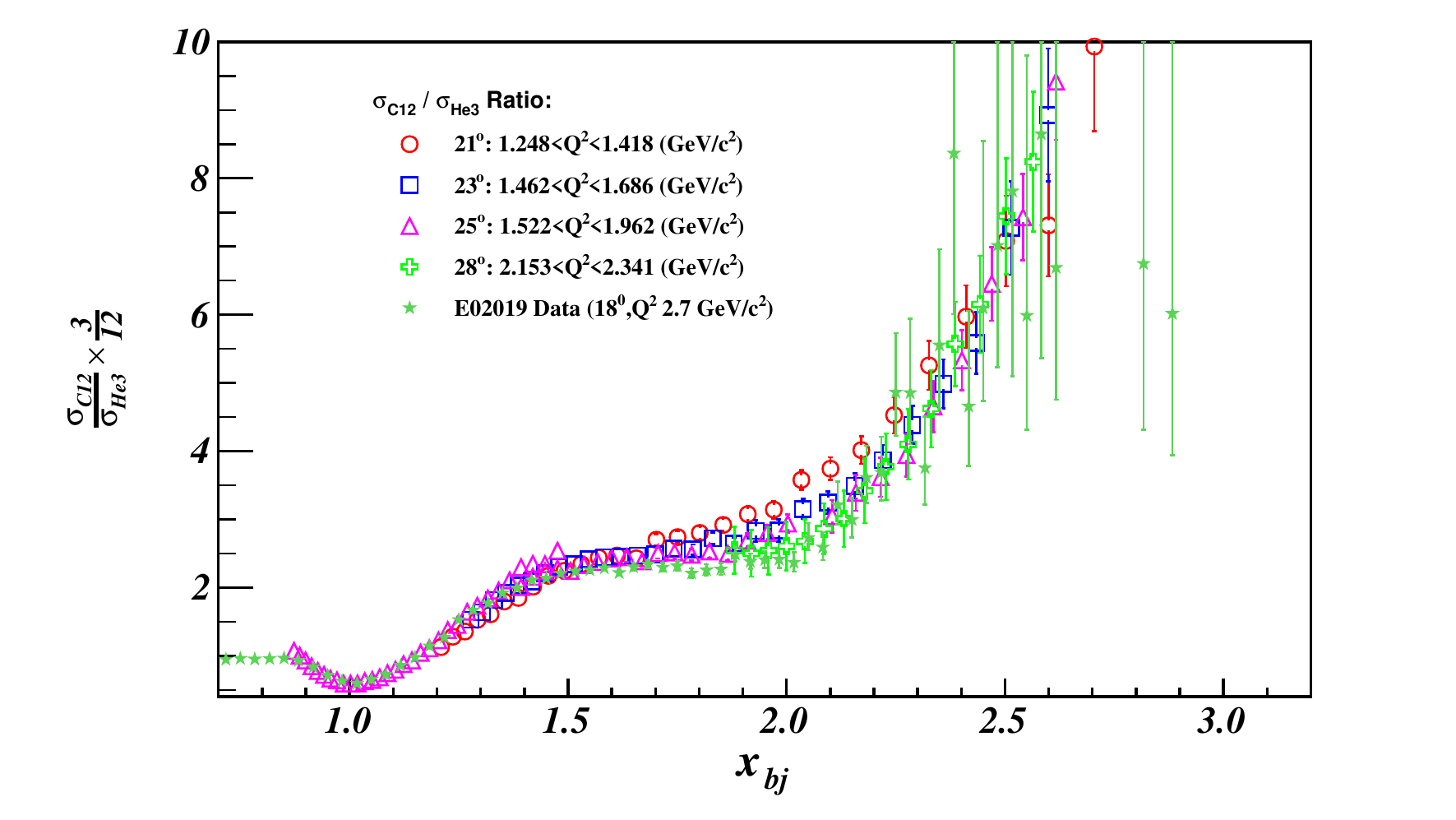}
    \caption[Cross section ratio of $\mathrm{^{12}C}$ to $\mathrm{^{3}He}$]{\footnotesize{Cross section ratio of $\mathrm{^{12}C}$ to $\mathrm{^{3}He}$, compared with the E02-019 preliminary results. Open symbols are the new E08-14 data at $21^{\circ}$ (circle), $23^{\circ}$ (box), $25^{\circ}$ (triangle) and $28^{\circ}$ (), respectively. The solid boxes are the E02-019 data. The $\mathrm{Q^{2}}$ value for each setting is given in the plot.}}
    \label{ratio_c12_he3}
  \end{center}
\end{figure}
Fig.~\ref{ratio_c12_he3} shows the preliminary cross section ratio of $\mathrm{^{12}C}$ to $\mathrm{^{3}He}$, as well as the results from the E02-019 data. Overall, both experiments have a good agreement in the region of the 2N-SRC, and the new result does not indicate a plateau of the 3N-SRC for $x_{bj}>2$.

 For each kinematic setting, a linear fit was performed for $1.55<x_{bj}<1.85$ to extract the ratio in the 2N-SRC region. For the E08-014 data, the values of the 2N-SRC plateau at $21^{\circ}$, $23^{\circ}$ and $25^{\circ}$  are 2.60$\pm$0.01, 2.46$\pm$0.01 and 2.43$\pm$0.01, respectively. For the E02-019 data, the value is 2.27$\pm$0.02. The 2N-SRC ratio decreases with increasing $\mathrm{Q^{2}}$, which suggests that FSI may still play an important role in these kinematic regions. 

\subsection{$\mathrm{^{40}Ca/^{2}H}$ and $\mathrm{^{48}Ca/^{2}H}$}
\begin{figure}[!ht]
  \begin{center}
    \includegraphics[type=pdf,ext=.pdf,read=.pdf,width=0.9\textwidth]{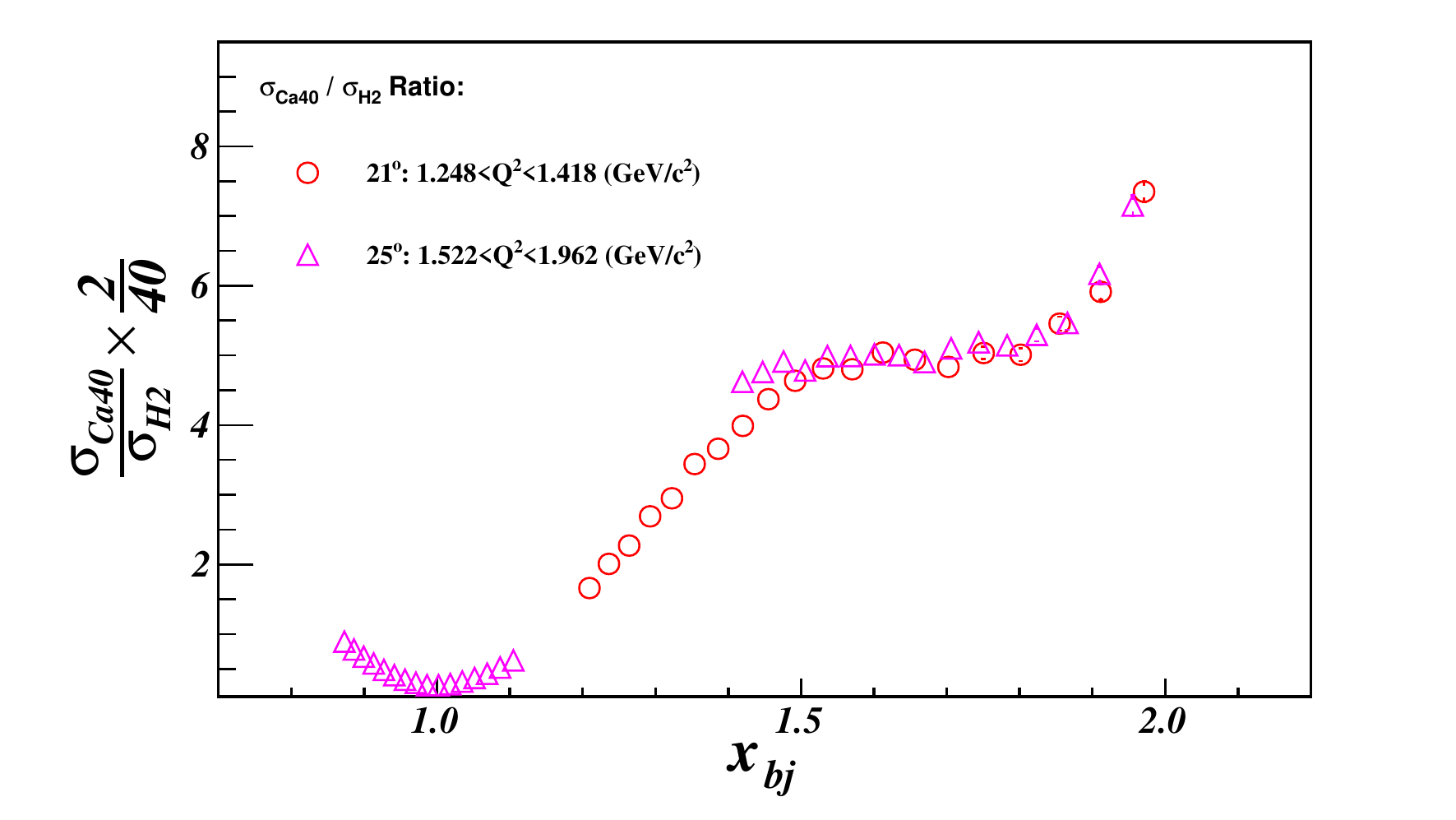}
   \caption[Cross section ratios of $\mathrm{^{40}Ca}$ to $\mathrm{^{2}H}$]{\footnotesize{Cross section ratios of $\mathrm{^{40}Ca}$ to $\mathrm{^{2}H}$.}}
    \label{ca40_h2_xgt2}
  \end{center}
\end{figure}

\begin{figure}[!ht]
  \begin{center}
    \includegraphics[type=pdf,ext=.pdf,read=.pdf,width=0.9\textwidth]{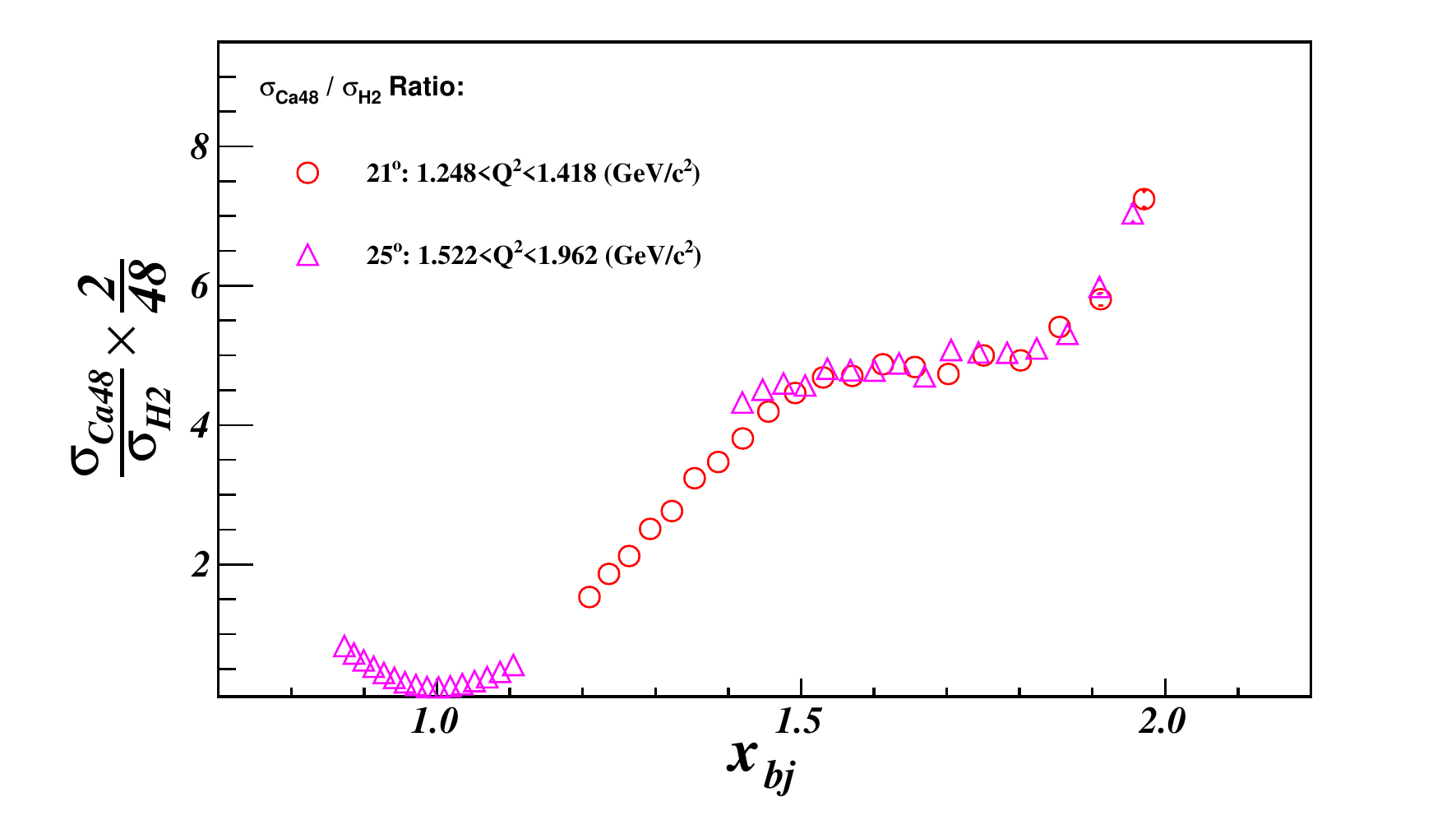}
   \caption[Cross section ratios of $\mathrm{^{48}Ca}$ to $\mathrm{^{2}H}$]{\footnotesize{Cross section ratios of $\mathrm{^{48}Ca}$ to $\mathrm{^{2}H}$.}}
    \label{ca48_h2_xgt2}
  \end{center}
\end{figure}

The 2N-SRC plateaus of the Calcium isotopes were firstly measured in this experiment, as shown in Fig.~\ref{ca40_h2_xgt2} and Fig.~\ref{ca48_h2_xgt2}. The values of $a_{2}$ for $\mathrm{^{40}Ca}$ and $\mathrm{^{48}Ca}$ are 4.987$\pm$0.018 and 4.863$\pm$0.016, respectively.  
\begin{figure}[!ht]
  \begin{center}
    \includegraphics[type=pdf,ext=.pdf,read=.pdf,width=1.\textwidth]{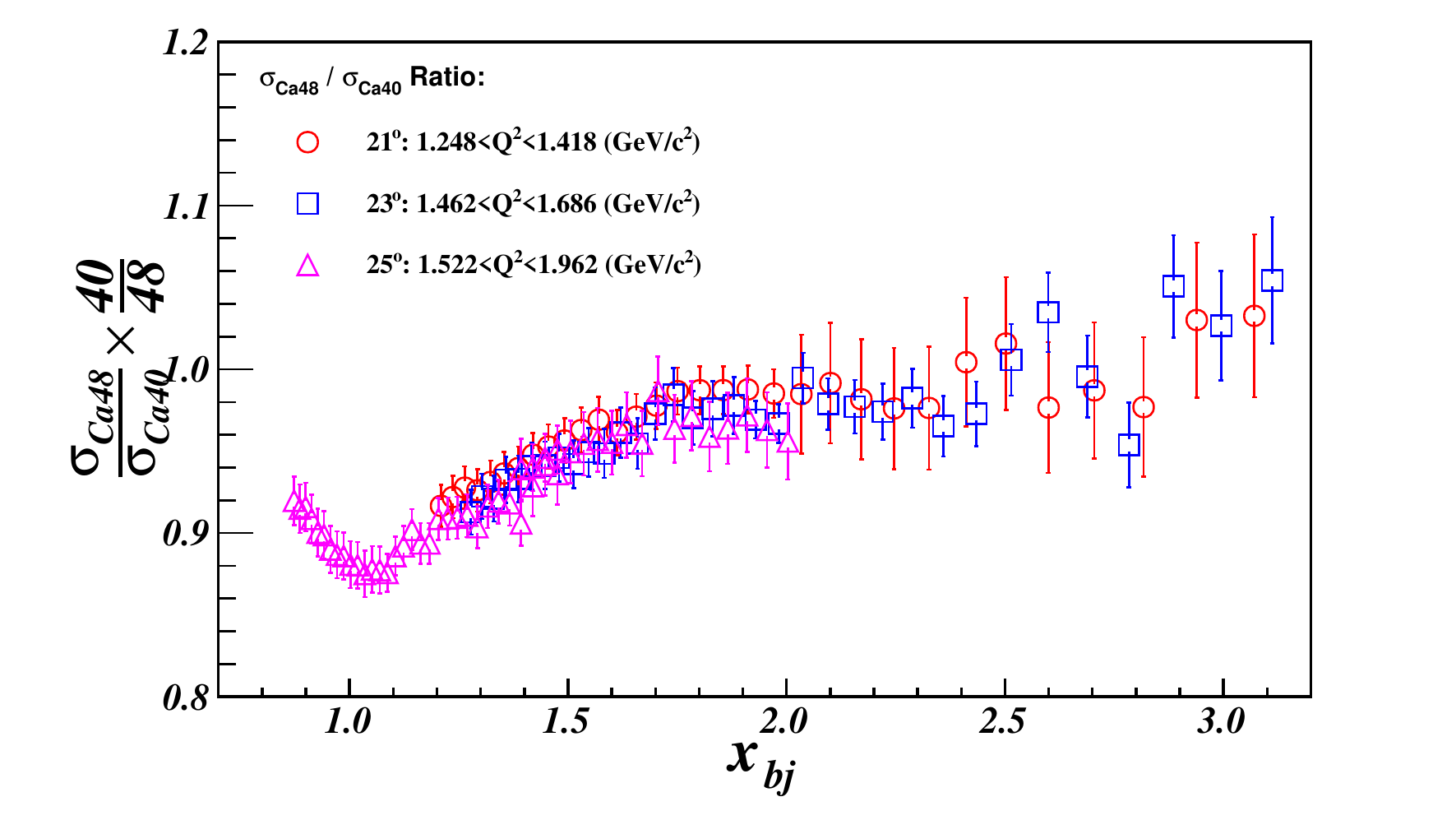}
    \caption[Cross section ratio of $\mathrm{^{48}Ca}$ to $\mathrm{^{40}Ca}$]{\footnotesize{Cross section ratio of $\mathrm{^{48}Ca}$ to $\mathrm{^{40}Ca}$}}
    \label{ratio_ca48_ca40}
  \end{center}
\end{figure}
\section{Isospin Effect Study with $\mathrm{^{48}Ca/^{40}Ca}$ Ratio}
 As discussed in Section 2.2.3, the cross section ratio of $\mathrm{^{48}Ca}$ to $\mathrm{^{40}Ca}$ is 0.916 with the isospin independence assumption. If one assumes the $np$ pairs dominate, the ratio becomes 1.17. Meanwhile, a theoretical prediction~\cite{PhysRevC.84.031302,PhysRevC.86.044619} claims that the ratio would be close to one. 

 \begin{figure}[!ht]
  \begin{center}
    \includegraphics[type=pdf,ext=.pdf,read=.pdf,width=1.0\textwidth]{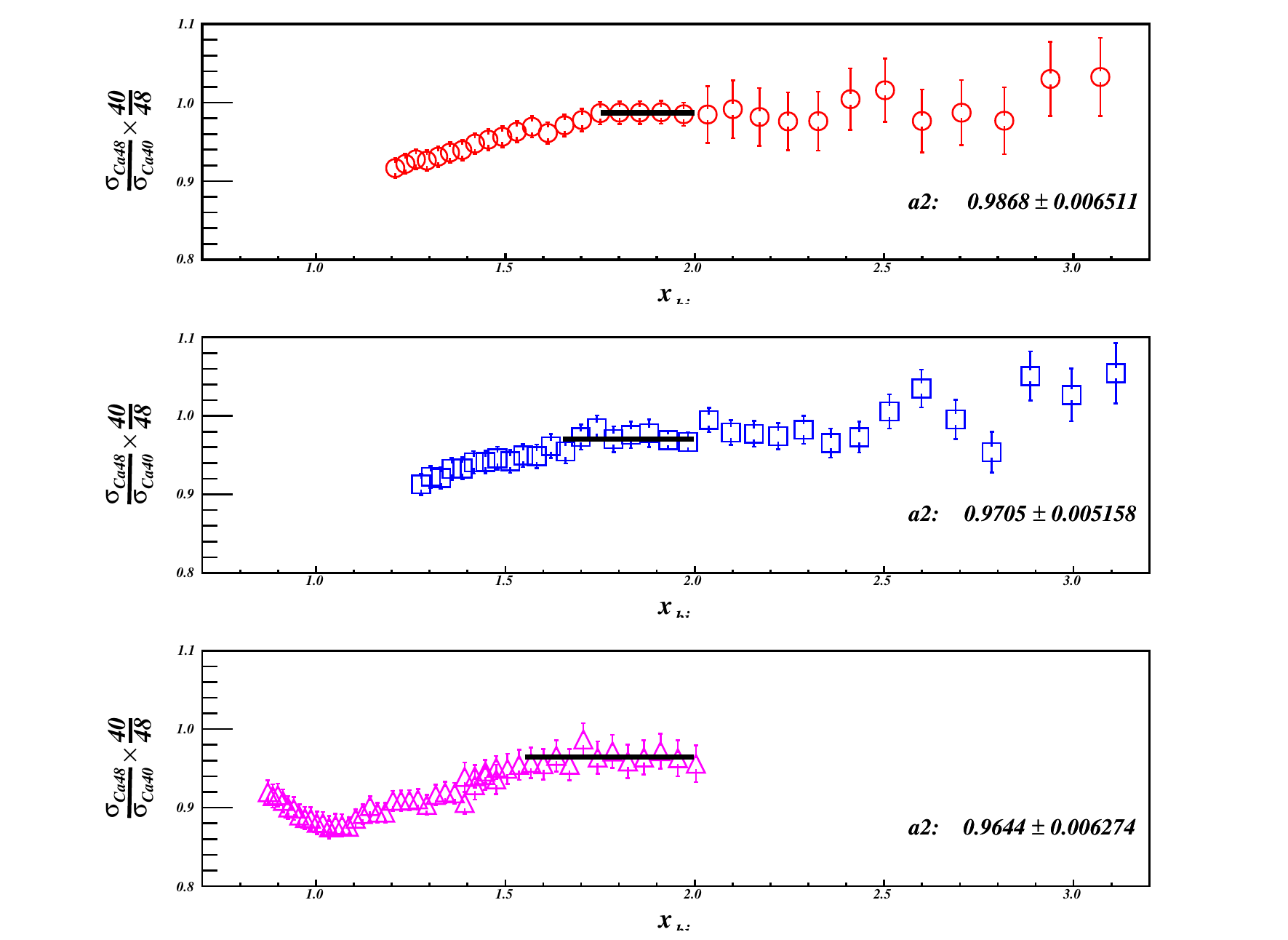}
    \caption[Fitting of $\mathrm{^{48}Ca}$ to $\mathrm{^{40}Ca}$ cross section ratio]{\footnotesize{Fitting of $\mathrm{^{48}Ca}$ to $\mathrm{^{40}Ca}$ cross section ratio for each kinematic setting, where each plot gives the result of each setting. The values of $\mathrm{a_{2}}$ are given in these plots. The results show that $\mathrm{a_{2}}$ is different in each setting.}}
    \label{ratio_ca48_ca40_fit}
  \end{center}
\end{figure} 

 The preliminary cross section ratio of $\mathrm{^{48}Ca}$ to $\mathrm{^{40}Ca}$ is presented in Fig.~\ref{ratio_ca48_ca40}. The ratio in the 2N-SRC region, $R_{2N-SRC}$, is fitted individually in each setting, as shown in Fig.~\ref{ratio_ca48_ca40_fit}. The average value of the ratios for three kinematics settings is 0.973$\pm$0.017. Note that there is a 3\% decrease from the lowest $\mathrm{Q^{2}}$ setting at $21^{\circ}$ to the highest one at $25^{\circ}$.
    
\section{SRC vs. EMC}
\begin{figure}[!ht]
  \begin{center}
    \includegraphics[type=pdf,ext=.pdf,read=.pdf,width=1.\textwidth]{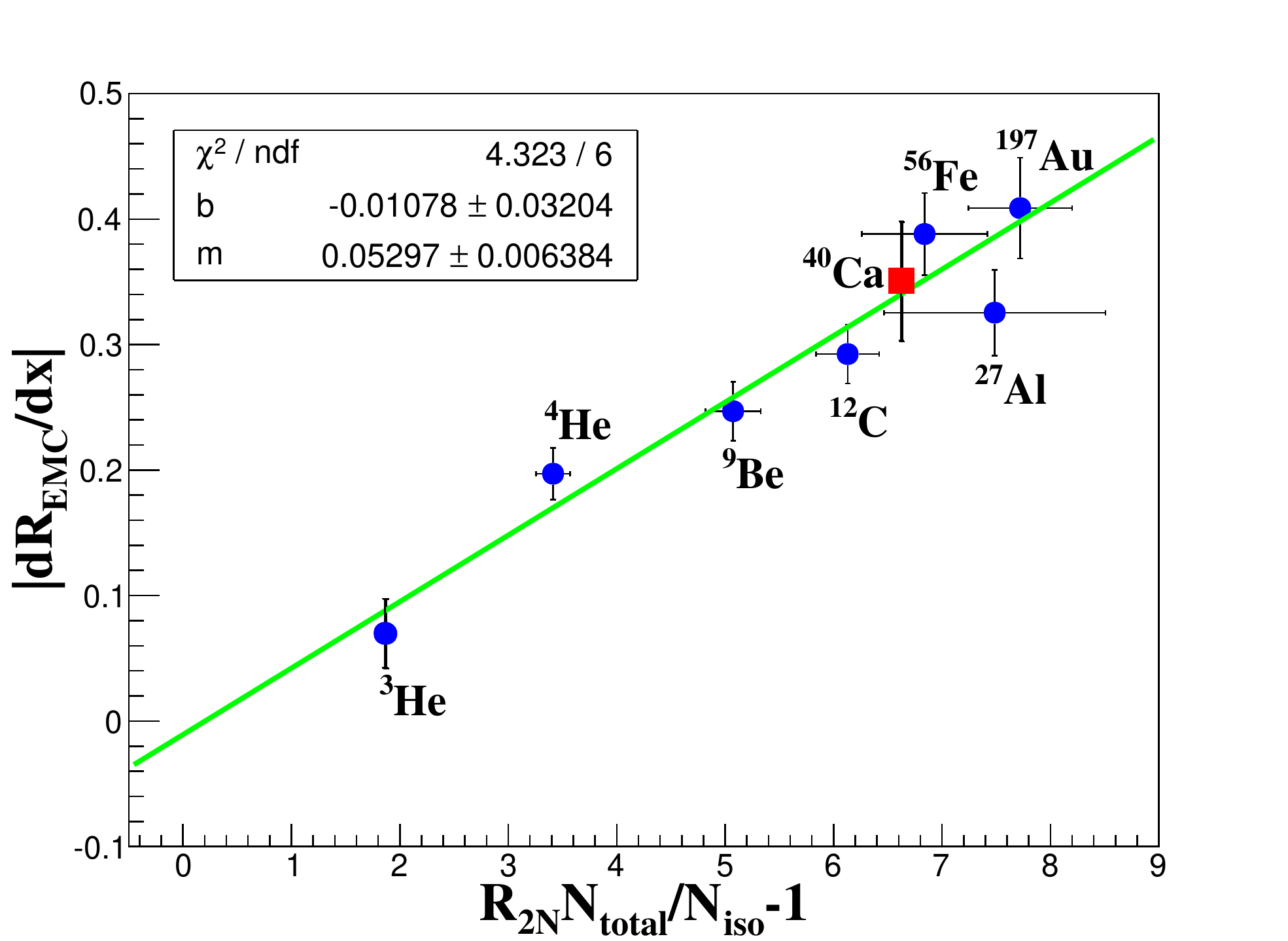}
    \caption[EMC vs. SRC with the new $\mathrm{^{40}Ca}$ measurement]{\footnotesize{EMC vs. SRC with the new $\mathrm{^{40}Ca}$ measurement. In the x-axis, $N_{total}=A(A-1)/2$ and $N_{iso}=(A-Z) Z$.}}
    \label{emc_vs_src_xgt2}
  \end{center}
\end{figure}

  As discussed in Section 2.3, the SRC and the EMC effect have a strong connection. As shown in Fig.~\ref{emc_vs_src}, the 2N-SRC plateau ($a_{2}$) is plotted against the slope of the EMC effect, and for all measured nuclei, the plot reveals a nearly linear correlation. The preliminary $a_{2}$ value of $\mathrm{^{40}Ca}$ is included in the study of the correlation between the EMC and the SRC, as shown in Fig.~\ref{emc_vs_src_xgt2}. The new data point falls onto the linear fit and supports the conclusion that the EMC effect and the SRC are strongly connected. Note that the error bar of the new data point will be larger after including all systematic errors.

\section{Discussion}
 Before drawing conclusions by comparing the new results with results from previous experiments  and theoretical calculations, there are several aspects needed to be discussed.
 
   First of all, during the cross section extraction of each cryo-target, the aluminium events from the target cell's two endcaps were removed by cutting out the peaks of both endcaps in the $z_{react}$ distributions. However, due to the optics reconstruction and multi-scattering, there were still some aluminium events mixed with the events from the cryo-target. At low $x_{bj}$, the residual aluminium events are negligibly smaller compared with events from $\mathrm{^{2}H}$, $\mathrm{^{3}He}$ or $\mathrm{^{4}He}$. With $x_{bj}$ getting larger, the yields of the cryo-target decrease dramatically, but the aluminium yields reduce much more slowly. For example, when $x_{bj}$ approaches to 3, the $\mathrm{^{3}He}$ cross section drops quickly toward zero, so its yields may be comparable with the yields from the endcaps. A study of aluminium contamination for cryo-targets is being undertaken at this time.
  
 Secondly, the cross section models in XEMC require careful examinations for different nuclei.  As discussed in Appendix B, the QE cross section model is based on the y-scaling function which requires a clean subtraction of the DIS contribution. The DIS model used in this experiment has been updated but it still has to be tested. For $\mathrm{^{3}He}$, its cross sections drop off quickly to zero in the model when $x_{bj}$ approaches to 3, but in reality the cross sections should decrease more slowly. A special treatment must be applied to model this target. The cross section models of $\mathrm{^{40}Ca}$ and $\mathrm{^{48}Ca}$ were first developed in this experiment and iterated based on the new data. However, the models could not be compared with pre-existing data since these two targets were firstly measured in this experiment in such a high $\mathrm{Q^{2}}$ range. 
 
 Moreover, this experiment ran at lower $\mathrm{Q^{2}}$ settings and the data might include contamination from processes other than the SRC at large $x_{bj}$. As discussed in Section 2.2.1, in performing a clean study of the SRC at high $\mathrm{Q^{2}}$, it is essential to look beyond the mean field contribution and other competing processes. Meanwhile, the effect of FSI becomes more significant at low $\mathrm{Q^{2}}$ (Section 2.3). Hence, this experimental data may be contaminated by the contributions from the mean field processes and the FSI. It still requires more theoretical studies to better understand these effects.
  
 Last but not least, due to the non-uniform target densities, the radiative correction of cryo-targets is more complicated (see Appendix D). Meanwhile, the current cross section model uses the peak-approximation method to calculate the radiative effect on the simplified density distributions, as discussed in Section D.4. This method may underestimate the radiative effect at large $x_{bj}$, and has to be carefully examined by more sophisticated methods. 

\chapter{Conclusion and Perspective}
 The E08-014 studied the short-distance properties of the NN interactions in the region of $1.3<x_{bj}<3$ where two experiments~\cite{PhysRevLett.96.082501,PhysRevLett.108.092502} showed different results in the 3N-SRC region. The preliminary results presented in this thesis confirm the 2N-SRC plateau at $1.3<x_{bj}<2$ as observed by previous experiments, but indicated no 3N-SRC plateau under the current analysis. The preliminary result of the $\mathrm{^{48}Ca/^{40}Ca}$ ratio agrees with the A(e,e'pN) measurement~\cite{Subedi:2008zz} which claims that the tensor force leads to the dominance of $np$ pairs in the 2N-SRC. There were still several analysis tasks before publishing the final results. For example, the aluminium contamination in the cryo-targets and the radiative corrections, must be carefully studied.

 New experiments have been approved in Hall-A and Hall-C at JLab, to study the isospin dependence in the SRC~\cite{E12_11_112_pr}, map-out the SRC and the EMC effect for a wide range of nuclei in different kinematic regions~\cite{E12_10_103_pr,E12_06_105_pr,E12_10_008_pr,E12_11_107_pr,E12_10_003_pr}, and systematically study the linear connection between these two effects. The results from this experiment will provide an important input for new experiments and for future theoretical developments.

%% file: append/append_trigger.tex
\chapter{Triggers in Data Analysis}
  This appendix is written specifically for people who analyze the Hall-A experiments or other experiments with similar trigger settings in Hall-A. It could also be useful for people who are interested in the event distributions in the trigger system and the trigger efficiency.
  
  When a scattered electron goes through the detectors located in the detector hut of each High Resolution Spectrometers (HRS), the signals created in specific detectors are used to form different triggers. The traditional single-arm production trigger, T1 for HRS-R or T3 for HRS-L, requires both the S1 and S2m scintillator planes to fire within a narrow time window. During the E08-014, a gas Cerenkov detector (GC) was also added into the production trigger in order to exclude most of pions events, and to reduce the total event rate as well as the dead-time. The new production triggers were the coincidence of logic signals from S1, S2m and GC. The original triggers were still used for the PID study but were assigned with different names, T6 for HRS-R and T7 for HRS-L, respectively. 

 Besides the main production triggers, there are two other important triggers, T2 for HRS-R and T4 for HRS-L, designed for the study of trigger efficiencies. Both T2 and T4 require only one of S1 and S2m logic signals to be coincident with the logic signal from a third detector plane, such as the GC in this experiment. The T2 and T4 triggers are generated by sending logic signals from S1, S2m and Cerenkov into a programmable module,called MLU~\cite{halla_daq}.

 Ideally, before the pre-scaling, T6 (T7) should be exactly the same as T1 (T3),if the GC has 100\% detection efficiency and there are no background events. However, T6 (T7) had much higher event rates than T1 (T3) mainly because of the pion contamination. During the data taking, the rates of  T1 and T3  were kept as high as possible until the dead time became high. T3, T4, T6 and T7 were prescaled to fix their rates no more than 50 $\sim$ 100Hz. T5, the coincident trigger of T1 and T3, was not used in this experiment and its rate was set to zero. T8 was the signal from the CPU clock and was also maintained at very low rate. 
\begin{table}[htbp]
 \begin{tabular}{lcccccccc}
 \toprule
 Trigger:       &    T1   &   T2   &   T3   &   T4   &   T5   &   T6   &   T7   &   T8\\
 \midrule
 TDC Channel:   &     1   &    2   &    3   &    4   &    5   &    6   &    7   &    8\\
 Decimal:       &     2   &   $\mathrm{2^{2}}$   &    $\mathrm{2^{3}}$   &   $\mathrm{2^{4}}$   &   $\mathrm{2^{5}}$   &  $\mathrm{2^{6}}$  &   $\mathrm{2^{7}}$  &   $\mathrm{2^{8}}$\\
 Hex:           &    0x02 &   0x04 &   0x08 &   0x10 &  0x20  &  0x40  &  0x80  &  0x100\\
 \bottomrule
 \end{tabular}
\caption{Triggers and their corresponding data types in data stream}
\label{trigger_table}
\end{table}

 All these trigger signals are sent to a 16-channel TDC port. Signals produced by an event can generate several types of triggers in a very narrow time window. Once one of the triggers is accepted by the DAQ system, all of the event's signals from detectors and other instruments are recorded by TDCs and ADCs. The trigger signals associated with this event are also stored. The analyzer decodes the TDC values of these triggers in Hex format and issues these values into a pointer-like variable in the \emph{\bf{T}} tree, "DBB.evtypebits". Table~\ref{trigger_table} lists the triggers and their corresponding values in different digital types.

 Based on this table, events belonging to the same trigger can be identified by applying cuts on the trigger variable. Note that an event can be affiliated with more than one trigger types. There are several kinds of trigger cuts used during data analysis, where differences are listed below:
 \begin{enumerate}
\item \textbf{DBB.evtypebits=0x02}: \\
    Selecting events which are associated with T1 trigger only. The cut returns a value of "1".
\item \textbf{(DBB.evtypebits\&0x02)==0x02}: \\
    Selecting events which are associated with T1 trigger and may also be associated with other triggers. The cut returns a value of "1".
\item \textbf{DBB.evtypebits$\gg1$\&1}: \\
    The same as (2).
\item \textbf{DBB.evtypebits \&(1$\gg$1)}: \\
    Exactly the same as (2) and (3), but returning a value of "2" instead of "1" (all non-zero values mean "TRUE")

\emph{The following two trigger cuts are not recommended:}
\item \textbf{DBB.evtype==1}: \\
     Selecting events only triggered by T1, and not by any other triggers coming within 5~ms window when the TS registers an event.This is almost the same as (1) except a slight difference caused by unknown reasons. 
\item \textbf{fEvtHdr.fEvtType==1}:\\
     Exactly the same as (5)
\end{enumerate}
\begin{figure}[!ht]
 \begin{center}
  \includegraphics[width=0.6\textwidth]{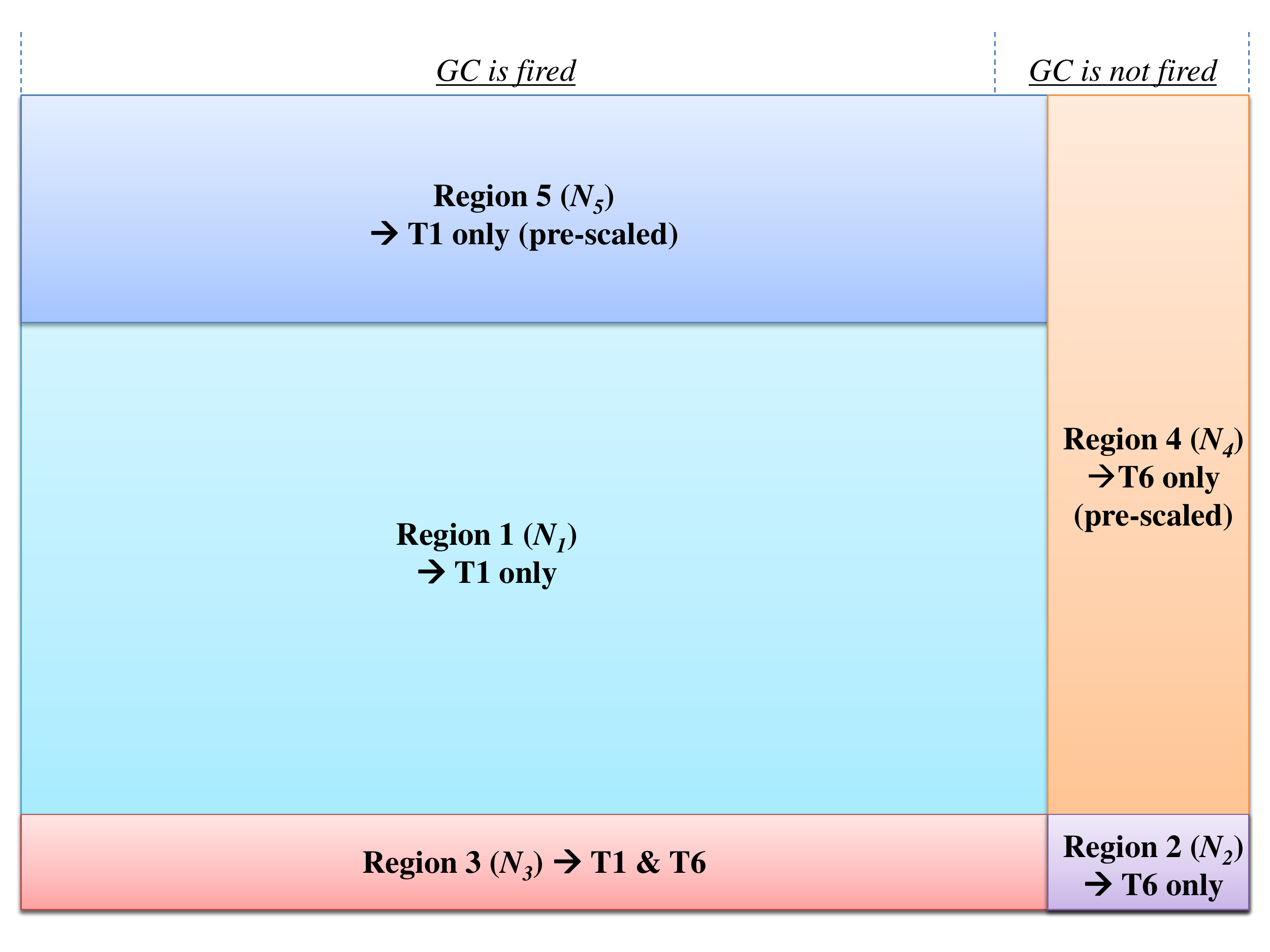}
  \caption[A scheme of events with different trigger cuts]{A scheme of events with different trigger cuts. Each box denotes the number of events associated with certain trigger types. The size of each box does not necessarily reflect the real distribution of events in the data.}
  \label{trig_region}
 \end{center}
\end{figure}

  Not all the scattered electrons arriving in the detector hut can be recorded by the DAQ system because the detectors do not have 100\% detection efficiencies. Meanwhile, a certain portion of detected events are skipped as a result of pre-scaling. For each run, the pre-scale factors for different trigger types are recorded in the raw data as well as in the log files created at the start and at the end of each run. The total number of events from a trigger has to be corrected by the efficiency of the trigger system (i.e. the trigger efficiency) including the electronic, the computer and the detectors (S1, S2m and GC for E08014). The procedure to extract the trigger efficiency has been discussed in Section 5.5.1. In the rest of this section, all variables related to the number of events are assumed to have been corrected by the trigger efficiency.
   
  Assuming the total number of the scattered electrons which fire both S1 and S2m on HRS-R is given as the big box in Fig.~\ref{trig_region}, the area of each small box represents the number of electrons (events) associated with different trigger types after applying the pre-scale factors. Region 1 gives the number of events ($\mathrm{N_{1}}$) from T1 only, and region 2 gives the total number of event ($\mathrm{N_{2}}$) from T6 only. Region 3 represents the events associated with both T1 and T6 ($\mathrm{N_{3}}$). The portions of events which are not recorded due to the pre-scaling are given as $\mathrm{N_{4}}$ in Region 4 for T6 and $\mathrm{N_{5}}$ in Region 5 for T1, respectively. $\mathrm{N_{2}+ N_{4}}$ denotes the number of events which are not detected by the GC, and it should be small since the GC has a very high detection efficiency. So the relationship between the number of events in those regions and the pre-scale factors can given as:
 \begin{equation}
 PS1 = \frac{N_{1}+N_{3}+N_{5}}{N_{1}+N_{3}},  PS6 = \frac{N_{1}+N_{2}+N_{3}+N_{4}+N_{5}}{N_{2}+N_{3}}=\frac{N_{2}+N_{4}}{N_{2}},
\end{equation}
 where $\mathrm{N_{1}}$, $\mathrm{N_{2}}$ and $\mathrm{N_{3}}$ can be extracted from data by applying Trigger cuts as listed in Table~\ref{trigger_cut_table}.
\begin{table}[htbp]
 \begin{tabular}{lcc}
\toprule
 Events  &  Cut\\
\midrule
$N_{1}$  &  \textbf{DBB.evtypebits$\gg1\&1$\&\&!(DBB.evtypebits$\gg6\&1$)} \\
$N_{2}$  &  \textbf{DBB.evtypebits$\gg6\&1$\&\&!(DBB.evtypebits$\gg1\&1$)} \\
$N_{3}$  &  \textbf{DBB.evtypebits$\gg1\&1$\&\&DBB.evtypebits$\gg6\&1$}  \\
$N_{1}+N_{3}$  &  \textbf{DBB.evtypebits$\gg1\&1$}  \\
$N_{2}+N_{3}$  &  \textbf{DBB.evtypebits$\gg6\&1$}  \\
\bottomrule
  \end{tabular}
  \caption[Events types with different Trigger cuts]{Events types with different Trigger cuts}
  \label{trigger_cut_table}
\end{table}

 If the pre-scale factors are known, the total number of trigger events in the box can be mathematically calculated:
\begin{equation}
 N_{0} = N_{1}+N_{2}+N_{3}+N_{4}+N_{5}=PS6\times(N_{2}+N_{3}).
\end{equation}
However, since the pre-scale factor of T6 was set to be large enough to keep the trigger rate around 50Hz, the value of $N_{2}+N_{3}$ should be very small and the statistical error in $\mathrm{N_{0}}$ would be very large. However, $\mathrm{N_{1}}$ has much more statistics since T1 was maintained to have big trigger rate. Combined with $\mathrm{N_{4}}$ and $\mathrm{N_{5}}$, it gives the total number of electrons in that box as:
\begin{equation}
 N_{0} = N_{1}+N_{2}+N_{3}+N_{4}+N_{5}=PS1\times(N_{1}+N_{3})+PS6\times N_{2}.
\label{event0_1}
\end{equation}
where the term, $PS1\times(N_{1}+N_{3})$, denotes the number of events which fired the GC, while $PS6\times N_{2}$ is the number of electrons which did not fire the detector. 

\begin{figure}[!ht]
 \begin{center}
  \includegraphics[width=0.6\textwidth]{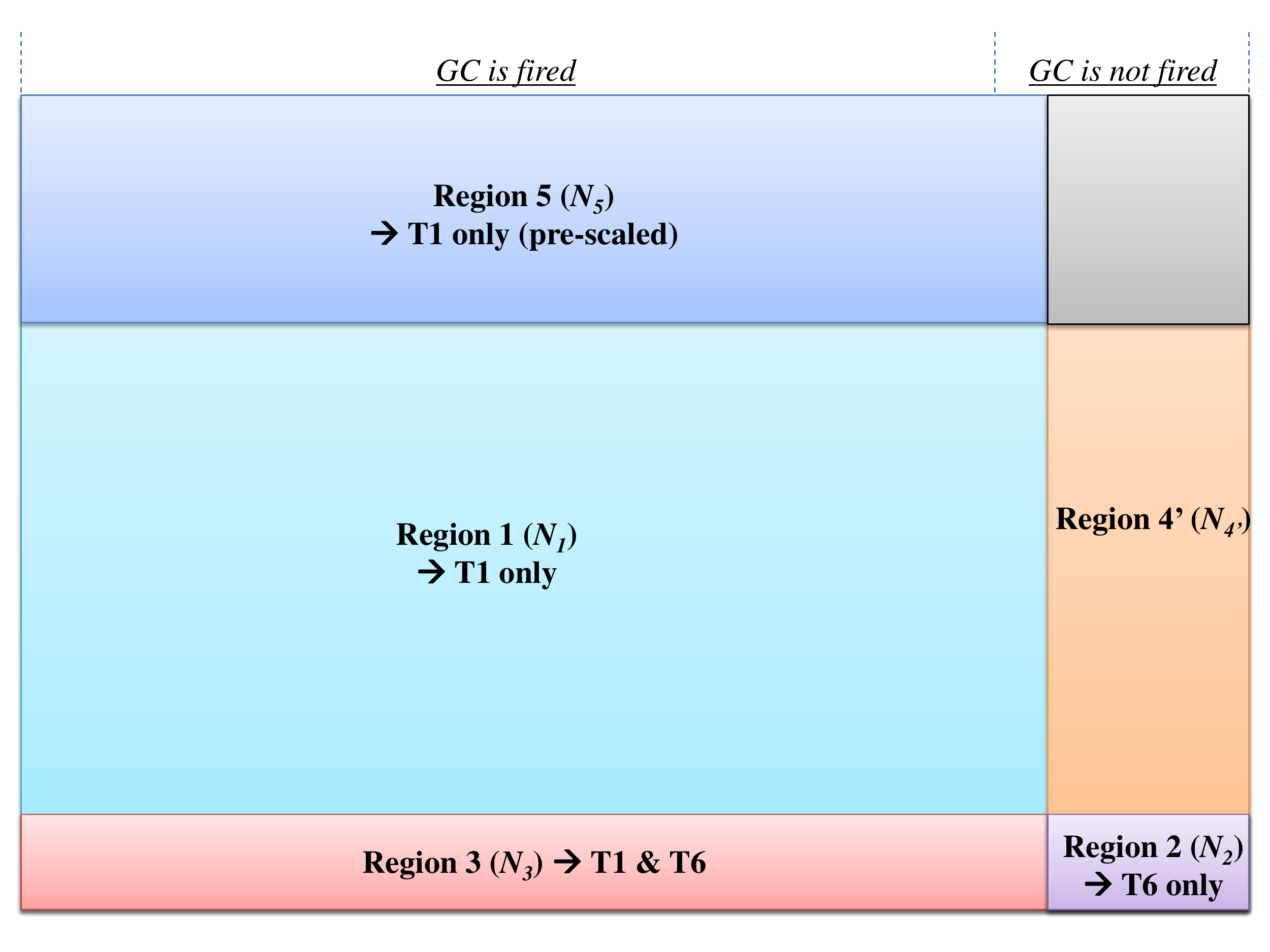}
  \caption[Another scheme of electrons with different trigger cuts]{Another scheme of electrons with different trigger cuts}
  \label{trig_region2}
 \end{center}
\end{figure}

Eq.~\ref{event0_1} can be further simplified. From Fig.~\ref{trig_region2}, a new region, called Region 4' ($N_{4'}$), can be defined:
\begin{equation}
\frac{N_{4'}}{N_{1}}=\frac{N_{2}}{N_{3}}=\frac{N_{2}+N_{4}}{N_{1}+N_{3}+N_{5}},
\end{equation}
which gives:
\begin{equation}
 N_{2}+N{4} = PS6\times N_{2} = PS1\times(N_{2}+N{4'}) = PS1\times(N_{2}+N_{1}N_{2}/N_{3}).
\end{equation}
and the relationship between PS1 and PS6 can be given by:
\begin{equation}
 PS6 = PS1(1+N_{1}/N_{3}).
\end{equation}

 So PS6 can be substituted by the formula above, and Eq.~\ref{event0_1} becomes:
\begin{equation}
 N_{0} = PS1\times(N_{1}+N_{3})\times \frac{N_{2}+N_{3}}{N_{3}}=\frac{PS1\times(N_{1}+N_{3})}{\epsilon},
\label{event0_2}
\end{equation}
where $\epsilon=\frac{N_{3}}{N_{2}+N_{3}}$ is the percentage of electrons firing GC when they pass through the detector, i.e. the exact definition of the GC detection efficiency. Since the number of events from T6 is very small, to reduce the statistical error, the typical way to get the detection efficiency of the GC is to select good electron samples from T1 events by applying a tight cut on the calorimeter and determining how many of them are detected by the GC (see Section 5.5.3): 
\begin{equation}
 \epsilon_{det}^{GC}=\frac{N^{GC}}{N^{Sample\_from\_Calo}}.
\end{equation}

 Based on the discussion above, $\mathrm{N_{0}}$ becomes straightforward: the total number of electrons passing through the HRS detectors is equal to the number of events triggered by S1, S2m and GC and corrected by the detection efficiency of the GC. It is important to emphasize that the value of $\mathrm{N_{0}}$ has to also be corrected by the trigger efficiency which is only related to the performance of S1 and S2m.
 
 The total number of trigger events from T3 on HRS-L can also be given in the same way.

%% file: append/append_xemc.tex
\chapter{XEMC: A Package for Inclusive Cross Section Models}

\section{Overview}
 XEMC is a stand-alone package written in C++ to calculate the inclusive cross section of electron-nucleus scattering. It is composed of three cross section models for the inelastic (DIS) process, three cross section models for the quasi-elastic (QE) process, and a radiative correction (RC) subroutine based on the peak approximation. In this model, the Born cross section ($\mathrm{\sigma^{Born}_{model}}$) is the sum of the inelastic cross section ($\mathrm{\sigma^{DIS}_{model}}$) and the QE cross section ($\mathrm{\sigma^{QE}_{model}})$. The RC subroutine calculates the radiative cross section ($\mathrm{\sigma^{rad}_{model}}$) from the Born cross section. The parameters of kinematic settings and target configurations are all defined in an external file.
 
 Cross section models are usually developed based on theoretical calculations, world data and additional corrections. Different models are designed for specific kinematic regions, depending on the physics processes and the final states. The inclusive cross section measured by the E08-014 was above the QE peak and can be well modelled by the y-scaling~\cite{West1975263,PhysRevC.41.R2474,Boffi19931,john_thesis}. The QE model was further iterated through comparing with experimental data at the similar kinematics. The DIS contribution to the total cross section was small and was calculated with the most updated DIS model~\cite{Bosted:2012qc}. 
  
  The basic structure of the package will be briefly introduced here, followed by a discussion of the cross section models. The results calculated with this code will be compared with experiment results. And a simple example of how to use this code is also given in the end. 
  
\section{Code Structure}
\begin{figure}[!ht]
 \begin{center}
  \includegraphics[angle=0,width=1.0\textwidth]{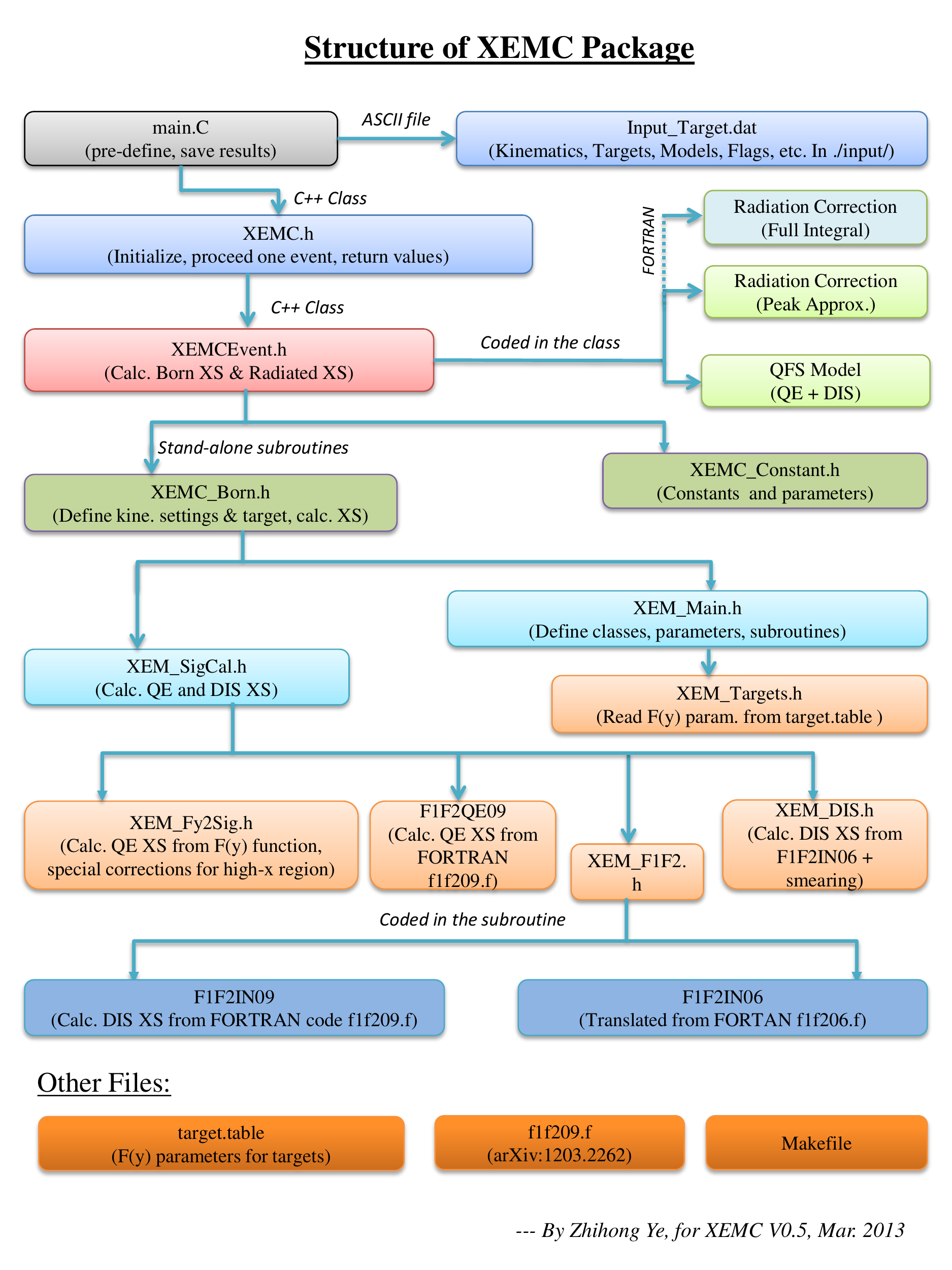}
  \caption[Structure of XEMC package]{Structure of XEMC package}
  \label{xemc_struct}
 \end{center}
\end{figure}
\begin{figure}[!ht]
 \begin{center}
  \includegraphics[angle=0,width=1.01\textwidth]{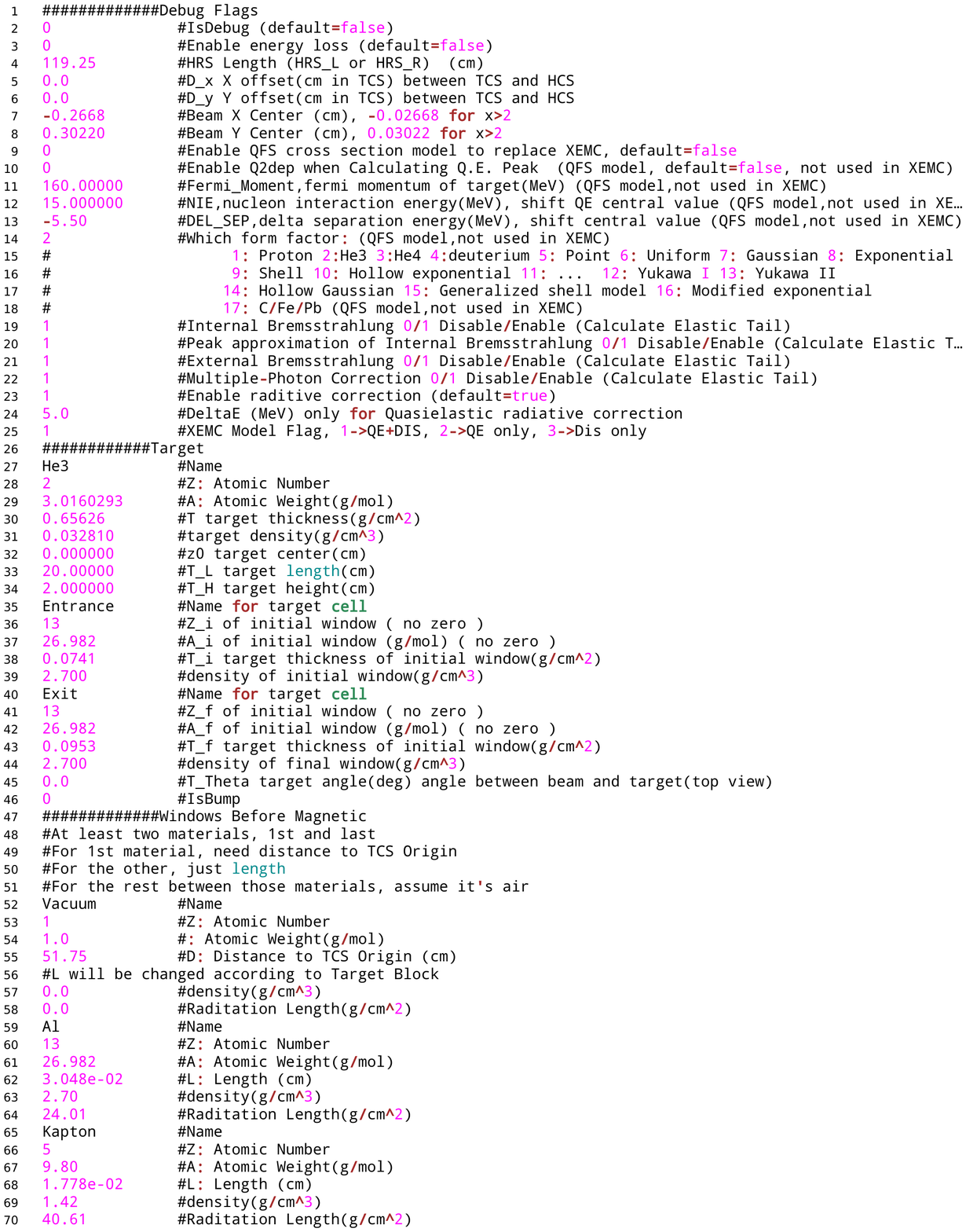}
 \caption[Input file for $\mathrm{^{3}He}$ target]{Input file for $\mathrm{^{3}He}$ target}
 \label{xemc_tgt_he3}
 \end{center}
\end{figure} 
Fig.~\ref{xemc_struct} shows the basic structure of the XEMC package. Outside the main code, the input file (Fig.~\ref{xemc_tgt_he3}) is defined to specify the choice of cross section models and any additional physics processes, such as the radiative correction. The reaction location can be corrected by giving the spectrometer center offset and beam position offset. The input file also includes the configuration of the target system, i.e. the target's name, mass and thickness. For cryo-targets, the materials of the target cell, the entrance and the exit of the target chamber are also given. Parameters in the input file are initialized only once in the code.

  A XEMC event has its specified values of the initial and scattered energies as well as the scattering angle. The Born cross section and radiated cross section of this event are calculated in \emph{XEMCEvent.h} where the QFS model is embedded by default.  Other Born cross section models are stored in an independent subroutine, \emph{$\mathrm{XEMC\_Born.h}$}, which will be introduced in next few sections. Once the target configuration and the kinematic setting are pre-defined, the RC subroutine in \emph{XEMCEvent.h} begins to calculate the radiated cross section. 

\section{Quasi-Elastic Cross Section Models}
 Three different QE cross section models, QE-XEM, QE-QFS, and QE-F1F209, are coded in this package. Each model will be introduced below.

 \subsection{QE-XEM} 
  \begin{figure}[!ht]
 \begin{center}
  \includegraphics[angle=0,width=1.0\textwidth]{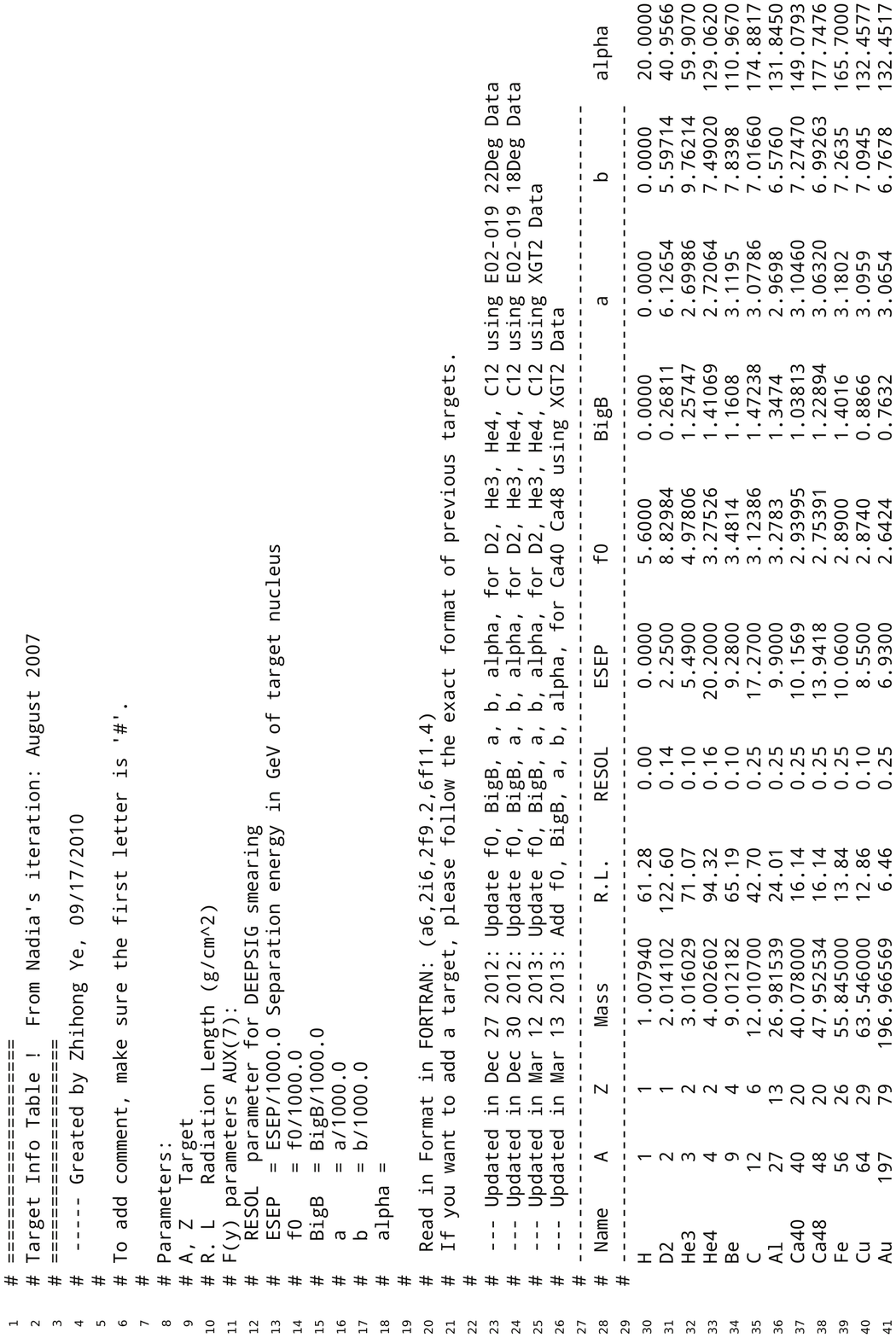}
  \caption[F(y) parameters for a list of targets]{F(y) parameters for a list of target. The file is called target.table which is in ASCII format. The values had been refitted with cross section results from the E02-019 and the E08-014. These values could be changed in the future.}
  \label{xemc_tgt_table}
 \end{center}
\end{figure} 

 QE-XEM was converted from the XEM cross section model, a FORTRAN package developed by the EMC collaboration in Hall-C at JLab~\cite{nadia_thesis,aji_thesis}. XEM includes a QE model (QE-XEM) based on y-scaling~\cite{West1975263,PhysRevC.41.R2474,Boffi19931,john_thesis}, a DIS model (DIS-XEM, see next section), and a RC subroutine. The entire subroutines have been converted into C++ (except the RC part) and coded in \emph{$XEMC\_Born.h$}. QE-XEM is the default QE model in the package.
 
  The scaling function, F(y) (Eq.~\eqref{fy_scaling_eq2} in Section 1.2.2), is directly fitted from experimental data. F(y) for $\mathrm{^{2}H}$ can be extracted from the function~\cite{PhysRevLett.56.1452}:
  \begin{equation}
  F(y) = (f_{0}-B)\frac{\alpha^{2}e^{-(\alpha y)^{2}}}{\alpha^{2}+y^{2}} + B e^{-b|y|},
  \label{fy_fit_func1}
  \end{equation}
where, $f_{0}$, $B$, $\alpha$, $a$ and $b$ are the parameters corresponding to the target. For heavy targets, the second term in the formula above is different:
  \begin{equation}
  F(y) = (f_{0}-B)\frac{\alpha^{2}e^{-(\alpha y)^{2}}}{\alpha^{2}+y^{2}} + B e^{-(by)^{2}}.
    \label{fy_fit_func2}
  \end{equation}

  For a list of targets, the parameters of $F(y)$ function ($f_{},~B,~\alpha,~a,~and~b$) are stored in an external ASCII file, called \emph{target.table}. To extract the parameters, one needs to  obtain the distribution of $F(y)$ from the experimental cross sections:
  \begin{equation}
   F(y) = \sigma_{EX}^{QE}\cdot\frac{1}{Z\sigma_{p}+N\sigma_{n}}\frac{q}{\sqrt{M^{2}+(y+q)^{2}}},
  \end{equation}
where $q=\sqrt{Q^{2}+\nu^{2}}$, and $y$ is the solution of the equation:
\begin{equation}
  M_{A}+\nu = \sqrt{M^{2}+q^{2}+y^{2}+2yq}+\sqrt{M_{A-1}^{2}+y^{2}},
\end{equation}
where $M$ is the mass of the struck nucleon, $M_{A}$ and $M_{A-1}$ are the masses of the target nucleus and the mass of the recoil system, respectively.

 The experimental QE cross sections, $\sigma_{EX}^{QE}$, can be extracted from the experimental Born cross sections subtracted by the DIS cross sections calculated from the model, i.e. $\sigma_{EX}^{QE}=\sigma_{EX}^{Born}-\sigma_{model}^{DIS}$. Hence, different DIS models yield different fitting values of the F(y) parameters. Fig.~\ref{xemc_tgt_table} gives a target table which lists the values of these parameters for all measured targets. The parameters have been determined from the the E02-019~\cite{nadia_thesis} and the E08-014 data with the DIS model, DIS-F1F209 (discussed in Section B.4.3). 
 
 \subsection{QE-QFS}
  QE-QFS is based on the QFS model, a phenomenological model~\cite{qfs_org,qfs_note} which has been used since 1960s. The model was designed to calculate both QE and DIS cross sections with the Plane-Wave Impulse Approximation (PWIA) and it works well at lower $Q^{2}$ region. The complete description of the QFS model can be found in Ref.~\cite{qfs_org,qfs_org2}. The subroutines of the model are coded in $XEMCEvent.h$ and were originally developed and maintained by the collaboration from Temple University~\cite{karl_thesis, hyao_thesis,whita}. 
   
 \subsection{QE-F1F209}
 QE-F1F209 is a part of the cross section model, F1F2QE09, which was developed by P. Bosted and V. Mamyan~\cite{Bosted:2012qc} based their work on empirical fit to electron-nucleus scattering. The model is coded in a stand-alone FORTRAN program, \emph{f1f209.f}. An external link is given in the XEMC package to call the subroutines in the FORTRAN code. To successfully compile the code, a library named \emph{libg2c.so} must be specified in the \emph{Makefile}.
 
\section{DIS Models}
 The DIS cross section model not only calculates the cross section of the deep inelastic scattering process but also includes other inelastic processes, such as resonance productions. There are three DIS models coded in the package. Since the kinematic settings of the E08-014 was well above the QE peak, the contribution from inelastic processes is relatively small, and these models were not iterated with the existing DIS data. 
 
\subsection{DIS-QFS}
 DIS-QFS is a part of the QFS subroutines~\cite{hyao_thesis}.  This model includes the following processes:
 \begin{itemize}
  \item Scattering from two interacting nucleons (MEC in Dip region between the QE peak and the resonances),
  \item Delta Electroproduction ($\Delta$),
  \item Resonance productions at 1500~MeV and 1700~MeV, and,
  \item Deep inelastic scattering (DIS).
 \end{itemize}
  
\subsection{DIS-XEM}
 DIS-XEM was specially designed for the XEM experiment based on P. Bosted's previous empirical fit, F1F2IN06~\cite{Bosted:2006}. To agree with the EMC data, the model included several corrections in different range of $0.8<x_{bj}<1.0$ , and the code became complicated and runs slowly, especially when performing radiative correction. The subroutines have been converted from FORTRAN into C++ and coded in \emph{$XEMC\_Born.h$}. 

\subsection{DIS-F1F209}
 DIS-F1F209 comes from F1F2IN09 and is coded in $f1f209.f$. It is the default DIS model in XEMC.

\section{Radiative Corrections}
\begin{figure}[!ht]
 \begin{center}
  \includegraphics[angle=0,width=0.8\textwidth]{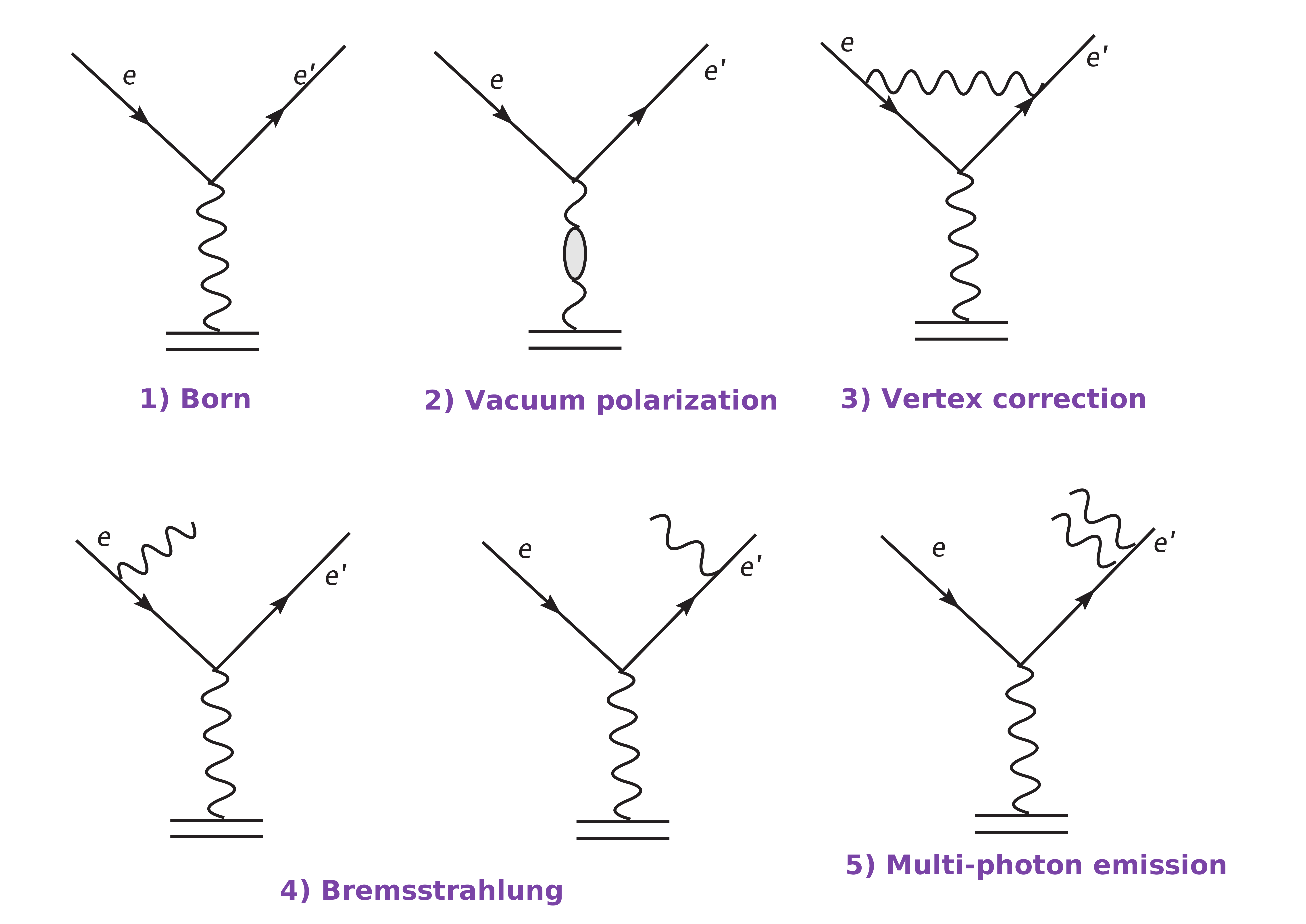}
  \caption[Feynman diagrams for radiation effect]{Feynman diagrams for radiation effect in inclusive lepton-nucleon scattering. Only the lowest orders are shown here.}
  \label{rad_feynm}
 \end{center}
\end{figure} 
 The electron-nucleon scattering process can be modelled by one-photon-exchange-approximation (OPEA), where the electron and the nucleon interact by exchanging one virtual photon. The inclusive cross section of the process is called the Born cross section. There are higher order processes, called radiative effects, contributing to the measured cross sections, as shown in Fig.~\ref{rad_feynm}. The experimental raw cross section is named as the radiated cross section, which has to be corrected to obtain the experimental Born cross section: 
 \begin{equation}
  \sigma^{EX}_{Born} = \frac{\sigma^{Model}_{Born}}{\sigma^{Model}_{rad}}\cdot\sigma^{EX}_{rad},
 \end{equation}
where $\sigma^{Model}_{Born}$ and $\sigma^{Model}_{rad}$ are the Born and radiated cross section calculated from the model, while $\sigma^{EX}_{Born}$ and $\sigma^{EX}_{rad}$ are the Born and radiated cross sections measured from the experiment. The ratio term is generally called the radiative correction factor. 
 
  The radiation effects contain the external radiation and the internal radiation. The external radiation, including external bremsstrahlung and ionization, happens when the incoming or the outgoing electron radiates a real photon when it interacts with the nuclear medium other than the target nucleon. This effect mainly depends on the material and thickness of the target. The internal radiation contains the soft processes, such as internal bremsstrahlung, and the hard processes, such as vacuum polarization, vertex corrections and multiple-photon exchange. The initial and final energies of the electron are modified during those processes, which causes the measured cross section to deviate from the Born cross section.

   The idea of radiative correction is carefully discussed in~\cite{mo_sai_rad, stein_radiation}, and a radiative correction package, RadCor, was developed based on this idea~\cite{karl_thesis,hyao_thesis}. Peak approximation method was used in the package to reduce the CPU time of the radiated cross section calculation. Important subroutines in this package have been migrated to XEMC. 
   
 \section{Performance} 
\begin{figure}[!h]
  \begin{center}
    \subfloat[$\mathrm{^{2}D}$ from Arrington 1995]{
      \includegraphics[type=pdf,ext=.pdf,read=.pdf,width=0.9\textwidth]{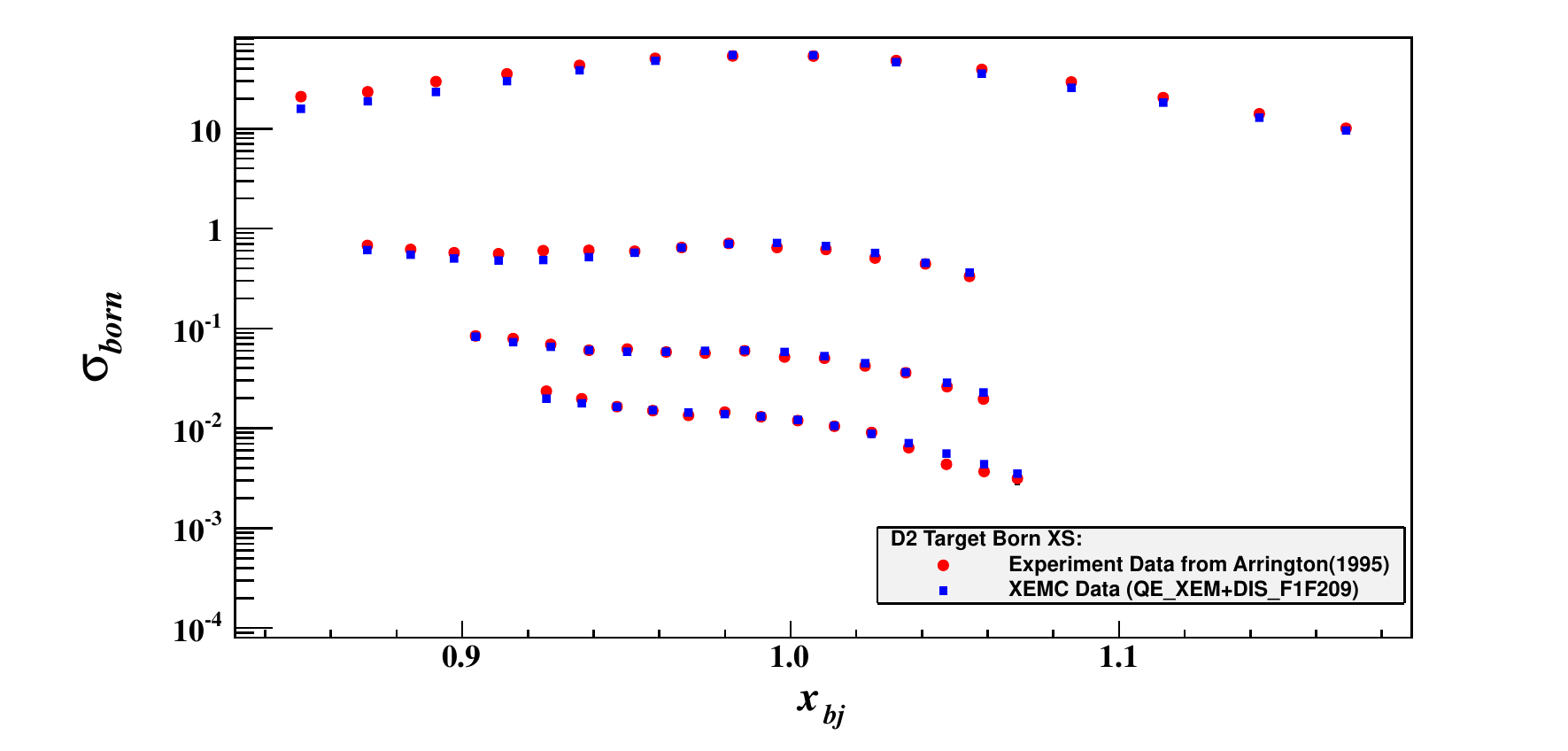}
    }
    \\
     \subfloat[$\mathrm{^{3}He}$ from Day 1979]{
      \includegraphics[type=pdf,ext=.pdf,read=.pdf,width=0.8\textwidth]{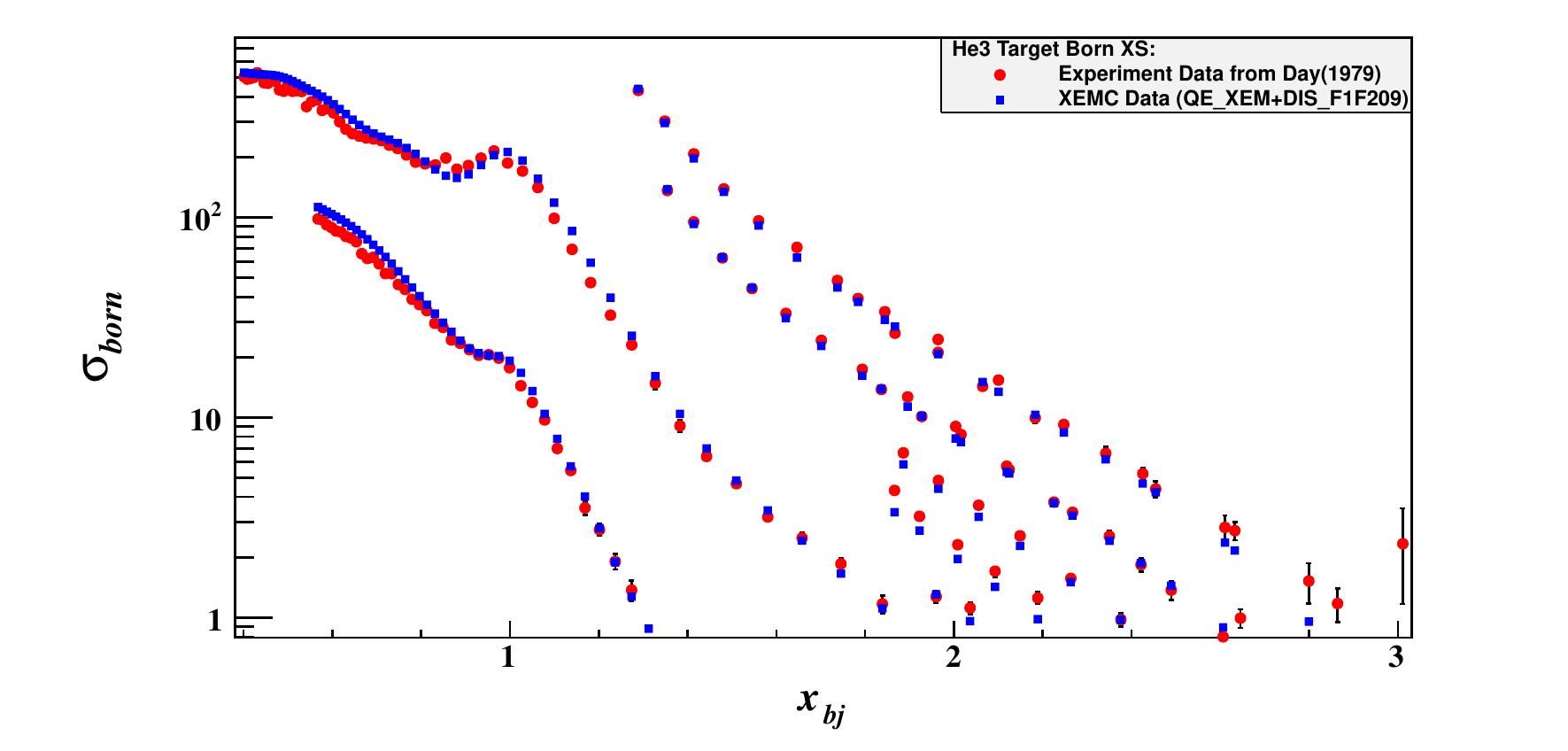}
    }
     \caption[Comparing XEMC models and experiment data for $\mathrm{^{2}D}$ and $\mathrm{^{3}He}$]{\footnotesize{Comparing XEMC models and experiment data for $\mathrm{^{2}D}$ and $\mathrm{^{3}He}$. Data is from QES-Archive~\cite{qe_donal}.}}
    \label{xs_achieve_com1}
  \end{center}
\end{figure}

\begin{figure}[!h]
  \begin{center}
     \subfloat[$\mathrm{^{4}He}$ from Meziani 1992]{
      \includegraphics[type=pdf,ext=.pdf,read=.pdf,width=0.9\textwidth]{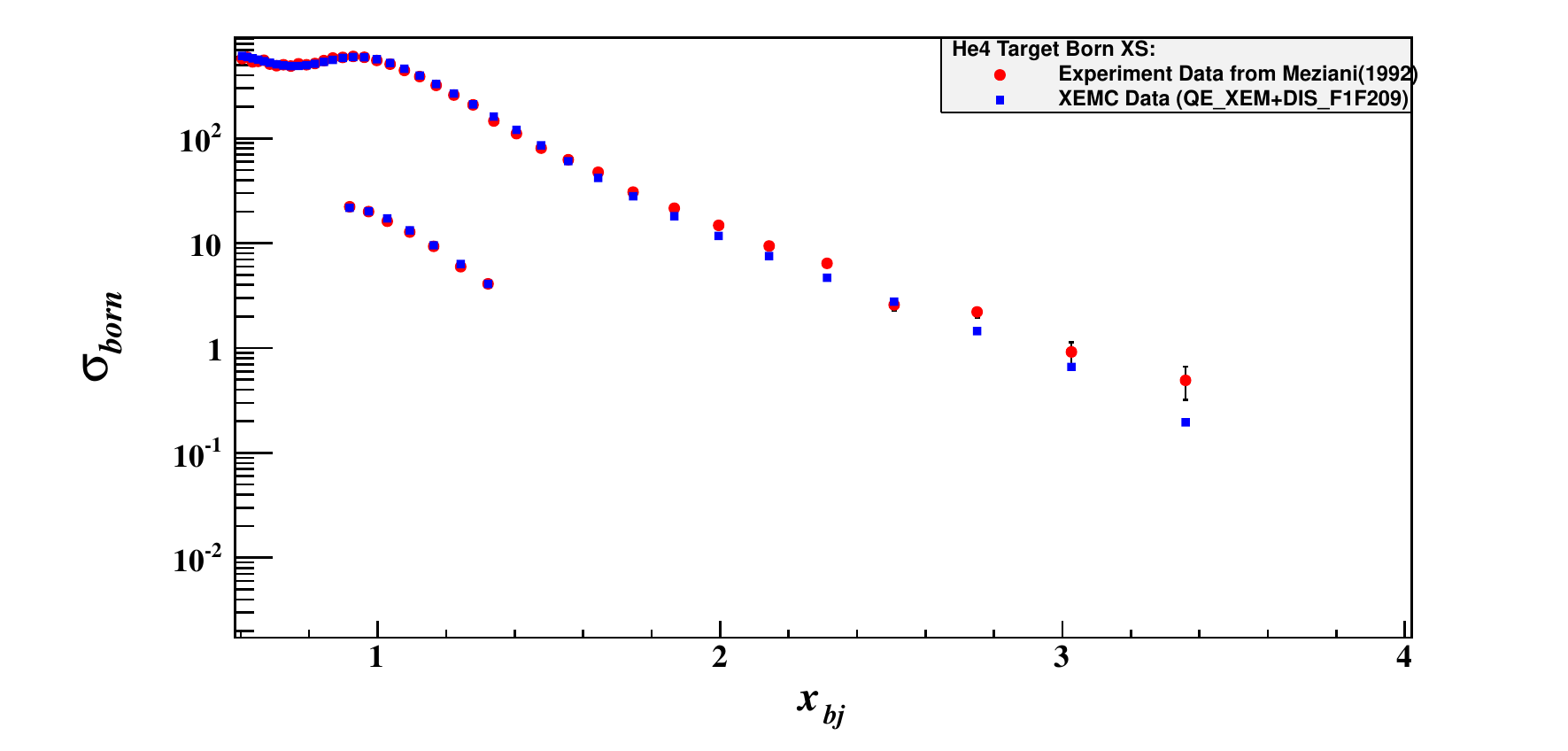}
      \label{target_run1}
    }
    \\
     \subfloat[$\mathrm{^{12}C}$ from Arrington 1998]{
      \includegraphics[type=pdf,ext=.pdf,read=.pdf,width=0.9\textwidth]{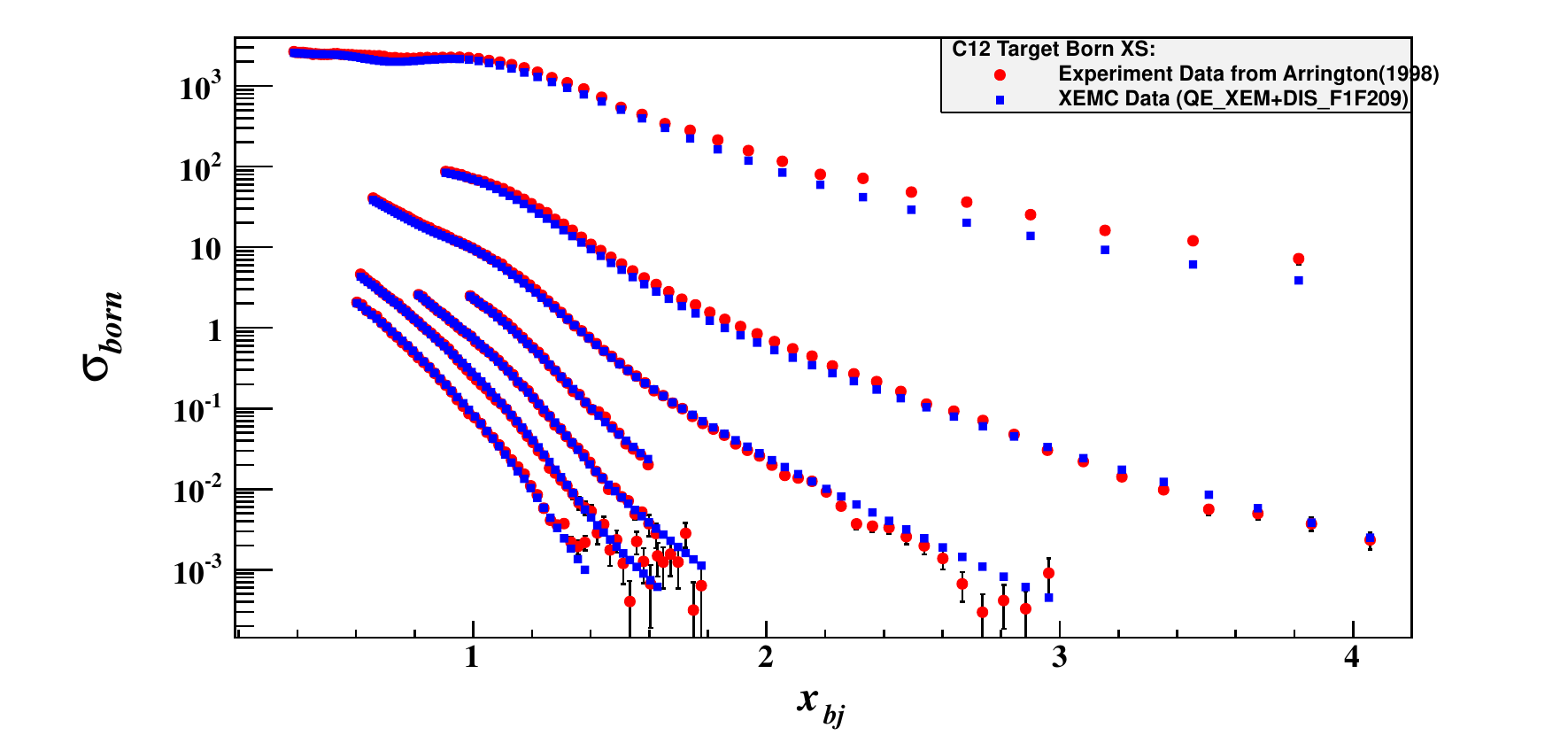}
    }
 \caption[Comparing XEMC models and experiment data for $\mathrm{^{4}He}$ and $\mathrm{^{12}C}$]{\footnotesize{Comparing XEMC models and experiment data for $\mathrm{^{4}He}$ and $\mathrm{^{12}C}$. Larger deviation can be seen in $\mathrm{^{12}C}$ data low $\mathrm{Q^{2}}$. Data is from QES-Archive~\cite{qe_donal}.}}   
  \label{xs_achieve_com2}
  \end{center}
\end{figure}

\begin{figure}[!h]
  \begin{center}
    \subfloat[$\mathrm{^{3}He}$ Born cross section]{
      \includegraphics[type=pdf,ext=.pdf,read=.pdf,width=0.9\textwidth]{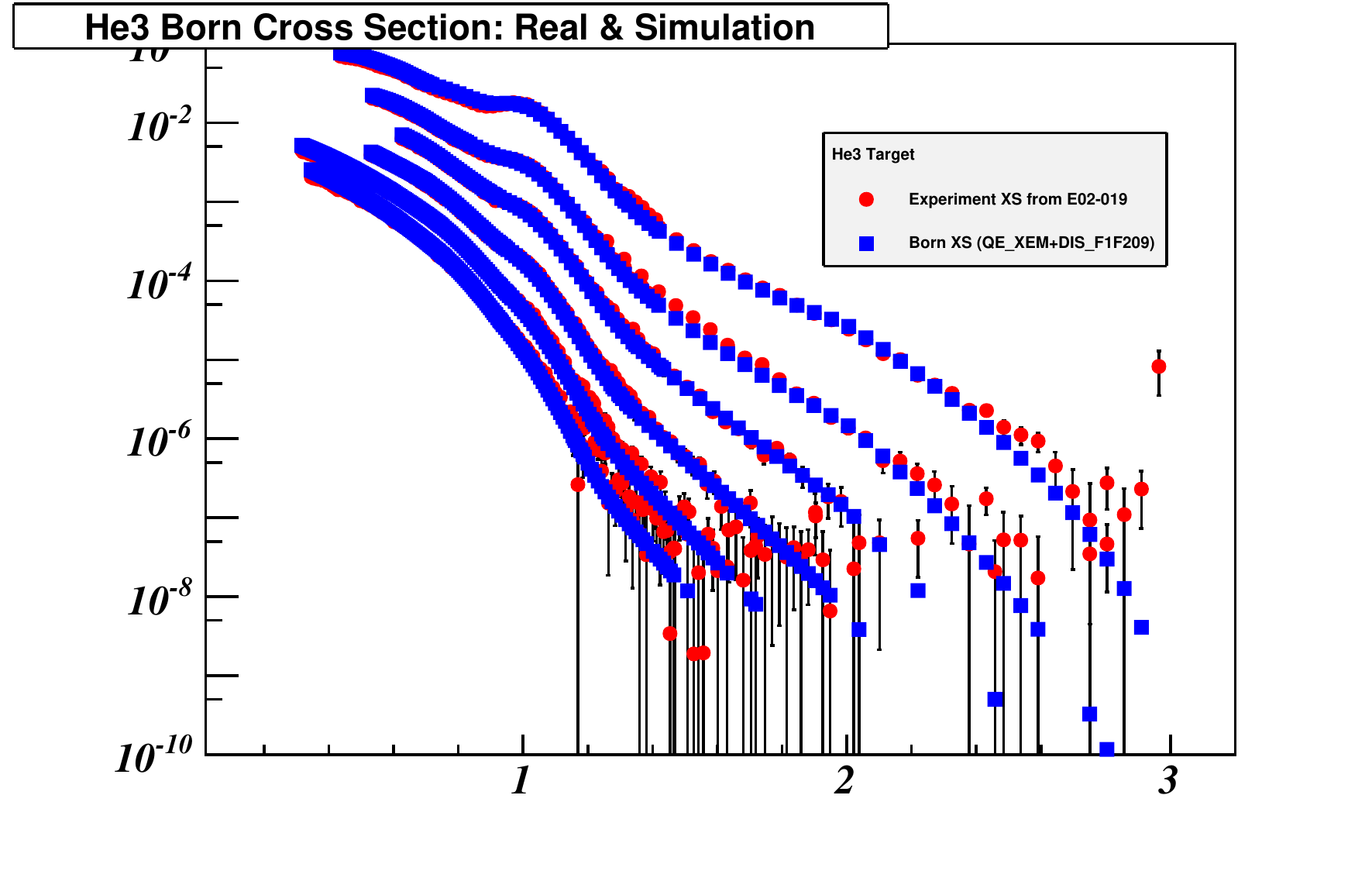}
      \label{xs_nadia_b1}
    }
    \\
       \subfloat[$\mathrm{^{3}He}$ radiated cross section]{
      \includegraphics[type=pdf,ext=.pdf,read=.pdf,width=0.9\textwidth]{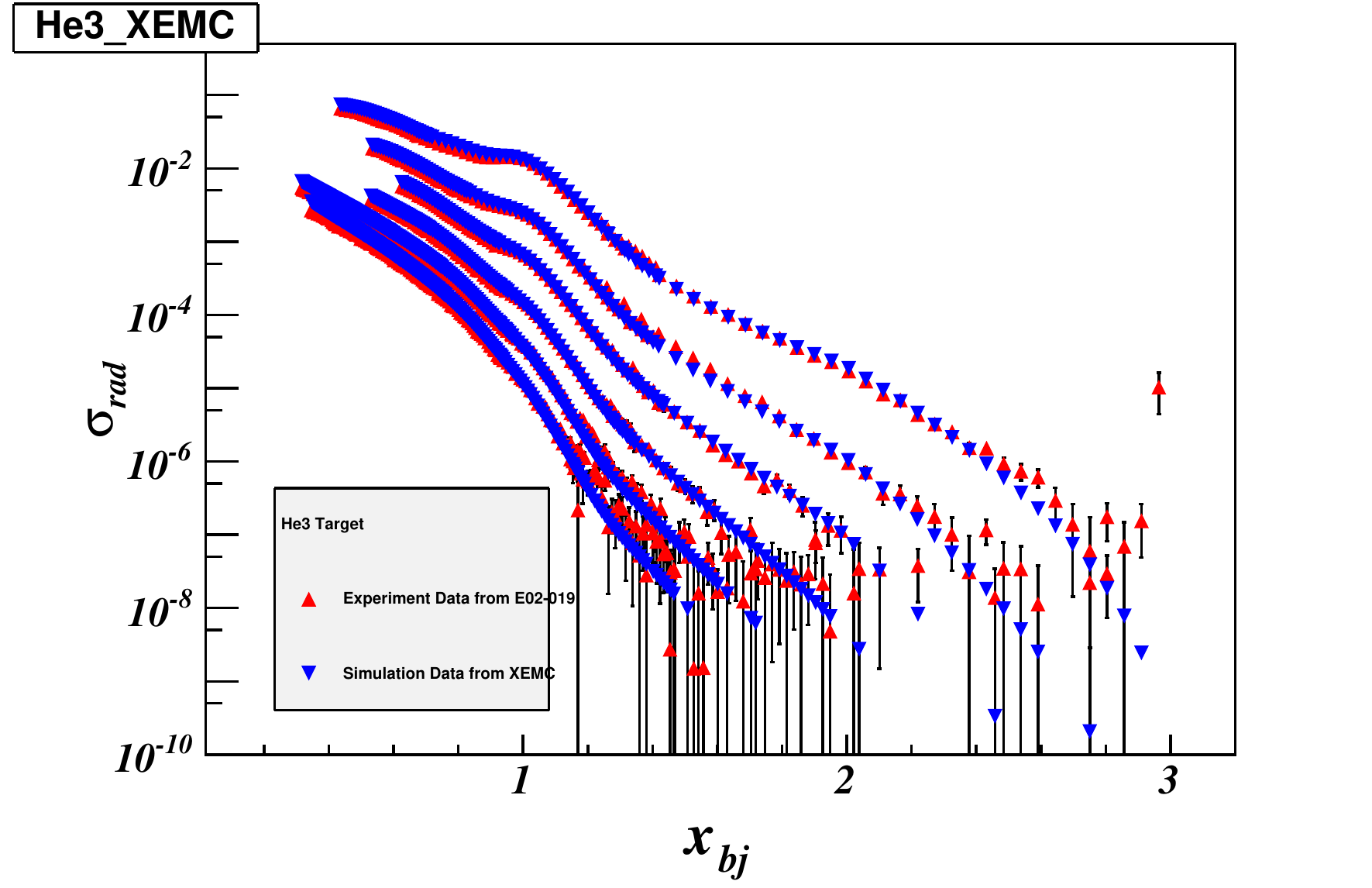}
      \label{xs_nadia_r1}
    }
    \caption[Comparing $\mathrm{^{3}He}$ cross sections from E02-019 and calculated in XEMC]{\footnotesize{Comparing $\mathrm{^{3}He}$ cross sections from E02-019~\cite{nadia_thesis} and calculated in XEMC, where top and bottom plots are the Born and radiated cross section, respectively. The thickness of the target and the configuration of the target system are included~\cite{nadia_thesis}.}}
    \label{xs_nadia_com1}
  \end{center}
\end{figure}

\begin{figure}[!h]
  \begin{center}
     \subfloat[$\mathrm{^{12}C}$ Born cross section]{
      \includegraphics[type=pdf,ext=.pdf,read=.pdf,width=0.9\textwidth]{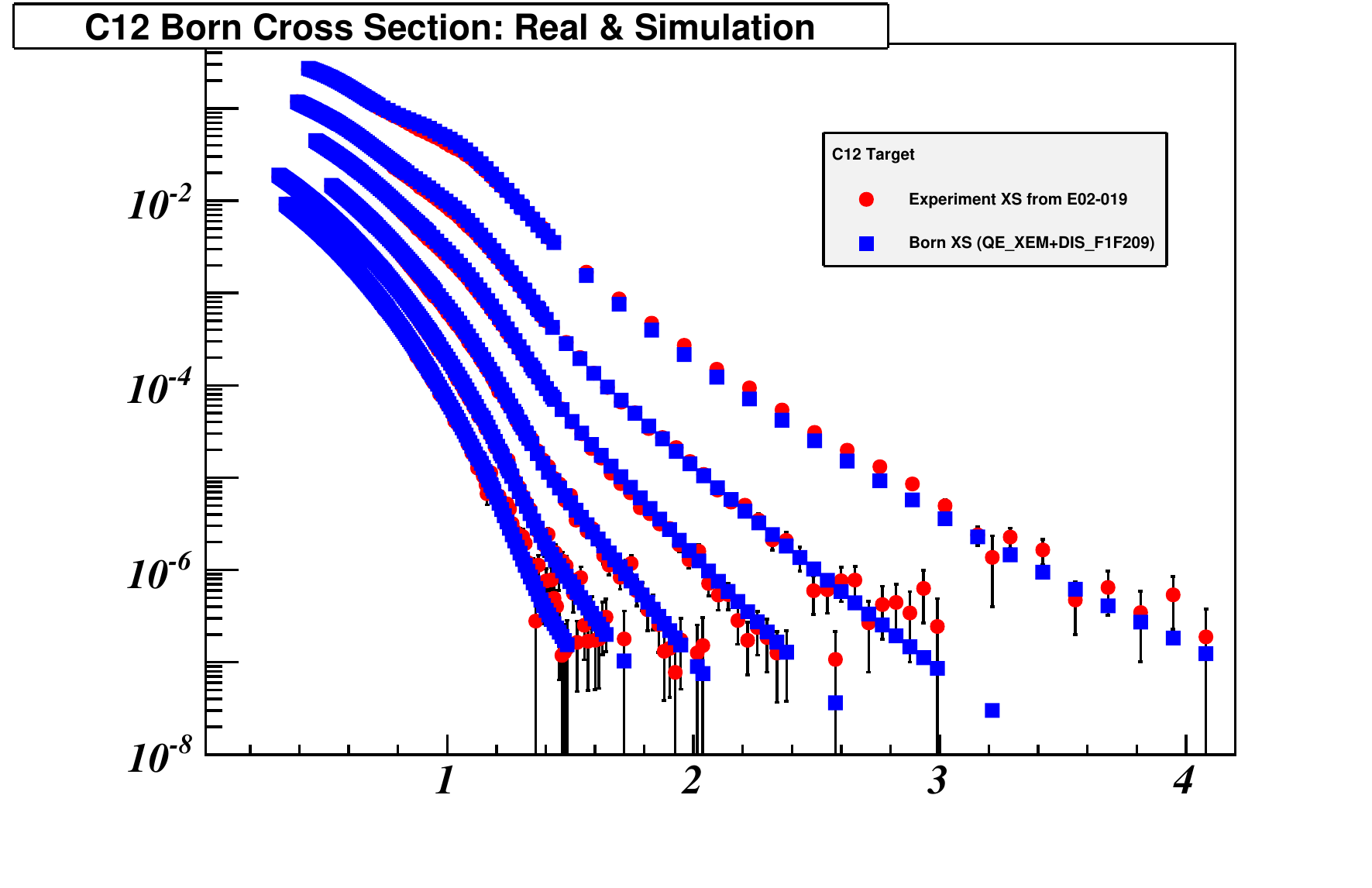}
      \label{xs_nadia_b2}
    }
    \\
     \subfloat[$\mathrm{^{12}C}$ radiated cross section]{
      \includegraphics[type=pdf,ext=.pdf,read=.pdf,width=0.9\textwidth]{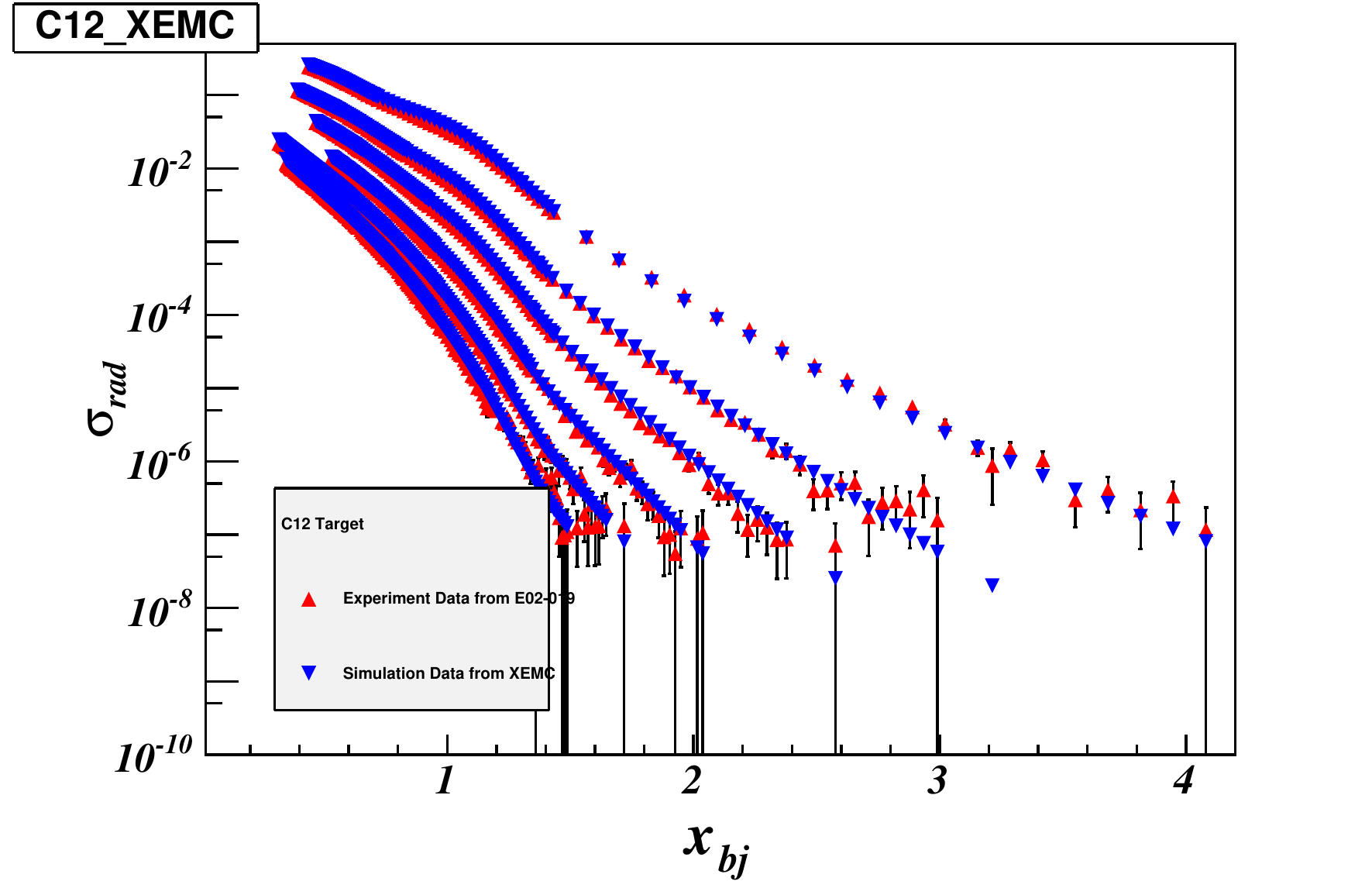}
      \label{xs_nadia_r2}
    }
    \caption[Comparing $\mathrm{^{12}C}$ cross sections from E02-019 and calculated in XEMC]{\footnotesize{Comparing $\mathrm{^{12}C}$ cross sections from E02-019~\cite{nadia_thesis} and calculated in XEMC, where top and bottom plots are the Born and radiated cross section, respectively. The thickness of the target and the configuration of the target system are included~\cite{nadia_thesis}.}}
    \label{xs_nadia_com2}
  \end{center}
\end{figure}

  In this section, the cross sections calculated from XEMC with the QE-XEM model and the DIS-F1F209 model are directly compared with previous experiment data stored in the QES-Archive~\cite{qe_donal} (Fig.~\ref{xs_achieve_com1} and Fig.~\ref{xs_achieve_com2}) as well as the E02-019 data (Fig.~\ref{xs_nadia_b1} and Fig.~\ref{xs_nadia_b2}). Overall, the model and the data agree nicely. The performance of the radiative correction was examined with the E02-019 data of which the target configurations were known. From Fig.~\ref{xs_nadia_r1} and Fig.~\ref{xs_nadia_r2}, the radiated cross sections from XEMC agree well with the data above the QE region. At $x_{bj}<1$ a small deviation can be seen due to the use of the peak approximation method. The E08-014 data is well above the QE peak so the deviation wasn't important. 

\newpage
\section{Examples}
 An example to use the XEMC package is given in this section. 
 \begin{lstlisting}
#include "XEMC.h"
int main(int argc,char** argv){
	XEMC* Event = new XEMC(); //Create a XEMC event
	//Target
	const int A = 12, Z = 6;    
	TString Target_Name = "C12";
    	Event->Init(Form("./input/%s_Input.dat",Target_Name.Data()));	
 	//Kinematic setting	
	double E0 = 3.356;    //GeV
	double Ep = 2.505;    //GeV/c
	double Theta = 25.00; //Degree
		
	//Calculate XS for the event
	int err = Event->Process(E0,Ep,Theta,A,Z);	
	if(err>=0){ //Return values
		double xs_rad  = Event->XS_Rad();
		double xs_qe   = Event->XS_QE();
		double xs_dis  = Event->XS_DIS();
		double xs_Born = Event->XS_Born();
	}
	//Print Out
	cerr<<Form("For %s Target, E0=%5.3f GeV, Ep=%5.3f GeV, Theta=%5.3f:",Target_Name.Data(), E0, Ep, Theta)<<endl;
	cerr<<Form("    XS_Born=%e, XS_QE=%e, XS_DIS=%e, XS_Rad=%e",
                    xs_Born, xs_qe, xs_dis, xs_rad)<<endl;	
	
	delete Event; //Release memory
}
\end{lstlisting}

%% file: append/append_deltap.tex
\chapter{Momentum Correction in HRS-R}
  Each HRS is composed of a dipole and three quadrupoles in the order of Q1, Dipole, Q2 and Q3. Four magnets typically have the same central momentum value. With the focal plane quantities provided by the VDC tracking, a HRS optics matrix reconstructs $\delta p$, $y_{tg}$, $\theta_{tg}$, and $\phi_{tg}$, the target plane quantities to describe an event at the reaction point. As discussed in Section 3.3, during the E08-014, the field of the Q3 magnet on HRS-R (RQ3) was scaled to 87.72\% of its normal value, and the transportation of particles in the HRS-R had been changed. While the matrix elements of $y_{tg}$, $\theta_{tg}$, and $\phi_{tg}$ have been properly optimized (see Section 4.3), the momentum matrix (the D-terms) could not be calibrated since the momentum calibration data was not available in the quasielastic region. It requires an additional correction to get the right value of $\delta p$. In this section, a method will be introduced to correct the $\delta p$ on HRS-R with the SAMC data and the SNAKE model~\cite{snack_lerose}.
 
 In the Hall-A Single Arm Monte Carlo tool (SAMC), each HRS magnet's transportation from the entrance to the exit is simulated in the SNAKE model as a series of forward transportation functions (FWDs). For example, the quantities at the Q1 entrance, $x_{Q1}^{en}$, $y_{Q1}^{en}$, $\theta_{Q1}^{en}$, $\phi_{Q1}^{en}$, and $l_{Q1}^{en}$ can be directly deduced from the target plane quantities via the linear transportation; and the quantities at the Q1 exit, $x_{Q1}^{ex}$, $y_{Q1}^{ex}$, $\theta_{Q1}^{ex}$, $\phi_{Q1}^{ex}$, and $l_{Q1}^{ex}$ can be calculated with their corresponding FWDs with the quantities at the entrance as inputs. These quantities at the Q1 exit are equal to the quantities at the dipole entrance and can be further used to calculate the quantities at the dipole exit; so on and so forth. The focal plane quantities, $x_{fp}$, $y_{fp}$,$\theta_{fp}$, and $\phi_{fp}$, are given by the FWDs of Q3. The focal plane quantities are then smeared with the resolution of the HRS VDCs defined in the simulation.
 
  Similar to the HRS optics matrix, a set of backward polynomial functions (BWDs) directly calculate each target plane quantity with the four focal plane quantities as inputs.
  
 To simulate the HRS-R setting during this experiment, new FWDs were generated for the RQ3 with the mis-matching field, named as $\mathrm{FWD^{Q3}_{mis}}$, and they replaced the FWDs with the normal field setting ($\mathrm{FWD^{Q3}_{norm}}$) in the simulation. Besides, two sets of new BWDs were also produced by SNAKE to describe the RQ3. The first set ($\mathrm{BWD_{mis}^{D}}$) has the correct reconstructions of all target plane quantities. It simulates the optics matrix on HRS-R with all terms being optimized. The second set ($\mathrm{BWD_{norm}^{D}}$) has also included the correct reconstructions of target plane quantities, except the one of $\delta p$ which was generated with the normal RQ3 field. This corresponds to the new optics matrix with the un-calibrated D-Terms.

 In SAMC, two groups of simulation events were generated with the same event seeds. The HRS-R in first group of events was simulated with $\mathrm{FWD^{Q3}_{mis}+BWD_{mis}^{D}}$, and the values of $\delta p$ in these events should be correctly reconstructed and are labelled as $\delta p_{cor}$. In the second group, the HRS-R was simulated with $\mathrm{FWD^{Q3}_{mis}+BWD_{norm}^{D}}$ which reconstructs incorrect values of $\delta p$, named as $\delta p_{in}$.
  
 In the real data, the error of the momentum reconstruction caused by using the un-calibrated D-terms can be studied by the difference of $\delta p_{cor}$ and $\delta p_{in}$ in the simulation data:
\begin{equation}
 \Delta\delta p = \delta p_{cor} - \delta p_{in},
\end{equation}
which can be specified by a correction function defined as:
\begin{eqnarray}
	 f(x_{fp}, \theta_{fp}, y_{fp}, \phi_{fp}) &=& \sum_{i=0}^{N_{A}}A_{i}x_{fp}^{i}+\sum_{j=0}^{N_{B}}B_{j}\theta_{fp}^{j}+\sum_{k=0}^{N_{C}}C_{k}y_{fp}^{k} \nonumber \\
                                   &+&\sum_{l=0}^{N_{D}}D_{l}\phi_{fp}^{l} +\sum_{m=0}^{N_{E}}E_{m}\delta p_{in}^{m},
\label{dp_corr_func}
\end{eqnarray}
where the first four terms are the polynomial functions of the focal plane quantities, and the last term is used to correct any high-order optics effects. The procedure to obtain the correction function from the simulation data is presented as follows. 

 First of all, the first term in Eq.~\eqref{dp_corr_func} is fitted with $x_{fp}$:
\begin{equation}
 \Delta\delta p(x_{fp}) = \delta p_{cor} - \delta p_{in} = \sum_{i=0}^{N_{A}}A_{i}x_{fp}^{i},
 \label{deltap_corr_xfp}
\end{equation}
which gives a new momentum value, $\delta p_{x_{fp}}=\delta p_{in}+\sum_{i=0}^{N_{A}}A_{i}x_{fp}^{i}$, and the residual error, $\Delta\delta p = \delta p_{cor}-\delta p_{x_{fp}}$, is further fitted with $\theta_{fp}$:
\begin{equation}
 \Delta\delta p(\theta_{fp}) = \delta p_{cor} - \delta p_{x_{fp}} = \sum_{j=0}^{N_{B}}B_{j}\theta_{fp}^{j},
 \label{deltap_corr_thetafp}
\end{equation}
which gives $\delta p_{\theta_{fp}}=\delta p_{x_{fp}}+\sum_{j=0}^{N_{B}}B_{j}\theta_{fp}^{j}$. Similar corrections are applied to $y_{fp}$, $\phi_{fp}$ and $\delta p_{in}$:
\begin{eqnarray}
 &&\Delta\delta p(y_{fp}) = \delta p_{cor} - \delta p_{\theta_{fp}} = \sum_{k=0}^{N_{C}}C_{k}y_{fp}^{k},\\
 &&\Delta\delta p(\phi_{fp})   = \delta p_{cor} - \delta p_{y_{fp}} = \sum_{l=0}^{N_{D}}D_{l}\phi_{fp}^{l},\\
 &&\Delta\delta p(\delta p_{in})   = \delta p_{cor} - \delta p_{\phi_{fp}} = \sum_{m=0}^{N_{E}}E_{m}\delta_{in}^{m}.
 \label{deltap_corr_deltap}
\end{eqnarray}
\begin{figure}[!ht]
  \begin{center}
    \subfloat[$\Delta\delta p(x_{fp})$ .vs. $x_{fp}$ and $\theta_{fp}$]{
      \includegraphics[type=pdf,ext=.pdf,read=.pdf,width=0.7\textwidth]{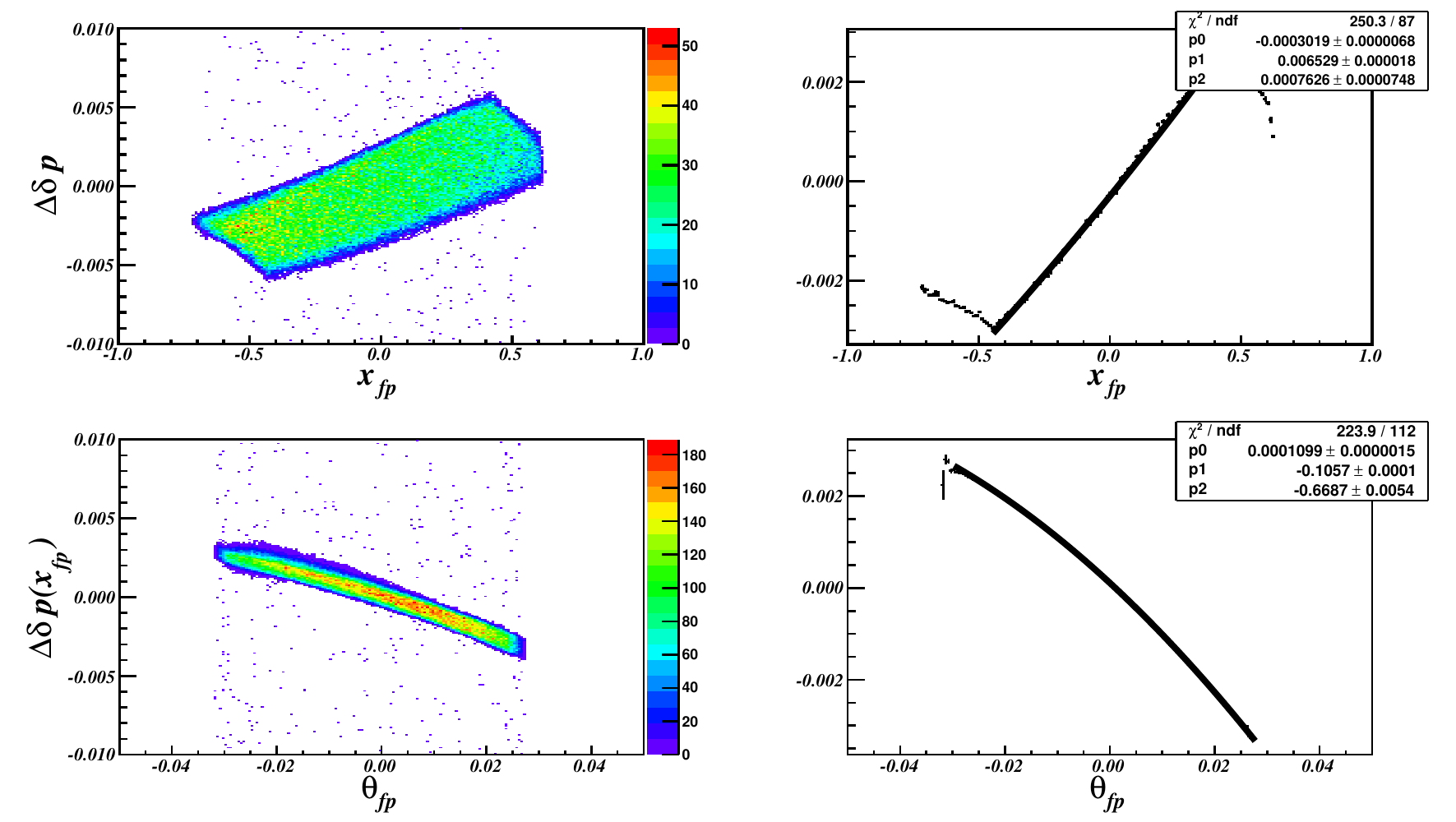} 
    }
    \hfill
     \subfloat[$\Delta\delta p(y_{fp})$ .vs. $y_{fp}$ and $\phi_{fp}$]{
      \includegraphics[type=pdf,ext=.pdf,read=.pdf,width=0.7\textwidth]{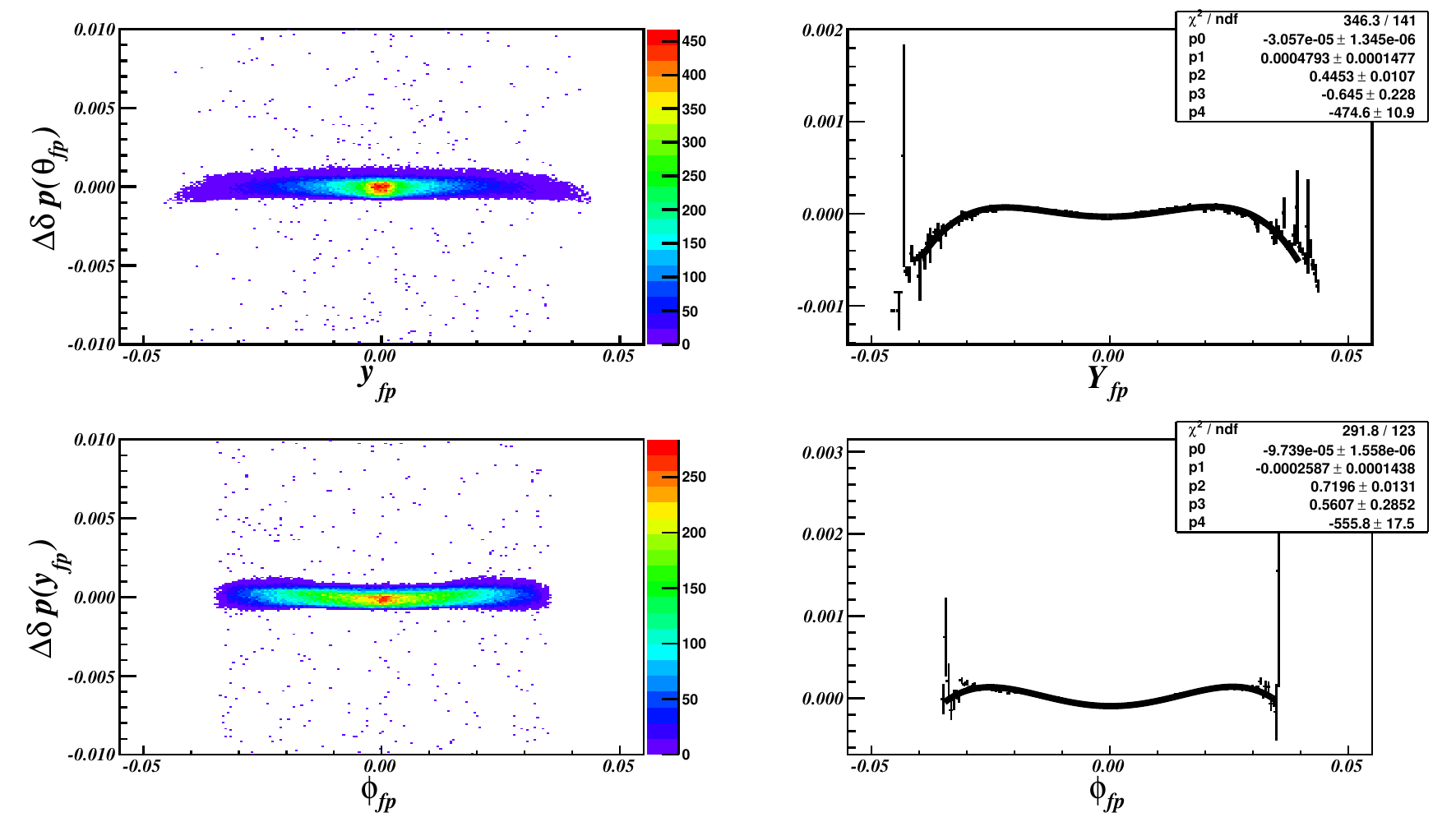} 
    }
    \hfill
   \subfloat[$\Delta\delta p(\phi_{fp})$ .vs. $\delta p_{in}$]{
      \includegraphics[type=pdf,ext=.pdf,read=.pdf,width=0.7\textwidth]{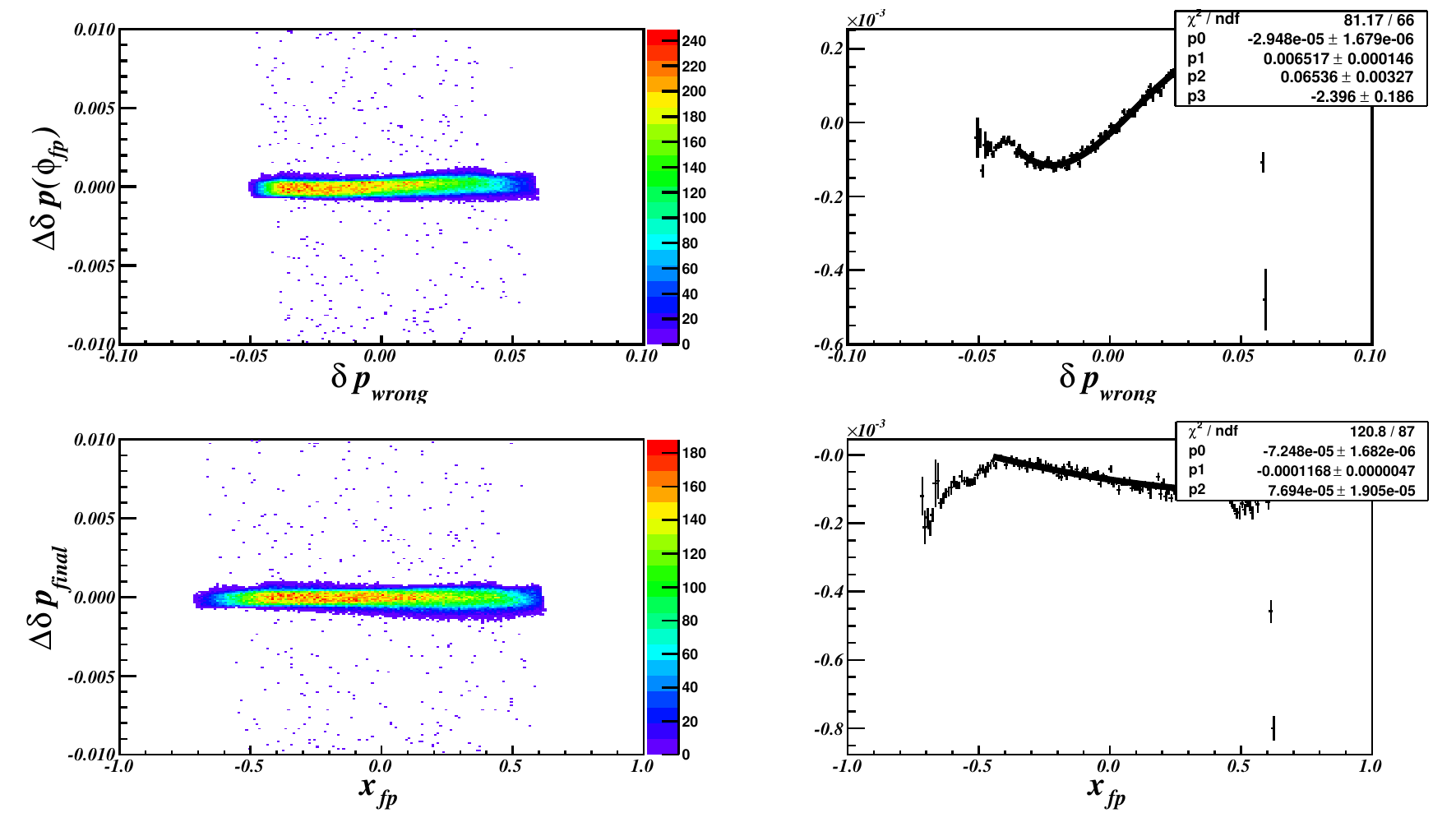}
    }
    
    \caption[$\delta p$ correction function fitting]{\footnotesize{$\delta p$ correction function fitting with the focal plane variables. In each raw, the left is the 2-D histogram of $\Delta\delta p$ versus each fitting variable, while the right plot is the profile of the 2-D histogram, which is fitted with the polynomial function defined in Eq.~\eqref{deltap_corr_all}. The bottom two plots give the final result after applying all corrections.}}
    \label{deltap_fitting}
  \end{center}
\end{figure}

 Combining equations from Eq.~\eqref{deltap_corr_xfp} to Eq.~\eqref{deltap_corr_deltap}, Eq.~\eqref{dp_corr_func} leads to:
\begin{equation}
  \delta p_{cor}= \delta p_{in}+ f(x_{fp}, \theta_{fp}, y_{fp}, \phi_{fp}),
   \label{deltap_corr_all}
\end{equation}

\begin{figure}[!ht]
 \begin{center}
  \includegraphics[type=pdf, ext=.pdf,read=.pdf,width=0.7\textwidth]{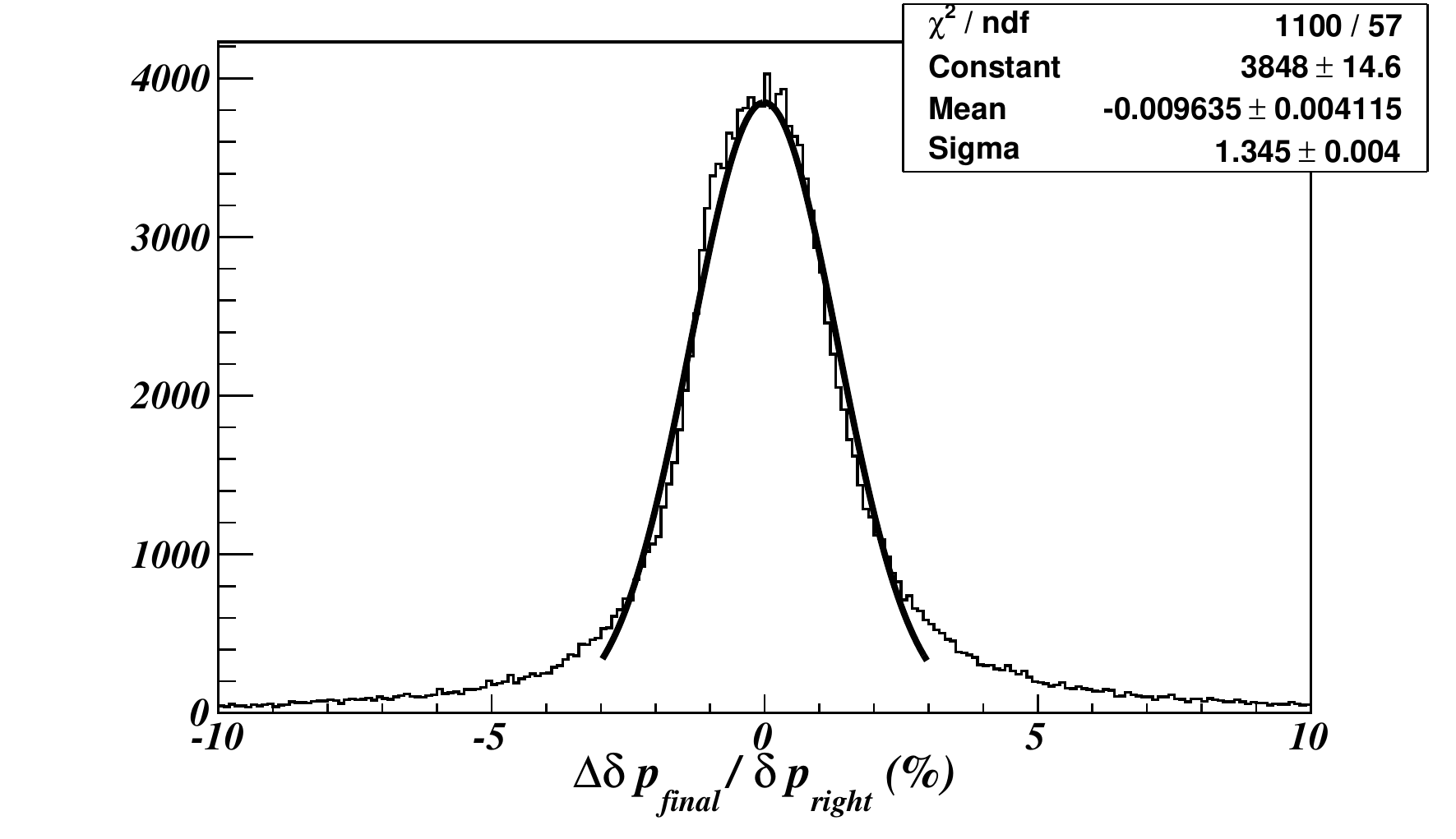}
  \caption[The residual error of $\delta p$ correction function]{The residual error of $\delta p$ correction function. The momentum reconstruction value can be corrected with the error around $1.3\%$.}
  \label{deltap_final}
 \end{center}
\end{figure}

 Fig.~\ref{deltap_fitting} shows the fitting result the distribution of $\Delta\delta p$ for each corrections. The final residual error, $\Delta\delta p(\delta p_{in})/\delta p_{cor}$, is close to $ 0.03\%$ (Fig.~\ref{deltap_final}) indicating that the mis-reconstructed momentum in the RQ3 has been corrected.
 \begin{figure}[!ht]
 \begin{center}
  \includegraphics[type=pdf, ext=.pdf,read=.pdf,width=1.0\textwidth]{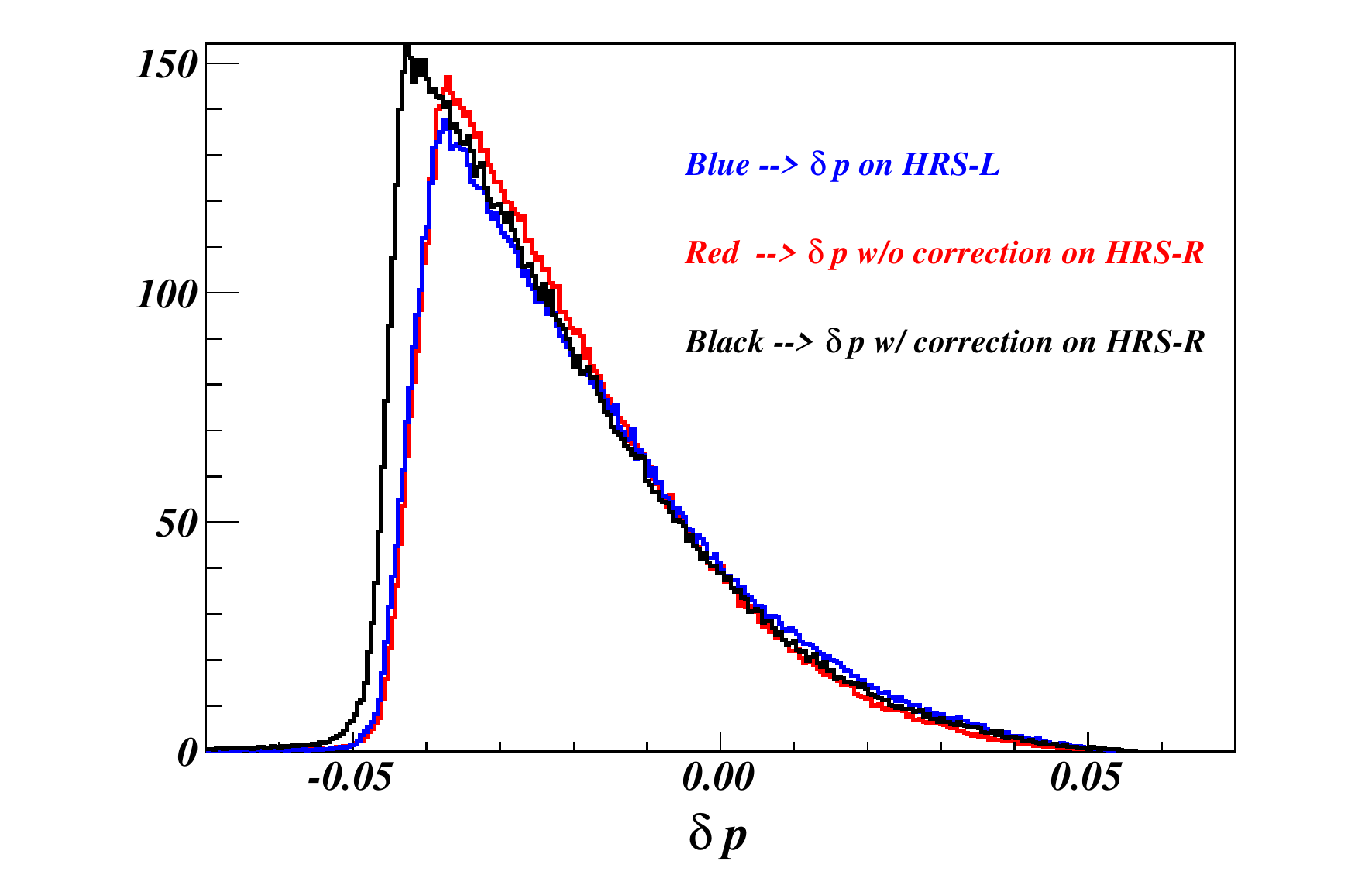}
  \caption[$\delta p$ correction function applying on real data]{$\delta p$ correction function applying on real data. The data was taken on the carbon target with both HRSs at Kin5.0. The momentum distribution on HRS-R without the correction is different from the one on HRS-L (blue), which, however, agrees with the momentum distribution on HRS-R after the correction (black) at the central region. The $\delta p$ acceptance range on HRS-R is wider than the one on HRS-L, which can be explained by the defusing effect of the RQ3.}
  \label{deltap_corr_real}
 \end{center}
\end{figure}

 Eq.~\ref{deltap_corr_all} can be applied to the experimental data to correct the value of $\delta p$ for each event which is firstly reconstructed by the un-calibrated D-Terms in the HRS-R optics matrix. The performance of the correction can be examined by comparing the momentum distribution of the data taken in two HRSs with the same setting. Shown in Fig.~\ref{deltap_corr_real}, the momentum distribution on HRS-R should be identical to the one on HRS-L after applying the correction function, but its acceptance would be slightly wider than the one on HRS-L because of the defocussing effect of the RQ3.

%% file: append/append_target.tex
\chapter{Non-Uniform Cryogenic Targets}
\begin{figure}[!ht]
  \begin{center}
    \subfloat[$\mathrm{^{2}H}$]{
      \includegraphics[type=pdf,ext=.pdf,read=.pdf,width=0.45\textwidth]{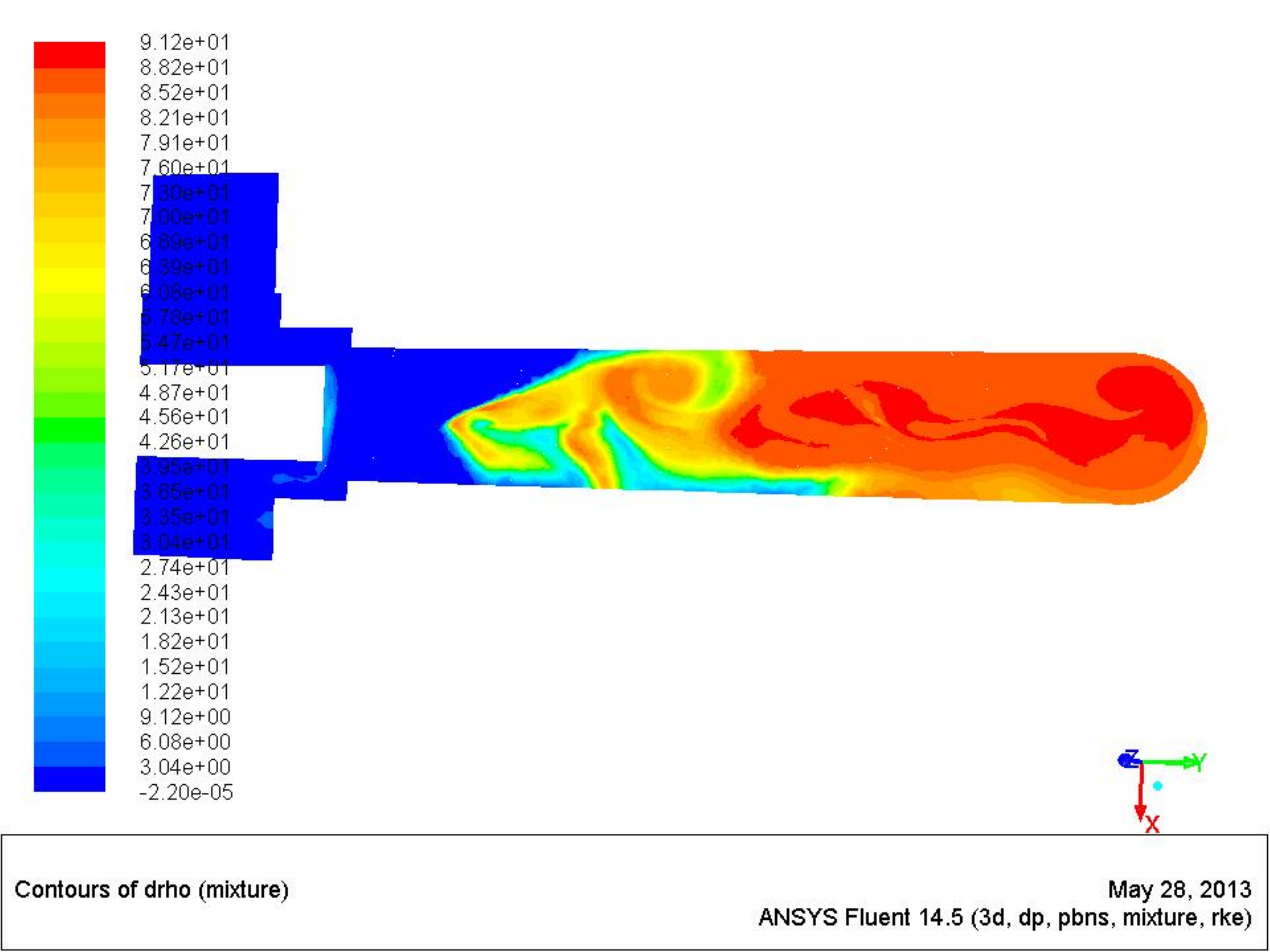} 
    }
    \hfill
     \subfloat[$\mathrm{^{4}He}$]{
      \includegraphics[type=pdf,ext=.pdf,read=.pdf,width=0.45\textwidth]{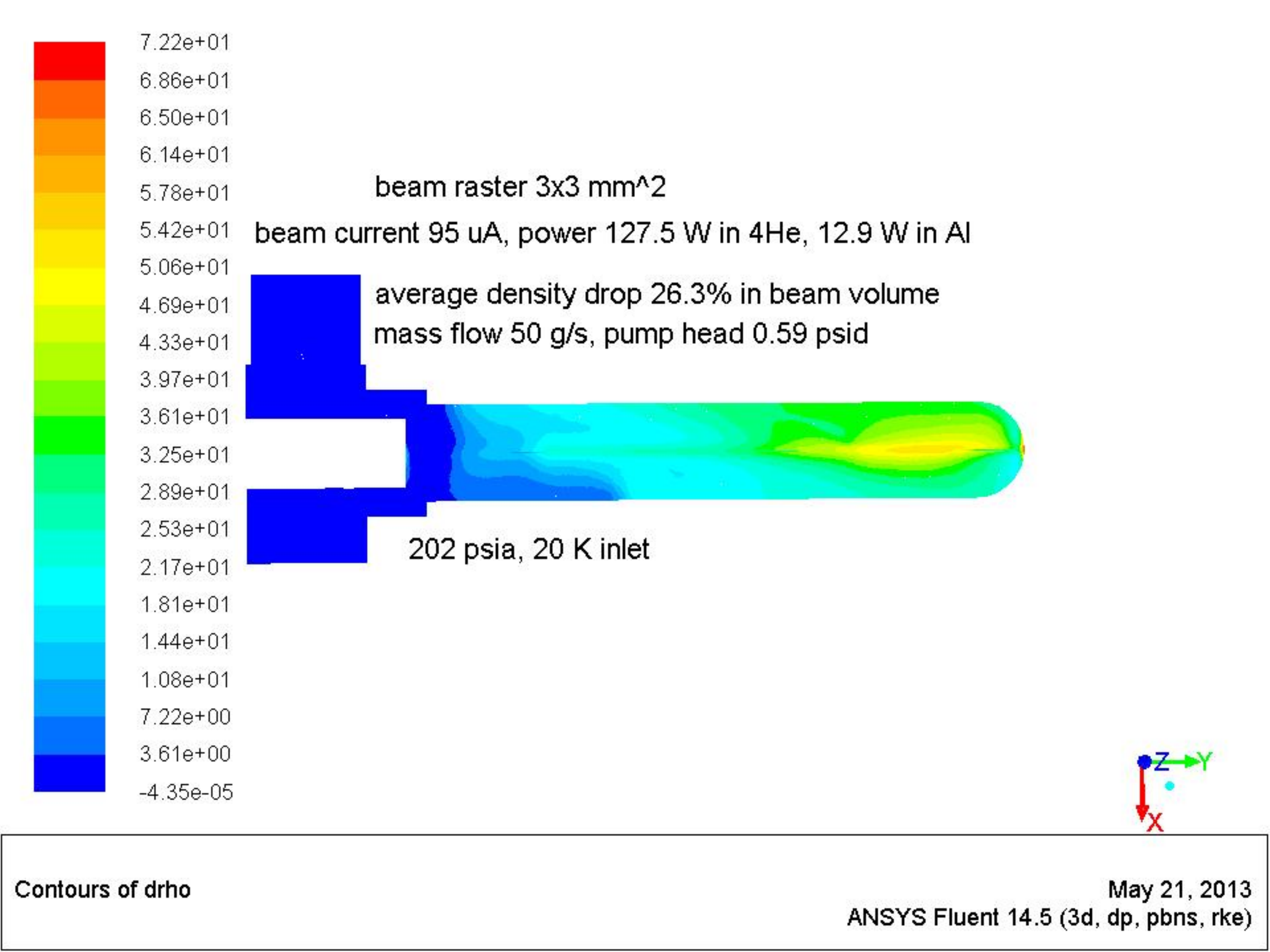} 
    }
    \caption[Cryo-target density profiles from simulation]{\footnotesize{Cryo-target density profiles from simulation. The color contour denotes the value of the target density. The left plot is for $\mathrm{^{2}H}$ and the right plot is for $\mathrm{^{4}He}$. The density profile of $\mathrm{^{3}He}$ is not shown here. Both plots present the fluctuations of the target density along the cell.}}
    \label{silviu_plots}
  \end{center}
\end{figure}

  As discussed in Section 5.4.2, the cryogenic targets (cryo-target) used in the E08-014, $\mathrm{^{2}H}$, $\mathrm{^{3}He}$ and $\mathrm{^{4}He}$, presented strange distributions in $z_{react}$. These distributions indicate that their densities were not uniform but instead, vary along the 20~cm cells. As proved by a Monte Carlo simulation of the cryogenic target system~\cite{silviu_target},  such non-uniform density profiles were caused by the poor design of the target cells and the direction of the cryogenic flow, as shown in Fig.~\ref{silviu_plots}.
 
  The absolute target density is required in order to extract the cross sections. The initial target density with the beam off can be determined before the experiment. When the beam is on, however, the density varies with the beam current because of the boiling effect. Such an effect is negligible for solid targets but it could be significant for cryo-targets. For a non-uniform cryo-target, the boiling effect differs along the target cell, and in addition, the radiation correction becomes more complicated since the radiation effect largely depends on the location and direction of the electron-scattering process. In this chapter, a detailed study of the boiling effect is given, followed by a discussion of extracting the cryo-target density distributions. A procedure to evaluate the radiation correction factors for non-uniform cryo-targets will be introduced in the end.

\section{Boiling Study}
 \begin{figure}[!ht]
  \begin{center}
    \subfloat[$\mathrm{^{12}C}$ on HRS-L]{
      \includegraphics[type=pdf,ext=.pdf,read=.pdf,width=0.45\textwidth]{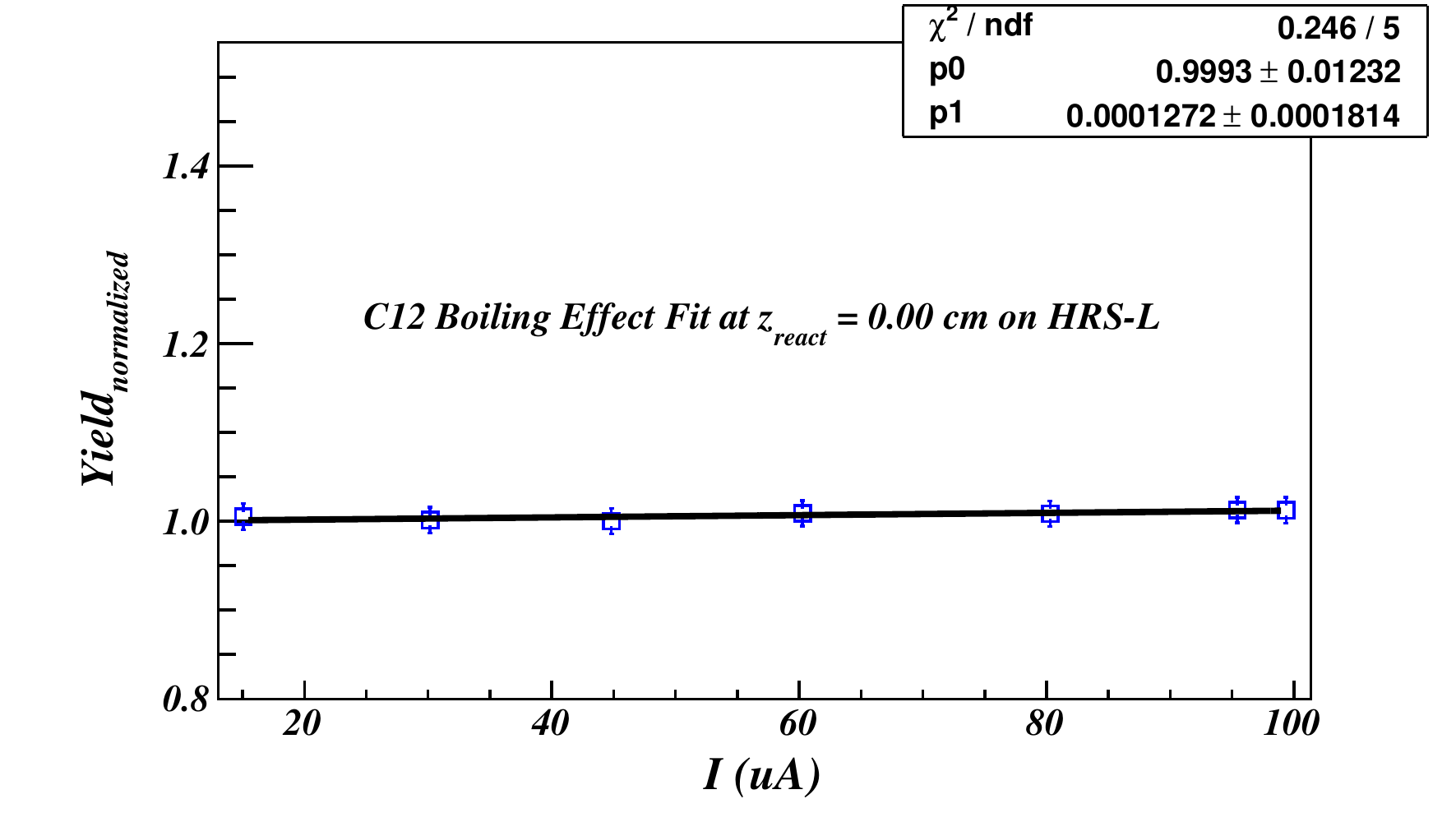} 
    }
    \hfill
    \subfloat[$\mathrm{^{12}C}$ on HRS-R]{
      \includegraphics[type=pdf,ext=.pdf,read=.pdf,width=0.45\textwidth]{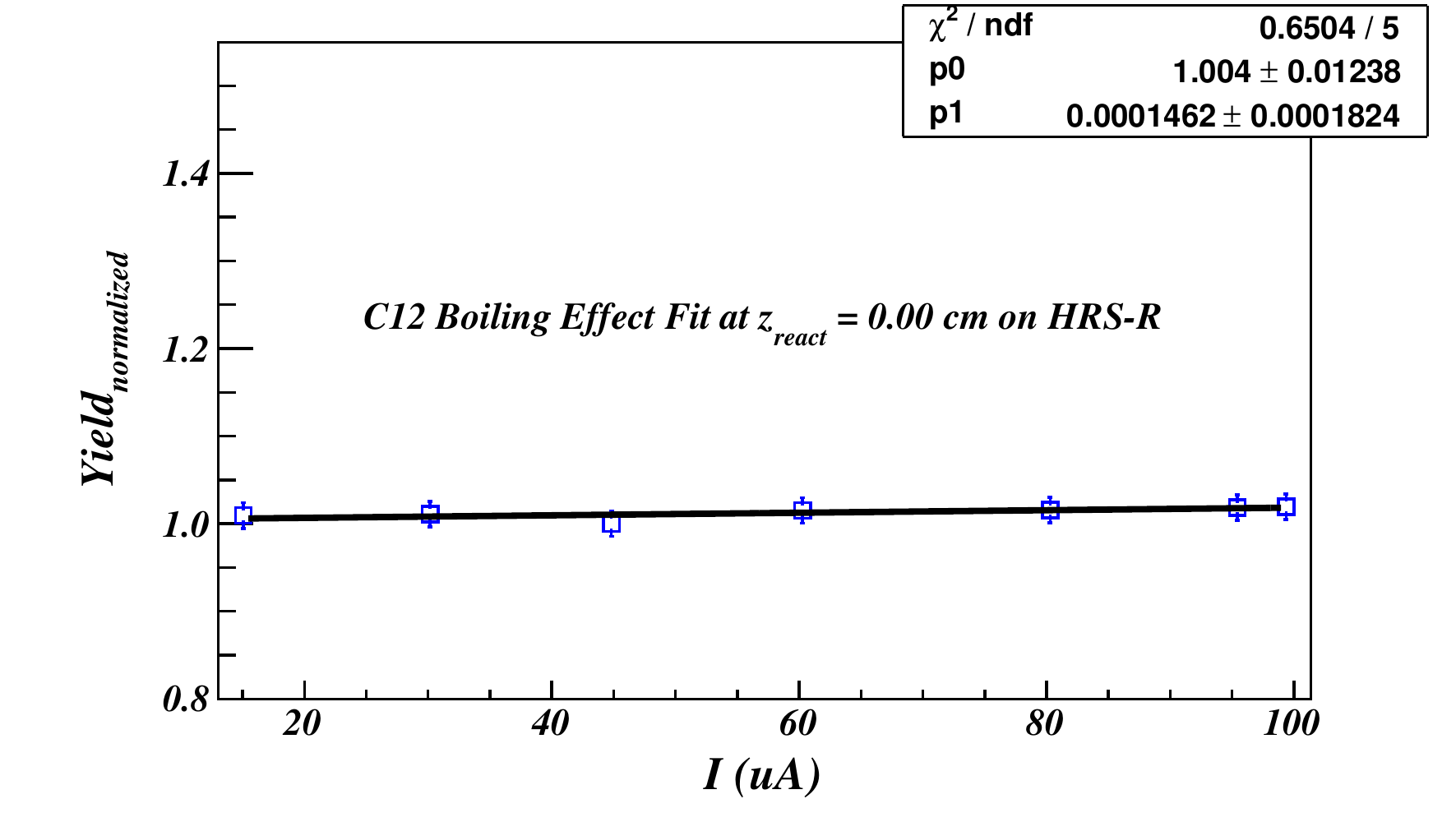} 
    }
    \caption[$\mathrm{^{12}C}$ boiling effect fitting]{\footnotesize{$\mathrm{^{12}C}$ boiling effect fitting. Since $\mathrm{^{12}C}$ should have very small boiling effect, this study is used to check any rate-dependence effect at different current settings. The yield values have been normalized by a common factor.}}
    \label{c12_boil_fit}
  \end{center}
\end{figure}
  During the E08-014, several boiling study runs were taken with different beam current on these cryo-targets, as well as on the $\mathrm{^{12}C}$ target which was used to check any rate-dependence effects.
  
  The experimental yield depends on the target density, the beam charge and the cross section of electron-nucleus scattering. While the average cross section shouldn't change for one kinematic setting, the yield normalized by the beam charge should be directly proportional to the target density:
\begin{equation}
  Y(I) = Y(0) + m\cdot I,
  \label{eq_yield_boiling}
\end{equation}  
where $Y(I)$, the yield for one run with the beam current $I$, is equal to the total number of experimental events after all necessary cuts divided by the the total accumulated charge. $Y(0)$ is the yield extrapolated to zero beam current and $m$ is the slop of $Y(I)$ on $I$. By fitting Eq.~\eqref{eq_yield_boiling} with the data of the boiling study runs, one can obtain the variation of the target density at different current, given as:
\begin{equation}
  \rho(I) = \rho(0) \cdot (1.0 + BF \cdot I /100),
  \label{eq_yield_rho}
\end{equation}
where $BF=Y(0)/m$ is the boiling factor of the target.

\begin{figure}[!h]
  \begin{center}
    \subfloat[$\mathrm{^{2}H}$ on HRS-L]{
      \includegraphics[type=pdf,ext=.pdf,read=.pdf,width=0.45\textwidth]{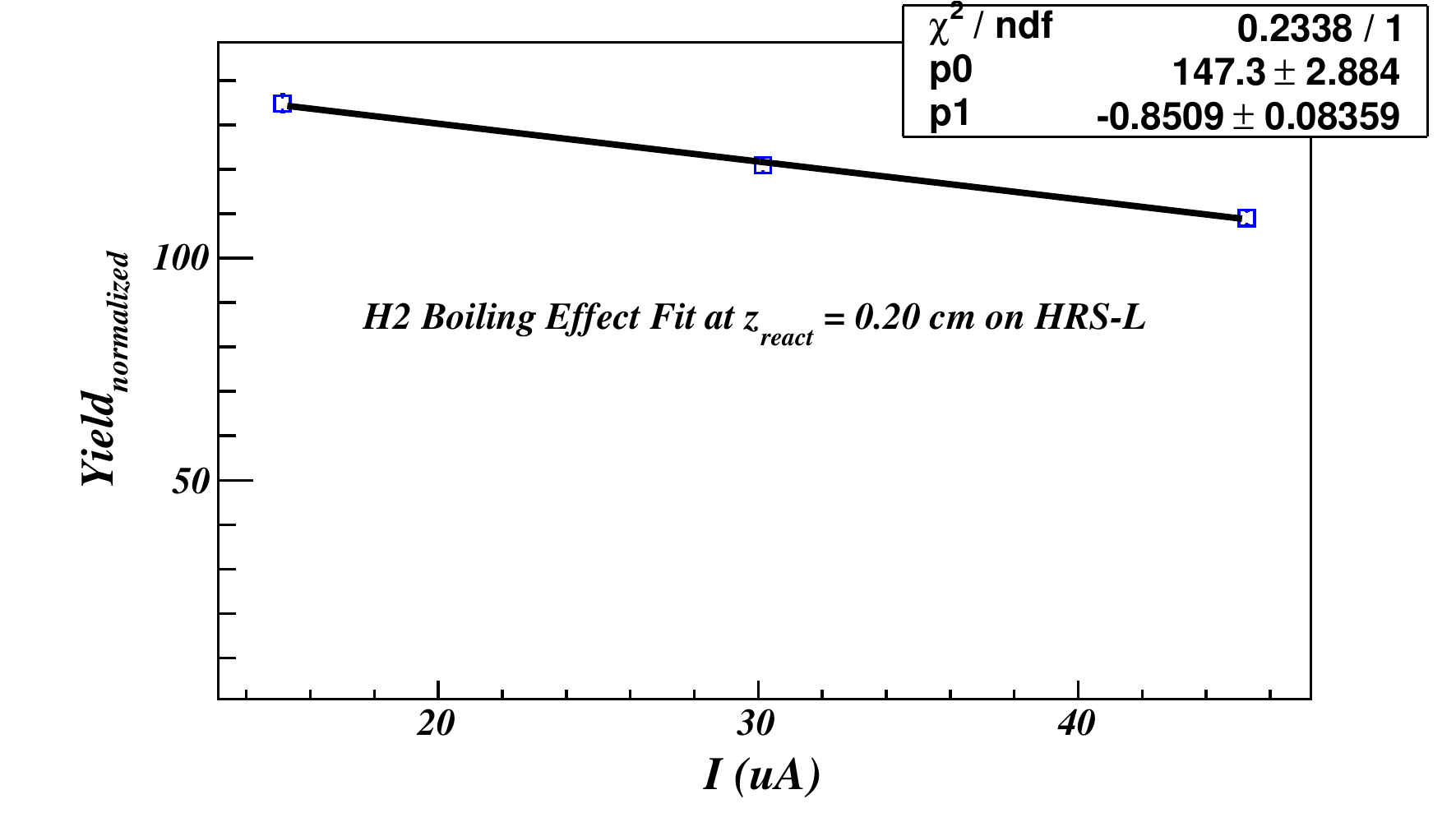} 
    }
    \hfill
    \subfloat[$\mathrm{^{2}H}$ on HRS-R]{
      \includegraphics[type=pdf,ext=.pdf,read=.pdf,width=0.45\textwidth]{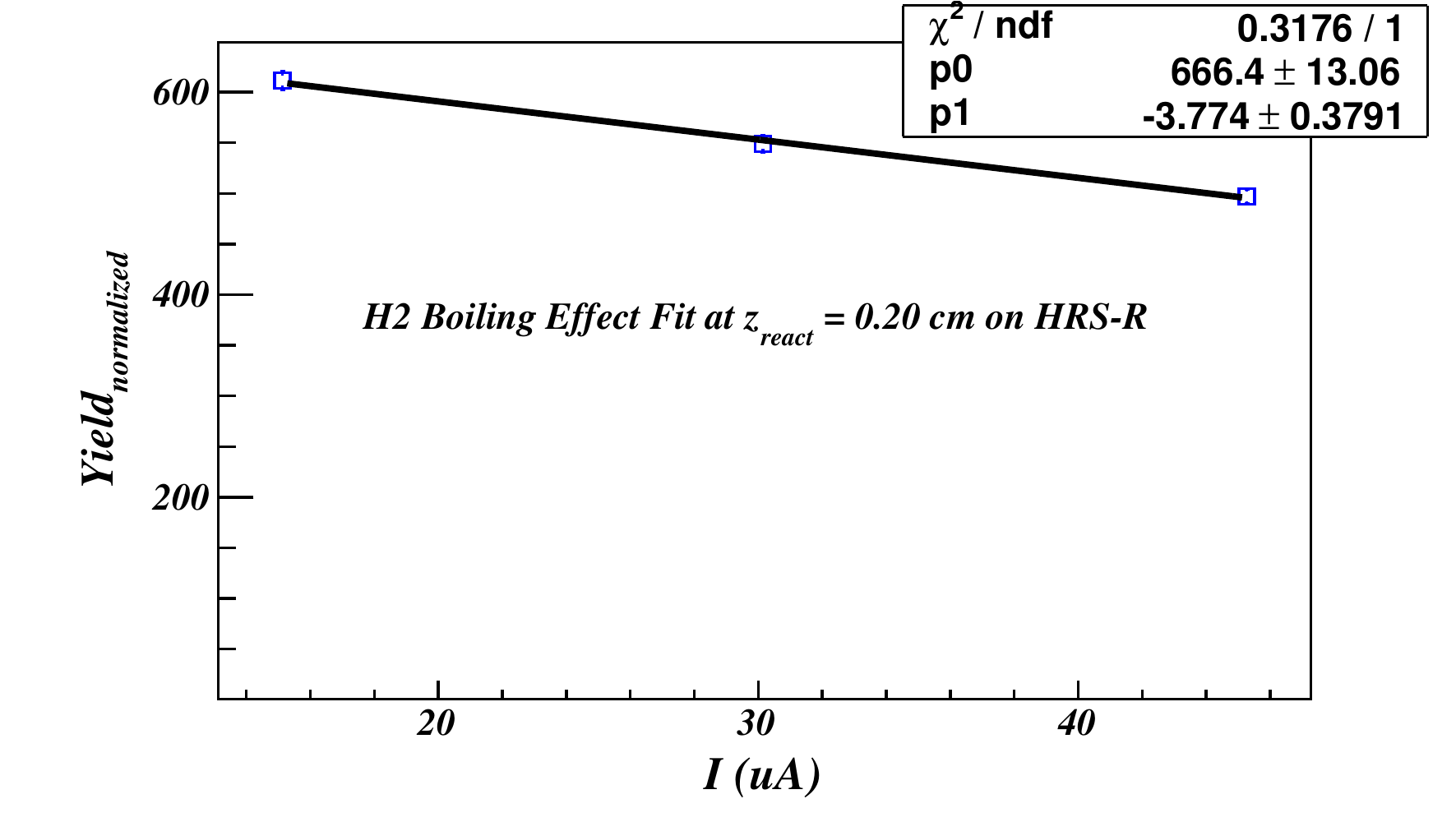} 
    }
\\
    \subfloat[$\mathrm{^{3}He}$ on HRS-L]{
      \includegraphics[type=pdf,ext=.pdf,read=.pdf,width=0.45\textwidth]{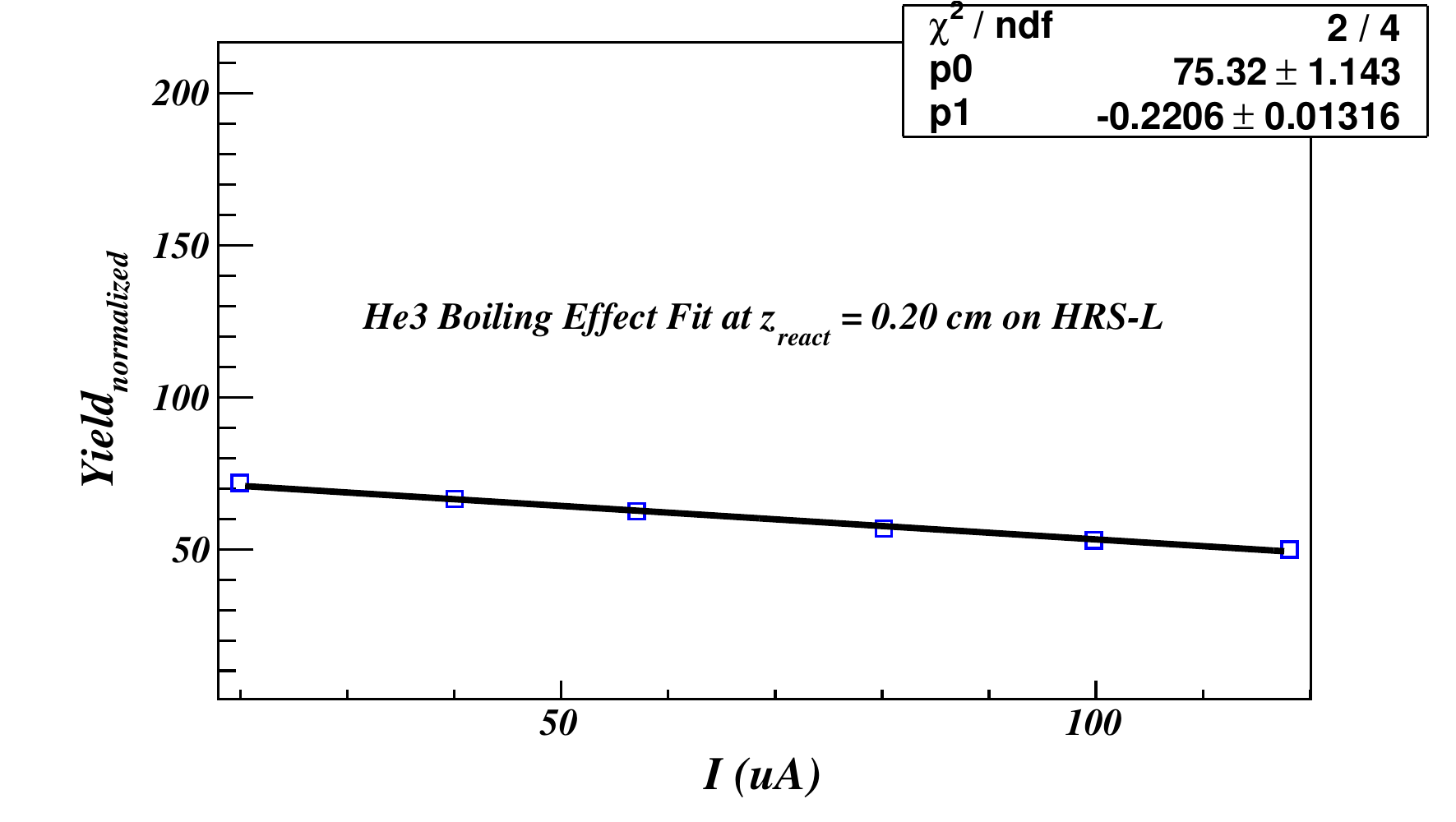} 
    }
    \hfill
    \subfloat[$\mathrm{^{3}He}$ on HRS-R]{
      \includegraphics[type=pdf,ext=.pdf,read=.pdf,width=0.45\textwidth]{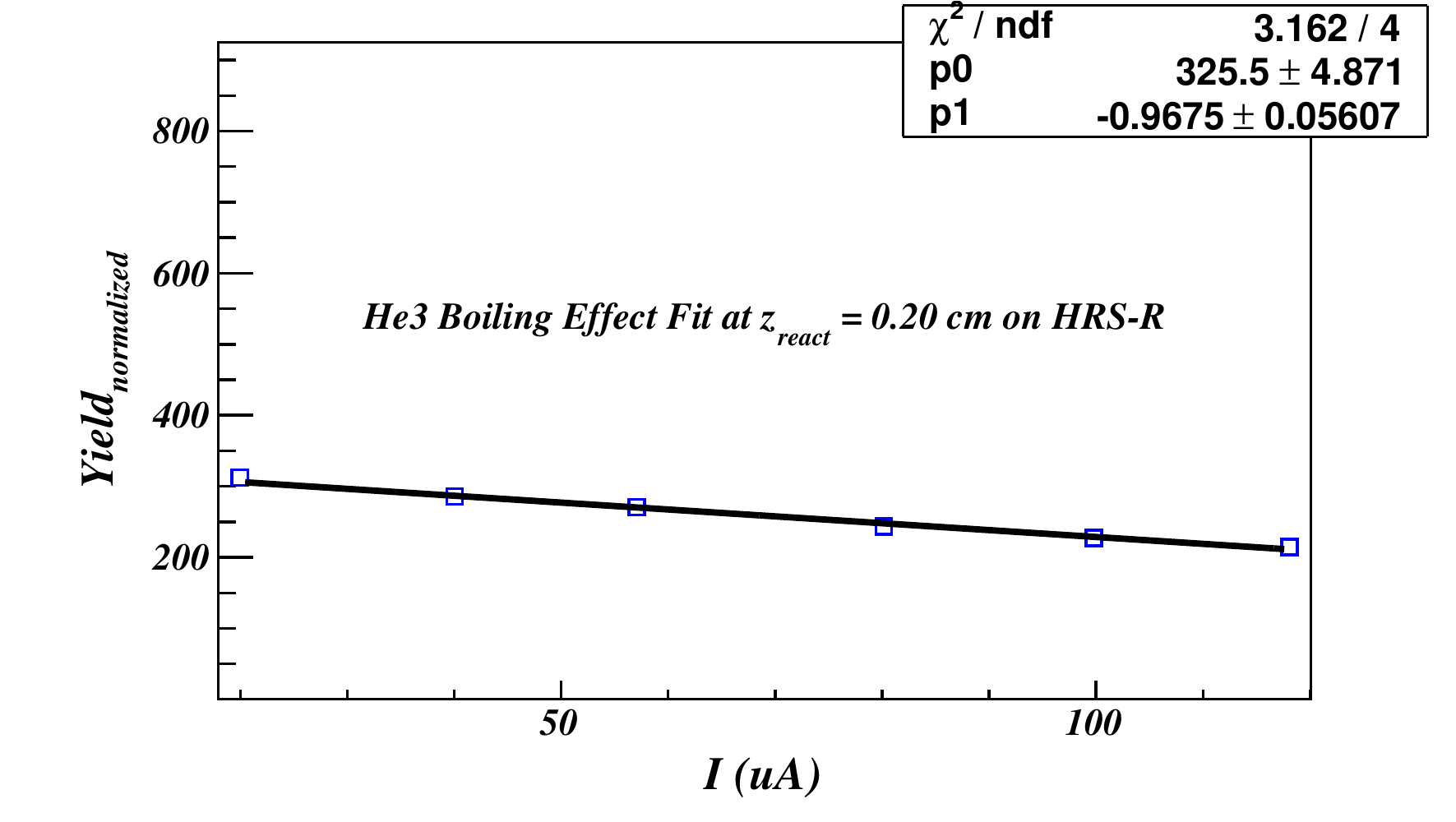} 
    }
\\
    \subfloat[$\mathrm{^{4}He}$ on HRS-L]{
      \includegraphics[type=pdf,ext=.pdf,read=.pdf,width=0.45\textwidth]{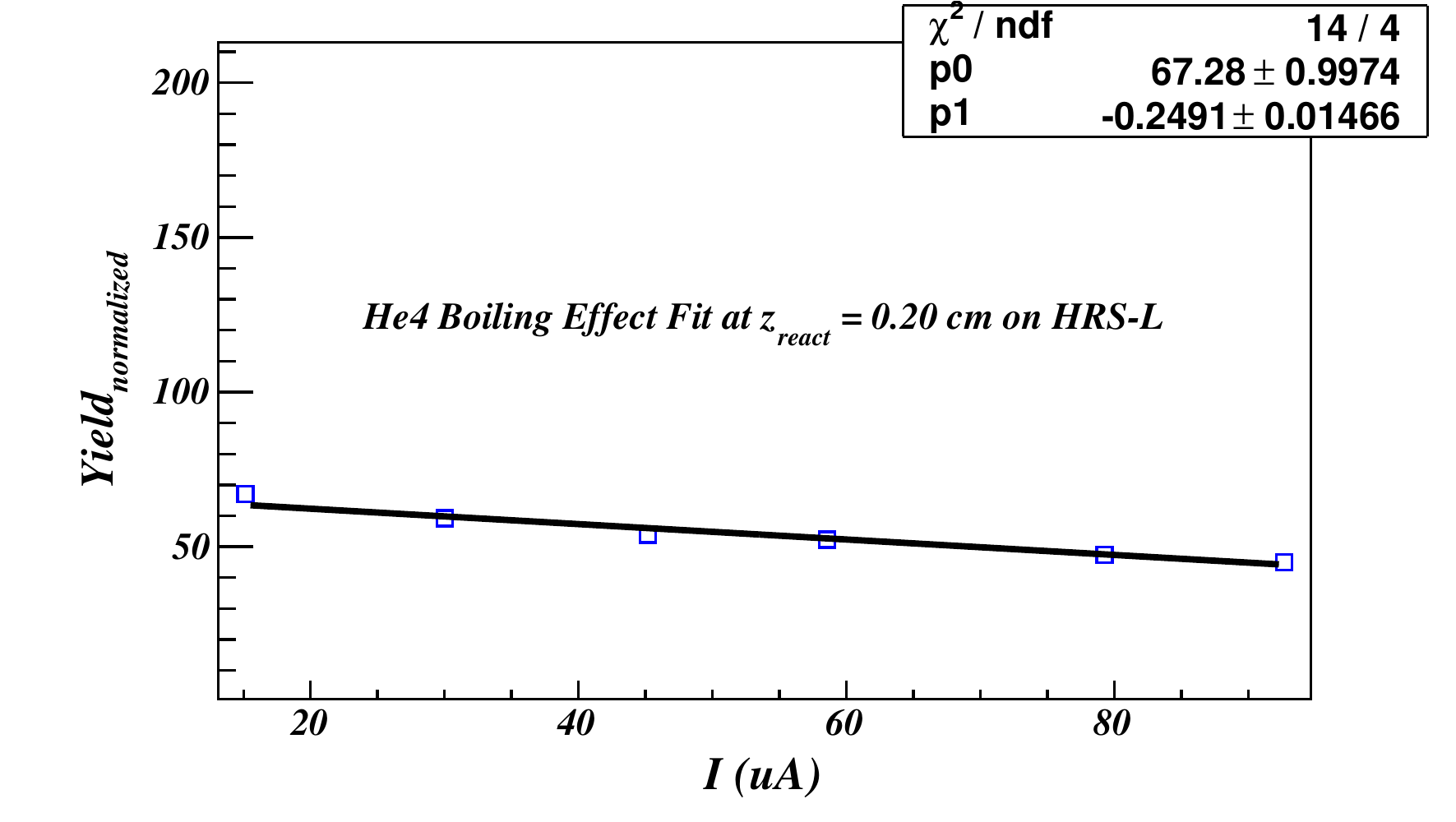} 
    }
    \hfill
    \subfloat[$\mathrm{^{4}He}$ on HRS-R]{
      \includegraphics[type=pdf,ext=.pdf,read=.pdf,width=0.45\textwidth]{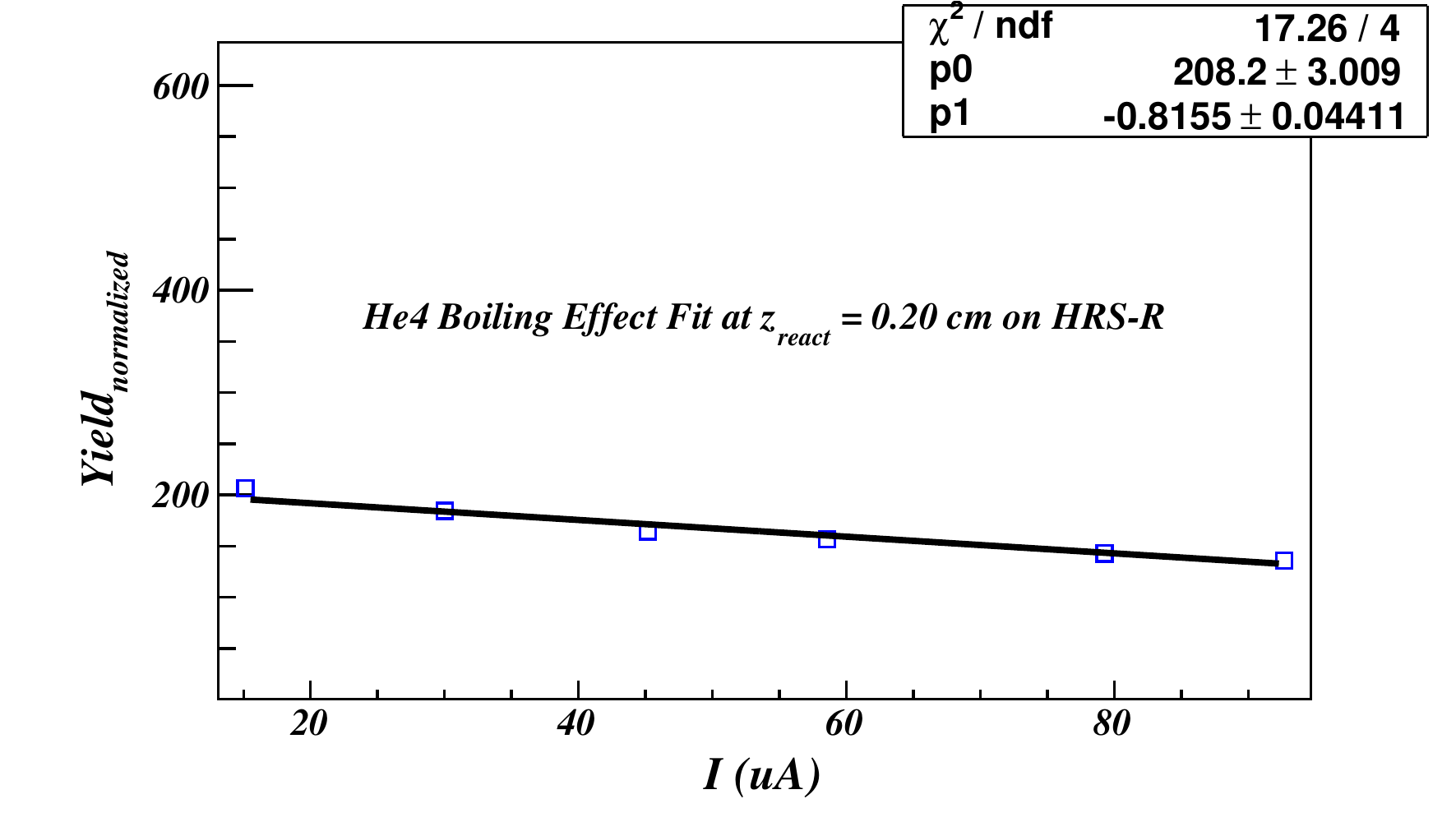} 
    }
    \caption[Cryo-targets boiling effect fitting]{\footnotesize{Cryo-targets boiling effect fitting. They are examples near the center of the targets. Each target was divided into 60 bins along the target cell, where the boiling factor was individually fitted. The yield values have been normalized by a common factor.}}
    \label{cryo_boil_fit}
  \end{center}
\end{figure}
\begin{figure}[!h]
  \begin{center}
    \subfloat[$\mathrm{^{2}H2}$]{
      \includegraphics[type=pdf,ext=.pdf,read=.pdf,width=0.7\textwidth]{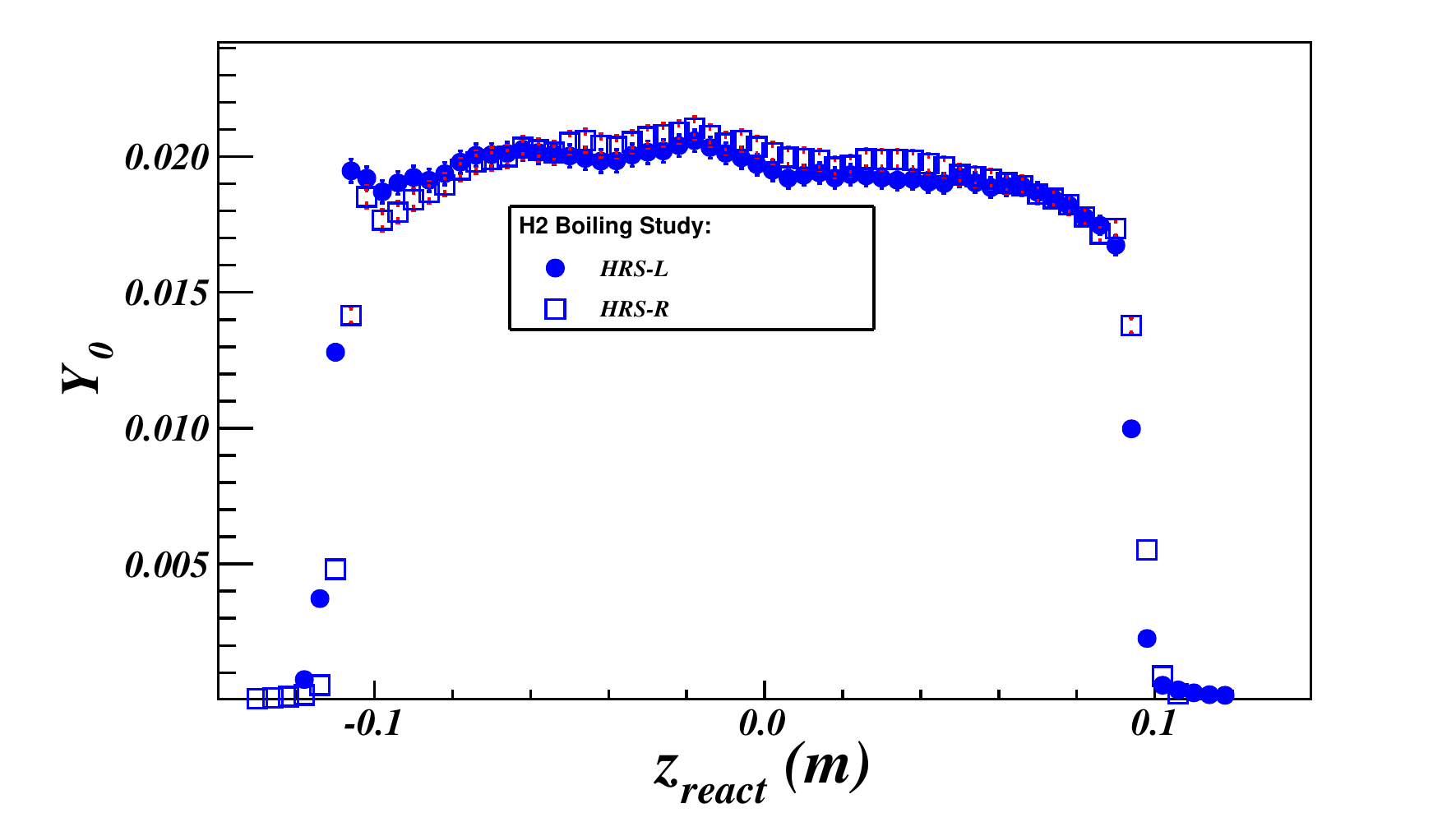} 
    }
    \hfill
    \subfloat[$\mathrm{^{3}He}$]{
      \includegraphics[type=pdf,ext=.pdf,read=.pdf,width=0.7\textwidth]{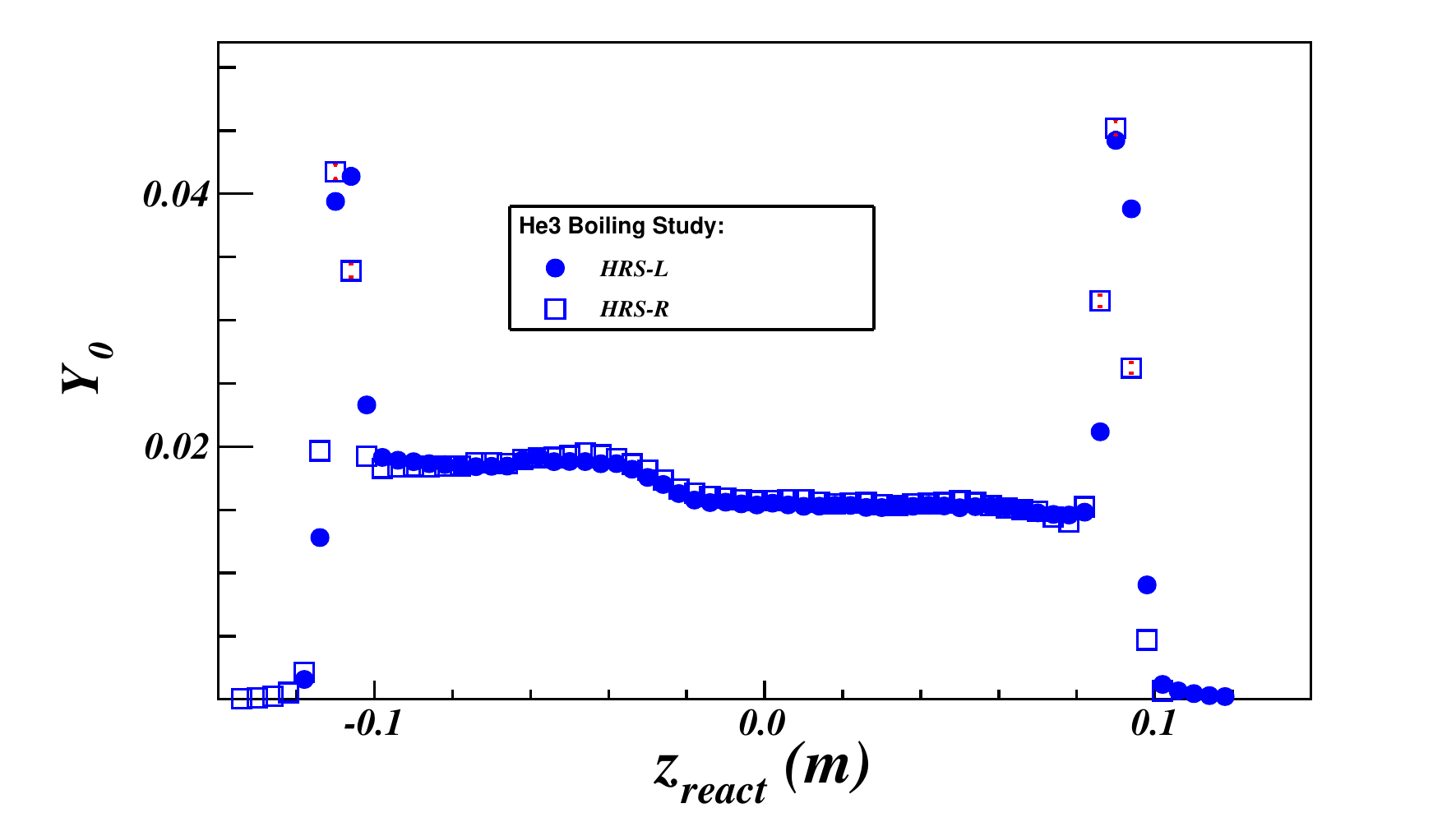} 
    }
     \hfill
    \subfloat[$\mathrm{^{4}He}$]{
      \includegraphics[type=pdf,ext=.pdf,read=.pdf,width=0.7\textwidth]{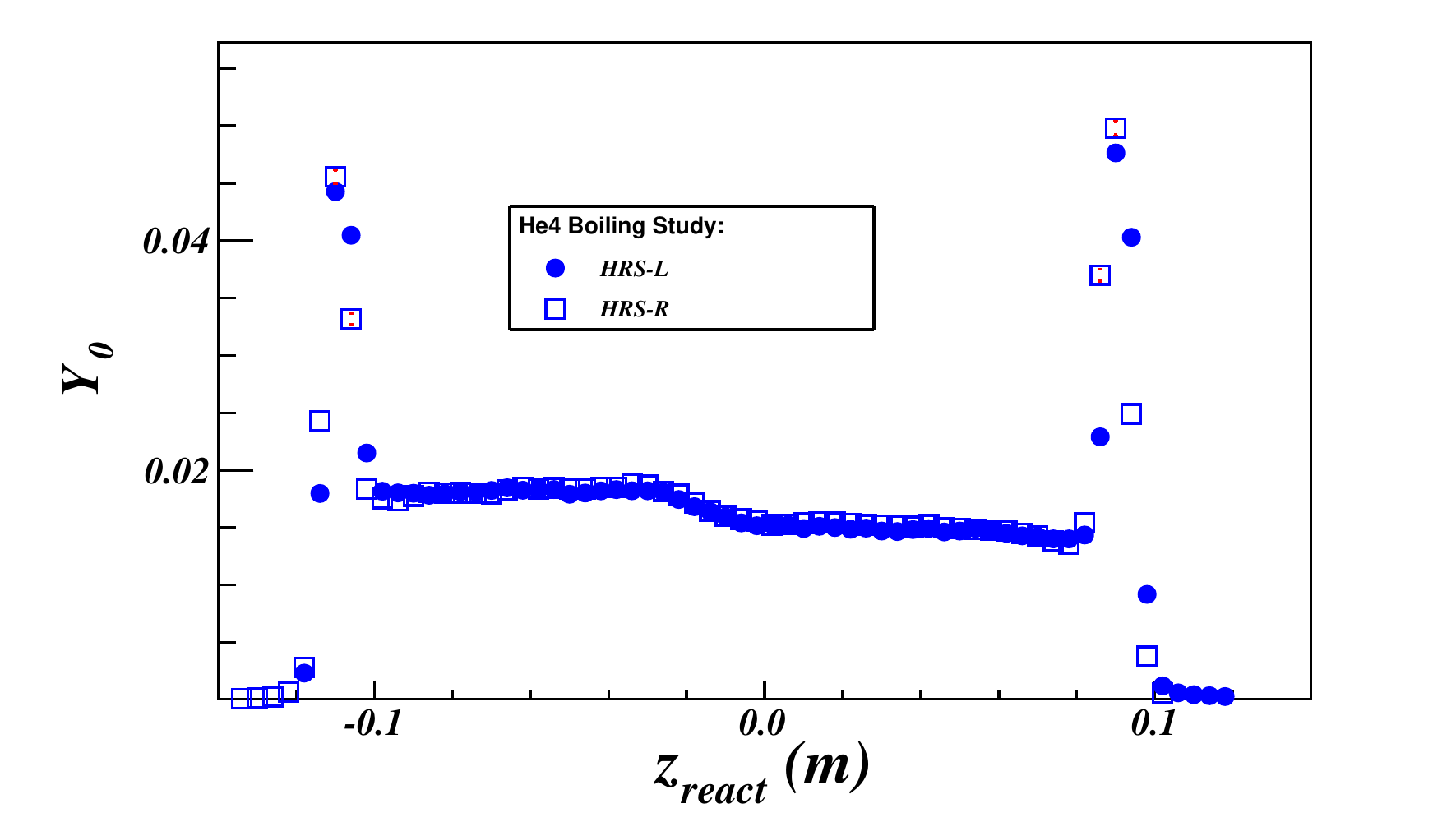} 
    }
    \caption[Cryo-target density profiles from the boiling study]{\footnotesize{Cryo-target density profiles from the boiling study. The results from both HRSs agree with each other for each target, and the peaks denote the contributions from the endcaps of the target cell.}}
    \label{cryo_boil_y0}
  \end{center}
\end{figure}

 Shown in Fig.~\ref{c12_boil_fit}, the fitting result of $\mathrm{^{12}C}$ indicates that for a fixed target density, the yield does not change at different current. For cryo-target, the data was binned in $z_{react}$ which was divided into 60 bins. In each bin, the yield was calculated and one can fit the boiling factor by the formula:
 \begin{equation}
  Y(I, z_{react}^{i}) = Y(0, z_{react}^{i}) + m(z_{react}^{i})\cdot I,,\quad where~i=1,\cdots,60,
  \label{eq_yield_rho_zbin}
\end{equation}
which gives the variation of the density in each bin:
 \begin{equation}
  \rho(I, z_{react}^{i}) = \rho(0, z_{react}^{i}) \cdot (1.0 + BF(z_{react}^{i}) \cdot I /100),
  \label{eq_rho_zbin}
 \end{equation}
where,
 \begin{equation}
   BF(z_{react}^{i})=\frac{Y(0, z_{react}^{i})}{m(z_{react}^{i})}. \\
     \label{eq_bf_zbin}
 \end{equation}

\begin{figure}[]
  \begin{center}
    \subfloat[$\mathrm{^{2}H}$]{
      \includegraphics[type=pdf,ext=.pdf,read=.pdf,width=0.7\textwidth]{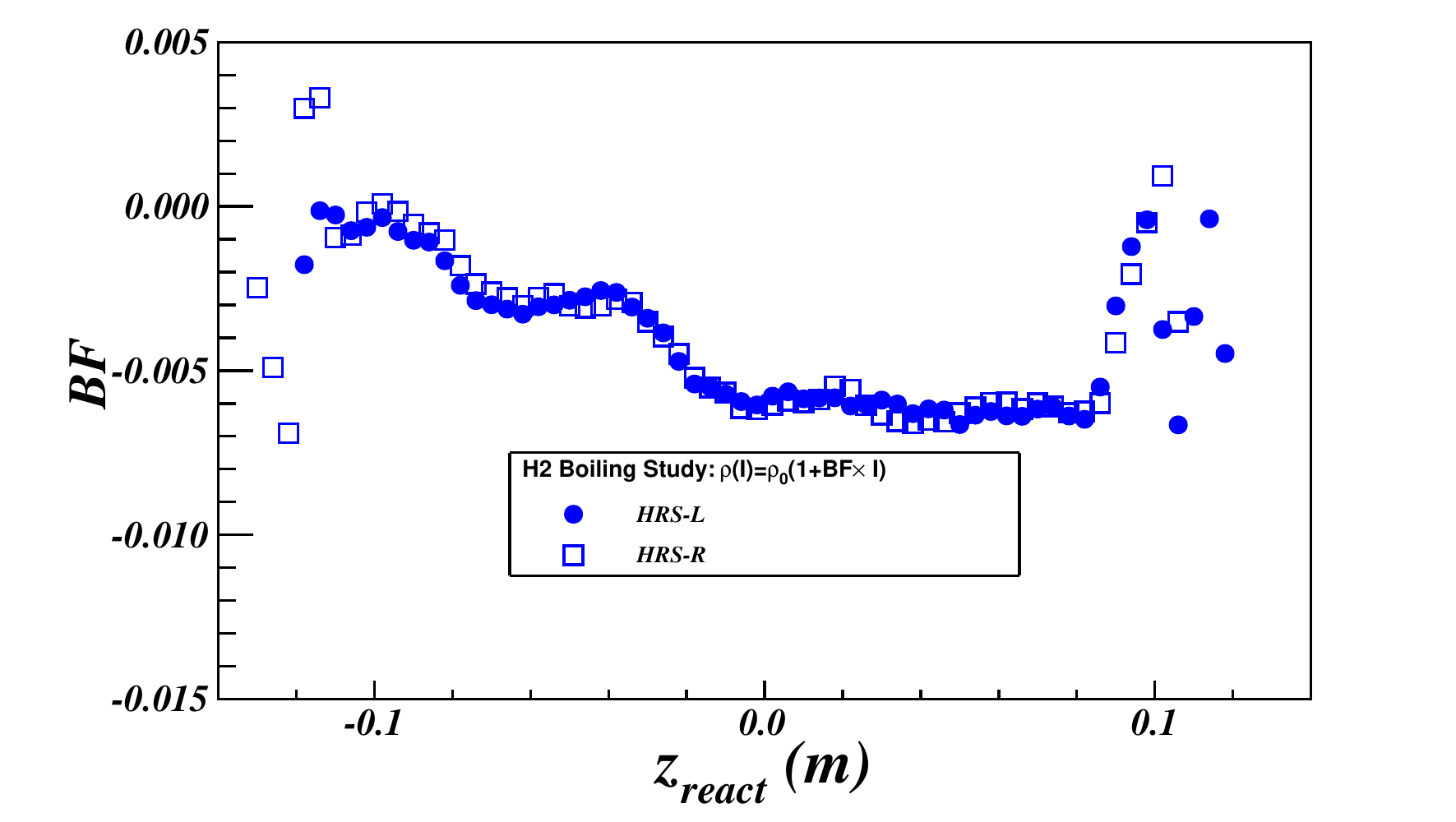} 
    }
    \hfill
    \subfloat[$\mathrm{^{3}He}$]{
      \includegraphics[type=pdf,ext=.pdf,read=.pdf,width=0.7\textwidth]{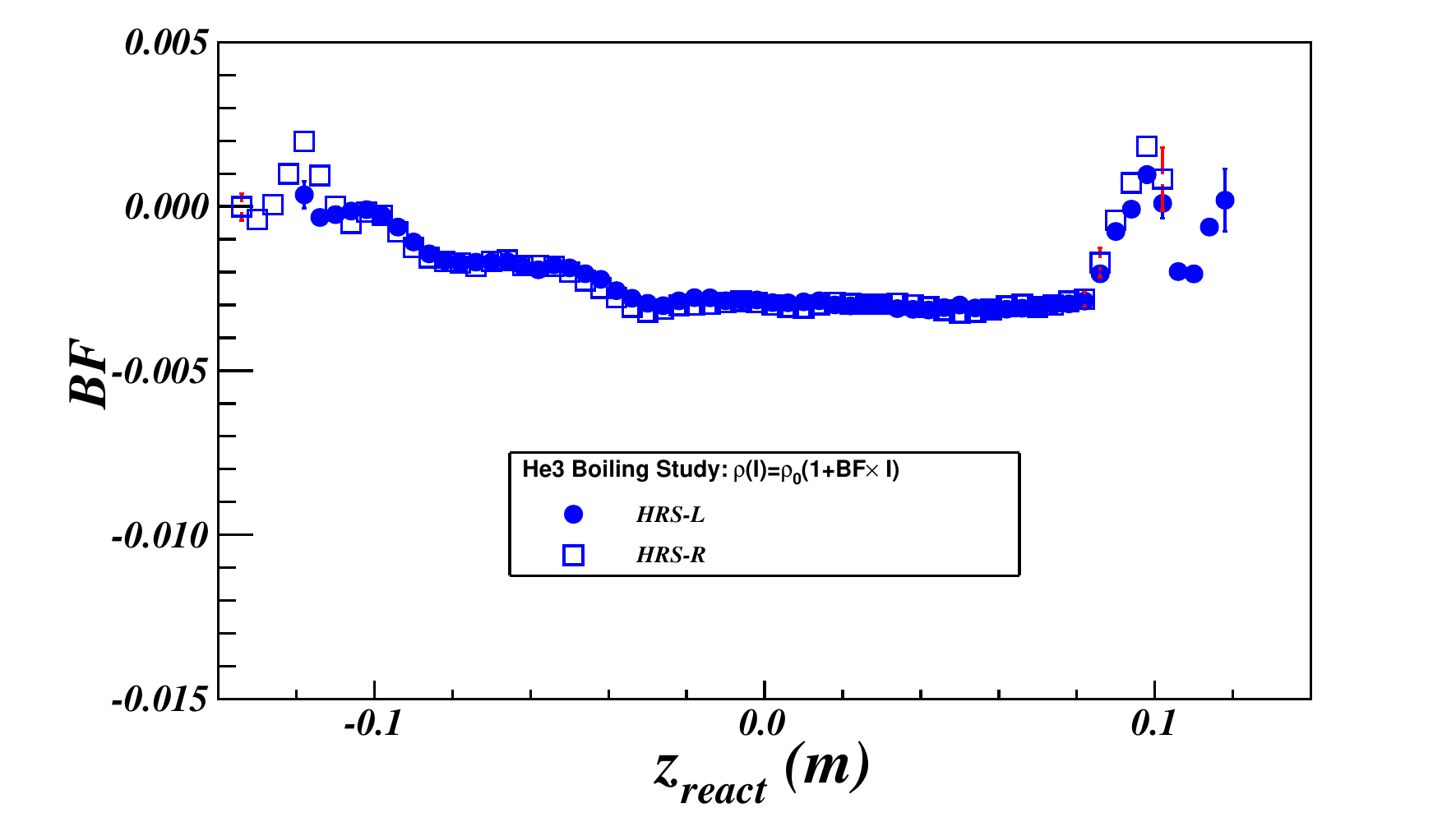} 
    }
     \hfill
    \subfloat[$\mathrm{^{4}He}$]{
      \includegraphics[type=pdf,ext=.pdf,read=.pdf,width=0.7\textwidth]{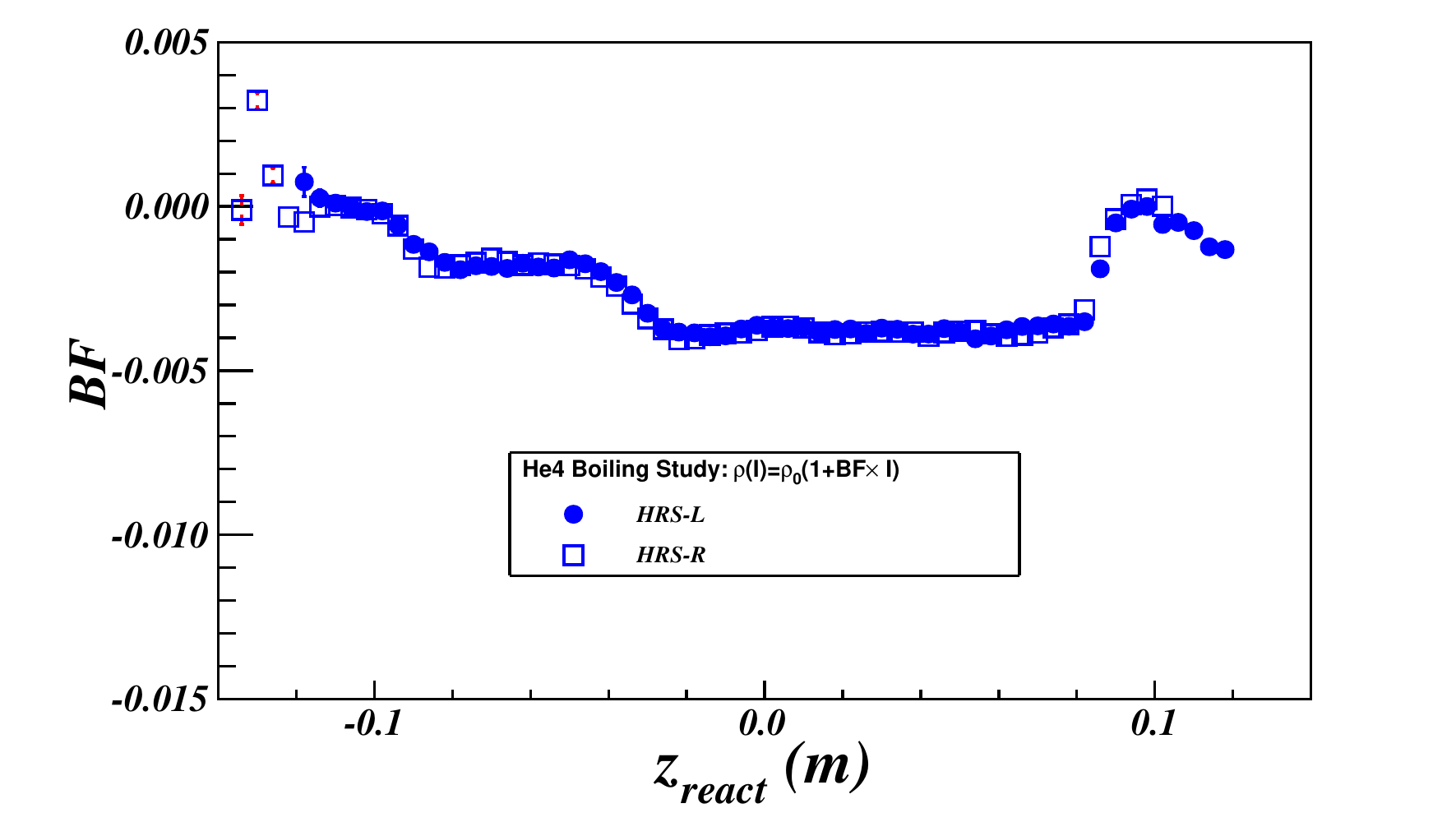} 
    }
    \caption[Cryo-target boiling factor distribution]{\footnotesize{Cryo-target boiling factor distribution. Each plot clearly shows that the boiling effect varies along the target cells. The boiling factors at the endcaps are reasonably close to zero. The studies from both HRSs give consistent results. The yield values had been normalized by a common factor.}}
    \label{cryo_boil_fact}
  \end{center}
\end{figure}

As examples, Fig.~\ref{cryo_boil_fit} shows the fitting results of boiling factors at the center of $z_{react}$ for three cryo-targets, where the curves are well fitted by linear functions. The curve of normalized $Y(0, z_{react}^{i})$ denotes the target density profile along the cell, as shown in Fig.~\ref{cryo_boil_y0}, where the peaks of endcaps can be clearly seen. The distribution of $BF(z_{react}^{i})$ for each target, given in Fig.~\ref{cryo_boil_fact}, shows the boiling effects at different $z_{react}$, and demonstrates that the non-uniform cryo-target densities were mainly caused by the highly localized boiling effects. In the plot, the values of $z_{react}$ at the positions of endcaps are close to zero which agree with the fact that the density of aluminium walls shouldn't change with beam current. Results from both HRSs were compared and found out to agree nicely with each other. 

\section{Extracting Density Distributions}
  From Eq.~\eqref{eq_yield_rho_zbin}, the target density profile can be obtained by extracting the distribution of $Y(0)$ during the boiling study. In this section, a different method is applied to extract the density distribution by with the experimental data and simulation data. 
  
  Since $z_{react}$ is along the incoming beam direction so as the orientation of the target cell, the $z_{react}$ distribution in the experimental data, $z_{react}^{EX}$, gives the distribution of yields in one current setting. Meanwhile, the yield for one $z_{react}^{EX}$ value is proportional to the density of the target in this location, so one expects to study the density distribution of the target with the $z_{react}^{EX}$ distribution. However, $z_{react}^{EX}$ should also contain the acceptance effect of the HRS and the cross section weighting effect. One can use the simulation data generated by SAMC which simulates the acceptance effect of HRSs, and plot the simulated $z_{react}$ distribution, $z_{react}^{MC}$. One can further weight $z_{react}^{MC}$ by the cross section values calculated from XEMC. When the target density distribution in the simulation data is uniform, $z_{react}^{MC}$ only carries the acceptance effect and the cross section effect. By plotting the histograms of $z_{react}^{EX}$ and $z_{react}^{MC}$ with the same range and bin-size, one takes to ratio of two histogram, which leads to a clean relative density distribution of the target at the current setting.
  
\begin{figure}[!ht]
  \begin{center}
    \subfloat[$\mathrm{^{2}H}$]{
      \includegraphics[type=pdf,ext=.pdf,read=.pdf,width=0.7\textwidth]{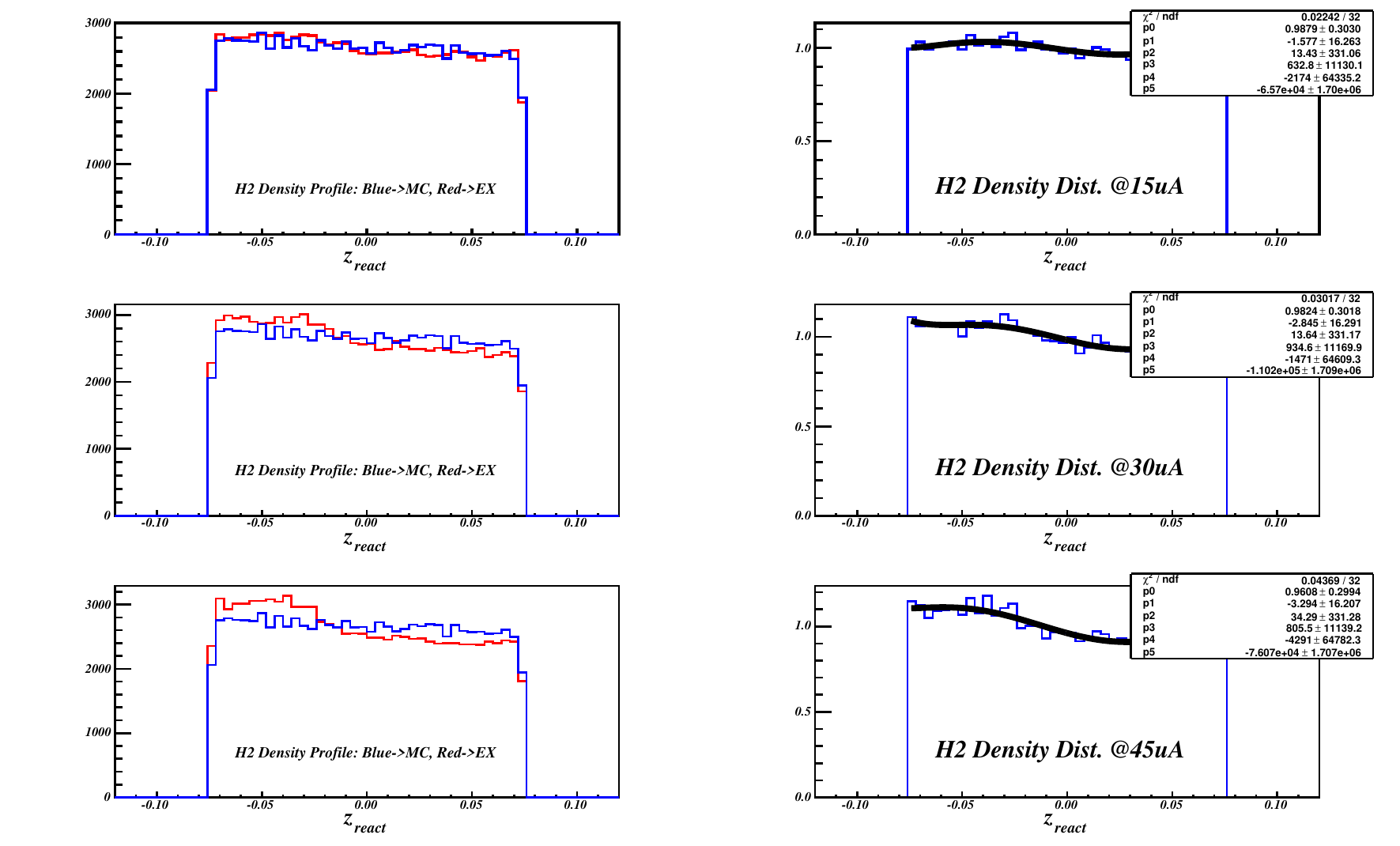} 
    }
    \hfill
    \subfloat[$\mathrm{^{3}He}$]{
      \includegraphics[type=pdf,ext=.pdf,read=.pdf,width=0.7\textwidth]{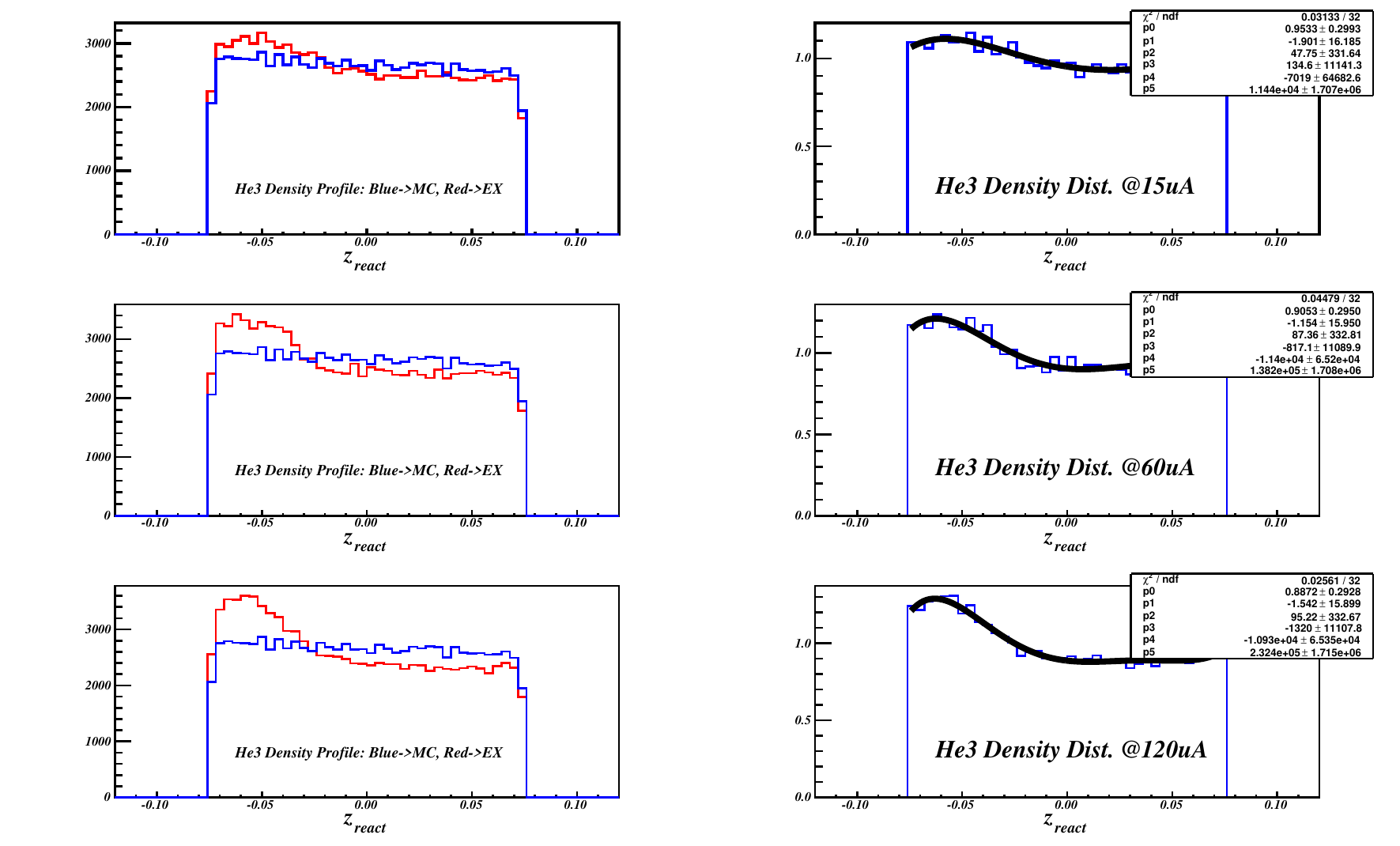} 
    }
     \hfill
    \subfloat[$\mathrm{^{4}He}$]{
      \includegraphics[type=pdf,ext=.pdf,read=.pdf,width=0.7\textwidth]{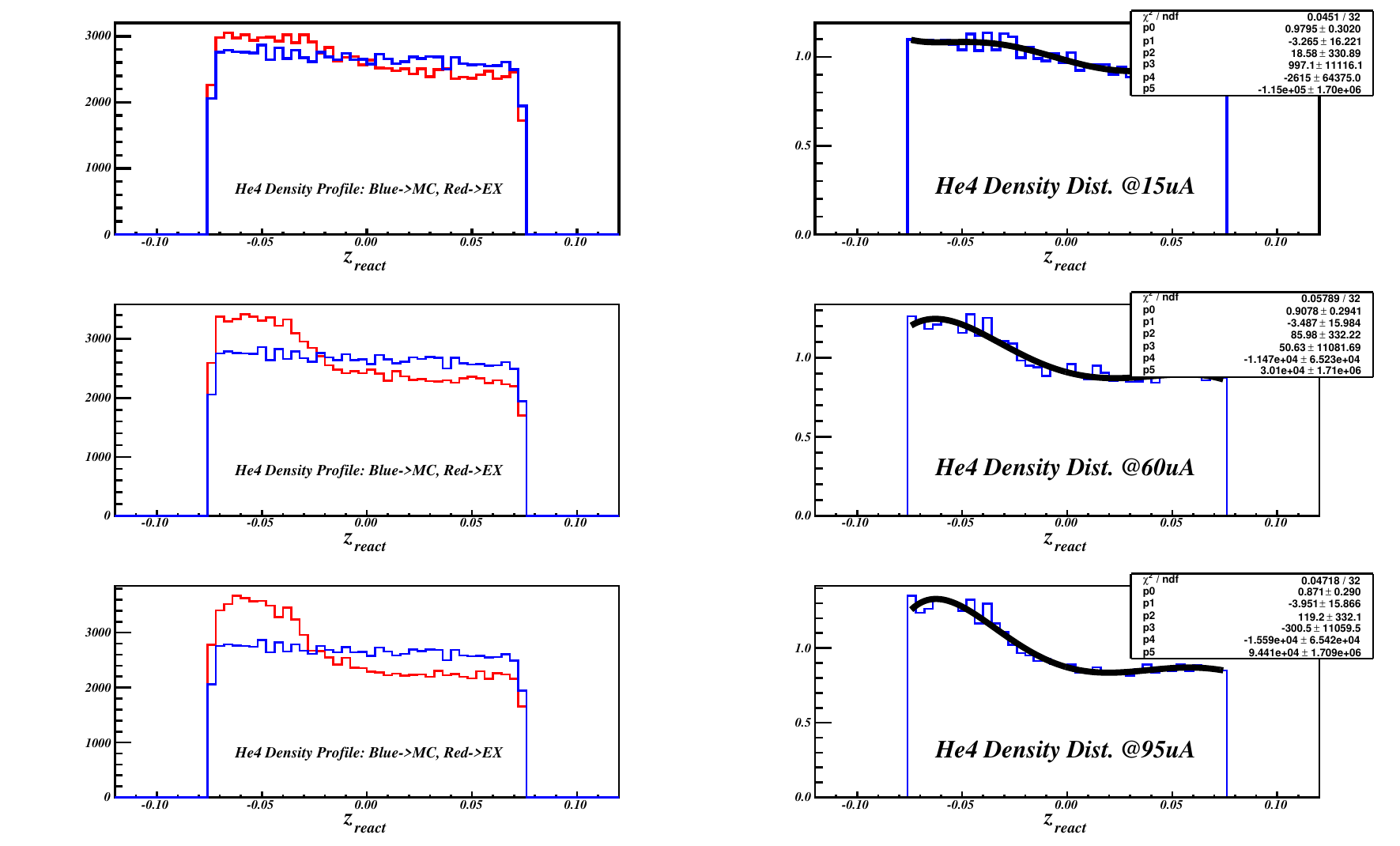} 
    }
    \caption[Cryo-target density distributions extracted from data]{\footnotesize{Cryo-target density distributions extracted from data. The density distribution was extracted by taking the histogram ratio of $z_{react}$ from experimental data (red lines in plots on the left panel) and from simulation data with flat density distribution (blue lines in plots on the left panel). For each target, the density distributions at the minimum, middle and maximum currents were individually extracted (on right panel). The current settings are given in the plots.}}
    \label{cryo_den_fit}
  \end{center}
\end{figure}

   The plots on the left hand side of Fig.~\ref{cryo_den_fit} show the distribution of $z_{react}^{EX}$ and $z_{react}^{MC}$ at three different current settings for each target, while the plots on the right hand side give the fitting results of the relative density distributions. A polynomial function is used for each fitting process. 

\begin{figure}[!ht]
  \begin{center}
    \subfloat[$\mathrm{^{2}H}$ at I=30 uA]{
      \includegraphics[type=pdf,ext=.pdf,read=.pdf,width=0.45\textwidth]{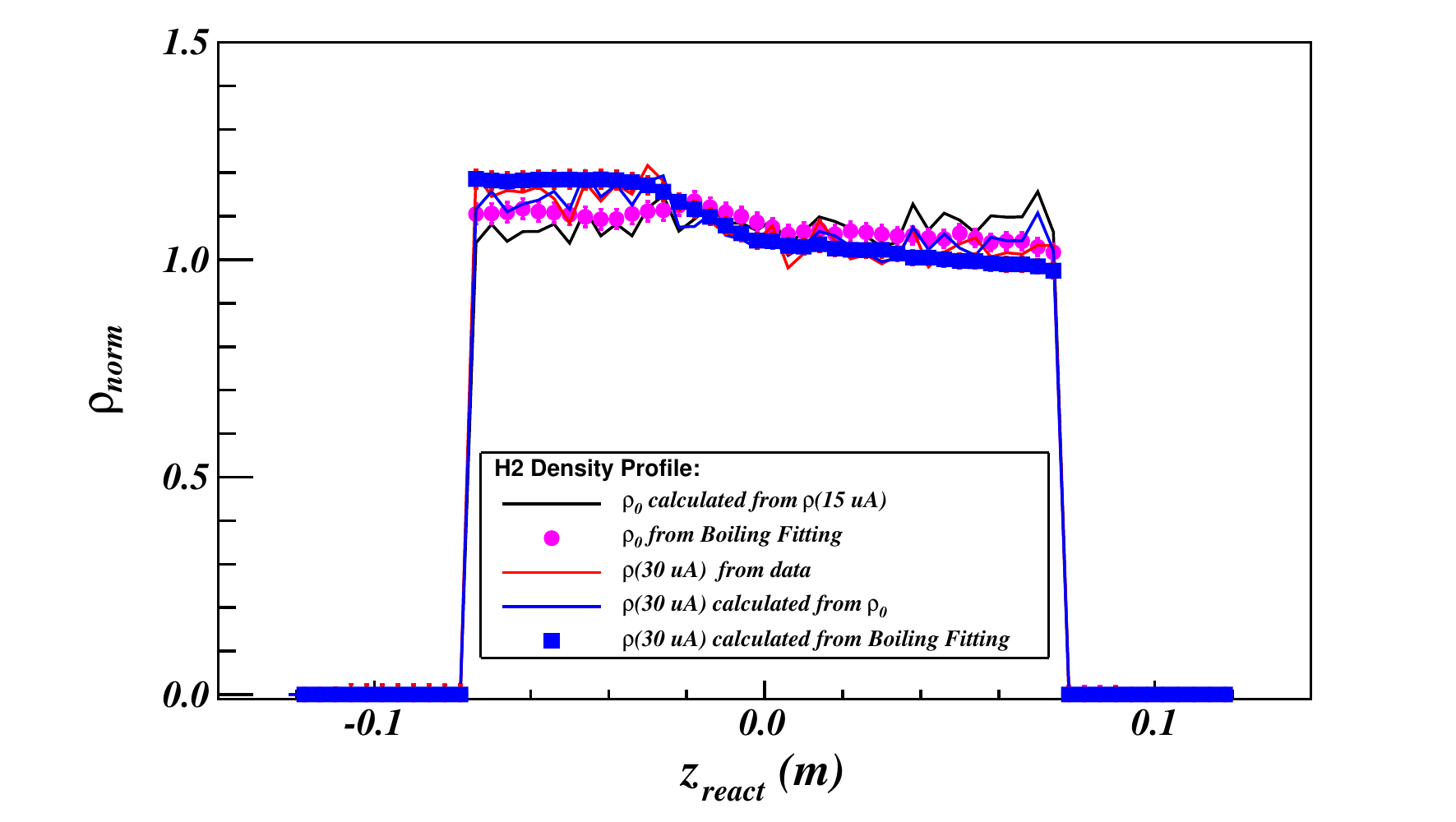} 
    }
    \hfill
    \subfloat[$\mathrm{^{2}H}$ at I=45 uA]{
      \includegraphics[type=pdf,ext=.pdf,read=.pdf,width=0.45\textwidth]{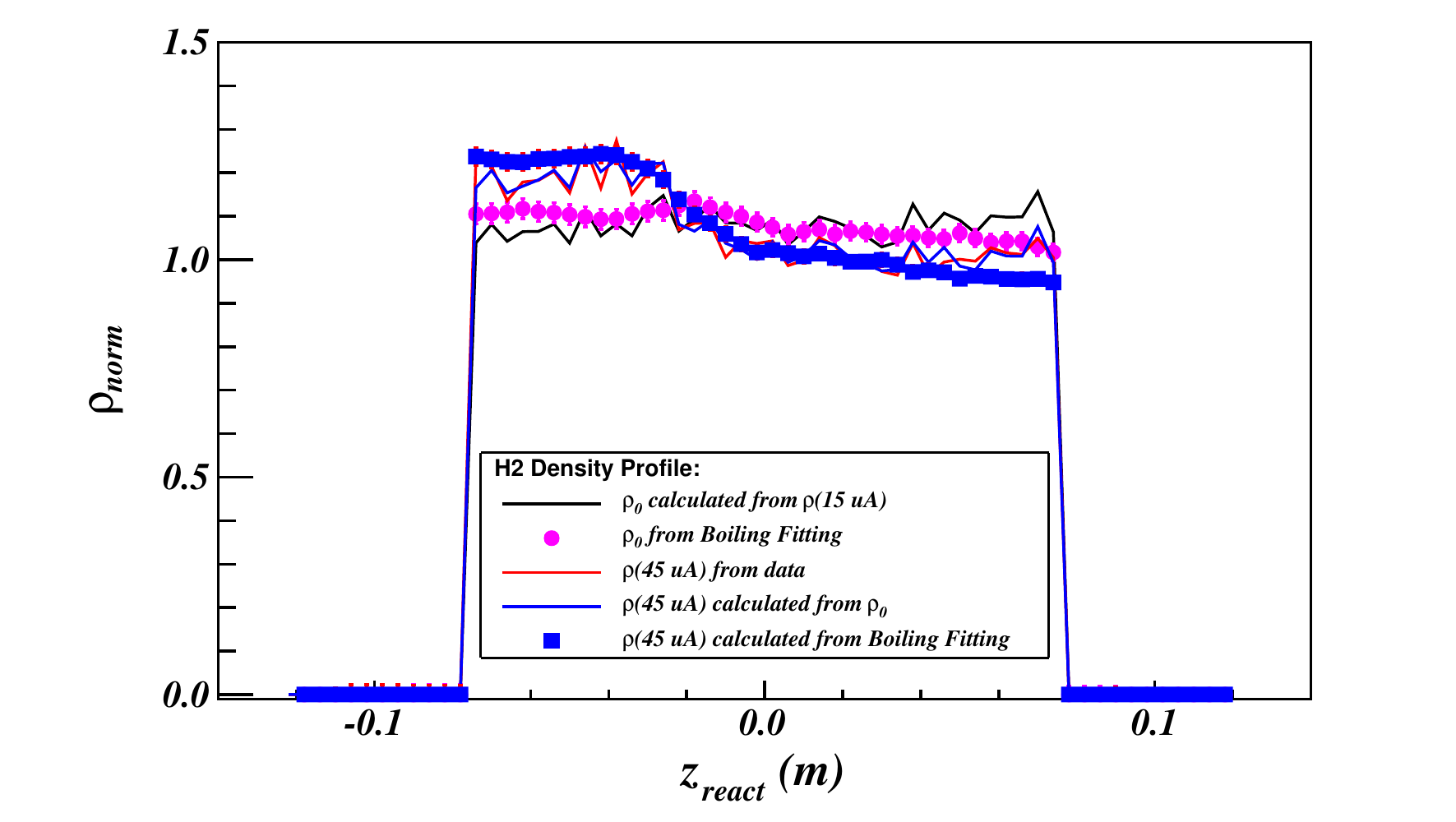} 
    }
\\
    \subfloat[$\mathrm{^{3}He}$ at I=60 uA]{
      \includegraphics[type=pdf,ext=.pdf,read=.pdf,width=0.45\textwidth]{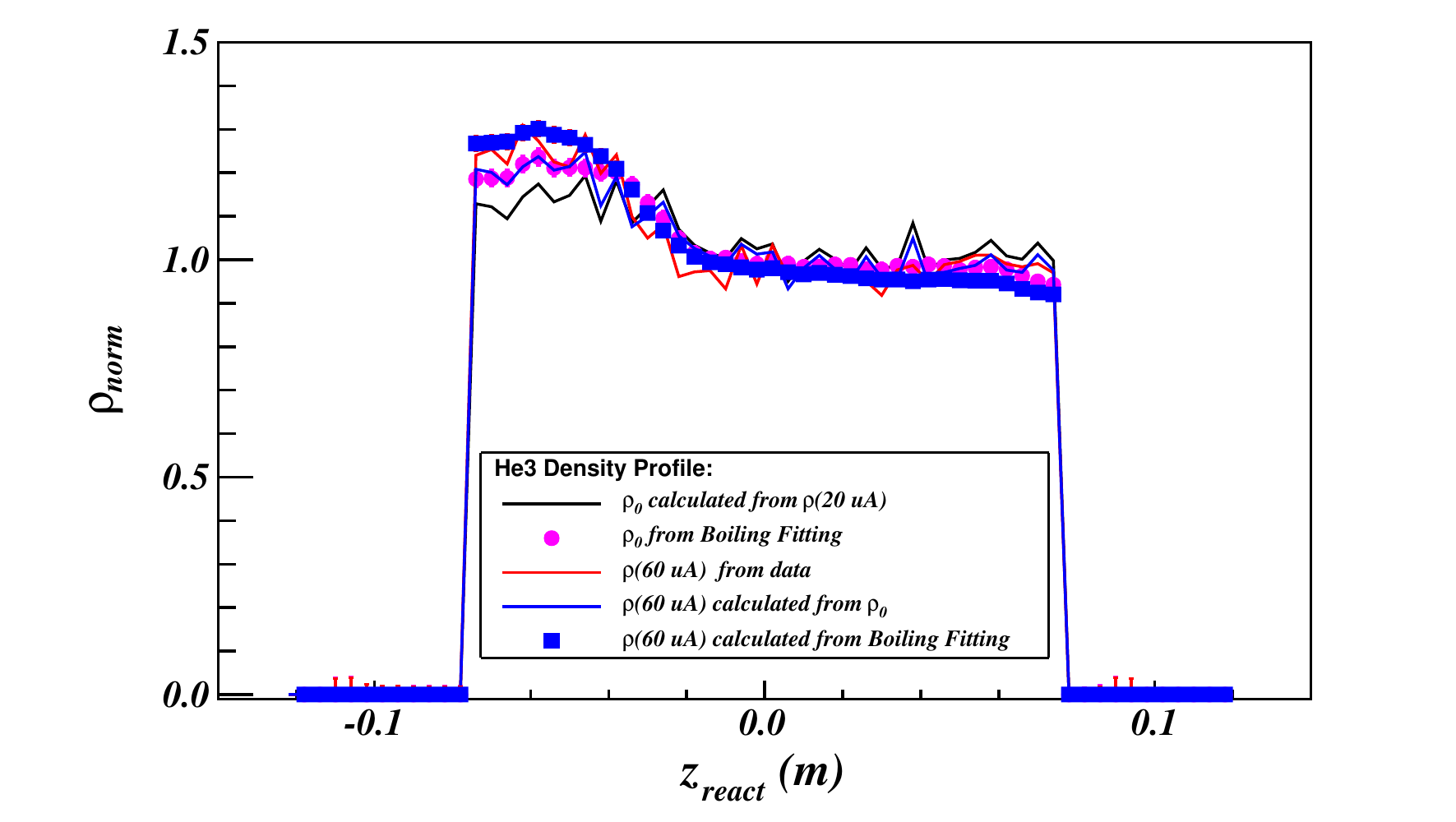} 
    }
    \hfill
    \subfloat[$\mathrm{^{3}He}$ at I=120 uA]{
      \includegraphics[type=pdf,ext=.pdf,read=.pdf,width=0.45\textwidth]{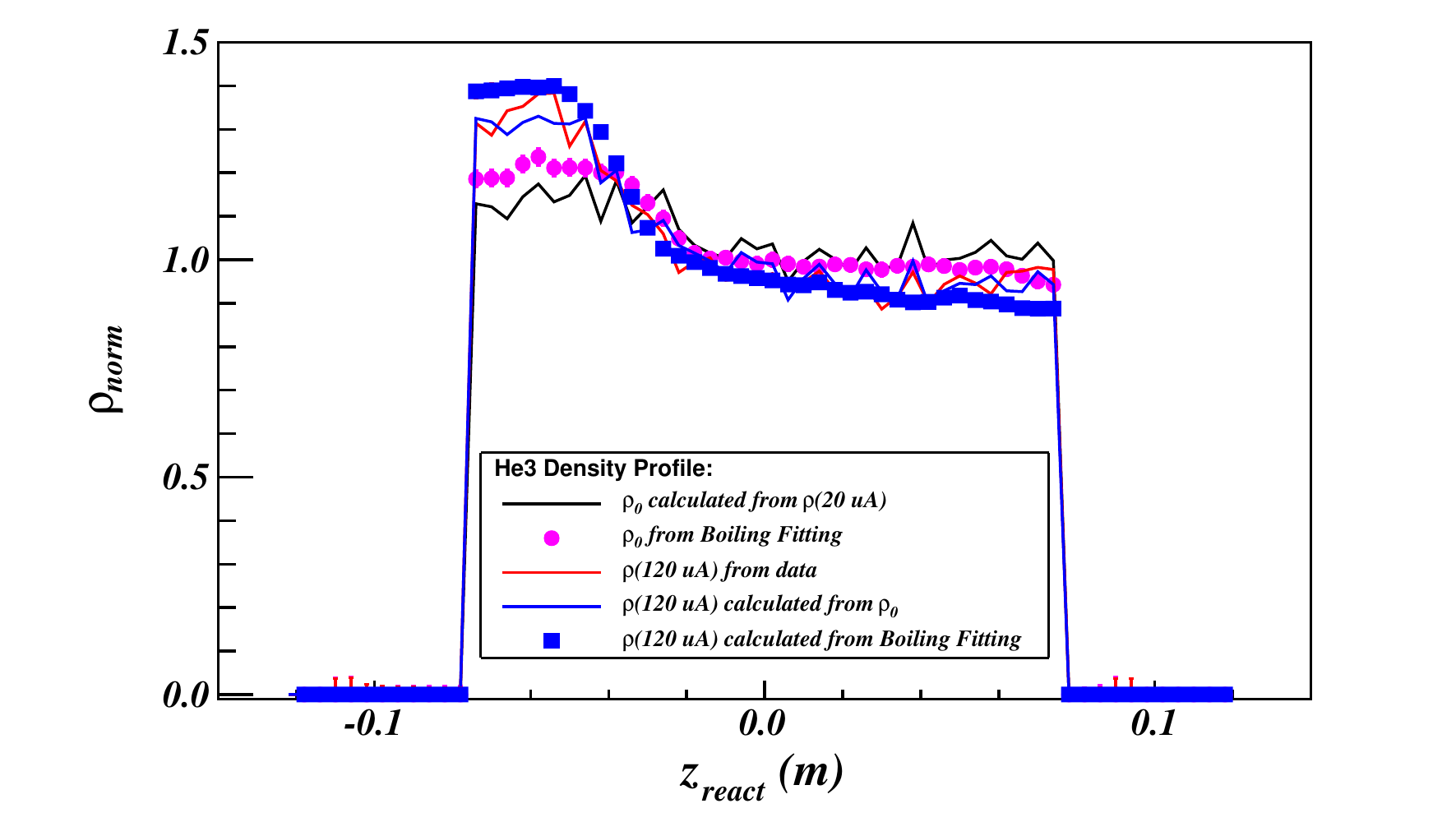} 
    }
\\
    \subfloat[$\mathrm{^{4}He}$ at I=60 uA]{
      \includegraphics[type=pdf,ext=.pdf,read=.pdf,width=0.45\textwidth]{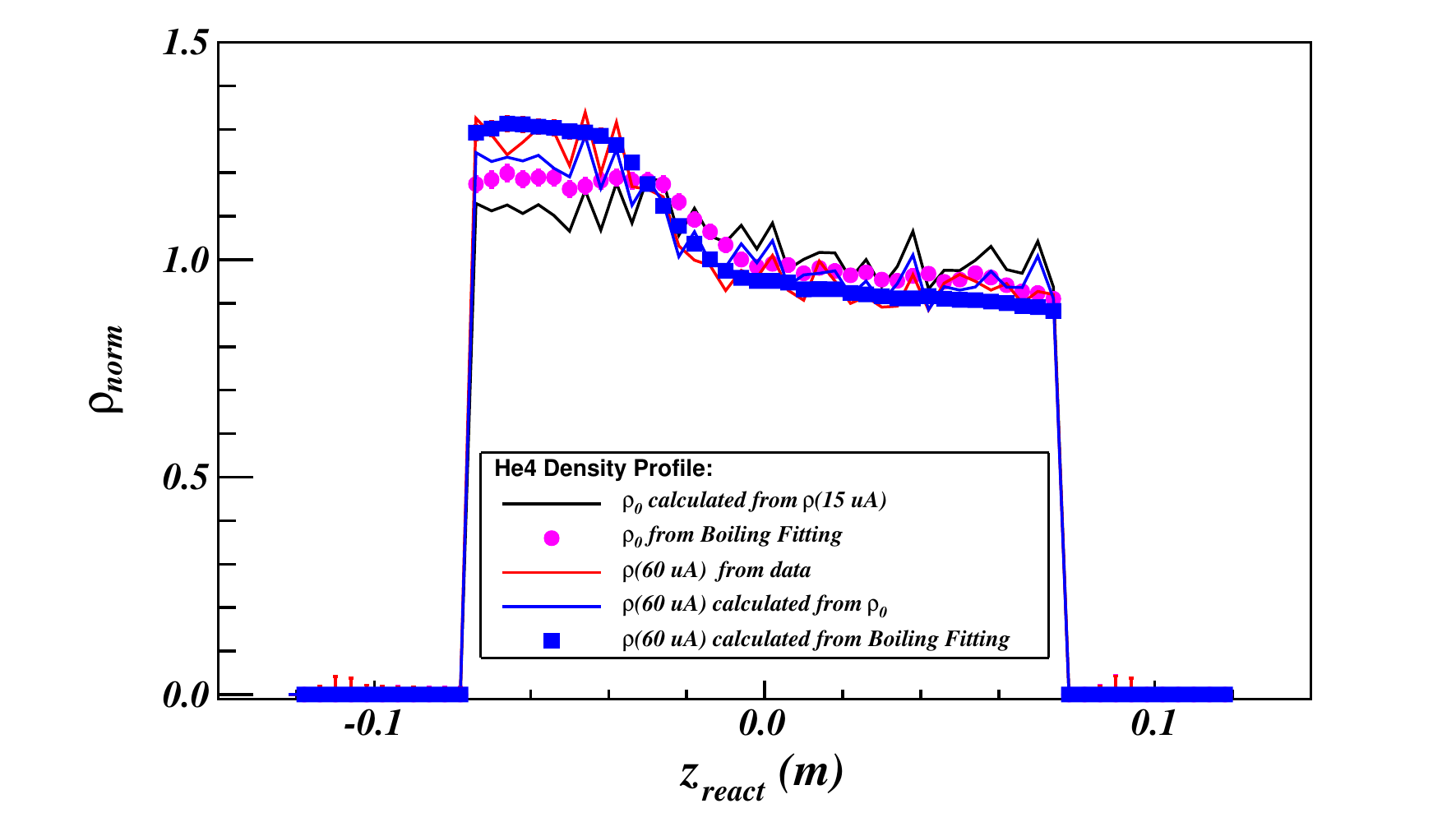} 
    }
    \hfill
    \subfloat[$\mathrm{^{4}He}$ at I=95 uA]{
      \includegraphics[type=pdf,ext=.pdf,read=.pdf,width=0.45\textwidth]{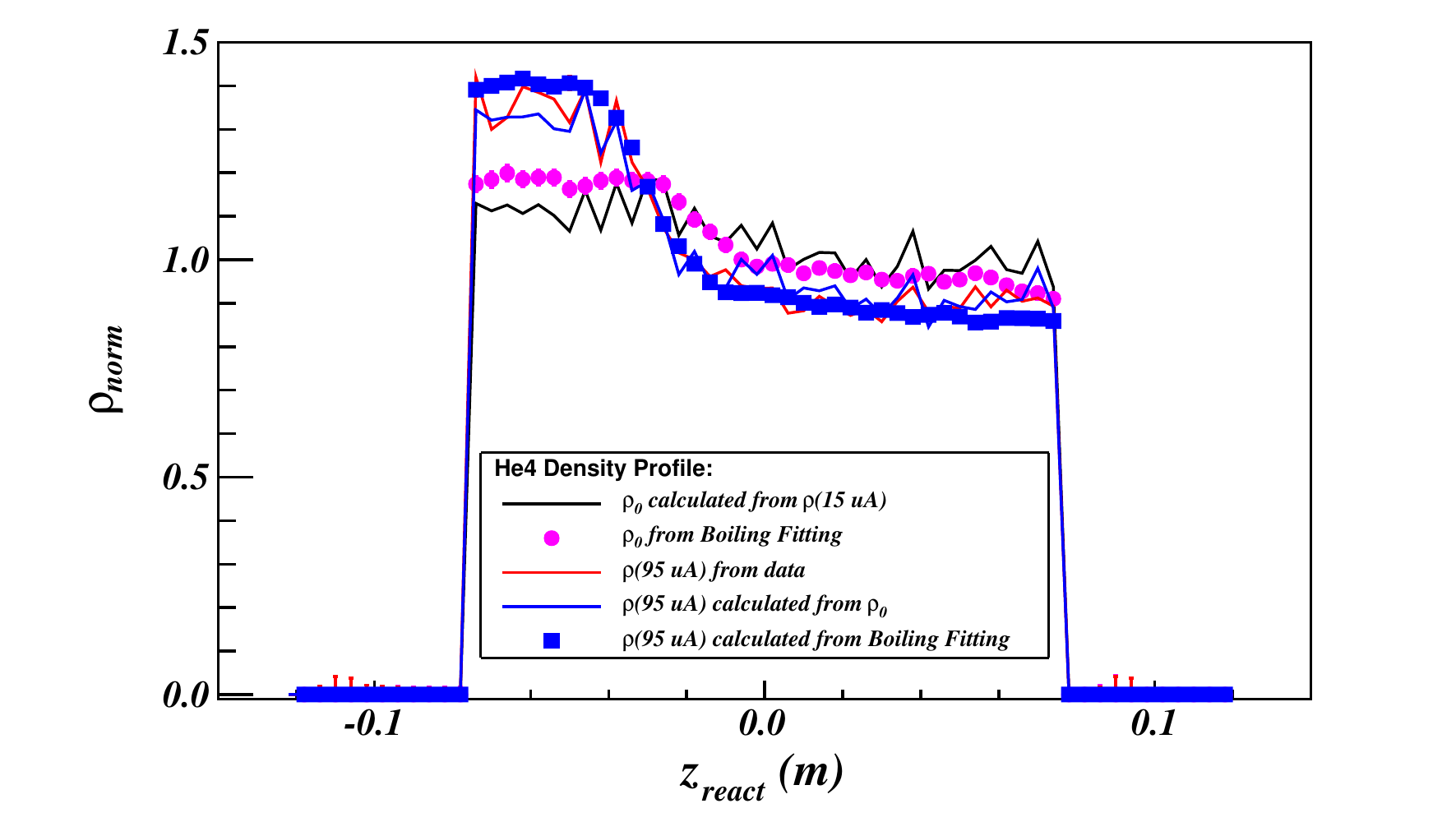} 
    }
    \caption[Cryo-targets relative density distribution]{\footnotesize{Cryo-targets relative density distribution extracted from data and corrected by the boiling factors. The density values were calculated in each $z_{react}$ bin and the peaks in these distribution were due to the statistical fluctuation. To remove the contribution of the endcaps, a cut was apply on the target length, $\mathrm{|z_{react}|\leq 7.5~cm}$}}
    \label{cryo_den_comp}
  \end{center}
\end{figure}

 One can use the density distributions at the minimum current (15 uA, 20 uA and 15 uA for $\mathrm{^{2}H}$, $\mathrm{^{3}He}$ and $\mathrm{^{4}He}$, respectively), and apply the boiling factors to calculate the density distribution at beam current equal to zero, $\rho(0)$. Then the density distribution at any current settings, $\rho_{Calc}(I)$ can be calculated with Eq.~\eqref{eq_rho_zbin}. To verify the boiling study results and the density distributions at different current settings, the distributions of $\rho_{Calc}(I)$ and $\rho(I)$ extracted in Fig.~\ref{cryo_den_fit} were compared, as shown in Fig.~\ref{cryo_den_comp}. Note that the contributions from the two endcaps were removed by applying the cut, $|z_{react}|\leq 7.5~cm$. The plots reveal that the results of boiling study successfully characterize the change of target density with different beam currents.

 Fig.~\ref{cryo_den_comp} also compares the distributions of the target density obtained from the boiling study and from the method discussed in this section. Ignoring the statistical fluctuations of the histograms, both methods give similar density profiles, and the small difference can be explained by the errors of the HRS acceptance simulation in SAMC and the cross section model in XEMC.
 
\section{Absolute Density}
 When the density distribution is uniform, the absolute density of a cryo-target can be calculated with the temperature and pressure readings from the target system with the fixed volume of the target cell. However, the calculation becomes impossible when the density is not uniform since the temperature and pressure fluctuate inside the target. While the relative density distribution has been extracted as discussed in previous section, one can obtain the absolute density distribution by calculating the density at the entrance of the target cell where the temperature and pressure were monitored.
 
  Whereas, the extracted relative density distribution at the entrance does not reflect the true density profile of the cryo-target due to the contamination of the aluminium endcaps during the boiling study, and an assumption has been made to assign the density value at the entrance to the value at $\mathrm{-10\leq z_{react} \leq -7.5~cm}$. The true value should not deviate too far away from this value since this location is very close to the entrance and the coolant flow should be able to maintain the same temperature as it at the entrance. The deviation can be corrected when comparing the experimental yield and the simulation yield, while the last one, yet, depends on the cross section model. To obtain the accurate density, one can utilize the 2N-SRC plateaus of cross section ratio of the carbon target to the cryo-targets~\cite{SLAC_Measurement_PRC.48.2451,PhysRevLett.96.082501,PhysRevLett.108.092502}, which have been well measured in previous experiments. Table~\ref{cryo_density_table} gives the densities of cryo-targets at the entrance and the yield-normalized density at $\mathrm{z_{react}=7.5~cm}$, where the values will be updated when they are further normalized by the 2N-SRC plateaus.
\begin{table}[htbp]
 \begin{tabular}{lcccccccc}
 \toprule
 Target:                        & $\mathrm{^{2}H}$  & $\mathrm{^{3}He}$-I  & $\mathrm{^{3}He}$-II & $\mathrm{^{4}He}$ \\
 \midrule
 $\rho_{entrance}~(g/cm^{3})$:   & 0.1676   & 0.0213      &  0.0296     & 0.0324 \\
 $\rho_{z_{react}=-7.5~cm}~(g/cm^{3})$:&  0.1906  & 0.0210 &  0.0292    & 0.0280 \\
 \bottomrule
 \end{tabular}
\caption[Cryo-targets densities]{\footnotesize{Cryo-targets densities, where two values of the $\mathrm{^{3}He}$ density refer to two different run periods. The values of $\rho_{entrance}$ are calculated from the temperature and pressure reading~\cite{target_report}. The values of $\rho_{z_{react}}=-7.5~cm$ are the values of $\rho_{entrance}$ normalized by the ratio of the experimental yield and the simulation yield and will be further corrected by comparing the 2N-SRC plateaus.}}
\label{cryo_density_table}
\end{table}

\section{Radiative Correction}
  The most essential parameter during the radiative correction is the radiation length of the target. For a uniform target, the radiation length is evaluated at the center of the target. For a non-uniform target, such an approximation has to be carefully examined. 

\begin{figure}[!ht]
 \begin{center}
  \includegraphics[angle=0,width=0.6\textwidth]{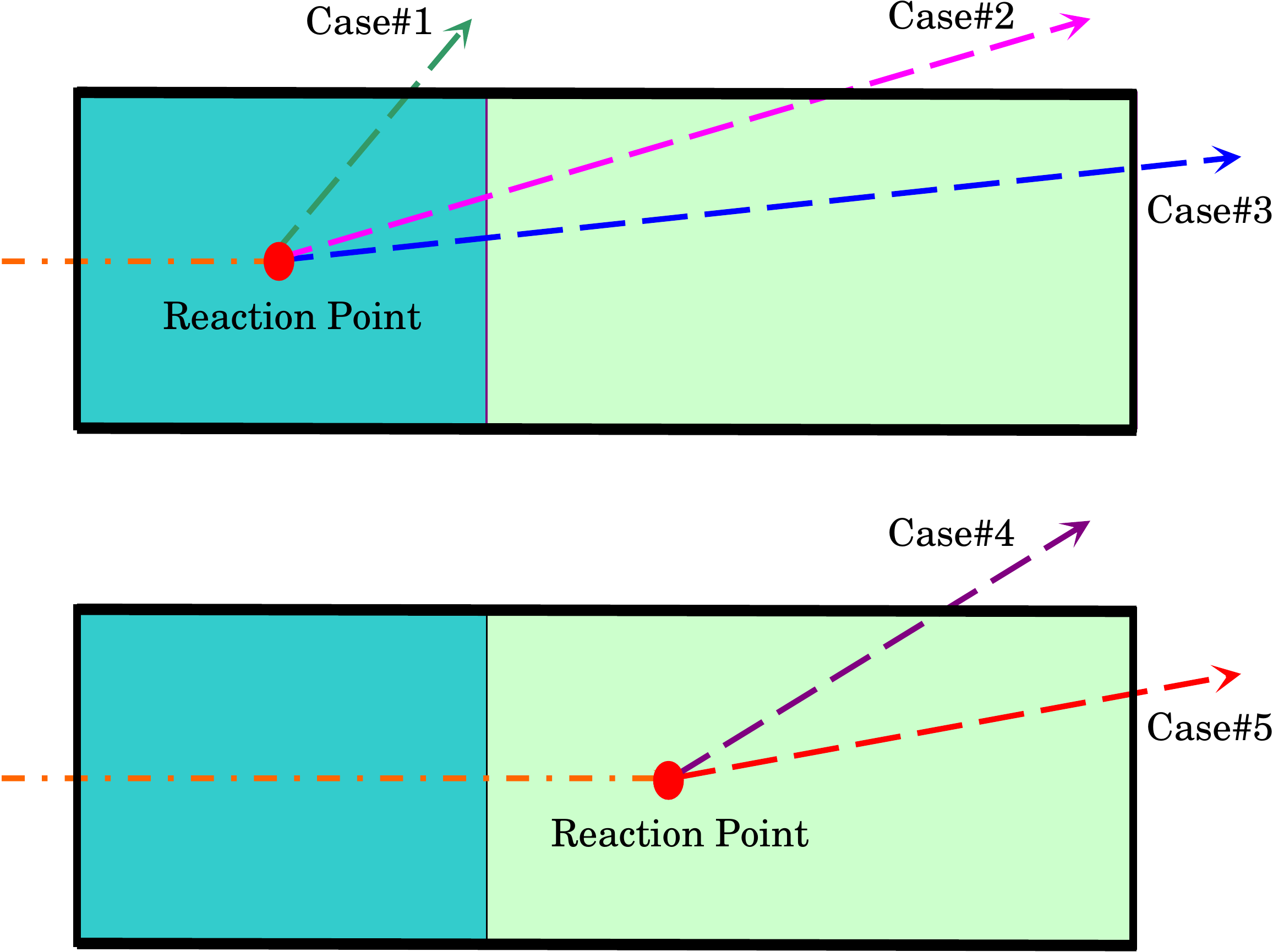}
  \caption[Different cases to calculate the radiation lengths]{\footnotesize{Different cases to calculate the radiation lengths, where the target has two parts with different density. The target cell's entrance, exit and wall (black lines ) also have different thickness. Depending on the reaction point in the forward scattering, there are five cases which give different radiation lengths.}}
  \label{radc_cases}
 \end{center}
\end{figure}
  In the radiative correction model in XEMC, the density distribution of a cryo-target is simplified as a step function, where the density is 30\% higher for -10~cm$\mathrm{\leq Z_{react}}\leq$-2~cm and is 20\% lower for the rest of the target. The value of radiation length in such a distribution depends on the reaction location and the scattering angle. Fig.~\ref{radc_cases} gives 5 different scattering paths which have been coded in the model. 
  
 \begin{figure}[!ht]
\begin{center}
  \includegraphics[angle=0,width=0.8\textwidth]{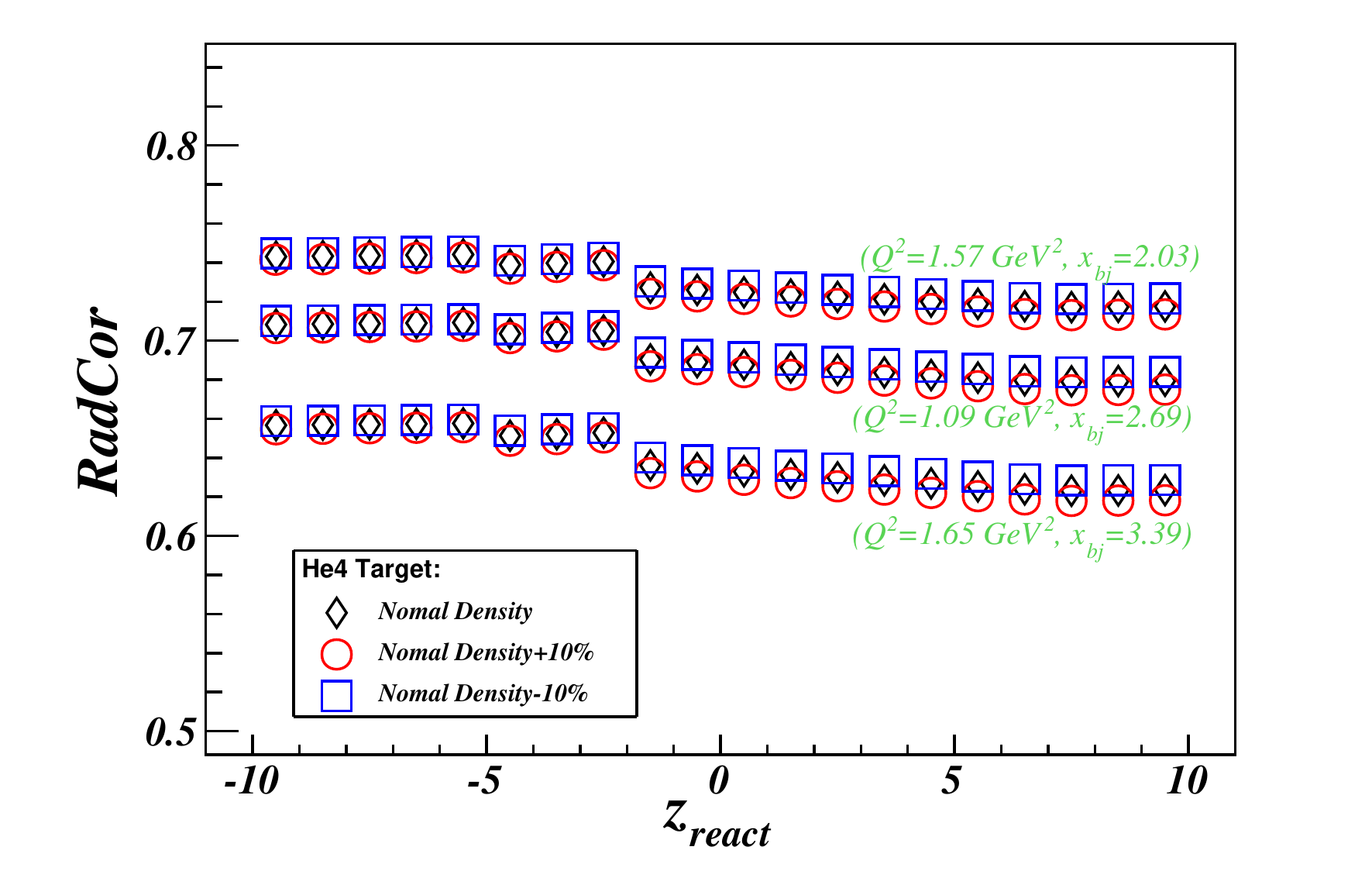}
  \caption[Position dependence of radiative effect on $\mathrm{^{4}He}$]{\footnotesize{Position dependence of radiative effect on $\mathrm{^{4}He}$. With a step function as the density distribution, $\mathrm{z_{react}}$ was divided into 20 bins where in each bin radiative correction factors were calculated at three different kinematic settings. To check the variation of the absolute density, the radiative effect was calculated in each $\mathrm{z_{react}}$ bin by changing the density by $\mathrm{\pm 10}$\%. The radiative effect clearly depends on the reaction location but has small dependence on the absolute density.}}
  \label{he4_rad_check}
 \end{center}
\end{figure}
  To study the position dependence of the radiative effect, the target was divided into 20 bins along the $\mathrm{z_{react}}$ distribution where the radiative correction factor (Eq.~\eqref{eq_radc_fact}) was calculated in each bin. Fig.~\ref{he4_rad_check} shows the distribution of the radiative correction factor as a function of $\mathrm{z_{react}}$. The downstream part of the target has stronger radiative effect since the incoming electron loses more energy while passing through the upstream part which has higher density. Overall, the variation of the radiative correction factors along the target length is less than 2\%. The distributions at different kinematic settings were also studied by changing the value of $E'$ by $\mathrm{\pm}$ 3\%. The results give the similar distribution which indicates that the $\mathrm{z_{react}}$-dependence of the radiative correction factors does not change much within one kinematics.
  
   The absolute target density varies with the beam current. Even though the beam current was set to a constant value for each target, it still had fluctuations depending on the stability of the beam. As discussed in the previous section, the absolute density for cryo-targets will be further normalized by the 2N-SRC ratio. In Fig.~\ref{he4_rad_check}, the radiative effect at different target densities was also examined and the results showed very small deviation when changing the density by  $\mathrm{\pm}$ 10\%.
   
    

%% file: append/append_xs.tex
\chapter{Cross Section Results}


\section{Cross Sectionss}
\begin{figure}[!ht]
  \begin{center}
    \subfloat[$\sigma^{^{2}H}_{born}$ .vs. $\nu$]{
      \includegraphics[type=pdf,ext=.pdf,read=.pdf,width=0.90\textwidth]{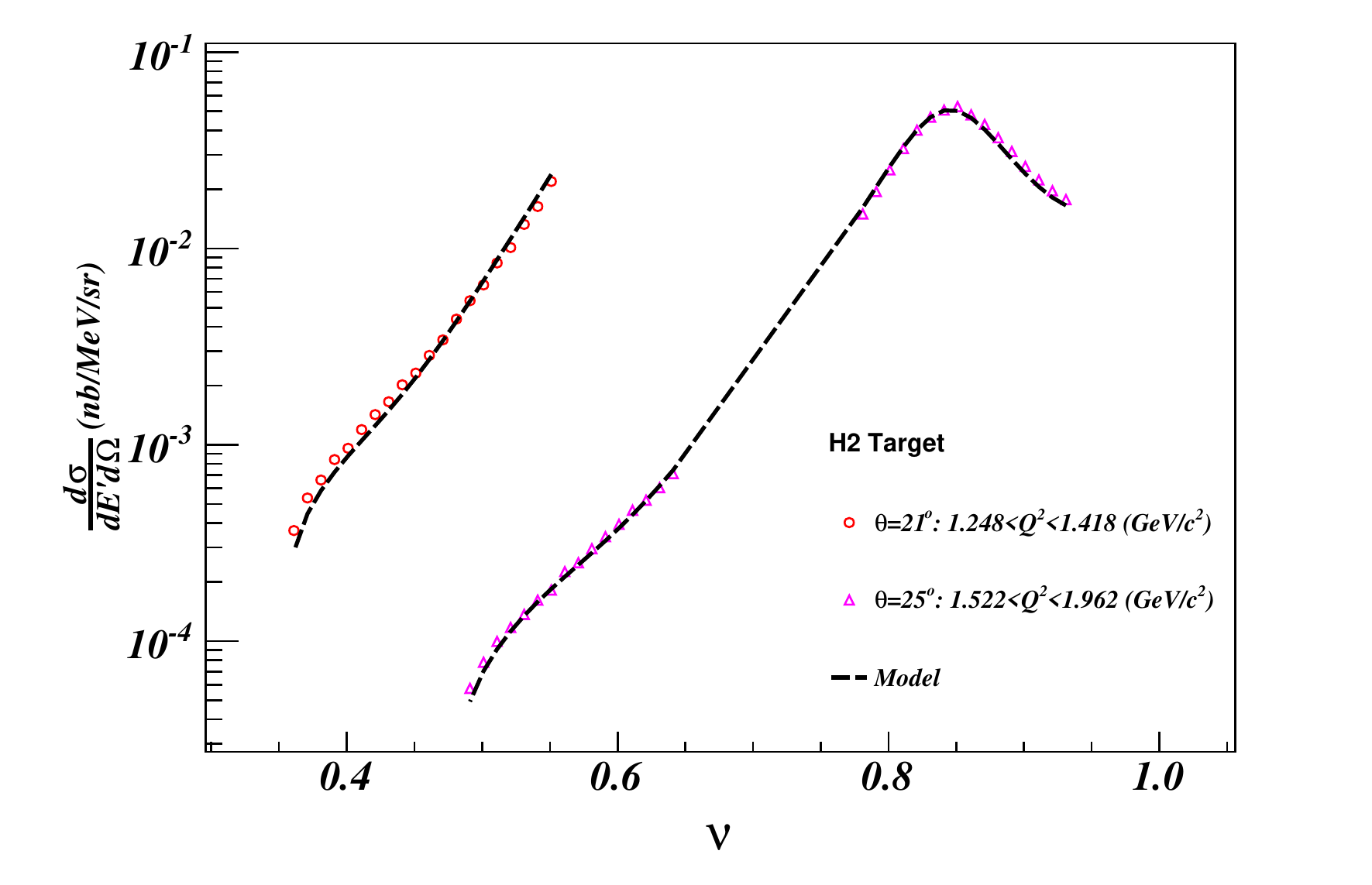}
    }
    \\
    \subfloat[$\sigma^{^{2}H}_{born}$ .vs. $x_{bj}$]{
      \includegraphics[type=pdf,ext=.pdf,read=.pdf,width=0.90\textwidth]{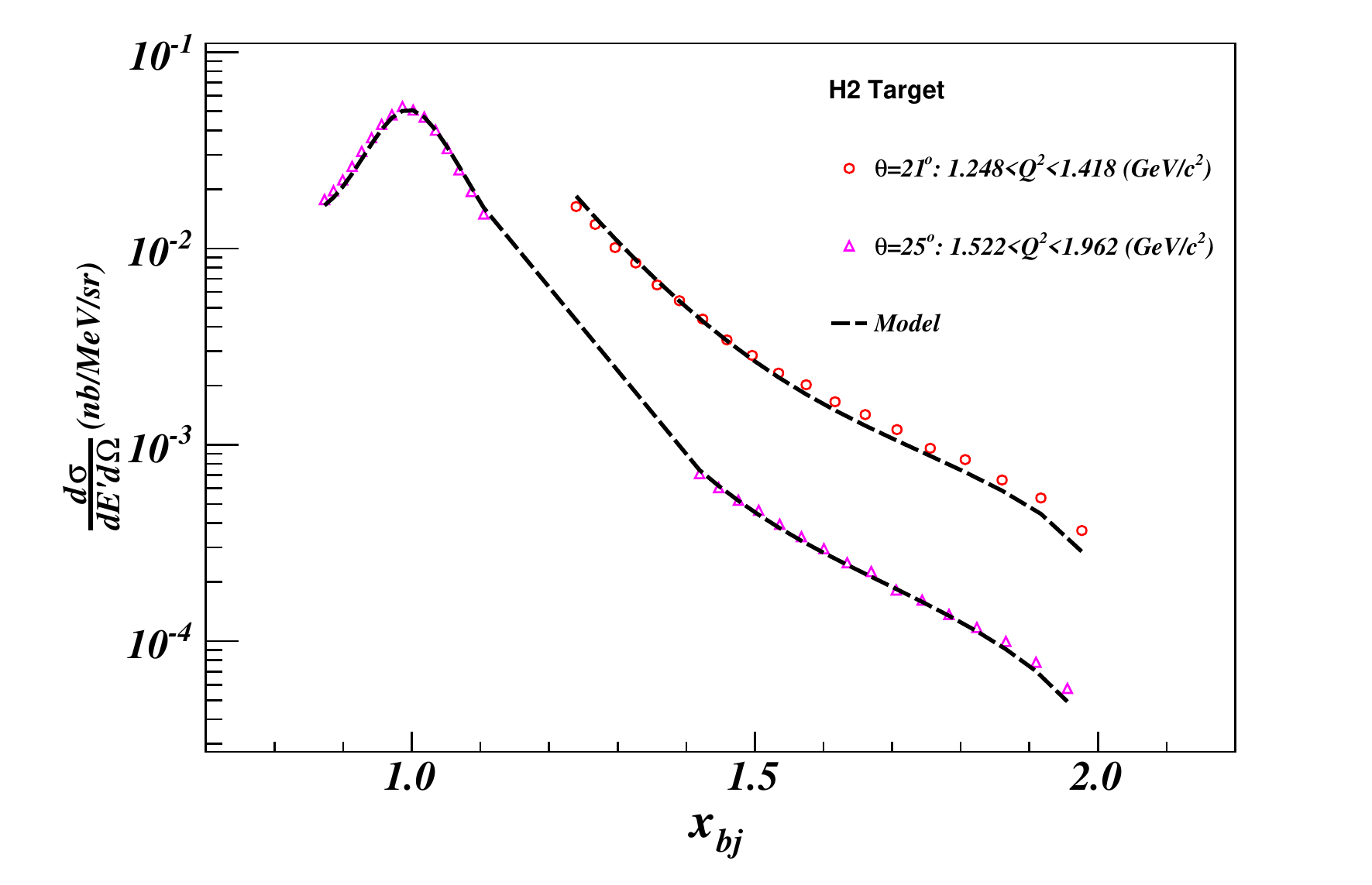}
    }
    \caption[$\mathrm{^{2}H}$ preliminary born cross sections]{\footnotesize{$\mathrm{^{2}H}$ preliminary born cross sections, where dots are from experimental results and lines are calculated from XEMC model}}
    \label{xs_born_h2}
  \end{center}
\end{figure}

\begin{figure}[!ht]
  \begin{center}
    \subfloat[$\sigma^{^{3}He}_{born}$ .vs. $\nu$]{
      \includegraphics[type=pdf,ext=.pdf,read=.pdf,width=0.90\textwidth]{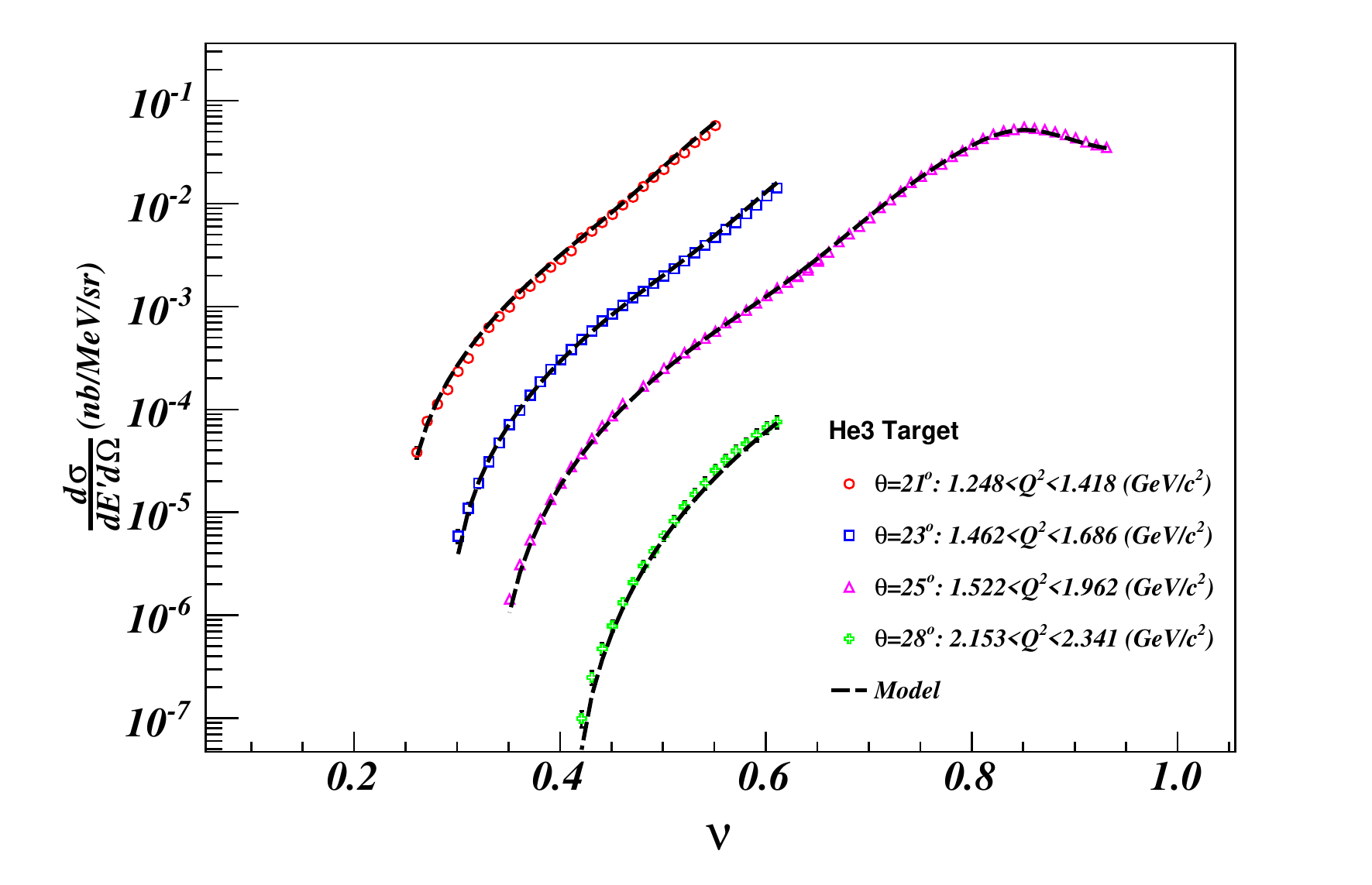}
    }
    \\
    \subfloat[$\sigma^{^{3}He}_{born}$ .vs. $x_{bj}$]{
      \includegraphics[type=pdf,ext=.pdf,read=.pdf,width=0.90\textwidth]{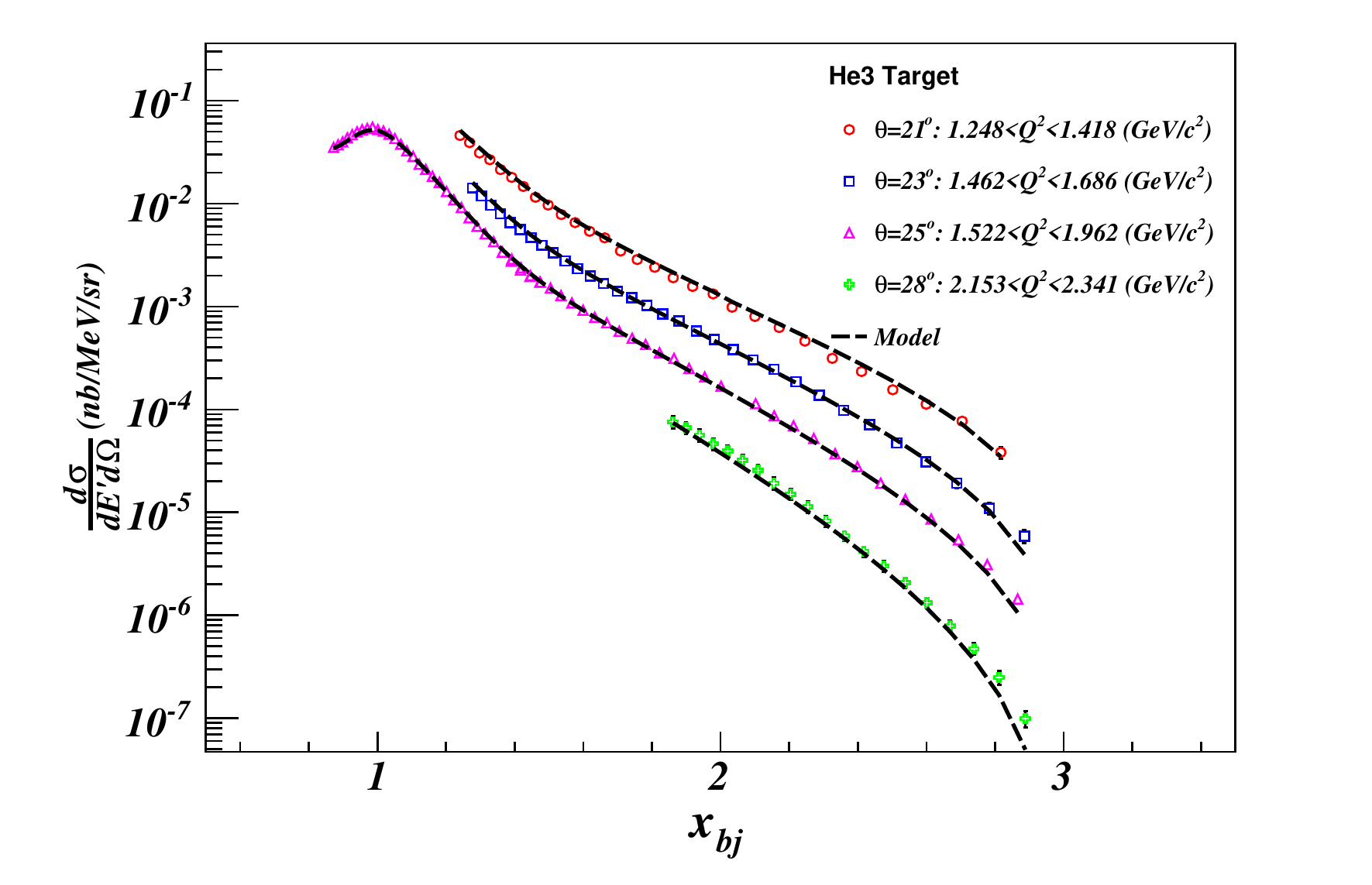}
    }
    \caption[$\mathrm{^{3}He}$ preliminary born cross sections]{\footnotesize{$\mathrm{^{3}He}$ preliminary born cross sections, where dots are from experimental results and lines are calculated from XEMC model}}
    \label{xs_born_he3}
  \end{center}
\end{figure}

\begin{figure}[!ht]
  \begin{center}
    \subfloat[$\sigma^{^{4}He}_{born}$ .vs. $\nu$]{
      \includegraphics[type=pdf,ext=.pdf,read=.pdf,width=0.90\textwidth]{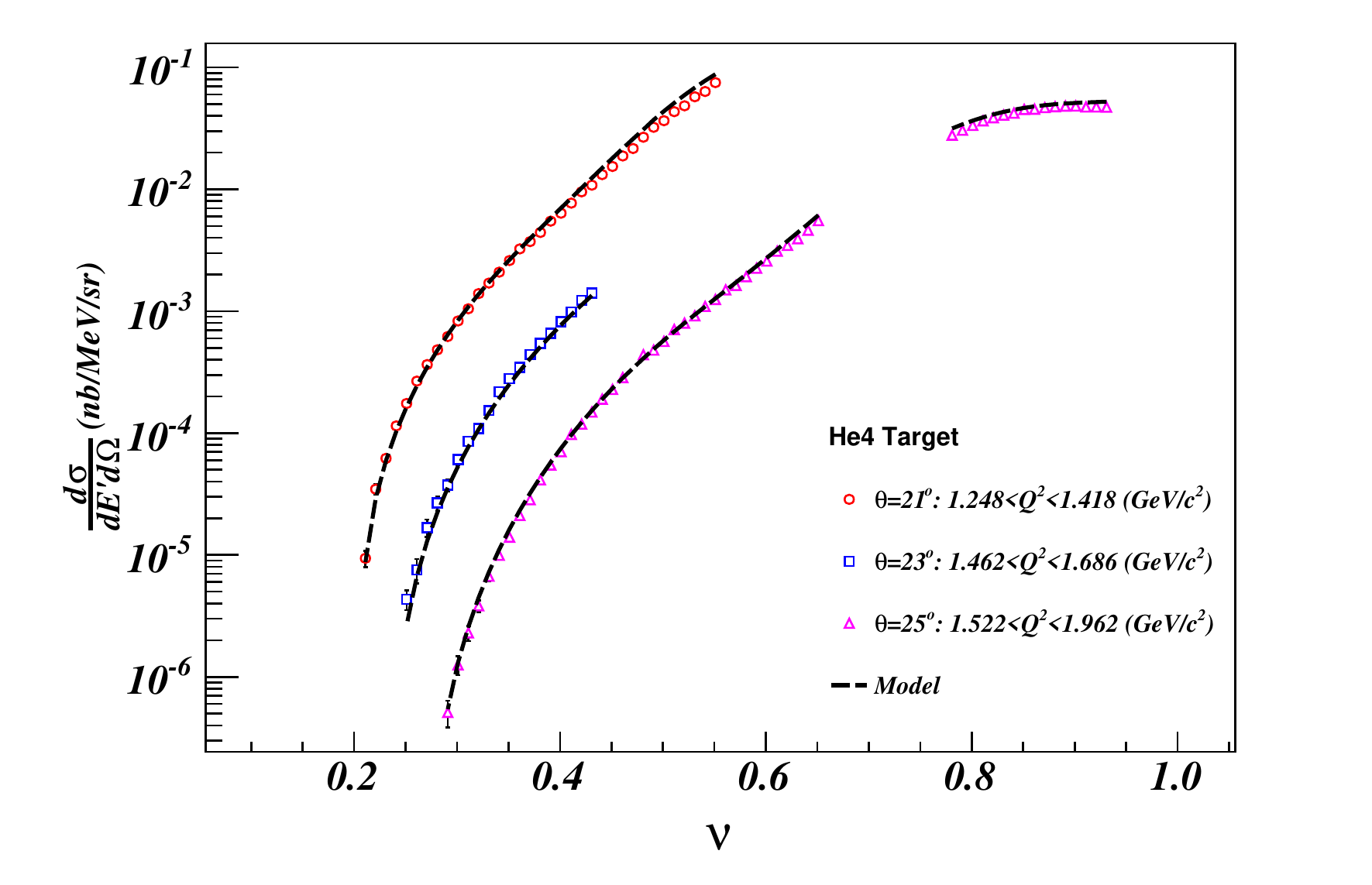}
    }
    \\
    \subfloat[$\sigma^{^{4}He}_{born}$ .vs. $x_{bj}$]{
      \includegraphics[type=pdf,ext=.pdf,read=.pdf,width=0.90\textwidth]{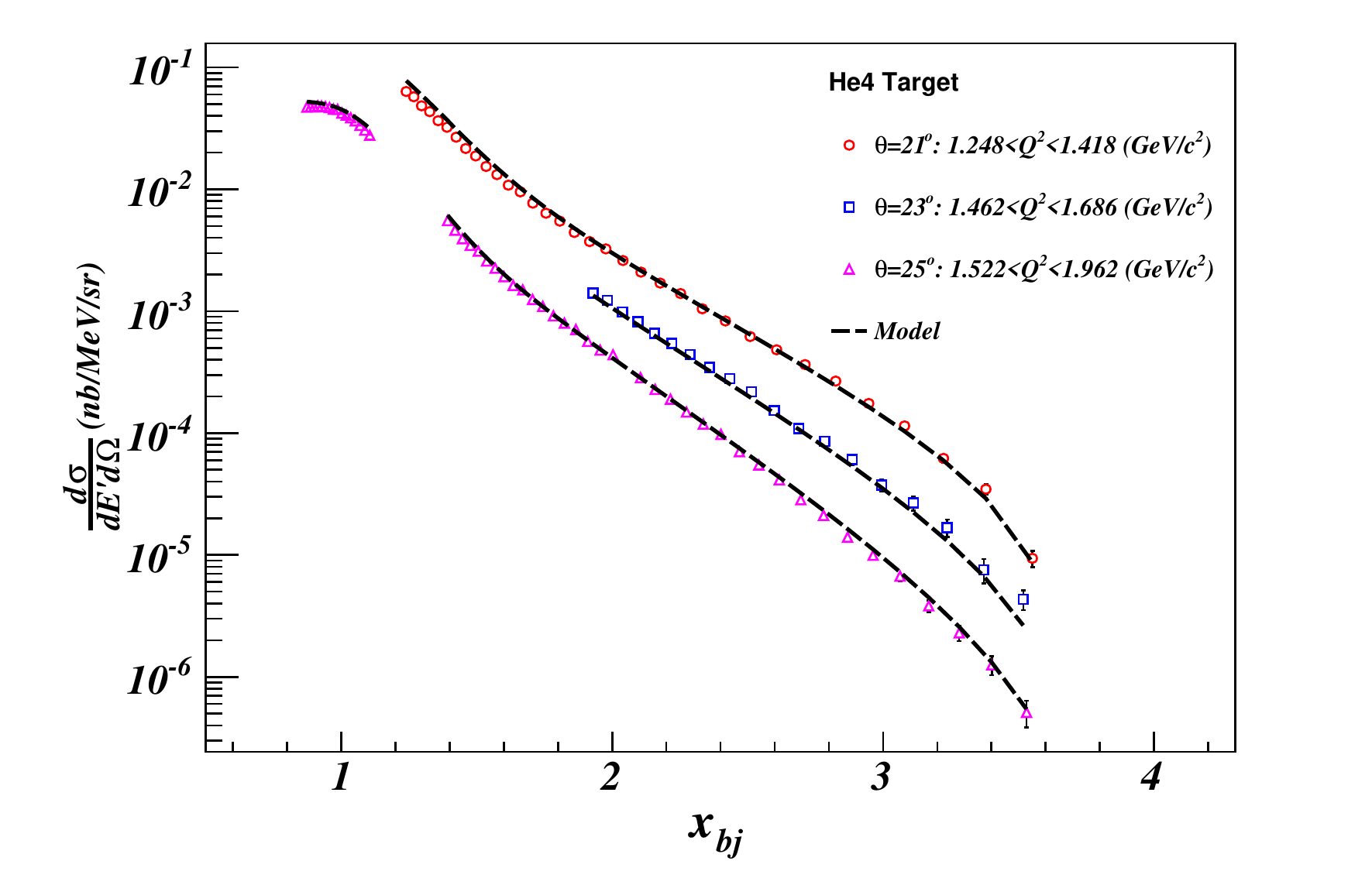}
    }
    \caption[$\mathrm{^{4}He}$ preliminary born cross sections]{\footnotesize{$\mathrm{^{4}He}$ preliminary born cross sections, where dots are from experimental results and lines are calculated from XEMC model}}
    \label{xs_born_he4}
  \end{center}
\end{figure}

\begin{figure}[!ht]
  \begin{center}
    \subfloat[$\sigma^{^{12}C}_{born}$ .vs. $\nu$]{
      \includegraphics[type=pdf,ext=.pdf,read=.pdf,width=0.90\textwidth]{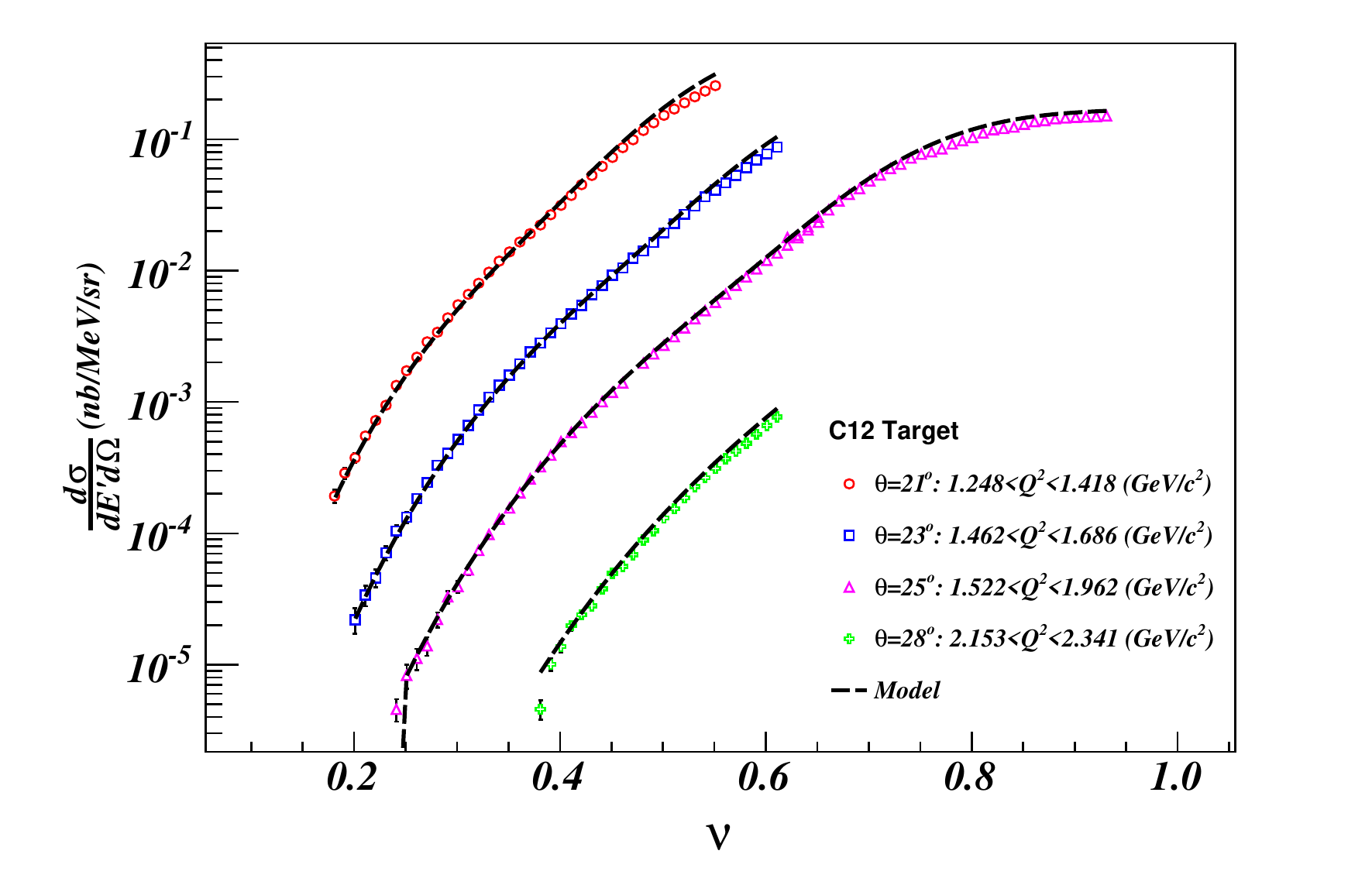}
    }
    \\
    \subfloat[$\sigma^{^{12}C}_{born}$ .vs. $x_{bj}$]{
      \includegraphics[type=pdf,ext=.pdf,read=.pdf,width=0.90\textwidth]{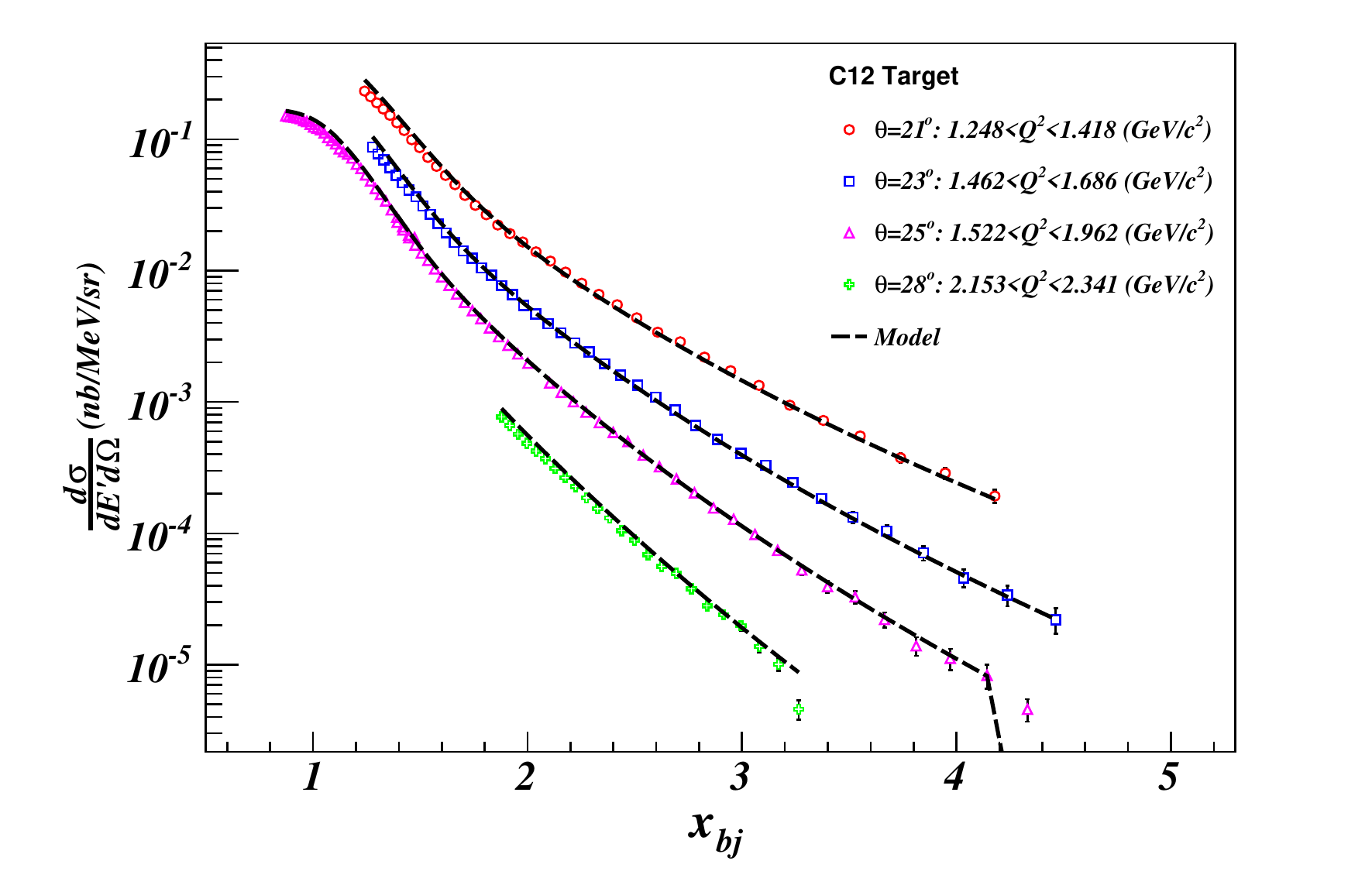}
    }
    \caption[$\mathrm{^{12}C}$ preliminary born cross sections]{\footnotesize{$\mathrm{^{12}C}$ preliminary born cross sections, where dots are from experimental results and lines are calculated from XEMC model}}
    \label{xs_born_c12}
  \end{center}
\end{figure}

\begin{figure}[!ht]
  \begin{center}
    \subfloat[$\sigma^{^{40}Ca}_{born}$ .vs. $\nu$]{
      \includegraphics[type=pdf,ext=.pdf,read=.pdf,width=0.90\textwidth]{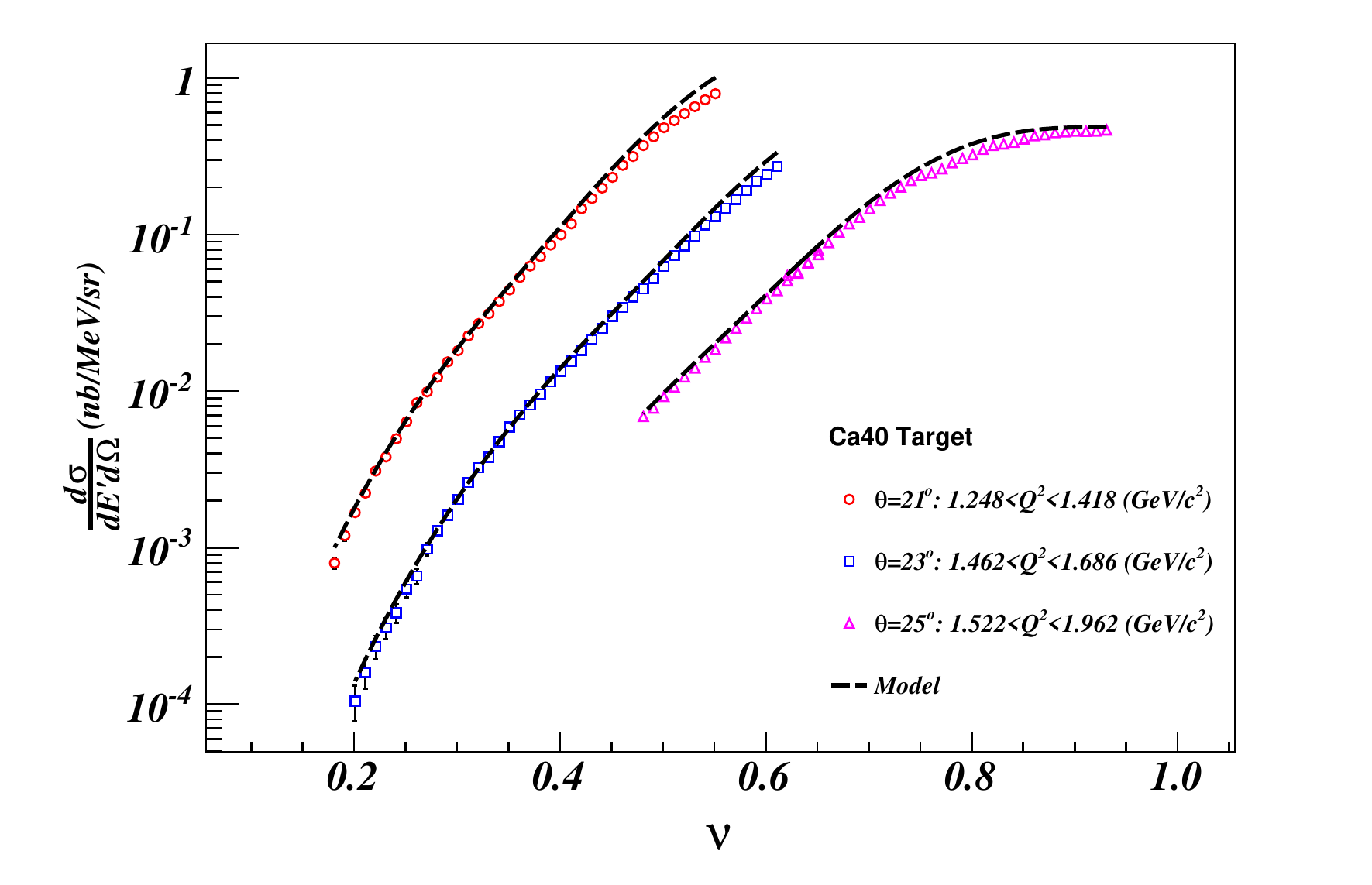}
    }
    \\
    \subfloat[$\sigma^{^{40}Ca}_{born}$ .vs. $x_{bj}$]{
      \includegraphics[type=pdf,ext=.pdf,read=.pdf,width=0.90\textwidth]{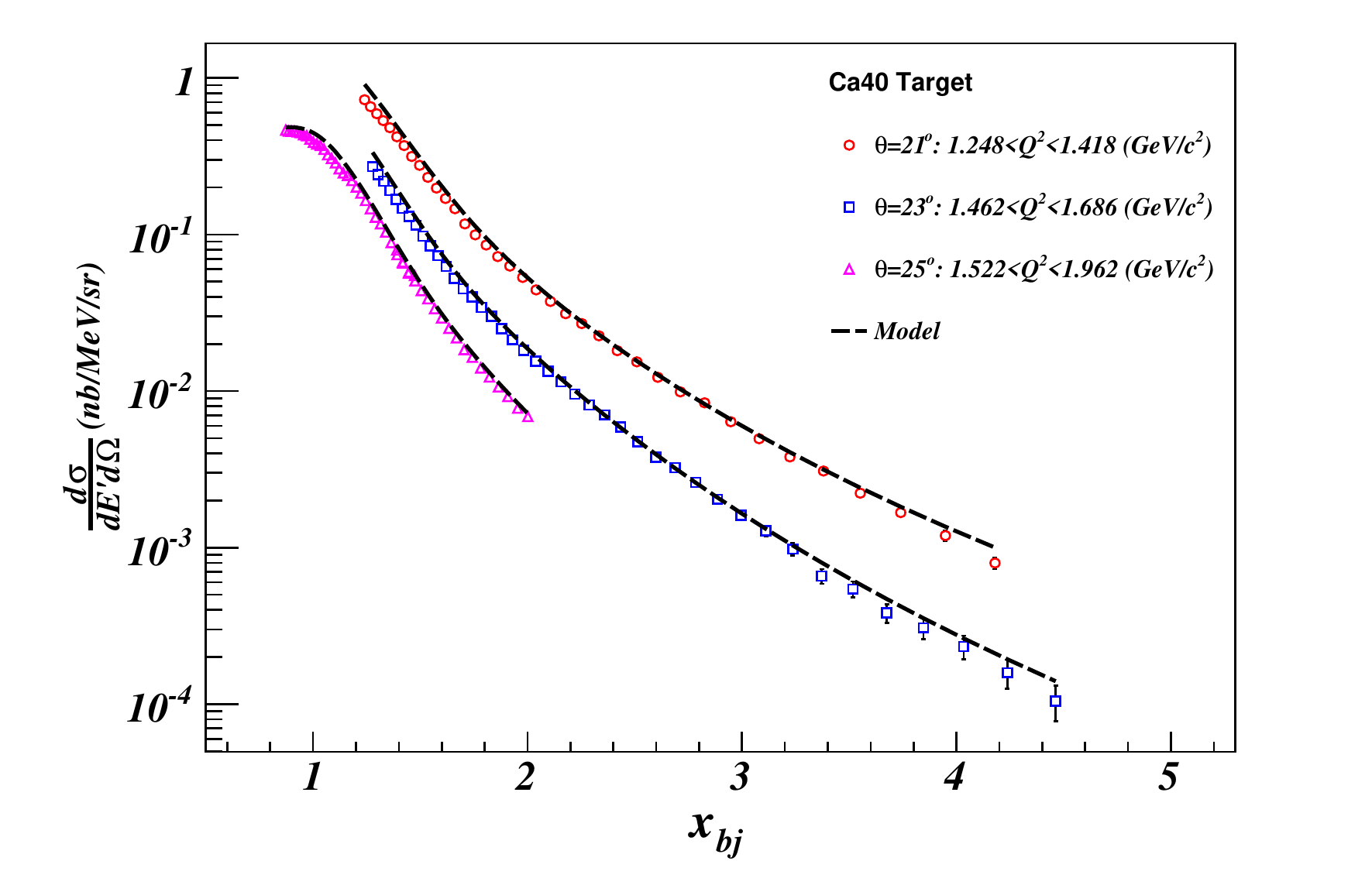}
    }
    \caption[$\mathrm{^{40}Ca}$ preliminary born cross sections]{\footnotesize{$\mathrm{^{40}Ca}$ preliminary born cross sections, where dots are from experimental results and lines are calculated from XEMC model}}
    \label{xs_born_ca40}
  \end{center}
  \end{figure}

\begin{figure}[!ht]
  \begin{center}
    \subfloat[$\sigma^{^{48}Ca}_{born}$ .vs. $\nu$]{
      \includegraphics[type=pdf,ext=.pdf,read=.pdf,width=0.90\textwidth]{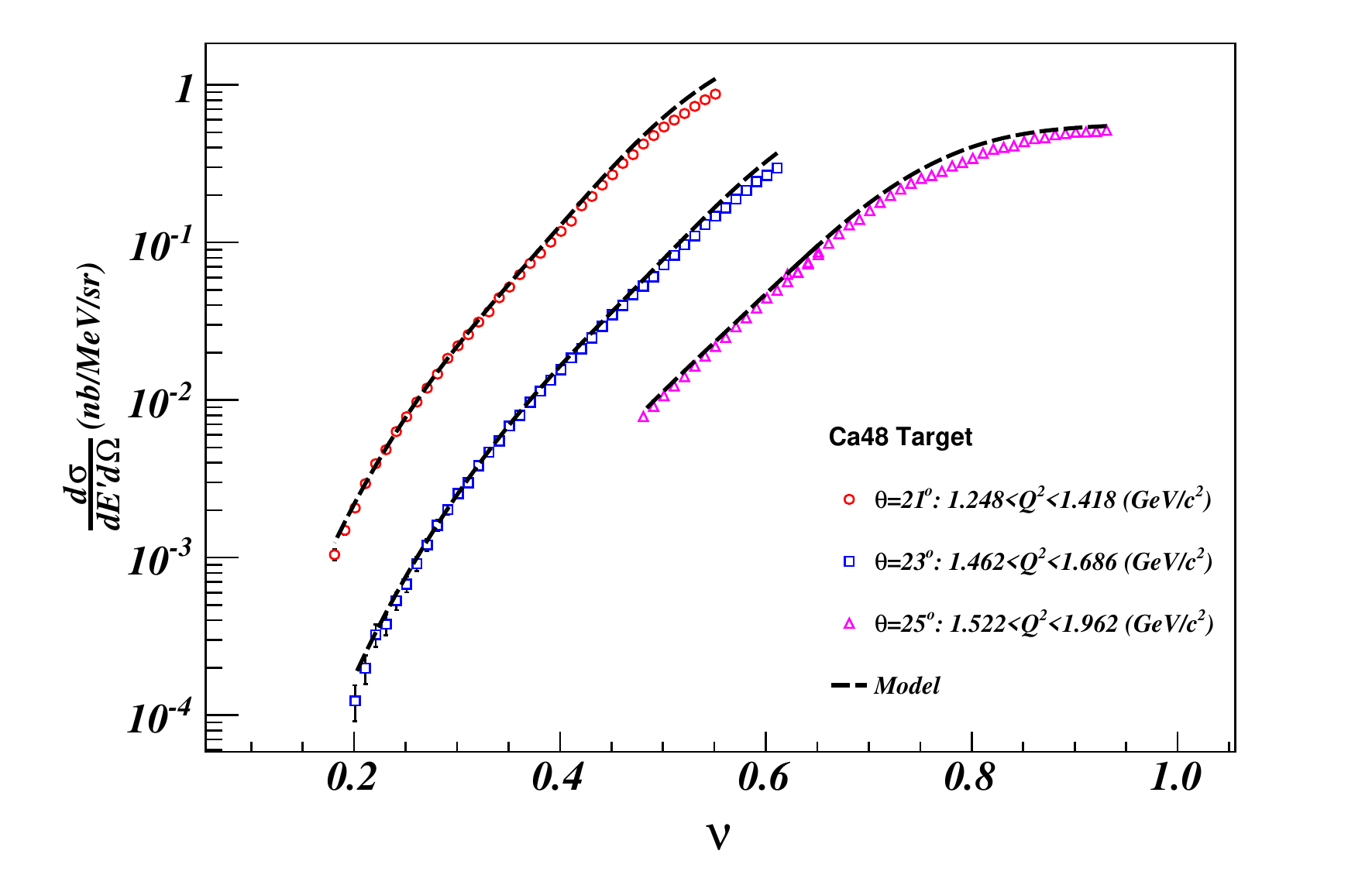}
    }
    \\
    \subfloat[$\sigma^{^{48}Ca}_{born}$ .vs. $x_{bj}$]{
      \includegraphics[type=pdf,ext=.pdf,read=.pdf,width=0.90\textwidth]{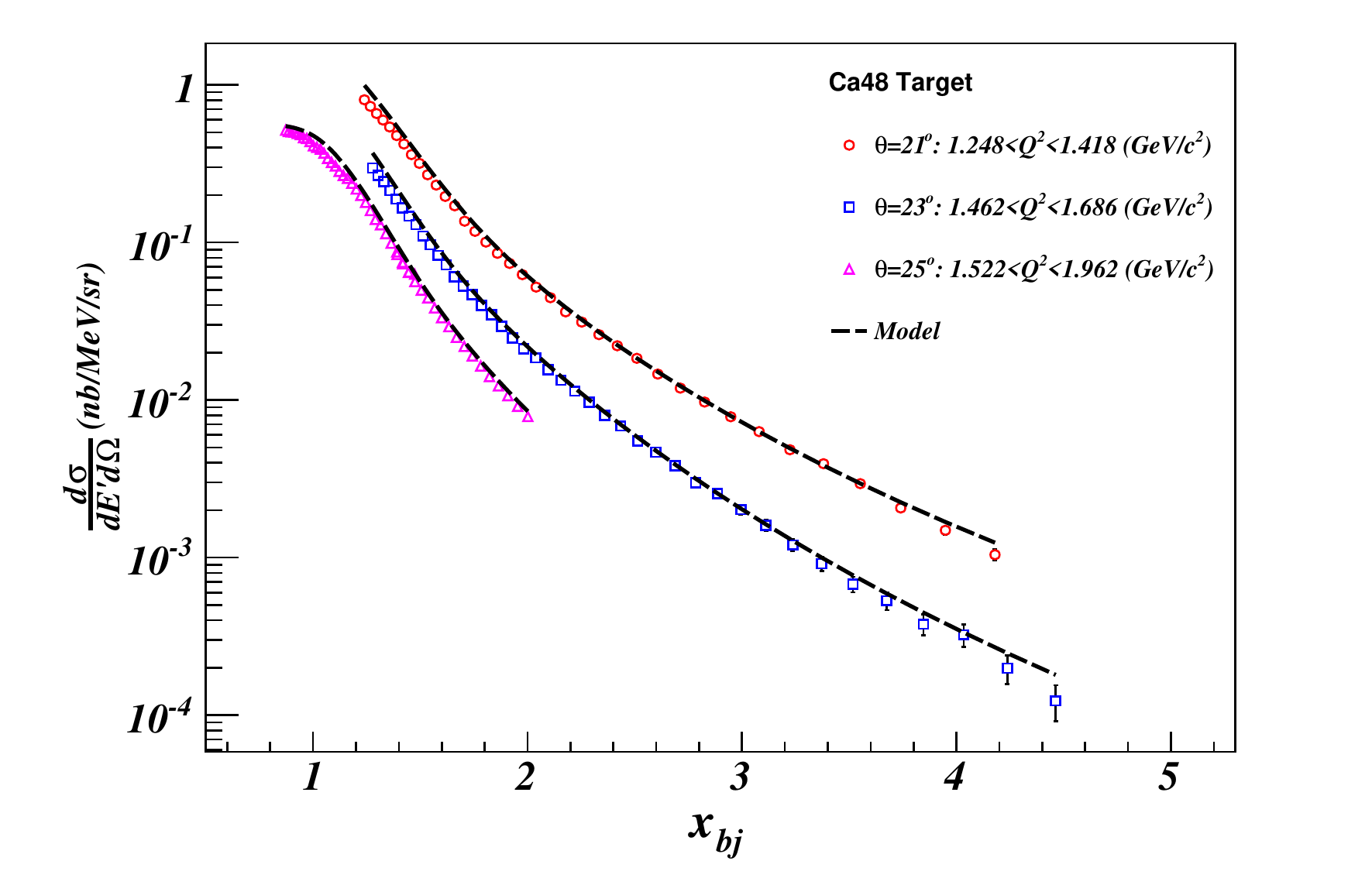}
    }
    \caption[$\mathrm{^{48}Ca}$ preliminary born cross sections]{\footnotesize{$\mathrm{^{48}Ca}$ preliminary born cross sections, where dots are from experimental results and lines are calculated from XEMC model}}
    \label{xs_born_ca48}
  \end{center}
\end{figure}

The measured born cross sections for $\mathrm{^{2}H}$, $\mathrm{^{3}He}$,$\mathrm{^{4}He}$,$\mathrm{^{12}C}$,$\mathrm{^{40}Ca}$,and $\mathrm{^{48}Ca}$ are shown in Fig.~\ref{xs_born_h2} through Fig.~\ref{xs_born_ca48}, respectively. Note that only the systematic errors from the detectors and the statistical errors are included. There are several other sources of systematic errors needed to be evaluated, e.g. the cryogenic target densities, acceptance correction, bin-centering correction and radiative corrections. etc..

 The detailed kinematic settings of two HRSs and list of targets measured are given in Table~\ref{kine_table_left} and Table~\ref{kine_table_right}. For each setting, if data is available from both arms, the cross section values are given as the average of individual cross sections extracted from these arms. 
\begin{table}[!ht]
  \centering
  \begin{tabular}{|c|c|c|c|c|}
    \hline
    Name      &  $\theta_{0} (^{o})$    & $P_{0}~(GeV/c)$ & $Q^{2}~(GeV^{2})$ &  Target \\
    \hline 
     Kin3.1   &21     & 2.905           & 1.295        & $^{2}H$,$^{3}He$, $^{4}He$,$^{12}C$,$^{40}Ca$,$^{48}Ca$ \\
    \hline                      
     Kin3.2   &21     & 3.055           & 1.362        &         $^{3}He$, $^{4}He$,$^{12}C$,                    \\
     \hline 
     Kin4.1   &23     & 2.855           & 1.523        &         $^{3}He$,          $^{12}C$,$^{40}Ca$,$^{48}Ca$ \\
      \hline 
     Kin4.2   &23     & 3.035           & 1.619        &         $^{3}He$, $^{4}He$,$^{12}C$,$^{40}Ca$,$^{48}Ca$ \\
     \hline 
     Kin5.0   &25     & 2.505           & 1.575        &         $^{3}He$, $^{4}He$,$^{12}C$,$^{40}Ca$,$^{48}Ca$ \\
     \hline 
     Kin5.05  &25     & 2.650           & 1.667        &         $^{3}He$,         $^{12}C$,$^{40}Ca$,$^{48}Ca$ \\
     \hline    
     Kin5.1   &25     & 2.795           & 1.758        &$^{2}H$, $^{3}He$, $^{4}He$,$^{12}C$,$^{40}Ca$,$^{48}Ca$ \\
     \hline    
     Kin5.2   &25     & 2.995           & 1.883        &$^{2}H$, $^{3}He$, $^{4}He$,$^{12}C$                     \\
    \hline 
     Kin6.5   &28     & 2.845           & 2.235        &          $^{3}He$,        $^{12}C$,                    \\
    \hline 
   \end{tabular}
  \caption[List of kinematic settings and target measured on HRS-L]{List of kinematic settings and targets measured on HRS-L}
  \label{kine_table_left}	
\end{table}
\begin{table}[!ht]
  \centering
  \begin{tabular}{|c|c|c|c|c|}
    \hline
    Name      &  $\theta_{0} (^{o})$    & $P_{0}~(GeV/c)$ & $Q^{2}~(GeV^{2})$ &  Target \\
    \hline 
     Kin3.1   &21     & 2.905           & 1.295   & $^{2}H$,          $^{4}He$,$^{12}C$,$^{40}Ca$,$^{48}Ca$ \\
    \hline                      
     Kin3.2   &21     & 3.055           & 1.362   &          $^{3}He$,$^{4}He$,$^{12}C$,$^{40}Ca$,$^{48}Ca$ \\
     \hline 
     Kin4.1   &23     & 2.855           & 1.523   &          $^{3}He$,         $^{12}C$,$^{40}Ca$,$^{48}Ca$ \\
      \hline 
     Kin4.2   &23     & 3.035           & 1.619   &          $^{3}He$,$^{4}He$,$^{12}C$,$^{40}Ca$,$^{48}Ca$ \\
     \hline 
     Kin5.0   &25     & 2.505           & 1.575   &          $^{3}He$,$^{4}He$,$^{12}C$,$^{40}Ca$,$^{48}Ca$ \\
     \hline 
     Kin5.05  &25     & 2.650           & 1.667   &                                                         \\
     \hline    
     Kin5.1   &25     & 2.795           & 1.758   & $^{2}H$, $^{3}He$,$^{4}He$,$^{12}C$,$^{40}Ca$,$^{48}Ca$ \\
     \hline    
     Kin5.2   &25     & 2.995           & 1.883   & $^{2}H$, $^{3}He$,$^{4}He$,$^{12}C$                     \\
    \hline 
     Kin6.5   &28                       & 2.845           & 2.235   &          $^{3}He$,        $^{12}C$,                    \\    
    \hline 
   \end{tabular}
  \caption[List of kinematic settings and target measured on HRS-R]{List of kinematic settings and target measured on HRS-R}
  \label{kine_table_right}	
\end{table}
  
 Fig.~\ref{xs_born_h2} shows the cross section of $\mathrm{^{2}H}$, where the Quasielastic (QE) peak can be clearly identified at $x_{bj}=1$ due to the relatively small Fermi motion of nucleons in the target. The results show good agreement with the calculation from XEMC model. $\mathrm{^{4}He}$ and $\mathrm{^{12}C}$ agree nicely with the model prediction (Fig.~\ref{xs_born_he4} and Fig.~\ref{xs_born_c12}). For $\mathrm{^{3}He}$ target, additional work is required on the XEMC model to correct the rapidly decreasing of cross section values when $x_{bj}\rightarrow 3$. Cross sections of $\mathrm{^{40}Ca}$ and $\mathrm{^{48}Ca}$ at high $\mathrm{Q^{2}}$ ($\mathrm{>1~GeV^{2}}$) are only available from this experiment and more iterations of the cross section models are necessary until the model and the data have solid agreement.
 

%% file: xgt2.bbl
\begin{thebibliography}{100}

\bibitem{PhysRevLett.87.172502}
P.~Navr\'atil {\em et~al.},
\newblock Phys. Rev. Lett. {\bf 87}, 172502 (2001).

\bibitem{PhysRevC.51.38}
R.~B. Wiringa, V.~G.~J. Stoks, and R.~Schiavilla,
\newblock Phys. Rev. C {\bf 51}, 38 (1995).

\bibitem{donal_prvt}
{D. Day},
\newblock {private communication}.

\bibitem{VanDerSteenhoven1988547}
G.~V.~D. Steenhoven {\em et~al.},
\newblock Nuclear Physics A {\bf 480}, 547  (1988).

\bibitem{review_mft}
D.~Lacroix,
\newblock Review of mean-field theory.

\bibitem{DeForest1983}
J.~D. Forest,
\newblock Nucl. Phys. {\bf A392}, 232 (1983).

\bibitem{hartree_fock_book}
N.~S. Szabo and A.~Ostlund,
\newblock Modern quantum chemistry.

\bibitem{Lapikas1993297}
L.~Lapikás,
\newblock Nuclear Physics A {\bf 553}, 297  (1993).

\bibitem{Kelly:1996hd}
J.~Kelly,
\newblock Adv. Nucl. Phys. {\bf 23}, 75 (1996).

\bibitem{qe_donal}
O.~Benhar, D.~Day, and I.~Sick,
\newblock Rev. Mod. Phys. {\bf 80}, 189 (2008).

\bibitem{john_thesis}
J.~Arrington,
\newblock {Ph.D Thesis, CalTech}, 1998.

\bibitem{West1975263}
G.~B. West,
\newblock Physics Reports {\bf 18}, 263  (1975).

\bibitem{day_arns}
D.~B. Day, J.~S. McCarthy, T.~W. Donnelly, and I.~Sick,
\newblock Annual Review of Nuclear and Particle Science {\bf 40}, 357 (1990).

\bibitem{PhysRevC.41.R2474}
C.~Ciofi~degli Atti, S.~Liuti, and S.~Simula,
\newblock Phys. Rev. C {\bf 41}, R2474 (1990).

\bibitem{Boffi19931}
S.~Boffi, C.~Giusti, and F.~Pacati,
\newblock Physics Reports {\bf 226}, 1  (1993).

\bibitem{PhysRevB.36.1208}
D.~Toma\ifmmode~\acute{n}\else \'{n}\fi{}ek and M.~A. Schluter,
\newblock Phys. Rev. B {\bf 36}, 1208 (1987).

\bibitem{PhysRevLett.108.092502}
N.~Fomin {\em et~al.},
\newblock Phys. Rev. Lett. {\bf 108}, 092502 (2012).

\bibitem{nadia_thesis}
N.~Fomin,
\newblock {Ph.D Thesis, University of Virginia}, 2008.

\bibitem{PhysRevC.49.2950}
V.~G.~J. Stoks, R.~A.~M. Klomp, C.~P.~F. Terheggen, and J.~J. de~Swart,
\newblock Phys. Rev. C {\bf 49}, 2950 (1994).

\bibitem{Lacombe1981139}
M.~Lacombe {\em et~al.},
\newblock Physics Letters B {\bf 101}, 139  (1981).

\bibitem{PhysRevC.46.1045}
C.~Ciofi~degli Atti, D.~B. Day, and S.~Liuti,
\newblock Phys. Rev. C {\bf 46}, 1045 (1992).

\bibitem{PhysRevC.87.024606}
O.~Benhar,
\newblock Phys. Rev. C {\bf 87}, 024606 (2013).

\bibitem{SLAC_Measurement_PRC.48.2451}
L.~L. Frankfurt, M.~I. Strikman, D.~B. Day, and M.~Sargsyan,
\newblock Phys. Rev. C {\bf 48}, 2451 (1993).

\bibitem{PhysRevLett.96.082501}
CLAS Collaboration, K.~S. Egiyan {\em et~al.},
\newblock Phys. Rev. Lett. {\bf 96}, 082501 (2006).

\bibitem{PhysRevLett.99.072501}
Jefferson Lab Hall A Collaboration, R.~Shneor {\em et~al.},
\newblock Phys. Rev. Lett. {\bf 99}, 072501 (2007).

\bibitem{Subedi:2008zz}
R.~Subedi {\em et~al.},
\newblock Science {\bf 320}, 1476 (2008), 0908.1514.

\bibitem{PhysRevC.53.1689}
C.~Ciofi~degli Atti and S.~Simula,
\newblock Phys. Rev. C {\bf 53}, 1689 (1996).

\bibitem{src_john}
J.~Arrington, D.~Higinbotham, G.~Rosner, and M.~Sargsian,
\newblock Progress in Particle and Nuclear Physics {\bf 67}, 898  (2012).

\bibitem{PhysRevLett.90.042301}
A.~Tang {\em et~al.},
\newblock Phys. Rev. Lett. {\bf 90}, 042301 (2003).

\bibitem{E07006_pr}
S.~Gilad, D.~Higinbotham, V.~Sulkosky, E.~Piasetzky, and J.~Watson,
\newblock {Studying Short-Range Correlations in Nuclei at the Repulsive Core
  Limit via the Triple Coincidence (e,e'pN) Reaction},
\newblock \url{http://hallaweb.jlab.org/experiment/E07-006/}, 2011.

\bibitem{PhysRevLett.94.082305}
Jefferson Lab Hall A Collaboration, F.~Benmokhtar {\em et~al.},
\newblock Phys. Rev. Lett. {\bf 94}, 082305 (2005).

\bibitem{Laget200549}
J.~Laget,
\newblock Physics Letters B {\bf 609}, 49  (2005).

\bibitem{Benhar:1994hw}
O.~Benhar, A.~Fabrocini, S.~Fantoni, and I.~Sick,
\newblock Nucl.Phys. {\bf A579}, 493 (1994).

\bibitem{Muther:1995bk}
H.~Muther, G.~Knehr, and A.~Polls,
\newblock Phys.Rev. {\bf C52}, 2955 (1995), nucl-th/9505038.

\bibitem{PhysRevLett.93.182501}
E97-006 Collaboration, D.~Rohe {\em et~al.},
\newblock Phys. Rev. Lett. {\bf 93}, 182501 (2004).

\bibitem{vanLeeuwe1998593}
J.~van Leeuwe {\em et~al.},
\newblock Nuclear Physics A {\bf 631}, 593  (1998),
\newblock Few-Body Problems in Physics.

\bibitem{Leeuwe20016}
J.~van Leeuwe {\em et~al.},
\newblock Physics Letters B {\bf 523}, 6  (2001).

\bibitem{Frankfurt1981215}
L.~Frankfurt and M.~Strikman,
\newblock Physics Reports {\bf 76}, 215  (1981).

\bibitem{Frankfurt_misak}
L.~Frankfurt, M.~Sargsian, and M.~Strikman,
\newblock International Journal of Modern Physics A {\bf 23}, 2991 (2008).

\bibitem{M_Sargsian_JPG_29_2003}
M.~Sargsian {\em et~al.},
\newblock Journal of Physics G: Nuclear and Particle Physics {\bf 29} (2003).

\bibitem{Frankfurt1988235}
L.~Frankfurt and M.~Strikman,
\newblock Physics Reports {\bf 160}, 235  (1988).

\bibitem{Arrington:2003tw}
J.~Arrington,
\newblock p. 567 (2003), nucl-ex/0306016.

\bibitem{Arrington:2006pn}
J.~Arrington,
\newblock (2006), nucl-ex/0602007.

\bibitem{e08014_pr}
J.~Arrington, D.~Day, D.~Higinbotham, and P.~Solvignon,
\newblock {Three-nucleon short range correlations studies in inclusive
  scattering for $0.8<Q^{2}<2.8 (GeV/c)^{2}$},
\newblock \url{http://hallaweb.jlab.org/experiment/E08-014/}, 2011.

\bibitem{PhysRevC.72.054310}
M.~Alvioli, C.~C.~d. Atti, and H.~Morita,
\newblock Phys. Rev. C {\bf 72}, 054310 (2005).

\bibitem{Pieper_Wiringa}
S.~C. Pieper and R.~B. Wiringa,
\newblock Annual Review of Nuclear and Particle Science {\bf 51}, 53 (2001).

\bibitem{PhysRevLett.98.132501}
R.~Schiavilla, R.~B. Wiringa, S.~C. Pieper, and J.~Carlson,
\newblock Phys. Rev. Lett. {\bf 98}, 132501 (2007).

\bibitem{PhysRevC.84.031302}
M.~Vanhalst, W.~Cosyn, and J.~Ryckebusch,
\newblock Phys. Rev. C {\bf 84}, 031302 (2011).

\bibitem{PhysRevC.86.044619}
M.~Vanhalst, J.~Ryckebusch, and W.~Cosyn,
\newblock Phys. Rev. C {\bf 86}, 044619 (2012).

\bibitem{E12_11_112_pr}
J.~Arrington, D.~Day, D.~Higinbotham, and P.~Solvignon,
\newblock {Precision measurement of the isospin dependence in the 2N and 3N
  short range correlation region},
\newblock \url{http://www.jlab.org/exp_prog/proposals/11/PR12-11-112.pdf}.

\bibitem{xiaohui}
X.~Zhan {\em et~al.},
\newblock Physics Letters B {\bf 705}, 59  (2011).

\bibitem{john_src_emc}
J.~Arrington {\em et~al.},
\newblock Phys. Rev. C {\bf 86}, 065204 (2012).

\bibitem{PhysRevLett.106.052301}
L.~B. Weinstein {\em et~al.},
\newblock Phys. Rev. Lett. {\bf 106}, 052301 (2011).

\bibitem{PhysRevLett.103.202301}
J.~Seely {\em et~al.},
\newblock Phys. Rev. Lett. {\bf 103}, 202301 (2009).

\bibitem{EMC_Review_1995}
D.~Geesaman, K.~Saito, and A.~Thomas,
\newblock Annual Review of Nuclear and Particle Science {\bf 45}, 337 (1995).

\bibitem{EMC_Review_2003}
P.~R. Norton,
\newblock Reports on Progress in Physics {\bf 66}, 1253 (2003).

\bibitem{EMC_first}
European Muon Collaboration, J.~Aubert {\em et~al.},
\newblock Phys.Lett. {\bf B123}, 275 (1983).

\bibitem{E12_10_103_pr}
J.~Annand, J.~Gomez, R.~Holt, G.~Petratos, and R,
\newblock {Measurement of the F2n/F2p, d/u Ratios and A=3 EMC Effect in Deep
  Inelastic Scattering off the Tritium and Helium Mirror Nuclei},
\newblock \url{http://www.jlab.org/exp_prog/proposals/10/PR12-10-103.pdf}.

\bibitem{E12_06_105_pr}
J.~Arrington, D.~Day, N.~Fomin, and P.~Solvignon,
\newblock {Inclusive Scattering from Nuclei at $x>1$ in the quasielastic and
  deeply inelastic regimes},
\newblock \url{http://www.jlab.org/exp_prog/proposals/06/PR12-06-105.pdf}.

\bibitem{E12_10_008_pr}
J.~Arrington, A.~Daniel, and D.~Gaskell,
\newblock {Detailed studies of the nuclear dependence of F2 in light nuclei},
\newblock \url{http://www.jlab.org/exp_prog/proposals/10/PR12-10-008.pdf}.

\bibitem{E12_11_107_pr}
O.~Hen, L.~Weinstein, and S.~Gilad,
\newblock {In Medium Nucleon Structure Functions, SRC, and the EMC effect},
\newblock \url{http://www.jlab.org/exp_prog/proposals/10/PR12-11-107.pdf}.

\bibitem{E12_10_003_pr}
W.~Boeglin and M.~Jones,
\newblock {Deuteron Electro-Disintegration at Very High Missing Momentum},
\newblock \url{http://www.jlab.org/exp_prog/proposals/10/PR12-10-003.pdf}.

\bibitem{Day:1987az}
D.~Day {\em et~al.},
\newblock Phys.Rev.Lett. {\bf 59}, 427 (1987).

\bibitem{halla_nim}
J.~Alcorn {\em et~al.},
\newblock Nucl. Instrum. Meth. {\bf A522}, 294 (2004).

\bibitem{halla_main}
Hall a main page,
\newblock \url{hallaweb.jlab.org/}.

\bibitem{bpm_cali}
B.~Reitz,
\newblock {Hall A BPM Calibration Procedure},
\newblock \url{http://hallaweb.jlab.org/podd/doc/bpm.html}.

\bibitem{bcm_patricia}
P.~Solvignon-Slifer,
\newblock {E08-014 BCM calibration report}, 2012.

\bibitem{beam_energy1}
P.~V. J.~Berthot,
\newblock Nucl. Phys. News {\bf 9}, 12 (1990).

\bibitem{beam_energy2}
O.~Ravel,
\newblock {Ph.D Thesis, University of Blaise Pascal}, 1997.

\bibitem{cryo_grp}
Jefferson lab cryogenic group,
\newblock \url{wwwwold.jlab.org/eng/cryo}.

\bibitem{target_report}
D.~Meekins,
\newblock E08-014 target report,
\newblock
  \url{http://hallaweb.jlab.org/experiment/E08-014/analysis/HallA_Target_Confi%
guration_Apr2011.pdf}, 2011.

\bibitem{R_Bock}
R.~Bock and A.Vasilescu,
\newblock {The Particle Detector Brief Book}.

\bibitem{pdg}
Particle Data Group, J.~Beringer {\em et~al.},
\newblock Phys. Rev. D {\bf 86}, 010001 (2012).

\bibitem{ts_tm}
E.~Jastrzembski,
\newblock Trigger supervisor transition module, 1997.

\bibitem{analyzer}
{ROOT/C++ Analyzer for Hall A}.

\bibitem{cern_root}
{ROOT - A Data Analysis Framework by CERN},
\newblock \url{http://root.cern.ch/drupal/}.

\bibitem{bpm_runs}
{E08-014 Harp scan results}, 2011.

\bibitem{shower_ak}
A.~Ketikyan {\em et~al.},
\newblock {About Shower Detector Software}, 1997.

\bibitem{shower_luhj}
H.~Lu {\em et~al.},
\newblock {Shower Calibration and Efficiency for E97-110},
\newblock \url{Shower Calibration and Efficiency for E97-110,
  http://hallaweb.jlab.org/experiment/E97-110/tech/shower\_calib.ps.gz"}, 2005.

\bibitem{nilanga_optics}
N.~Liyanage,
\newblock {Optics Calibration of the Hall A High Resolution Spectrometers using
  the new optimizer},
\newblock
  \url{http://hallaweb.jlab.org/publications/Technotes/files/2002/02-012.pdf},
  2002.

\bibitem{survey_dvcs_2010}
{Hall A survey report 1343},
\newblock \url{http://www.jlab.org/eng/survalign/documents/dthalla/A1343.pdf},
  2010.

\bibitem{survey_pvdis_2009}
{Hall A survey report A1249},
\newblock \url{http://www.jlab.org/eng/survalign/documents/dthalla/A1239.pdf},
  2009.

\bibitem{survey_a1n_2009}
{Hall A survey report A1239},
\newblock \url{http://www.jlab.org/eng/survalign/documents/dthalla/A1249.pdf},
  2009.

\bibitem{jinge_thesis}
G.~Jin,
\newblock {Ph.D Thesis, University of Virginia (2011)}, 2011.

\bibitem{jin_huang_optics}
J.~Huang,
\newblock {HRS Optics Optimizer}, 2009.

\bibitem{A_Duer}
A.~Duer,
\newblock {Single Arm Monte Carlo for Polarized 3He experiments in Hall A
  v.0.2},
\newblock
  \url{http://hallaweb.jlab.org/publications/Technotes/files/2001/01-004.pdf},
  2002.

\bibitem{hyao_thesis}
H.~Yao,
\newblock {Ph.D Thesis, Temple University }, 2011.

\bibitem{snack_lerose}
J.~LeRose,
\newblock {HRS transportation functions}.

\bibitem{vince_thesis}
V.~Sulkosky,
\newblock {Ph.D Thesis, College of William and Mary}, 2007.

\bibitem{halla_daq}
Hall a daq setup,
\newblock \url{hallaweb.jlab.org/equipment/daq/trigsetup_2003.html}.

\bibitem{Bosted:2012qc}
P.~Bosted and V.~Mamyan,
\newblock {Empirical Fit to electron-nucleus scattering}, 2012, arXive
  1203.2262.

\bibitem{aji_thesis}
A.~Daniel,
\newblock {Ph.D Thesis, Houston University}, 2007.

\bibitem{PhysRevLett.56.1452}
R.~D. McKeown,
\newblock Phys. Rev. Lett. {\bf 56}, 1452 (1986).

\bibitem{qfs_org}
J.~S. O'Connell {\em et~al.},
\newblock Phys. Rev. C {\bf 35}, 1063 (1987).

\bibitem{qfs_note}
{J. Bevelacqua},
\newblock FIZIKA {\bf 1}, 129 (1992).

\bibitem{qfs_org2}
J.~W. Lightbody and J.~S. O'Connell,
\newblock Comput. Phys. {\bf 2}, 57 (1988).

\bibitem{karl_thesis}
K.~Slifer,
\newblock {Ph.D Thesis, Temple University }, 2003.

\bibitem{whita}
W.~Armstrong,
\newblock {InSane - QFS model},
\newblock
  \url{http://quarks.temple.edu/~whit/code/InSANE++/html/da/dfc/group__xsectio%
ns.html}.

\bibitem{Bosted:2006}
P.~E. Bosted and M.~E. Christy,
\newblock Phys. Rev. C {\bf 77}, 065206 (2008).

\bibitem{mo_sai_rad}
L.~W. Mo and Y.~S. Tsai,
\newblock Rev. Mod. Phys. {\bf 41}, 205 (1969).

\bibitem{stein_radiation}
S.~Stein {\em et~al.},
\newblock Phys. Rev. D {\bf 12}, 1884 (1975).

\bibitem{silviu_target}
S.~Covrig,
\newblock {E08-014 Cryogenic Target System Simulation}, 2013.

\end{thebibliography}
